%% file: full_thesis.tex
\documentclass[a4paper,12pt]{uclphd}
\usepackage{setspace}
\usepackage{graphicx}
\usepackage{subfigure}
\usepackage{caption}
\usepackage{color}
\usepackage{amsmath}
\usepackage{amssymb}
\usepackage{caption}
\usepackage{bm}
\usepackage{float}
\usepackage{fancyhdr}
\DeclareGraphicsExtensions{.epsi,.eps,.ps}

\def\mathbi#1{\textbf{\em #1}}

\title{Entanglement and Quantum Information Transfer in Arrays of Interacting Quantum Systems}
\author{Martina Avellino}
\advisor{Professor Andrew J. Fisher}
\date{February 2009}

\begin{document}

% -------------------------------------------------------------
% 	TITLE PAGE
% -------------------------------------------------------------

\maketitle

\mainmatter
\pagestyle{plain}
I declare that the work presented in this thesis is my own. Where information has been obtained from other sources, I declare this has been clearly indicated in the thesis.

\newpage

% -------------------------------------------------------------
% 	ABSTRACT
% -------------------------------------------------------------
\pagestyle{plain}
\begin{abstract}
Quantum Information Technology (QIT) promises faster, more secure means of data manipulation by making use of the quantum properties of matter. In recent years, QIT has experienced a steady transition from the realm of the theoretical to that of the practical, and seemingly abstract properties, such as \emph{entanglement}, can now be observed and, to an extent, manipulated. However, current research is still investigating means to control these elusive quantum properties, which seldom survive the transition from the microscopic to the macroscopic. This thesis examines some of the more fundamental requirements of a successful quantum computation, namely the ability to transmit quantum information with maximum efficiency, and the creation of entanglement. I focus specifically on neutron entanglement, showing that the spins of two or more distinct neutrons can be measurably entangled by forward-scattering from an isotropic medium. This result is of great interest, as inter-neutron entanglement has never yet been observed. Furthermore, a mathematical treatment of neutron scattering puts forth intriguing questions regarding the interpretation of `time' in scattering experiments. Though the problem is by no means new, a unique resolution is outstanding. I present a simple treatment based on the Heisenberg $S$-matrix, from which it emerges that in certain situations the quantum-mechanical time parameter appearing in the effective time-evolution operator for the spin system has an intuitive physical interpretation. The final part of the thesis deals with quantum information transfer in arrays of permanently coupled dipolar systems. It is shown that spin chains with dipolar couplings offer high fidelity long-distance state transmission, but transfer times in unmodulated chains are unfeasibly long. Possible optimization methods are discussed, concluding with a review of recent achievements in this field. Finally, I give a brief summary of each chapter, and discuss future extensions of this work.

\end{abstract}
\newpage
\setcounter{page}{3}

% -------------------------------------------------------------
% 	COPYRIGHT
% -------------------------------------------------------------
%\makecopyright

% -------------------------------------------------------------
% 	ACKNOWLEDGEMENTS
% -------------------------------------------------------------
\begin{acknowledgements}
I wish to express my deepest, most heartfelt thanks to my supervisors Professor Andrew Fisher and Professor Sougato Bose. It has been my immense privilege to work with two such outstanding minds, who have shown me nothing but guidance, support and plain old-fashioned kindness. I cannot speak highly enough of them, and hope the future will bring many more opportunities for fruitful collaborations. Very special thanks also go to colleagues, friends and family who have lent much help and support over the past years. I especially remember Christian Ruegg, Tom Fennell, Des McMorrow, Steve Bramwell and Duc Le, who had the grace not to laugh when I came to them with the hare-brained schemes of a naive theoretician. Che Gannarelli, for programming suggestions, proof-reading and general advice. Kyle Rogers and Andrew Gormanly, for always being at hand to resolve technical difficulties. Lorenzo Stella, Wisdom Beyhum, Hoi-Yu Tang, Mose Bevilacqua, Emily Milner, Ana Sofia Vila Verde and Jennifer Brookes, for many laughs and much tea. Lisa Harris, Seema Babbar, Mischa Stocklin and Alan Wright for many laughs and slightly more varied beverages. Giusi Cerbone, for being my sister in all but name.

I would like to dedicate a very special mention to my family. I admire them immensely for their achievements and experiences, and will never find the words to thank them enough for having strived to provide me with opportunities and choices. They have taught me the importance of dignity and honesty, and ensured I would never want for comfort or warmth. I sincerely hope they might understand and feel for me the love and pride I feel for them, and that I may always return the care and security I have received.

Finally, a person who has enriched, challenged and sustained me, for and with whom I wish to be the highest true self I can be. To Ben, for giving me light, and nurturing my soul.
\end{acknowledgements}

%\begin{document}

%
%\include{intro}
%\include{ent_review}
%\include{neutron_proposal2}
%\include{s_matrix1}
%\include{two_neutron_results}
%\include{many_neutrons}
%\include{spin_chain_review}
%\include{dipolar_spin_chains}
%\include{conclusions}

\pagestyle{fancy}
\fancyhead{}
\fancyfoot{}
\renewcommand{\chaptermark}[1]{%
\markboth{\chaptername
 \ \thechapter.\ #1}{}}
\renewcommand{\sectionmark}[1]{\markright{\thesection.\ #1}}
\renewcommand{\headrulewidth}{0.1pt}
\fancyhead[RE]{\slshape \rightmark}
\fancyhead[LO]{\slshape \leftmark}
\fancyhead[RO,LE]{\thepage}
\fancypagestyle{plain}{%
\fancyhf{}
\fancyfoot[C]{\thepage}
\renewcommand{\headrulewidth}{0pt}
\renewcommand{\footrulewidth}{0pt}}
\tableofcontents
%here goes your chapter content
\input{intro.tex}

\input{ent_review.tex}

\input{neutron_proposal2.tex}

\input{s_matrix1.tex}
\input{two_neutron_results.tex}
\input{many_neutrons.tex}
\input{spin_chain_review.tex}
\input{dipolar_spin_chains.tex}

\input{conclusions.tex}

% ----------------------------------------------------------------
%	APPENDIX
% ----------------------------------------------------------------

\newpage
\appendix

\chapter{The $S$-Matrix Method}\label{app_a}

\pagestyle{fancy}
\fancyhf{}
\fancyhead[L]{\sf Appendix \thechapter}
\fancyhead[R]{\sf \rightmark}
\fancyfoot[LE,RO]{\thepage}

The $S$-matrix formalism was originally proposed by Heisenberg in the early 1940s, as an attempt to reformulate quantum field theory in terms of observable quantities \cite{cushing}. This project, which became known as the $S$-matrix program, was abandoned soon after, but laid the foundations for the future developments of scattering theory at the hand of, among others, Gell-Mann and Goldberger \cite{gell-mann53}, and Lippmann and Schwinger. The treatment I present here is that of \cite{gell-mann53}, which describes the emergence of an operator akin to the Heisenberg $S$-matrix from the interaction representation.

Consider a scattering system represented by a Hamiltonian $H=K+V$, where $K$ represents the kinetic energy of the system, and $V$ represents the scattering potential. Let us denote with $\Phi_i\left(t\right)=\phi_ie^{-iE_it}$ the stationary state solutions to the Schr\"{o}dinger equation (SE) in the absence of the potential, where the $\phi_i$ are solutions of the time-independent SE. The full SE takes eigenstates $\Psi(t)$, such that
\begin{equation}
i\frac{\partial \Psi(t)}{\partial t}=\left(K+V\right)\Psi(t).
\end{equation}
We work in the approximation that interaction timescales are negligible with respect to measurement timescales. Therefore, the initial and measured states of the system coincide with eigenstates of $K$. Suppose, then, we wished to calculate the cross-section for scattering from initial state $\Phi_j$ to final state $\Phi_i$ at some convenient time $t=0$. To describe the history of the system up until the moment $t=0$, one can represent the initial state $\Psi_j(0)$ as a superposition of $\Phi_j$ states, i.e. as a `train' of waves released at some time $T$ in the distant past and fed into the sample over a prolonged period of time. Hence
\begin{equation}\label{psi j 0}
\Psi_j(t)=\frac{\int_{-\tau}^0 e^{\epsilon T}e^{-iH\left(t-T\right)}\Phi_j(T)\mathrm{d}T}{\int_{-\tau}^0\mathrm{d}T},
\end{equation}
where $\tau=\epsilon^{-1}$ will eventually be taken to the limit of $+\infty$, and the factor of $e^{\epsilon T}$ is included to ensure the integral converges in this limit.

Let us now define an interaction representation wavefunction $\Psi^{\prime}$ by
\begin{equation}\label{psi prime t}
\Psi^{\prime}(t)=e^{iKt}\Psi(t),
\end{equation}
such that $\Psi^{\prime}(0)=\Psi(0)$. The SE for $\Psi^{\prime}(t)$ reads
\begin{equation}
i\frac{\partial \Psi^{\prime}(t)}{\partial t}=V\left(t\right)\Psi^{\prime}(t),
\end{equation}
with
\begin{equation}\label{v of t}
V\left(t\right)=e^{iKt}Ve^{-iKt}.
\end{equation}
$\Psi^{\prime}(t)$ can be written in terms of $\Psi^{\prime}(t_0)$ at some time $t_0$ with the help of the time-evolution operator:
\begin{equation}\label{u tt0}
\Psi^{\prime}(t)=U\left(t,t_0\right)\Psi^{\prime}(t_0).
\end{equation}
Knowing that $\Psi(t)=e^{-iHt}\Psi(t_0)$, equations \eqref{psi prime t} and \eqref{u tt0} then give
\begin{equation}\label{u tt0a}
U\left(t,t_0\right)=e^{iKt}e^{-i\left(K+V\right)\left(t-t_0\right)}e^{-iKt_0}.
\end{equation}
By differentiating $U\left(t,t_0\right)$ with respect to $t$ and substituting back the form of \eqref{u tt0a} and \eqref{v of t}, one finds a SE for the time-evolution operator:
\begin{equation}
i\frac{\partial U\left(t,t_0\right)}{\partial t}=V\left(t\right)U\left(t,t_0\right).
\end{equation}
Integrating both sides from $t_0$ to $t$ then gives
\begin{equation}\label{u tt0b}
U\left(t,t_0\right)=1-i\int_{t_0}^t V\left(t^{\prime}\right)U\left(t^{\prime},t_0\right)\mathrm{d}t^{\prime}.
\end{equation}
By repeating the same process, but with respect to $t_0$, we obtain
\begin{equation}\label{u tt0c}
U\left(t,t_0\right)=1+i\int_t^{t_0} U\left(t,t^{\prime}\right)V\left(t^{\prime}\right)\mathrm{d}t^{\prime}.
\end{equation}
With the help of Dyson's ordering operation \cite{dyson49}, the solutions to equations \eqref{u tt0b} and \eqref{u tt0c} can be expressed as
\begin{align}
U\left(t,t_0\right)&=\left(\exp{\left[-i\int_{t_0}^t V\left(t^{\prime}\right)\mathrm{d}t^{\prime}\right]}\right)_+,\label{dyson1}\\
U\left(t,t_0\right)&=\left(\exp{\left[i\int_t^{t_0} V\left(t^{\prime}\right)\mathrm{d}t^{\prime}\right]}\right)_-,\label{dyson2}
\end{align}
where the $+$ ($-$) subscripts indicate that terms in the series expansion should be written from left to right in order of decreasing (increasing) times. The Heisenberg $S$-matrix is often written as $U\left(\infty,-\infty\right)$. Therefore, we must now show that for the scattering process yielding the state \eqref{psi j 0}, it is meaningful to extrapolate equation \eqref{u tt0a} to the limit of $t,t_0=\pm\infty$.

Let us transform \eqref{psi j 0} with the help of \eqref{psi prime t}, letting $\tau\rightarrow\infty$:
\begin{equation}\label{psi j}
\Psi_j^{\prime}(t)=e^{iKt}e^{-iHt}\epsilon\int_{-\tau}^0 e^{\epsilon T}e^{iHT}e^{-iKT}\phi_j\mathrm{d}T,
\end{equation}
which becomes, from \eqref{u tt0a}
\begin{equation}\label{psi j}
\Psi_j^{\prime}(t)=\epsilon\int_{-\infty}^0 e^{\epsilon T}U\left(t,T\right)\phi_j\mathrm{d}T.
\end{equation}
We take
\begin{align}
U\left(t,-\infty\right)&=\mathrm{lim}_{\epsilon\rightarrow0^+}\epsilon\int_{-\infty}^0 e^{\epsilon T}U\left(t,T\right)\mathrm{d}T,\label{rules 1}\\
U\left(\infty,t\right)&=\mathrm{lim}_{\epsilon\rightarrow0^+}\epsilon\int_0^{\infty} e^{-\epsilon T}U\left(T,t\right)\mathrm{d}T\label{rules 2}.
\end{align}
By thus defining our limits, it is possible to show that equations \eqref{u tt0b} and \eqref{u tt0c} do indeed reduce to
\begin{align}
U\left(t,-\infty\right)&=1-i\int_{-\infty}^t V\left(t^{\prime}\right)U\left(t^{\prime},-\infty\right)\mathrm{d}t^{\prime},\label{u tt0d}\\
U\left(t,\infty\right)&=1+i\int_t^{\infty} U\left(t,t^{\prime}\right)V\left(t^{\prime}\right)\mathrm{d}t^{\prime}.\label{u tt0e}
\end{align}
We therefore conclude that
\begin{equation}\label{s1}
S\equiv U\left(-\infty,\infty\right)=1-i\int_{-\infty}^{\infty} V\left(t^{\prime}\right)U\left(t^{\prime},-\infty\right)\mathrm{d}t^{\prime}.
\end{equation}

Now to justify the form of equation \eqref{u tt0a} as $t_0\rightarrow -\infty$. We know the scattered states of the system in the asymptotic regime must satisfy the Lippmann-Schwinger equation \eqref{lipp schw eq}, such that
\begin{equation}
\psi_j=\phi_j+G\left(E_j\right)V\psi_j.
\end{equation}
where the $\psi$ are eigenstates of the total Hamiltonian, satisfying
\begin{equation}
H\psi_j=E_j\psi_j,
\end{equation}
and $G\left(E_j\right)$ is the retarded Green Function of the free system evaluated at energy $E_j$:
\begin{equation}
G\left(E_j\right)=\frac{1}{E_j-K+i\epsilon}.
\end{equation}
From equations \eqref{u tt0a} and \eqref{rules 1}
\begin{equation}\label{u0 minfty a}
U\left(t,-\infty\right)=e^{iKt}e^{-i\left(K+V\right)t}\:\mathrm{lim}_{\epsilon\rightarrow0^+}\frac{\epsilon}{i\epsilon+i\left(H-E_j\right)}\:|\phi_j\rangle\langle\phi_j|,
\end{equation}
with $\sum_j |\phi_j\rangle\langle\phi_j|=1$.
One can show that
\begin{equation}\label{u0 minfty b}
U\left(0,-\infty\right)\phi_j=\psi_j,
\end{equation}
meaning $U\left(0,-\infty\right)$ takes an incident wave $\phi_j$, which is a stationary state of the interaction-free Hamiltonian, and produces an initial state $\psi_j$ which is an eigenstate of the \emph{total} Hamiltonian. Similarly, $U\left(0,+\infty\right)$ takes $\phi_i$ to an eigenstate of the total Hamiltonian $\psi_i$ corresponding to an outgoing wave $\phi_i$. These operators are found to obey the unitarity condition
\begin{align}
U\left(0,-\infty\right)U\left(-\infty,0\right)&=1,\\
U\left(0,+\infty\right)U\left(+\infty,0\right)&=1.
\end{align}
Then, using \eqref{u0 minfty a} and \eqref{u0 minfty b}
\begin{equation}\label{u0 minfty c}
U\left(t,-\infty\right)=e^{iKt}\sum_j  e^{-iE_jt}|\psi_j\rangle\langle\phi_j|,
\end{equation}
giving
\begin{equation}\label{u0 minfty d}
V\left(t\right)U\left(t,-\infty\right)=\sum_{i,\:j} |\phi_i\rangle\langle\phi_i| e^{i\left(E_i-E_j\right)t}V|\psi_j\rangle\langle\phi_j|.
\end{equation}
Finally, substituting into \eqref{s1}:
\begin{equation}
S=\mathbb{I}-2\pi i\sum_{i,\:j} |\phi_i\rangle\langle\phi_j| \delta\left(E_i-E_j\right)R_{ij},
\end{equation}
with
\begin{equation}
R_{ij}=\langle \phi_i\left|V\right|\phi_j+\frac{V\phi_j}{E_j-H+i\epsilon}\rangle.
\end{equation}
To first order in the scattering potential, the $S$-matrix is then
\begin{equation}\label{1st order s}
S^{(1)}=\mathbb{I}-2\pi iV\delta\left(E_i-E_j\right),
\end{equation}
where $E_j$ and $E_i$ are the initial and final eigenenergies of the free Hamiltonian, respectively.

\chapter{The PF2 Source of UCN}\label{app_b}
\begin{figure}[H]
\renewcommand{\captionfont}{\footnotesize}
\renewcommand{\captionlabelfont}{}
\begin{center}
\subfigure{\label{reactor}\includegraphics[width=6.5cm]{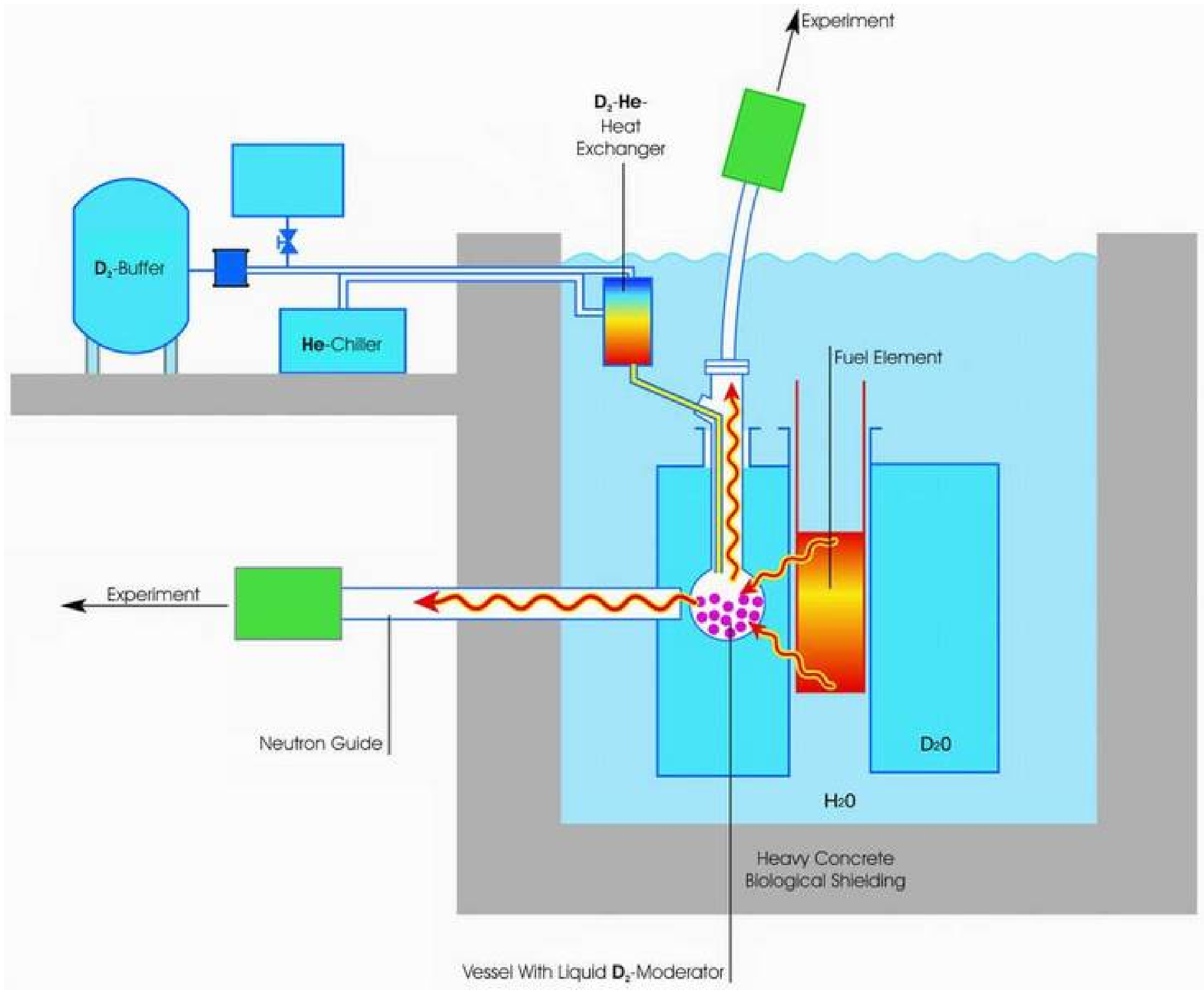}}
\hspace{0.3cm}
\subfigure{\label{ucnb}\includegraphics[width=6.5cm]{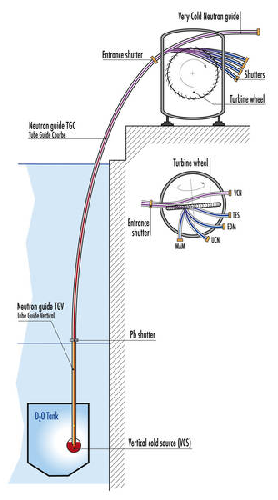}}
\end{center}
\caption{A schematic of the Vertical Cold Source (left) and of the PF2 instrument at the ILL, Grenoble. The image of the VCS is reproduced from the website of the Technischen Universit\"{a}t M\"{u}nchen, which operates the cold neutron source FRM II, whose design is extremely similar to that of the VCS. Image available at http://www.frm2.tum.de/en/technik/secondary-sources/cold-source/index.html.}\label{appb 1}
\end{figure}

The PF2 source produces UCNs by further cooling neutrons that emerge from the Vertical Cold Source (VCS), currently operating at the Insitut Laue-Langevin in Grenoble. The structure of the VCS comprises a fuel element and liquid deuterium moderator immersed in a pool of heavy water, all encased in concrete shielding. As the neutrons interact with the moderator, the deuterium is heated to boiling point. The vapour is then recondensed in a heat exchanger, and flows back into the moderator vessel. The moderated neutrons are directed into a nickel guide, the lower half of which is filled with helium to increase cooling. They emerge in a turbine house with five exit ports, part of which is occupied by a device known as a \emph{Garching turbine}. Half the beam feeds the turbine, which slows the neutrons to the meV regime by total reflection along semi-circular nickel blades. These neutrons emerge as ultra-cold beams, whose cross section is determined by the exit port to which they are distributed. The remaining part of the beam bypasses the turbine, and emerges through the upper of the five exit ports at a velocity of approximately $40$ ms$^{-1}$. A schematic of the reactor and of the PF2 source is shown in figure \ref{appb 1}, below.

\chapter{The Energy Spectrum of a Heisenberg Ring}\label{app_c}
The Heisenberg nearest-neighbour interaction is expressed by the Hamiltonian of equation \eqref{h0}:
\begin{equation}\label{h0_app}
H_0=-\sum_{\langle ij \rangle}J_{ij}\:\bm{\sigma^i \cdot \sigma^j}.
\end{equation}
The eigenenergies of this Hamiltonian in a periodic system are obtained by calculating the expectation values $\langle \bm{m}|H_0|\bm{m}\rangle$ between eigenstates of the form
\begin{equation}\label{ring estates app}
|\bm{m}\rangle=\frac{1}{\sqrt{N}}\sum_{j=1}^Ne^{ik_mj}|\bm{j}\rangle,
\end{equation}
where $j$ represents the location of the spin flip. From equations \eqref{h0_app} and \eqref{ring estates app}, one therefore obtains
\begin{align}
E^H_m&=\langle \bm{m}|H_0|\bm{m}\rangle=\frac{1}{N}\sum_{\left(j,j^{\:\prime}\neq j\right)\:=1}^Ne^{ik_m\left(j-j^{\:\prime}\right)}\langle \bm{j}^{\prime}|H_0|\bm{j}\rangle=\nonumber\\
&=-\frac{J}{N}\sum_{\left(j,j^{\:\prime}\neq j\right)\:=1}^N e^{ik_m\left(j-j^{\:\prime}\right)}\left(\delta_{j^{\:\prime},\:j+1}+\delta_{j^{\:\prime},\:j-1}\right)=\nonumber\\
&=-\frac{J}{N}\sum_{j=1}^N\left(e^{ik_m}+e^{-ik_m}\right)=-2J\cos{k_m}.
\end{align}
This is the result used to compute the fidelities and timescales shown in figure \ref{Fidtime}.

\backmatter

\pagestyle{fancy}
\fancyhf{}
\fancyhead[L]{}
\fancyhead[R]{\sf \rightmark}
\fancyfoot[LE,RO]{\thepage}

\addcontentsline{toc}{chapter}{References}

\include{bibliography}
\end{document}

%% file: intro.tex
%\documentclass[a4paper,12pt]{article}
%\usepackage{setspace}
%\usepackage{graphicx}
%\usepackage{subfigure}
%\usepackage{caption}
%\usepackage{color}
%\usepackage{amsmath}
%\usepackage{amssymb}
%\usepackage{caption}
%\usepackage{bm}
%\DeclareGraphicsExtensions{.epsi,.eps,.ps}
%
%\onehalfspacing
%%\title{Entanglement and Quantum Information Transfer in Arrays of Interacting Quantum Systems}
%%\author{M. Avellino}
%%\def\mathbi#1{\textbf{\em #1}}
%%\def\hh{{\mathfrak h}}
%\begin{document}
%%\tableofcontents
%%\maketitle
%%\newpage
\chapter{Introduction}\label{intro}
In the past four decades, the evolution of computer hardware has borne out a prediction commonly known as Moore's Law. In its most recent formulation \cite{moore97}, Moore's Law decrees that the number of transistors on an integrated circuit doubles roughly every two years. In a keynote speech delivered earlier this year at the Intel developer forum in Shanghai, the senior vice president of Intel technologies stated the trend is set to continue for at least another twenty years. By that time, a microprocessor no bigger than a few square nanometres will have come to house as many as $10^{12}$ transistors \cite{gelsinger,intel}. Already in the early 1970s, scientists began to realize the
implications of this continued miniaturization of computer hardware: microchip transistors would soon be
so densely packed as to experience non-negligible intrinsic quantum
mechanical effects. The idea of a computer
whose functioning would rely in a significant way on quantum
processes aroused scientists' curiosity: what if \emph{all} the
internal processing of a computer were to follow quantum laws? Would
such a machine have any advantage over a classical computer? How
could it be built?

One of the first to address these issues was Richard Feynman.
During a conference at Massachusetts Institute of Technology (MIT) in 1981, Feynman observed that a
classical computer could not efficiently simulate the evolution of a
quantum system, and proposed a basic model of quantum ``simulator''
that would overcome this limitation. Feynman's ideas were
formalized by David Deutsch, who in the seminal paper ``Quantum
Theory, the Church-Turing Principle and the Universal Quantum
Computer''\cite{deutsch85} provided a more comprehensive description
of a quantum computer's features and capabilities. In comparing a
quantum computer to a classical machine, Deutsch stated:\\
\begin{quote}
... [the universal quantum computer] admits a further class of
programs which evolve computational basis states into linear
superpositions of each other.\\
\end{quote}
This is, in fact, the defining feature of a quantum computer: the
ability to manipulate data that is not in an eigenstate of the computational basis. Whereas
classical bits can exist only in the states 0 or 1, quantum bits (or
\textit{qubits}), can be found in superpositions of the form
$\alpha |0 \rangle + \beta |1 \rangle$, where $\alpha$ and $\beta$
are real or complex coefficients satisfying the condition
$|\alpha|^2+|\beta|^2=1$. Consequently, a quantum computer of
\textit{N} registers can exist in a superposition of $2^N$ possible
states, which greatly increases its computing power compared to a
classical computer of the same size. Consider, for example, a
function of many variables, $f(x_1,x_2... x_n)$, with $x_n=[0,1]$.
This function is defined constant if it is single-valued for all
permutations of $\{x_1,x_2... x_n\}$, and balanced if its value is 0
for half of the input domain and 1 for the other half. Classically,
the characterization of the function might require $2^{(n-1)}+1$
evaluations. With a quantum computer, just one evaluation is enough
to give an answer that is always correct. The quantum algorithm that
achieves this is known as the Deutsch-Josza algorithm, and was
proposed by Deutsch, in collaboration with Richard Josza, in 1992
\cite{deutschjozsa92}.

Two years later, Peter Shor of Bell Labs discovered a remarkable
algorithm that would allow large integers to be factored very
quickly, thus invalidating even the most secure public-key cryptosystems available at the time \cite{shor95}. Arriving as it did in a period of intense debate on the individual's right to absolute privacy of communication, Shor's discovery sparked a huge interest in quantum computation and kick-started some truly outstanding progress
in the field. The very next year, a group at NIST (Colorado)
succeeded in making the first quantum logic gate with trapped ions,
according to a proposal advanced previously by Cirac and Zoller of
Innsbruck University \cite{zc95,monroe95}. In 1996 Lov Grover of
Bell Labs developed an algorithm that would allow the quadratic speed
up of a database search \cite{grover96}, and 1997 brought the first publications on quantum simulation using liquid-state nuclear magnetic resonance (NMR) techniques, by Cory, Chuang \textit{et al.} of MIT \cite{coryprop97}. The following year, two
independent groups at MIT and Oxford University succeeded in
building the first two-qubit NMR quantum computers, and executed the Deutsch-Josza algorithm and Grover's algorithm
for a database with four entries \cite{jones98g,chuang98,jones98dj}.
The new millennium saw the construction of five- and seven-qubit NMR computers,
and in 2001 researchers at IBM's Almaden Research Centre and
Stanford University implemented Shor's algorithm to factor the
number 15 on a 7-qubit NMR computer \cite{steffen01}.

The appearance of a working quantum computer less than
fifteen years after its `official' conception was a truly impressive
achievement, greatly owed to a backup of almost five
decades of work in the field of magnetic resonance. However, it was soon realized that NMR quantum computing would become unfeasible for systems of more than 10 qubits, which is considerably fewer than the number required for any truly interesting application. As a result, the focus of the scientific community largely shifted to other QC strategies, and some of the most popular proposals currently include trapped ions, superconducting Josephson junctions, charge states of quantum dots, and atoms trapped in optical potentials, usually referred to as optical lattices. The past few years have yielded numerous achievements in all these fields, but current research is still grappling with issues such as qubit generation, manipulation and stability, as well as reliable data storage and transfer.

This thesis proposes to examine some of the more fundamental requirements of a successful quantum computation, namely the ability to transmit quantum information with maximum efficiency, and the creation of \emph{entanglement}, a stronger-than-classical quantum correlation that many believe embodies `quantumness' itself. I will begin with a discussion of this intriguing phenomenon, describing in a mainly qualitative fashion its principal properties and its role in quantum computation and communication. Experimental achievements will be reviewed, and possible `gaps in the market' will be identified. Chapter \ref{neutron proposal} attempts to fill one such gap by proposing an original scheme to generate entanglement between distinct neutrons, which has never been observed to date. The entanglement arises from the sequential scattering of a neutron pair from a macroscopic substrate, which is described using a fully time-dependent approach based on unitary evolution. This somewhat atypical treatment of the scattering problem raises the question of the physical interpretation of the quantum-mechanical time parameter, and a detailed comparison with standard scattering theory reveals interesting, and often pleasingly simple results. Following a discussion of this point in chapter \ref{s matrix}, the performance of the protocol will be assessed in chapter \ref{two ns}, together with experimental requirements and feasibility in light of methods and instrumentation currently at our disposal. Chapter \ref{many ns} then describes an extension of the scheme, which allows for greater experimental flexibility. In chapter \ref{spin chain review}, the focus shifts slightly from entanglement generation to entanglement distribution by transmission of arbitrary quantum states. The discussion will hinge on quantum communication using spin chains, remarkable yet simple systems identified some years ago as potentially faithful carriers of quantum information. I will review the original proposal for quantum state transfer through nearest-neighbour-coupled Heisenberg spin chains, and present a new model for near-perfect spin chain communication based on long-range dipolar interactions in chapter \ref{dipolar chains}. Finally, chapter \ref{conclusions} is dedicated to a short summary and discussion of the main results of this thesis, together with a vision for improvements, extensions and future avenues of investigation.

%% file: ent_review.tex
\chapter{The Role of Entanglement}\label{ent review}
\textit{This chapter introduces quantum entanglement, a property identified by Schr\"{o}dinger in the mid 1930's as the very embodiment of `quantumness' itself. After a brief historical note, I will illustrate the main properties of entangled states and the concept of an `entanglement measure', a theoretical tool used to quantify entanglement and study its dynamics. Entanglement generation schemes will be reviewed, with particular emphasis on long-distance entanglement via ancillary mediators. I will discuss detection of entangled states using \emph{witnesses}, and current experimental achievements, concluding with a few personal remarks on the ontological value of understanding what Einstein called `spooky action at a distance'.}

\section{Introduction}
The act of communication between two parties can be interpreted as a movement of information from one to the other. This information somehow links the two parties, hence a successful communication fulfils the objective of establishing a correlation between a sender and a receiver. In the quantum world, this correlation has rather peculiar properties, and upon close analysis raises questions that stray from the realm of physics into that of philosophy. We are speaking, of course, of \emph{entanglement}.

From an operational viewpoint, entanglement is a physical resource, which enables faithful communication of quantum information and is believed by many to fuel the power of quantum computation (\cite{josza03,lloyd99,ding07} and references therein). Entangled quantum systems share a stronger-than-classical correlation, such that a measurement on one will
instantaneously project the other into a state conditional on the outcome of
that measurement, even if they are separated by a space-like interval. In the words of Schr\"{o}dinger himself
\cite{schrodinger35}:
\begin{quote}
When two systems, of which we know the states by their respective
representatives, enter into temporary physical interaction due to
known forces between them, and [...] separate again, then they can
no longer be described [...] by endowing each of them with a
representative of its own [...] By the interaction the two
representatives have become entangled.

[...] even though we restrict the disentangling measurements to one
system, the representative obtained for the other system is by no
means independent of the particular choice of observations which we
select for that purpose and which by the way are entirely arbitrary.
It is rather discomforting that the theory should allow a system to
be steered or piloted into one or the other type of state at the
experimenter's mercy in spite of his having no access to it.
\end{quote}

Schr\"{o}dinger's discomfort stemmed from the apparent conflict between the predictions of quantum mechanics and the reigning opinion that a complete physical theory should be local and realistic \cite{EPR}. According to such a theory, (i) a measurement made at one point in space should not be influenced by anything outside its past light cone; (ii) physical properties should have a well-defined value whether they are measured or not. It is believed Einstein once said, ``I like to think that the moon is there even if I am not looking at it'' \cite{seelig56}. The idea that observations made on one party of a entangled pair could instantaneously project the other into a well-defined quantum state contradicts both these axioms, suggesting (i) that faster than light communication is possible, and (ii) that physical quantities lack objective reality. Quantum mechanics was therefore deemed incomplete, and local `hidden variable' (LHV) models were proposed to explain the manifestation of this `spooky action at a distance'. After many years of heated debate, British physicist John Bell concluded that empirical evidence alone would resolve the controversy, and in 1966 published details of an experiment designed to test the deterministic world view \cite{bell64}. The key result of this publication was a measurable inequality, derived independently of any type of mathematical formalism, which should be respected by any LHV. A more general version of Bell's inequality was measured for the first time in the 1980s in a series of experiments by Aspect \textit{et al.}, whose results conclusively precluded the existence of local hidden variables, therefore ascribing to quantum mechanics the characteristics of a non-local theory \cite{aspect}. The overwhelming majority of Bell's test experiments carried out since have reached similar conclusions \cite{bell's test}; however, there remain shadows of uncertainty owing to the possibility of `loopholes', which arise from the need to make assumptions about the experimental setup \cite{loopholes}\footnote{An example is the `fair sampling' assumption, which relates to optical realizations of the experiment, and states that the number of photons reaching a detector is representative of the number of photons emitted by the source.}.

Entangled states have many interesting properties. It has recently been shown that entangled particles may signal to each other more than 10000 times faster than the speed of light \cite{gisin08}. Luckily for special relativity, this does not imply super-luminal communication, because the signal cannot be used to send a message. For this to be possible, control over a measurement outcome is required - but this outcome is always probabilistic. Indeed, the information carried by a maximally entangled state relates to the whole rather than the components, which are individually in maximally mixed states. This property can be phrased in terms of entropy, in that the state of the whole is always more ordered than that of the subsystems \cite{horodeki94a}. The operational capabilities of entangled states can then be expressed as entropic inequalities which show, for example, that if two parties share a maximally entangled state and a classical channel, each party can arrive at full knowledge of the state without any further exchange of qubits \cite{horodeki05}.

Entanglement is perhaps best appreciated in the context of quantum computation. It is believed that entanglement is, at least partly, responsible for the exponential speedup of a quantum process over a classical one. It has been shown that an arbitrary quantum computation requires exponentially higher resource overheads if entanglement is \emph{not} involved \cite{josza03,lloyd99,ding07}. This view is upheld by the fact that many quantum algorithms, such as Shor's factorization and Grover's search, rely on the production of highly entangled intermediate states \cite{ekert98,orus04}. Sophisticated quantum search (SQS) algorithms capable of outperforming a classical algorithm \emph{without} recourse to entanglement have been suggested; however, it has also been observed that equally efficient classical versions of SQS exist, therefore no true advantage over classical computation can be claimed \cite{meyer00,wojcik05}. Simulations of quantum systems also rely on entanglement, because not all quantum processes can be reproduced by a classical computer. In some exceptional cases, highly entangled quantum computations can be efficiently simulated by classical means \cite{gottesmann99,barlett02,terhal02}. However, as a general rule, this is not possible unless the amount of entanglement present in the system does not exceed $k\:\log n$, where $k>0$ and $n$ is the number of input qubits \cite{josza03,vidal03}.

Entanglement is by no means a comprehensively understood phenomenon, however despite its more esoteric characteristics it is now well established as a real physical resource. Unfortunately, entanglement is often somewhat fragile, and can be diminished or destroyed if the entangled systems experience contact with a noisy environment. Much of today's entanglement theory comes from attempts to preserve the precious correlations, by determining how entanglement can be created, manipulated and quantified. Bi-partite entanglement is well understood and, now, accessible from an experimental viewpoint also. The field of multi-partite entanglement characterization is currently one of ever expanding horizons, strongly motivated by recent observations of three- to eight-party entangled states \cite{haffner05}.

\section{Fundamental Properties}
The main property
of an entangled state is that its component subsystems are no longer
independent. Formally, this means an entangled $n$-party state $\rho$ cannot
be written as a mixture of separable states, i.e.
\begin{equation}
\rho^{A,B...N} \neq \sum_ip_i\rho_i,
\end{equation}
with
\begin{equation}
\rho_i=\rho_i^A\bigotimes \rho_i^B\bigotimes \cdots \rho_i^N.
\end{equation}
It follows that all tensor product (or separable) states contain no entanglement, while all non-separable states are entangled, and
therefore allow some tasks to be achieved better than by local
operations and classical communication (LOCC) alone \cite{masanes05,horodeki07}. Separable states can evolve into entangled states, however this requires non-local operations such as two-qubit gates or interaction with ancillary systems. Indeed, the total entanglement of an $n$-party state cannot increase under LOCC alone \cite{bennett96}. This goes back to entropy, as there exist theorems stating that a pure state $\psi$ can only be transformed into another $\phi$ if the subsystems of $\psi$ are less ordered than those of $\phi$ \cite{horodeki07,nielsen99}. Therefore, as the entanglement of a state is directly proportional to the entropy of its subsystems, a locally ordered, separable state cannot be transformed into a locally disordered entangled state. LOCC can however change the distribution of entanglement within a many-particle system; this property enables protocols such as quantum teleportation and quantum dense coding, which allow Alice to transmit a message to Bob by performing local measurements on her half of a previously shared entangled state \cite{bennett93,bennett92}. It has also been shown there exists a special class of entangled states that increase the scope of LOCC if employed as ancillae. These are called \emph{bound} entangled states, sometimes referred to as `activators', owing to their ability to facilitate entanglement-boosting protocols such as distillation. An interesting result is that any bipartite pure state can be transformed into another by LOCC alone, provided one has access to bound-entangled states \cite{masanes05,horodeki98,horodeki99b,ishizaka04}.

The invariance of entanglement under LOCC can be related to a property known as \textit{entanglement monogamy} \cite{wootters00}. According to this property, there is a limit on the amount of information a particle can share with others. If a particle is maximally entangled with another, its entanglement with any other system must be zero. This is owed to the fact that a maximally entangled state contains all the information available on that state; there is none left to share with other parties. If, on the other hand, two particles are only partly entangled, correlations with other parties are allowed. The bound on the amount of entanglement between any pair of particles of a multiparty state has recently been formalized in terms of an entanglement measure known as the \textit{tangle}, which will be introduced in the following section.

\section{Measures}\label{ent measures}
Analyzing the entanglement of a quantum state can mean several things. On the one hand, one could be content with establishing whether the state is in fact entangled or separable. For bipartite systems whose dimensions $d$ do not exceed 6, this is easily decided via a criterion known as the Positive Partial Transpose (PPT) \cite{peres96,horodeki96}. The criterion states that if the eigenvalue spectrum of a state $\rho_{AB}$ becomes negative under partial transposition with respect to either system, the state is entangled. This does not mean a state with positive partial transpose is necessarily separable; the aforementioned activator states, for example, are entangled states with PPT. The situation becomes more complicated for multipartite states; in this case, the PPT criterion can tell us whether any two parties are entangled, but not whether genuine multipartite entanglement exists. It is, therefore, often simpler to quantify the entanglement of a state via \textit{ad hoc} functions known as \textit{entanglement measures}. Not all measures have the same operational meaning, therefore different entanglement measures can afford different state ordering. Calculating a spectrum of entanglement measures for a given state can then provide a useful means of characterization.

Entanglement measures $E$ satisfy the following fundamental
axioms:
\begin{enumerate}
\item $E(\rho)=0$ if $\rho$ is separable;
\item $E(\rho)$ does not increase over LOCC, and is therefore an \textit{entanglement monotone}.
\end{enumerate}
As regards \emph{pure} states, the best known and most commonly used entanglement measure is the \textit{Von Neumann Entropy}, defined as
\begin{equation}\label{vn entropy}
S(\rho)=-\mathrm{Tr}_{A\:(B)}\:\left[\rho_{A\:(B)}\log_2\rho_{A\:(B)}\right].
\end{equation}
This is the quantum equivalent of the Shannon entropy, which measures the uncertainty associated with a classical probability distribution \cite{nielsen and chuang}. In the context of pure-state entanglement, $S\left(\rho\right)$ represents something of a `reference', as all entanglement measures $E(\rho)$ must reduce to $S(\rho)$ if $\rho$ is pure.

In more general terms, any function that adheres to the two conditions listed above is a valid entanglement measure. Therefore, the set of measures is in principle infinite \cite{vidal00}. However, for reasons of mathematical convenience, it is often useful to choose functions which satisfy more specific conditions \cite{horodeki07}, for example:
\renewcommand{\labelenumii}{\alph{enumii}}
\begin{enumerate}
\item \textit{Normalization}: $E(\rho)=1$ if $\rho$ is a Bell state, or related to a Bell state by a local unitary transformation;
\item \textit{Convexity}: For a mixed state $\rho=\sum_ip_i\rho_i$, where the $p_i$ are the weights of the different $\rho_i$ components, an entanglement measure is convex if
    \begin{equation}
    E\left(\sum_ip_i\rho_i\right)\leq\sum_ip_iE\left(\rho_i\right).
    \end{equation}
\end{enumerate}

Some entanglement measures have an operational definition, others are constructed on the basis of the above axioms, and not all have a physical interpretation. The brief summary that follows is intended as a largely qualitative overview rather than an exhaustive and mathematically detailed discussion, as I have focussed on measures that are either relevant to the present work or have an intuitive physical interpretation. I refer the reader to \cite{horodeki07,plenio07} and references therein for a more comprehensive and fully mathematical treatment.

\begin{description}
\item[\textbf{The Distillable Entanglement}] Entanglement
distillation (or concentration) is an entanglement purification protocol whereby Alice and Bob share $n$ copies of an entangled state $\rho$, perform LOCC and obtain $k<n$ copies of a Bell state \cite{brassard96}. The \textit{distillable entanglement} is defined as the optimal ratio $k/n$ yielded by this process in the limit of large $n$. It is interesting to note that all entangled bipartite states can be distilled, however this may require the help of an activator state. Activator states, on the other hand, cannot be distilled; their entanglement is inaccessible, hence the term `bound entanglement'.
\item[\textbf{The Entanglement Cost}] represents the number of singlet states $n$ one has to share to distill an arbitrary state, such that errors become infinitesimal in the limit of large $n$ \cite{brassard96}. It is therefore defined in relation to a process which is the opposite of distillation.
\item[\textbf{The Entanglement of Formation}] was introduced in the context of entanglement purification, with the principal objective of quantifying the entanglement of mixed states, i.e. states which are a mixture of entangled pure states \cite{bennett96}. The entanglement of formation is then defined as the minimal
average entanglement over all pure state decompositions. In a two-qubit system, this can be calculated exactly from a second quantity, the
\textit{concurrence} \cite{wootters98}, defined as
\begin{equation}
C(\rho)=max\{0,\lambda_1-\lambda_2-\lambda_3-\lambda_4\}.
\end{equation}
Taking $\rho$ to be the density matrix of the two qubits, the
$\lambda_i$ are the eigenvalues in decreasing order of the matrix
$R=\sqrt{\sqrt{\rho}\tilde{\rho}\sqrt{\rho}}$, which has
$\tilde{\rho}=(\sigma_y\bigotimes\sigma_y)\rho^*(\sigma_y\bigotimes\sigma_y)$
and
\[ \sigma_y = \left( \begin{array}{cc}
0 & -i \\
i & 0\end{array} \right),\]
where $\rho^*$ is the
element-by-element complex conjugate of $\rho$.
\item [\textbf{The Logarithmic Negativity}] A bipartite state is entangled if it has a negative partial transpose.
The magnitude of the negative eigenvalue can be related to the degree of entanglement using the \textit{logarithmic negativity} \cite{zyczkowski98,vidal02}, defined as
\begin{equation}
E_N(\rho)=\log_2||\rho^{T_B}||
\end{equation}
where $||X||=Tr\sqrt{X^{\dagger}X}$, and $T_B$ indicates the partial
transpose with respect to qubit B.
\item [\textbf{The Robustness of Entanglement}] This quantifies how
much noise it is possible to mix in with a given entangled
state before it becomes separable \cite{vidal99}.
\item[\textbf{The Squashed Entanglement}] is related to the \textit{intrinsic information}, which measures the correlations between two discrete random variables whose value is conditioned by a third \cite{christandl04a}. It is proportional to the minimal quantum intrinsic information over all states $\rho_{ABC}$ satisfying $\mathrm{Tr}_C\:\rho_{ABC}=\rho_{AB}$:
\begin{equation}
E_{sq}=\mathrm{inf}_{\lbrace\rho_{ABC}\rbrace}\frac{1}{2}\left[S(AC) + S(BC) - S(ABC) - S(C)\right],
\end{equation}
where $S\left(\cdot\right)$ is the Von-Neumann entropy. The squashed entanglement is related to the distillable entanglement and the entanglement of formation, as $E_D\leq E_{sq}\leq E_F$.
\end{description}

Strictly, the above quantities relate to bipartite entanglement only, though multipartite generalizations can sometimes be derived. An example is the `global entanglement' \cite{meyer01}, which is simply the sum of the pairwise concurrences of a many qubit state. The squashed entanglement can also be extended to many particle states by replacing the bipartite mutual information with its equivalent for multipartite systems \cite{horodeki94}.

A commonly used measure is the \textit{tangle} $\tau$ \cite{wootters00}, defined for three qubits as
\begin{equation}
\tau(A,B,C)=\mathcal{C}^2_{A,\:BC}-\mathcal{C}^2_{A,\:B}-\mathcal{C}^2_{A,\:C},
\end{equation}
where $\mathcal{C}$ is the concurrence. It has been shown that $\mathcal{C}^2_{A,\:BC}=4\:det\:\rho_A$, where $\rho_A$ is the reduced density matrix of the first subsystem. This is owed to the fact that the pair $BC$ can be viewed as a `unit'; therefore, the four-dimensional subspace of $BC$ effectively becomes two-dimensional. The tangle is then the entanglement between system $A$ and the unit $BC$ which is not accounted for by the separate entanglements of $A$ with $B$ and $C$ individually. This is sometimes referred to as the `residual entanglement'. In its original context, the tangle was used to define the monogamy inequality for three qubits
\begin{equation}
\mathcal{C}^2_{A,\:BC}\geq\mathcal{C}^2_{A,\:B}+\mathcal{C}^2_{A,\:C}.
\end{equation}
Recently, this has been generalized to $m$-qubit systems \cite{osborne06}. One then has
\begin{equation}
\mathcal{C}^2_{1,\:2}+\cdots+\mathcal{C}^2_{1,\:m}\leq\mathcal{C}^2_{1,\:\left(2,\cdots m\right)},
\end{equation}
hence the entanglement of one qubit with all the others individually cannot exceed the entanglement of that qubit with the others taken as a block. The quantity $\mathcal{C}^2_{1,\:\left(2,\cdots m\right)}$ is defined in analogy to $\mathcal{C}^2_{A,\:BC}$ via the so called convex roof construction \cite{osborne07}. Taking qubits $\lbrace2,m\rbrace$ to represent a $2^{\left(m-1\right)}$-dimensional system $B$, one finds
\begin{equation}
\mathcal{C}^2_{1,\:B}\triangleq \mathrm{inf}_{\lbrace p_x,\:\psi_x\rbrace}\sum_x p_x S(\mathrm{Tr}_B\:\rho_{\psi_x}),
\end{equation}
where the infimum runs over all pure state decompositions of $\rho_{1\:B}=\sum_x p_x\rho_{\psi_x}$ yielded by the pure state $\psi_x$.

Another interesting measure, showing some formal similarities to the bipartite entanglement of formation, has been defined in the context of spin chains. This is the \textit{localisable entanglement} \cite{verstraete04}, and represents the maximum bipartite entanglement one can obtain by choosing two subsystems of a multipartite entangled state and performing LOCC to optimize their correlation. For three qubit states, the localisable entanglement is equivalent to the so-called entanglement of assistance, i.e. the amount of entanglement Alice and Bob can establish between their qubits if Charlie measures his and communicates the outcome \cite{divincenzo98}.

For reasons of relevance and brevity I shall stop here, although the above is by no means an exhaustive list. For a comprehensive discussion, I once more refer the reader to \cite{horodeki07,plenio07} and references therein.

\section{Witnesses}
Entanglement measures provide a theoretical characterization of entanglement dynamics, but cannot be measured directly because they do not represent observables. To verify the entanglement of a system in an experimental setting, it is therefore useful to define a set of hermitian operators termed \textit{entanglement witnesses} \cite{horodeki96,terhal00}, whose expectation values represent average values of measurable quantities, i.e. observable properties of a physical systems. A valid entanglement witness must satisfy
\begin{equation}\label{witness_ent_rev}
\langle W\rangle=\mathrm{Tr}\:\left[W\rho_{AB}\right]> 0
\end{equation}
if $\rho_{AB}$ is separable, hence the expectation value of a witness must be positive on all product states. A witness therefore provides an experimentally accessible means of establishing whether a state is entangled or not.

The estimation of the entanglement properties of a system from expectation value measurements bears an interesting relation to the Jaynes maximum entropy principle \cite{jaynes57}. It is known that the state of a system is completely characterized by the ensemble of the expectation values of all commuting observables relating to that state. Suppose, however, that one disposed of an incomplete data set which did not allow a unique characterization; the Jaynes principle affirms that, of all possible states of the system, the most probable one is that which maximizes entropy. Let us now imagine we hold an incomplete data set relating to some quantum state; what could we infer about its entanglement properties? According to the Jaynes principle, if the data set is compatible with a range of entanglements $\epsilon\in[\epsilon_-,\epsilon_+]$, one should always assume $\epsilon=\epsilon_+$. This is due to the correspondence between entanglement and entropy discussed above. However, it has been shown that such a maximization is not always `truthful'; in other words, by applying the Jaynes principle to quantum states one runs the risk of overestimating the amount of entanglement at one's disposal \cite{horodeki99a}. It has therefore been suggested that in a quantum setting one should apply \textit{minimization} of entanglement under experimental constraints. In other words, if an incomplete data set is compatible with the range of entanglements $\epsilon\in[\epsilon_-,\epsilon_+]$, one should always assume the true entanglement of the system is represented by the lower bound.

Witness operators can be ordered by the entanglement they detect. A witness is \textit{finer} than another if it detects more entanglement, and is termed \textit{optimal} if no finer witness exists, i.e. if it detects all the entanglement of a given state. Witnesses are usually represented by a combination of observables, rather than a single one. As the measurement of each component of the witness typically requires a different setting of the laboratory apparatus, much effort is dedicated to finding optimal witnesses that can be measured with a minimal number of device settings. This problem is sometimes referred to as `finding the optimal decomposition' \cite{guhne02,guhne03}. Constructing a witness capable of detecting the entanglement of an arbitrary state is generally an arduous task. However, the problem simplifies greatly if one has some \textit{a priori} knowledge of the structure of the target state. In this case, it is possible to define witnesses that are measurable with as few as three device settings \cite{guhne02}.

In practise, the spectrum of measurable witnesses for a given system is restricted by experimental constraints. However, witnesses find application as theoretical tools also, because for any valid witness $W$, condition (\ref{witness_ent_rev}) provides a separability criterion. Conditions such as (\ref{witness_ent_rev}) are known as \textit{linear} separability criteria, because only the first power of the expectation value of the witness appears. \textit{Non-linear} criteria feature higher powers of $\langle W\rangle$. Typical examples involve the uncertainties associated with measurements of local variables, expressed in terms of the variance, which is a quadratic function of the expectation value \cite{giovannetti03,hofmann03}.

\section{Entanglement Generation}
A vast proportion of all research in quantum computation and communication relates to entanglement generation and control, with a view to
find experimentally feasible ways
of producing robust entanglement on demand. Nearby qubits are best entangled via two-qubit gates; however, gating operations become inefficient on distant qubits, whose direct coupling is
usually extremely weak. This problem represents one of the main obstacles to upscaling
current devices, and has therefore been looked into by several
authors. Many existing proposals for creating long distance
entanglement rely on single- or two-photon interference effects
\cite{cabrillo99,browne03,feng03,plenio03,simon03,matsukevich06}. In
these types of schemes, distant qubits are entangled via the detection of photons
resulting from their radiative decay\footnote{The entanglement arises because it is impossible to trace the provenance of the individual
photons, hence the qubits are left in a superposition of ground and excited state.}. Recently, however, the focus seems to have shifted somewhat
to the alternative scenario of employing \emph{entanglement
mediators}. These are ancillary qubits, made to interact successively with two non-communicating qubits A and B; the entanglement between A and B arises when measurements on these ancillae are made.
Various ways to implement this idea have been proposed
\cite{compagno04,yuasa05,yuasa06,costa06}. In \cite{compagno04}, for example, the authors describe an experiment to entangle two optical cavities by repeated interaction with a mediator. The success of the scheme is shown to depend on the interaction times and
strengths, and on the measurement of the mediator. Yuasa
and Nakazato \cite{yuasa05} build on this scheme to demonstrate that
maximally entangled states can be created by three successive
interactions of identical mediators with the cavities, provided one
assumes rotating wave couplings of a specific strength, acting for
specific times. In later work, the same authors find that repeated
resonant scattering of a mediator also generates maximal
entanglement, particularly if the momentum of the mediator can be
fixed to a desired value \cite{yuasa06}. Costa \textit{et al.} have
analyzed the more complex scenario of an injected electron being
spatially scattered by two spin-$\frac{1}{2}$ magnetic impurities in
a one-dimensional, non-magnetic, metallic chain. In this case, it
was found that the electron could almost perfectly entangle the
impurities, conditional to a favourable (and reasonably probable)
outcome of a measurement on the electron's spin.

Very recently, some authors have been investigating the issue of
entanglement on a macroscopic scale
\cite{dechiara06,cunha06,cavalcanti06}, with a view to underline
that:
\begin{quote}
... entanglement is not an artificial mathematical property but [..]
can be extracted and therefore used for quantum information
processing in the same way as heat can be extracted and used for
work in thermodynamics \cite{dechiara06}.
\end{quote}
The situation described in \cite{dechiara06} is that of two probe
particles simultaneously scattering from two entangled spins in a
solid. It is shown that the amount of entanglement extractable from
the solid depends mainly on the interaction Hamiltonian and the
joint state of the spins.

Another interesting proposal is that described by Cunha and Vedral
\cite{cunha06}. This builds on previous work by the same authors, in
which the Pauli principle is invoked as a means to create
entanglement between fermions in a degenerate Fermi gas. Because
fermions only have spatial and spin degrees of freedom, it is argued
that if two fermions were simultaneously detected at the same
position, their spin degrees of freedom would necessarily be
entangled. The authors show this entanglement persists provided the
fermions remain within a certain distance of each other. The
\emph{entanglement distance} is found to be inversely proportional
to the Fermi wavevector $k_F$, which imposes constraints on the type
of fermion one could use for an experimental verification of this
scheme. The setup described involves a vessel
containing ultracold neutrons, with a small hole. The authors
suggest that provided certain precautions are observed in preparing
the Fermi gas, the neutrons emitted from this hole within a given
time interval of each other will be entangled. The scheme is simple and efficient, however an experimental realization would be extremely challenging, owing to the difficulty of preparing a Fermi gas of neutrons in a laboratory environment.

The Pauli principle is used as a means to create entanglement also
in \cite{cavalcanti06}. Here, the authors consider two electrons
with opposite momenta and spin, trapped in a potential well inside
an optical cavity. Two independent photons with orthogonal polarizations are fired into the cavity, creating two excited electrons with different spin and momenta. The electrons are made to interact with some
external energy absorbing reservoir, which ``drags'' them into the
same zero-momentum state, causing their spins to become entangled.
The entangled electrons then decay according to
well-defined selection rules, emitting spin-polarized entangled photons. Interestingly, the dissipative interaction with the environment is essential to the success of the scheme, because the photons become entangled only if the electrons' momenta are equal. This is ensured by the dissipative interaction with the environment, hence decoherence is quite surprisingly viewed as a `pro' rather than a `con'.

\section{Experimental Achievements}
The study of entanglement was in origin a largely theoretical enterprise. The advancement of experimental prowess in the past few years now means the scientific community is poised to mount a two-pronged attack on the subject: theory tells us what we \emph{should} see, experiment tells us what we \emph{do} see. Attempting to reconcile the two produces a positive feedback, whence an understanding of the relationship between the microscopic and macroscopic world is born.

Recent experimental studies of entanglement generation have pursued two main objectives. On the one hand, great effort has been dedicated to the realization of many-particle entangled states. These are essential to measurement-based quantum computation with graph or cluster states \cite{cluster}, which provides an alternative to the more widespread gate model. Achievements in this area include the realization of four- to eight-ion W-states \cite{haffner05}, four-ion \cite{sackett00}, five-photon \cite{zhao04} and three-atom \cite{rauschenbeutel00} GHZ states .

Much research is also directed at sourcing novel systems to entangle. Indeed, entangled particles are not only at the heart of quantum computation and communication, but also find application in more fundamental areas, such as investigating the non-locality and non-contextuality of quantum mechanics. Photons, ions, atoms, quantum dots \cite{bayer01} and superconducting qubits \cite{steffen06} are all demonstrably feasible possibilities, but neutrons, for example, have never been entangled \cite{hasegawa03,sponar08} - an issue this thesis will attempt to address. An area of rapidly growing interest is that of single-particle entanglement \cite{tan91}, which involves entangling distinct degrees of freedom of the same particle. Systems of this type clearly cannot exhibit non-locality, though non-contextual behaviour has been observed for both single neutrons and single photons \cite{hasegawa03,kim03}. One of the most impressive achievements to date involves the creation of a three-way entangled $GHZ$-type state of a single neutron, the three `qubits' being the spatial and spin parts of the neutron's wavefunction, and its energy \cite{sponar08}.

\section{Conclusions}
The study of entanglement occupies a prime position in the field of quantum information processing. From a practical viewpoint, it is clear that generating and preserving entanglement in a controllable fashion are paramount to building quantum computers that rival existing classical machines. In addition, entanglement is a probe of nature, given its potential to reveal whether the world around us has objective reality or whether, to paraphrase slightly, reality is in the eye of the beholder. Finally, entanglement is a probe of `quantumness', because it is destroyed by contact with the macroscopic. I believe the practical and fundamental aspects are closely interwoven: to use a resource to its full potential, one must first understand its properties. However, if cracking an RSA-encrypted message and understanding `life, the universe and everything' truly are two sides of the same coin, I'd place strong odds on achieving the former before the latter.

%% file: neutron_proposal2.tex
\chapter{A Novel Proposal for Neutron Entanglement}\label{neutron proposal}
\textit{In this chapter, I put forward a proposal for a scattering experiment aimed at achieving entanglement between two distinct neutrons. This has never been accomplished to date; indeed, neutron entanglement has only been observed between distinct degrees of freedom of single particles. The realization of entangled states of two neutrons is therefore a highly desirable objective, and would open up the possibility of testing the non-local nature of quantum mechanics with an as yet untapped resource. After a general overview of neutron scattering as an experimental technique, I will describe the protocol, and discuss the assumptions and approximations underlying it. In the concluding part of the chapter, I outline the mathematical formalism employed to model the experiment, and comment on its relationship to the traditional approach to scattering problems, which will be treated in detail in the next chapter.}

\section{Neutrons and Scattering}\footnote{The majority of this material was obtained from \cite{brukner07}.}
A great deal of our current understanding of the structure and dynamical properties of matter comes from neutron scattering. This technique, originally developed to further the study of condensed matter physics and crystallography, now finds application in the most diverse areas of science, including biology, nanotechnology and planetary science.

Neutrons are chargeless particles which, together with protons, form atomic nuclei. The standard model of particle physics classifies them as baryons, the subclass of hadrons composed of three quarks - two down and one up in this specific case. Free neutrons are unstable, and have a typical lifetime of the order of 15 minutes, after which they decay into a proton, an electron and a neutrino. Scattering events are usually concluded well within this time frame. Owing to the absence of charge, neutrons are largely insensitive to stray electric fields and surface defects, and do not themselves create fields capable of disturbing the environment of the scatterer. Consequently, neutrons are excellent non-destructive probes both of surface and bulk properties of matter, and can be used to study samples immersed in complex environments \cite{zaliznyak05}.

Typical scattering experiments are aimed at the characterization of a sample. For reasons that go back to wave-particle duality, this is achieved by analyzing the momentum distribution of the outgoing neutrons. Indeed, the neutron scatters as a wave but is detected as a particle. The De Broglie wavelength of a neutron is determined by its energy; so-called thermal neutrons have wavelengths comparable to interatomic spacings, and can therefore penetrate bulk materials. The energy of a thermal neutron is of order $\mathcal{O}\left(10^{-3}\right)$ eV, which is similar to the energy scales of elementary acoustic or magnetic excitations (phonons and magnons). Momentum changes of scattered neutrons can then be used to map dispersion curves for lattice vibrations and spin fluctuations. Elastic scattering, on the other hand, provides information on the structural properties of the sample, and, on a more fundamental level, on the optical potential that couples nucleons to nuclei and to each other. Finally, if the resolution of the experiment is sufficient to distinguish elastic and inelastic processes, scattering can also identify whether or not the sample is entangled \cite{cowley03}.

\begin{figure}[H]
\renewcommand{\captionfont}{\footnotesize}
\renewcommand{\captionlabelfont}{}
\begin{center}
 \includegraphics[width=15cm]{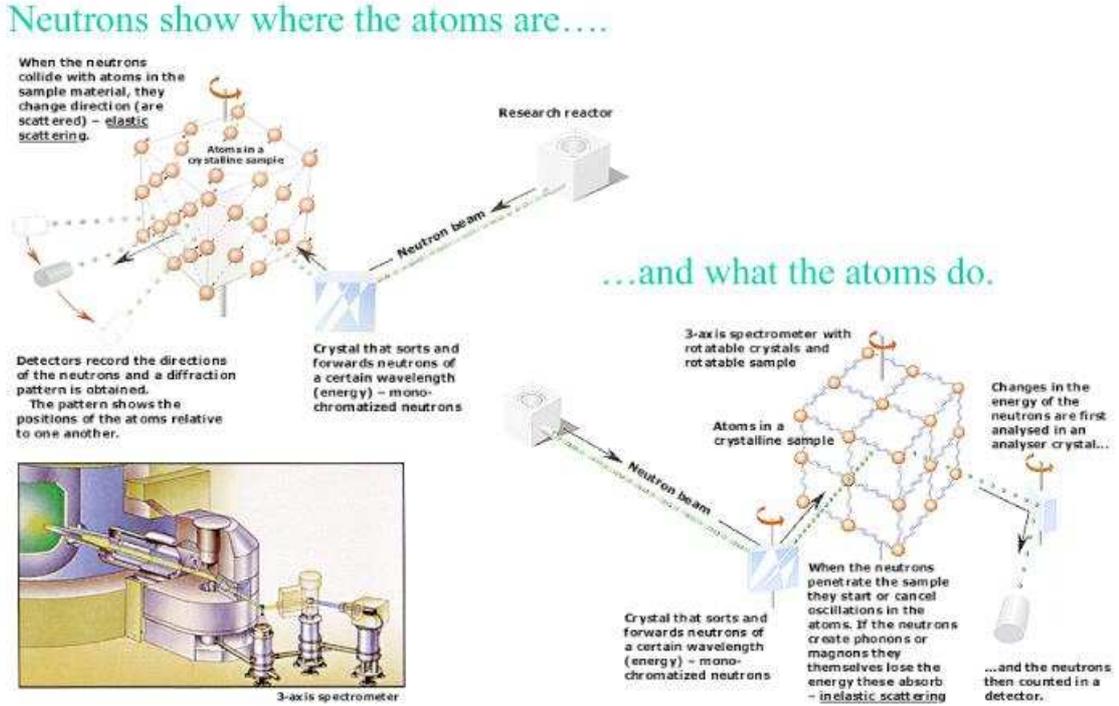}
 \end{center}
 \caption{A visual description of neutron scattering as an experimental technique. Image reproduced with kind permission of Professor Sunil K. Sinha, University of California San Diego and Los Alamos National Laboratory.} \label{scatt1}
\end{figure}

The interaction between the neutron and the sample is determined mainly by two mechanisms: the strong interaction, which couples the neutrons to the nuclei, and the magnetic dipole interaction, which couples the neutrons to the sample's magnetic moment \cite{squires78}. The latter mechanism is owed to the neutron's non-zero magnetic moment, given by
\begin{equation}
\mu=g_N\mu_N\frac{\bm{s}}{\hbar},
\end{equation}
where $g_N= -1.913$ is the neutron gyromagnetic ratio, $\bm{s}$ is the neutron spin operator and $\mu_N$ is the nuclear magneton, whose magnitude is approximately 1000 times smaller than the Bohr magneton. The potential experienced by a neutron during a scattering event is then a combination of a short-range interaction, arising from the nuclei, and a long-range potential, arising from the sample's unpaired electrons. Both interactions are typically rather weak, hence it is common in mathematical models of experimental data to treat scattering as a first order process \cite{lefmann06}.

Neutron scattering is unfortunately a costly technique, which must be carried out in apposite facilities. These come under two categories: fission research reactors, and spallation sources. Fission reactors, also known as continuous sources, produce a steady stream of neutrons via a controlled chain reaction. When heavy elements such as uranium or plutonium absorb a slow neutron, the nucleus splits into two lighter components, releasing two or three fast neutrons in the process. These are referred to as \emph{prompt neutrons}, because they are the first products of the reaction. They are subsequently followed by \emph{delayed} neutrons, produced by the decay of the fission fragments. Each of these neutrons can trigger a new fission event, giving rise to a chain reaction. This is kept in check by removing excess neutrons, which are slowed by passage through a \emph{moderator} and absorbed by \emph{control rods}. The moderator is a large volume of material such as water or solid graphite, that absorbs weakly but scatters strongly. The rods, on the other hand, are typically boron-rich, as boron has a high neutron capture cross section.

It is sometimes desirable to produce pulses of neutrons rather than a continuous stream. This offers a means to distinguish the processes contributing to the scattering event, which do not typically occur on the same timescales\footnote{Inelastic and fast-neutron processes occur shortly after the pulse, whereas neutron capture processes follow some time after.}. Fission reactors can be made to operate as pulsed sources with the help of a \emph{chopper}. Essentially, this is a shutter positioned downstream from the source, which is opened and closed periodically so that only neutrons with well-defined velocities pass through. Pulsed beams can also be produced by \emph{spallation}, which involves bombarding a heavy nucleus with a high-energy proton beam, usually produced in a particle accelerator. As a result of the collision, several neutrons are knocked out of the atom, while others boil off as the nucleus heats up. Spallation typically produces 20 or 30 neutrons per collision, but requires more energy than it delivers, therefore chain reactions cannot occur.

Once the neutrons have been produced, they are typically directed away from the source by a neutron guide. This is a channel with high reflectivity at small incidence angles, often curved to reduce the line of sight from the source to the sample, hence avoiding contamination of the scattered signal by stray high-energy neutrons. Upon emerging from the guide, the neutrons often undergo some kind of velocity selection, after which their fate is determined by the purpose of the experiment.

Finally, once scattering has taken place, the outgoing neutrons are detected. This is not a trivial task, because most currently available detection techniques are designed for charged particles. High-efficiency detection can be achieved using hydrogen-rich scintillators, because hydrogen is a strong scatterer. Otherwise, detection occurs via a two-step process, typically involving a reaction that produces charged by-products \cite{lefmann06,keimer04}.

The field of neutron scattering is incredibly vast; the above is intended to provide only the most basic overview. Where necessary, I will go into more detail with regard to instrumentation and experimental techniques. The formal description of neutron scattering will be the subject of a separate discussion.

\section{Notation}
Having reached this point, it is convenient to define some `shorthand' notation, which will be employed throughout this thesis. Unless otherwise specified, quantum states will be expressed in accordance with the Bloch sphere convention, which labels the spin-up and down eigenstates of $\sigma_z$ with the kets $|0\rangle$ and $|1\rangle$, respectively. The wavefunction of a many-body system fully polarized in the positive (negative) $\bm{\hat{z}}$ direction will be represented as $|\bm{0}\rangle$ ($|\bm{1}\rangle$). Depending on the magnitude and orientation of an external magnetic field along $\bm{\hat{z}}$, either $|\bm{0}\rangle$ or $|\bm{1}\rangle$ may correspond to the system's ground state. Excitations from the ground state are obtained by rotating individual spins through an angle $\pi$. A particle whose spin is anti-aligned with the ground state configuration will be referred to as a `spin flip'. A fully aligned state with a single spin flip at site $j$ will be represented as $|\mathbi{j}\rangle$; a fully aligned state with spin flips at sites $j$ and $l$ will be represented as $|\mathbi{jl}\rangle$, and so on.

For reasons of mathematical simplicity, the majority of the analysis carried out in this thesis is concerned with pure states. Often, these will be described in terms of a density operator, conventionally labeled $\rho$. For a state vector $|\Psi\rangle$, this takes the form
\begin{equation}
\rho=|\Psi\rangle\langle\Psi|.
\end{equation}
If $\rho$ represents a many-particle system, the density operator of a subsystem is obtained by performing a partial trace. For a system of two qubits $A$ and $B$, one then has
\begin{equation}
\rho_{A\:(B)}=\mathrm{Tr}_{A\:(B)}\left[\rho_{AB}\right],
\end{equation}
where $\rho_{AB}$ represents the composite density matrix.

Variables and operators will usually be assigned a set of qualifying indices; however, these will be omitted whenever the context allows, to make for more compact notation.

\section{Approximations}\label{approximations}
To provide a more complete picture of the scope and interpretation of this work, it is useful to head the description of the protocol with an explanation of the approximations that underlie it. These have been chosen according to the twin criteria of analytical computability and experimental feasibility, and much care has been taken to strike a balance between the two, as the ultimate goal is to design a feasible experiment.

As regards the operating conditions, it will be assumed the sample is at zero temperature and completely isolated from surrounding noise. This removes the risk of thermally induced spin flips or other forms of decoherence. The sample and the neutrons are immersed in a static, uniform magnetic field aligned with the $\mathbf{\hat{z}}$ direction. This is chosen as the quantization axis. The effect of the field on the neutrons will be neglected, as the neutron magnetic moment is $\mathcal{O}(10^3)$ times smaller than the electron magnetic moment. The neutrons themselves are produced by an ultra-cold source capable of imparting an arbitrary, user-defined polarization. Neutron velocities are chosen not to exceed $10$ ms$^{-1}$, with corresponding energies of order $10^{-7}$ eV. Sources of this type have been in operation for some years now, an example being the PF2 source at the Institute Laue-Langevin in Grenoble (see Appendix \ref{app_b} and \cite{steyerl86}). It is assumed the intensity of the source is low, so that very few neutrons interact with the sample at any chosen moment. This should not be a problem, as scattering facilities go to great lengths to \emph{avoid} producing dilute beams. High-intensity sources could also be adapted to serve our purpose, provided they were operated in pulsed mode - indeed, when a continuous beam is chopped, a large proportion of the flux is lost. In addition, neutrons are weakly interacting particles with an extremely low scattering cross section; therefore, the probability of many neutrons scattering at once is small.

The sample from which the neutrons scatter is modeled as a magnetically ordered crystal, in which the unpaired electrons are localized near the equilibrium positions of the ions in the lattice. This is known as the Heitler-London model, and provides a good description of ferromagnetic insulators such as EuO, $\mathrm{LaMnO_{3.12}}$ and few other materials \cite{vanvleck61,lopes06,wei08}. Magnetic scattering is provided by the long range dipolar force between the neutron and the unpaired electrons, while nuclear scattering is provided by the short-range strong force. I assume $LS$ coupling, i.e. that orbital and spin angular momenta of the individual ions can be summed to yield total orbital and spin angular momenta, defined by the quantum numbers $L$ and $S$. I work in the limit of $L=0$; this could describe a lattice of s-shell electrons, and also a more general situation in which the resultant orbital angular momentum is either zero, or quenched by the internal electric field of the crystal. Magnetic scattering of the neutrons will be teated as a purely electronic effect; the nuclear spins will be completely neglected. This is reasonable, as the strength of magnetic scattering from nuclear dipole moments is approximately two orders of magnitude weaker than that from electronic dipole moments \cite{sears86}.

The electrons in the sample are coupled by a translationally invariant potential which conserves the total spin quantum number $S_z$. A reasonable example of such a potential is Coulomb nearest-neighbour exchange, given by the Heisenberg Hamiltonian
\begin{equation}\label{h0}
H_0=-\sum_{\langle ij \rangle}J_{ij}\bm{\sigma^i \cdot \sigma^j},
\end{equation}
where the notation $\langle ij \rangle$ indicates the sum over all nearest-neighbouring pairs, and $J_{ij}$ represents the coupling strength between nearest neighbours. For simplicity, I will take this to be constant and positive, so as to describe a ferromagnetic sample. The full Hamiltonian of the system then reads
\begin{equation}\label{h0 tot}
H_0=-J\sum_{\langle ij \rangle}\bm{\sigma^i \cdot \sigma^j}+B_z\sum_{j=1}^N\mathbf{\sigma}_z^j,
\end{equation}
where $B_z>0$ is the strength of the external field and $N$ is the number of spins in the sample. Each spin is represented by the operator $\bm{\sigma}^j=\left(\sigma_x^j,\sigma_y^j,\sigma_z^j\right)$,
where the $\sigma_{\alpha}^j$ are the Pauli matrices for
spin $j$.

\begin{figure}[H]
\renewcommand{\captionfont}{\footnotesize}
\renewcommand{\captionlabelfont}{}
\begin{center}
 \includegraphics[width=8cm]{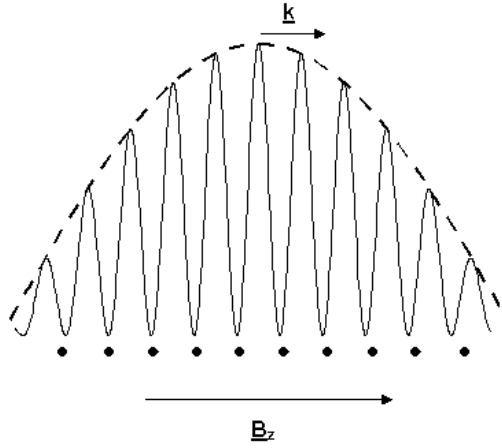}
 \end{center}
 \caption{A one-dimensional snapshot of the scattering geometry. The sample sees the neutron as a wavepacket, the peak of which travels with momentum $\mathbf{k}_0$.} \label{scatt geom}
\end{figure}

The sample itself is a macroscopic, periodic system experiencing negligible boundary effects. Each neutron scatters from the sample as a wavepacket of spatial width $\delta_x$ and momentum width $\delta_p$, whose relationship is determined by the uncertainty principle
\begin{equation}\label{u princ}
\delta_x\delta_p\geq\frac{\hbar}{2}.
\end{equation}
It will be assumed $\delta_x$ is comparable to the size of the sample, or, equivalently, that the neutron interacts with all the sample spins at once (figure \ref{scatt geom}). The validity of this assumption will be discussed in section \ref{exp feasibility}.

The scattering events are described as sequential, time-dependent interactions with the potential occurring such that no direct interaction between the neutrons takes place at any time. Only neutrons scattered into a specific Bragg peak will be detected. Upon emerging from the source, the neutrons are uncorrelated and polarized in the same (arbitrary) direction, specified by the coefficients $\alpha$ and $\beta$, which can take on any real or complex value satisfying $|\alpha|^2+|\beta|^2=1$. To begin with, I will consider a simplified model in which only two neutrons scatter from the sample. The initial state of the system as a whole is then a product of the states of the neutrons and the sample. It is convenient to represent this as
\begin{equation}\label{psi0}
|\psi_0\rangle=\left(\alpha |0\rangle+\beta |1\rangle\right)_2\:\left(\alpha |0\rangle+\beta |1\rangle\right)_1\:|\psi^s\rangle,
\end{equation}
where $|\psi^n\rangle_m=\left(\alpha |0\rangle+\beta |1\rangle\right)_m$ is the state of neutron $m$, and $s$ stands for `sample'. Hereafter, I will refer to the system of neutrons plus sample simply as `the system'. With these details in place, I now describe the entanglement scheme.

\section{The Protocol}\label{the protocol}
The protocol can be thought of in terms of four phases:
\begin{enumerate}
\item The first period of free evolution of the sample, before the arrival of the first neutron pulse;
\item The interaction between the first neutron and the sample;
\item The second period of free evolution of the sample, before the arrival of the second neutron pulse;
\item The interaction between the second neutron and the sample.
\end{enumerate}
From a qualitative viewpoint, the first scattering event has the effect of depositing the `spin information' carried by the first neutron in the sample. As we will see in chapter \ref{two ns}, this signature can remain intact during the second period of free evolution. The sample therefore behaves as an `entanglement safety-deposit box', which retains the information deposited by the first neutron until the second neutron comes to claim it. However, the entanglement between the neutrons is conditional on the second scattering event being able to `undo' -in part or in full- the transformation on the sample spin precipitated by the the arrival of the first neutron (see section \ref{log negativity}).

The protocol begins by initializing the sample and the neutrons to a known state, chosen such that $|\psi_0\rangle$ contains at most two spin flips. The motivation for this restriction is mathematical rather than physical, and stems from the hope of attaining an analytically solvable model by reducing the size of the working Hilbert space (see section \ref{symm and op rep}). Two possible combinations of $|\psi^n\rangle$ and $|\psi^s\rangle$ are considered here, and I label these $|\psi_0^A\rangle$ and $|\psi_0^B\rangle$. In state $|\psi_0^A\rangle$, the neutrons are polarized parallel to the positive $\hat{\bm{z}}$-axis and the sample is prepared in a single spin-flip state. In $|\psi_0^B\rangle$, the neutrons have arbitrary polarization while the sample is prepared in a fully aligned state. Hence
\begin{eqnarray}
|\psi_0^A\rangle&=&|0\rangle_2|0\rangle_1\left[\frac{1}{\sqrt{N}}\sum_{j=1}^N|\mathbi{j}\:\rangle\right],\label{psi0a}\\
|\psi_0^B\rangle&=&\left(\alpha |0\rangle+\beta |1\rangle\right)_2\:\left(\alpha |0\rangle+\beta |1\rangle\right)_1\:|\mathbf{0}\rangle\label{psi0b}.
\end{eqnarray}
From equations (\ref{psi0}) and (\ref{psi0a}), for $|\psi_0^A\rangle$ we have
\begin{eqnarray}
\alpha&=&1,\\
\beta&=&0,\\
|\psi^n\rangle_1&=&|\psi^n\rangle_2=|0\rangle,\\
|\psi^s\rangle&=&\frac{1}{\sqrt{N}}\sum_{j=1}^N|\mathbi{j}\:\rangle\label{psiAs}.
\end{eqnarray}
The state $|\psi_0^A\rangle$ contains a single spin flip, which is localized on the sample. This is the only realistic means to create a single spin flip state, as the alternative would involve polarizing the neutrons in opposite directions while taking $|\psi^s\rangle=|\mathbf{0}\rangle$. While the latter is certainly feasible, the former is decidedly not. The form of $|\psi^s\rangle$ is familiar from linear spin wave theory \cite{vanvleck58}; indeed, equation (\ref{psiAs}) represents the zero-momentum single magnon eigenstate of the Hamiltonian (\ref{h0 tot}). In zero field, this is degenerate with the ground state, but a finite field breaks the degeneracy.

Now, comparing equations (\ref{psi0}) and (\ref{psi0b}) shows that, for $|\psi_0^B\rangle$
\begin{eqnarray}
|\psi^n\rangle&=&\alpha |0\rangle+\beta |1\rangle,\\
|\psi^s\rangle&=&|\mathbf{0}\:\rangle\label{psiBs}.
\end{eqnarray}
In its most general form, this state contains zero, one and two-spin-flip components. Both spin flips are localized on the neutrons, while the sample is in a fully polarized state. In zero field, this is one of the degenerate ground states of the system. In a finite field, it's energy is determined by the relative magnitude of $J$ and $B_z$.

In the time interval between state initialization and the arrival of the first pulse, the sample undergoes free evolution under the effect of (\ref{h0 tot}). When the pulse reaches the sample, all neutrons but one pass through without scattering. The neutron in question remains coupled to the sample for a time $\tau$, and the interaction generates entanglement between the neutron and the sample\footnote{The mechanism whereby this occurs will be explained in more detail in chapter \ref{two ns}.}. When the neutron departs, the sample undergoes a second period of free evolution, which ends with the arrival of a second pulse. The second scattering event is identical to the first, and has the effect of redistributing the entanglement in the system. As a result, part of the entanglement between the first neutron and the sample is swapped to the second neutron.

The next step involves establishing whether the two scattered neutrons have acquired a correlation as a result of the scattering process.
This can be achieved by measuring a witness operator \textit{W}, as defined in chapter \ref{ent review}. Let us call the density operator of the scattered neutrons $\rho_n$; if this state is entangled, the eigenspectrum of its partial transpose should contain a negative eigenvalue $\lambda_-$, corresponding to eigenvector $|e^-\rangle$. The witness operator is then defined by
\begin{equation}\label{witness}
W=|e^-\rangle\langle e^-|,
\end{equation}
such that $\mathrm{Tr}\left[W^{T_A}\rho_n\right]=\mathrm{Tr}\left[W\rho_n^{T_A}\right]<0$. It will be shown in chapter \ref{two ns} that, when the protocol has the desired outcome, this witness can be measured with as few as three device settings.

The protocol is deemed successful if the entanglement measured by the witness exceeds some threshold value $\mathcal{E}$. In general, this depends on two
factors: (i) creating at least $\mathcal{E}$ units of entanglement between the first neutron
and the sample during the first scattering event, and (ii) swapping the state of the second neutron and the state of the sample during the second scattering event. Point (i) follows from theorems on entanglement monogamy
\cite{wootters00}, according to which the total entanglement
of the first neutron with the rest of the system must remain
constant throughout the protocol. Therefore, if the two neutrons are found to share
\textit{X} ebits of entanglement after the second scattering
event, the first neutron must have shared that same amount of
entanglement with the sample prior to this event. Point (ii) implies the second scattering event must change the distribution of entanglement within the
system. In an ideal scenario, the second neutron and the sample would experience a complete role reversal; in practise only a partial swap takes place, because it is not possible to tune neutron interaction times separately (see section \ref{haroche}). Therefore, the entanglement between the neutrons is typically less than \textit{X}, owing both to the residual correlation between the first neutron and the sample, and to the creation of correlations between the second neutron and the sample.

\section{The Hamiltonian}\label{neutron h}
During each phase of the protocol described in section \ref{the protocol}, the evolution of the system as a whole is governed by a specific Hamiltonian. In phases 1 and 3, this is just the free Hamiltonian of the sample, given in equation (\ref{h0 tot}). Phases 2 and 4, on the other hand, require a Hamiltonian that describes both the intrinsic dynamics of the sample and its interaction with the incoming neutron. If the initial state of the system is fully separable, the Hamiltonians of phases 2 and 4 must not commute, else no correlation can be established between the scattered neutrons \cite{brennen02}. Therefore, the sample needs to `know' which neutron arrived first; in other words, the sample must have some memory.

Let us label the Hamiltonians associated with phases 2 and 4 with $\mathcal{H}_1$ and $\mathcal{H}_2$, to indicate the neutron involved in that particular phase. Then
\begin{eqnarray}
\mathcal{H}_1=\left(H_0+\mathcal{V}_1\right)\theta_D \left(\mathbf{r}\right),\label{hcal1}\\
\mathcal{H}_2=\left(H_0+\mathcal{V}_2\right)\theta_D \left(\mathbf{r}\right),
\end{eqnarray}
where $\mathcal{V}_1$ and $\mathcal{V}_2$ represent the coupling between the neutron and the sample, and $\theta_D(\mathbf{r})$ is a heaviside function defined to be unity inside the sample and zero elsewhere. This accounts for the finite size of the sample. The dominant terms in the potential are the short-range, strong interaction with the nuclei, and the long-range magnetic dipole interaction with the unpaired electrons. The potential experienced by neutron $m$ can then be written as
\begin{equation}\label{vm}
\mathcal{V}_m=\theta_D \left(\mathbf{r}\right)\left[\sum_{j=1}^{\infty}\frac{2\pi\hbar^2b_j}{m}\:\delta\left(\mathbf{r}-\mathbf{r}_j\right)-\bm{\mu}_m\cdot\mathbf{B}(\mathbf{r})\right],
\end{equation}
The first term represents the point-like, spin-independent coupling of the neutron to the nucleus at position $\mathbf{r}_j$, often referred to as the \emph{Fermi pseudopotential}, and the second term is the spin-dependent interaction of the neutron's dipole moment $\bm{\mu}_m$ with the magnetic field created by the sample electrons. The quantity $b_j$ is known as the \emph{scattering length}. This measures the strength of nuclear scattering, and must be determined empirically for each nuclear species. For the remaining terms, one has
\begin{eqnarray}
\bm{\mu}_m&=&g_N\mu_N\frac{\bm{s}_m}{\hbar},\\
\bm{B}&=& \frac{g_e\mu_0\mu_B}{4\pi\hbar}\sum_{j=1}^{\infty}\nabla \times\left(\frac{\bm{s}_j\times\hat{\mathbf{R}_j}}{R_j^2}\right)\label{b field},
\end{eqnarray}
where $\bm{s}_m$ is the spin of neutron $m$, $\bm{s}_j$ is the spin of the electron at position $\mathbf{r}_j$, and $\mathbf{R}_j=\mathbf{r}-\mathbf{r}_j$ is the displacement of the neutron from this electron \cite{squires78}.

The state of the incoming neutron can be represented as a plane wave with momentum $\mathbf{k}$ carrying spin $\bm{\sigma}_m$
\begin{equation}\label{neutron wave}
|\psi^n_k\rangle=|k\sigma_m\rangle\equiv\frac{1}{\sqrt{L^3}}\:e^{i\mathbf{k}\cdot\mathbf{r}}|\psi^n\rangle,
\end{equation}
where $|\psi^n\rangle$ is the spin state of the neutron, as defined in the previous section, and $L$ is an arbitrary normalization length. As we only consider neutrons scattered into a specific Bragg peak, it is convenient to integrate the potential over the spatial coordinates of the neutron. This gives a spin-dependent spatial matrix element of the potential, $\mathcal{V}_m^s$. Note this does \emph{not} in itself represent the action of the potential on the spin state of the system. The true form of the spin potential warrants a separate discussion, and will be derived in the following chapter.

Neglecting for the moment the subscript $m$, for a cubic sample of side $D$ centred at the origin one finds \cite{squires78,fisher93}
\begin{align}
\mathcal{V}^s&=\frac{1}{L^3}\int_{all\:space} e^{-i\mathbf{k}_f\cdot\mathbf{r}}\mathcal{V}e^{i\mathbf{k}_i\cdot\mathbf{r}}\mathrm{d}\mathbf{r}\nonumber\\
&=\frac{D^3}{L^3}\left[\frac{2\pi\hbar^2}{m}\sum_{j=1}^{N}b_j e^{i\mathbf{Q}\cdot\mathbf{r}_j}+\Lambda\:\sum_{j=1}^{N}\bm{s}
\cdot\lbrack \mathbf{\hat{Q}}\times \left(\bm{s}_j\times
\mathbf{\hat{Q}}\right)\rbrack \:e^{i \mathbf{Q}\cdot\mathbf{r}_j}\right]\star \mathrm{Sa}\left[\frac{\mathbf{Q}\cdot\mathbf{D}}{2}\right]\nonumber\\
&=\frac{D^3}{L^3}\left[\frac{2\pi\hbar^2}{m}\sum_{j=1}^{N}b_j +\Lambda\:\sum_{j=1}^{N}\bm{s}
\cdot\lbrack \mathbf{\hat{Q}}\times \left(\bm{s}_j\times
\mathbf{\hat{Q}}\right)\rbrack\right]\label{vs} ,
\end{align}
where $\mathbf{\hat{Q}}$ is the direction of the
neutron scattering wavevector $\mathbf{Q}=\mathbf{k}_i-\mathbf{k}_f$, $\mathrm{Sa}\left[\cdot\right]$ is the hyperbolic sine function
\begin{equation}\label{sinc}
\mathrm{Sa}\left[\frac{\mathbf{Q}\cdot\mathbf{D}}{2}\right]=\mathrm{sinc}\left[\frac{Q_xD}{2}\right]\mathrm{sinc}\left[\frac{Q_yD}{2}\right]\mathrm{sinc}\left[\frac{Q_zD}{2}\right],
\end{equation}
and $\Lambda$ is related to the magnetic coupling strength, whose precise value will be discussed in the following chapter. The last step of equation (\ref{vs}) follows from the Bragg condition, which states that peaks in the intensity of the scattered beam are obtained when the momentum transfer is a reciprocal lattice vector.

If the sample is composed of a single nuclear species, the short-range term in the scattering potential can be approximated by a constant. Therefore, $\mathcal{V}^s$ depends only on the direction of the momentum transfer. Expanding the cross product in equation (\ref{vs}) shows the component in the direction of $\mathbi{Q}$ must disappear. Assuming for simplicity that $\mathbi{Q}$ is parallel to the $\mathbf{\hat{z}}$-direction, one finds
\begin{equation}\label{h xy1}
\mathcal{V}^{s_{xy}}=\frac{D^3}{L^3}\left[V_0+\Lambda\: \left(s_m^x\sum_{j=1}^Ns_j^x+s_m^y\sum_{j=1}^Ns_j^y\right)\right],
\end{equation}
where $\sum_{j=1}^N s_j^{\chi}$ is the $\chi$-component of the total spin of the sample, and $V_0=\frac{2\pi\hbar^2}{m}\sum_{j=1}^Nb_j$. Note this is simply a spatial matrix element of the total interaction, \emph{not} the spin potential acting on the system at finite momentum transfer.

Special care must be taken if the momentum transfer is zero, because the spin-dependent part of $\mathcal{V}^s$ is undefined. To account for the rapid variation of $\mathbi{Q}$ in the vicinity of $\mathbi{Q}=0$, one can replace the double cross product in equation (\ref{vs}) with its average value over a small sphere \textsl{S} of radius $\epsilon \ll r^{-1}_j$, centred on $\mathbi{Q}=0$
\begin{equation}
\langle\mathbf{\hat{Q}}\times \left(\bm{s}_j\times
\mathbf{\hat{Q}}\right)\rangle=\frac{\int_S \mathbf{\hat{Q}}\times \left(\bm{s}_j\times
\mathbf{\hat{Q}}\right)\,d\mathbf{Q}}{ \int_S\,d\mathbf{Q}}.
\end{equation}
By carrying out the substitutions
\begin{eqnarray}
\mathbf{\hat{Q}}\times \left(\bm{s}_j\times
\mathbf{\hat{Q}}\right)&=&\bm{s}_j-|\bm{s}_j|\cos{\theta}\mathbf{\hat{Q}},\\
d\mathbf{Q}&=&dQ\:\sin{\theta}\:d\theta\: d\phi,
\end{eqnarray}
with $|\mathbf{\hat{Q}}|^2=1$, $Q\in[0,\epsilon]$, $\theta\in[0,\pi]$, and $\phi\in[0,2\pi]$, it can be shown that
\begin{equation}
\langle\mathbf{\hat{Q}}\times \left(\bm{s}_j\times
\mathbf{\hat{Q}}\right)\rangle=\bm{s}_j-\frac{1}{3}|\bm{s}_j|\bm{\hat{z}}=\frac{2}{3}\:\bm{s}_j.
\end{equation}
The matrix element of the potential at $\mathbf{Q}=0$ can therefore be approximated by the expression
\begin{equation}\label{h ex1}
\mathcal{V}^{s_{ex}}=\frac{D^3}{L^3}\left[V_0+\frac{2\Lambda}{3}\:\left(s_m^x\sum_{j=1}^Ns_j^x+s_m^y\sum_{j=1}^Ns_j^y+s_m^z\sum_{j=1}^Ns_j^z\right)\right].
\end{equation}
As we will see (cfr. section \ref{time at 0 mtm trans}), this gives a true spin potential $V^{s_{ex}}$ of the form
\begin{equation}\label{h ex}
V^{s_{ex}}=V_0+\frac{2\Lambda}{3}\:\left(s_m^x\sum_{j=1}^Ns_j^x+s_m^y\sum_{j=1}^Ns_j^y+s_m^z\sum_{j=1}^Ns_j^z\right),
\end{equation}
which represents an exchange interaction between the neutron and the sample. Physically, $\mathbi{Q}=0$ describes forward scattering, i.e. scattering in the direction of the incoming momentum.

Equation \eqref{h ex} represents an interaction capable of creating spin excitations in the sample. This process costs energy, which in zero external field is detracted from the kinetic energy of the neutron. It may then seem counterintuitive that scattering at zero momentum transfer - that is, elastic scattering - creates spin flips. To resolve the controversy, we recall that $V^{s_{ex}}$ is an average over a \emph{range} of $\mathbf{Q}$ around $\mathbf{Q}=0$. The averaging procedure accounts for the fact that a Bragg peak is not infinitely sharp, an effect owed to the finite size of the sample. Averaging over a range of momenta centred at a certain $\mathbf{Q}$ is therefore equivalent to detecting all neutrons scattered into a Bragg peak of finite width about $\mathbf{Q}$. The width of the peak is of order $D^{-1}$, as expressed by the range of the sinc function of equation (\ref{vs}). The variation of the Hamiltonian over this range is negligible, hence the effective Hamiltonian is well-approximated by its value at the centre of the peak.

\section{Quantum State Evolution}
The final point I would like to make regards the method I use to derive the scattered state of the system. As the initial state $|\psi_0\rangle$ is pure, the scattered state can be obtained by straightforward time-evolution using the canonical operator $U(\mathcal{H},t)=\exp{\lbrack-i\mathcal{H}t}\rbrack$. This operator must reflect the four stages of the protocol, hence $U(\mathcal{H},t)$ can be expressed as a product of four terms
\begin{equation}\label{teo}
U(\mathcal{H},t)=U(\mathcal{H}_2,\tau)U(H_0,\tau_f^{\prime})U(\mathcal{H}_1,\tau)U(H_0,\tau_f),
\end{equation}
describing phases 4, 3, 2 and 1, respectively. None of the terms commute (save, trivially, the second and fourth), so $U(\mathcal{H},t)$ truly represents a specific sequence of events. It is furthermore evident that, in zero field, the first and second periods of free evolution only contribute a phase to the scattered state. During phase 1, this is true in finite fields also, because both $|\psi_0^A\rangle$ and $|\psi_0^B\rangle$ are eigenstates of $H_0$. The scattered state $|\psi_f^{A,B}\rangle$ is then, modulo a phase factor
\begin{equation}\label{qtm psi}
|\psi_f^{A,B}\rangle=U(\mathcal{H}_2,\tau)U(H_0,\tau_f^{\prime})U(\mathcal{H}_1,\tau)|\psi_0^{A,B}\rangle.
\end{equation}
From here, the state $\rho_n$ of the neutrons can be obtained by calculating the system density matrix $\rho=|\psi_f^{A,B}\rangle\langle\psi_f^{A,B}|$ and tracing out the sample.

Ultimately, the outcome of the protocol is determined by the neutron polarization $(\alpha,\beta)$ and the interaction time $\tau$. It is therefore essential to relate these parameters to the experimental settings of our equipment. The physical interpretation of  $\alpha$ and $\beta$ is immediate, as they are simply the direct numerical expression of the real property that is the neutron polarization. For $\tau$, on the other hand, the situation is a little more hazy, because the interaction time between the neutron and the sample is not in itself a well-defined quantity. In fact, scattering problems are not typically solved in terms of state vectors. Conventionally, one works with \emph{cross sections}, which represent the probability that a certain scattering event will result in neutrons being detected in a specific region of space, and with a certain polarization. The wavefunctions $\Psi$ of the scattered neutrons satisfy the Lippmann-Schwinger equation \cite{sears89}
\begin{equation}\label{lipp schw eq}
\Psi=\Phi+GV\Psi,
\end{equation}
where $\Phi$ is a solution to the time-independent Schr\"{o}dinger equation, $V$ is the scattering potential and $G$ is known as the \emph{retarded Green function}. To obtain these solutions, one acts on the initial state of the system with an operator known as the $S$-matrix, given by
\begin{equation}
S=\mathbb{I}-2\pi i T \delta\left(E^{\:\prime}-E\right),
\end{equation}
where $T$ is the \emph{transition matrix}, related to the scattering potential as $T=V\left(1-GV\right)^{-1}$.

The $S$-matrix is universally acknowledged as providing the correct description for the observable outcome of a scattering experiment. It follows that any results obtained with a quantum-mechanical treatment should measure up to the predictions of the $S$-matrix method. Comparing the outcomes of the two techniques may therefore provide an indication of the relationship between $\tau$ and real, physical quantities one might be able to measure and tune. I will explore this possibility in the following chapter.

\section{Conclusions}
In this chapter, I have described an original protocol aimed at creating entanglement between two distinct neutrons. The success of the protocol would provide the quantum information processing community with a novel source of entanglement, as well as a playground for the test of quantum effects with neutrons.

The present scheme bears some superficial similarities with a proposal for \emph{entanglement extraction} from a solid, advanced some years ago by De Chiara \textit{et al.} \cite{dechiara06}. As the name suggests, the protocol involves sending two uncorrelated probe neutrons to interact separately with two entangled spins in a bulk solid. It is shown that, if the neutrons and the sample spins are coupled by an interaction similar to (\ref{h ex}), for some values of the interaction time it is possible to transfer the entanglement of the sample spins to the probes. In addition, in the limit of multiple collisions, the entanglement between the probes approaches unity. The authors proceed to describe a more realistic situation, in which each probe neutron, being a wavepacket of finite width, interacts with a subset of spins in the sample, which is prepared in a state analogous to (\ref{psiAs}). In this case, the probes are entangled only if the separate spin subsets are entangled; in other words, the success of the protocol relies on the presence of entanglement between different parts of the sample.

Three main points distinguish this protocol from the scheme I have suggested. First, the interaction of the neutrons with the sample is sequential, meaning the respective interaction Hamiltonians do not commute. This is not the case in \cite{dechiara06}, because the probes interact simultaneously with spatially separated regions of the sample. Second, I assume the incoming neutron couples to \emph{all} the sample spins, rather than a subset. This creates a three-body entangled state of the neutrons and the sample, which for certain values of the interaction time is close to being separable with respect to the neutrons/sample bi-partition. Finally, and most important of all, no prior entanglement is necessary for the protocol to succeed: the scattered neutrons can become entangled even if the sample is prepared in a fully separable state. This allows for greater flexibility in state initialization, and in general renders the protocol a more versatile and realistic alternative to that of \cite{dechiara06}.

%% file: s_matrix1.tex
\chapter{The Formal Theory of Neutron Scattering}\label{s matrix}
\textit{In this chapter, I illustrate the relationship between time-dependent and time-independent treatments of scattering as applied to the entangling protocol discussed previously. The time-independent counterpart of the scattered state $|\psi_f^{A,B}\rangle$ will be calculated for both $\mathbi{Q}=0$ and $\mathbi{Q}\neq 0$ scattering geometries. This form will be compared to an expansion of equation (\ref{qtm psi}) in terms of the acting potentials to yield the physical significance of the time parameter $\tau$ in this context. A review of the traditional, time-independent scattering formalism developed by Heisenberg, often referred to as the $S$-matrix method, can be found in Appendix \ref{app_a} and follows the work of Gell-Mann and Goldberger \cite{gell-mann53}.}

\section{A Preliminary Study}
\begin{figure}[H]
\renewcommand{\captionfont}{\footnotesize}
\renewcommand{\captionlabelfont}{}
\begin{center}
 \includegraphics[width=14cm]{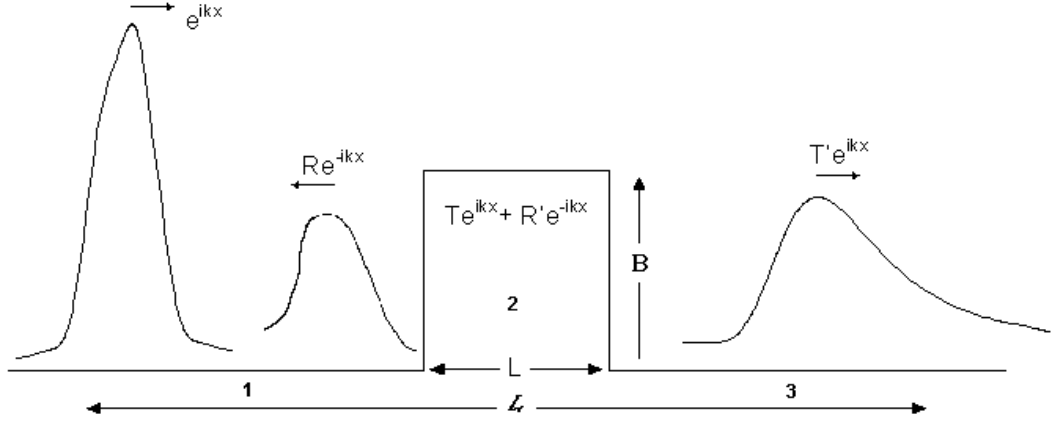}
 \end{center}
 \caption{Transmission and reflection of a wavepacket of incident energy $E$ from a barrier of height $B<E$.} \label{smat1}
\end{figure}

To first order in the scattering potential, the $S$-matrix representing scattering between initial and final eigenstates of the free Hamiltonian of the system is
\begin{equation}\label{1st order sa}
S^{(1)}=\mathbb{I}-2\pi iV\delta\left(E_i-E_j\right),
\end{equation}
where $E_j$ and $E_i$ are the initial and final eigenenergies of the free Hamiltonian, respectively. In the case of forward scattering, the derivation of scattered states according to the $S$-matrix formalism reduces to solving the familiar problem of a particle of energy $E$ traversing a potential barrier of height $V<E$. Let us then consider a one-dimensional Gaussian wavepacket of width $w$, centred on position $x_0$ and momentum $k_0$, traversing a region of uniform magnetic field of strength $B_z$ oriented along $\hat{\bm{z}}$, as shown in figure \ref{smat1}.

At time $t$, the state of the wavepacket at any point along the $\hat{\bm{x}}$ axis is given by
\begin{equation}
|\psi\left(x,x_0,k_0,w,t\right)\rangle=\sum_{\sigma}\sum_k \left(\frac{\pi}{2w^2}\right)^{-\frac{1}{4}}e^{-i\left(k-k_0\right)x_0}e^{-\left(k-k_0\right)^2w^2}e^{-iE_kt/\hbar}\phi_{k,\sigma}(x),
\end{equation}
where the $E_k$ represent the energies of each $k$-component, corresponding to space and spin eigenfunctions $\phi_{k,\sigma}$. The $\phi_{k,\sigma}$ are easily (if somewhat laboriously) derived by expressing the wavefunctions outside and inside the barriers in terms of incident, transmitted and reflected waves, and matching wavefunctions and derivatives at the boundaries. One has
\begin{align}
\phi^1_{k,\sigma}&=\left(e^{ikx}+R_{\sigma}e^{-ikx}\right)|\sigma\rangle,\\
\phi^2_{k,\sigma}&=\left(T_{\sigma}e^{iq_{\sigma}x}+R_{\sigma}^{\prime}e^{-iq_{\sigma}x}\right)|\sigma\rangle,\\
\phi^3_{k,\sigma}&=T_{\sigma}^{\prime}e^{ikx}|\sigma\rangle,
\end{align}
with
\begin{eqnarray}
R_{\sigma}&=&\frac{\left(k^2-q_{\sigma}^2\right)\sin{q_{\sigma}L}}{\left(k^2+q_{\sigma}^2\right)\sin{q_{\sigma}L}+2iq_{\sigma}k},\\
T_{\sigma}&=&-\frac{2k\left(k+q_{\sigma}\right)}{e^{2iq_{\sigma}L}\left(k^2-q_{\sigma}^2\right)-\left(k^2+q_{\sigma}^2\right)},\\
R_{\sigma}^{\prime}&=&\frac{2ke^{2iq_{\sigma}L}\left(k-q_{\sigma}\right)}{e^{2iq_{\sigma}L}\left(k^2-q_{\sigma}^2\right)-\left(k^2+q_{\sigma}^2\right)},\\
T_{\sigma}^{\prime}&=&-\frac{4e^{-iL\left(k-q_{\sigma}\right)}kq_{\sigma}}{e^{2iq_{\sigma}L}\left(k^2-q_{\sigma}^2\right)-\left(k^2+q_{\sigma}^2\right)},
\end{eqnarray}
where $L$ is the extent of the field region, $R_{\sigma}$ and $R_{\sigma}^{\prime}$ are the spin-dependent reflection coefficients in regions 1 and 2, $T_{\sigma}$ and $T_{\sigma}^{\prime}$ are the spin-dependent transmission coefficients in regions 2 and 3, and $q_{\sigma}$ is the spin-dependent wavevector inside the barrier. In natural units, this is given by
\begin{align}
q_{\uparrow}&=\left(k^2-2B_z\right)^{\frac{1}{2}},\label{q up}\\
q_{\downarrow}&=\left(k^2+2B_z\right)^{\frac{1}{2}},\label{q down}
\end{align}
for particles of spin up and down, respectively. Values of $k$ can be fixed by imposing periodic boundary conditions over some length $\mathcal{L}\gg L$. For a set of integers $m$, one then has
\begin{equation}
k_m=m\:\frac{2\pi}{\mathcal{L}}.
\end{equation}
The group velocity of each $k_m$ component is a linear function of $k_m$; therefore, the wavepacket spreads as it travels, as shown in figure \ref{smat1}.
%\begin{figure}[H]
%\renewcommand{\captionfont}{\footnotesize}
%\renewcommand{\captionlabelfont}{}
%\begin{center}
% \includegraphics[width=14cm]{smat1.eps}
% \end{center}
% \caption{Transmission and reflection of a wavepacket of incident energy $E$ from a barrier of height $B<E$.} \label{smat1}
%\end{figure}
Let us assume that, at some time $t=0$, the wavepacket is located at a point $x_{in}$ to the left-hand side of the barrier. After a time $t_s$, we measure the position of its peak and find it to be at $x_{out}$, on the right-hand side of the barrier. $t_s$ then defines the time required for the wavepacket to travel a distance $|x_{in}-x_{out}|$. Provided $x_{in}$ is sufficiently far from the barrier, the spin component of the wavepacket is initially oriented along the positive $\hat{\bm{x}}$ axis. By the time we reach $x_{out}$, the spin has been rotated to some new orientation, which can be estimated by calculating the expectation values of the Pauli operators $\sigma_x$, $\sigma_y$ and $\sigma_z$ from the density operator of the system at time $t_s$
\begin{equation}\label{sigma alpha}
\langle\sigma_{\alpha}\rangle=\mathrm{Tr}\left[\sigma_{\alpha}\rho\left(t_s\right)\right],
\end{equation}
with $\alpha\in\lbrace x,y,z\rbrace$. The square modulus of the wavefunction at $t_s$ therefore defines the probability of finding the wavepacket at position $x_{out}$ with a certain spin orientation $\left(\theta,\phi\right)$, defined by \ref{sigma alpha}.

Now, let us put aside the spatial part of the wavepacket's evolution, and focus on spin. Changes in the $\phi$ coordinate of the spin presumably took place while the wavepacket was passing through the field. We can estimate the approximate duration of this phase from the wavepacket's average velocity inside the barrier
\begin{equation}
\bar{v}=\frac{\left(q_{\uparrow}+q_{\downarrow}\right)}{2},
\end{equation}
giving
\begin{equation}
t_B\approx\frac{L}{\bar{v}}.
\end{equation}
Is it then valid to identify $t_B$ with the time taken by the spin to rotate from $\left(\pi/2,0\right)$ to $\left(\theta,\phi\right)$, or is there some other timescale associated with spin rotation? In other words, do the `space' and `spin' clocks tick at the same rate? To answer this question, I calculate how much a single spin initially oriented along the positive $\hat{\bm{x}}$ axis precesses about $\hat{\bm{z}}$ under the effect of the field $B_z\hat{\bm{z}}$. If this value is compatible with the $\phi$ angles given by the $\langle\sigma_{\alpha}\rangle$, there is good reason to assume the `space' and `spin' clocks measure time in the same way.

\begin{figure}[H]
\renewcommand{\captionfont}{\footnotesize}
\renewcommand{\captionlabelfont}{}
\begin{center}
\subfigure{\label{smat2a}\includegraphics[width=6.5cm]{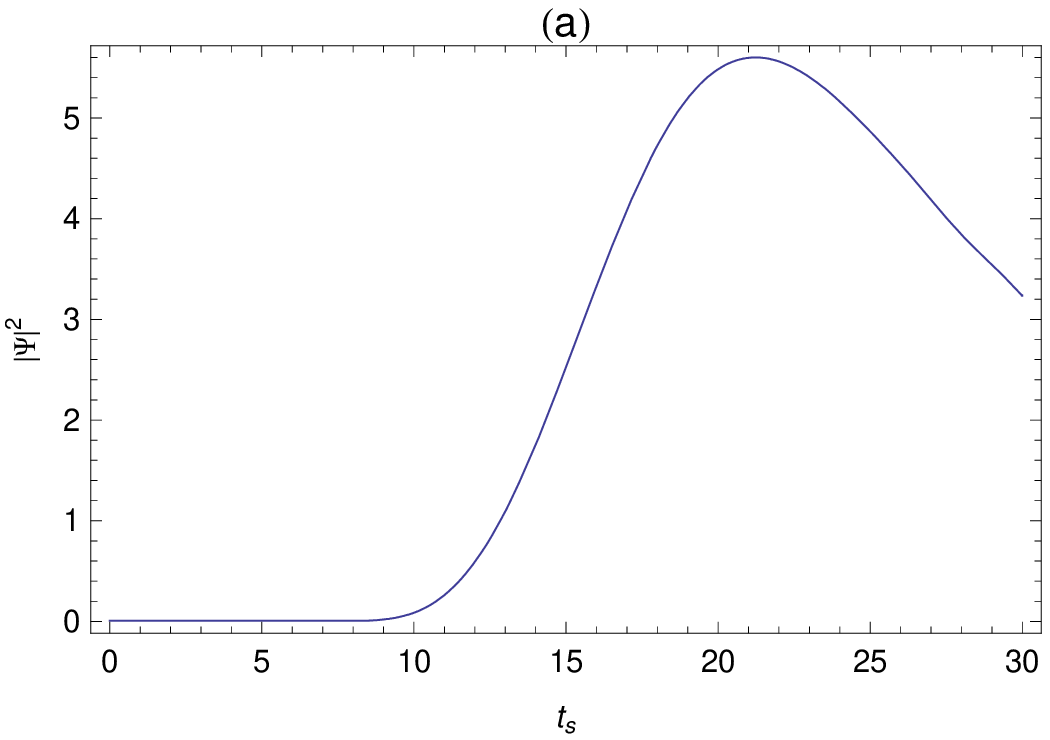}}
\hspace{0.3cm}
\subfigure{\label{smat2b}\includegraphics[width=5.5cm]{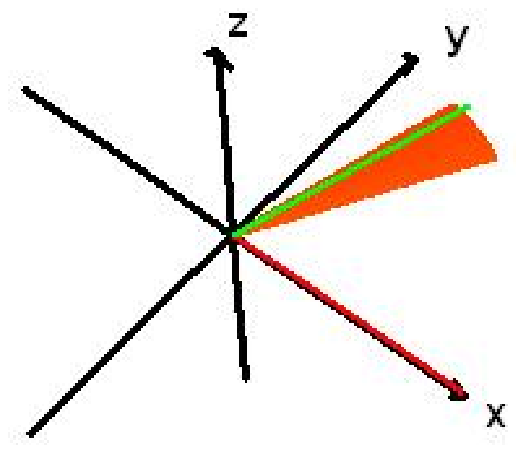}}
\hspace{0.3cm}
\subfigure{\label{smat2c}\includegraphics[width=6.5cm]{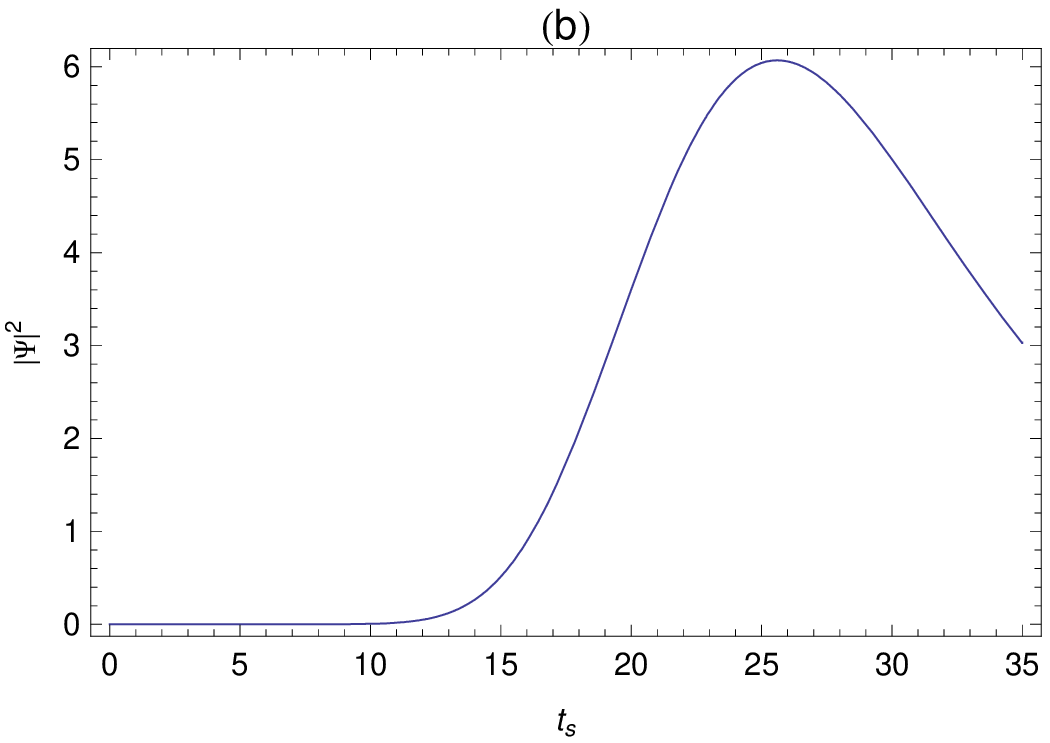}}
\hspace{0.3cm}
\subfigure{\label{smat2d}\includegraphics[width=5.5cm]{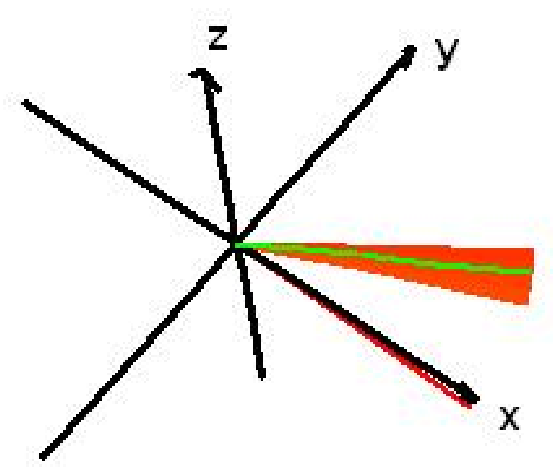}}
\end{center}
\caption{A comparison between the `space' and `spin' clocks, for the situation shown in figure \ref{smat1}. In all cases $\mathcal{L}=40$, $L=10$, $x_0=-8$, $k_0=\frac{10\pi}{40}$. It is assumed the barrier originates at position $x=0$. (a) The un-normalized probability distribution $|\Psi|^2$ as a function of travel time $t_s$ at position $x_{out}=16$ for $w=2$ and $B_z=0.05$. To the right, the distribution of measured spin states over the range $\Delta t_s=[19,25]$, compared with the initial spin state (red line) and the expected final spin state (green line). (b) The un-normalized probability distribution $|\Psi|^2$ as a function of travel time $t_s$ at position $x_{out}=19$ for $w=3$ and $B_z=0.02$. To the right, the distribution of measured spin states over the range $\Delta t_s=[21,31]$, compared with the initial spin state (red line) and the expected final spin state (green line). All quantities are expressed in natural units.}
\label{smat2}
\end{figure}

Figure \ref{smat2} shows a few examples at fixed $\mathcal{L}$, $L$, $k_0$ and $x_0$. Here, the three axes represent the eigenstates of $\sigma_x$, $\sigma_y$ and $\sigma_z$ in the Bloch sphere, and the red vector indicates the spin orientation of the wavepacket at time $t=0$. The green vector shows the expected precession around $\hat{\bm{z}}$ over a time $t_B$, and the orange segment illustrates the range of possible spin orientations over the width of the peak. Note that the states of the system at time $t_s$ are not always pure, hence the magnitude of the vectors forming the orange segment can be less than unity. It is clear from this figure that for the chosen values of the input parameters one finds good agreement between the `space' and `spin' clocks of the system.

What remains now is to verify whether this holds in a more general context also. Therefore, I return to the entangling protocol, and to a more formal approach, by deriving the scattered state of the system according to the $S$-matrix formalism. For simplicity, I will assume elastic scattering at both zero and finite momentum transfer. As discussed at the end of section \ref{neutron h}, under these circumstances spin flips can still be occasioned by neutrons scattered slightly away from the centre of the Bragg peak.

\section{Time at Zero Momentum Transfer}\label{time at 0 mtm trans}
I will derive the scattered states of the system using only the first-order expansion of the $S$-matrix, as expressed by equation \eqref{1st order sa}. At zero momentum transfer, the $S$-matrix representing the first scattering event is given by
\begin{equation}
S=\mathbb{I}-2 \pi i \mathcal{V}_1\delta\left(E^{\:\prime}-E\right),
\end{equation}
where $\mathcal{V}_1$ is the scattering potential of equation (\ref{vm}), and $E$ is the energy of the initial state of the system. For incoming neutrons with momentum $\mathbf{k}_0$, given the dispersion relation $E=\frac{\hbar^2k^2}{2m}$, this can be written as
\begin{equation}\label{smat 0mtm}
S=\mathbb{I}-2 \pi i \mathcal{V}_1\frac{\delta(k-k_0)}{k},
\end{equation}
where $k_0$ and $k$ are the magnitudes of the neutron's initial and final momenta, respectively. The general state of a neutron with momentum $\mathbi{k}$ and spin $\sigma_m$ is given by equation (\ref{neutron wave}). Neglecting for the moment the subscript $m$, the orthonormality of the neutron states obeys the relation
\begin{equation}\label{orthonorm}
\langle k^{\:\prime}\sigma^{\:\prime}| k\sigma\rangle=\left(\frac{2\pi}{L}\right)^3\delta(\mathbi{k}^{\:\prime}-\mathbi{k})\delta_{\sigma\:\sigma^{\:\prime}},
\end{equation}
where $\delta(\mathbi{k}^{\:\prime}-\mathbi{k})$ is a three-dimensional delta function in momentum and $\left(\frac{L}{2\pi}\right)^3$ is the number of states per unit volume of momentum space. The orthonormality condition of equation (\ref{orthonorm}) also applies to the states of the system as a whole, provided one now takes the ket $|\sigma\rangle$ to label the overall spin state of neutron plus sample. Indeed, as the spatial part of the sample's wavefunction is not changed by the scattering event, the sample is entirely described by its spin state.

Following section \ref{neutron h}, we focus on neutrons scattered into the Bragg peak at $\mathbf{Q}=0$ and calculate the spin state produced by the action of the $S$ matrix by integrating the scattered state over the spatial coordinates of the neutron. Recall, however, that the peak has finite width, due to the finite size of the sample. Our intention is to collect \emph{all} neutrons scattered into the peak, not just those precisely obeying $\mathbf{k}_i=\mathbf{k}_f$. As measurements will be made after both neutrons have scattered, it is in our interest to maintain a certain amount of uncertainty in these measurements, because the spatial and spinor parts of the wavefunction of scattered neutrons are usually entangled \cite{rauch book}. Therefore, an exact measurement of momentum would destroy any indeterminacy in the spin state of the neutron, proving counterproductive to the success of the entangling protocol. To account for this effect, we take the outgoing spin state of the system to be an average over all outgoing spatial states $\mathbf{q}$ consistent with scattering into the $\mathbf{Q}=0$ peak
\begin{equation}
|\psi_f^{\sigma}\rangle=\left(\frac{L}{2\pi}\right)^3\int_{q}\langle \mathbf{q}|S|\mathbf{k}\sigma \rangle\:d\mathbf{q},
\end{equation}
which becomes, with the help of (\ref{smat 0mtm})
\begin{equation}\label{psi fin s mat}
|\psi_f^{\sigma}\rangle=|\sigma \rangle - \frac{iL^3}{4\pi^2}\int_q\langle \mathbf{q}|\mathcal{V}_1|\mathbf{k}\sigma \rangle \frac{\delta(q-k_0)}{q}\:d\mathbf{q}.
\end{equation}
Note the function $\langle q|\mathcal{V}_1|k\sigma \rangle$ differs from the matrix element of equation (\ref{vs}) because it is evaluated at a generic value of $\mathbf{Q}$ within the Bragg peak, rather than at its centre.

\begin{figure}[H]
\renewcommand{\captionfont}{\footnotesize}
\renewcommand{\captionlabelfont}{}
\begin{center}
 \includegraphics[width=15cm]{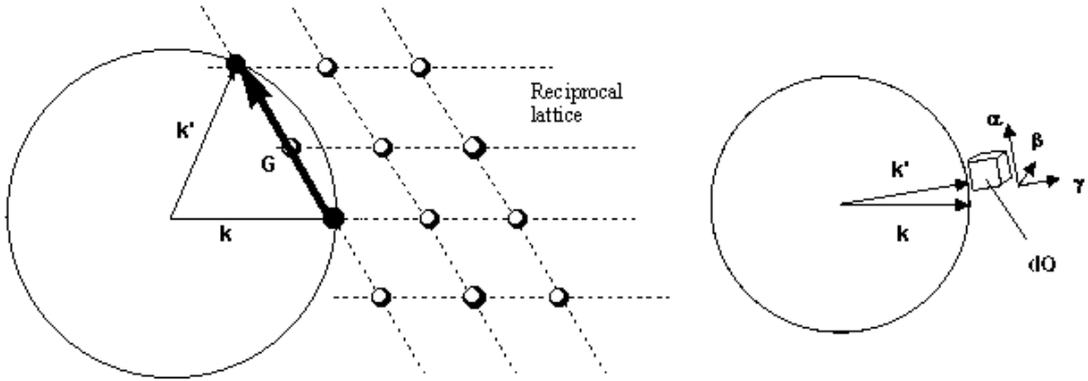}
 \end{center}
 \caption{The Ewald construction in two dimensions (left), and the set of integration coordinates $\alpha$, $\beta$ and $\gamma$ (right). The left-hand diagram is adapted from a figure taken from http://perso.fundp.ac.be/~jwouters/DRX/ewald1.GIF} \label{ewald}
\end{figure}

Let us assume the scattering length $b_j$ is a constant for all nuclei $j$, and set
\begin{equation}\label{f of q}
\frac{2\pi\hbar^2}{m}\sum_{j=1}^{N}b_j e^{i\mathbf{Q}\cdot\mathbf{r}_j}+\Lambda\:\sum_{j=1}^{N}\bm{s}
\cdot\lbrack \mathbf{\hat{Q}}\times \left(\bm{s}_j\times
\mathbf{\hat{Q}}\right)\rbrack \:e^{i \mathbf{Q}\cdot\mathbf{r}_j}=f\left(\mathbf{Q}\right)\delta\left(\mathbf{Q}-\mathbf{G}\right),
\end{equation}
where $\mathbf{G}$ is a reciprocal lattice vector. The three-dimensional delta function $\delta\left(\mathbf{Q}-\mathbf{G}\right)$ accounts for the fact that the right hand side of equation (\ref{f of q}) is small unless the momentum transfer coincides with a reciprocal lattice vector. From equations (\ref{hcal1}) and (\ref{vs}), the matrix element $\langle q|\mathcal{V}_1|k\sigma \rangle$ then becomes
\begin{align}\label{v qk}
\langle q|\mathcal{V}_1|k\sigma \rangle&=\frac{D^3}{L^3}\left[\left.f\left(\mathbf{Q}\right)\delta\left(\mathbf{Q}-\mathbf{G}\right)\star \mathrm{Sa}\left(\frac{\mathbf{Q}\cdot\mathbf{D}}{2}\right)\right]\right|_{\mathbf{Q}\approx0}\nonumber\\
&=\frac{D^3}{L^3}\:f\left(\mathbf{G}\right)\mathrm{Sa}\left[\frac{\left(\mathbf{Q}-\mathbf{G}\right)\cdot\mathbf{D}}{2}\right]
\end{align}
with $\mathrm{Sa}\left[\cdot\right]$ given by equation \eqref{sinc}, and
\begin{equation}
f\left(\mathbf{G}\right)=\left[V_0 +\Lambda\:\sum_{j}\bm{s}
\cdot\lbrack \mathbf{\hat{G}}\times \left(\bm{s}_j\times
\mathbf{\hat{G}}\right)\rbrack\right].
\end{equation}
To within a constant, the function $f\left(\mathbf{G}\right)$ represents the spatial matrix element of the potential evaluated at the centre of the Bragg peak. At $\mathbf{G}=0$, the form of this matrix element is known from equation (\ref{h ex1}). With hindsight, let us then set
\begin{equation}\label{f of 0}
f\left(0\right)\equiv V_0+\frac{2\Lambda}{3}\:\left(s_m^x\sum_{j=1}^Ns_j^x+s_m^y\sum_{j=1}^Ns_j^y+s_m^z\sum_{j=1}^Ns_j^z\right)=V^{s_{ex}}.
\end{equation}
Substituting equations (\ref{v qk}) and (\ref{f of 0}) into (\ref{psi fin s mat}), one finds
\begin{equation}\label{psi fin 1}
|\psi_f^{\sigma}\rangle=|\sigma \rangle - \frac{iD^3}{4\pi^2}V^{s_{ex}}\int_q \mathrm{Sa}\left[\frac{\mathbf{Q}\cdot\mathbf{D}}{2}\right]\frac{\delta(q-k_0)}{q}\:d\mathbf{q}.
\end{equation}
At fixed $\mathbf{k}_i$, the integral of equation (\ref{psi fin 1}) can be written in terms of the momentum transfer as
\begin{equation}\label{int1}
\int_q \mathrm{Sa}\left[\frac{\mathbf{Q}\cdot\mathbf{D}}{2}\right]\frac{\delta(Q_0)}{k_0-Q}\:d\mathbf{Q},
\end{equation}
where
\begin{align}
Q_0&=q-k_0\nonumber,\\
q&=k_0-Q\nonumber.
\end{align}
To define a convenient and visually descriptive set of integration coordinates, I make use of a construction known as the \emph{Ewald sphere}, shown in two dimensions in figure \ref{ewald}. This is a sphere of radius $k_0$, centred at the origin of reciprocal space. If the scattering process is elastic, the tip of the final wavevector $\mathbf{k}_f$ must lie somewhere on the surface of the sphere, as $k_0=k_f$. The momentum transfer $\mathbf{Q}$ is the vector joining $\mathbf{k}_0$ and $\mathbf{k}_f$. The integration region $d\mathbf{Q}$ can then be approximated by a small cube of volume $dQ_{\alpha}dQ_{\beta}dQ_{\gamma}$ centred on the tip of the vector $\mathbf{k}_f$, where the directions $\lbrace\alpha,\beta\rbrace$ and $\gamma$ lie parallel and perpendicular to the surface of the sphere, respectively (fig. \ref{ewald}). In these coordinates, the integral of equation (\ref{int1}) at fixed $\mathbf{k}_0$ becomes
\begin{equation}\label{ints} \int\mathrm{Sa}\left[\frac{Q_{\alpha}D}{2}\right]\mathrm{Sa}\left[\frac{Q_{\beta}D}{2}\right]\mathrm{Sa}\left[\frac{Q_{\gamma}D}{2}\right]\frac{\delta(Q_0)}{k_0-Q}\:dQ_{\alpha}dQ_{\beta}dQ_{\gamma}.
\end{equation}
The integration with respect to $Q_{\gamma}$ is immediately resolved by the delta function
\begin{equation}\label{ints2}
\int\mathrm{Sa}\left[\frac{Q_{\gamma}D}{2}\right]\frac{\delta(Q_0)}{k_0-Q_{\gamma}}\:dQ_{\gamma}=\frac{1}{k_0}.
\end{equation}
The remaining terms must be integrated over the range of the sinc function, which gives
\begin{equation}\label{sincs}
\int_{-\infty}^{\infty}\int_{-\infty}^{\infty}\mathrm{Sa}\left[\frac{Q_{\alpha}D}{2}\right]\mathrm{Sa}\left[\frac{Q_{\beta}D}{2}\right]\:dQ_{\alpha}dQ_{\beta}=\left(\frac{2 \pi}{D}\right)^2.
\end{equation}
Finally, from equations (\ref{psi fin 1}), (\ref{ints2}) and (\ref{sincs}), the scattered spin state of the system is
\begin{equation}\label{psi fin2}
|\psi_f^{\sigma}\rangle= \left(\mathbb{I}-iV^{s_{ex}}\frac{D}{k_0}\right)|\sigma \rangle.
\end{equation}

Let us now consider the scattering event in a time-dependent frame. If the neutron is scattered by a spin-dependent potential $V$ in a time $\tau$, the spin state of the system at time $\tau$ is described by
\begin{equation}\label{psi fin3}
|\psi_f^{\sigma}\rangle=e^{-iV\tau}|\sigma \rangle\approx\left(\mathbb{I}-iV\tau\right)|\sigma \rangle,
\end{equation}
with $\hbar=1$. Comparing equations (\ref{psi fin2}) and (\ref{psi fin3}) shows that first-order agreement between the time-independent and time-dependent pictures is achieved when
\begin{align}
V&=V^{s_{ex}},\label{vs ex}\\
\tau&=\frac{D}{k_0}.
\end{align}
The form of $V^{s_{ex}}$ is known from equation (\ref{f of 0}); the quantity $\frac{D}{k_0}$ is simply the free-flight time of the neutron through the sample. These results lead us to conclude that:
\begin{enumerate}
\item The potential that determines the evolution of the spin state of the system is, from equations (\ref{h0 tot}), (\ref{hcal1}) and (\ref{f of 0})
    \begin{align}
    \mathcal{H}=-J\sum_{\langle ij \rangle}\bm{\sigma^i \cdot \sigma^j}&+B_z\sum_{j=1}^N\mathbf{\sigma}_z^j+\nonumber\\
    &+V_0+\frac{2\Lambda}{3}\:\left(s_m^x\sum_{j=1}^Ns_j^x+s_m^y\sum_{j=1}^Ns_j^y+s_m^z\sum_{j=1}^Ns_j^z\right)\label{h ex tot s mat}.
    \end{align}
\item To first order in the scattering potential, the interaction time of a forward-scattered neutron coincides with its free-flight time through the sample. This can be tuned by simply adjusting the neutron momentum, thus providing a realistic means of setting the interaction time to the desired value.
\end{enumerate}

\section{Time at Finite Momentum Transfer}\label{time finite mtm}
The derivation of the scattered state at finite momentum transfer $\mathbf{Q}=Q_z\hat{\bm{z}}$ is virtually identical to that outlined in the previous section. What changes is the form of the scattering matrix, now given by
\begin{equation}\label{smat}
S=-2 \pi i \mathcal{V}_1\frac{\delta(k-k_0)}{k}.
\end{equation}
The disappearance of the first term reflects the absence of forward scattering; it is now assumed all neutrons scatter away from the incident direction. The scattered spin state, averaged over the width of the Bragg peak, is then
\begin{equation}\label{psi fin xy}
|\psi_f^{\sigma}\rangle=- \frac{iL^3}{4\pi^2}\int_q\langle \mathbf{q}|\mathcal{V}_1|\mathbf{k}\sigma \rangle \frac{\delta(q-k_0)}{q}\:d\mathbf{q}.
\end{equation}
Let us repeat the integration of equation (\ref{v qk}). To within a constant, the form of $f\left(\mathbf{G}\right)$ is known from equation (\ref{h xy1}). It is therefore convenient to set
\begin{equation}\label{f of qz}
f\left(Q_z\right)=V_0+\Lambda\: \left(s_m^x\sum_{j=1}^Ns_j^x+s_m^y\sum_{j=1}^Ns_j^y\right)=V^{s_{xy}}.
\end{equation}
Substituting equations (\ref{v qk}) and (\ref{f of qz}) into (\ref{psi fin xy}), one finds
\begin{equation}\label{psi fin 1a}
|\psi_f^{\sigma}\rangle=- \frac{iD^3}{4\pi^2}V^{s_{xy}}\int_q \mathrm{Sa}\left[\frac{\mathbf{Q}\cdot\mathbf{D}}{2}\right]\frac{\delta(q-k_0)}{q}\:d\mathbf{q}.
\end{equation}
This can be resolved as before, with the help of equations (\ref{int1}), (\ref{ints}), (\ref{ints2}) and (\ref{sincs}), to give
\begin{equation}\label{psi fin xy1}
|\psi_f^{\sigma}\rangle= -iV^{s_{xy}}\frac{D}{k_0}|\sigma \rangle=-i\left(V_0+\Lambda\: s_m^x\sum_{j=1}^Ns_j^x+s_m^y\sum_{j=1}^Ns_j^y\right)\frac{D}{k_0}\:|\sigma \rangle.
\end{equation}
Owing to the lack of a non-imaginary component proportional to the identity, this cannot be immediately traced back to an expansion of the time evolution operator to any order in the scattering potential. One is therefore left with a series of possibilities, among which:
\begin{enumerate}
\item The neutron interaction time is not a well-defined quantity for this process;
\item The correct interpretation of the interaction time emerges from a higher order expansion of the $S$ matrix;
\item The present model is too simplistic; the presence of an apparently complex-valued time parameter may then result from failing to take into account additional processes, such as decoherence.
\end{enumerate}
Related to this final point, the significance of imaginary time in quantum mechanics, and particularly in scattering theory, is a topic of intense debate, which has yet to find a unique resolution. It has been suggested that the complex part of the time parameter arises from changes in the scattering probability with potential. Specifically, Im$\left[\tau\right]$ is thought to describe the variation in the number of particles scattered per unit time, per unit change of the scattering potential \cite{solokovski87}. The derivation of this result is based on a spin-independent potential, therefore it may not be of strict relevance to the present discussion. However, it is interesting that qualitatively similar findings should crop up also in the context of tunneling \cite{buttiker82,gasparian98}.

We can conclude that scattering at finite momentum transfer cannot immediately be described in a physically intuitive time-dependent frame. Consequently, it is difficult to see how parameters such as the neutron interaction time may be tuned in an experimental setting. It may be that the entangling protocol of the previous chapter translates into a feasible scheme at zero momentum transfer only. Therefore, I will restrict subsequent analysis to realizations of the protocol at $\mathbf{Q}=0$.

\section{The Magnetic Coupling Strength}\label{lambda derivation}
We conclude from sections \ref{time at 0 mtm trans} and \ref{time finite mtm} that in the limit of weak scattering ($T\approx V$), one can ascribe a well-defined physical meaning to the time-evolution of the spin part of the system wavefunction only in the case of forward scattering. The Hamiltonian representing this process resembles an exchange interaction, whose strength we define as $\Lambda_E$. To complete our understanding of the scattering process, I will now discuss the value of this parameter.

The Hamiltonian of \eqref{h ex tot s mat} is derived by taking a spatial matrix element of the potential of equation \eqref{vm}. Let us isolate the magnetic part of this potential, writing
\begin{equation}
\mathcal{V}_B=-\bm{\mu}_m\cdot\mathbf{B}(\mathbf{r})\:\theta_D \left(\mathbf{r}\right).
\end{equation}
We then have
\begin{align}
\langle k|\mathcal{V}_B|k\rangle&=-\frac{1}{L^3}\int_{all\:space} e^{-i\mathbf{k}_f\cdot\mathbf{r}}\left[\bm{\mu}_m\cdot\mathbf{B}(\mathbf{r})\:\theta_D \left(\mathbf{r}\right)\right]e^{i\mathbf{k}_i\cdot\mathbf{r}}\mathrm{d}\mathbf{r}\nonumber=\\
&=-\frac{\bm{\mu}_m}{L^3}\cdot\int e^{i\mathbf{Q}\cdot\mathbf{r}}\mathbf{B}(\mathbf{r})\:\theta_D \left(\mathbf{r}\right)\mathrm{d}\mathbf{r}.
\end{align}
At zero momentum transfer, the exponential term reduces to unity, leaving
\begin{equation}
\langle k|\mathcal{V}_B|k\rangle=-\frac{\bm{\mu}_m}{L^3}\cdot\int_{all\:space}\mathbf{B}(\mathbf{r})\:\theta_D \left(\mathbf{r}\right)\mathrm{d}\mathbf{r}.
\end{equation}
The integral on the right-hand side of this equation is simply the average magnetic field of the sample multiplied by its volume. One can then write
\begin{equation}\label{vb s mat}
\langle k|\mathcal{V}_B|k\rangle=-\frac{D^3}{L^3}\left(\bm{\mu}_m\cdot\bar{\mathbf{B}}\right).
\end{equation}
Comparing equations \eqref{vs}, \eqref{h ex}, \eqref{h ex tot s mat} and \eqref{vb s mat}, it becomes clear that for $\mathbf{Q}=0$
\begin{equation}\label{mbar1}
-\bm{\mu}_m\cdot\bar{\mathbf{B}}=\frac{2\Lambda}{3}\:\left(s_m^x\sum_{j=1}^Ns_j^x+s_m^y\sum_{j=1}^Ns_j^y+s_m^z\sum_{j=1}^Ns_j^z\right).
\end{equation}
Expanding the left-hand side of this equation in terms of the average magnetization $\bar{\mathbf{M}}$ gives
\begin{equation}
-\bm{\mu}_m\cdot\bar{\mathbf{B}}=-\frac{g_N\mu_N\mu_0}{\hbar}\left(\bm{s}_m\cdot\bar{\mathbf{M}}\right).
\end{equation}
Then, recalling the magnetization of the sample is defined as the magnetic moment per unit volume, we obtain
\begin{equation}\label{mbar2}
-\bm{\mu}_m\cdot\bar{\mathbf{B}}=-\frac{g_N\mu_N\mu_0g_e\mu_B}{\hbar^2D^3}\left(\bm{s}_m\cdot\sum_{j=1}^{N}\bm{s}_j\right).
\end{equation}
It follows from equations \eqref{mbar1} and \eqref{mbar2} that
\begin{equation}\label{big lambda}
\Lambda=-\frac{3}{2}\:\frac{g_N\mu_N\mu_0g_e\mu_B}{\hbar^2D^3}.
\end{equation}
The action of the spin Hamiltonian $V^{s_{ex}}$ on the combined spin state of the interacting neutron and the sample produces terms proportional to $\frac{\hbar^2}{4}$. The coupling energy $\Lambda_E$ can therefore be defined as
\begin{equation}\label{big lambda e}
\Lambda_E=-\frac{3}{8}\:\frac{g_N\mu_N\mu_0g_e\mu_B}{a_0^3N},
\end{equation}
where the sample volume $D^3$ has been replaced with the number of spins $N$ multiplied by the volume of the unit cell $a_0^3$. We finally conclude that the Hamiltonian describing the evolution of the system spin state during forward scattering of a neutron is given by
\begin{equation}
V^{s_{ex}}=\frac{3\lambda}{8N}\:\frac{4}{\hbar^2}\:\left(s_m^x\sum_{j=1}^Ns_j^x+s_m^y\sum_{j=1}^Ns_j^y+s_m^z\sum_{j=1}^Ns_j^z\right),
\end{equation}
with
\begin{equation}\label{lambda}
\lambda=-\frac{g_N\mu_N\mu_0g_e\mu_B}{a_0^3},
\end{equation}
where the value of $\lambda$ can be interpreted as the strength of the magnetic interaction of a neutron with an electron at a distance of one lattice constant.

\section{Symmetries and Operator Representation}\label{symm and op rep}
The full Hilbert space of a system comprising $N$ sample spins and two neutrons has dimensions $d=2^{N+2}$. Fortunately, the need to deal with exponentially growing matrices is removed by exploiting the symmetries of equation (\ref{h ex tot s mat}), and the periodic nature of the sample. Let us define the operators $\bm{S}_z$ and $\mathbb{S}_z$, which represent the $z$-components of the total spin of the sample and of the system, respectively. It is easily shown that
\begin{equation}\label{comms}
\lbrack H_0,\mathbb{S}_z\rbrack=\lbrack H_0,\bm{S}_z\rbrack=\lbrack \mathcal{V}^s,\mathbb{S}_z\rbrack=0,
\end{equation}
therefore both the free Hamiltonian of the sample and the interaction potential have excitation-number conserving symmetries. Bearing in mind our chosen initial states contain at most two spin flips, this allows us to define a reduced computational basis, comprising all possible zero-, single- and double-spin-flip states. The dimensions of the reduced basis can be calculated by summing the binomial coefficients $_{(N+2)}C_0$, $_{(N+2)}C_1$ and $_{(N+2)}C_2$, to obtain
\begin{equation}
d=\frac{N^2}{2}+\frac{5N}{2}+4.
\end{equation}
This grows quadratically rather than exponentially with $N$, but still blows up as $N$ becomes large.

A further reduction is possible if one abandons the canonical basis. The single- and double- spin-flip eigenstates of a periodic sample can be expressed in terms of Bloch vectors $|s^1_k\rangle$ and $|s^2_k\rangle$
\begin{eqnarray}
|s^1_k\rangle&=&\frac{1}{\sqrt{N}}\sum_{j=1}^Ne^{i\mathbf{k}\cdot\mathbf{r}_j}|\mathbi{j}\rangle,\\
|s^2_k\rangle&=&\sqrt{\frac{2}{N(N-1)}}\sum_{\lbrace
j,\:l\neq
j\rbrace=1}^Ne^{i\mathbf{k}\cdot\left(\mathbf{r}_j+\mathbf{r}_l\right)}|\mathbi{j\:l}\rangle.
\end{eqnarray}
As the initial states of the system are translationally invariant, and neither $H_0$ nor $\mathcal{V}^s$ contain any spatial dependence, coupling to states $|s^1_k\rangle$ and $|s^2_k\rangle$ is null unless $\mathbf{k}=0$. The Hilbert space of the sample is, therefore, fully specified by the vectors $|\bm{0}\rangle$, $|s^1_0\rangle$ and $|s^2_0\rangle$. Combining these with the four canonical two-neutron basis states, it is possible to define a complete 8-dimensional basis for the system, comprising the following eigenvectors
\begin{eqnarray}\label{basis states}
|1\rangle&=&|00\mathbf{0}\rangle,\label{st1}\\
|2\rangle&=&|001_s\rangle,\\
|3\rangle&=&|002_s\rangle,\\
|4\rangle&=&|01\mathbf{0}\rangle,\\
|5\rangle&=&|011_s\rangle,\\
|6\rangle&=&|10\mathbf{0}\rangle,\\
|7\rangle&=&|101_s\rangle,\\
|8\rangle&=&|11\mathbf{0}\rangle\label{st8},
\end{eqnarray}
where $|00\mathbf{0}\rangle\equiv|0\rangle_2|0\rangle_1|\mathbf{0}\rangle$, and so on. Unless otherwise stated, from this point onwards equations (\ref{st1})-(\ref{st8}) will be my computational basis. For ease of reference, I call this the \textit{Bloch basis}, and drop all indices to simplify notation. Henceforth, the values inside bras and kets should be interpreted as labeling $|$\textit{second neutron}, \textit{first neutron}, \textit{sample}$\rangle$, respectively.

The dimensions of the Bloch basis are fixed, manageable, and independent of the size of the sample. As a result, many calculations can be done analytically and operator representation is greatly simplified. The initial states of the system, for instance, can now be written as
\begin{eqnarray}
|\psi_0^A\rangle&=&|2\rangle,\\
|\psi_0^B\rangle&=&\alpha^2|1\rangle+\alpha \beta\:\left(|4\rangle+|6\rangle\right)+\beta^2|8\rangle.
\end{eqnarray}
Also, the zero field free Hamiltonian becomes a multiple of the identity
\begin{equation}\label{h0a}
H_0=-JN\mathbb{I},
\end{equation}
owing to the fact that the sample basis states $|\bm{0}\rangle$, $|s^1_0\rangle$ and $|s^2_0\rangle$ are degenerate eigenstates of $H_0$ with eigenvalue $-J N$. Equation (\ref{h0a}) underlines a very important point: the internal coupling of the spins in the sample plays no meaningful
role in the dynamics of the system. Hence, both the numerical value of $J$
and its sign are irrelevant. This warrants two observations.
First, provided one could prepare the same initial states, ferromagnetic and antiferromagnetic samples would produce exactly the same outcome. Second, in zero field the evolution of the system between scattering events is trivial. Therefore, if one did succeed in producing a correlation between the neutrons, this correlation would not depend on the time elapsed between the departure of the first neutron and the arrival of the second. Qualitatively, this means the `information' deposited in the sample by
the first neutron can survive indefinitely; in other words, the sample functions as a perfect quantum
memory between scattering events. Note this is a specific feature of the present model, and should not be interpreted as a result of general validity.

\section{Conclusions}
The scattering of a neutron from a sample is usually described as a time-independent, probabilistic process. In the case of forward scattering, it is possible to recast the problem in a temporal frame, thus showing that the interaction time of a neutron with a scatterer is, in the limit of weak scattering, the time taken by the neutron to traverse the sample. At finite momentum transfer, the interpretation of $\tau$ remains unclear. It emerges from a first-order calculation that $\tau$ may be a complex-valued, or even completely imaginary quantity. This possibility has been the focus of myriad studies, but no unique resolution has been found to date. Indeed, a \emph{unique} resolution may not exist, as the details of the scattering geometry or the scattering potential may be a factor. Some answers may lie in a higher order treatment of the problem; alternatively, it may be necessary to look into the \emph{dynamical} theory of neutron scattering, which accounts for the possibility of the incident neutron being re-scattered within the sample before emerging \cite{rauch book}. These options are not explored in the present thesis, but may become the topic of future work.

%\end{document} 

%% file: two_neutron_results.tex
%\documentclass[a4paper,12pt]{article}
%\usepackage{setspace}
%\usepackage{graphicx}
%\usepackage{subfigure}
%\usepackage{caption}
%\usepackage{color}
%\usepackage{amsmath}
%\usepackage{amssymb}
%\usepackage{caption}
%\usepackage{bm}
%\usepackage{float}
%\DeclareGraphicsExtensions{.epsi,.eps,.ps}
%
%\onehalfspacing
%\def\mathbi#1{\textbf{\em #1}}
%\begin{document}
%\tableofcontents
\chapter{Two-Neutron Entanglement via Scattering}\label{two ns}
\textit{This chapter considers two alternative realizations of the neutron entanglement scheme outlined in chapter \ref{neutron proposal}. In order of discussion, these are: (i) scattering at zero momentum transfer from a sample in a single magnon state; (ii) scattering at zero momentum transfer from a sample containing no spin excitations. In each case, the performance of the system in zero and finite field is analyzed. After describing the quantum state of the system at each stage of the protocol, I will discuss the behaviour of the entanglement between the neutrons as a function of the input parameters. It will emerge that, for certain combinations of these parameters, the concurrence attains high values. Furthermore, the concurrence proves independent of the time of free evolution $\tau_f^{\:\prime}$ between scattering events. To investigate this property further, I will compare the time-evolution of the concurrence to that of the logarithmic negativity, which is not independent of $\tau_f^{\:\prime}$. I then discuss parallels with the optical entanglement scheme proposed by Haroche \textit{et al.} \cite{haroche97}, and evaluate the robustness of the protocol and its experimental feasibility in light of current neutron scattering facilities. Finally, I propose a simple model to describe how sample anisotropies might affect the performance of the protocol.}

\section{Scattering from a Single-Magnon State}\label{p0a zero mtm}
Let us first consider forward scattering from a single-magnon state, taking the input wavefunction of the system to be $|\psi_0^A\rangle=|001\rangle\equiv|2\rangle$. One sees from equation (\ref{qtm psi}) that the first non-trivial step of the protocol is the first scattering event. It was shown in sections \ref{time at 0 mtm trans} and \ref{lambda derivation} of the previous chapter that the Hamiltonian describing this process is
\begin{align}
\mathcal{H}_1=-J\sum_{\langle ij \rangle}\bm{\sigma^i \cdot \sigma^j}&+B_z\sum_{j=1}^N\mathbf{\sigma}_z^j\nonumber\\
&+V_0+\frac{3\lambda}{2N\hbar^2}\:\left(s_1^x\sum_{j=1}^Ns_j^x+s_1^y\sum_{j=1}^Ns_j^y+s_1^z\sum_{j=1}^Ns_j^z\right)\label{h ex tot}.
\end{align}
It is clear from this equation that both $V_0$ and the exchange coupling between the sample spins simply map the state of the system onto itself. In the interaction picture, $\mathcal{H}_1$ can then be written as
\begin{equation}\label{h1prime}
\mathcal{H}_1^{\prime}=H_0^B+\mathcal{V}_1^{\sigma},
\end{equation}
with
\begin{align}
H_0^B&=B_z\sum_{j=1}^N\bm{\sigma}_z^j,\\
\mathcal{V}_1^{\:\sigma}&=\frac{3\lambda}{8N}\:\left[\sigma_m^x\sum_{j=1}^N\sigma_j^x+\sigma_m^y\sum_{j=1}^N\sigma_j^y+\sigma_m^z\sum_{j=1}^N\sigma_j^z\right].
\end{align}
Absorbing the factor of $3/8$ into the value of $\lambda$, the non-zero matrix elements of $\mathcal{H}^{\prime}_1$ are
\begin{eqnarray}
\langle1|\mathcal{H}^{\prime}_1|1\rangle&=&\langle6|\mathcal{H}^{\prime}_1|6\rangle=NB_z+\lambda,\label{v1}\\
\langle4|\mathcal{H}^{\prime}_1|4\rangle&=&\langle8|\mathcal{H}^{\prime}_1|8\rangle=NB_z-\lambda,\\
\langle2|\mathcal{H}^{\prime}_1|2\rangle&=&\langle7|\mathcal{H}^{\prime}_1|7\rangle=\lambda \left(1-\frac{2}{N}\right)+B_z(N-2),\\
\langle3|\mathcal{H}^{\prime}_1|3\rangle&=&\lambda \left(1-\frac{4}{N}\right)+B_z(N-4),\\
\langle5|\mathcal{H}^{\prime}_1|5\rangle&=&\lambda \left(\frac{2}{N}-1\right)+B_z(N-2),\\
\langle2|\mathcal{H}^{\prime}_1|4\rangle&=&\langle4|\mathcal{H}^{\prime}_1|2\rangle=\frac{2\lambda}{\sqrt{N}},\\
\langle3|\mathcal{H}^{\prime}_1|5\rangle&=&\langle5|\mathcal{H}^{\prime}_1|3\rangle=2\lambda\sqrt{\frac{2}{N}\left(1-\frac{1}{N}\right)},\label{v7}
\end{eqnarray}
The time evolution of state $|\psi_0^A\rangle$ is dominated by only two of the eight eigenstates of this potential. We write these as $|\mathfrak{1}\rangle$ and $|\mathfrak{2}\rangle$:
\begin{eqnarray}
|\mathfrak{1}\rangle&=&c|2\rangle+d|4\rangle,\label{evec1 ex}\\
|\mathfrak{2}\rangle&=&d|2\rangle-c|4\rangle\label{evec2 ex},
\end{eqnarray}
with
\begin{eqnarray}
\phi&\equiv&\phi^{ex}(N,\lambda,B_z)=\sqrt{B_z^2 - 2 B_z \left(1-\frac{1}{N}\right) \lambda + \left(1+\frac{1}{N}\right)^2 \lambda^2},\label{phi}\\
\varphi&\equiv&\varphi^{ex}(N,\lambda,B_z)=\lambda \left(\frac{1}{N}-1\right)+B_z+\phi,\label{varphi}\\
c&=&-\frac{\sqrt{N}\varphi}{\sqrt{4\lambda^2+N\varphi^2}}\label{eq c},\\
d&=&\frac{2\lambda}{\sqrt{4\lambda^2+N\varphi^2}}\label{eq d}.
\end{eqnarray}
The scattered states of the system can then be derived by expanding the `effective' time-evolution operator $U\left(\mathcal{H}_1^{\prime},\tau\right)$ in terms of its eigenstates $|i\rangle$ and eigenvalues $E_i$, as
\begin{equation}
U\left(\mathcal{H}_1^{\prime},\tau\right)=\sum_{i}e^{-iE_i\tau}|i\rangle\langle i|.
\end{equation}
To within a global phase, the first scattered state is therefore
\begin{equation}\label{psi2exch}
|\psi_2^A\rangle=e^{-i\Lambda\tau}\lbrack (c^2+d^2e^{-2i\phi\tau})|2\rangle+cd(1-e^{-2i\phi\tau})|4\rangle\rbrack,
\end{equation}
where $\Lambda=-\frac{\lambda}{N}+B_z(N-1)-\phi\label{l1}$.

During the second period of free evolution, each component of $|\psi_2^A\rangle$ simply acquires a phase proportional to the corresponding eigenvalue of $H_0^B$, thus yielding the state
\begin{equation}\label{psi3exch}
|\psi_3^A\rangle=e^{-i\Lambda\tau}e^{-iB_zN\tau_f^{\prime}}\lbrack e^{2iB_z\tau_f^{\prime}}(c^2+d^2e^{-2i\phi\tau})|2\rangle+cd(1-e^{-2i\phi\tau})|4\rangle\rbrack.
\end{equation}
The final stage of the protocol is the second scattering event. As before, this is governed by an effective Hamiltonian $\mathcal{H}^{\prime ex}_2=H_0^B+\mathcal{V}_{2}^{\sigma}$, which is formally similar to $\mathcal{H}^{\prime}_1$ but for the placement of signs and off-diagonal matrix elements. The three eigenstates of $\mathcal{H}^{\prime }_2$ contributing to the evolution of $|\psi_3^A\rangle$ are
\begin{eqnarray}
|\mathfrak{4}\rangle&=&|4\rangle,\\
|\mathfrak{5}\rangle&=&c|2\rangle+d|6\rangle,\\
|\mathfrak{6}\rangle&=&d|2\rangle-c|6\rangle,
\end{eqnarray}
hence the final state of the system $|\psi_f^A\rangle$ becomes
\begin{eqnarray}\label{psi4exch}
|\psi_f^A\rangle&=&e^{-i(2\Lambda\tau+B_zN\tau_f^{\prime})}\lbrack e^{2iB_z\tau_f^{\prime}}(c^2+d^2e^{-2i\phi\tau})^2|2\rangle
+ cde^{-i\tau y}(1-e^{-2i\phi\tau})|4\rangle\nonumber\\
&+&cde^{2iB_z\tau_f^{\prime}}(1-e^{-2i\phi\tau})(c^2+d^2e^{-2i\phi\tau})|\:6\rangle\rbrack,
\end{eqnarray}
with
\begin{equation}
y=\lambda\left(1+\frac{1}{N}\right)+B_z+\phi.
\end{equation}
For brevity, let us label the coefficients of $|\psi_f^A\rangle$ by $\Gamma$, $\Delta$ and $\Theta$, respectively. The reduced density matrix of the neutrons in the canonical basis is calculated by performing a partial trace over the system, which gives
\begin{eqnarray}
\rho_n&=&\mathrm{Tr}_s\lbrack|\psi_f^A\rangle\langle\psi_f^A|\:\rbrack=|\Gamma|^2|00\rangle\langle00|+|\Delta|^2|01\rangle\langle01|\nonumber\\
&+&|\Theta|^2|10\rangle\langle10|+\Delta\Theta^*|01\rangle\langle10|+\Delta^*\Theta|10\rangle\langle01|.
\end{eqnarray}
The concurrence of the neutrons is then calculated from the eigenvalues of the spin-flipped density matrix described in section \ref{ent measures}. If $\rho=\rho_n$, the only non-zero eigenvalue of this matrix is $\tilde{\Lambda}=4|\Delta|^2|\Theta|^2$. Substituting the values of $\Delta$ and $\Theta$, one obtains
\begin{equation}\label{conc psiA ex}
C(N,\lambda,B_z,\tau)=\frac{8\sqrt{2}\lambda^2\sin^2{\phi\tau}}{N\phi^{3}\varphi}\sqrt{\left[\phi\varphi-\frac{2\lambda^2}{N}\right]\left[\phi^2-\frac{4\lambda^2}{ N}\sin^2{\phi\tau}\right]}.
\end{equation}
Equation (\ref{conc psiA ex}) immediately reveals several important characteristics. First, the concurrence does not depend on the time between scattering events $\tau_f^{\prime}$. This is essential for the purpose of experimental implementation, because $\tau_f^{\prime}$ cannot be tuned. In zero field, this invariance follows from the fact that the sample Hamiltonian $H_0$ is a multiple of the identity. In finite field the interpretation is more complex, as we will see.

Second, the concurrence is an oscillating function of time, whose periodicity is determined by the cosine function $\cos(2\phi\tau)$:
\begin{equation}\label{T ex}
T_{\phi}(N,\lambda,B_z)=\frac{2\pi}{2\phi}=\frac{\pi}{\sqrt{B_z^2 - 2 B_z \lambda  \left(1-\frac{1}{N}\right)+ \left(1+\frac{1}{N}\right)^2 \lambda^2}}.
\end{equation}
The quantity $2\phi$ is the energy splitting of the eigenstates of $\mathcal{H}^{\prime}_m$ corresponding to the spin-flip being shared between the interacting neutron and the sample. Therefore, during each scattering event the system oscillates between states $|2\rangle$ and $|4\rangle$ ($|6\rangle$), in which the spin flip is localized on the sample or the first (second) neutron, respectively. This is a promising beginning, as it suggests there will be times at which some entanglement is present in the system. To verify how this entanglement is distributed, however, we must focus in detail on the behaviour of the concurrence as a function of time, spin density and magnetic field.

\subsection{Zero-Field Evolution}\label{p0a zero mtm zero field}
\begin{figure}[H]
\renewcommand{\captionfont}{\footnotesize}
\renewcommand{\captionlabelfont}{}
\begin{center}
\subfigure{\label{c0fieldpsiA}\includegraphics[width=6.5cm]{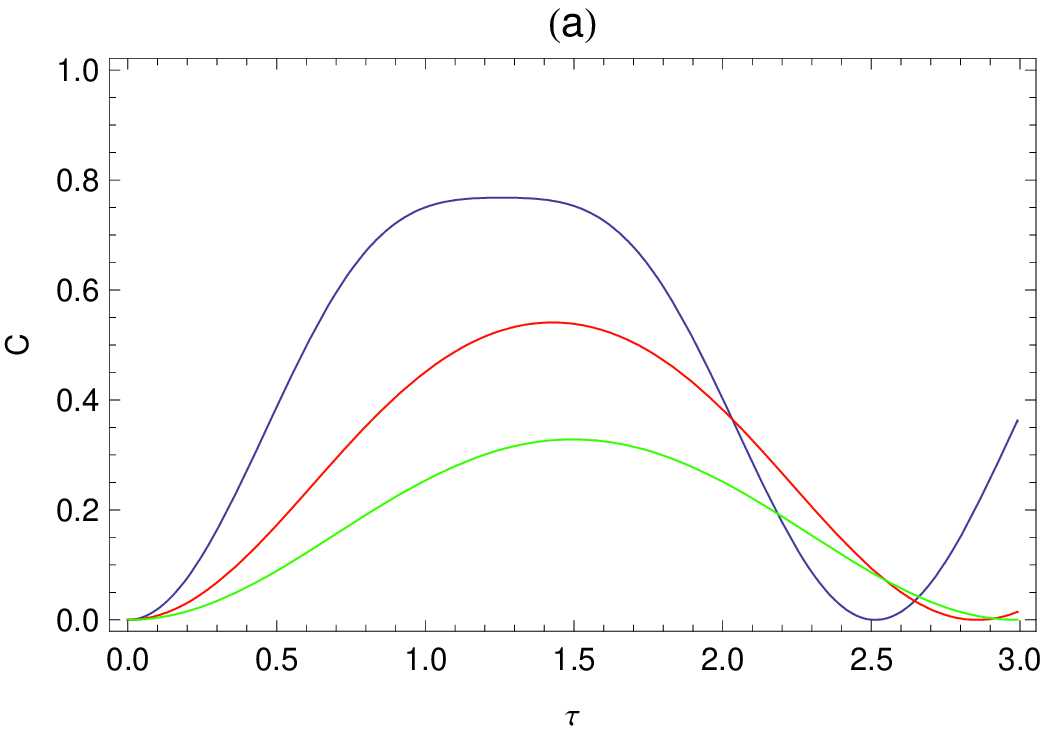}}
\hspace{0.3cm}
\subfigure{\label{cp0fieldpsiA}\includegraphics[width=6.5cm]{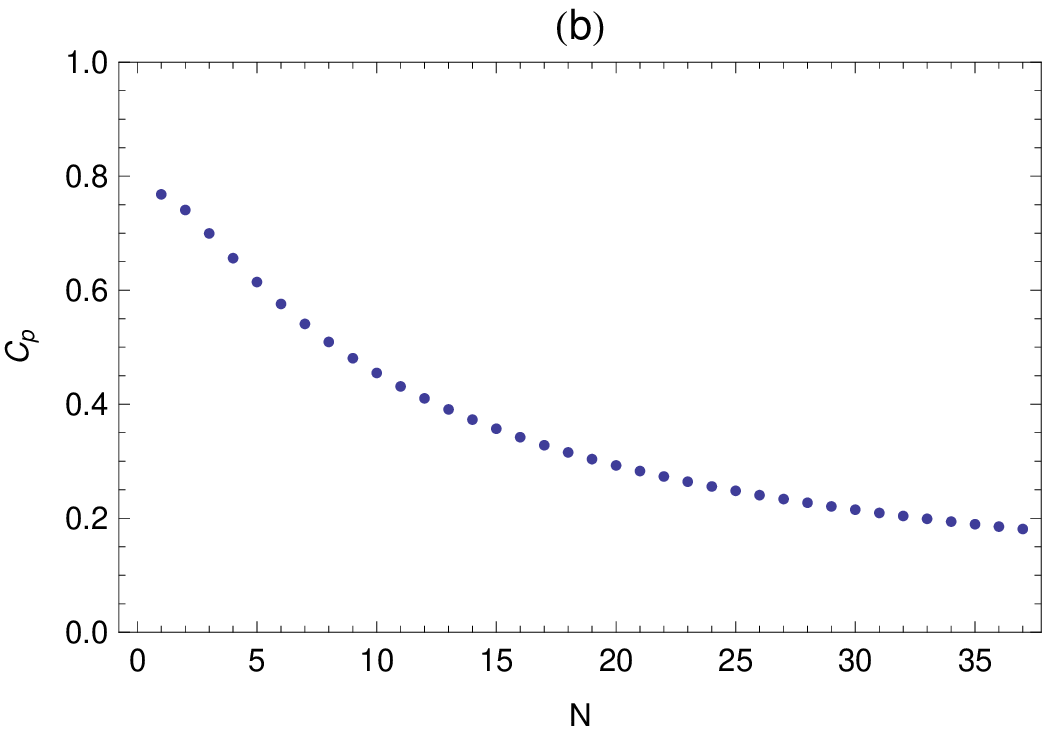}}
\end{center}
\caption{Input state $|\psi_0^A\rangle$, $\bm{Q}=0$, $B_z=0$, $\lambda=1$. (a) The evolution of the concurrence as a function of interaction time $\tau$ for $N=4$, $N=10$ and $N=20$ (blue, red and green curves, respectively). (b) The peak concurrence as a function of $N$. All quantities are expressed in natural units.}
\label{c0fieldpsiAtot}
\end{figure}

If the applied field is zero, equation (\ref{conc psiA ex}) becomes
\begin{equation}\label{conc psiA ex 0b}
C(N,\lambda,\tau)=\frac{8N\sin^2\left[{\lambda\left(1+\frac{1}{N}\right)\tau}\right]\sqrt{N^2+1+2N\cos\left[{2\lambda\left(1+\frac{1}{N}\right)\tau}\right]}}{(N+1)^3}.
\end{equation}
At fixed $N$ and $\lambda$, the only free parameter in this expression is the interaction time. Maximizing with respect to $\tau$ and neglecting un-physical solutions yields
\begin{equation}
\tau^*=\frac{\pi}{2\lambda\left(1+\frac{1}{N}\right)}\equiv\frac{T_{\phi}(N,\lambda,0)}{2},
\end{equation}
hence in zero field the optimal interaction time is half the oscillation period. The peak concurrence $C_p$ is then found to be
\begin{equation}\label{cp 0 field}
C_p(N,\lambda,\tau^*)=\frac{8N(N-1)}{(N+1)^3}.
\end{equation}
It is evident from equation (\ref{cp 0 field}) that $C_p$ falls as $N$ increases, converging to a value $8/N$ as $N\rightarrow\infty$ (fig. \ref{c0fieldpsiAtot}). This phenomenon can be explained in terms of the concurrence of the first neutron and the sample prior to the second scattering event, which is
\begin{equation}\label{c1 0 field}
C_1(N,\lambda,\tau)=\frac{4\sqrt{N}|\sin[{\lambda\left(1+\frac{1}{N}\right)\tau}]|\sqrt{N^2+1+2N\cos[{2\lambda\left(1+\frac{1}{N}\right)\tau}]}}{(N+1)^2}.
\end{equation}
The maximum value of $C_1$ drops with $N$ as $N^{-1}$. The behaviour of $C_p$ therefore stems from the fact that the total amount of `available' entanglement falls as the number of spins in the sample becomes large.

\subsection{Finite-Field Evolution}\label{p0a zero mtm finite field}
We conclude from the previous section that if no field is applied there is a limit to the amount of information the sample can store; but is this true in finite field also? And if not, is there an optimal field value? To answer this question, let us consider the first scattered state. The success of the protocol relies on the system being as entangled as possible before the second scattering event. From equation (\ref{psi2exch}), the state $|\psi_2^A\rangle$ is maximally entangled if
\begin{equation}\label{c cond ex}
2c^2(1-c^2)(1-\cos{2\phi\tau})=\frac{1}{2}.
\end{equation}
If this is true, the derivative of equation (\ref{c cond ex}) with respect to $c$ must vanish. Therefore
\begin{equation}
4c^2(1-2c^2)(1-\cos{2\phi\tau})=0.
\end{equation}
The only reasonable solution to this equation is $c=\pm\frac{1}{\sqrt{2}}$, hence $d=\pm\frac{1}{\sqrt{2}}$ also. Therefore, the concurrence of the neutrons is maximized when the eigenstates of the interaction Hamiltonian in the single-excitation sector correspond to the spin-flip being equally shared between the interacting neutron and the sample. From equations (\ref{phi})-(\ref{eq d}), one finds
\begin{equation}\label{bz star}
B_z=\lambda\left(1-\frac{1}{N}\right)\equiv B_z^*.
\end{equation}
It follows from equation (\ref{lambda}) that, in the large $N$ limit, the optimal field represents the interaction of a neutron with an electron at a distance of one lattice constant. Substituting the value $B_z^*$ into $\mathcal{H}^{\prime}_1$, the difference between the diagonal elements $\langle2|\mathcal{H}^{\prime}_1|2\rangle$ and $\langle4|\mathcal{H}^{\prime}_1|4\rangle$ becomes zero, which ensures the states $|\mathfrak{1}\rangle$ and $|\mathfrak{2}\rangle$ take on the form of equal superpositions. As we will see, $B_z^*$ represents the optimal field value for input state $|\psi_0^B\rangle$ also, because in both cases the first scattered state can be written in terms of eigenstates formally equivalent to $|\mathfrak{1}\rangle$ and $|\mathfrak{2}\rangle$.

\begin{figure}[H]
\renewcommand{\captionfont}{\footnotesize}
\renewcommand{\captionlabelfont}{}
\begin{center}
 \includegraphics[width=8cm]{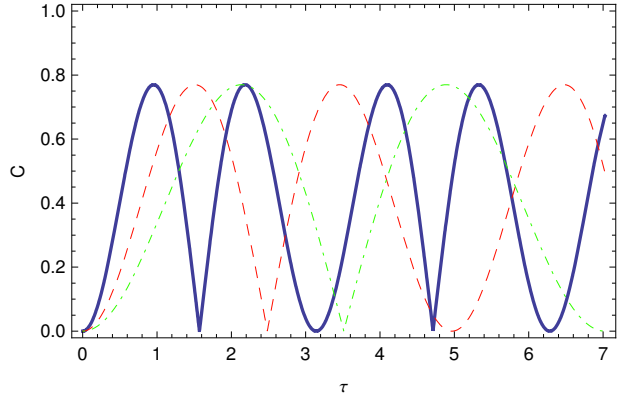}
 \end{center}
 \caption{Input state $|\psi_0^A\rangle$, $\bm{Q}=0$, $\lambda=1$. The evolution of the concurrence with interaction time at optimal field $B_z^*$ for $N=4$, $N=10$ and $N=20$ (thick blue, dashed red and dot-dashed green curves, respectively). All quantities are expressed in natural units.} \label{cOptfieldpsiA}
\end{figure}

Substituting $B_z=B_z^*$ into equation (\ref{conc psiA ex}), the concurrence takes on a particularly simple form
\begin{equation}\label{cp opt field}
C_p(N,\lambda,B_z^*,\tau)=2\sin^2{\left(\frac{2\lambda\tau}{\sqrt{N}}\right)}\left|\cos{\left(\frac{2\lambda\tau}{\sqrt{N}}\right)}\right|.
\end{equation}
This is now easily maximized with respect to $\tau$ to yield
\begin{equation}\label{tau star}
\tau^*=\frac{\sqrt{N}}{4\lambda}\cos^{-1}\left(-\frac{1}{3}\right).
\end{equation}
For all values of $N$, the peak concurrence between the neutrons is now fixed at 0.77 (fig. \ref{cOptfieldpsiA}). This can be expressed in units of entanglement by making use of the relationship between the concurrence $C$ and the entanglement of formation $\mathcal{E}$:
\begin{eqnarray}
\mathcal{E}&=&h\left(\frac{1-\sqrt{1-C^2}}{2}\right);\\
h(x)&=&-x\log_2 x-(1-x)\log_2 (1-x).
\end{eqnarray}

\begin{figure}[H]
\renewcommand{\captionfont}{\footnotesize}
\renewcommand{\captionlabelfont}{}
\begin{center}
 \includegraphics[width=8cm]{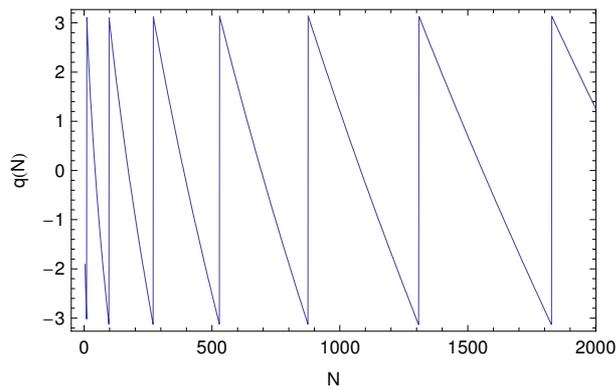}
 \end{center}
 \caption{Input state $|\psi_0^A\rangle$, $\bm{Q}=0$, $\lambda=1$. The argument of the coefficient $\nu\xi^*$ of state $\rho_n^A$ as a function of the number of spins in the sample, N.} \label{arg}
\end{figure}

Hence, for a concurrence $C_p=0.77$ the neutrons share approximately 0.68 ebits of entanglement, and are found to be in state
\begin{eqnarray}\label{rho n A}
\rho_n^A&=&|\mu|^2|00\rangle\langle00|+|\nu|^2|01\rangle\langle01|+|\xi|^2|10\rangle\langle10|\nonumber\\
&+&\nu\xi^*|01\rangle\langle10|+\nu^*\xi|10\rangle\langle01|,
\end{eqnarray}
with $|\mu|^2=\frac{1}{9}$, $|\nu|^2=\frac{2}{3}$, $|\xi|^2=\frac{2}{9}$, $|\nu\xi^*|=|\nu^*\xi|\approx 0.385$, and $\mathrm{Arg}\left[\nu\xi^*\right]=q(N)$, where the function $q(N)$ is shown in figure \ref{arg}.

\begin{figure}[H]
\renewcommand{\captionfont}{\footnotesize}
\renewcommand{\captionlabelfont}{}
\begin{center}
\subfigure{\label{cBpmpsiA}\includegraphics[width=6.5cm]{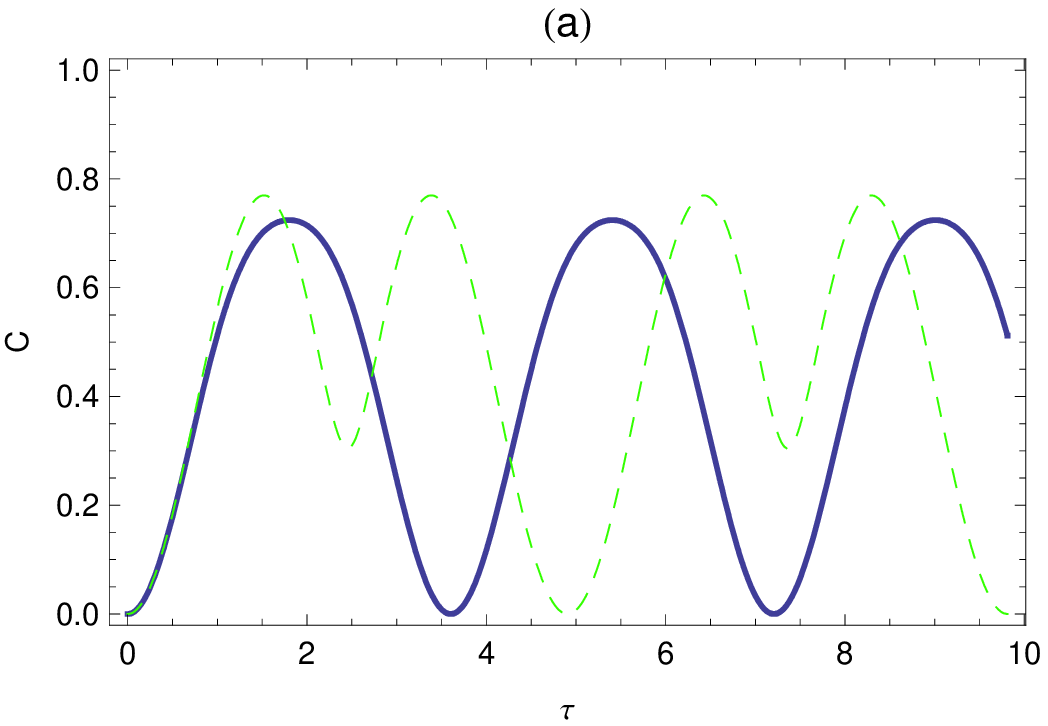}}
\hspace{0.3cm}
\subfigure{\label{intBpmpsiA}\includegraphics[width=6.5cm]{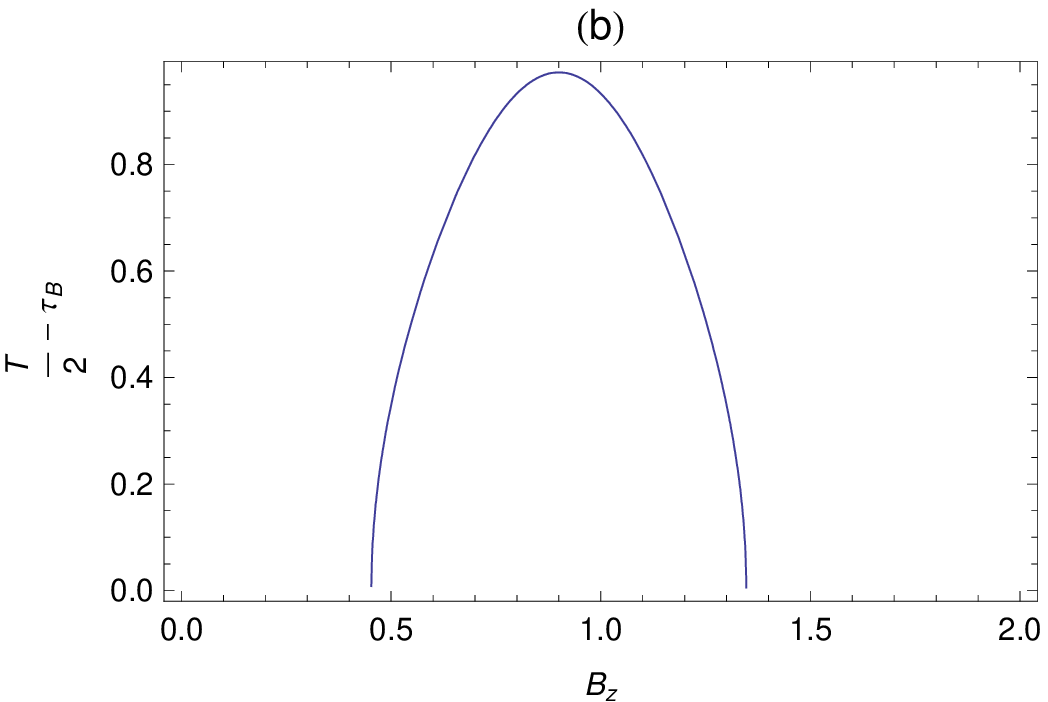}}
\hspace{0.3cm}
\subfigure{\label{cIntpsiA}\includegraphics[width=6.5cm]{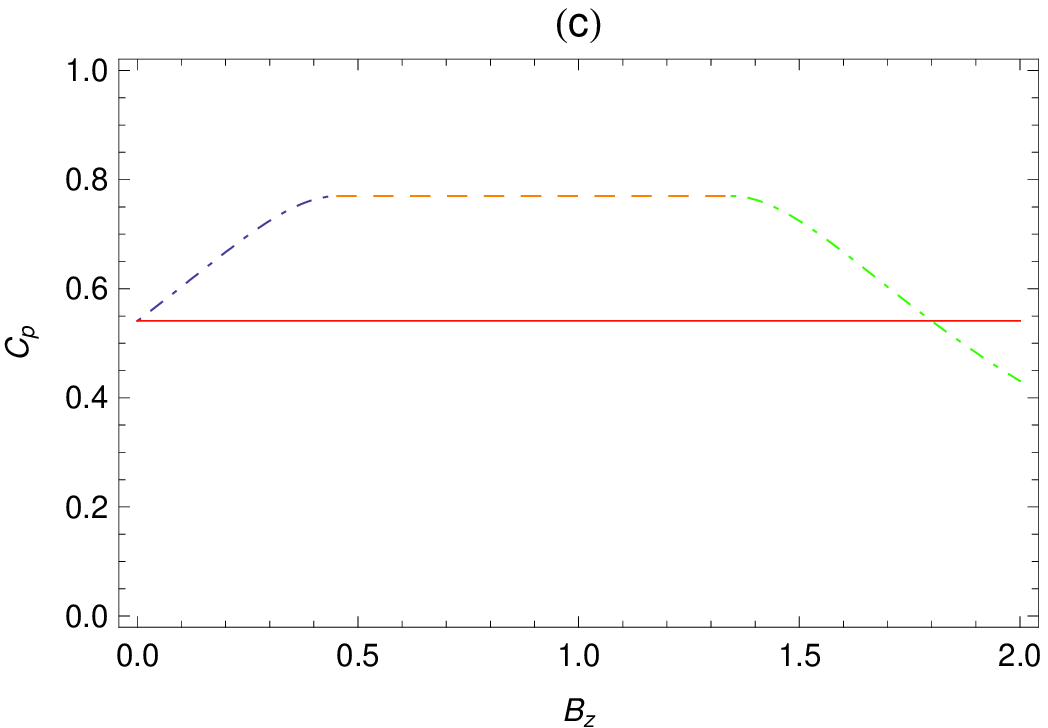}}
\end{center}
\caption{Input state $|\psi_0^A\rangle$, $\bm{Q}=0$, $N=10$, $\lambda=1$. (a) The evolution of the concurrence in time for $B_z=0$ (solid blue line) and $B_z=1$ (dashed green line), showing the development of a double-peaked structure when $B_{z-}<B_z<B_{z+}$. (b) The discrepancy between the optimal interaction times, and the range of the interval $[B_{z-},B_{z+}]$. (c) The peak concurrence outside this interval (blue and green dot-dashed lines) and inside this interval (orange dashed line) as a function of the applied field. The solid red line represents the peak concurrence at $B_z=0$, and shows the existence of the limiting field $B_z=2\lambda\left(1-\frac{1}{N}\right)$. All quantities are expressed in natural units.}
\label{cIntpsiA}
\end{figure}

Let us now consider the evolution of the concurrence for a generic value of $B_z$. By maximizing equation \eqref{conc psiA ex} with respect to $\tau$, one finds the optimal interaction time as a function of applied field to be
\begin{equation}\label{tau b}
\tau_B=\frac{1}{\phi}\sin^{-1}{\left(\frac{\phi\sqrt{N}}{\lambda\sqrt{6}}\right)}.
\end{equation}
If $B_z$ lies outside a certain range, $\tau_B$ is equivalent to half the oscillation period $T_{\phi}$; hence, the concurrence oscillates sinusoidally with time [fig. \ref{cBpmpsiA}]. Inside this range, however, $\tau_B$ and $T_{\phi}/2$ no longer coincide and the concurrence develops a double-peaked structure, showing a higher frequency oscillation modulated by an envelope with period $T_{\phi}$. The limits of this range are found by calculating when the equality $\tau_B=T_{\phi}/2$ ceases to apply. This yields the interval $B_z\in\:[B_{z-},B_{z+}]$ shown in figure \ref{intBpmpsiA}, with
\begin{eqnarray}
B_{z-}&=&\lambda\left(1-\frac{1}{N}-\sqrt{\frac{2}{N}}\:\right)\label{bz-},\\
B_{z+}&=&\lambda\left(1-\frac{1}{N}+\sqrt{\frac{2}{N}}\:\right)\label{bz+}.
\end{eqnarray}
The peak concurrence outside and inside this interval then takes on the form
\begin{eqnarray}
C_o(N,\lambda,B_z)&=&\frac{8\sqrt{2}\lambda^2|\varphi-\phi|\sqrt{\phi\varphi-\frac{2\lambda^2}{N}}}{N\phi^3\varphi}\label{c out ex},\\
C_i(N,\lambda,B_z)&=&\frac{4}{3\sqrt{3}}\label{c in ex},
\end{eqnarray}
Equations (\ref{cp 0 field}), (\ref{c out ex}) and (\ref{c in ex}) then show the peak concurrence is always improved by applying a field, provided this does not exceed an upper limit of $B_z=2\lambda\left(1-\frac{1}{N}\right)$ (fig. \ref{cIntpsiA}).

To summarize: if the system is initialized to a single magnon state, with the spin-flip localized on the sample, the concurrence of the scattered neutrons is an oscillating function of time, which ranges from zero to a maximum of 0.77. In zero applied field, the oscillation is sinusoidal, and its amplitude falls as the spin density of the sample increases. The same result applies if the field lies outside a certain interval $[B_{z-},B_{z+}]$. Conversely, if the field lies within this interval, it is always possible to find an interaction time for which the concurrence attains its optimal value.

\subsection{Concurrence in the Thermodynamic Limit}
The expressions derived so far to describe the behaviour of the concurrence in different parameter regimes are valid at all $N$. In a macroscopic sample, $N$ may be of order $10^{23}$. Consequently, it is useful to note the functional form of these expressions, and other relevant quantities, as $N$ becomes large. We find
\begin{align}
&\phi\rightarrow B_z-\lambda\\
&\varphi\rightarrow 2\left(B_z-\lambda\right)\\
&C(N,\lambda,B_z,\tau)\rightarrow\frac{8\lambda^2\sin^2\phi\tau}{N\left(B_z-\lambda\right)^2}\\
&T_{\phi}(N,\lambda,B_z)\rightarrow\frac{\pi}{B_z-\lambda}\\
&C(N,\lambda,0,\tau)\rightarrow\frac{8\sin^2\lambda\tau}{N^2}\\
&B_z^*\rightarrow\lambda\\
&B_{z\pm}\rightarrow\lambda\\
&C_o\rightarrow\frac{8\lambda^2}{N\left(B_z-\lambda\right)^2}.
\end{align}
The only form to remain unchanged is that of $C(N,\lambda,B_z^*,\tau)$; hence equation \eqref{cp opt field} describes the time-evolution of the concurrence in the thermodynamic limit also.

\section{Scattering from a fully Polarized Sample}\label{p0b 0mtm}
The derivation of the system wavefunctions for initial state $|\psi_0^B\rangle$ is rather elaborate, because it involves the zero-, one- and two-spin-flip sectors of the Hilbert space. Let us write $|\psi_0^B\rangle$ explicitly in terms of the basis states:
\begin{equation}
|\psi_0^B\rangle=\alpha^2|1\rangle+\alpha\beta|4\rangle+\alpha\beta|6\rangle+\beta^2|8\rangle.
\end{equation}
There are now six eigenstates of $\mathcal{H}^{\prime }_1$ that have non-zero overlap with $|\psi_0^B\rangle$:
\begin{eqnarray}
|\mathfrak{1}\rangle&=&|1\rangle,\\
|\mathfrak{2}\rangle&=&|6\rangle,\\
|\mathfrak{3}\rangle&=&c|7\rangle+d|8\rangle,\\
|\mathfrak{4}\rangle&=&c|2\rangle+d|4\rangle,\\
|\mathfrak{5}\rangle&=&d|7\rangle-c|8\rangle,\\
|\mathfrak{6}\rangle&=&d|2\rangle-c|4\rangle,
\end{eqnarray}
where $\phi$, $c$ and $d$ are given by equations \eqref{phi}, (\ref{eq c}) and (\ref{eq d}), respectively. The state of the system after the first neutron has scattered is
\begin{eqnarray}
|\psi_2^B\rangle&=&e^{-i\Lambda\tau}\lbrack \alpha^2|1\rangle+\alpha\beta cd(e^{ix\tau}-e^{iy\tau})|2\rangle\nonumber\\
&+&\alpha\beta(d^2e^{ix\tau}+c^2e^{iy\tau})|4\rangle+\alpha\beta|6\rangle\nonumber\\
&+&\beta^2 cd(e^{ix\tau}-e^{iy\tau})|7\rangle+\beta^2(d^2e^{ix\tau}+c^2e^{iy\tau})|8\rangle\rbrack\label{psi2b exch},
\end{eqnarray}
with
\begin{eqnarray}
\Lambda&=&\lambda+B_zN,\\
x&=&B_z+\lambda\left(1+\frac{1}{N}\right)+\phi,\\
y&=&B_z+\lambda\left(1+\frac{1}{N}\right)-\phi.
\end{eqnarray}
The second period of free evolution simply introduces extra phases between the components of $|\psi_2^B\rangle$:
\begin{eqnarray}
|\psi_3^B\rangle&=&e^{-i\Lambda\tau}e^{iB_z(2-N)\tau^{\prime}_f}\lbrack e^{-2iB_z\tau^{\prime}_f}\alpha^2|1\rangle+\alpha\beta cd(e^{ix\tau}-e^{iy\tau})|2\rangle\nonumber\\
&+&e^{-i2B_z\tau^{\prime}_f}\alpha\beta(d^2e^{ix\tau}+c^2e^{iy\tau})|4\rangle+e^{-2iB_z\tau^{\prime}_f}\alpha\beta|6\rangle\nonumber\\
&+&\beta^2 cd(e^{ix\tau}-e^{iy\tau})|7\rangle+e^{-2iB_z\tau^{\prime}_f}\beta^2(d^2e^{ix\tau}+c^2e^{iy\tau})|8\rangle\rbrack\label{psi2b exch}.
\end{eqnarray}
Hence, the final state of the system is found to be
\begin{eqnarray}\label{psifb}
|\psi_f^B\rangle&=&e^{-2i\Lambda\tau}e^{iB_z(2-N)\tau^{\prime}_f}\lbrack e^{-2iB_z\tau^{\prime}_f}\alpha^2|1\rangle\nonumber\\
&+&cd\alpha\beta(e^{ix\tau}-e^{iy\tau})(c^2e^{ix\tau}+d^2e^{iy\tau}+e^{-2iB_z\tau^{\prime}_f})|2\rangle\nonumber\\
&+&cdfg\beta^2(e^{ix\tau}-e^{iy\tau})(e^{iz\tau}-e^{iw\tau})|3\rangle\nonumber\\
&+&e^{-2iB_z\tau^{\prime}_f}\alpha\beta(d^2e^{ix\tau}+c^2e^{iy\tau})|4\rangle\nonumber\\
&+&e^{-2iB_z\tau^{\prime}_f}cd\beta^2(d^2e^{ix\tau}+c^2e^{iy\tau})(e^{ix\tau}-e^{iy\tau})|5\rangle\nonumber\\
&+&\alpha\beta c^2d^2(e^{ix\tau}-e^{iy\tau})^2+\alpha\beta e^{-2iB_z\tau^{\prime}_f}(d^2e^{ix\tau}+c^2e^{iy\tau})|6\rangle\nonumber\\
&+&cd\beta^2 (e^{ix\tau}-e^{iy\tau})(g^2e^{iz\tau}+f^2e^{iw\tau})|7\rangle\nonumber\\
&+&e^{-2iB_z\tau^{\prime}_f}\beta^2(d^2e^{ix\tau}+c^2e^{iy\tau})^2|8\rangle\rbrack,
\end{eqnarray}
with
\begin{eqnarray}
\gamma&\equiv&\gamma^{ex}(\lambda,N,B_z)=\sqrt{B_z^2 - 2 B_z \left(1-\frac{3}{N}\right) \lambda + \left(1 + \frac{1}{N}\right)^2 \lambda^2},\\
\vartheta&\equiv&\vartheta^{ex}(\lambda,N,B_z)=\lambda \left(\frac{3}{N}-1\right)+B_z+\gamma,\\
f&=&-\frac{\vartheta N}{\sqrt{8\lambda^2(N-1)+N^2\vartheta^2}},\\
g&=&\frac{2\lambda\sqrt{2(N-1)}}{\sqrt{8\lambda^2(N-1)+N^2\vartheta^2}},\\
w&=&3 B_z + \lambda\left(1 + \frac{1}{N}\right) - \gamma,\\
z&=&3 B_z + \lambda\left(1 + \frac{1}{N}\right) + \gamma.
\end{eqnarray}
The form of $|\psi_f^B\rangle$ is unfortunately too complex to yield a manageable analytical expression for the concurrence, so much of the analysis must be carried out using numerical simulation. Fortunately, the results of the previous section provide a useful starting point and somewhat reduce the magnitude of the task.

\subsection{Zero-Field Evolution}\label{p0b zero mtm zero field}
\begin{figure}[H]
\renewcommand{\captionfont}{\footnotesize}
\renewcommand{\captionlabelfont}{}
\begin{center}
\subfigure{\label{c0fieldpsiB}\includegraphics[width=6.5cm]{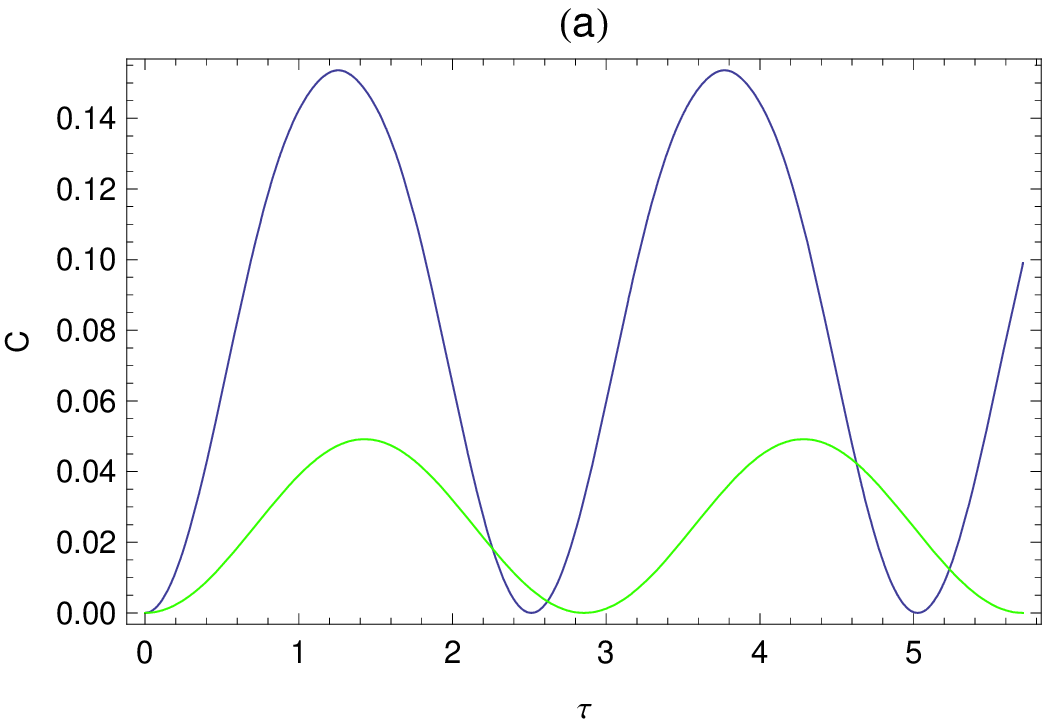}}
\hspace{0.3cm}
\subfigure{\label{cfntfppsiB}\includegraphics[width=6.5cm]{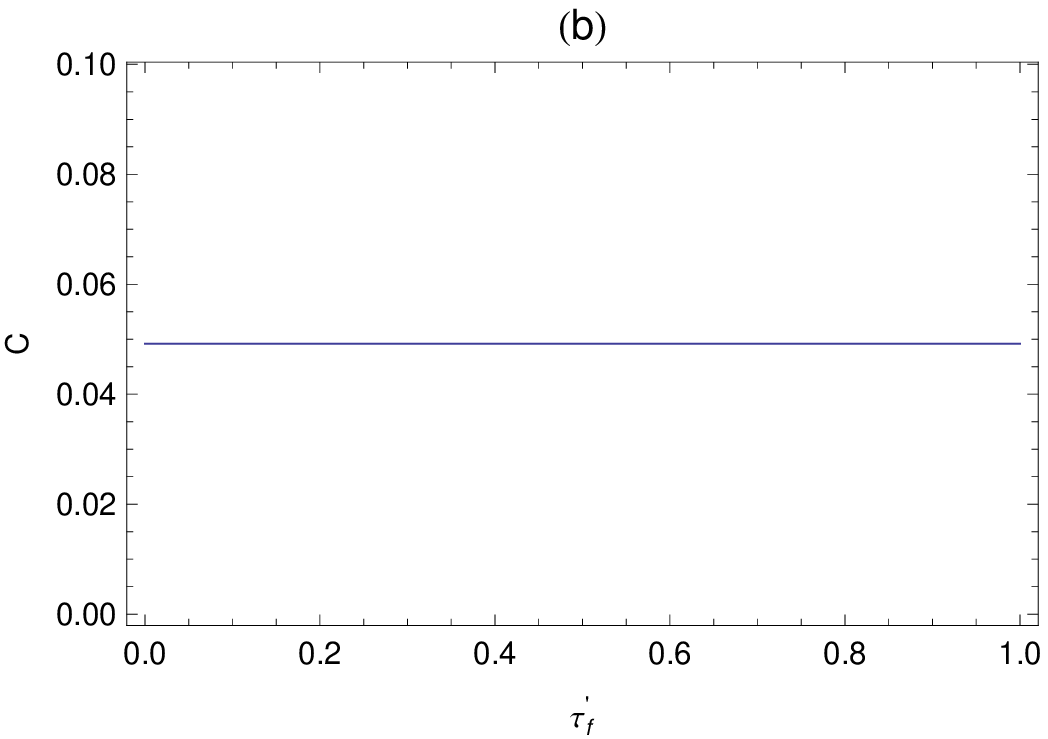}}
\hspace{0.3cm}
\subfigure{\label{cFnNpsiB}\includegraphics[width=6.5cm]{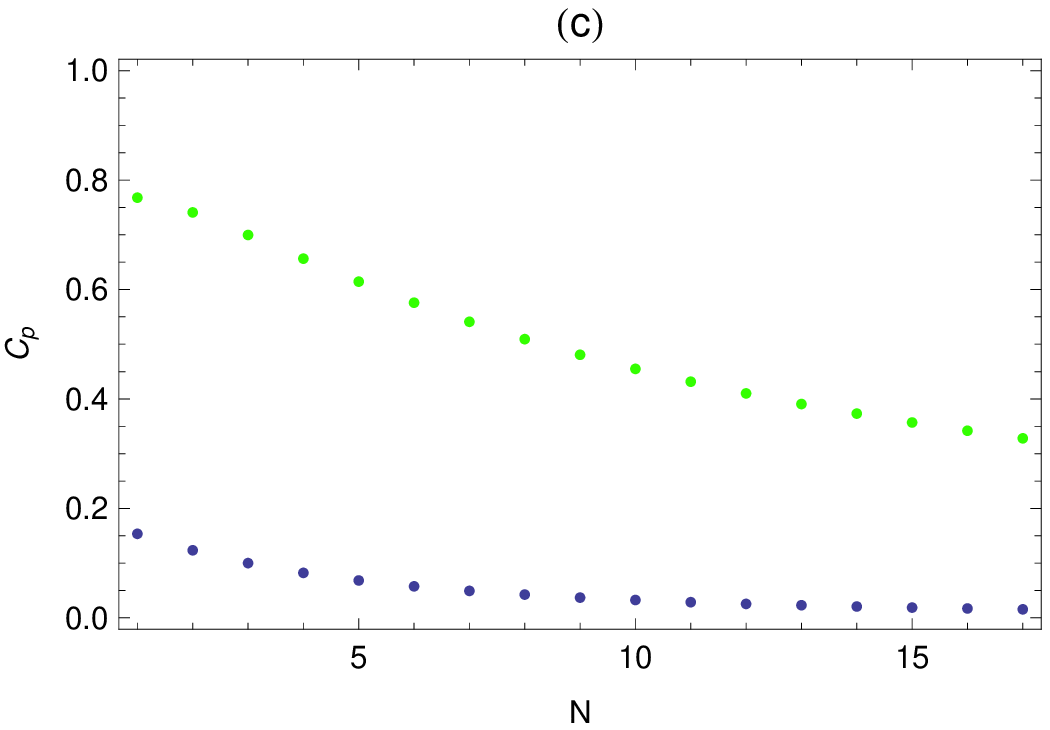}}
\end{center}
\caption{Input state $|\psi_0^B\rangle$, $\bm{Q}=0$, $B_z=0$, $\alpha=\beta=\frac{1}{\sqrt{2}}$, $\lambda=1$. (a) The evolution of the concurrence as a function of interaction time for $N=4$ and $N=10$ (blue and green curves, respectively). (b) The concurrence as a function of the time between scattering events for $N=10$. (c) The peak concurrence as a function of $N$ (blue dots). To ease comparison, the same curve for $|\psi_0^A\rangle$ is also shown (green dots). All quantities are expressed in natural units.}
\label{cexpsiB}
\end{figure}

The main features of zero-field time evolution of the concurrence are illustrated in figure \ref{cexpsiB}. We note that:
\begin{enumerate}
\item The concurrence does not depend on the duration of the period of free evolution between scattering events.
\item The concurrence is an oscillating function of interaction time.
\item The period of the oscillation is defined by the energy splitting of the single-excitation eigenstates that correspond to the spin flip being shared between the interacting neutron and the sample.
\item The amplitude of the oscillation decays as the density of the sample increases.
\end{enumerate}
Therefore, the behaviour of the system mirrors that observed for input state $|\psi_0^A\rangle$, except the peak concurrence for input state $|\psi_0^B\rangle$ is considerably lower for all $N$ and seems to decay more slowly as the number of spins in the sample increases.

\begin{figure}[H]
\renewcommand{\captionfont}{\footnotesize}
\renewcommand{\captionlabelfont}{}
\begin{center}
\subfigure{\label{aldep1a}\includegraphics[width=6.5cm]{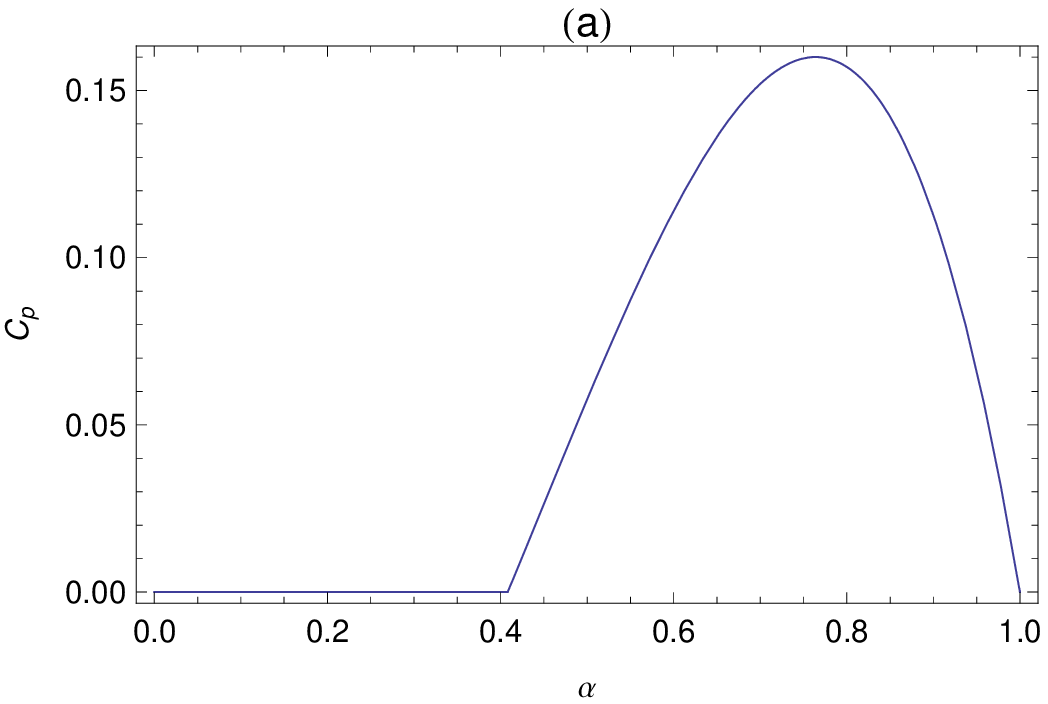}}
\hspace{0.3cm}
\subfigure{\label{aldep1b}\includegraphics[width=6.5cm]{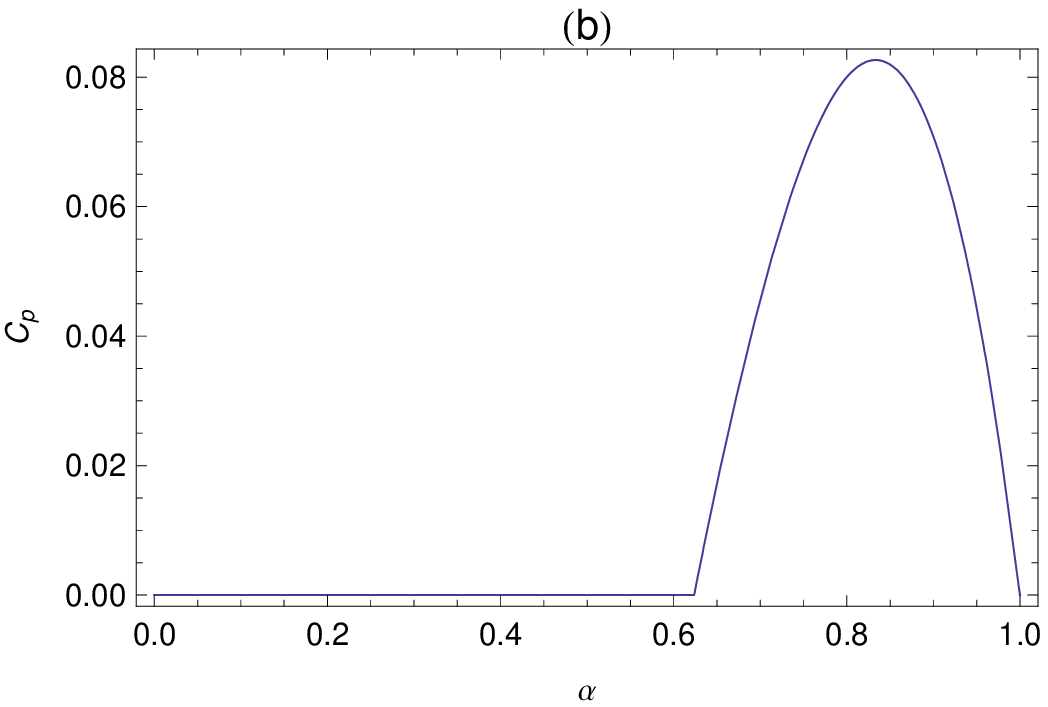}}
\hspace{0.3cm}
\subfigure{\label{aldep1c}\includegraphics[width=6.5cm]{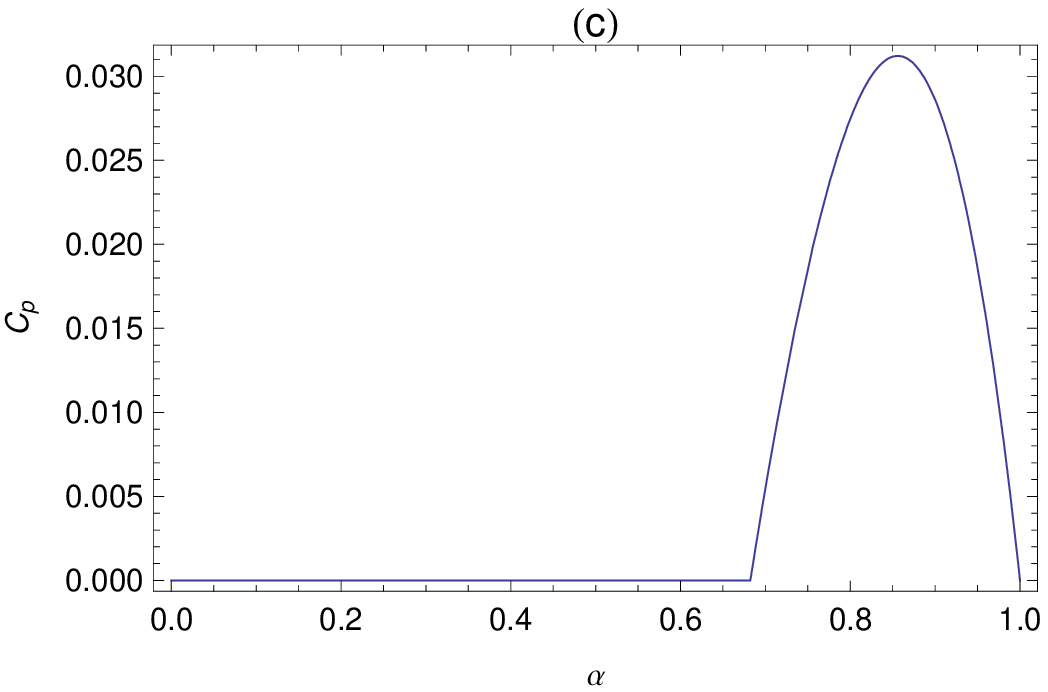}}
\end{center}
\caption{Input state $|\psi_0^B\rangle$, $\bm{Q}=0$, $B_z=0$, $\lambda=1$. The peak concurrence as a function of $\alpha$ for (a) $N=4$, (b) N=10, and (c) N=30. All quantities are expressed in natural units.}\label{aldep1}
\end{figure}

\begin{figure}[H]
\renewcommand{\captionfont}{\footnotesize}
\renewcommand{\captionlabelfont}{}
\begin{center}
\subfigure{\label{aldep2a}\includegraphics[width=6.5cm]{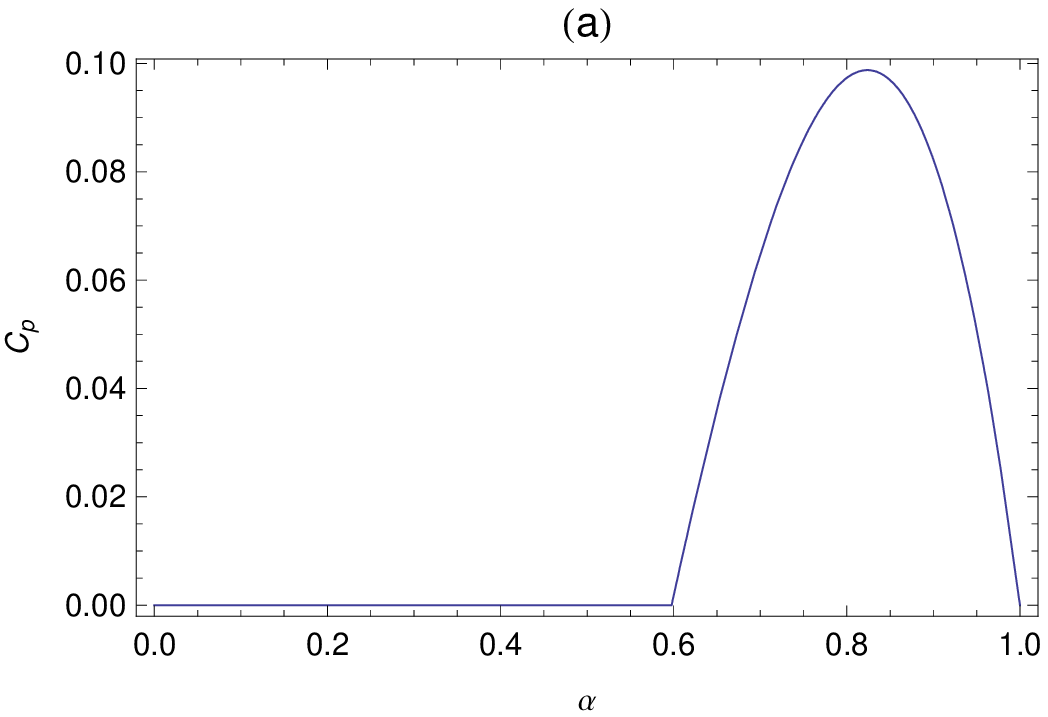}}
\hspace{0.3cm}
\subfigure{\label{aldep2b}\includegraphics[width=6.5cm]{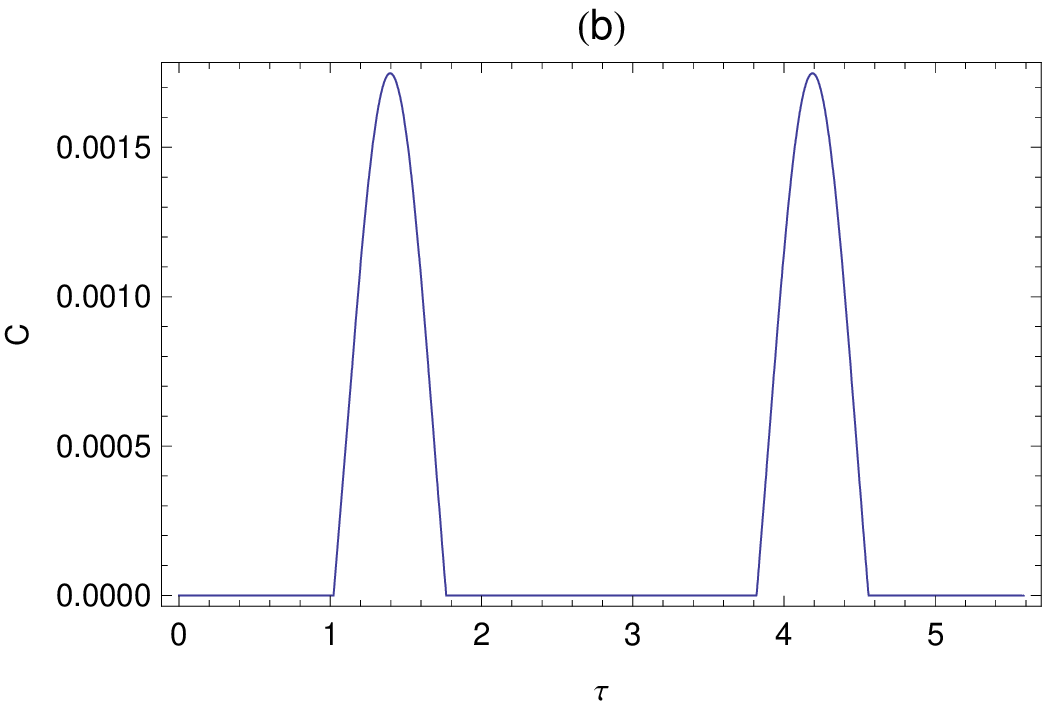}}
\hspace{0.3cm}
\subfigure{\label{aldep2c}\includegraphics[width=6.5cm]{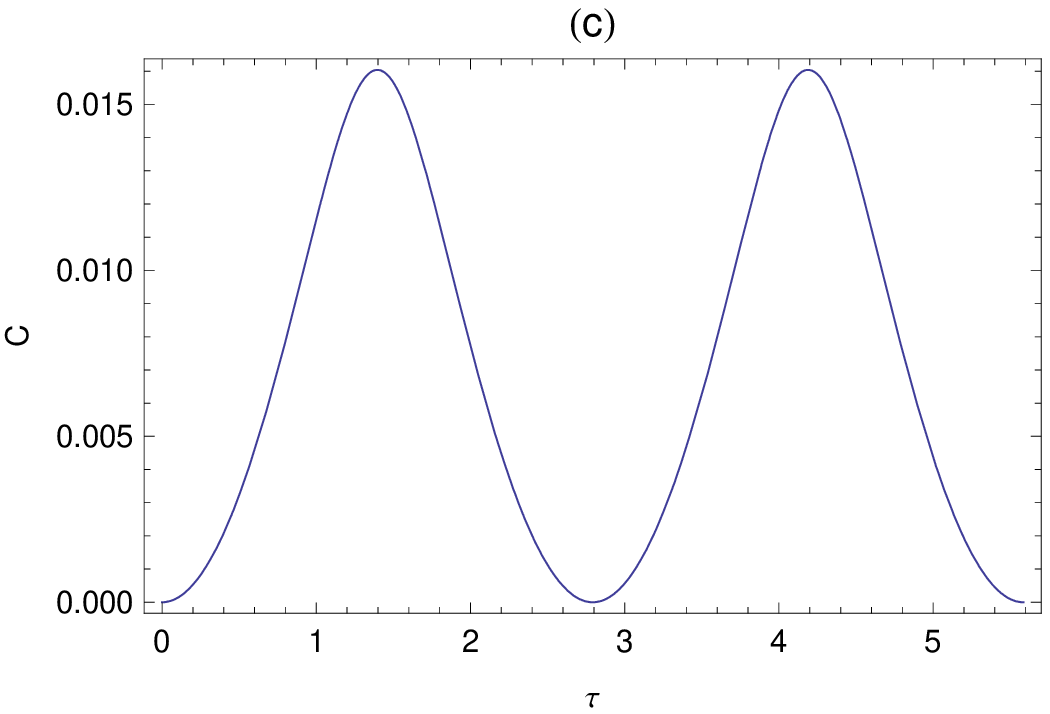}}
\end{center}
\caption{Input state $|\psi_0^B\rangle$, $\bm{Q}=0$, $B_z=0$, $N=8$, $\lambda=1$. (a) The peak concurrence as a function of $\alpha$, showing $\alpha_-(N=8)\approx 0.6$. (b) The concurrence as a function of interaction time for $\alpha=0.6$. (c) The concurrence as a function of interaction time for $\alpha=0.62$. Note the non-analytical behaviour of the curve at $\alpha=\alpha_-$. All quantities are expressed in natural units.}\label{aldep2}
\end{figure}

\begin{figure}[H]
\renewcommand{\captionfont}{\footnotesize}
\renewcommand{\captionlabelfont}{}
\begin{center}
\subfigure{\label{aldep3a}\includegraphics[width=6.5cm]{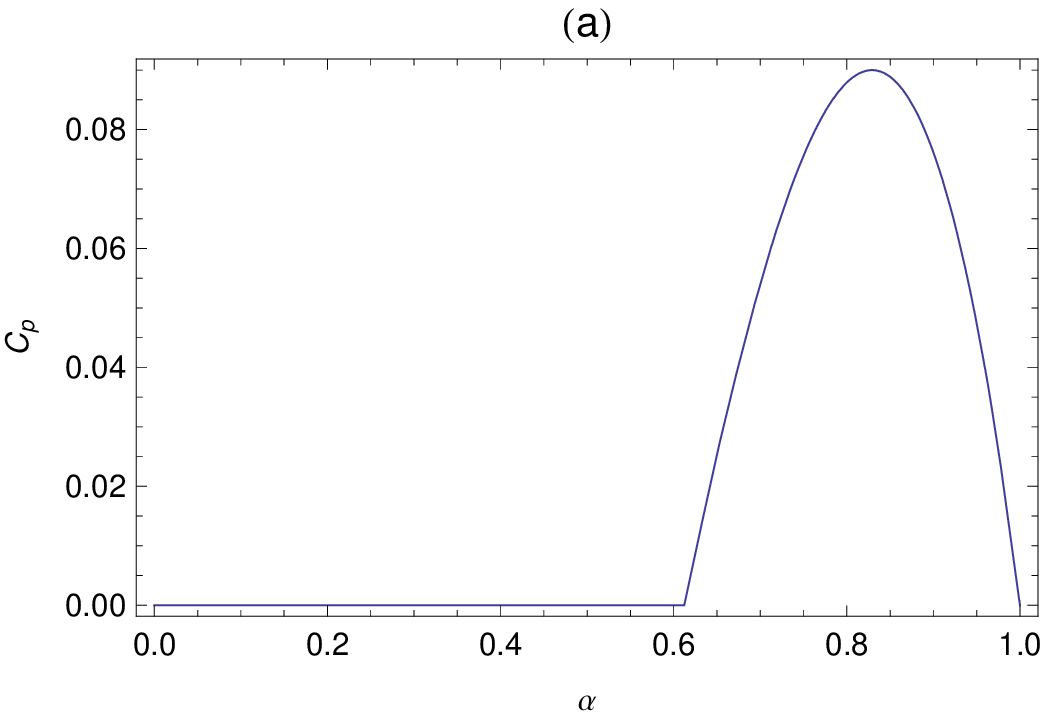}}
\hspace{0.3cm}
\subfigure{\label{aldep3b}\includegraphics[width=6.5cm]{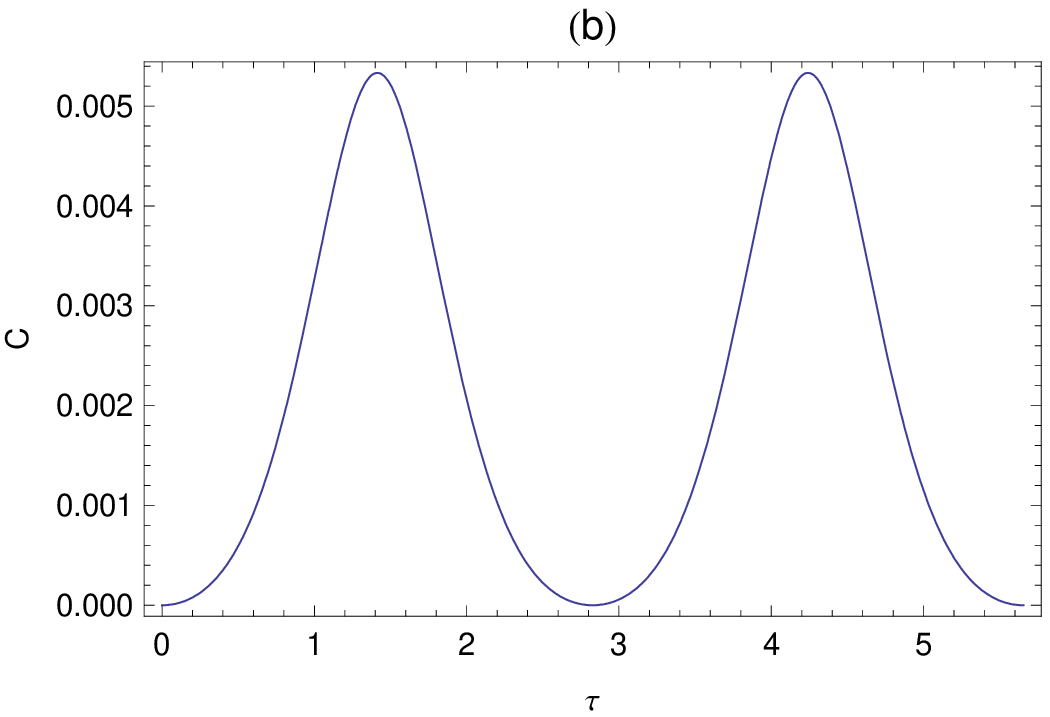}}
\hspace{0.3cm}
\subfigure{\label{aldep3c}\includegraphics[width=6.5cm]{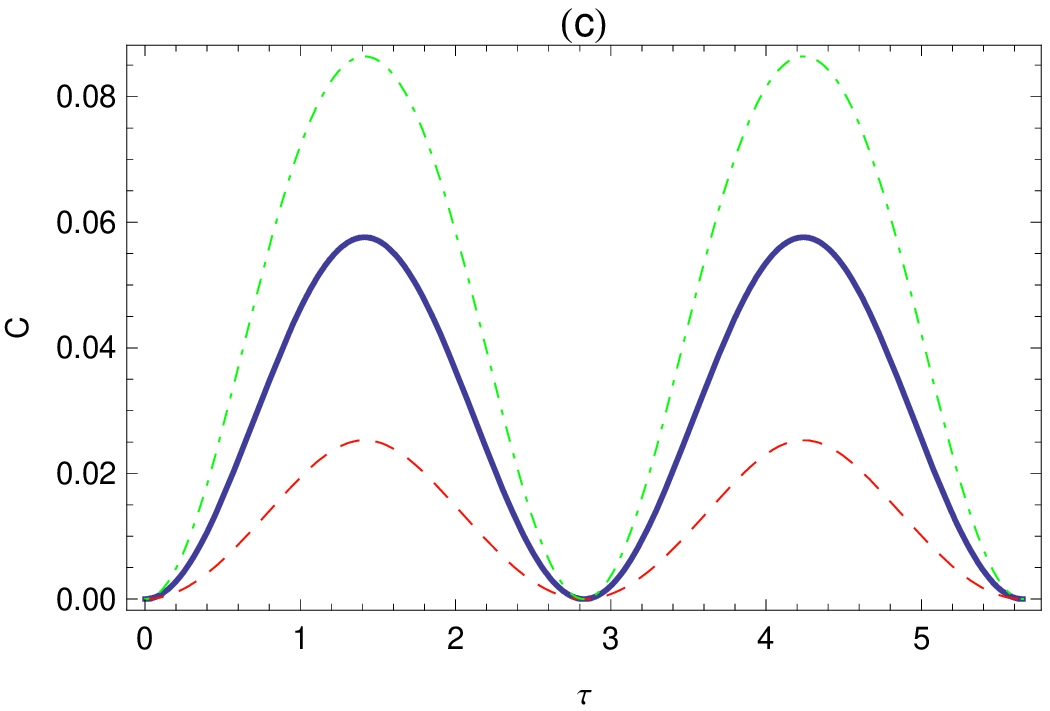}}
\end{center}
\caption{Input state $|\psi_0^B\rangle$, $\bm{Q}=0$, $B_z=0$, $N=9$, $\lambda=1$. (a) The peak concurrence as a function of $\alpha$, showing $\alpha_-(N=9)\approx 0.62$. (b) The concurrence as a function of interaction time for $\alpha=\alpha_-$. Note the concurrence is no longer discontinuous at this value of $\alpha$. (c) The concurrence as a function of interaction time for $\alpha=\frac{1}{\sqrt{2}}$ (solid blue curve), $\alpha=0.65$ (dashed red curve) and $\alpha=\frac{\sqrt{3}}{2}$ (dot-dashed green curve). The optimal neutron polarization is the latter. All quantities are expressed in natural units.}\label{aldep3}
\end{figure}

The only outstanding question is then the relationship between the concurrence and the neutron polarization. Figure \ref{aldep1} shows the peak concurrence as a function of $\alpha$ for different values of $N$, and clearly illustrates that, for each $N$, the concurrence is non-zero for a specific range of $\alpha$. The upper boundary of this range is fixed at $\alpha_{+}=1$; the lower boundary $\alpha_{-}$ increases with $N$ between approximately 0.4 and $\frac{1}{\sqrt{2}}$, converging to $\frac{1}{\sqrt{2}}$ at large $N$. The value of $\alpha$ that maximizes the peak concurrence varies with $N$, and tends to $\alpha\approx \frac{\sqrt{3}}{2}$ in the large $N$ limit. The peak concurrence decreases as $\alpha$ tends to unity, finally returning to zero at $\alpha=1$ (figs. \ref{aldep1}, \ref{aldep2} and \ref{aldep3}).

The absence of entanglement when $\alpha=1$ is evidently due to the fact that when the neutron spin is parallel to the positive $z$-axis the input state is an eigenstate of the interaction Hamiltonian. To explain why similar behaviour is observed at $\alpha=0$, it is useful to examine the state of the system at the different stages of the protocol. The first scattered state can be written as $|\psi_2^B\rangle=\Gamma|7\rangle+\Delta|8\rangle$, and for the protocol to succeed this state must be entangled. The concurrence of the first neutron and the sample can be expressed in closed form as
\begin{equation}
C_1=2|\Gamma\Delta|=4\:\frac{\sqrt{N}|\sin{[\lambda \left(1+\frac{1}{N}\right)\tau]}|\sqrt{N^2+1+2N\cos{[2\lambda \left(1+\frac{1}{N}\right)\tau]}}}{(N+1)^2},
\end{equation}
which is once again a regularly oscillating function of $\tau$ with period $T_{\phi}$. For $N>5$, $C_1$ peaks at time $\tau=T_{\phi}/2$ with a maximum value $4\sqrt{N}(N-1)(N+1)^{-2}$, while for $N\leq5$ the peak concurrence reaches unity. We can verify this is true by showing there exist times at which the square moduli of the coefficients of state $|\psi_2^B\rangle$ are equal to $\frac{1}{2}$, imposing for instance
\begin{equation}
|cd(e^{ix\tau}-e^{iy\tau})|^2=\frac{1}{2}.
\end{equation}
Substituting the values of $c$ and $d$ and solving for $\tau$ then yields
\begin{equation}
\tau=\frac{1}{\lambda}\frac{N}{N+1}\:\sin^{-1}{\left(\frac{N+1}{2\sqrt{2N}}\right)},
\end{equation}
which has real solutions for $N\leq5$. Hence, the concurrence of the first neutron and the sample does not reach unity for $N>5$, but can assume non-zero values for all $N$.

Evidently, then, it is not the first stage of the protocol that is at fault. Let us therefore consider the state of the system after the second scattering event; this takes the form $|\psi_f^B\rangle=\Theta|3\rangle+\Upsilon|5\rangle+\Xi|7\rangle+\Omega|8\rangle$. The concurrence of the neutrons can be expressed analytically as $C=2\left(\left|\Upsilon\Xi\right|-\left|\Theta\Omega\right|\right)$, which substituting the zero-field values of the coefficients takes the rather complicated form
\begin{equation}\label{co psib zero field}
C=\frac{8G(N)\sin^2{\left[\lambda \left(1+\frac{1}{N}\right)\tau\right]}}{(N+1)^4}\:\left[NF(N)-\sqrt{2N(N-1)}G(N)\right],
\end{equation}
with
\begin{eqnarray}
F(N)&=&\sqrt{N^2+1+2N\cos{\left[2\lambda \left(1+\frac{1}{N}\right)\tau\right]}},\\
G(N)&=&\sqrt{N^2-2N+5+4\left(N-1\right)\cos{\left[2\lambda \left(1+\frac{1}{N}\right)\tau\right]}}.
\end{eqnarray}
The pre-factor to equation \eqref{co psib zero field} is always positive. In the limit of large $N$, the function enclosed in square brackets tends to $N^2\left(1-\sqrt{2}\right)$, which is always negative. Therefore, the concurrence between the neutrons must always be null.

In summary, if the applied field is zero the concurrence of the neutrons is determined by the spin density of the sample and the neutron polarization. The peak concurrence is maximal when the neutrons are polarized at approximately $60^{\circ}$ to the quantization axis, but this maximum never exceeds 0.16 and falls roughly as $N^{-1}$.

\subsection{Finite Field Evolution}\label{p0b zero mtm finite field}
\begin{figure}[H]
\renewcommand{\captionfont}{\footnotesize}
\renewcommand{\captionlabelfont}{}
\begin{center}
\subfigure{\label{Bdep0a}\includegraphics[width=6.5cm]{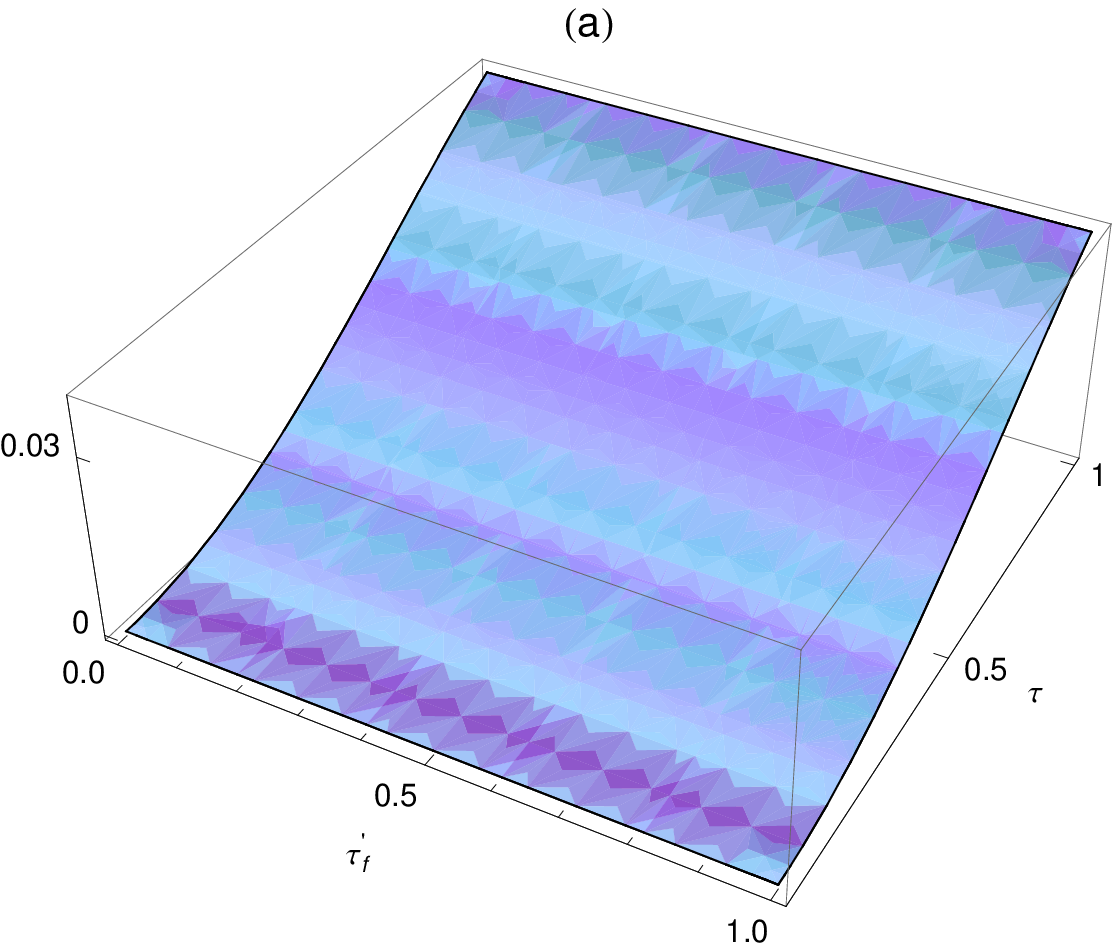}}
\hspace{0.3cm}
\subfigure{\label{Bdep0b}\includegraphics[width=6.5cm]{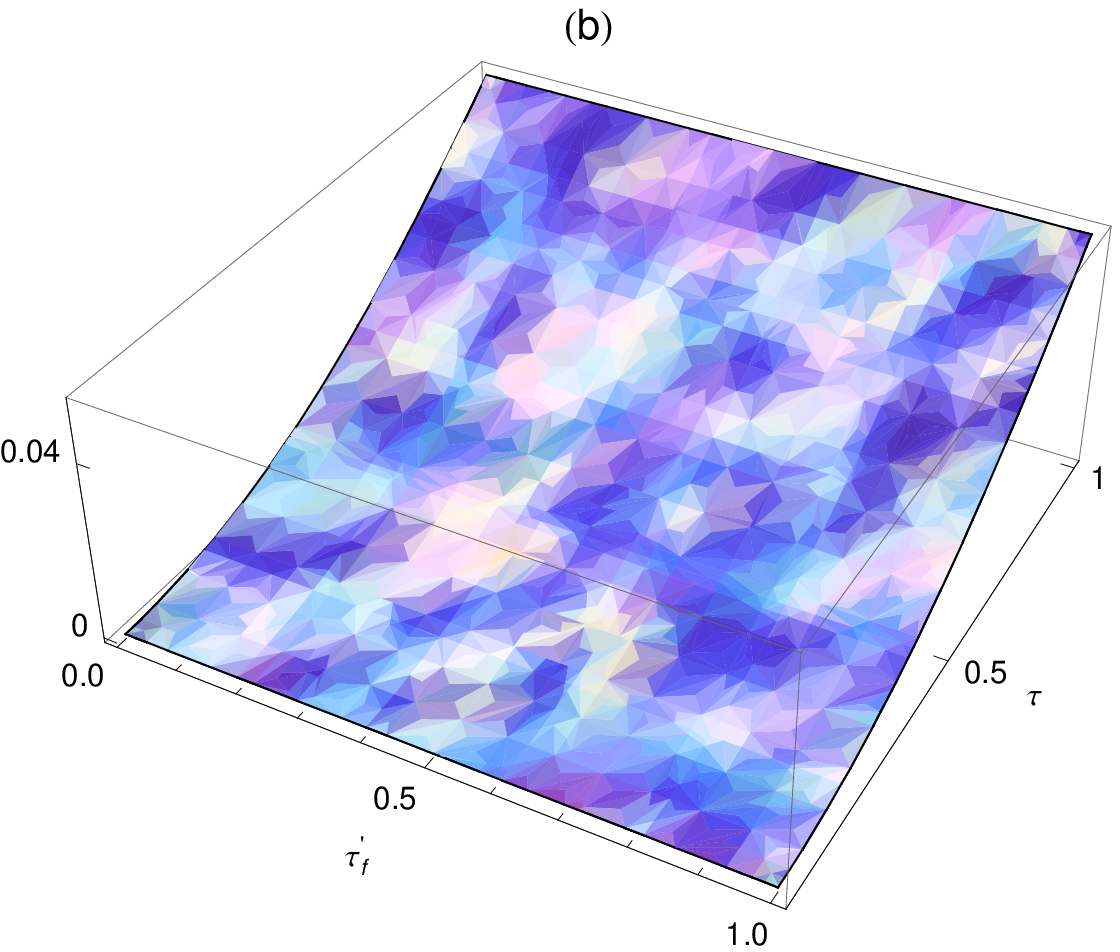}}
\hspace{0.3cm}
\subfigure{\label{Bdep0c}\includegraphics[width=6.5cm]{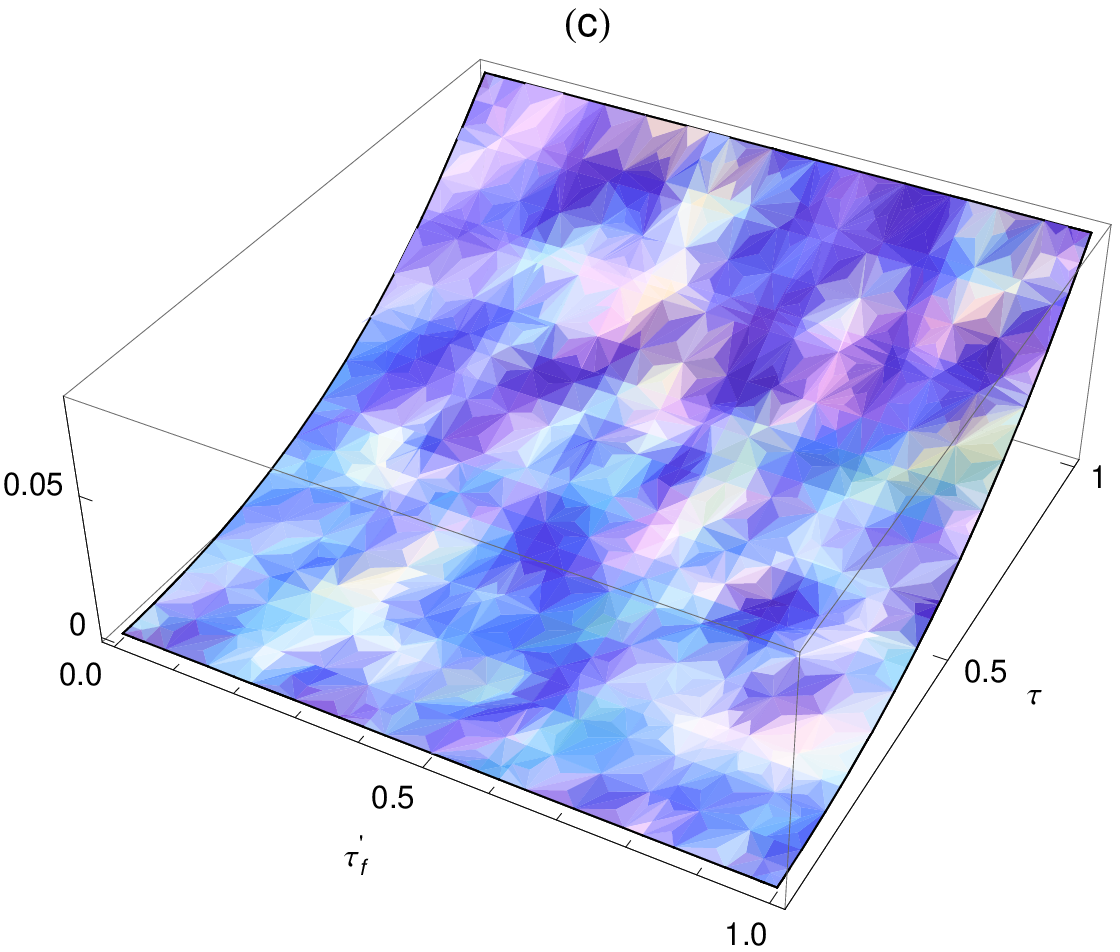}}
\hspace{0.3cm}
\subfigure{\label{Bdep0d}\includegraphics[width=6.5cm]{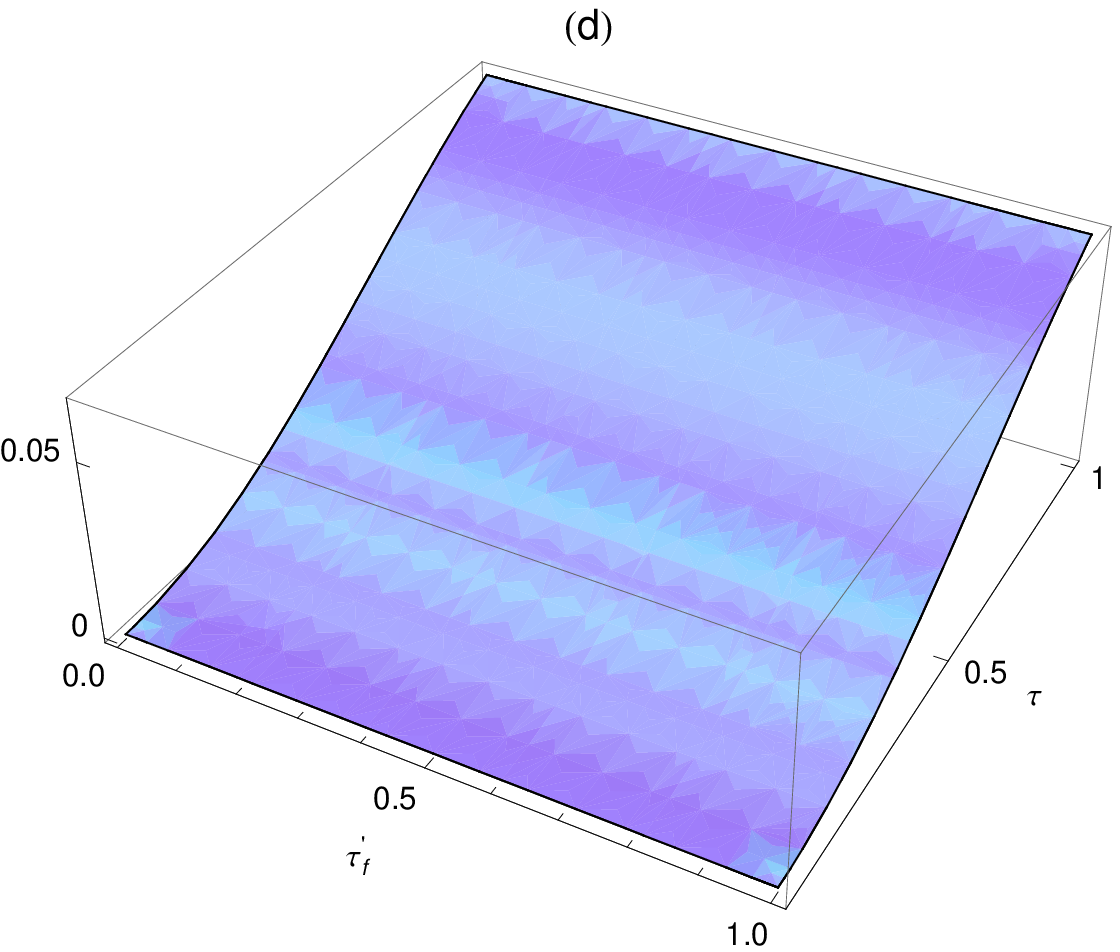}}
\hspace{0.3cm}
\subfigure{\label{Bdep0e}\includegraphics[width=6.5cm]{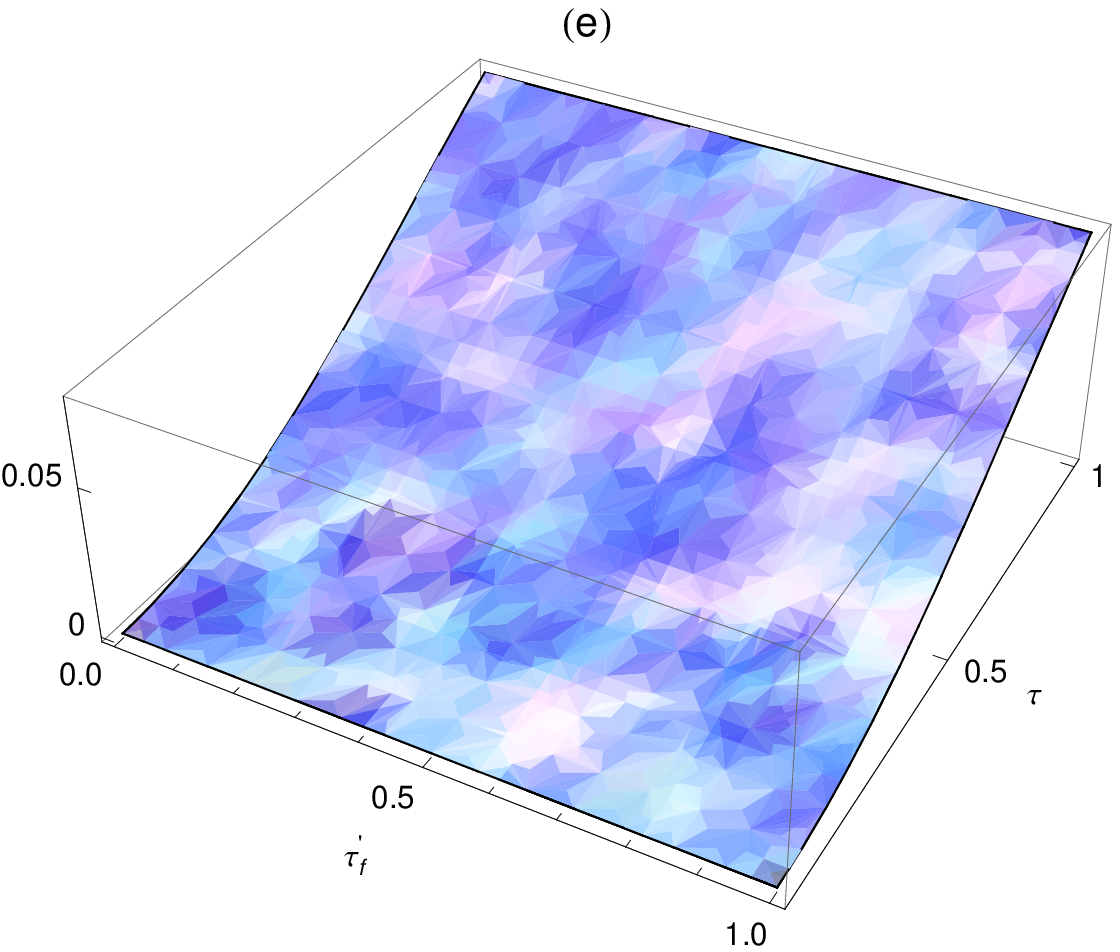}}
\hspace{0.3cm}
\subfigure{\label{Bdep0f}\includegraphics[width=6.5cm]{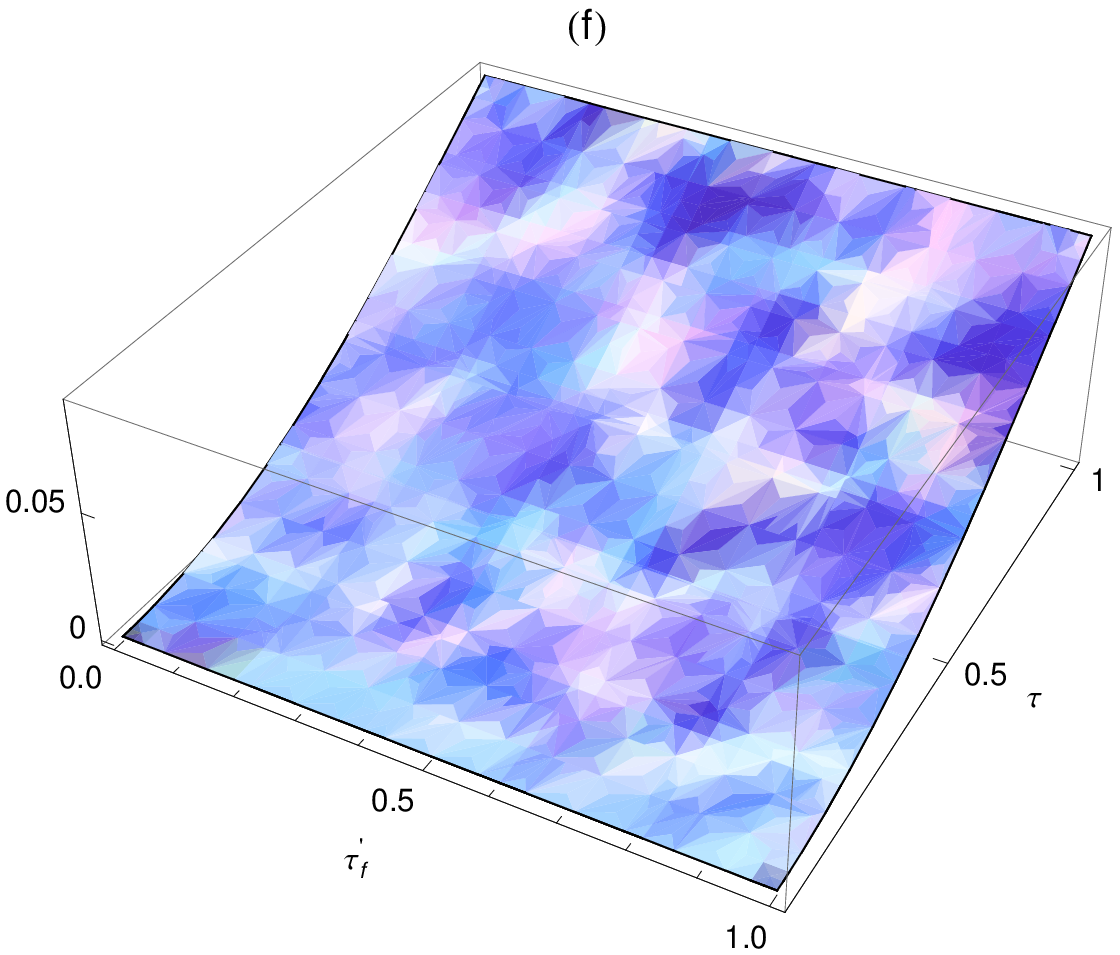}}
\end{center}
\caption{Input state $|\psi_0^B\rangle$, $\bm{Q}=0$, $N=10$, $\lambda=1$. The evolution of the concurrence as a function of $\tau$ and $\tau_f^{\:\prime}$ for different field strengths and neutron polarizations. (a) $B_z=0$, $\alpha=\frac{1}{\sqrt{2}}$, (b) $B_z=0.3$, $\alpha=\frac{1}{\sqrt{2}}$, (c) $B_z=1$, $\alpha=\frac{1}{\sqrt{2}}$, (d) $B_z=0$, $\alpha=\frac{\sqrt{3}}{2}$, (e) $B_z=0.3$, $\alpha=\frac{\sqrt{3}}{2}$, (f) $B_z=1$, $\alpha=\frac{\sqrt{3}}{2}$. All quantities are expressed in natural units.}\label{Bdep0}
\end{figure}

\begin{figure}[H]
\renewcommand{\captionfont}{\footnotesize}
\renewcommand{\captionlabelfont}{}
\begin{center}
\subfigure{\label{Bdep1a}\includegraphics[width=6.5cm]{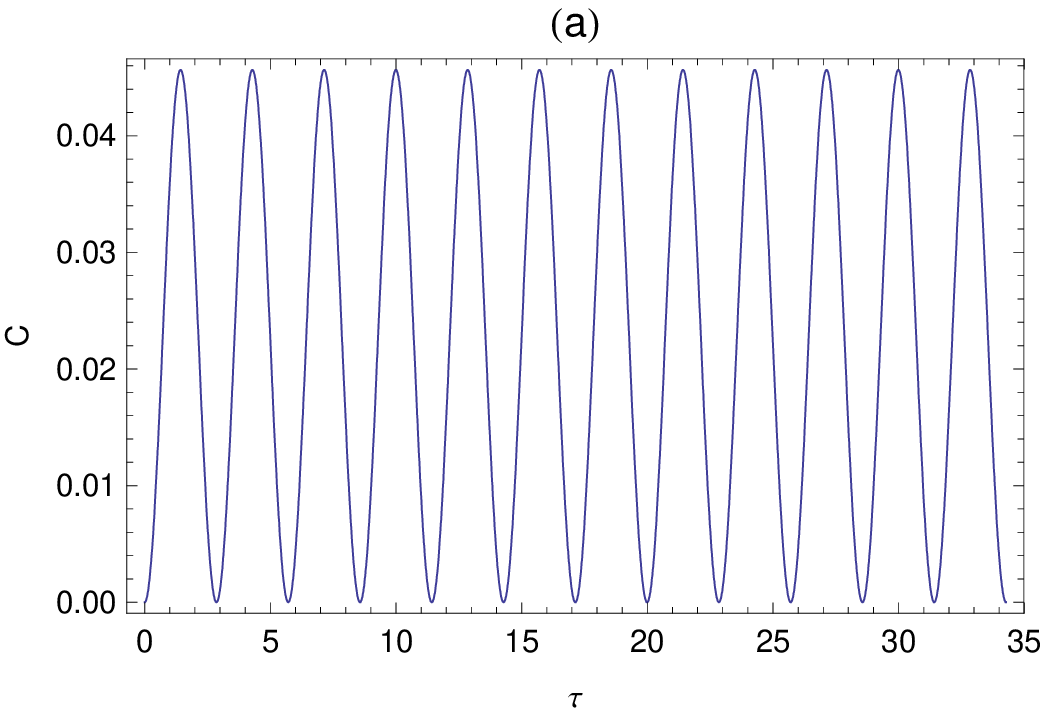}}
\hspace{0.3cm}
\subfigure{\label{Bdep1b}\includegraphics[width=6.5cm]{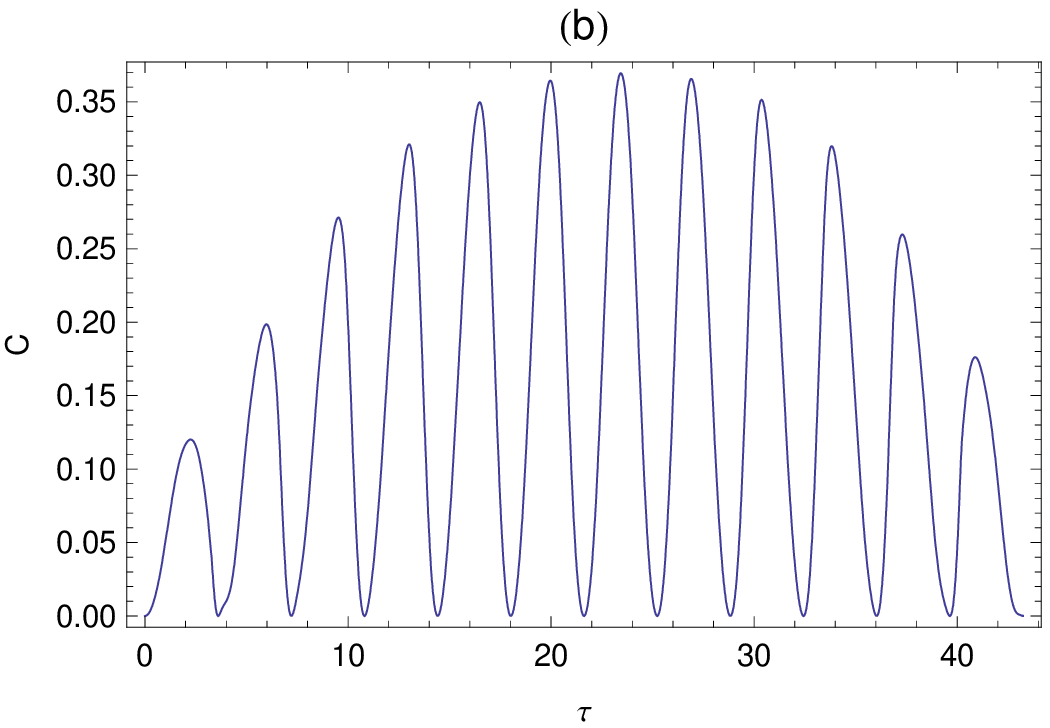}}
\hspace{0.3cm}
\subfigure{\label{Bdep1c}\includegraphics[width=6.5cm]{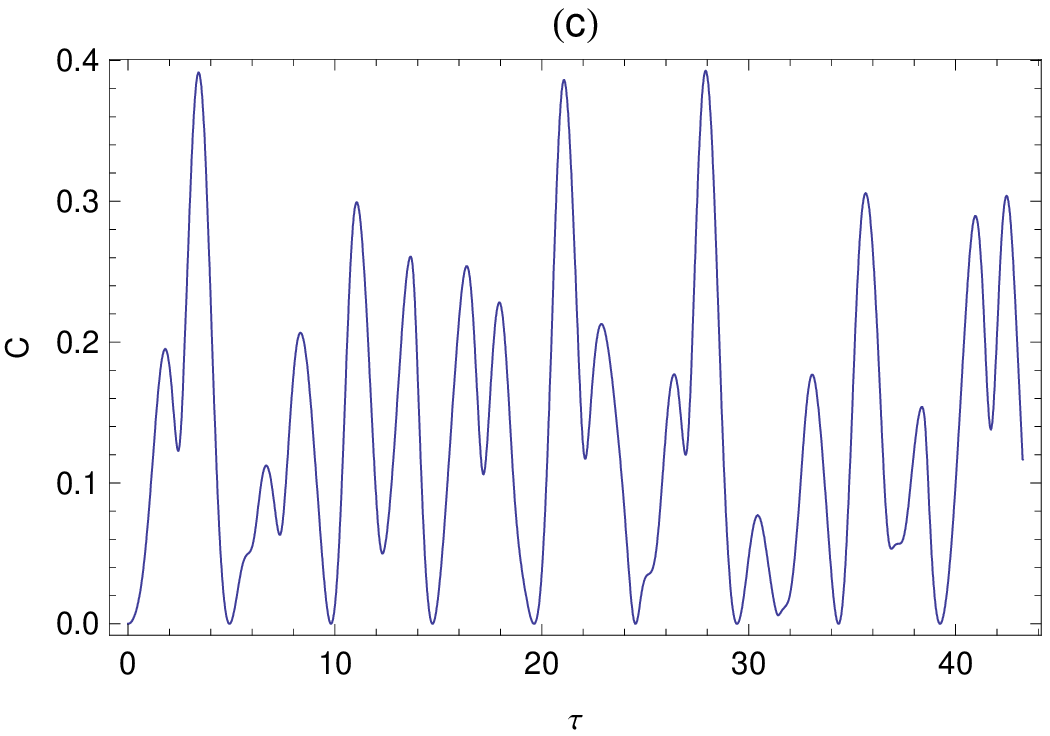}}
\hspace{0.3cm}
\subfigure{\label{Bdep1d}\includegraphics[width=6.5cm]{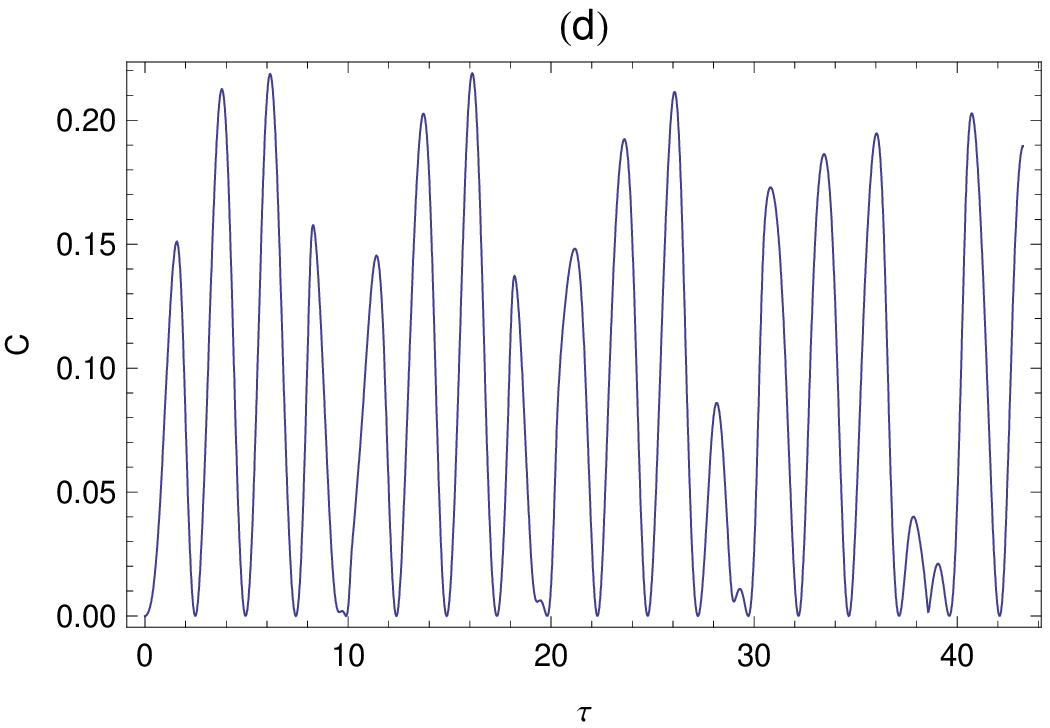}}
\hspace{0.3cm}
\subfigure{\label{Bdep1e}\includegraphics[width=6.5cm]{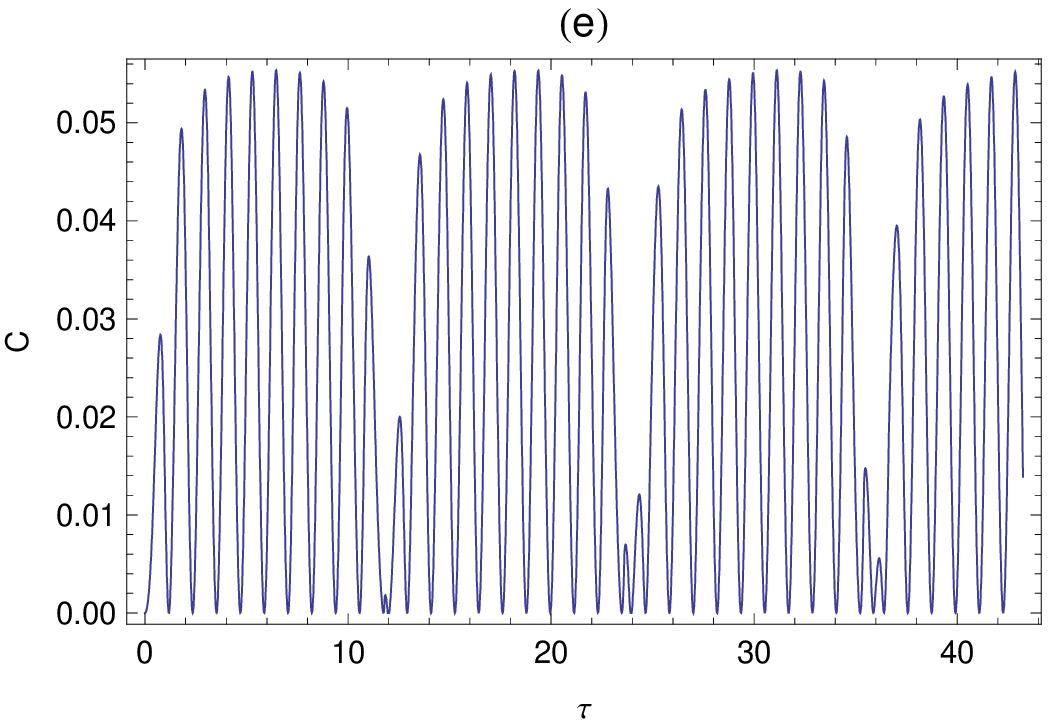}}
\end{center}
\caption{Input state $|\psi_0^B\rangle$, $\bm{Q}=0$, $\alpha=0.7$, $N=10$, $\lambda=1$. The evolution of the concurrence with interaction time at different $B_z$, showing the existence of a limiting field of approximately $3\lambda$. (a) $B_z=0$, (b) $B_z=0.3$, (c) $B_z=1$, (d) $B_z=2$, (e) $B_z=3.5$. All quantities are expressed in natural units. }\label{Bdep1}
\end{figure}

\begin{figure}[H]
\renewcommand{\captionfont}{\footnotesize}
\renewcommand{\captionlabelfont}{}
\begin{center}
\subfigure{\label{Bdep2a}\includegraphics[width=6.5cm]{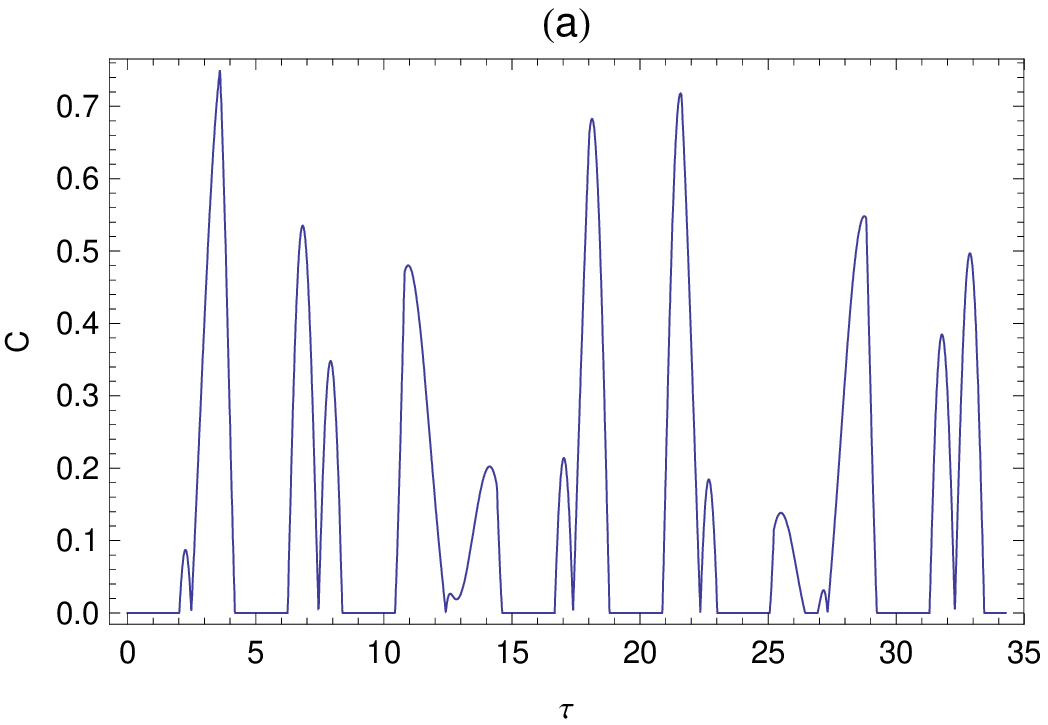}}
\hspace{0.3cm}
\subfigure{\label{Bdep2b}\includegraphics[width=6.5cm]{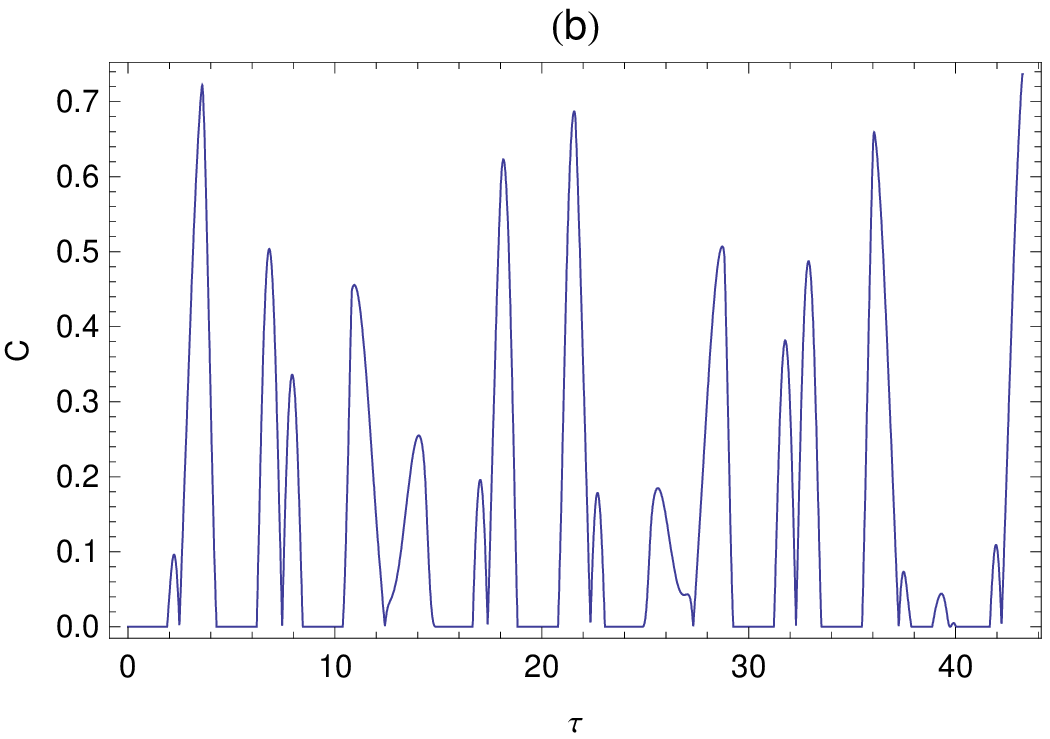}}
\hspace{0.3cm}
\subfigure{\label{Bdep2c}\includegraphics[width=6.5cm]{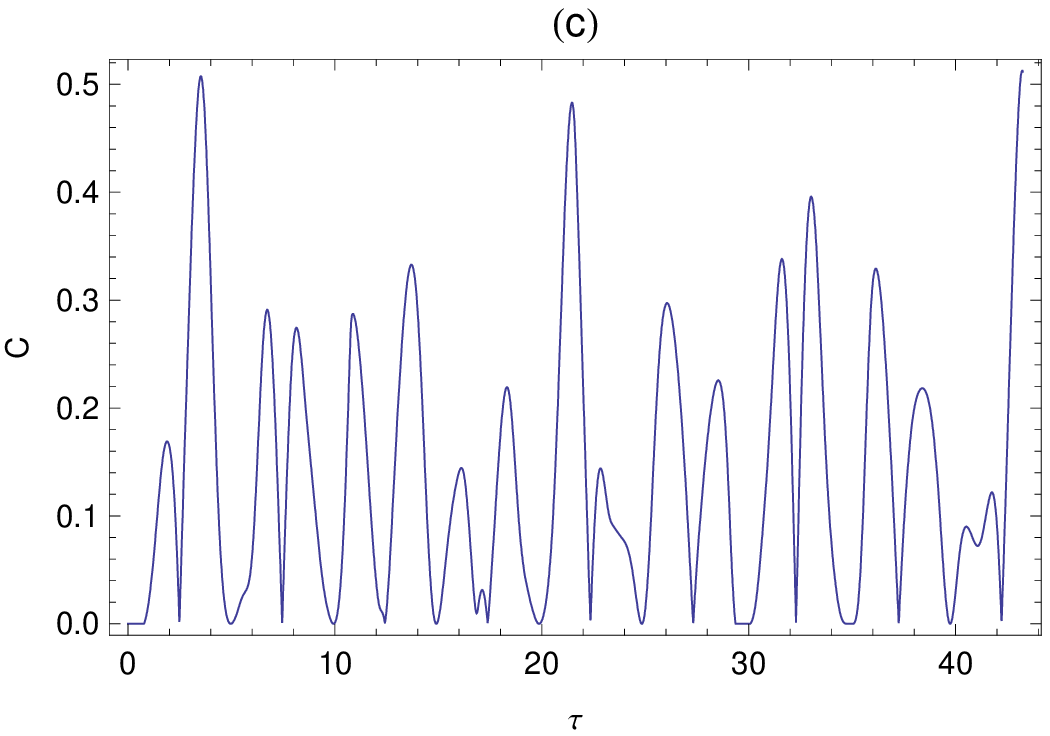}}
\hspace{0.3cm}
\subfigure{\label{Bdep2d}\includegraphics[width=6.5cm]{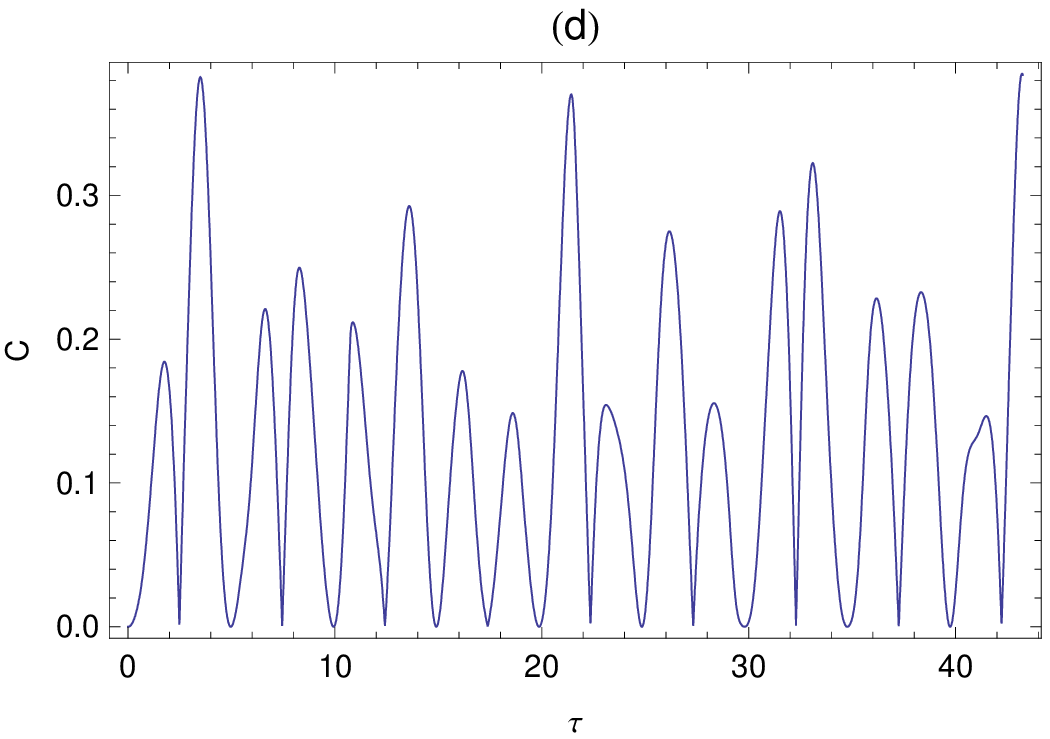}}
\hspace{0.3cm}
\subfigure{\label{Bdep2e}\includegraphics[width=6.5cm]{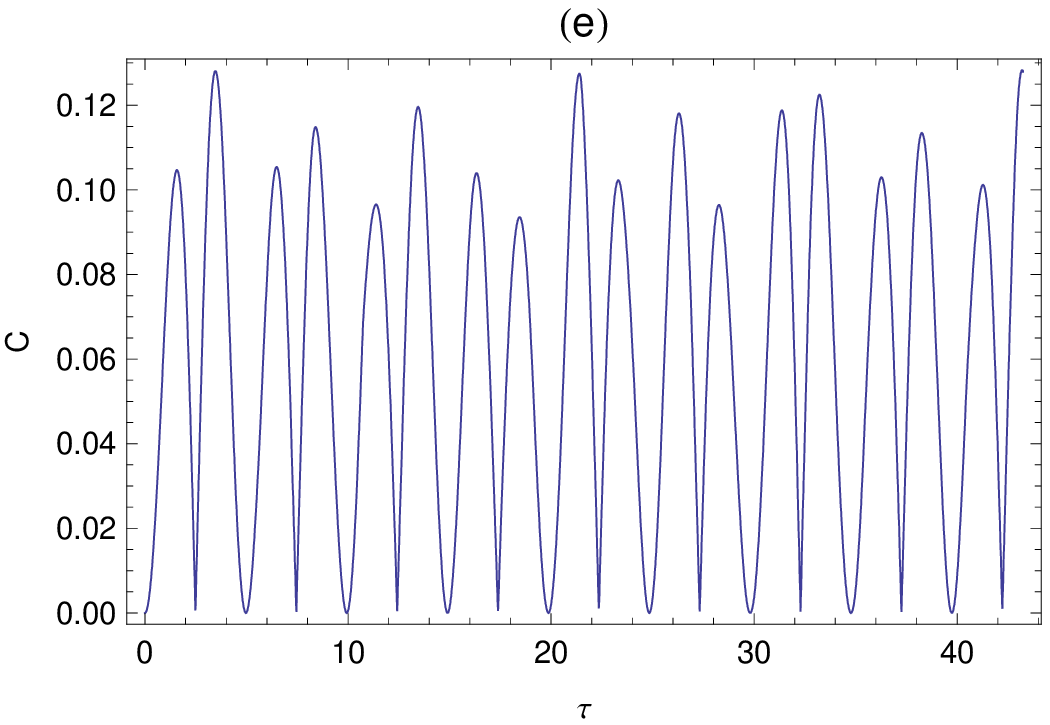}}
\hspace{0.3cm}
\subfigure{\label{Bdep2f}\includegraphics[width=6.5cm]{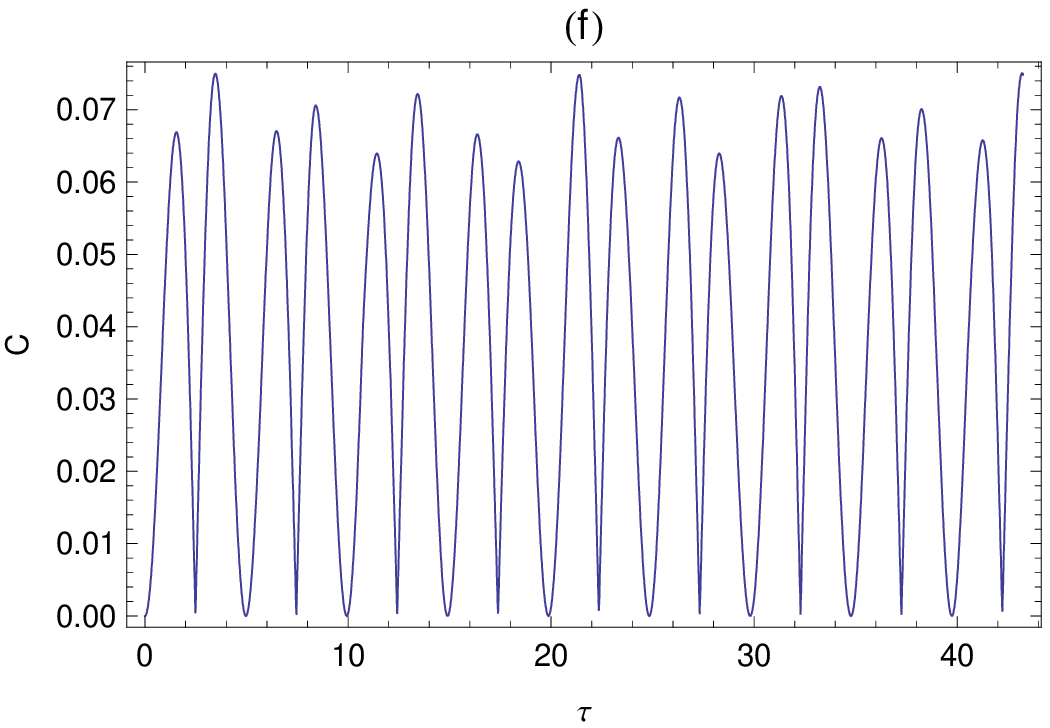}}
\end{center}
\caption{Input state $|\psi_0^B\rangle$, $\bm{Q}=0$, $\lambda=1$, $B_z=\frac{9}{10}$, $N=10$. The evolution of the concurrence with interaction time at different $\alpha$ for (a) $\alpha=0$, (b) $\alpha=0.2$, (c) $\alpha=\frac{1}{\sqrt{3}}$, (d) $\alpha=\frac{1}{\sqrt{2}}$, (e) $\alpha=\frac{\sqrt{5}}{\sqrt{6}}$, and (f) $\alpha=0.95$. Note the concurrence is finite for all $\alpha\neq1$. All quantities are expressed in natural units.}\label{Bdep2}
\end{figure}

Let us now consider the effect of applying a finite field. Bearing in mind the results of sections \ref{p0a zero mtm finite field} and \ref{p0b zero mtm zero field}, there are several features we might expect to see:
\begin{enumerate}
\item Independence of the concurrence from $\tau_f^{\prime}$;
\item Discontinuities and null concurrence when $\alpha<\alpha_{-}$;
\item Increase or decrease of the peak concurrence according to the applied field and the neutron polarization;
\item Maximal concurrence when $B_z=B_z^*$;
\end{enumerate}
The first point holds for all $N$ and $B_z$, hence the concurrence is once again independent of the time between scattering events (see fig. \ref{Bdep0}). It is also undoubtedly true that if $\alpha=1$ the neutrons must remain unentangled, because the initial state of the system is an eigenstate of the interaction potential. In addition, we know the final point must apply, owing to to structure of the potential. The remaining items are easily verified from figures \ref{Bdep1} and \ref{Bdep2}, which illustrate the evolution of the concurrence in time for different field strengths and polarizations. We observe that, for fixed $\alpha$, the peak concurrence $\mathcal{C}_p$ improves when a field is applied, provided the strength of the field does not exceed a rough upper limit of $3\lambda$. Surprisingly, however, the concurrence in finite field is now non-zero for all $\alpha\neq1$.

\begin{figure}[H]
\renewcommand{\captionfont}{\footnotesize}
\renewcommand{\captionlabelfont}{}
\begin{center}
\subfigure{\label{Bdep3a}\includegraphics[width=6.5cm]{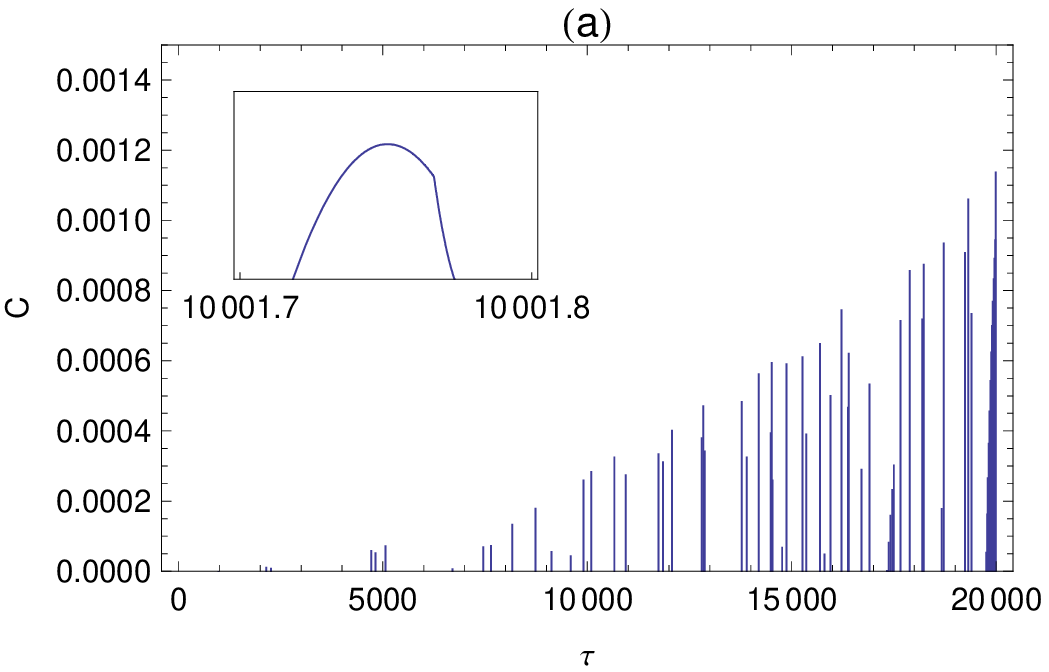}}
\hspace{0.3cm}
\subfigure{\label{Bdep3b}\includegraphics[width=6.5cm]{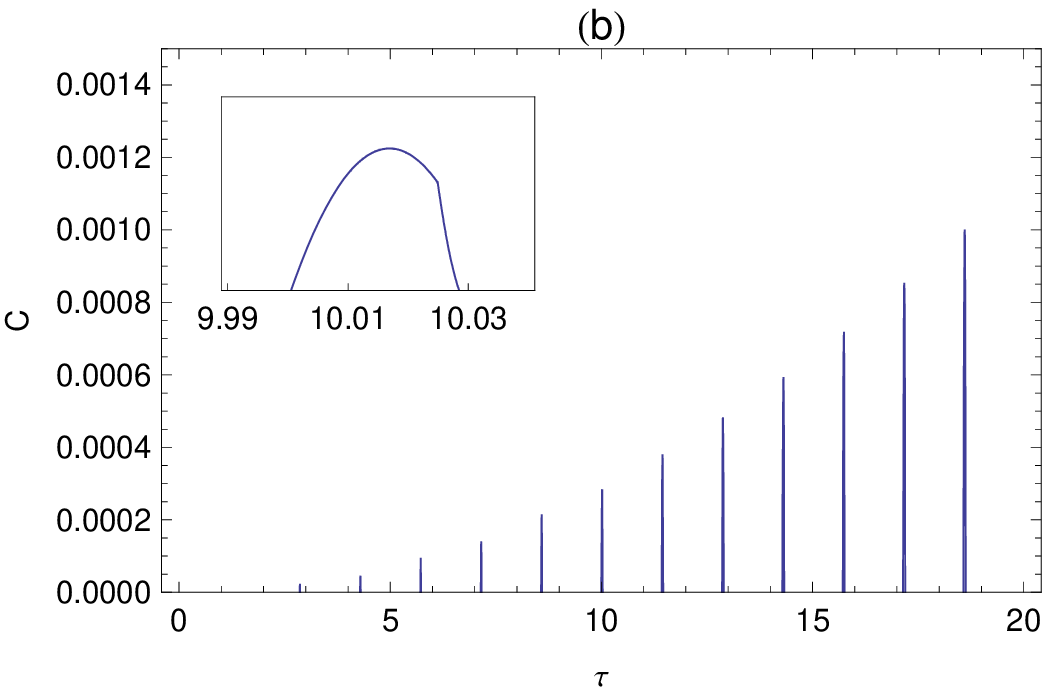}}
\hspace{0.3cm}
\subfigure{\label{Bdep3c}\includegraphics[width=6.5cm]{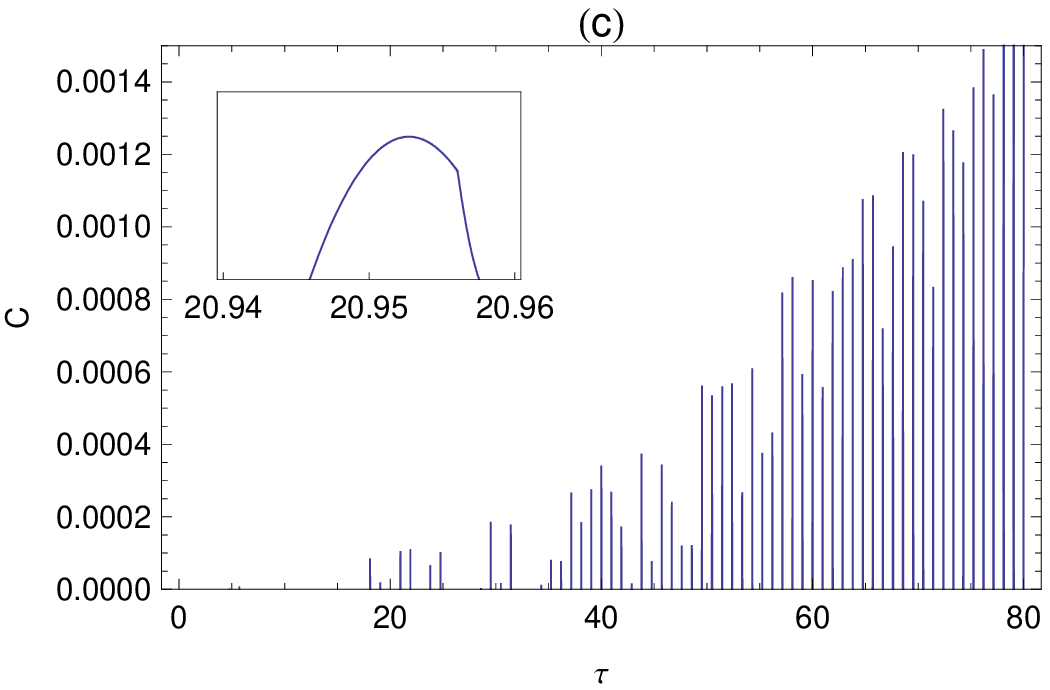}}
\end{center}
\caption{Input state $|\psi_0^B\rangle$, $\bm{Q}=0$, $\alpha=0$, $N=10$. The development of a non-zero concurrence as $B_z\tau$ exceeds the limiting value $0.1\lambda$. (a) $B_z=10^{-5}\lambda$, $\lambda=1$, (b) $B_z=0.01\lambda$, $\lambda=2$, (c)  $B_z=0.003\lambda$, $\lambda=3$. The insets show that, on a sufficiently small timescale, the `spikes' pictured have finite width. All quantities are expressed in natural units.}\label{Bdep3}
\end{figure}

\begin{figure}[H]
\renewcommand{\captionfont}{\footnotesize}
\renewcommand{\captionlabelfont}{}
\begin{center}
 \includegraphics[width=8cm]{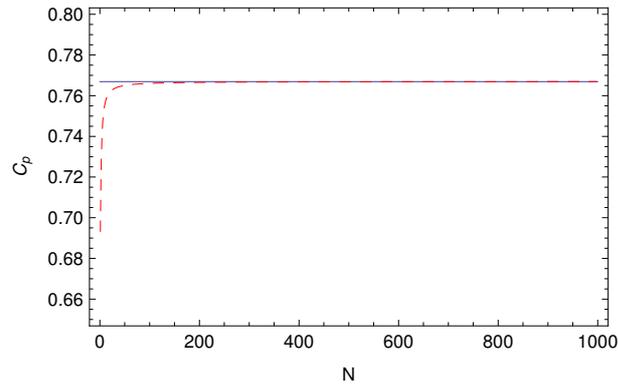}
 \end{center}
 \caption{Input state $|\psi_0^B\rangle$, $\bm{Q}=0$, $\alpha=0$, $B_z=B_z^*$, $\lambda=1$. The peak concurrence as a function of $N$ (red dashed line), which at large $N$ converges to 0.77 (blue solid line). All quantities are expressed in natural units.} \label{Bdep4}
\end{figure}

In fact, $\mathcal{C}_p$ is maximal when $\alpha=0$. To explain this behaviour, it is instructive to analyze the state of the system at each stage of the protocol. From equation (\ref{psifb}), neglecting the state of the second neutron, the first scattered state $|\psi_2^B\rangle$ can be written as
\begin{equation}
|\psi_2^B\rangle=\alpha|00\rangle+\beta cd\left(e^{ix\tau}-e^{iy\tau}\right)|01\rangle+\beta\left(d^2e^{ix\tau}+c^2e^{iy\tau}\right)|10\rangle.
\end{equation}
%with:
%\begin{eqnarray}
%x&=&2\lambda\sqrt{N}(\sqrt{N}+1),\\
%y&=&2\lambda\sqrt{N}(\sqrt{N}-1).
%\end{eqnarray}
Evidently, for a given $c$ and $d$ the entanglement of this state is directly proportional to $\beta$, and is maximal if $\beta=1$. Therefore, to optimize the first stage of the protocol the neutron spin must be anti-aligned with the $z$-axis. Let us then assume $\beta=1$ and consider the state of the system after the second scattering event; this has the form $|\psi_f^B\rangle=\Theta^{\prime}|3\rangle+\Upsilon^{\prime}|5\rangle+\Xi^{\prime}|7\rangle+\Omega^{\prime}|8\rangle$. As before, the concurrence of the neutrons is $C=2\left(\left|\Upsilon^{\prime}\Xi^{\prime}\right|-\left|\Theta^{\prime}\Omega^{\prime}\right|\right)$, hence substituting the values of the coefficients at finite $B_z$ yields
\begin{eqnarray}\label{c psib field 1}
C&=&-\frac{32N^3\lambda^2\sqrt{\varepsilon}}{(4\lambda^2+N\varphi^2)^3\left[8\lambda^2(N-1)+N^2\vartheta^2\right]}\bigg\lbrack\sqrt{2\varepsilon N(N-1)}|\varphi\vartheta\sin{\phi\tau}\sin{\gamma\tau}|\nonumber\\
&-&\left.N\varphi^2\sin^2{\phi\tau}\sqrt{\left[\frac{8\lambda^2\left(N-1\right)}{N^2}+\vartheta^2\right]^2-32\left(N-1\right)\vartheta^2\frac{\lambda^2}{N^2}\sin^2{\gamma\tau}}\right],
\end{eqnarray}
with
\begin{equation}
\varepsilon=\frac{16\lambda^4}{N^2}+\varphi^4+\frac{8\lambda^2}{N}\varphi^2\cos{2\phi\tau}.
\end{equation}
Both terms in the square brackets are always positive. The concurrence will therefore be finite if there exist values of $B_z$ and $\tau$ for which the magnitude of the second term is greater than that of the first. Let us set a lower limit for the concurrence of $C_{\mathrm{min}}=10^{-3}$. For any arbitrarily small value $B_z<B_z^*$, $C$ exceeds $C_{\mathrm{min}}$ at certain values of $\tau$ satisfying $B_z\tau\gtrsim 0.1\lambda$. Not all values of $\tau$ described by this condition yield a finite concurrence, because $C$ is not a smooth function of time (fig. \ref{Bdep3}). Equation (\ref{c psib field 1}) peaks when the difference between the bracketed terms is a maximum, which for $B_z=B_z^*$ first occurs at time
\begin{equation}
\tau^*=\frac{\pi N}{2\lambda\sqrt{2N-1}}.
\end{equation}
Hence, as for initial state $|\psi_0^A\rangle$, setting $\tau=\tau^*$ the peak concurrence converges to 0.77 in the limit of large $N$ (fig. \ref{Bdep4}).

\begin{figure}[H]
\renewcommand{\captionfont}{\footnotesize}
\renewcommand{\captionlabelfont}{}
\begin{center}
\subfigure{\label{Bdep5a}\includegraphics[width=6.5cm]{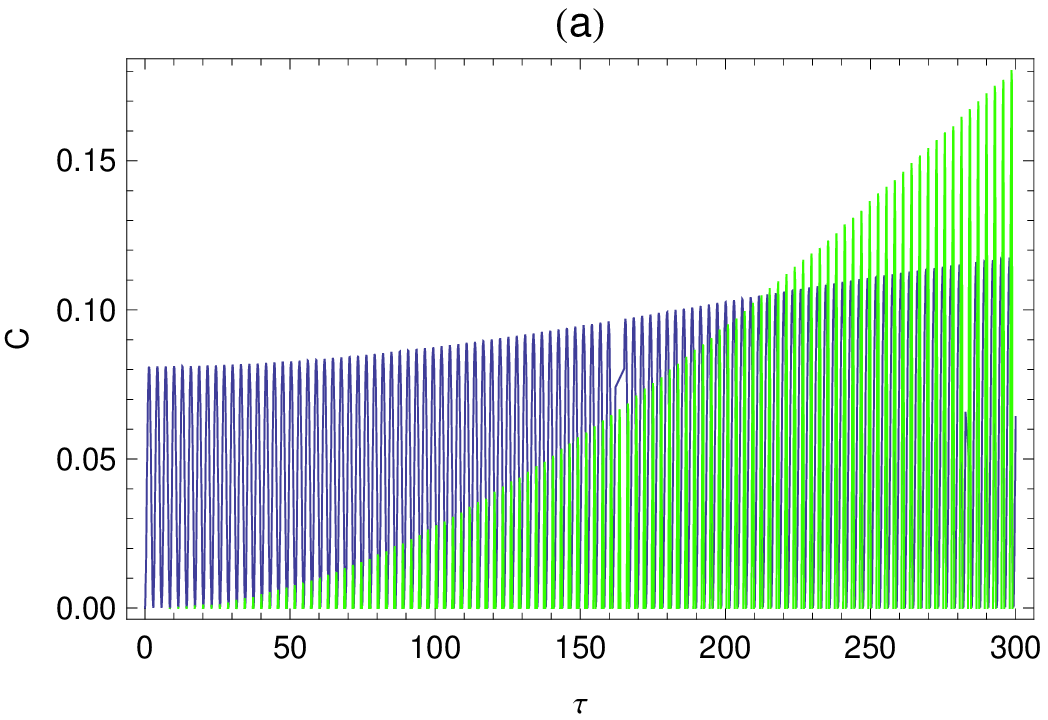}}
\hspace{0.3cm}
\subfigure{\label{Bdep5b}\includegraphics[width=6.5cm]{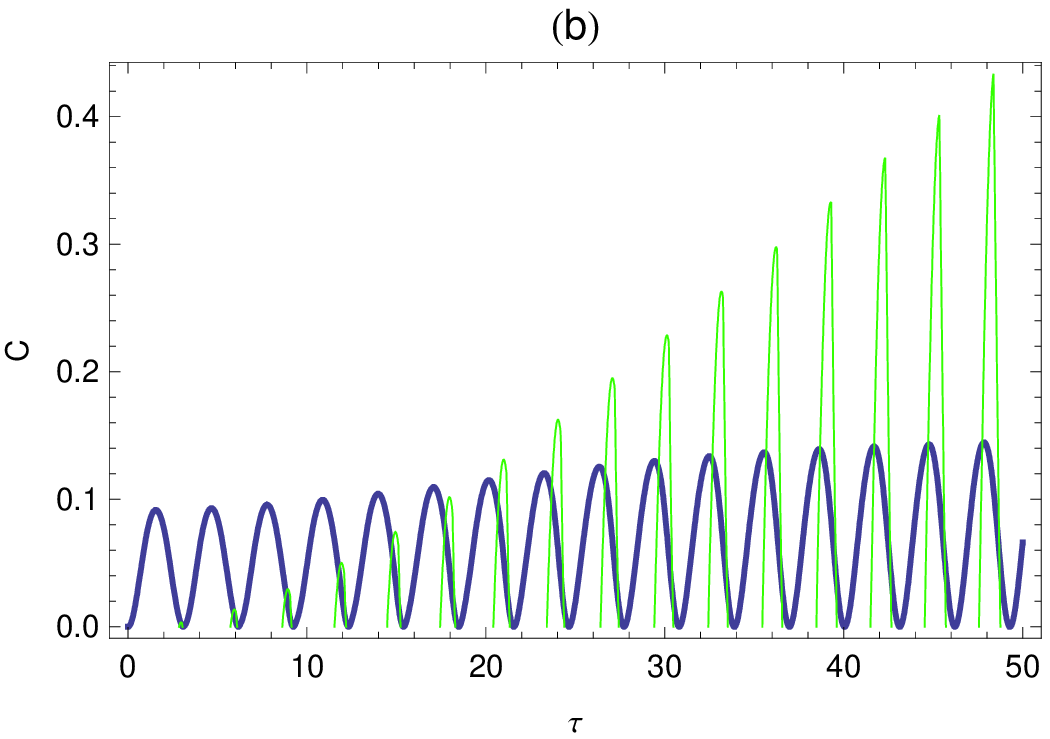}}
\hspace{0.3cm}
\subfigure{\label{Bdep5c}\includegraphics[width=6.5cm]{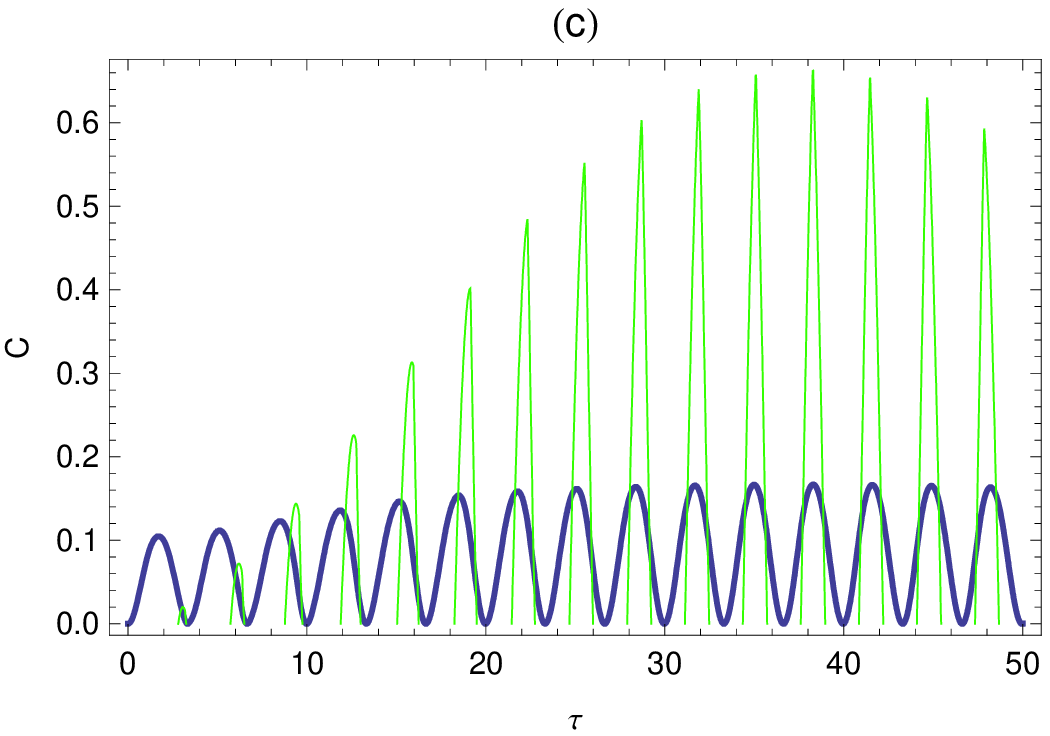}}
\end{center}
\caption{Input state $|\psi_0^B\rangle$, $\bm{Q}=0$, $N=10$, $\lambda=1$. The evolution of the concurrence at short times for $\alpha=\frac{\sqrt{3}}{2}$ (blue curve) and $\alpha=0$ (green curve) at (a) $B_z=0.01$, (b) $B_z=0.1$, and (c) $B_z=0.2$. All quantities are expressed in natural units.}\label{Bdep5}
\end{figure}

We conclude that, from a purely mathematical viewpoint, the performance of the system is optimal when the neutrons are polarized in the negative $z$-direction. However, in practical terms this may not be the most advantageous choice, because when the field is weak the concurrence at short times is poor. Comparing the evolution of the concurrence for different field strengths at $\alpha=0$ and $\alpha=\frac{\sqrt{3}}{2}$ illustrates that, for $B_z\tau\lesssim2\lambda$, the optimal spin orientation is in fact the latter (fig. \ref{Bdep5}).

\begin{figure}[H]
\renewcommand{\captionfont}{\footnotesize}
\renewcommand{\captionlabelfont}{}
\begin{center}
\subfigure{\label{Bdep6a}\includegraphics[width=6.5cm]{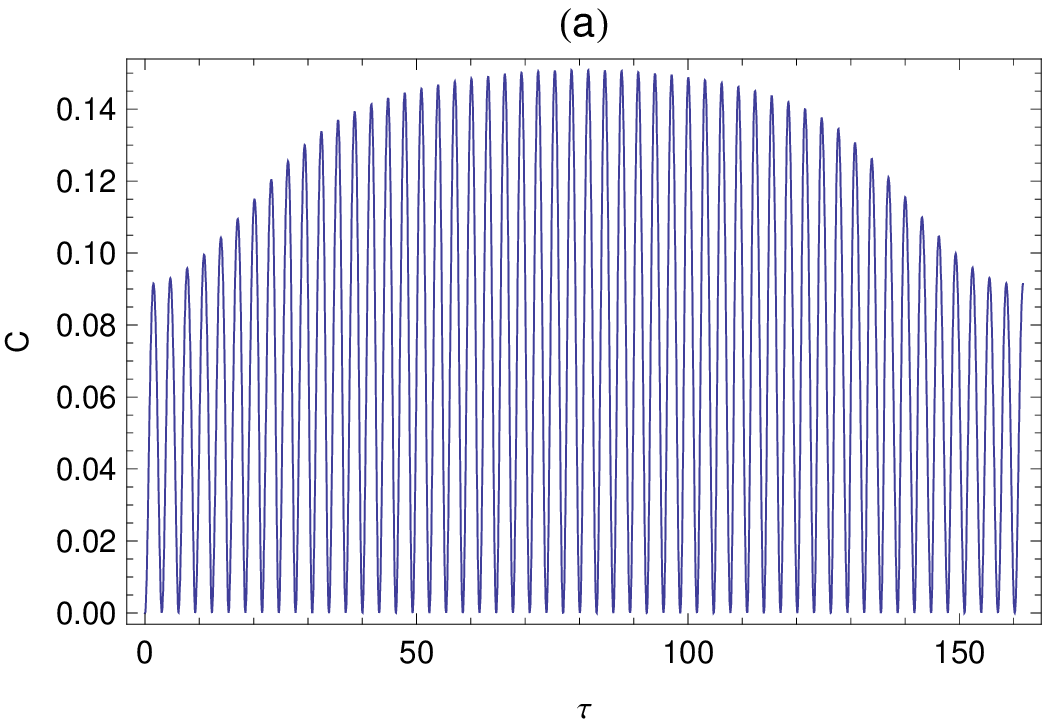}}
\hspace{0.3cm}
\subfigure{\label{Bdep6b}\includegraphics[width=6.5cm]{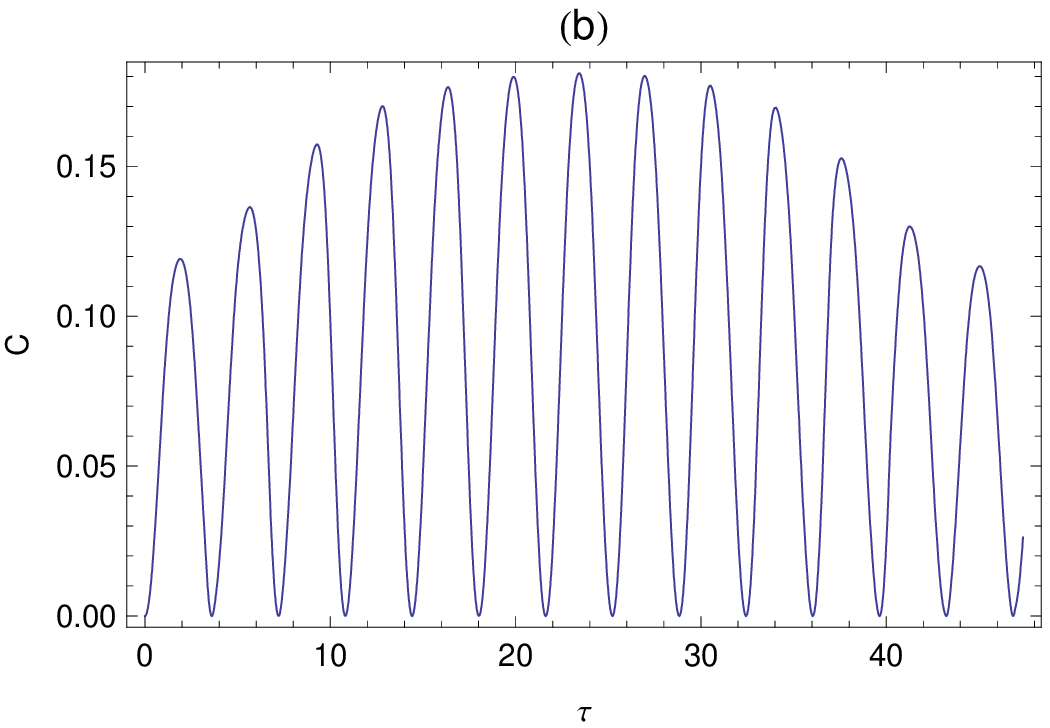}}
\hspace{0.3cm}
\subfigure{\label{Bdep6c}\includegraphics[width=6.5cm]{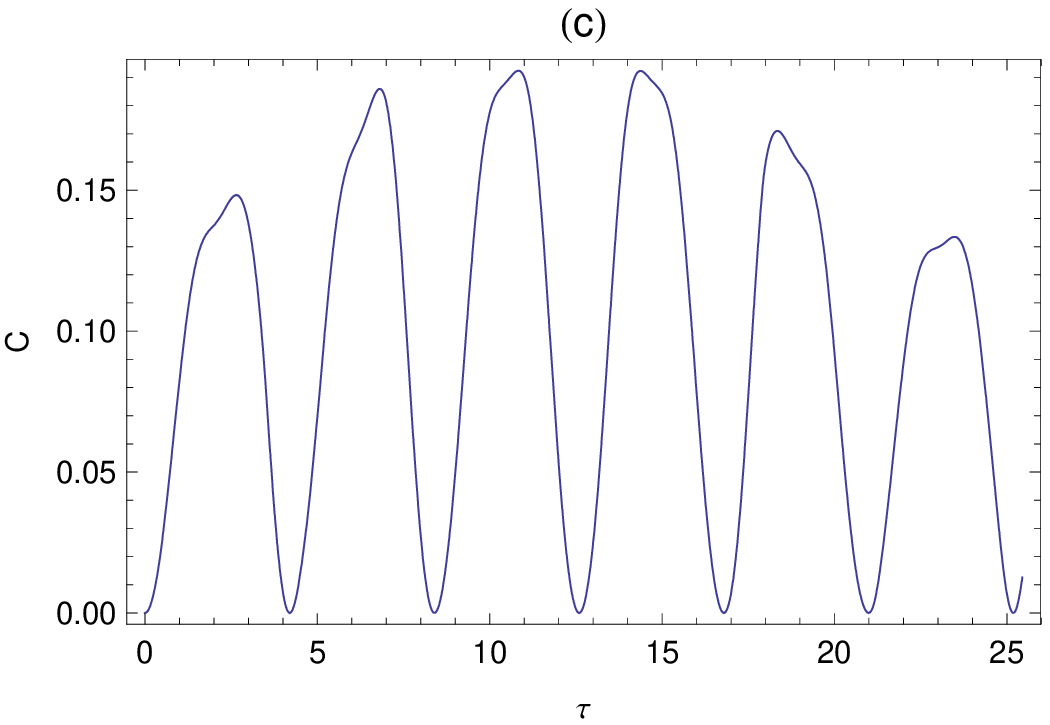}}
\hspace{0.3cm}
\subfigure{\label{Bdep6d}\includegraphics[width=6.5cm]{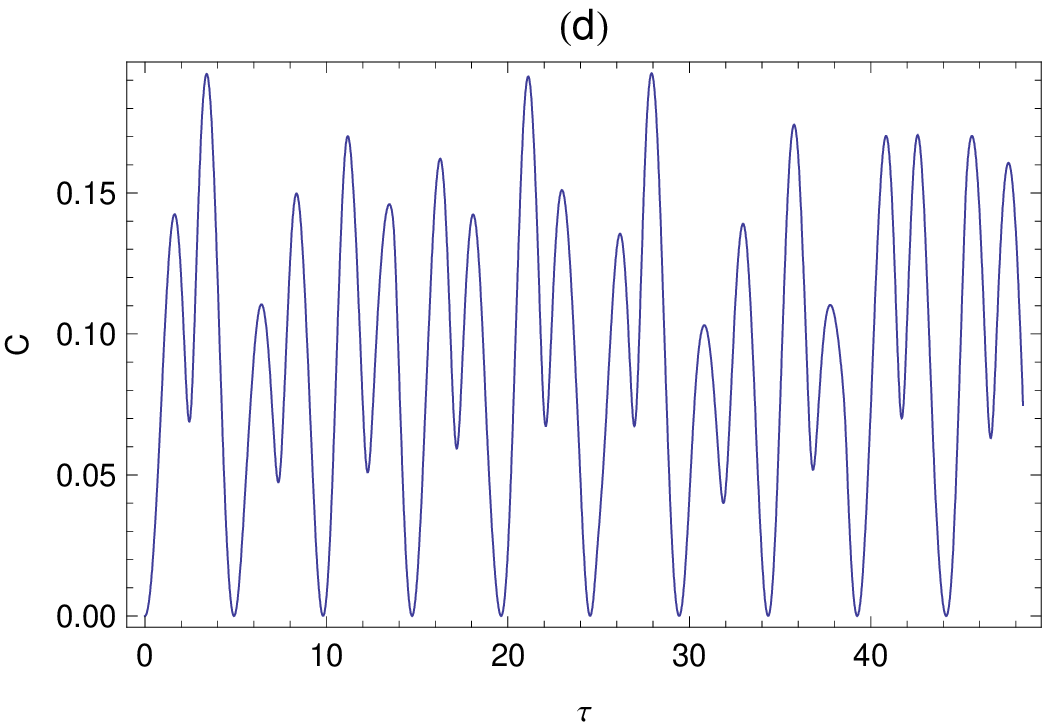}}
\hspace{0.3cm}
\subfigure{\label{Bdep6e}\includegraphics[width=6.5cm]{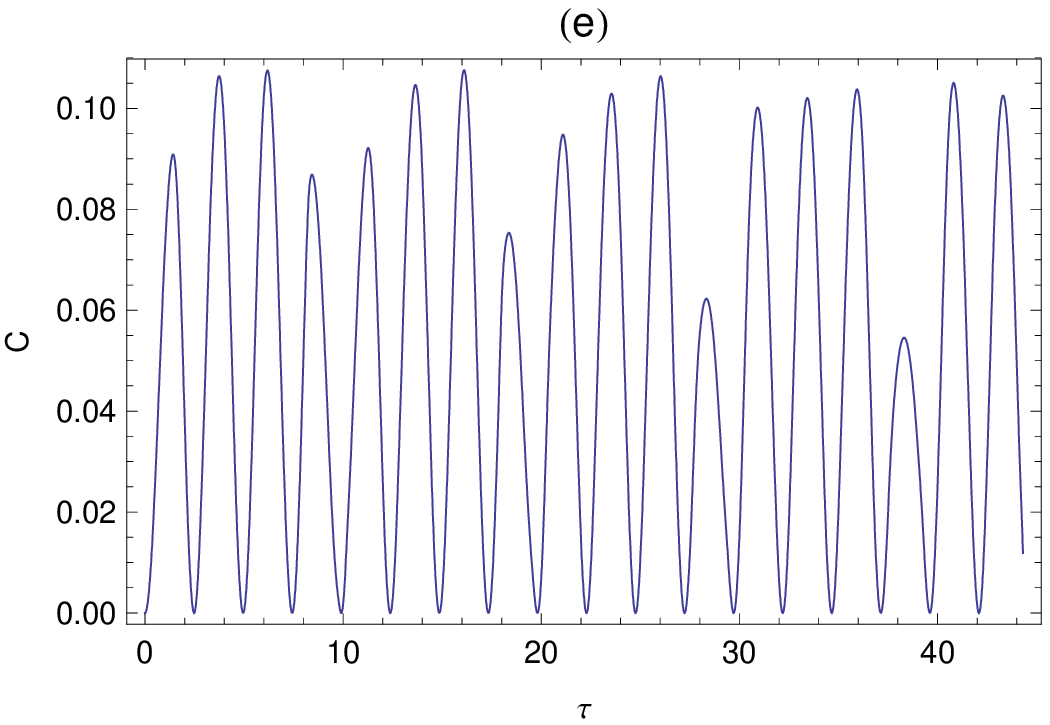}}
\hspace{0.3cm}
\subfigure{\label{Bdep6f}\includegraphics[width=6.5cm]{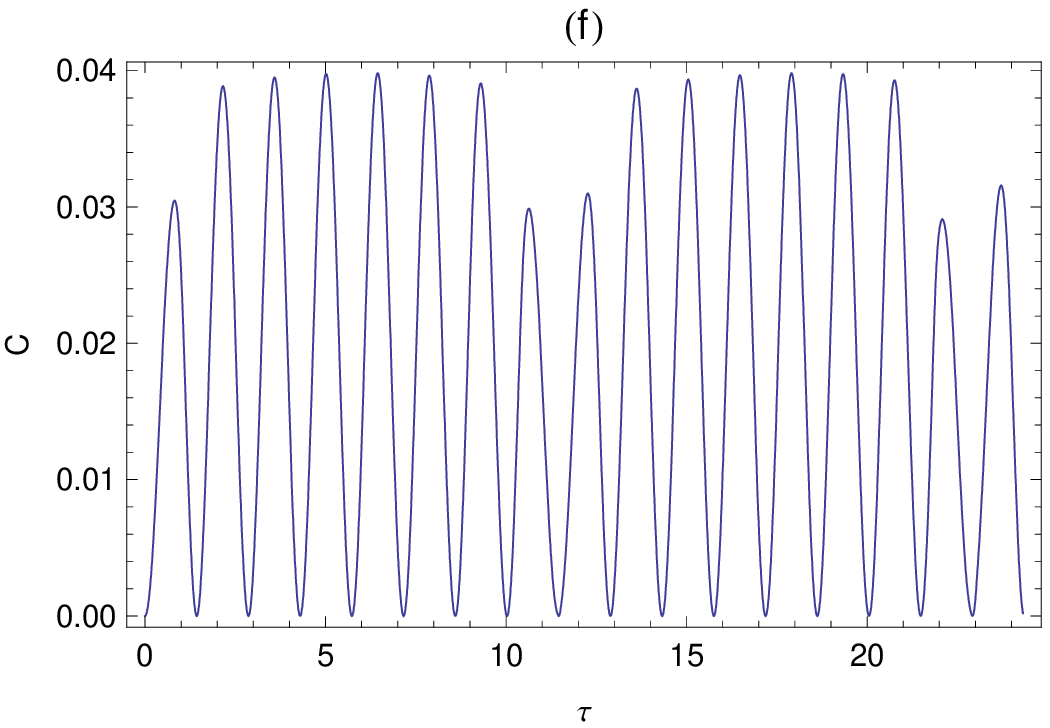}}
\end{center}
\caption{Input state $|\psi_0^B\rangle$, $\bm{Q}=0$, $\alpha=\frac{\sqrt{3}}{2}$, $N=10$, $\lambda=1$. The emergence of a beating effect in the evolution of the concurrence as a function of interaction time. In figures (a) to (f), the applied field sweeps through the range $B_z<B_{z-}$ to $B_z>B_{z+}$, with (a) $B_z=0.1$, (b) $B_z=0.3$, (c) $B_z=0.5$, (d) $B_z=1$, (e) $B_z=2$, (f) $B_z=3$. All quantities are expressed in natural units.}\label{Bdep6}
\end{figure}

Figure \ref{Bdep5c} shows evidence of a rather interesting beating effect. To explore this further, let us examine the concurrence for generic $\alpha$ and $B_z$; this is illustrated in figure \ref{Bdep6}. If the field lies within an interval $[B_{z-},B_{z+}]\approx\left[\frac{\lambda}{2},2\lambda\left(1+\frac{1}{N}\right)\right]$, no evident systematic behaviour is observed. Conversely, if $B_z$ is outside this interval there emerges a pattern which resembles a high frequency oscillation modulated by a low frequency envelope. The period of the rapidly oscillating component is the familiar quantity $T_{\phi}$. The period of the envelope function is defined by
\begin{equation}\label{tgamma}
T_{\gamma}=\frac{\pi}{|\phi-\gamma|},
\end{equation}
where $\gamma\equiv\gamma^{ex}(N,\lambda,B_z)$. The behaviour of the system is therefore determined by the beating of four eigenstates, of which two correspond to the spin-flip being shared between the first neutron and the sample, and two correspond to the spin-flip being shared between the second neutron and the sample. The optimal interaction time can then be estimated by tracking the evolution of the peaks of the envelope function.

In summary, when a field is applied the peak concurrence always increases, provided the field does not exceed an upper limit. This limit depends on the neutron polarization, but is generally of order $4\lambda$. For a given field, $\mathcal{C}_p$ is a maximum when the neutron spin is anti-aligned with the quantization axis. If the field lies outside the interval $[B_{z-},B_{z+}]$ this maximum occurs at approximately $\tau=T_{\gamma}/2$. However, if the field is weak then $T_{\gamma}$ is large, hence at short times the system performs better if both $\alpha$ and $\beta$ are non-zero. If the field lies within the interval $[B_{z-},B_{z+}]$ the behaviour of the concurrence is more difficult to predict, but setting $B_z=B_z^*$ and $\alpha=0$ the concurrence takes on a well-defined analytical form and peaks at predictable times. For $N\geq5$ the peak value exceeds 0.7, and converges to 0.77 as $N$ becomes large. The performance of the system at zero momentum transfer for input state $|\psi_0^B\rangle$ therefore matches in efficiency that observed for input state $|\psi_0^A\rangle$, although its behaviour is overall more complex and less predictable.

\section{Concurrence and Negativity}\label{log negativity}
One of the most interesting properties of the neutron concurrence is the invariance with respect to the second period of free evolution. In zero field, this arises from the form of the free Hamiltonian $H_0$. In finite field, however, $H_0$ is no longer a multiple of the identity, nor does it commute with $\mathcal{H}_1$ or $\mathcal{H}_2$; therefore the explanation is not as evident. To gain insight on the origin of this behaviour, it is useful to investigate its dependence on the structure of the interaction potential. Specifically, let us focus on the second scattering event. One finds that, provided the interaction Hamiltonian is hermitian and displays an excitation number conserving symmetry, the concurrence is always independent of the second period of free evolution. It would therefore seem that the sample behaves as an `entanglement safety-deposit box', which holds the spin information deposited by the first scattering event until a second neutron bearing the right `key' comes along to claim it. In other words, the entanglement between the neutrons is conditional on the second scattering event being able to `undo' -in part or in full- the transformation on the sample spin precipitated by the the arrival of the first neutron.

\begin{figure}[H]
\renewcommand{\captionfont}{\footnotesize}
\renewcommand{\captionlabelfont}{}
\begin{center}
\subfigure{\label{Neglpota}\includegraphics[width=6.5cm]{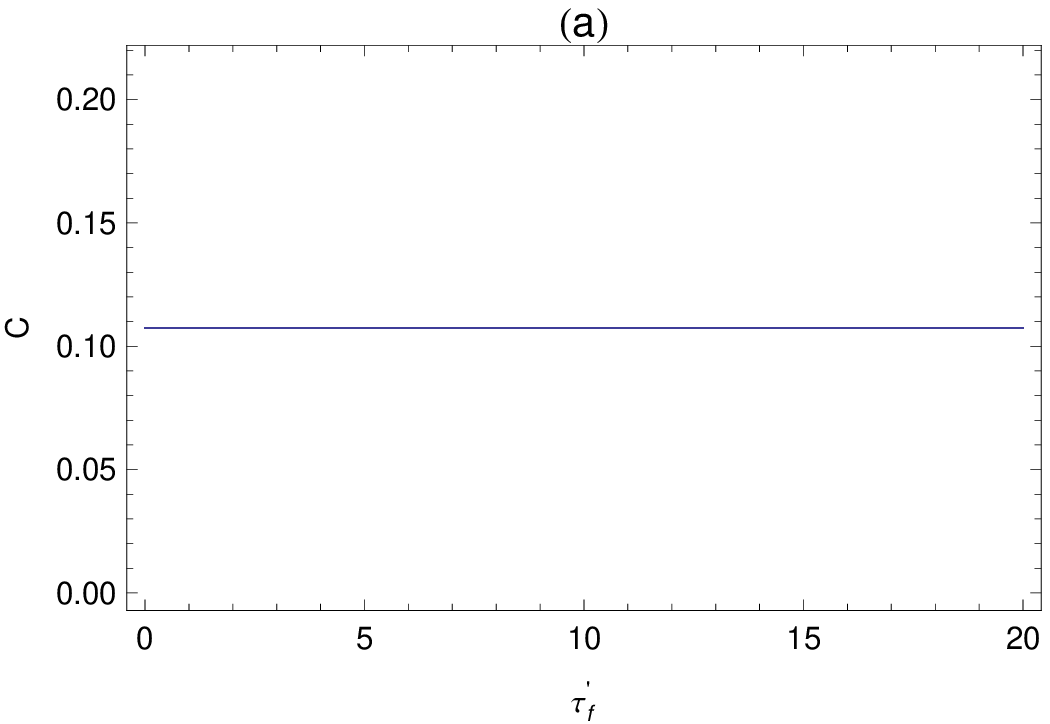}}
\hspace{0.3cm}
\subfigure{\label{Negplotb}\includegraphics[width=6.5cm]{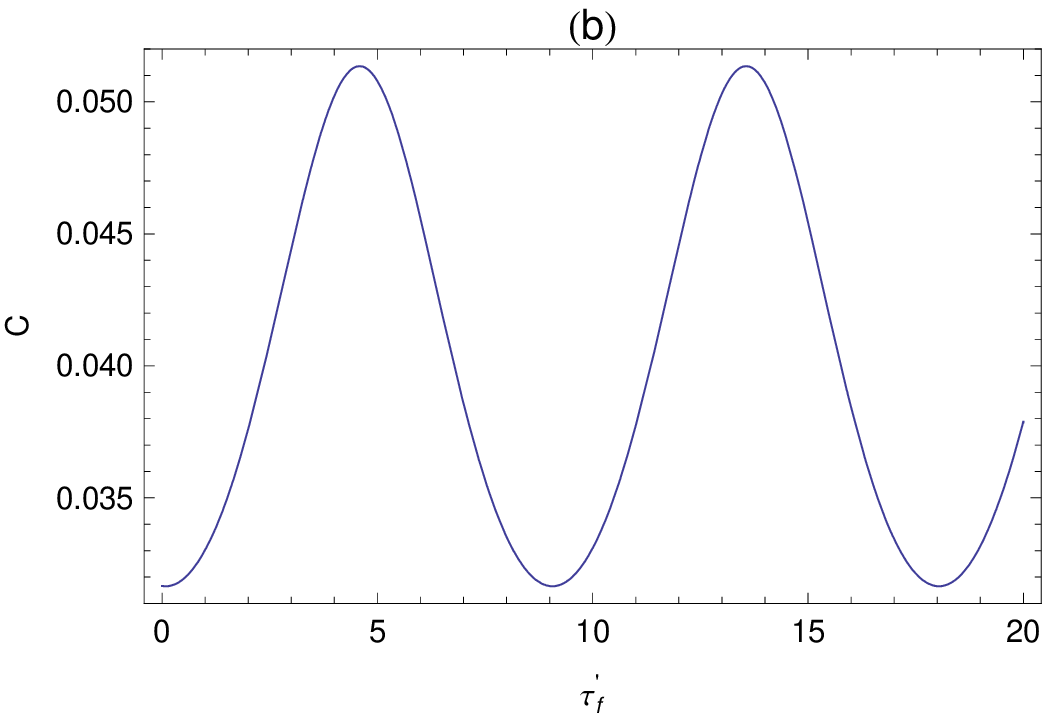}}
\hspace{0.3cm}
\subfigure{\label{Negplotc}\includegraphics[width=6.5cm]{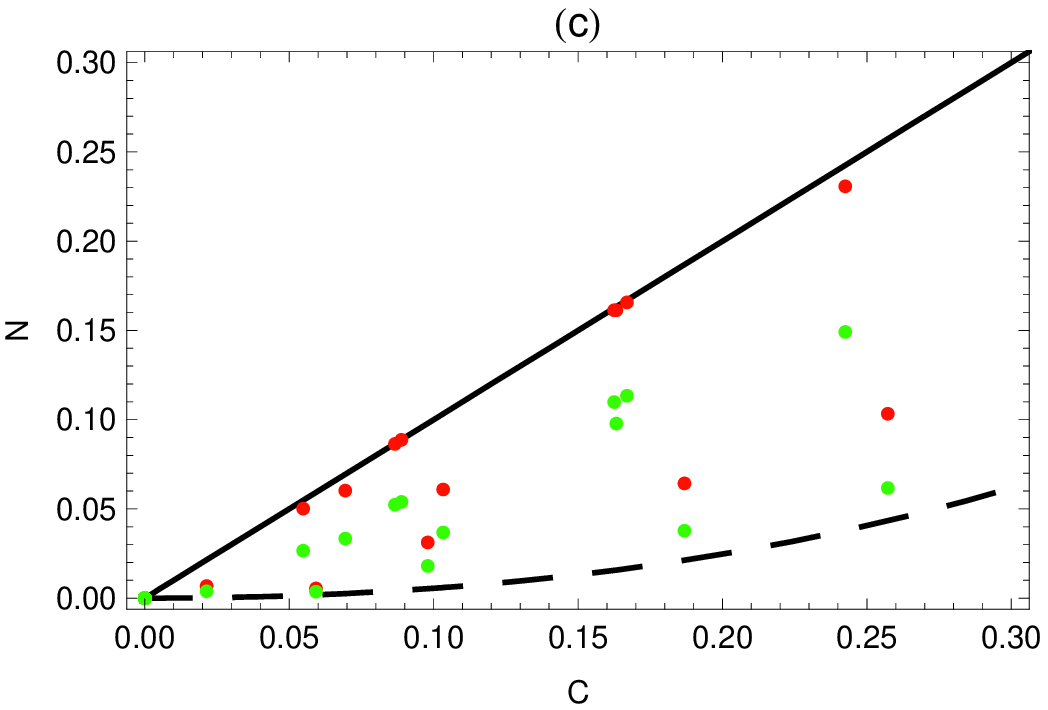}}
\end{center}
\caption{Input state $|\psi_0^B\rangle$, $\bm{Q}=0$, $B_z=0.35$, $N=10$, $\alpha=\frac{1}{\sqrt{6}}$, $\lambda=1$. Figure (a) shows the concurrence of the neutrons remains constant during the second period of free evolution. Figure (b), conversely, shows the negativity of the neutron state varies with $\tau_f^{\:\prime}$. Figure (c) shows the spread of negativity values for states with a given concurrence, sampled at random values of $\tau$. The
red and green points are maximum and minimum negativity values, respectively.
The solid and dashed lines are the bounds for the maximum and minimum negativity of
a state with a given concurrence \cite{miranowicz04}. All quantities are expressed in natural units.}\label{neg1}
\end{figure}

It is interesting to ask whether the invariance with $\tau_f^{\prime}$ extends to all entanglement measures. Figure \ref{neg1} illustrates the evolution of the logarithmic negativity during the second period of free evolution for input state $|\psi_0^B\rangle$, and reveals a definite variation with $\tau_f^{\prime}$. The discrepancy between the concurrence and the negativity may arise from the difference in the physical significance of the two quantities: given an
arbitrarily large number of maximally entangled states, the
concurrence quantifies the entanglement cost of generating an
arbitrary state $\rho$ under local operations and classical
communication, whereas the negativity quantifies the entanglement
cost of generating an arbitrary state $\rho$ under
positive-partial-transpose-preserving operations. The dependence of the negativity on $\tau_f^{\prime}$ then becomes a useful indicator of the inequivalence of the two classes of operations.

%\begin{figure}[t]
%\renewcommand{\captionfont}{\footnotesize}
%\renewcommand{\captionlabelfont}{}
%\begin{center}
%\subfigure{\label{Neglpota}\includegraphics[width=6.5cm]{Neglpota.eps}}
%\hspace{0.3cm}
%\subfigure{\label{Negplotb}\includegraphics[width=6.5cm]{Negplotb.eps}}
%\hspace{0.3cm}
%\subfigure{\label{Negplotc}\includegraphics[width=6.5cm]{Negplotc.eps}}
%\end{center}
%\caption{Anisotropic Sample: input state $|\psi_0^B\rangle$, $\bm{Q}=0$, $B_z=0.35$, $N=10$, $\alpha=\frac{1}{\sqrt{6}}$. Figure (a) shows the concurrence of the neutrons remains constant during the second period of free evolution. Figure (b), conversely, shows the negativity of the neutron state varies with $\tau_f^{\:\prime}$. Figure (c) shows the spread of negativity values for states with a given concurrence, sampled at random values of $\tau$. The
%red and green points are maximum and minimum negativity values, respectively.
%The solid and dashed lines are the bounds for the maximum and minimum negativity of
%a state with a given concurrence \cite{miranowicz04}.}\label{neg2}
%\end{figure}

\section{Parallels with Optical Entanglement Generation}\label{haroche}
The entanglement of massive particles via successive interactions
with a macroscopic mediator is by no means a novel idea. Almost ten
years ago, a group led by Haroche of the Ecole Normale
Superieure (ENS) reported the generation of EPR pairs of atoms via
the exchange of a single photon in a high-\textit{Q} cavity according to the following scheme
\cite{haroche97}. A single atom is prepared in an excited state $|e\rangle$
and is made to interact with a cavity in its ground state. Assuming
the cavity and the atom exchange photons at a rate $\Omega$, after a
period of time $\tau_1=\frac{\pi}{2\Omega}$ the combined state of
the atom and the cavity becomes
\begin{equation}
|\psi_1\rangle=\frac{1}{\sqrt{2}}\left(|e0\rangle+|g1\rangle\right),
\end{equation}
where the right-most number refers to the photon number inside the
cavity. After a delay time $\tau_2$, a second atom, this time
prepared in its ground state $|g\rangle$, is made to interact with
the cavity for a period $\tau_3$, which is twice the duration of $\tau_1$. If the first atom has
emitted a photon, the second atom becomes excited; otherwise it
remains in $|g\rangle$. As a result, the composite state of the
system is
\begin{equation}
|\psi_2\rangle=\frac{1}{\sqrt{2}}\left(|ge0\rangle+|eg0\rangle\right),
\end{equation}
which is a maximally entangled state of the atoms.

\begin{figure}[H]
\renewcommand{\captionfont}{\footnotesize}
\renewcommand{\captionlabelfont}{}
\begin{center}
\subfigure{\label{uneqtimes1}\includegraphics[width=6.5cm]{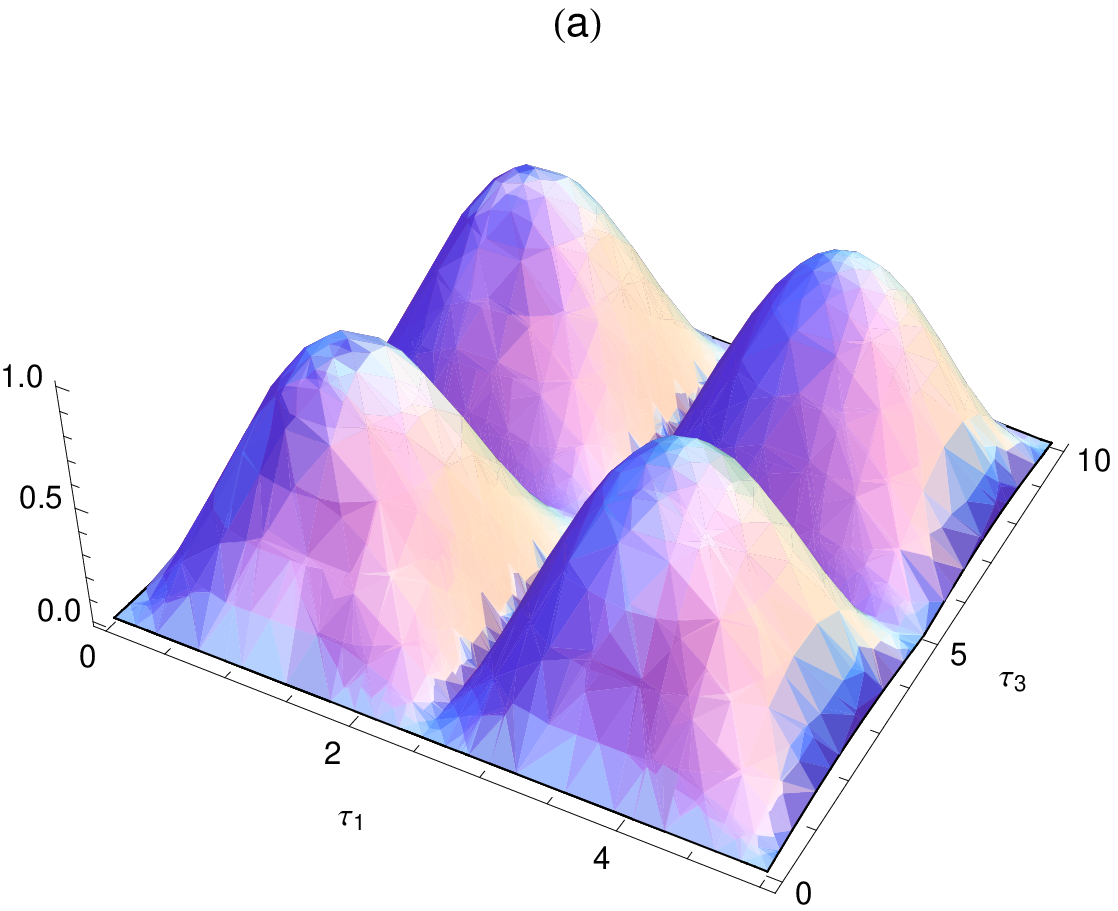}}
\hspace{0.3cm}
\subfigure{\label{uneqtimes2}\includegraphics[width=6.5cm]{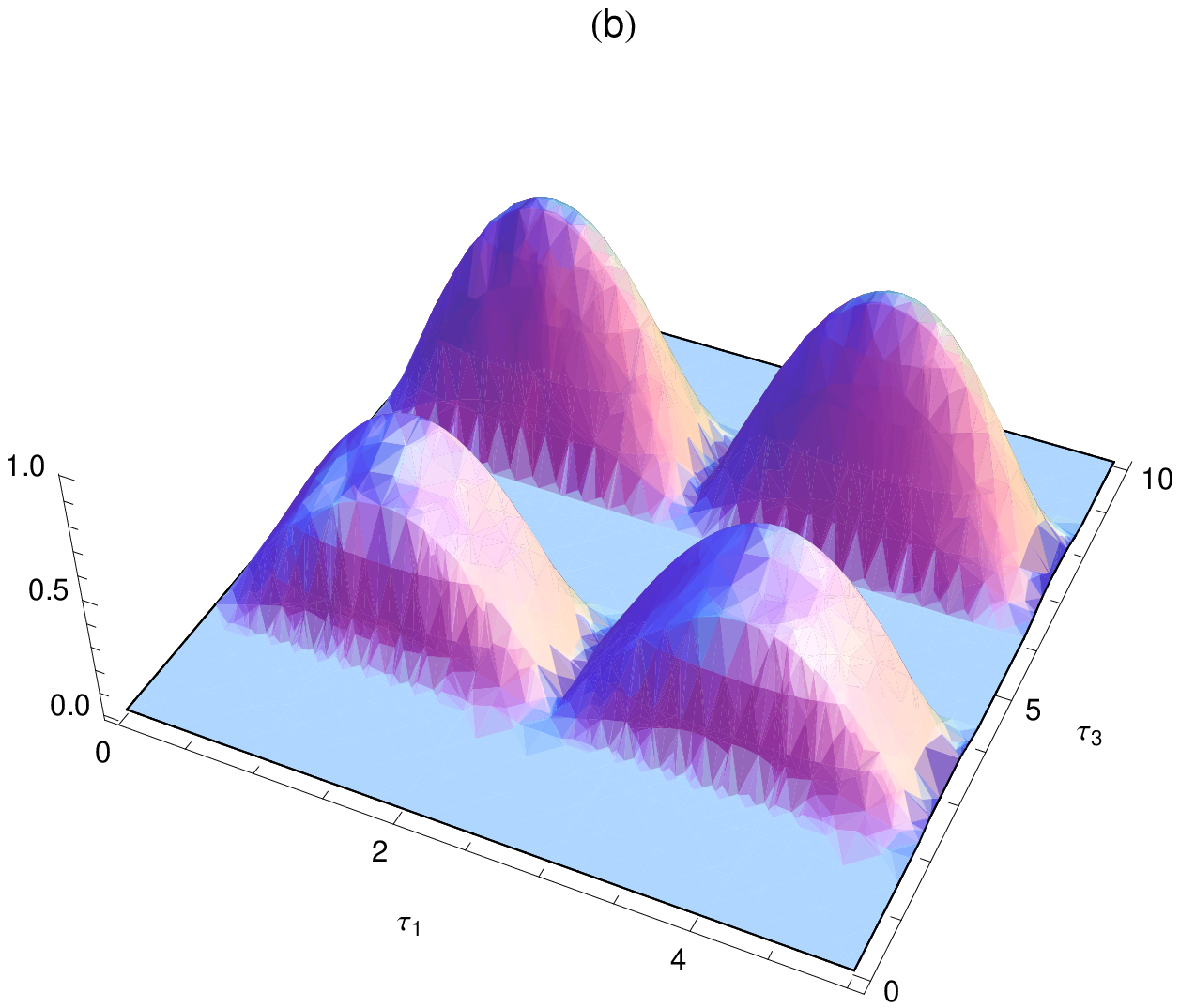}}
\end{center}
\caption{The time-evolution of the neutron concurrence assuming independently tuneable interaction times, for $N=10$, $B_z=B_z^*$ and $\lambda=1$. Figure (a) relates to input state $|\psi_0^A\rangle$, for which optimal concurrence is first achieved at $\tau_3=2\tau_1$. Figure (b) relates to input state $|\psi_0^B\rangle$, with $\alpha=0$ and $\beta=1$. All quantities are expressed in natural units.}\label{uneqtimes}
\end{figure}

It is shown in figure \ref{uneqtimes} that if the ENS protocol were exactly mapped to the scattering scheme I have described, it would be possible to generate fully entangled neutron states. Indeed, if we restrict ourselves to the single excitation subspace by working with input state $|\psi_0^A\rangle$, near unit concurrence is achieved when $\tau_3=2\tau_1$ [figure \ref{uneqtimes1}]. However, such a mapping requires the possibility of separately tuning the interaction times of consecutive neutrons, which is not currently possible. Therefore, to compare the ENS protocol and the scattering scheme on a more even footing, it is useful to examine the evolution of state $|ee0\rangle$, which is contextually equivalent to state $|8\rangle\equiv|11\bm{0}\rangle$ of the Bloch basis, for equal interaction times $\tau_1=\tau_3$. In this case, the concurrence of the atoms is given by
\begin{equation}
C_h=2\left(\cos^2{\Lambda\tau}|\sin{\sqrt{2}\Lambda\tau}\sin{\Lambda\tau}|-\sin^2{\Lambda\tau}|\cos{\sqrt{2}\Lambda\tau}\cos{\Lambda\tau}|\right).
\end{equation}
This expression peaks at a value 0.77 for $\tau\approx\frac{3\pi}{\Lambda}$, and, as we will see, greatly resembles the evolution of the concurrence in optimal field for input state $|8\rangle$, when scattering takes place from an anisotropic sample [see equation \eqref{c psib field xy} and section \ref{p0b finite mtm}]. The ENS scheme and the scattering protocol are therefore comparable in all but the timescale.

\section{Experimental Feasibility}\label{exp feasibility}
From the results presented so far in this chapter, it would seem that the scattering protocol of chapter \ref{neutron proposal} offers a valid means of creating entanglement between distinct neutrons. The question to ask is therefore: can it be done? Let us review the quantitative requirements of the protocol:
\begin{enumerate}
\item The spin lattice relaxation time ($T_1$) must be longer than the first period of free evolution, which roughly coincides with the time taken by the neutrons to reach the sample;
\item The phase coherence time ($T_2$) must be longer than the period of free evolution between scattering events;
\item The neutron coherence volume must be comparable to the size of the sample;
\item The optimal field $B^*$ and the optimal interaction time $\tau^*$ must be attainable values;
\item The protocol must be robust against experimental uncertainties;
\item The final state of the neutrons must yield a measurable witness.
\end{enumerate}
To address these points, I take as a reference the technical specifications of the PF2 source of UCNs at the Institut Laue-Langevin (ILL) in Grenoble (see Appendix \ref{app_b} and \cite{steyerl86}).

%\begin{figure}[H]
%\renewcommand{\captionfont}{\footnotesize}
%\renewcommand{\captionlabelfont}{}
%\begin{center}
% \includegraphics[width=10cm]{ucn.eps}
% \end{center}
% \caption{The layout of the PF2 source at the ILL (Grenoble). The four ultra-cold neutron beams (TES, EDM, UCN and MAM) have cross-sections ranging from $16-140$ cm$^2$, with a total flux of $2.6 \cdot 10^{-8}$ m$^{-2}$s$^{-1}$ for $v<6.2$ ms$^{-1}$. Experiments are conducted at a distance of approximately 4 metres from the turbine house. The operation of this source is described in \ref{app b}. Image available at http://www.ill.eu/pf2/ .} \label{scatt1}
%\end{figure}

For UCNs with velocity $v=7$ ms$^{-1}$ and a flight path of $10^{-2}-1$ m, the time taken to reach the sample is of order $10^{-2}-1$ s. $T_1$ times consistent with these figures are achievable in materials such as phosphorus-doped silicon or N@C$_{60}$ \cite{tyrishkin03,spaeth96}. Furthermore, for a neutron flux $F=10^8$ m$^{-2}$s$^{-1}$ and a sample area of, say, $10^{-2}$ m$^{2}$, the time between scatterings might be of order $10^{-6}$ s. $T_2$ times similar to this are also attainable in these systems at low temperature \cite{tyrishkin03}.

Next, we require the neutron coherence volume to be comparable to the size of the sample. This condition is expressed by the uncertainty relation of equation (\ref{u princ}), which in terms of the neutron velocity becomes
\begin{equation}
\Delta v=\frac{\hbar}{m_N\Delta L}.
\end{equation}
For a sample of length $\Delta L=10$ cm, $\Delta v$ is of order $10^{-7}$ ms$^{-1}$. For neutrons traveling at $7$ ms$^{-1}$, therefore
\begin{equation}
\frac{\Delta v}{v}\approx10^{-8}.
\end{equation}
Currently, this is somewhat unrealistic. Resolutions of order $10^{-3}$ have been achieved in the field of neutron spin-echo spectroscopy \cite{bayrakci06}, but this yields coherence lengths of only $\Delta L\approx 10^{-6}-10^{-5}$ m. The situation could be improved if the neutron velocity were reduced, which might be achieved by prolonging or improving the efficiency of the cooling stage described in Appendix \ref{app_b}, or by allowing the neutrons to slow down under the effect of gravity (see below).

For the remainder of this discussion, I will consider scattering from a sample in a single magnon state. To estimate working values of $B^*$ and $\tau^*$, it is useful to re-write equations (\ref{bz star}) and (\ref{tau star}) in SI units and take the large $N$ limit:
\begin{align}
B_z^*&\approx\left(g_e\mu_B\right)^{-1}\lambda=\frac{g_N\mu_N\mu_0}{a_0^3},\\
\tau^*&=\frac{\hbar\sqrt{N}}{4\lambda}\approx\frac{a_0^3\hbar \sqrt{N}}{4g_N\mu_N\mu_0g_e\mu_B}.
\end{align}
Hence
\begin{align}
B_z^*&\approx10^{-32}\left(\frac{a_0}{\mathrm{m}}\right)^{-3} \mathrm{T},\label{bSI}\\
\tau^*&\approx10^{20}\left(\frac{a_0}{\mathrm{m}}\right)^{3}N^{\frac{1}{2}}\:\mathrm{s}\label{tauSI}.
\end{align}
Assuming a lattice constant $a_0=10^{-10}$ m and a macroscopic number of scatterers $N=10^{23}$ gives an optimal field $B_z^*=10^{-2}$ T and a sample volume $D^3=10^{-7}$m$^3$. For these parameters, the neutron velocity is $v\approx10^{-4}$ms$^{-1}$, corresponding to an interaction time $\tau^*\approx100$s. Currently, such low values of $v$ are not attainable. This is not a fundamental constraint, but a technical limitation. In principle, UCN could be slowed to this energy regime by causing them to decelerate in a gravitational potential \cite{golub91}: a UCN with $v=7$ms$^{-1}$ would stop at a height of approximately $2.5$m, having converted all its kinetic energy to gravitational potential energy. Unfortunately, there are several problems with this approach. First, the neutron absorption cross-section is inversely proportional to the neutron velocity. Therefore, the slower the neutrons, the greater the chance they will be absorbed by the scatterer or indeed by any other component of the experimental setup. Second, even if it were possible to produce such a slow neutron beam, one would expect its intensity to be extremely low. Therefore, it is difficult to see how one could produce enough neutrons in the required energy range to carry out a statistically significant measurement of the witness (see discussion below on how to measure the witness).

It was stated in chapter \ref{neutron proposal} that the protocol would be deemed successful if the neutrons were found to share an amount $\mathcal{E}$ of entanglement, corresponding to a concurrence $C_{\mathcal{E}}$. I now set $C_{\mathcal{E}}=\frac{2}{3}$, which implies we still consider the protocol successful if the concurrence deviates from its optimal value by $0.1$. For simplicity, the possibility of simultaneous errors in the field and in the timing will be excluded. In other words, I assume we operate either at optimal field, or at optimal time.

\begin{figure}[H]
\renewcommand{\captionfont}{\footnotesize}
\renewcommand{\captionlabelfont}{}
\begin{center}
\subfigure{\label{dtau1}\includegraphics[width=6.5cm]{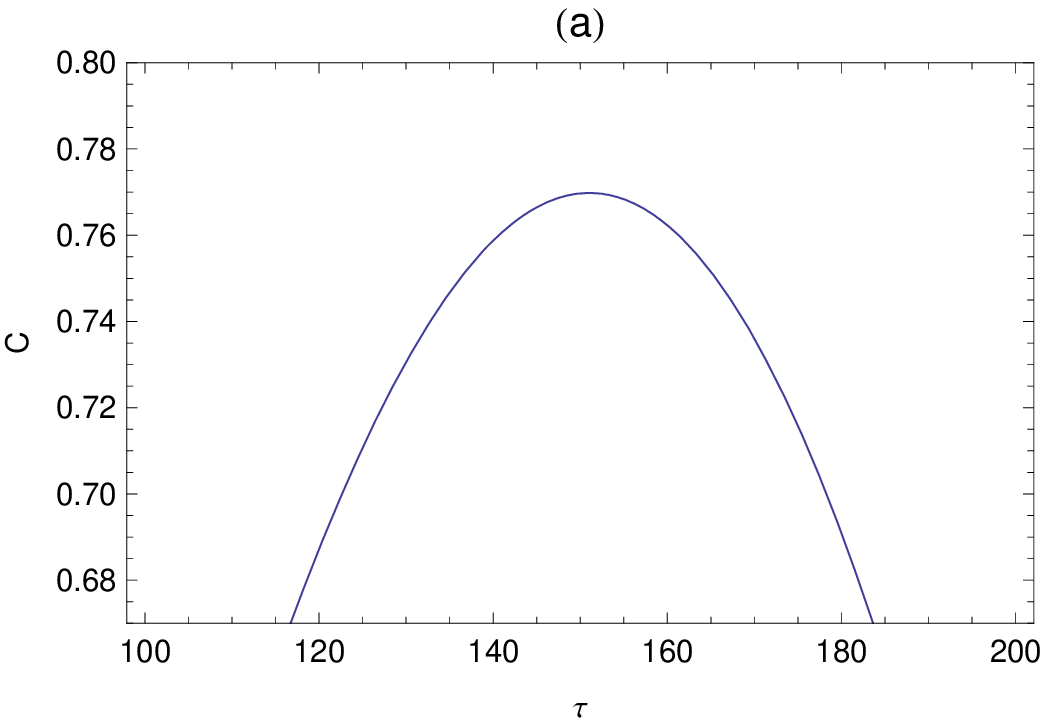}}
\hspace{0.3cm}
\subfigure{\label{db1}\includegraphics[width=6.5cm]{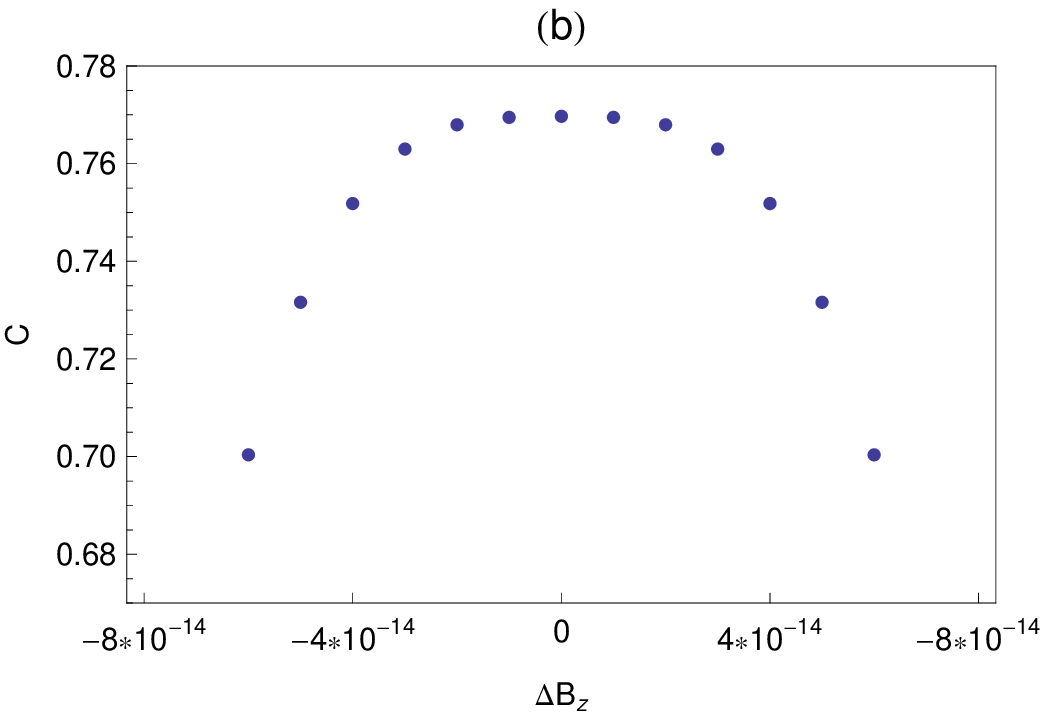}}
\end{center}
\caption{Input state $|\psi_0^A\rangle$, $\bm{Q}=0$. (a) The spread in concurrence around $\tau=\tau^*$ at $B_z=B_z^*$. The origin of the ordinate axis is set at $C_{\mathcal{E}}=0.67$ to better illustrate the tolerance interval $\Delta\tau$. (b) The spread in concurrence around $B_z=B_z^*$ at $\tau=\tau^*$. }\label{dtau and db}
\end{figure}

At optimal field, the time evolution of the concurrence is described by equation (\ref{cp opt field}). An acceptable timing error can then be estimated by measuring the half-width of this function in the vicinity of $\tau^*$, at $C=C_{\mathcal{E}}$. For the set of parameters defined above, one finds $\tau^*\approx150$ s and $\Delta\tau\approx\pm70$ s [see fig. \ref{dtau1}], hence
\begin{equation}
\frac{\Delta\tau}{\tau^*}=\frac{\Delta v}{v}\approx0.5.
\end{equation}
Note this value does not take into account the restrictions on the neutron coherence volume.

Let us now fix $\tau=\tau^*$, and consider the effect of errors in field calibration. The concurrence as a function of applied field is given by equation (\ref{conc psiA ex}). As before, the allowed spread in $B_z$ can be estimated from the half-width of the curve at $C=C_{\mathcal{E}}$ around $B_z=B_z^*$. From figure \ref{db1}:
\begin{align}
\Delta B_z&\approx10^{-13},\\
\frac{\Delta B_z}{B_z^*}&\approx10^{-11}.
\end{align}
To recap, for a completely successful realization of the protocol we require
\begin{align}
\frac{\Delta v}{v}&\approx10^{-8},\\
\frac{\Delta B_z}{B_z^*}&\approx10^{-11}.
\end{align}
The second condition may be attainable, but the first is currently not \cite{li01,aynajian08,baciak03}.

The final point regards the measurement of the witness. We saw in section \ref{p0a zero mtm finite field} that a successful realization of the protocol yields the scattered state $\rho_n^A$ of equation (\ref{rho n A}). As expected, the partial transpose of $\rho_n^A$ is negative. The eigenvector corresponding to the negative eigenvalue has the form
\begin{equation}
|e^-\rangle=a|00\rangle+b|11\rangle,
\end{equation}
where both $a$ and $b$ are approximately equal to $\frac{1}{\sqrt{2}}$. It has been shown that witness operators constructed from states of this form can be decomposed as follows \cite{guhne02}:
\begin{eqnarray}\label{wit}
W&=&a^2|z^+z^+\rangle\langle z^+z^+|+b^2|z^-z^-\rangle\langle z^-z^-|\\
&+&ab\left(|x^+x^+\rangle\langle x^+x^+|+|x^-x^-\rangle\langle x^-x^-|-|y^+y^-\rangle\langle y^+y^-|-|y^-y^+\rangle\langle y^-y^+|\right)\nonumber,
\end{eqnarray}
where $|x^{\pm}\rangle$, $|y^{\pm}\rangle$ and $|z^{\pm}\rangle$ are the spin-up and down eigenstates of the Pauli matrices $\sigma_x$, $\sigma_y$ and $\sigma_z$, respectively. Such a witness could be measured with as few as three device settings, provided one could detect all outgoing neutrons and measure each component of their spin. A Stern-Gerlach experiment in an arbitrary direction would achieve this objective. Bearing in mind the stochasticity of existing neutron sources, $\langle W \rangle$ could then be evaluated on pairs of neutrons detected in rapid succession following a gap long enough to reset the sample state. This is conceptually possible, though very challenging from a technical viewpoint.

\section{Scattering from an Anisotropic Medium in a Single Magnon State}\label{p0a finite mtm}
We have assumed so far that the sample is an isotropic medium. This is reflected in the form of the interaction Hamiltonian of equation \eqref{h ex tot}, which is proportional to an electronic $g$-factor with equal components $g_e$ along all three axes. In general terms, the $g$-factor is a tensor with diagonal elements $g_x$, $g_y$ and $g_z$, whose magnitude can vary in space according to the chemical nature of the environment, or the presence of external fields. This property is of great relevance in the field of spintronics, which exploits $g$-tensor anisotropies to manipulate electron spins using electrical signals \cite{salis01}. It is then interesting to explore how the scattering scheme might perform if the sample were anisotropic. To describe this situation accurately, one would have to account for the possible breakdown of the $LS$ model. However, as a very basic approximation, we can assume $LS$ coupling still holds, and simply choose a $g$-tensor with unequal components along the three spatial directions. Let us choose an extreme example: $g_x=g_y=g_e$, and $g_z=0$. In this limit, one might imagine a spin potential of the form
\begin{equation}\label{h xy}
V^{s_{xy}}=V_0+\Lambda\: \left(s_m^x\sum_{j=1}^Ns_j^x+s_m^y\sum_{j=1}^Ns_j^y\right).
\end{equation}
As before, this can be expressed in terms of an `effective' interaction potential similar to equation (\ref{h1prime}), provided one takes the correct form for $V_m$. Quantitatively, $V^{s_{xy}}$ differs from $V^{s_{ex}}$ because it lacks a diagonal component; hence the diagonal elements of $\mathcal{H}^{\prime}_m$ are simply those of $H_0^B$. As a result, the equations describing the states of the system in the different stages of the protocol are identical to (\ref{psi2exch}), (\ref{psi3exch}) and (\ref{psi4exch}), provided one sets
\begin{eqnarray}
\phi&\equiv&\phi^{xy}(N,\lambda,B_z)=\sqrt{B_z^2+\frac{4\lambda^2}{N}}\label{phi xy},\\
\Lambda&=&B_z(N-1)-\phi\label{lambda1 xy},\\
c&=&\sqrt{\frac{1}{2}+\frac{B_z}{2\phi}},\label{c xy}\\
d&=&-\sqrt{\frac{1}{2}-\frac{B_z}{2\phi}},\label{d xy}\\
y&=&B_z+\phi\label{y xy}.
\end{eqnarray}
The concurrence between the neutrons is then
\begin{equation}\label{conc psiA xy}
C(N,\lambda,B_z,\tau)=\frac{8\lambda^2\sin^2{\phi\tau}}{N\phi^3}\sqrt{B_z^2+\frac{4\lambda^2}{N}\cos^2{\phi\tau}}.
\end{equation}
There are striking similarities between this and equation (\ref{conc psiA ex}). First, (\ref{conc psiA xy}) is an oscillating function of the interaction time, and independent of the time between scattering events $\tau_f^{\prime}$. The period of the oscillation has the form of equation (\ref{T ex}), with the value of $\phi$ given by equation (\ref{phi xy}). Second, (\ref{conc psiA xy}) reduces to the form of (\ref{cp opt field}) if the field is set to zero, suggesting the concurrence is maximal when there is no applied field. Indeed, as discussed in section \ref{p0a zero mtm finite field}, at $B_z=0$ the single excitation eigenstates of $\mathcal{H}^{\prime}_1$ are reduced to equally weighted superpositions.

\begin{figure}[H]
\renewcommand{\captionfont}{\footnotesize}
\renewcommand{\captionlabelfont}{}
\begin{center}
\subfigure{\label{cxypsiA}\includegraphics[width=6.5cm]{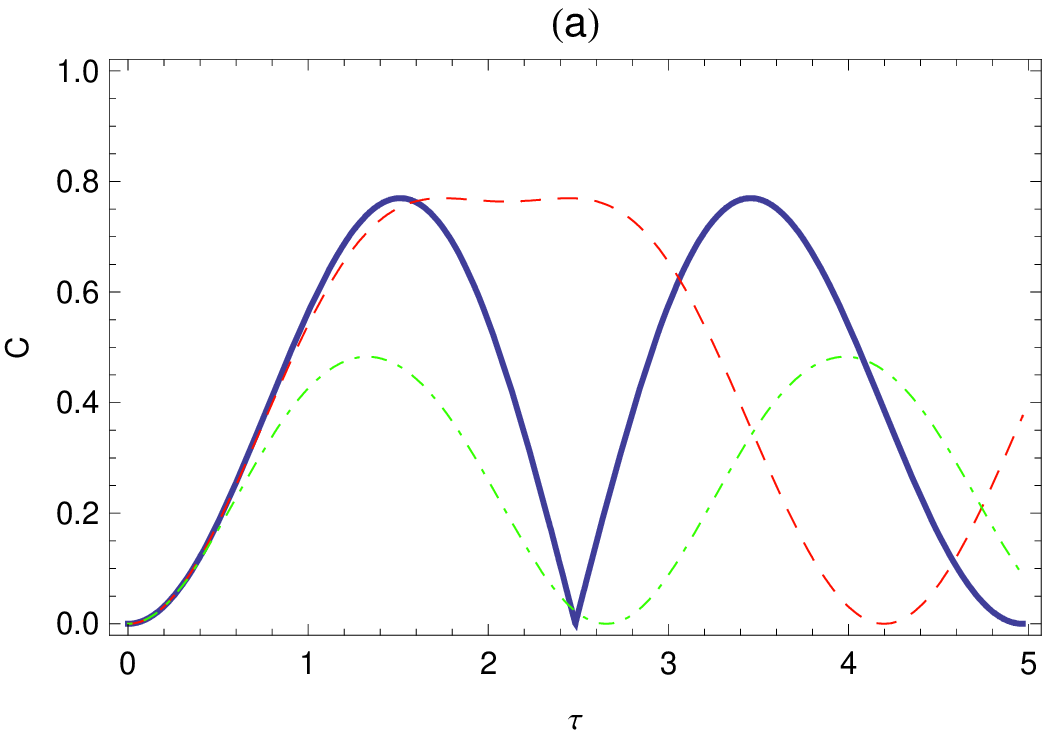}}
\hspace{0.3cm}
\subfigure{\label{xyIntpsiA}\includegraphics[width=6.5cm]{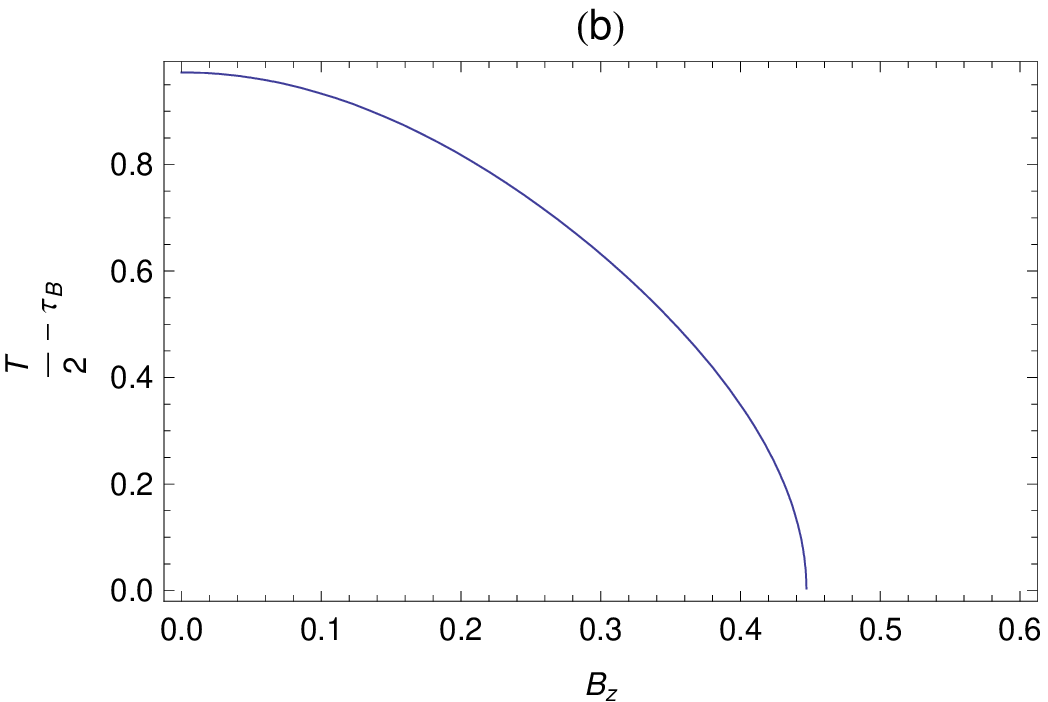}}
\hspace{0.3cm}
\subfigure{\label{CpeakxyIntpsiA}\includegraphics[width=6.5cm]{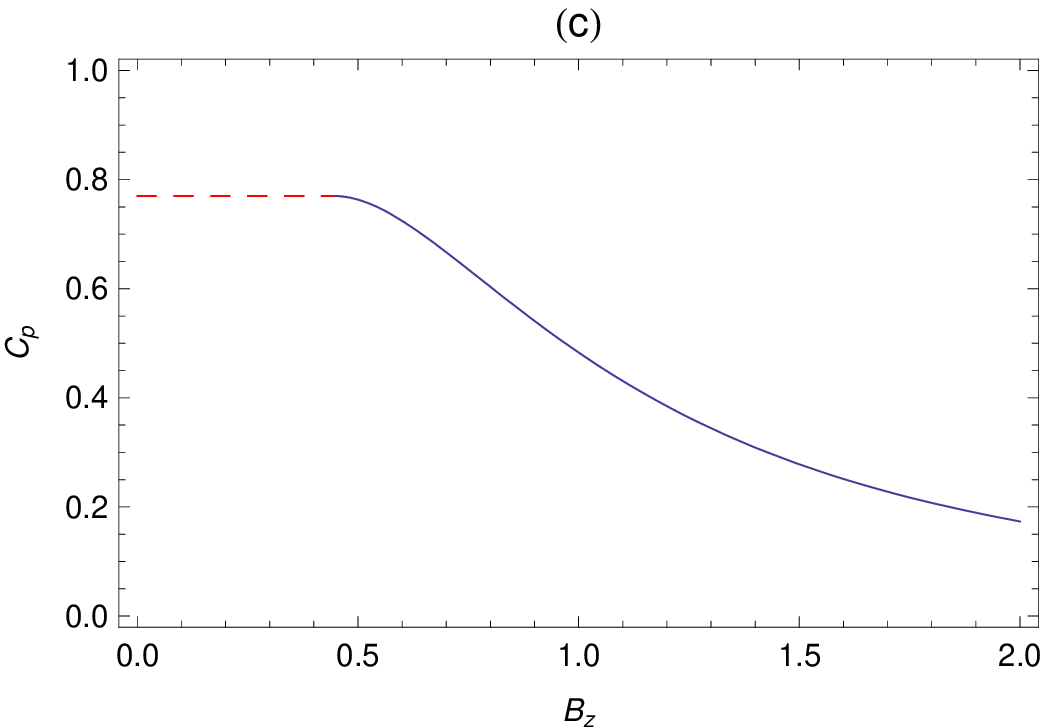}}
\hspace{0.3cm}
\subfigure{\label{CoutxypsiA}\includegraphics[width=6.5cm]{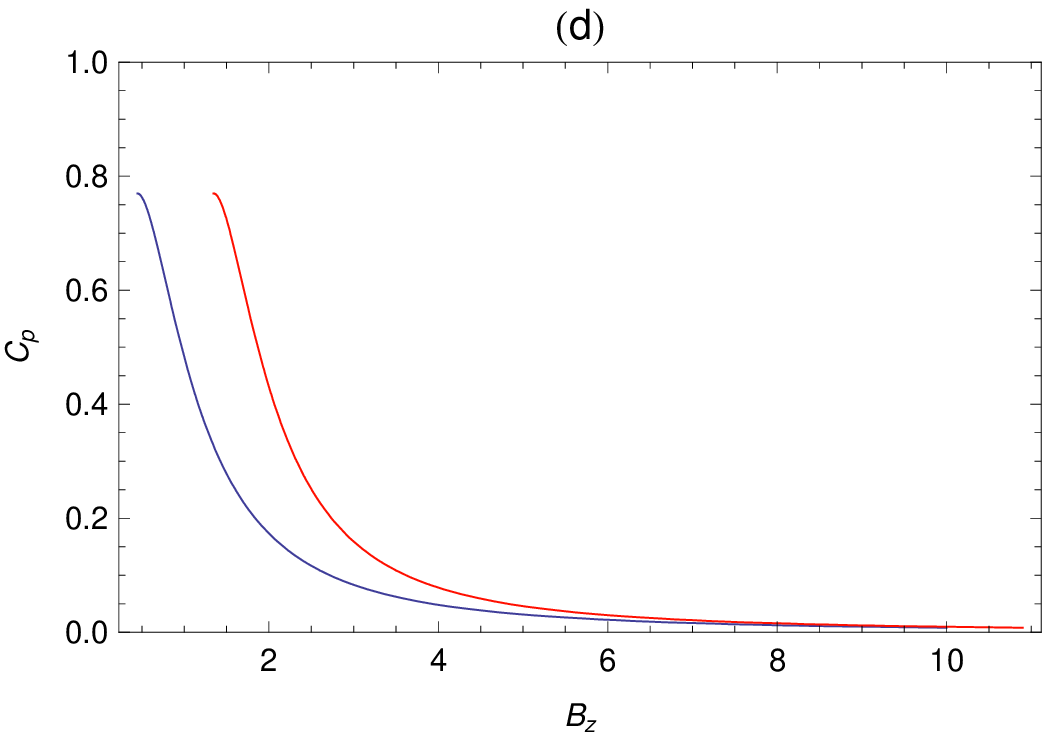}}
\end{center}
\caption{Input state $|\psi_0^A\rangle$, $\bm{Q}=0$, $N=10$, $\lambda=1$. (a) The evolution of the concurrence as a function of interaction time for $B_z=0$, $B_z=0.1$ and $B_z=1$ (solid blue, dashed red and dot-dashed green curves, respectively). (b) The discrepancy between the optimal interaction times as a function of applied field, showing the existence of a threshold field $B_t$. (c) The peak concurrence as a function of $B_z$, for $B_z<B_t$ and $B_z>B_t$ (red dashed and blue solid lines, respectively). (d) The convergence of $C_+$ (blue curve) and $C_o$ (red curve) in the limit of large $N$. All quantities are expressed in natural units.}\label{xy1}
\end{figure}

In zero field, the evolution of the concurrence in time shows the familiar double-peaked structure observed in figure \ref{cOptfieldpsiA}. This structure disappears when the applied field exceeds a certain threshold, to be replaced with a sinusoidal oscillation [fig. \ref{cxypsiA}]. To determine the threshold field we repeat the analysis of section \ref{p0a zero mtm finite field}, maximizing equation (\ref{conc psiA xy}) with respect to $\tau$ and imposing equality with half the oscillation period $T_{\phi}/2$ [fig. \ref{xyIntpsiA}]. One finds
\begin{equation}\label{taub xy}
\tau_B=\frac{1}{2\phi}\cos^{-1}{\left[-\left(\frac{NB_z^2+\lambda^2}{3\lambda^2}\right)\right]},
\end{equation}
which reduces to the form of (\ref{tau star}) when $B_z=0$. From equations \eqref{phi xy}, (\ref{taub xy}) and (\ref{T ex}), the threshold field $B_t$ is then
\begin{equation}
B_t=\frac{\sqrt{2}\lambda}{\sqrt{N}}.
\end{equation}
Substituting into equation (\ref{conc psiA xy}), the peak concurrence above and below $B_z=B_t$ can therefore be expressed by the following:
\begin{eqnarray}
C_+(N,\lambda,B_z)&=&\frac{8B_z\lambda^2}{N\phi^3}\label{c+ xy},\\
C_-(N,\lambda,B_z)&=&\frac{4}{3\sqrt{3}}\label{c- xy}.
\end{eqnarray}
This is illustrated in figure \ref{CpeakxyIntpsiA}. The trends described by (\ref{c+ xy}) and (\ref{c- xy}) are qualitatively similar to those observed for an isotropic sample. Apart from the obvious equivalence of $C_i$ and $C_-$ at optimal field, the behaviour of $C_+$ strongly resembles that of $C_o$ for $B_z>B_{z+}$, and both tend to the same limit as $B_z\rightarrow\infty$ [fig. \ref{CoutxypsiA}].

To summarize: if the system is initialized to a single magnon state, with the spin-flip localized on the sample, the concurrence of the scattered neutrons is an oscillating function of time, which ranges from zero to a maximum of 0.77. Above a threshold field $B_t$, the oscillation is sinusoidal, and its amplitude falls as with $N$. Below this threshold, the concurrence can always be boosted to the peak value. The only salient difference between scattering from an isotropic or anisotropic sample therefore lies in the value of the optimal field, which is finite in the former case, but zero in the latter.

\section{Scattering from an Anisotropic Medium in a fully Polarized State}\label{p0b finite mtm}
\begin{figure}[H]
\renewcommand{\captionfont}{\footnotesize}
\renewcommand{\captionlabelfont}{}
\begin{center}
\subfigure{\label{Bdep0xya}\includegraphics[width=6.5cm]{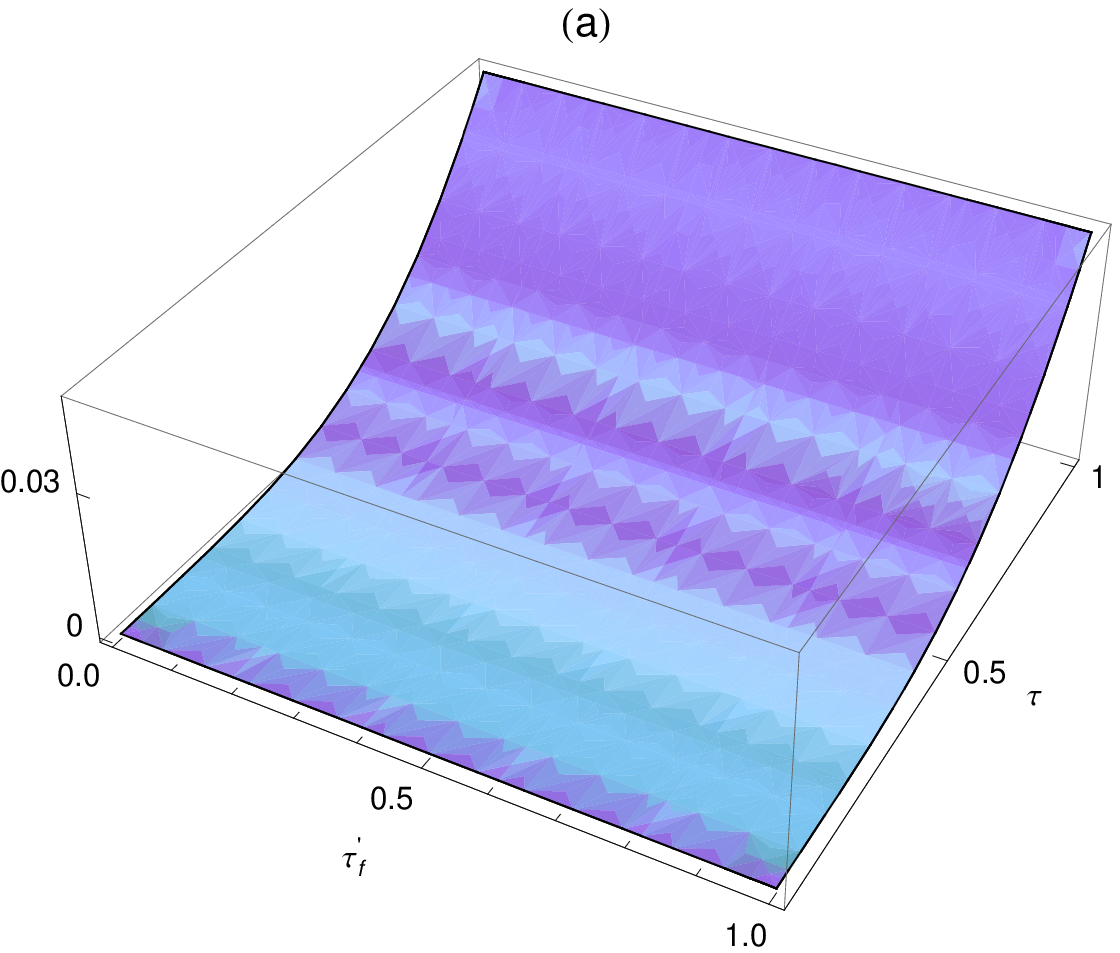}}
\hspace{0.3cm}
\subfigure{\label{Bdep0xyb1}\includegraphics[width=6.5cm]{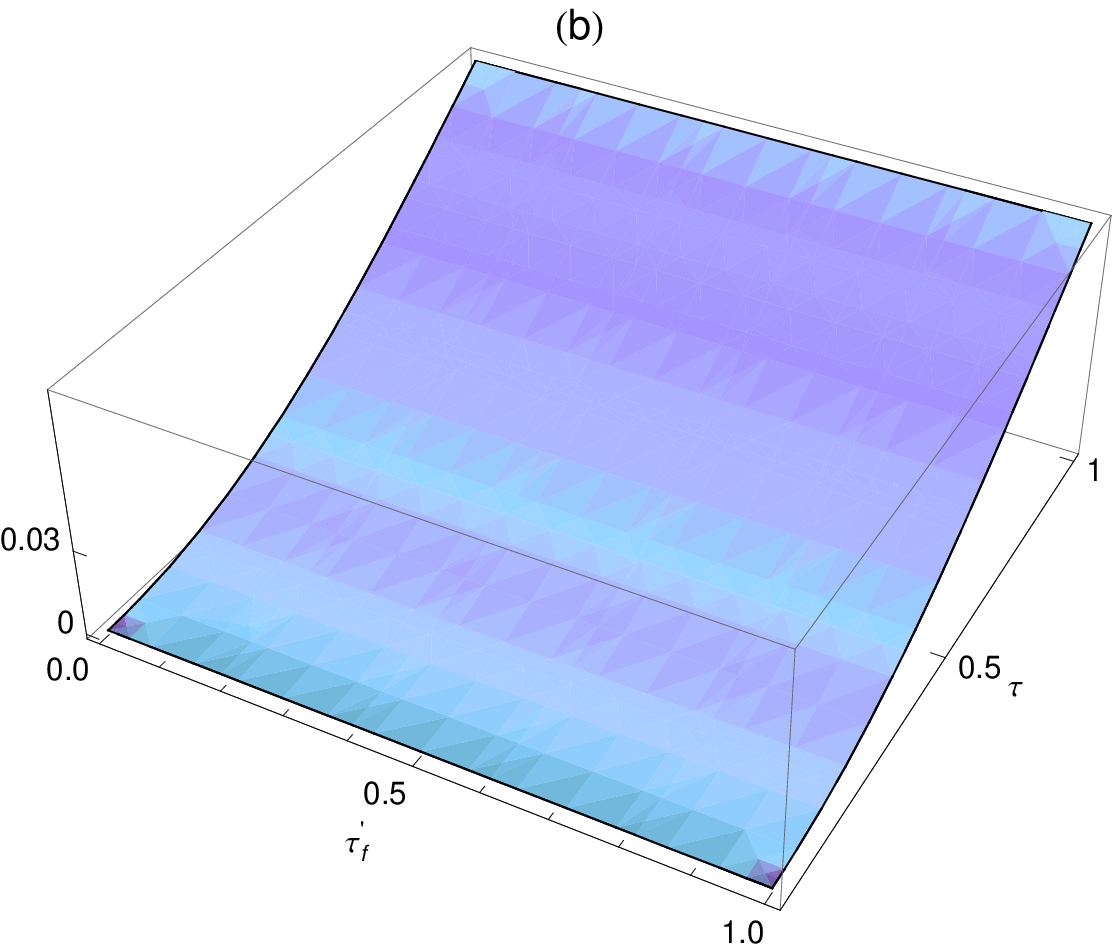}}
\end{center}
\caption{Input state $|\psi_0^B\rangle$, $\bm{Q}=0$, $B_z=0$, $N=10$. The evolution of the concurrence as a function of $\tau$ and $\tau_f^{\:\prime}$ for different neutron polarizations. (a) $\alpha=\frac{1}{\sqrt{2}}$, (b) $\alpha=\frac{\sqrt{3}}{2}$. }\label{Bdep0xy}
\end{figure}

\begin{figure}[H]
\renewcommand{\captionfont}{\footnotesize}
\renewcommand{\captionlabelfont}{}
\begin{center}
\subfigure{\label{Bdep1xya}\includegraphics[width=6.5cm]{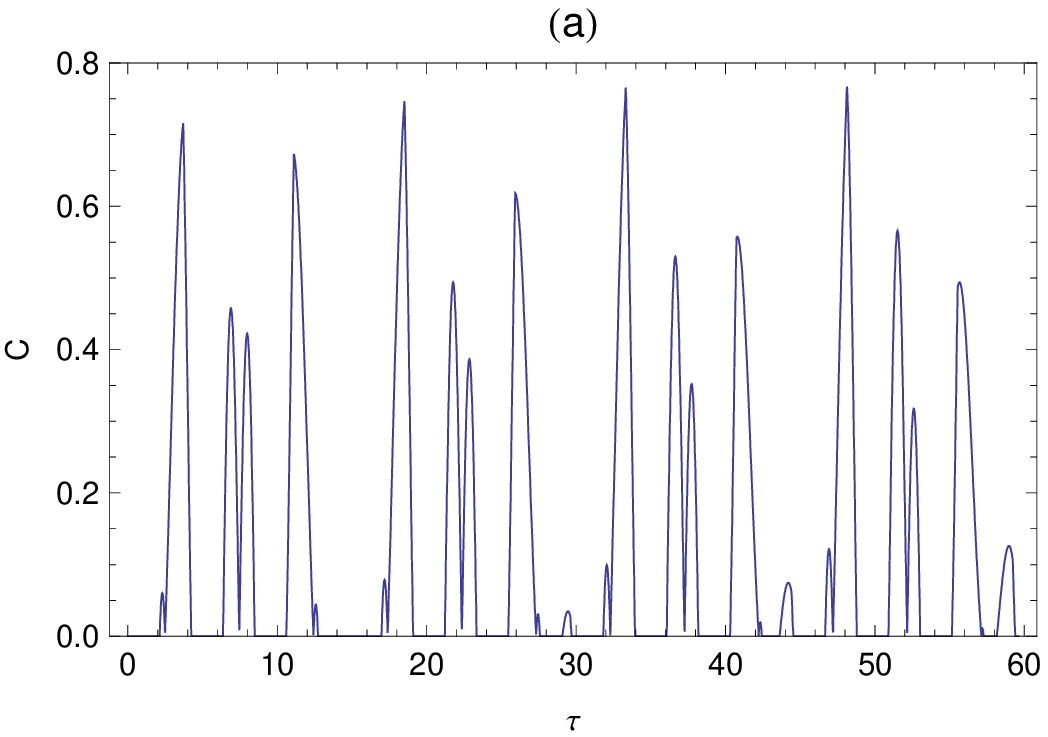}}
\hspace{0.3cm}
\subfigure{\label{Bdep1xyb}\includegraphics[width=6.5cm]{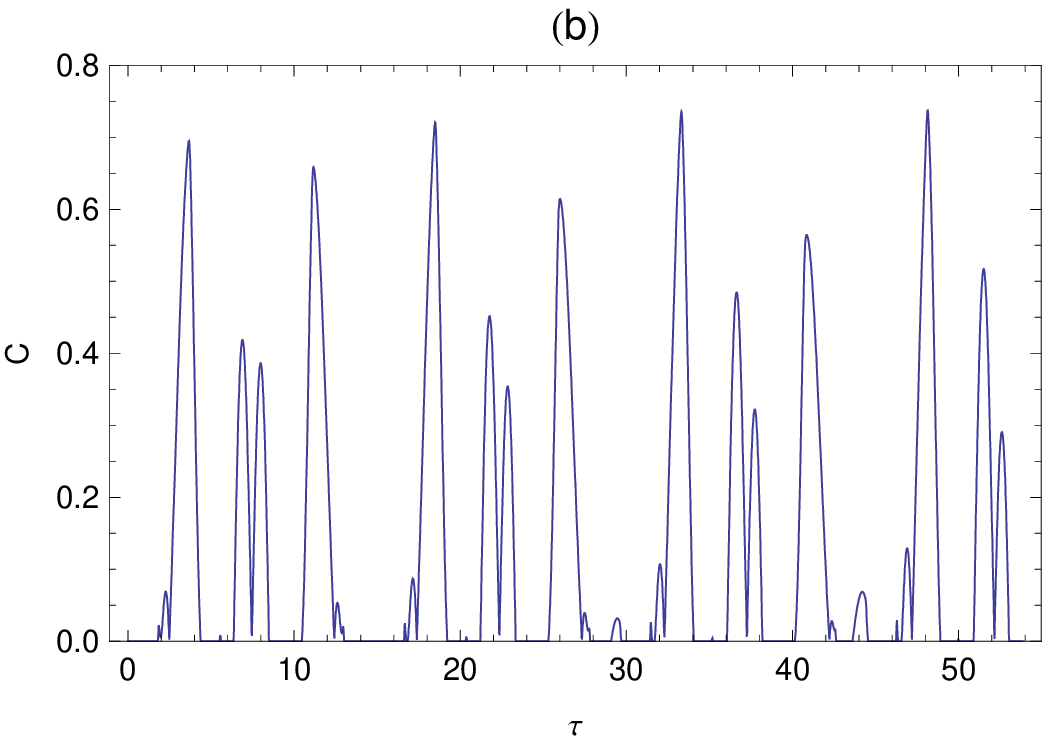}}
\hspace{0.3cm}
\subfigure{\label{Bdep1xyc}\includegraphics[width=6.5cm]{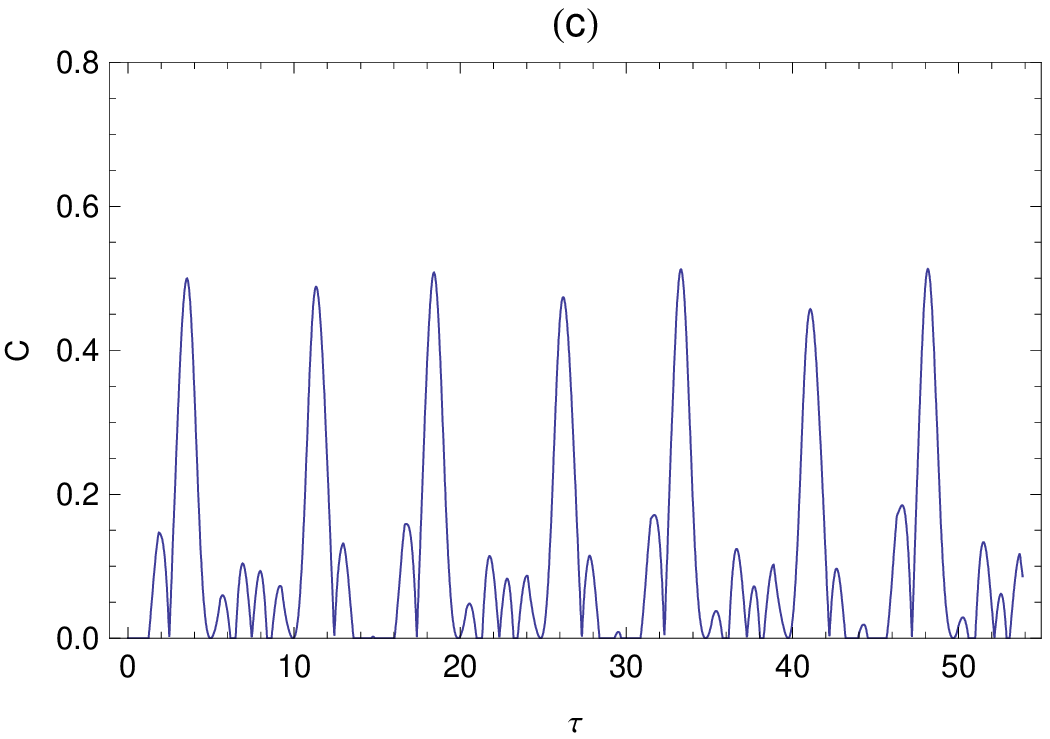}}
\hspace{0.3cm}
\subfigure{\label{Bdep1xyd}\includegraphics[width=6.5cm]{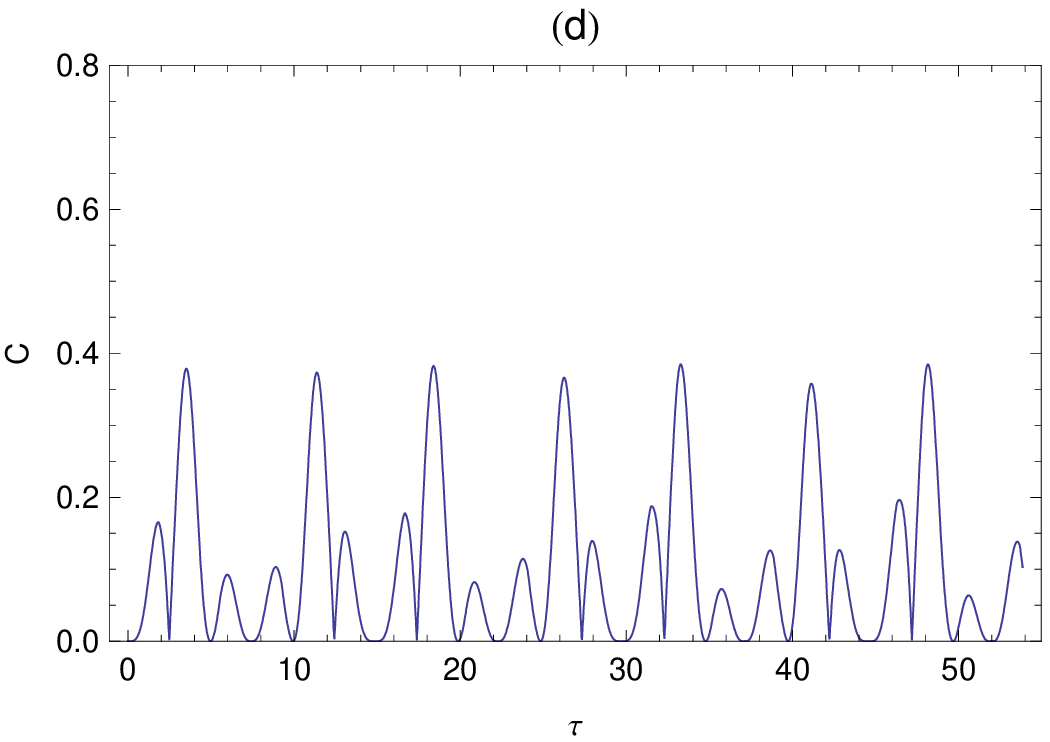}}
\hspace{0.3cm}
\subfigure{\label{Bdep1xye}\includegraphics[width=6.5cm]{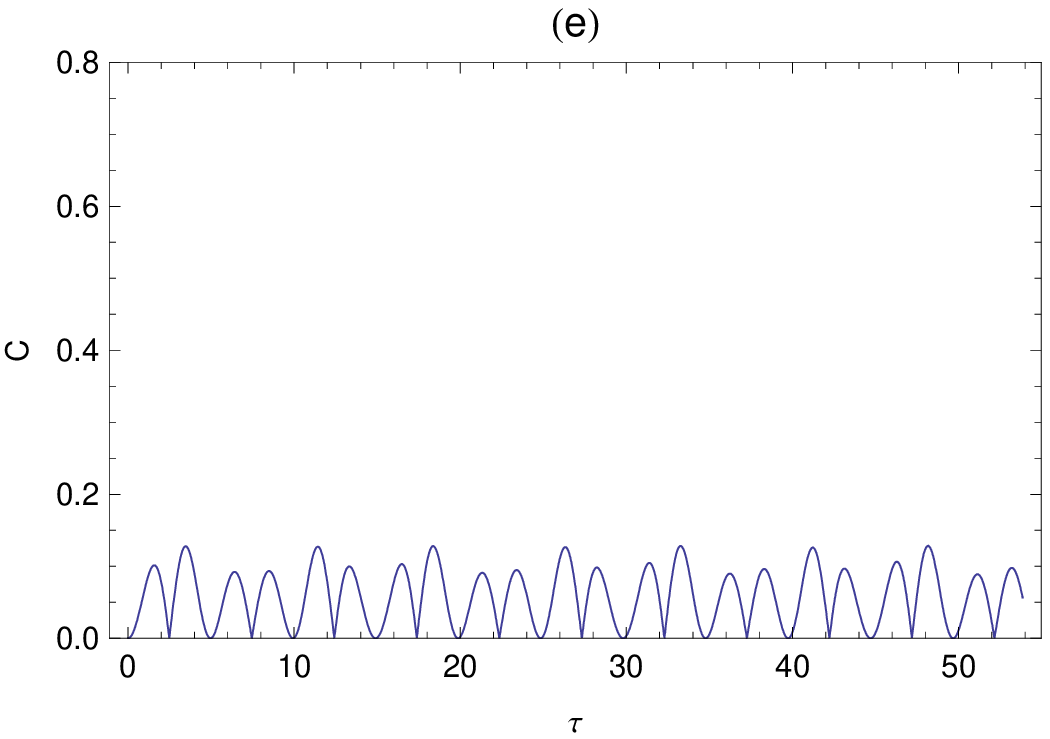}}
\hspace{0.3cm}
\subfigure{\label{Bdep1xyf}\includegraphics[width=6.5cm]{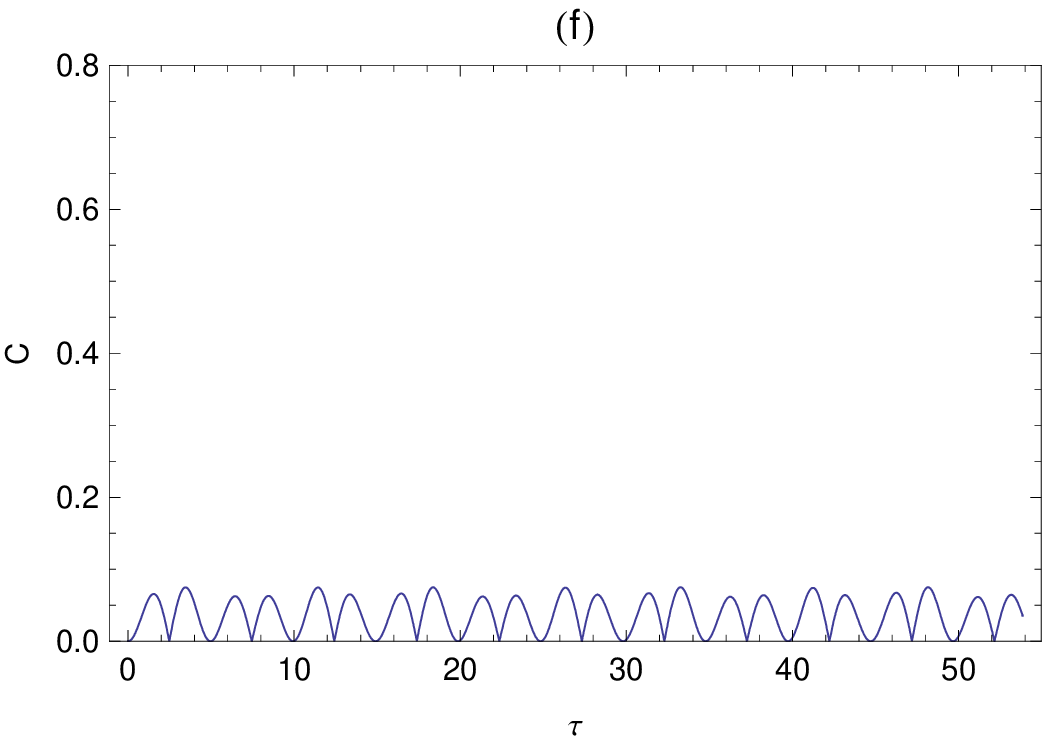}}
\end{center}
\caption{Input state $|\psi_0^B\rangle$, $\bm{Q}=0$, $B_z=0$, $N=10$, $\lambda=1$. The evolution of the concurrence with interaction time at different $\alpha$ for (a) $\alpha=0$, (b) $\alpha=0.2$, (c) $\alpha=\frac{1}{\sqrt{3}}$, (d) $\alpha=\frac{1}{\sqrt{2}}$, (e) $\alpha=\frac{\sqrt{5}}{\sqrt{6}}$, and (f) $\alpha=0.95$. Note the concurrence is finite for all $\alpha\neq1$. All quantities are expressed in natural units.}\label{Bdep1xy}
\end{figure}

The derivation of the system states for initial state $|\psi_0^B\rangle$ is identical to that outlined in the previous section, hence to avoid unnecessary repetition I will simply quote the results of the calculation. The first scattered state can be written as
\begin{eqnarray}
|\psi_2^B\rangle&=&e^{-i\Lambda\tau}\lbrack \alpha^2e^{-iy\tau}|1\rangle+\alpha\beta cd(1-e^{-2i\phi\tau})|2\rangle\nonumber\\
&+&\alpha\beta(d^2+c^2e^{-2i\phi\tau})|4\rangle+\alpha\beta e^{-iy\tau}|6\rangle\nonumber\\
&+&\beta^2 cd(1-e^{-2i\phi\tau})|7\rangle+\beta^2(d^2+c^2e^{-2i\phi\tau})|8\rangle\rbrack\label{psi2b exch},
\end{eqnarray}
where $\phi\equiv\phi^{xy}(\lambda,N,B_z)$, $\Lambda=B_z(N-1)-\phi$, and $c$, $d$ and $y$ are defined by equations (\ref{c xy}), (\ref{d xy}) and (\ref{y xy}), respectively. The second period of free evolution simply adds phases:
\begin{eqnarray}
|\psi_3^B\rangle&=&e^{-i\Lambda\tau}e^{-iB_zN\tau^{\prime}_f}\lbrack e^{-iy\tau}\alpha^2|1\rangle+\alpha\beta cde^{2iB_z\tau^{\prime}_f}(1-e^{-2i\phi\tau})|2\rangle\nonumber\\
&+&\alpha\beta(d^2+c^2e^{-2i\phi\tau})|4\rangle+e^{-iy\tau}\alpha\beta|6\rangle\nonumber\\
&+&\beta^2 e^{2iB_z\tau^{\prime}_f}cd(1-e^{-2i\phi\tau})|7\rangle+\beta^2(d^2+c^2e^{-2i\phi\tau})|8\rangle\rbrack\label{psi2b exch}.
\end{eqnarray}
The final state of the system is therefore
\begin{eqnarray}
|\psi_f^B\rangle&=&e^{-2i\Lambda\tau}e^{-iB_zN\tau^{\prime}_f}\lbrack e^{-2iy\tau}\alpha^2|1\rangle\nonumber\\
&+&cd\alpha\beta e^{2iB_z\tau^{\prime}_f}(1-e^{-2i\phi\tau})(c^2+d^2e^{-2i\phi\tau}+e^{-2iB_z\tau^{\prime}_f}e^{-iy\tau})|2\rangle\nonumber\\
&+&cdfg\beta^2e^{2iB_z\tau^{\prime}_f}(1-e^{-2i\phi\tau})(e^{-iw\tau}-e^{-iz\tau})|3\rangle\nonumber\\
&+&e^{-iy\tau}\alpha\beta(d^2+c^2e^{-2i\phi\tau})|4\rangle\nonumber\\
&+&cd\beta^2(d^2+c^2e^{-2i\phi\tau})(1-e^{-2i\phi\tau})|5\rangle\nonumber\\
&+&\alpha\beta c^2d^2e^{2iB_z\tau^{\prime}_f}(1-e^{-2i\phi\tau})^2+\alpha\beta e^{-iy\tau}(d^2+c^2e^{-2i\phi\tau})|6\rangle\nonumber\\
&+&cd\beta^2 e^{2iB_z\tau^{\prime}_f}(1-e^{-2i\phi\tau})(f^2e^{-iz\tau}+g^2e^{-iw\tau})|7\rangle\nonumber\\
&+&\beta^2(d^2+c^2e^{-2i\phi\tau})^2|8\rangle\rbrack\nonumber,
\end{eqnarray}
with
\begin{eqnarray}
\gamma&\equiv&\gamma^{xy}(\lambda,N,B_z)=\sqrt{B_z^2+\frac{8\lambda^2}{N^2}\left(N-1\right)},\\
f&=&\sqrt{\frac{1}{2}-\frac{B_z}{2\gamma}},\\
g&=&\sqrt{\frac{1}{2}+\frac{B_z}{2\gamma}},\\
w&=&-2 B_z + \gamma + \phi,\\
z&=&-2 B_z - \gamma + \phi.
\end{eqnarray}
The results of section \ref{p0a finite mtm} show the behaviour of the system in the presence of sample anisotropy is similar to that observed for an isotropic sample, offset by a field $B_z^*$; this is true here also. The evolution of the concurrence in time at zero field can therefore be summarized in the following observations: (i) the concurrence is independent of $\tau_f^{\prime}$; (ii) for $\alpha\neq0$, the concurrence is not a periodic function of the interaction time; (iii) the peak concurrence is finite for all $\alpha\neq1$, but maximal when $\alpha=0$, in which case it takes the form
\begin{align}\label{c psib field xy}
C&=4\left|\sin{\frac{4\lambda\tau}{\sqrt{N}}}\left(\cos{\frac{2\lambda\tau}{\sqrt{N}}}\sin{\frac{2\lambda\sqrt{2(N-1)}\tau}{N}}-\sin{\frac{2\lambda\tau}{\sqrt{N}}}\cos{\frac{2\lambda\sqrt{2(N-1)}\tau}{N}}\right)\right|.\nonumber\\
&\:
\end{align}
This peaks above 0.76 for all $N$, converging to a value 0.77 as $N\rightarrow\infty$. These characteristics are illustrated in figures \ref{Bdep0xy} and \ref{Bdep1xy}.

\begin{figure}[H]
\renewcommand{\captionfont}{\footnotesize}
\renewcommand{\captionlabelfont}{}
\begin{center}
\subfigure{\label{Bdep2xya}\includegraphics[width=6.5cm]{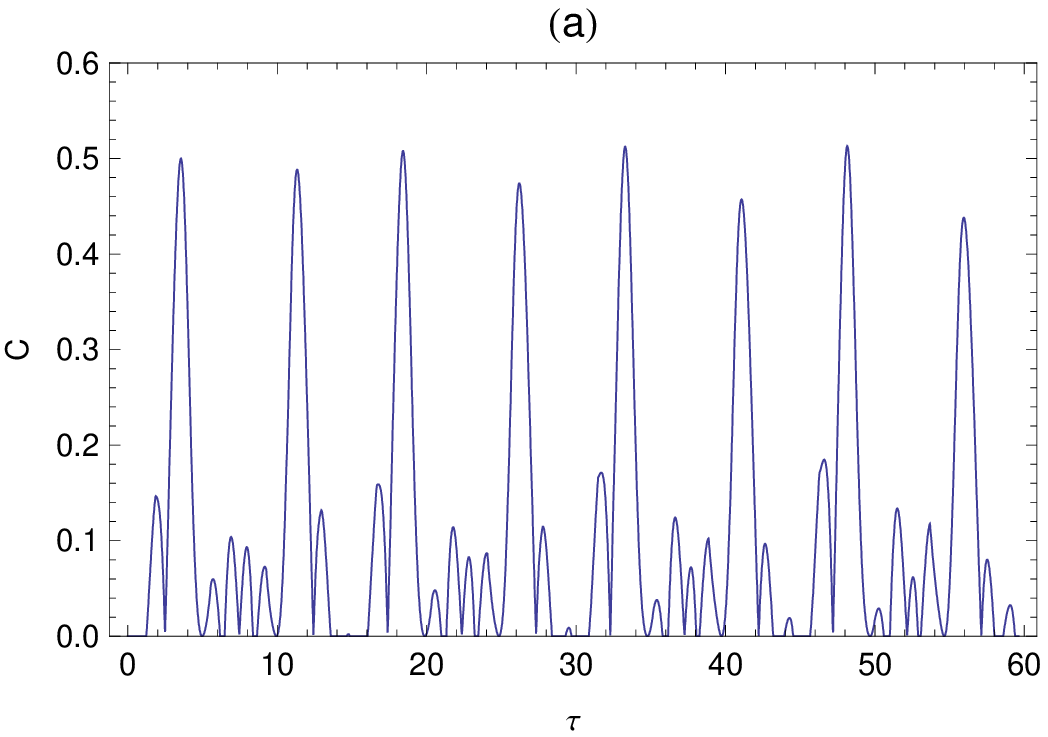}}
\hspace{0.3cm}
\subfigure{\label{Bdep2xyb}\includegraphics[width=6.5cm]{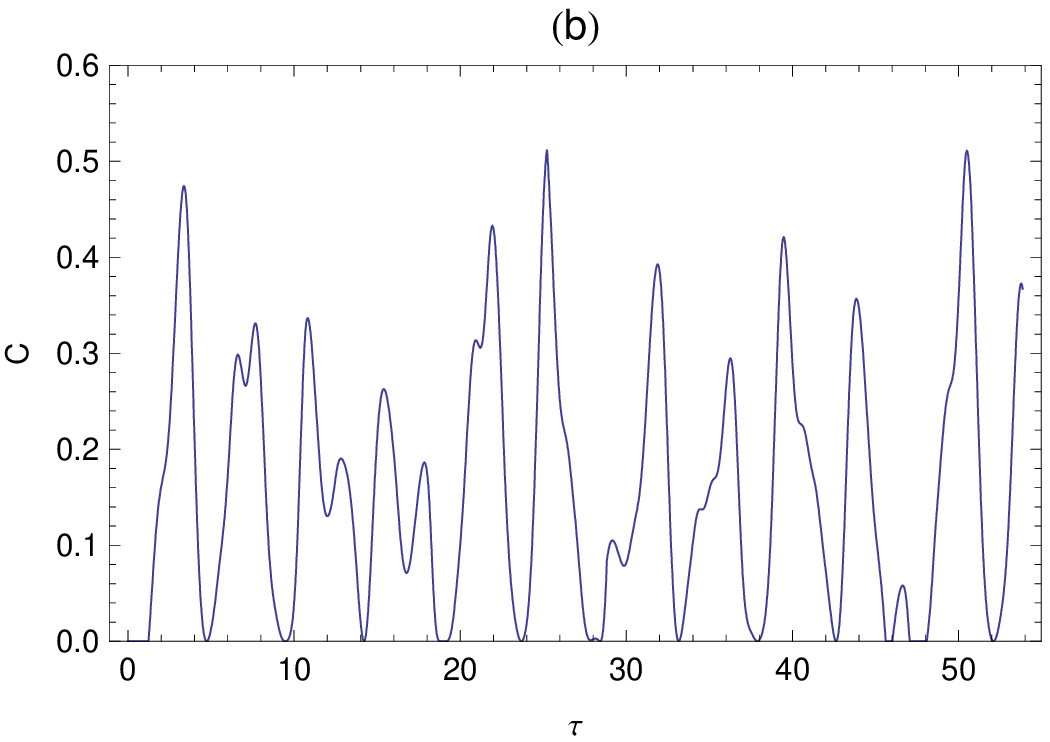}}
\hspace{0.3cm}
\subfigure{\label{Bdep2xyc}\includegraphics[width=6.5cm]{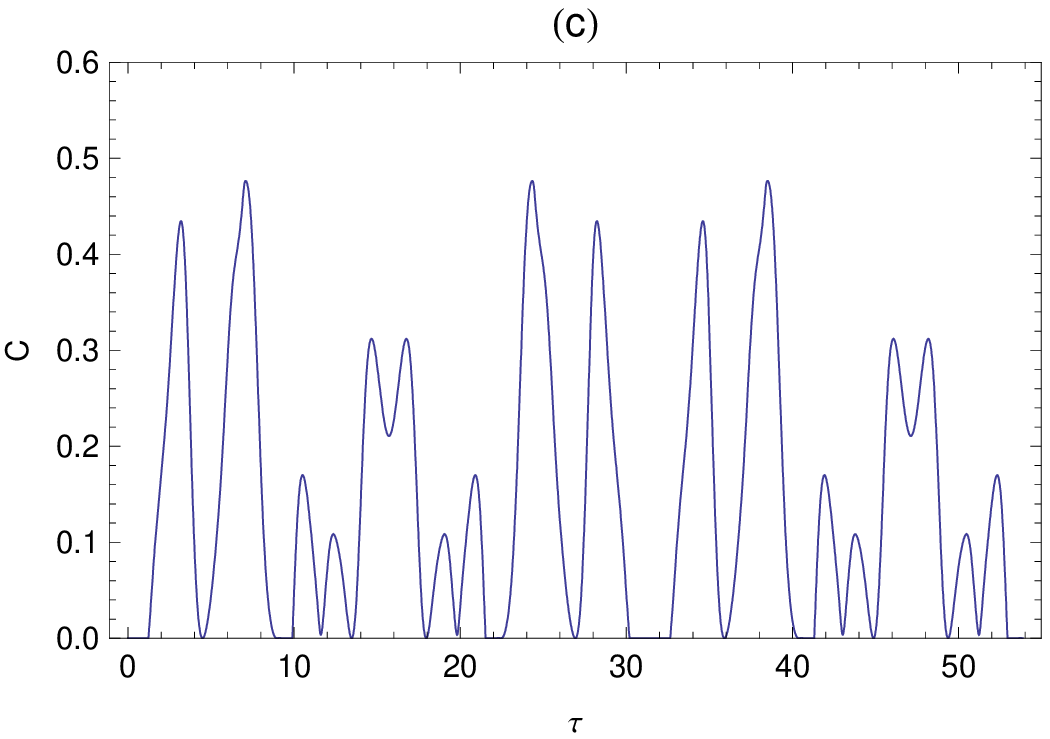}}
\hspace{0.3cm}
\subfigure{\label{Bdep2xyd}\includegraphics[width=6.5cm]{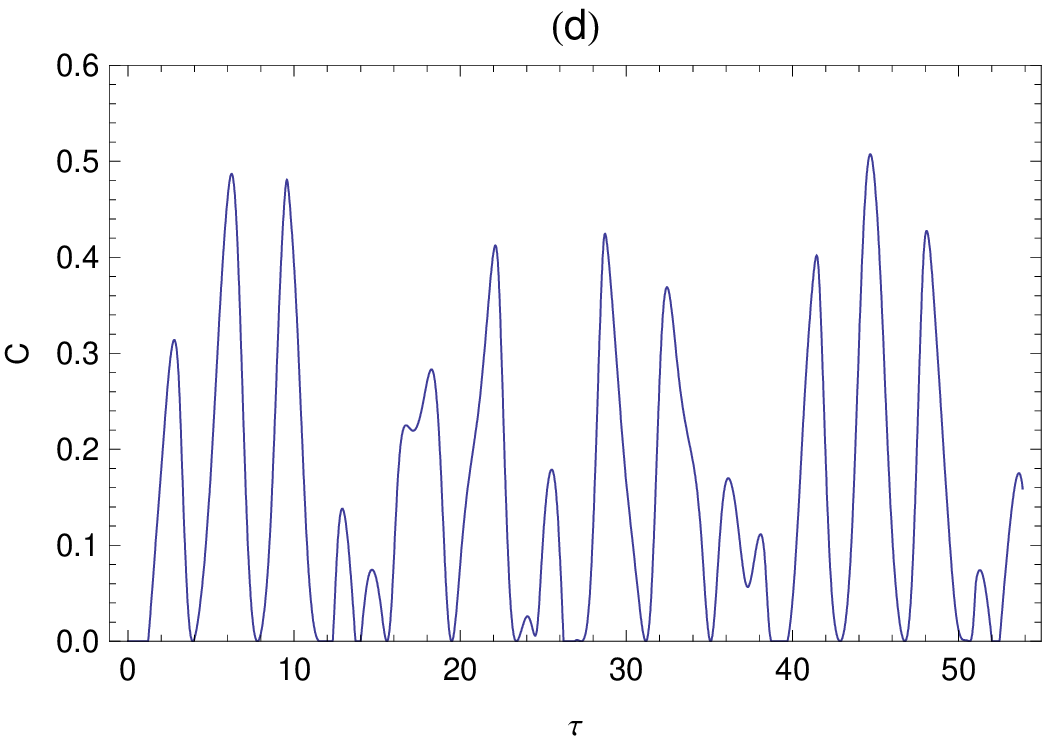}}
\hspace{0.3cm}
\subfigure{\label{Bdep2xye}\includegraphics[width=6.5cm]{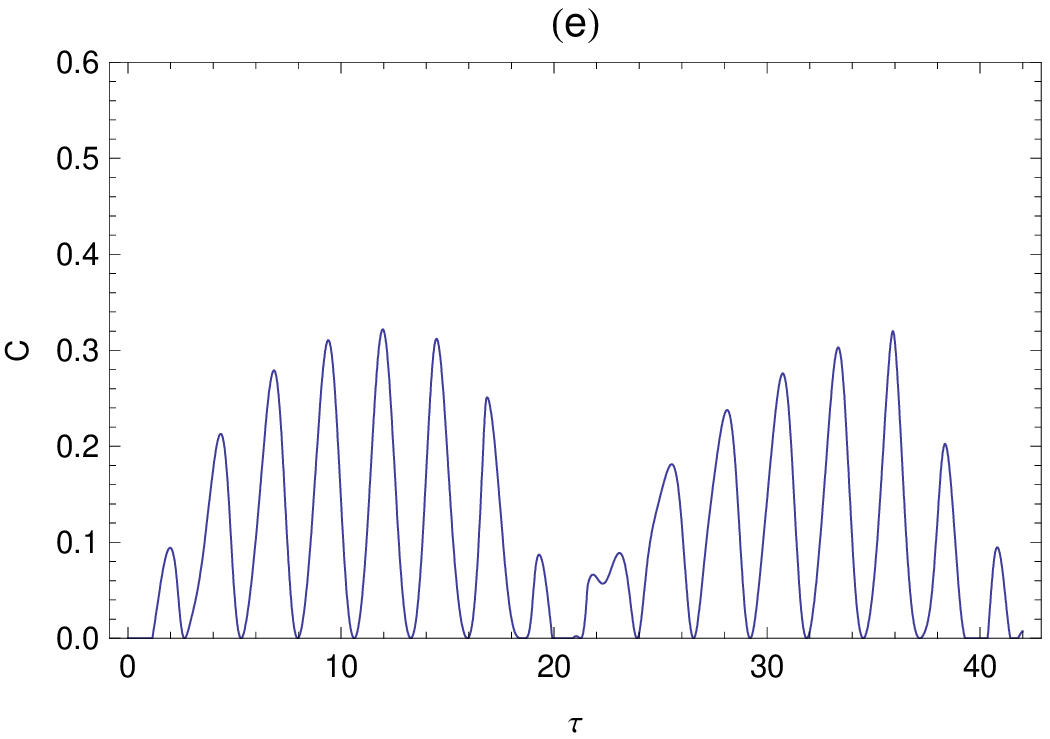}}
\hspace{0.3cm}
\subfigure{\label{Bdep2xyf}\includegraphics[width=6.5cm]{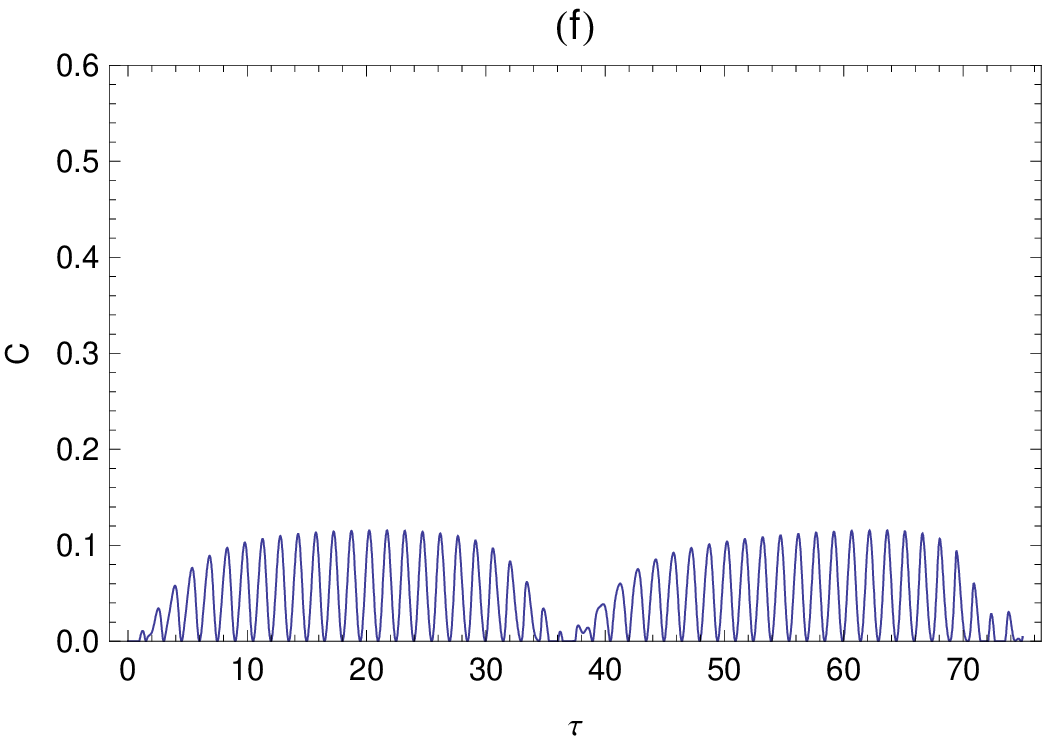}}
\end{center}
\caption{Input state $|\psi_0^B\rangle$, $\bm{Q}=0$, $\alpha=\frac{1}{\sqrt{3}}$, $N=10$, $\lambda=1$. The evolution of the concurrence with interaction time at different $B_z$ for (a) $B_z=0$, (b) $B_z=0.2$, (c) $B_z=0.3$, (d) $B_z=0.5$, (e) $B_z=1$, and (f) $B_z=2$. Note that, for $B_z<\lambda$, the peak concurrence remains roughly constant with $B_z$. All quantities are expressed in natural units.}\label{Bdep2xy}
\end{figure}

\begin{figure}[H]
\renewcommand{\captionfont}{\footnotesize}
\renewcommand{\captionlabelfont}{}
\begin{center}
\subfigure{\label{Bdep3xya}\includegraphics[width=6.5cm]{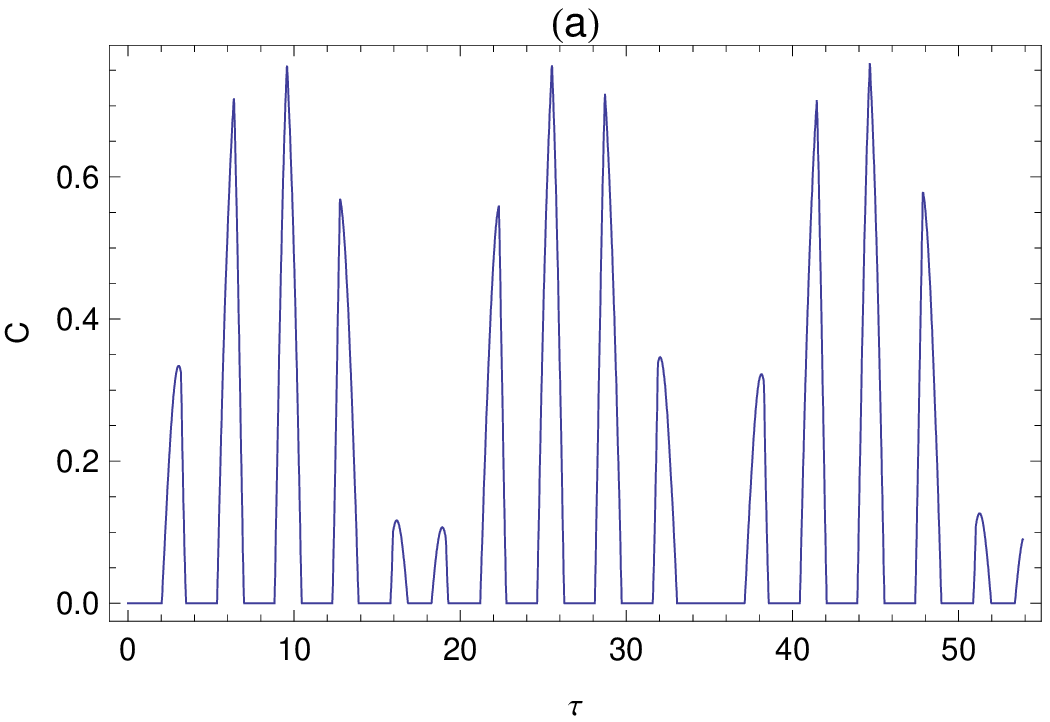}}
\hspace{0.3cm}
\subfigure{\label{Bdep3xyb}\includegraphics[width=6.5cm]{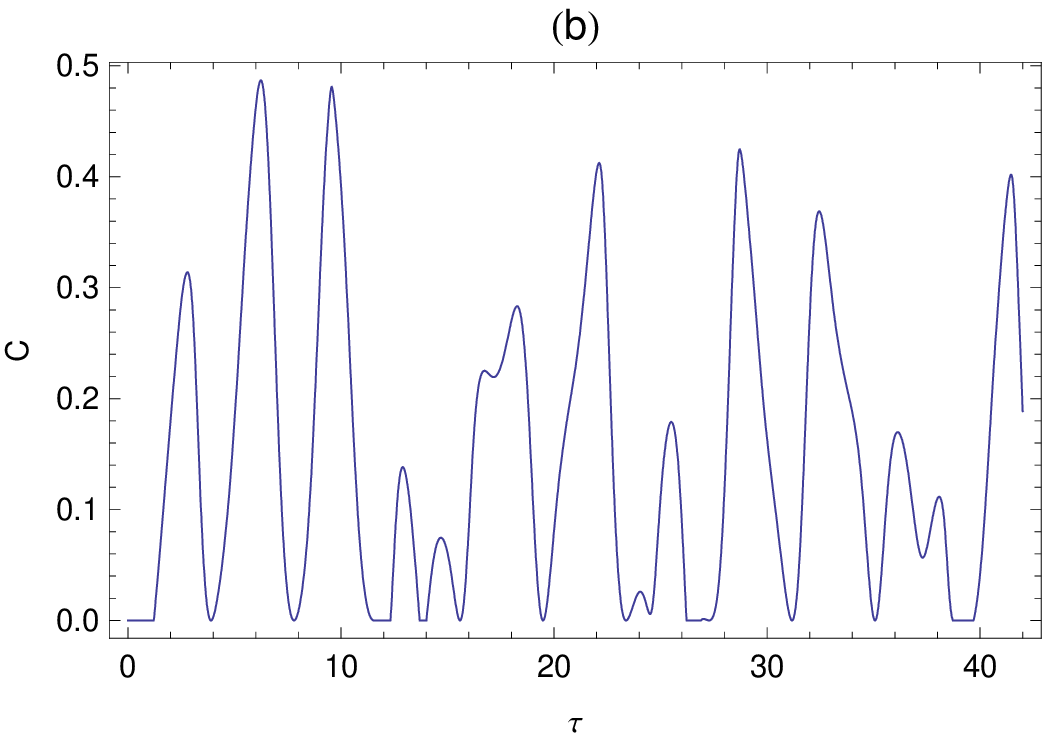}}
\hspace{0.3cm}
\subfigure{\label{Bdep3xyc}\includegraphics[width=6.5cm]{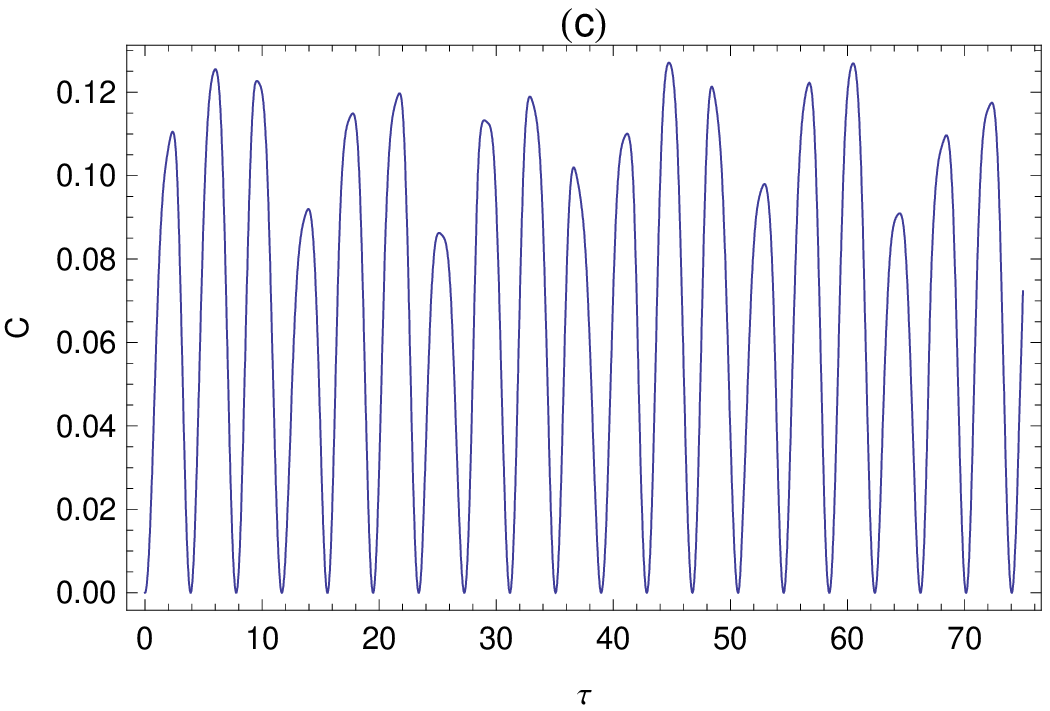}}
\end{center}
\caption{Input state $|\psi_0^B\rangle$, $\bm{Q}=0$, $B_z=0.5$, $N=10$, $\lambda=1$. The evolution of the concurrence with interaction time at different $B_z$ for (a) $B_z=0$, (b) $B_z=0.2$, (c) $B_z=0.3$, (d) $B_z=0.5$, (e) $B_z=1$, and (f) $B_z=2$. Note that, for $B_z<\lambda$, the peak concurrence remains roughly constant with $B_z$. All quantities are expressed in natural units.}\label{Bdep3xy}
\end{figure}

The behaviour of the concurrence in finite field is more complex. For all $\alpha$, $N$ and $B_z$ the concurrence is an oscillating function of time. At $B_z\lesssim \lambda$, the oscillation is irregular and lacks an easily discernible periodicity, but the peak concurrence remains roughly constant with $B_z$. Conversely, for $B_z\gtrsim\lambda$, there emerges a beating effect similar to that observed in figure \ref{Bdep6} for $B_z>B_{z+}$ and $B_z<B_{z-}$ (see figure \ref{Bdep2xy}). The timescales involved are once again $T_{\phi}$ (rapid oscillation) and $T_{\gamma}$ (envelope function), with $\phi\equiv \phi^{xy}$ and $\gamma\equiv \gamma^{xy}$. In this range of $B_z$, the concurrence drops rapidly with increasing field (figure \ref{Bdep3xy}). Bearing in mind the form of $c$ and $d$, this is due to the fact that, as the field is raised, the single-excitation eigenstates of the potential deviate increasingly from being equal superpositions. Indeed, by the same argument the concurrence at fixed $B_z$ increases with increasing $N$. For all $B_z$, the peak concurrence is strongly related to the neutron polarization, and attains a maximum when $\alpha=0$. The concurrence in this case takes the form
\begin{eqnarray}
C&=&\frac{8\lambda^2\sqrt{B_z^2+\frac{2\lambda^2}{N}\left[1+\cos{2\phi\tau}\right]}}{N^2\gamma\phi^3}\left[N\sin^2{\phi\tau}-\right.\nonumber\\
&-&\left.\sqrt{2N(N-1)\left[B_z^2+\frac{2\lambda^2}{N}\left(1+\cos{2\phi\tau}\right)\right]}|\sin{\phi\tau}\sin{\gamma\tau}|\right].
\end{eqnarray}

In summary, with the exception of a few quantitative details, the performance of the protocol is not significantly affected by a strong sample anisotropy. As observed previously, if the sample is in a fully polarized state the behaviour of the system becomes more complex and less predictable, but no less effective.

\section{Conclusions}
Two distinct neutrons can be entangled with respect to their spin degree of freedom if they are made to scatter sequentially from a macroscopic sample having the properties of a ferromagnetic insulator. I have considered forward scattering from an isotropic or strongly anisotropic sample. In both cases, the system must be initialized to a well-defined state, chosen such that the sample contains at most one spin excitation. More freedom is permitted in the choice of the neutron states, however there does exist an optimal polarization, which is determined by the initial state of the sample and the magnitude of the external magnetic field. It must also be possible to tune the time each neutron spends interacting with the sample. This can be done by regulating the neutron velocity. Provided these conditions can be met, the neutrons can come to share as much as 0.68 ebits of entanglement.

In light of present-day achievements across the range of disciplines employing the tools and methods I have discussed, a realization of the entanglement scheme in its present form does not seem feasible. However, future progress in neutron cooling and manipulation may change this state of affairs.

Leaving aside for a moment the problem of producing neutrons with sufficiently large coherence volumes, the most difficult part of the protocol is the entanglement detection stage. Evaluating the witness of equation (\ref{wit}) involves measuring expectation values of the polarizations of the scattered neutrons along all three axes. Currently operating neutron spectrometers are capable of full XYZ polarization analysis \cite{stewart00}, but with two limitations. First, for a given scattering event, it is only possible to measure the spin of the outgoing neutrons along one axis - either X, Y or Z. Second, only neutrons with spin in one direction along the chosen axis are detected. For example, suppose one had a beam of scattered neutrons containing a mixture of particles with spin up or down along the $\hat{\bm{z}}$-direction, and wished to measure the $z$-component of the spin of each neutron. One could either set the analyzer to detect neutrons with spin up \emph{or} to detect neutrons with spin down. If the setting were chosen as spin up, none of the neutrons with spin down would be detected. Instead, the relative occurrence of spin up and down in the beam would be inferred by relating the measured flux to the total expected scattered flux. Now, for the purposes of measuring the witness of equation (\ref{wit}), this is not a problem until one comes to the term in $|y^+y^-\rangle\langle y^+y^-|+hc$, which can only be measured reliably if both the scattered neutrons are detected. It may then be necessary to employ more than one measurement strategy to evaluate the witness. One might, for example, measure the first four components using a spectrometer such as D7, currently in operation at the ILL \cite{stewart00}, and measure the final two by routing the scattered neutrons through a rotated Stern-Gerlach apparatus. This would certainly be extremely challenging, but is not \textit{a priori} impossible.

%% file: many_neutrons.tex
\chapter{Many-Neutron Scattering}\label{many ns}
\textit{In this chapter, I extend the protocol described in chapter \ref{neutron proposal} to study the concurrence of an arbitrary pair of neutrons scattered from the sample. Once again, I consider sequential scattering at zero momentum transfer from a sample in one of two possible states - the polarized, and single-magnon states of equations \eqref{psiAs} and \eqref{psiBs}. I find that, under optimal circumstances, pairs of neutrons can share a finite concurrence even if they are not scattered in succession. It is therefore possible to create entangled states of many neutrons; in fact, for certain choices of input parameters, the amount of entanglement of any neutron with any other becomes a slowly-varying function of the neutron indices. Consequently, the probability of creating an entangled neutron pair is high, provided the sample can be periodically reset to its initial state. In addition, measurement of the witness is no longer subject to the constraint of detecting the first two scattered neutrons.}

\section{Introduction}
I showed in the previous chapter that, conditional on some limitations, the first two neutrons to scatter from a sample prepared in a specific eigenstate of its internal Hamiltonian can be measurably entangled. For an optimal realization of the protocol, this entanglement is detected by the witness operator of equation \eqref{wit}, whose expectation value is calculated on the states of the two outgoing neutrons. As the neutron scattering cross-section is small, if the beam is sufficiently dilute and the sample state is periodically reset, it is reasonable to assume the first two neutrons detected in rapid succession following a reset pulse are indeed the first two neutrons to scatter from the sample.

But what if this were not the case? Equation \eqref{wit} represents a particularly simple decomposition of a generic witness operator capable of detecting the entanglement of quantum states having some overlap with the state $p|01\rangle+q|10\rangle$. This decomposition is valid provided the eigenvector corresponding to the negative eigenvalue of the partially-transposed density matrix of the neutrons has the form $a|00\rangle+b|11\rangle$. The state $\rho^A$ of equation \eqref{rho n A} falls into this category, however the state $\rho_{m,n}$ of two arbitrary scattered neutrons $m$ and $n$ may not. This means the witness in its present form is no longer guaranteed to detect the entanglement of $\rho_{m,n}$. In addition, detection of the wrong neutron pair could lead to an inaccurate assessment of the performance of the protocol. Suppose, for example, that $\rho_{1,2}$ is entangled, but $\rho_{2,3}$ is not. By accidentally detecting the second and third neutrons, rather than the first and the second, we might conclude the protocol has failed, whereas in fact it has succeeded.

It may then be useful to know exactly how much rests on the ability to detect the first two neutrons. To this end, I study the concurrence $C_{m,n}$ of an arbitrary neutron pair $m$ and $n$, and calculate the average concurrence over all detected neutron pairs $\lbrace m,n\rbrace$.

\section{Scattering from a Sample in a Single Magnon State}\label{mn psi0a}
We saw in section \ref{p0a zero mtm} that when the system is initialized to $|\psi_0^A\rangle$, quantum state evolution is restricted to the single excitation sector of the Hilbert space. This allows the concurrence of the scattered neutrons, and the behaviour of the system in general, to be described in terms of analytical expressions, making clear how the performance of the protocol is affected by changing the input parameters. The same argument applies to the situation we now wish to describe. By deriving the state of the system after the first few scattering events, with the help of equation \eqref{teo}, there quickly emerges a pattern in the coefficients of the scattered state. This pattern allows us to extrapolate the form of the wavefunction, hence of the reduced density matrix of a chosen neutron pair, after an arbitrary number of scattering events. I will spare the reader a detailed account of this derivation, as it is completely straightforward but very lengthy. Suffice it to say that, omitting global phase factors, one finds the first four scattered states of the system as a whole to be
\begin{align}
|\psi_1\rangle&=P|01\rangle+Q|10\rangle;\\
|\psi_2\rangle&=P^2|001\rangle+Qe^s|010\rangle+R|100\rangle;\\
|\psi_3\rangle&=P^3|0001\rangle+Qe^{2s}|0010\rangle+Re^s|0100\rangle+QP^2|1000\rangle;\\
|\psi_4\rangle&=P^4|00001\rangle+Qe^{3s}|00010\rangle+Re^{2s}|00100\rangle\nonumber\\
&\:+QP^2e^s|01000\rangle+QP^3e^s|10000\rangle;
\end{align}
with
\begin{align}
s&=\lambda\left(1+\frac{1}{N}\right)+B_z+\phi,\\
P&=c^2+d^2e^{-2i\phi\tau},\\
Q&=cd\left(1-e^{-2i\phi\tau}\right),\\
R&=c^3d\left(1-e^{-4i\phi\tau}\right).
\end{align}
The basis I have used is an extension of the Bloch basis, where the right-most qubit represents the state of the sample and all qubits to the left of this label the neutrons, disposed in order of arrival from right to left. It then becomes clear that scattering a fifth neutron from the sample will produce the state
\begin{align}
|\psi_5\rangle&=P^5|000001\rangle+Qe^{4s}|000010\rangle+Re^{3s}|000100\rangle\nonumber\\
&\:+QP^2e^{2s}|001000\rangle+QP^3e^s|010000\rangle+QP^4|100000\rangle.
\end{align}
Therefore, the general state $|\psi_N\rangle$ can be written as
\begin{equation}
|\psi_N\rangle=P^N|1\rangle+e^{\left(N-2\right)s}R|4\rangle+\sum_{i=1,\:i\neq 2}^N QP^{i-1}e^{\left(N-i\right)s}|2^i\rangle,
\end{equation}
where the number inside the ket represents the binary value of the state in a single-spin-flip basis.

In theory, knowing the form of the $|\psi_i\rangle$ provides a means to calculate the reduced density matrix $\rho_{m,n}$ of any neutron pair, hence the concurrence $C_{m,n}$ of that pair. At first glance, this appears to be quite a task, however things are much simpler than they seem. One finds from numerical simulations that the concurrence of any neutron and any other can be related to the concurrence of the first neutron and those scattered after. Therefore, one can obtain from the set of density matrices $\rho_{1,n}$ all the information necessary to provide a full characterization of the concurrence dynamics of any neutron pair.

It can be shown that, for all $n$, the structure of $\rho_{1,n}$ in the canonical basis is of the form
\[ \rho_{1,n} = \left( \begin{array}{cccc}
p & 0&0&0 \\
0 & q&r&0\\
0&r^*& s&0\\
0&0&0&0\end{array} \right),\]
where the coefficients $p$, $q$, $r$ and $s$ are obtained from the $|\psi_i\rangle$ by tracing over all neutrons save the first and the $n^{th}$. From the eigenvalues of the spin-flipped counterpart of $\rho_{1,n}$, the concurrence $C_{1,n}$ is
\begin{equation}
C_{1,n}=2\sqrt{qs},
\end{equation}
with $qs=\left|P^{n-1}\right|^2\left|Q\right|^4$. Therefore
\begin{equation}\label{c1n}
C_{1,n}\left(N,\lambda,B_z,\tau\right)=2\left|P^{n-1}\right|\left|Q\right|^2,
\end{equation}
as $P$ and $Q$ are in general functions of $N$, $\lambda$, $B_z$ and $\tau$. Finally, from numerical studies, one finds for the concurrence of an arbitrary neutron pair $\lbrace m,n\rbrace$
\begin{equation}\label{c mn}
C_{m,n}=C_{1,m+n-1}.
\end{equation}
This equation reveals two interesting properties of our system. First, at fixed $m$, the concurrence $C_{m,n}$ will always fall as $|m-n|$ increases. Therefore, neutrons separated by long time gaps can never be more entangled than neutrons detected in rapid succession. Second, the concurrence of a given pair of neutrons \emph{not} scattered in succession is optimal if one of those neutrons is the first.

Let us now consider how $C_{m,n}$ is affected by the neutron interaction time and the strength of the applied field.

\subsection{Zero Field Evolution}\label{0f psia mn}
\begin{figure}[H]
\renewcommand{\captionfont}{\footnotesize}
\renewcommand{\captionlabelfont}{}
\begin{center}
\subfigure{\label{psia0f1a}\includegraphics[width=6.5cm]{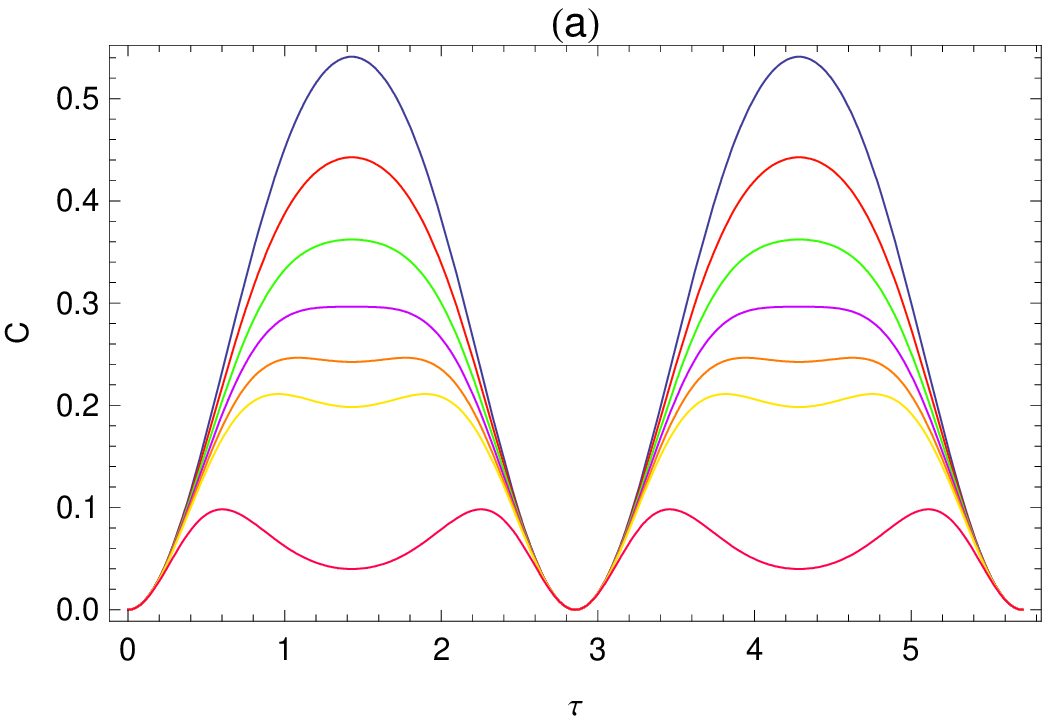}}
\hspace{0.3cm}
\subfigure{\label{psia0f1b}\includegraphics[width=6.5cm]{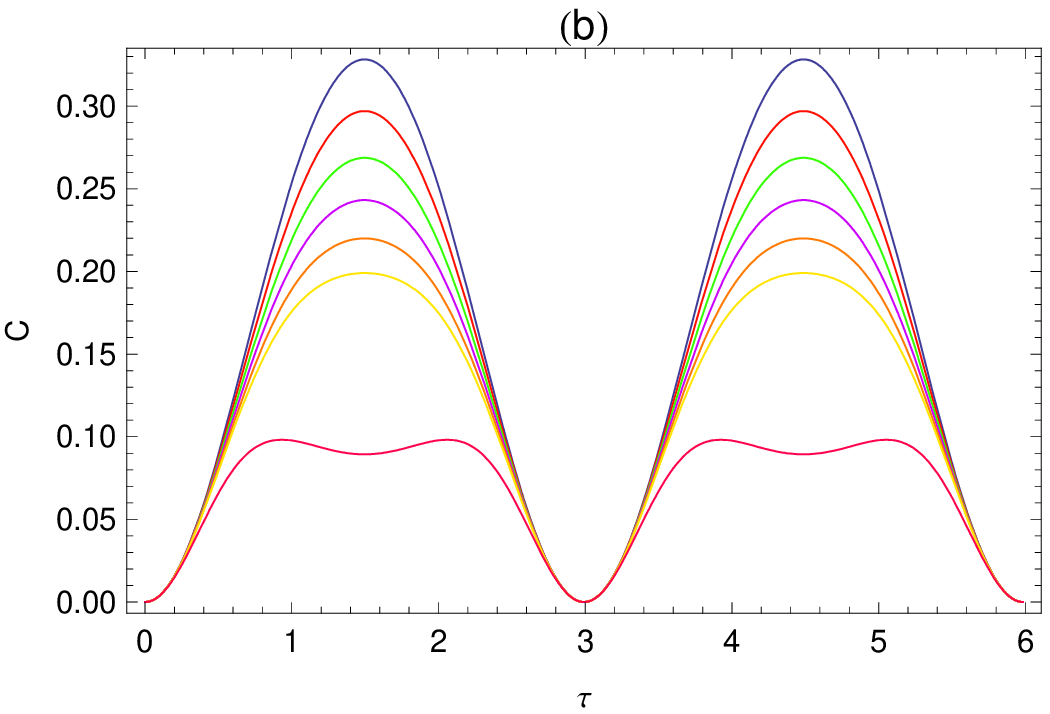}}
\hspace{0.3cm}
\subfigure{\label{psia0f1c}\includegraphics[width=6.5cm]{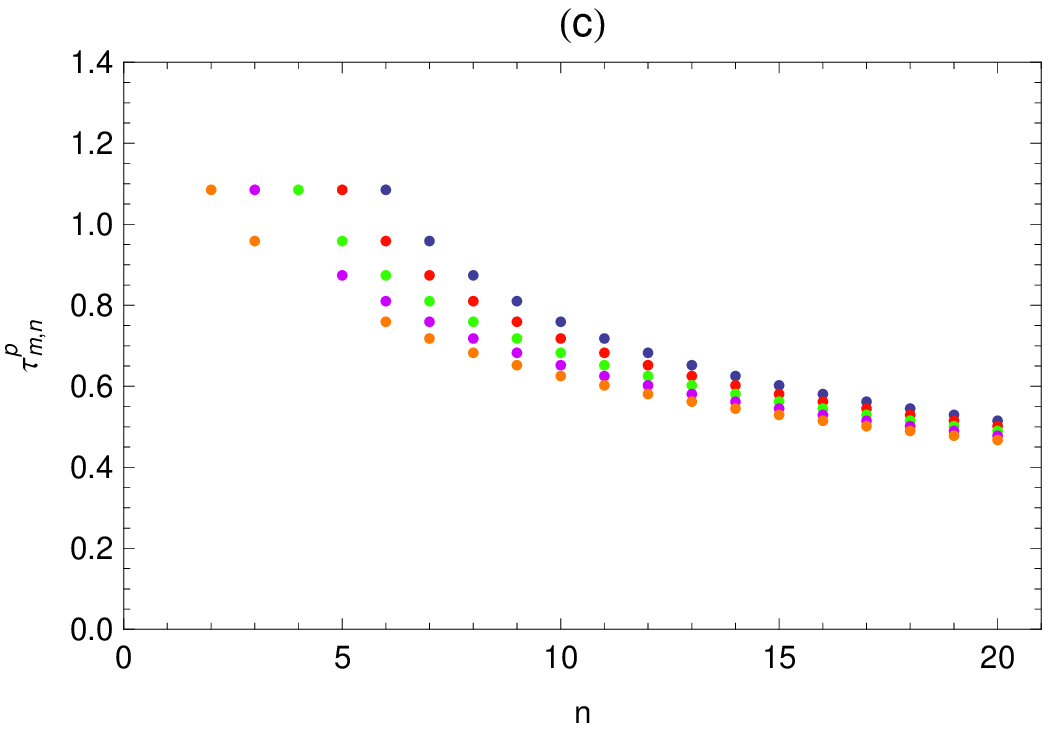}}
\hspace{0.3cm}
\subfigure{\label{psia0f1d}\includegraphics[width=6.5cm]{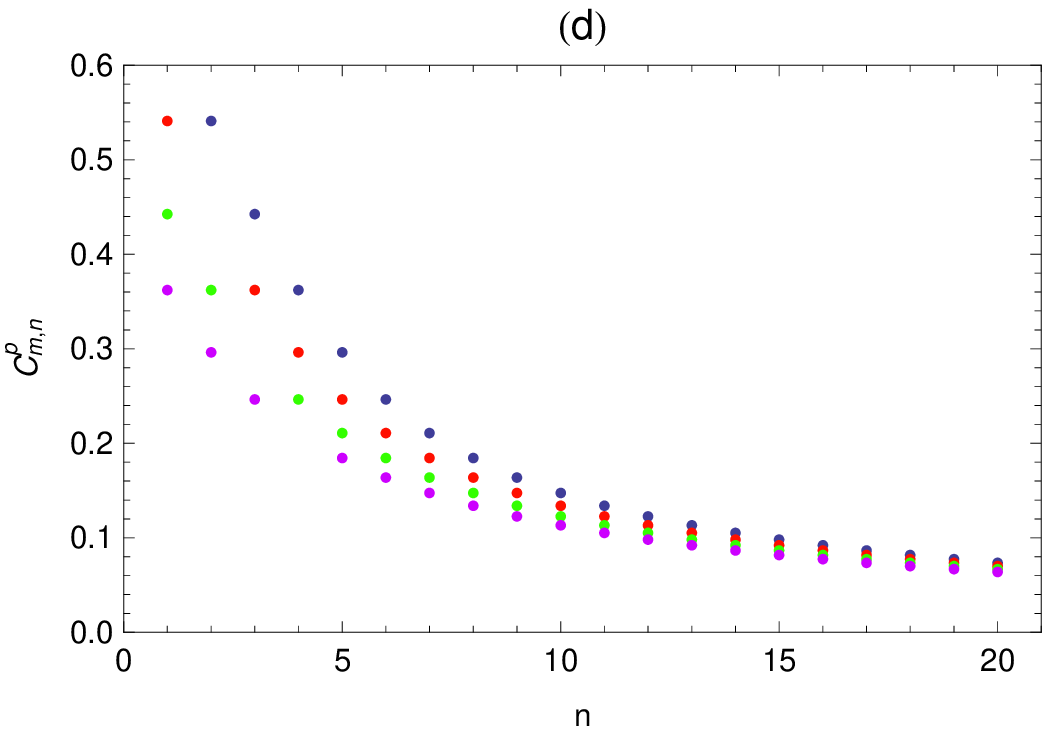}}
\end{center}
\caption{(a) The concurrence $C_{1,n}$ for $n=\lbrace2,3,4,5,6,7,15\rbrace$ and $N=10$. Increasing values of $n$ are indicated by the blue, red, green, purple, orange, yellow and magenta curves, respectively. Note $\zeta=6.05$, hence the relative structure of the purple and orange curves. (b) As figure (a) for $N=20$. Note in this case $\zeta=11.03$, hence the relative structure of the yellow and magenta curves. (c) The optimal interaction time $\tau^0_{m,n}$ at $N=10$ and large $\zeta$, for $m=1$ (blue dots), $m=2$ (red dots), $m=3$ (green dots), $m=4$ (purple dots) and $m=5$ (orange dots). (d) The peak concurrence as a function of $n$ at $N=10$ for $m=1$ (blue dots), $m=2$ (red dots), $m=3$ (green dots) and $m=4$ (purple dots). All quantities are expressed in natural units, with $\lambda=1$.}
\label{psia0f1}
\end{figure}

In zero field, equation \eqref{c1n} becomes
\begin{align}
C_{1,n}\left(N,\lambda,0,\tau\right)=8N\left(N+1\right)^{-n-1}&\sin^2{\left[\left(1+\frac{1}{N}\right)\lambda\tau\right]}\nonumber\\
&\:\bigg\lbrace N^2+1+2\cos{\left[2\left(1+\frac{1}{N}\right)\lambda\tau\right]}\bigg\rbrace^{\frac{n-1}{2}}.
\end{align}
From \eqref{c mn}, the concurrence of an arbitrary neutron pair is then
\begin{align}\label{c mn 0f}
C_{m,n}\left(N,\lambda,0,\tau\right)=8N\left(N+1\right)^{-m-n}&\sin^2{\left[\left(1+\frac{1}{N}\right)\lambda\tau\right]}\nonumber\\
&\:\bigg\lbrace N^2+1+2\cos{\left[2\left(1+\frac{1}{N}\right)\lambda\tau\right]}\bigg\rbrace^{\frac{m+n-2}{2}}.
\end{align}
Generalizing the remarks of section \ref{p0a zero mtm zero field}, we now see that the concurrence of any two neutrons is roughly proportional to $N^{-1}$. For all $m$ and $n$, $C_{m,n}$ is an oscillating function of time, with periodicity $T_{\phi}(N,\lambda,0)$ [see equation \eqref{T ex}]. The mode of these oscillations is determined by the relative values of $N$ and $m+n$. Let us define the quantity $\zeta=m+n$. If
\begin{equation}\label{zeta}
\zeta<\frac{\left(N+1\right)^2}{2 N},
\end{equation}
$C_{m,n}$ is sinusoidal, and peaks at time $\tau_0=T_{\phi}/2$. Otherwise, the concurrence develops a double-peaked structure, whose maxima shift further apart as $\zeta$ increases [figures \ref{psia0f1a} and \ref{psia0f1b}]. The position of the first peak is determined by the relation
\begin{equation}
\tau^0_{m,n}=\frac{N}{2\lambda\left(N+1\right)}\sec^{-1}\left[\frac{N\zeta}{1+N^2-N\left(\zeta-2\right)}\right],
\end{equation}
where $\tau^0_{m,n}$ at fixed $m$ is shown in figure \ref{psia0f1c}. The peak concurrence of an arbitrary neutron pair is, therefore
\begin{equation}\label{cp0fp}
\mathcal{C}^{p+}=4\sqrt{\frac{\left(\zeta-2\right)^{\zeta-2}}{\zeta^{\zeta}}} ,
\end{equation}
if equation \eqref{zeta} is satisfied, or
\begin{equation}\label{cp0fm}
\mathcal{C}^{p-}=\frac{8N\left(N-1\right)^{\zeta-2}}{\left(N+1\right)^{\zeta}},
\end{equation}
if it is not. The transition between the two regimes can be observed by comparing the form of the purple and orange curves of figure \ref{psia0f1a}. Interestingly, $\mathcal{C}^{p+}$ is independent of $N$, suggesting that as $\zeta$ becomes large, the maximum achievable entanglement between any two neutrons depends only on the number of scattering events which have taken place. The peak concurrence at fixed $m$ for $N=10$ is shown in figure \ref{psia0f1d}.

\begin{figure}[H]
\renewcommand{\captionfont}{\footnotesize}
\renewcommand{\captionlabelfont}{}
\begin{center}
\subfigure{\label{psia0f2a}\includegraphics[width=6.5cm]{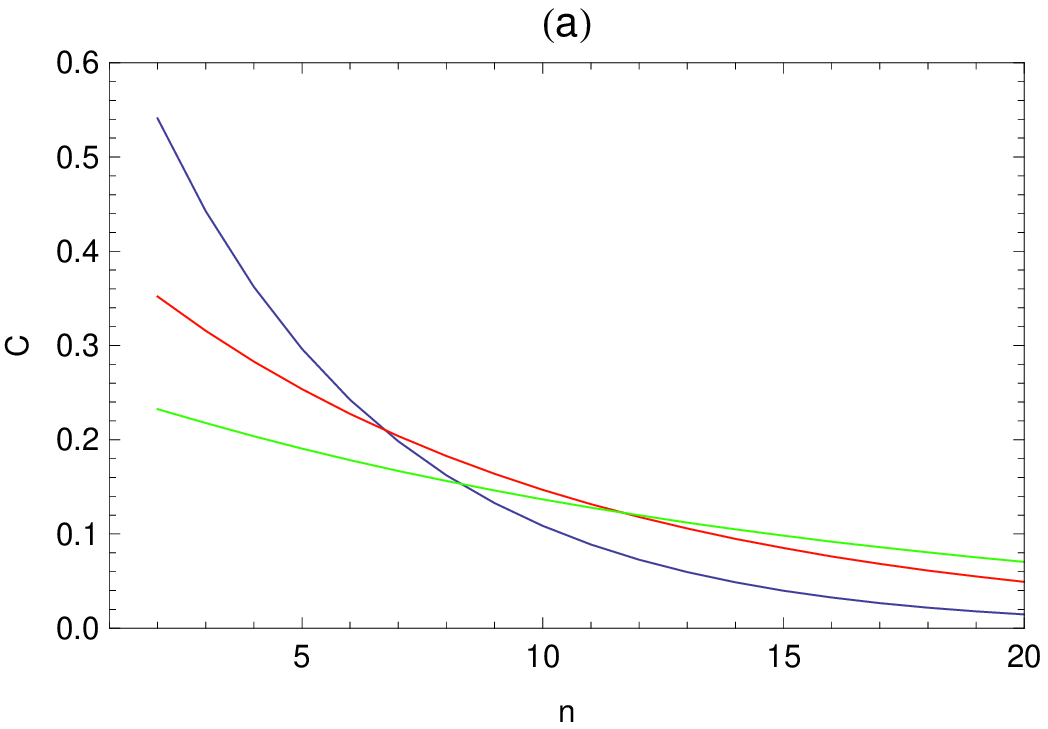}}
\hspace{0.3cm}
\subfigure{\label{psia0f2b}\includegraphics[width=6.5cm]{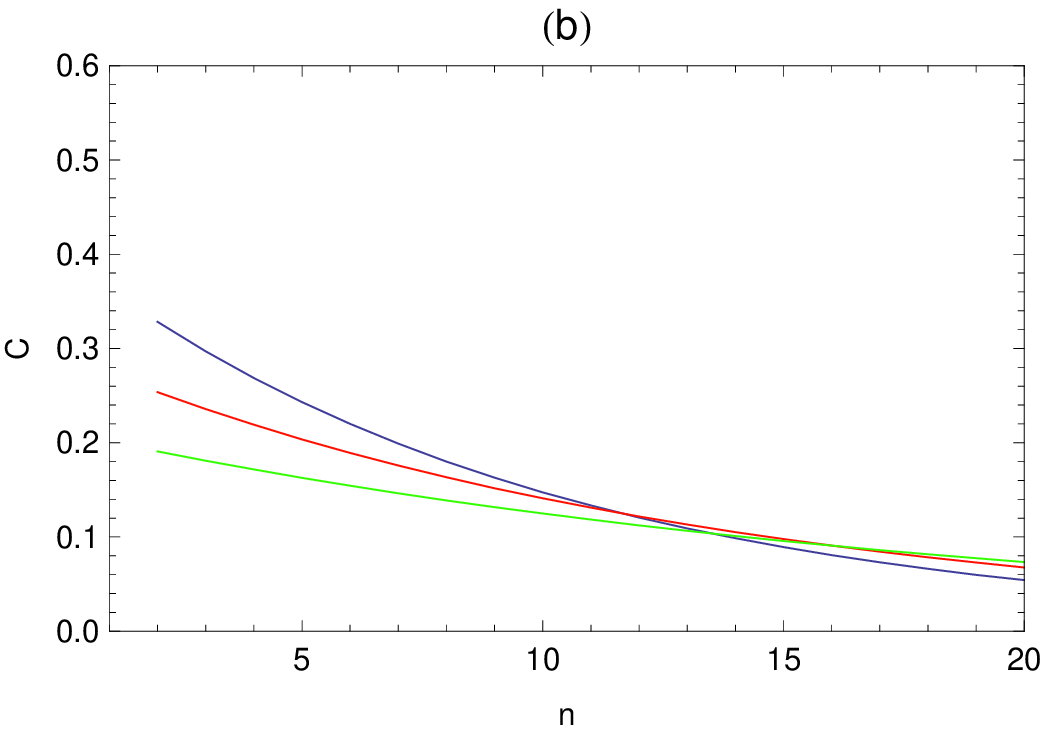}}
\end{center}
\caption{(a) The concurrence $C_{1,n}$ for $N=10$ at $\tau=T_{\phi}/2$, $\tau=0.8$ and $\tau=0.6$ (blue, red and green curves, respectively). (b) $C_{1,n}$ for $N=20$ at $\tau=T_{\phi}/2$, $\tau=1$ and $\tau=0.8$ (blue, red and green curves, respectively). All quantities are expressed in natural units, with $\lambda=1$.}
\label{psia0f2}
\end{figure}

In general, it is not possible to maximize the concurrence of every neutron pair. Equations \eqref{cp0fp} and \eqref{cp0fm} describe the results one might obtain \emph{if} it were possible to tune the interaction time $\tau$ of individual neutrons. Of course, this cannot be done; as discussed in the previous chapter, $\tau$ is determined by the neutron momentum and the size of the sample, and remains constant throughout the realization of the protocol. From figures \ref{psia0f1a} and \ref{psia0f1b}, it is clear that an optimal choice of $\tau$ depends on $m$ and $n$, that is, on the neutron pair we feel most certain of detecting. Setting $\tau=\tau_0$, for example, will maximize the concurrence of $\lbrace m,n \rbrace$ pairs $\lbrace1,2 \rbrace$, $\lbrace1,3 \rbrace$, $\lbrace1,4 \rbrace$, $\lbrace1,5 \rbrace$, $\lbrace2,3 \rbrace$, and $\lbrace2,4 \rbrace$, but is far from ideal for $\lbrace1,15 \rbrace$, $\lbrace6,10 \rbrace$, etc. This makes us heavily reliant on detecting either the first or the second neutron, together with a neutron scattered very soon after. On the other hand, going back to figure \ref{psia0f1a}, by moving towards the outer edge of the curves the concurrence may not be so high, but $C_{m,n}$ falls much less sharply with $m$ and $n$ (see figure \ref{psia0f2}). Given the form of $\zeta$, a similar result can also be obtained by increasing $N$.

We conclude from figure \ref{psia0f2} that, provided $m$ and $n$ remain small, the chances of detecting an entangled neutron pair are reasonably fair. This entanglement may be low, but by repeating the protocol a sufficient number of times one could produce enough copies of the scattered state to perform distillation. However, this would rely on detecting the same neutron pair with every run of the experiment and, as we know, such conditions are difficult to meet.

\subsection{Finite Field Evolution}
\begin{figure}[H]
\renewcommand{\captionfont}{\footnotesize}
\renewcommand{\captionlabelfont}{}
\begin{center}
\subfigure{\label{psiaf1a}\includegraphics[width=6.5cm]{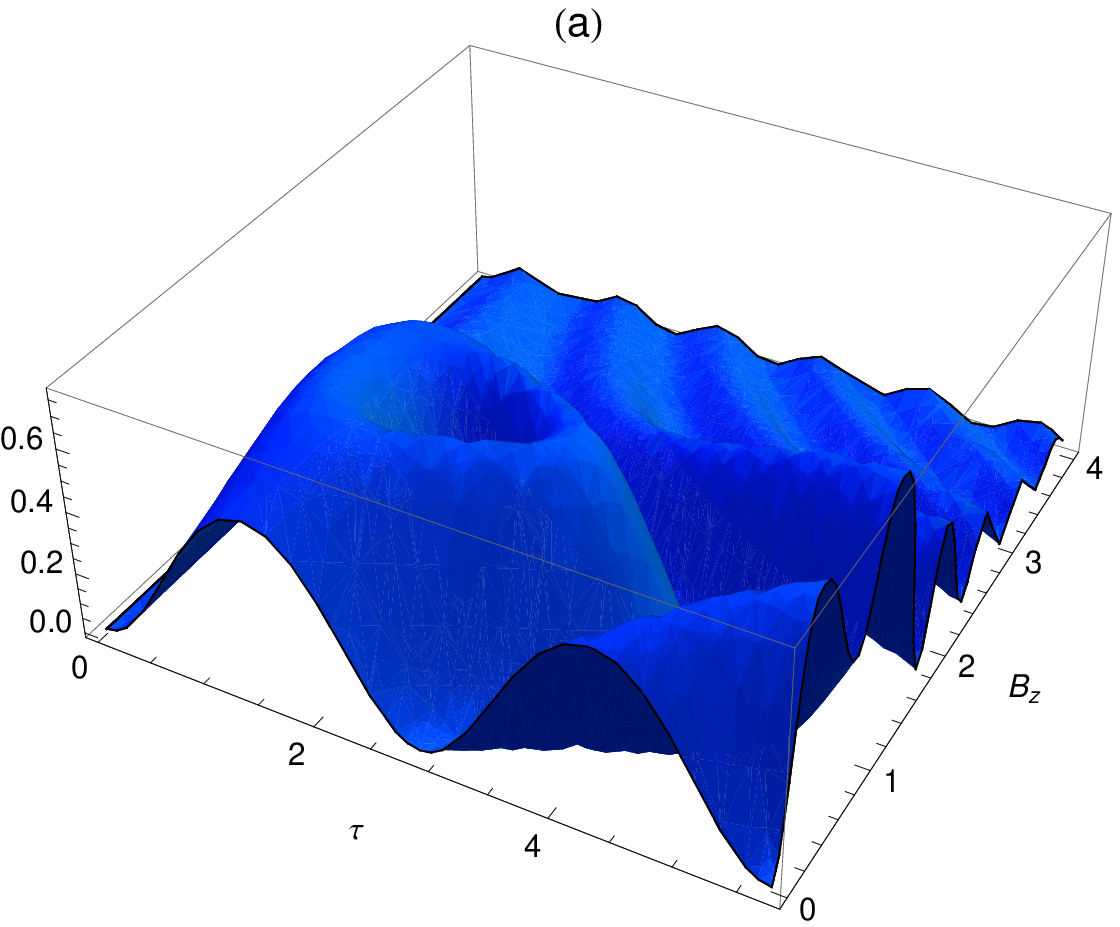}}
\hspace{0.3cm}
\subfigure{\label{psiaf1b}\includegraphics[width=6.5cm]{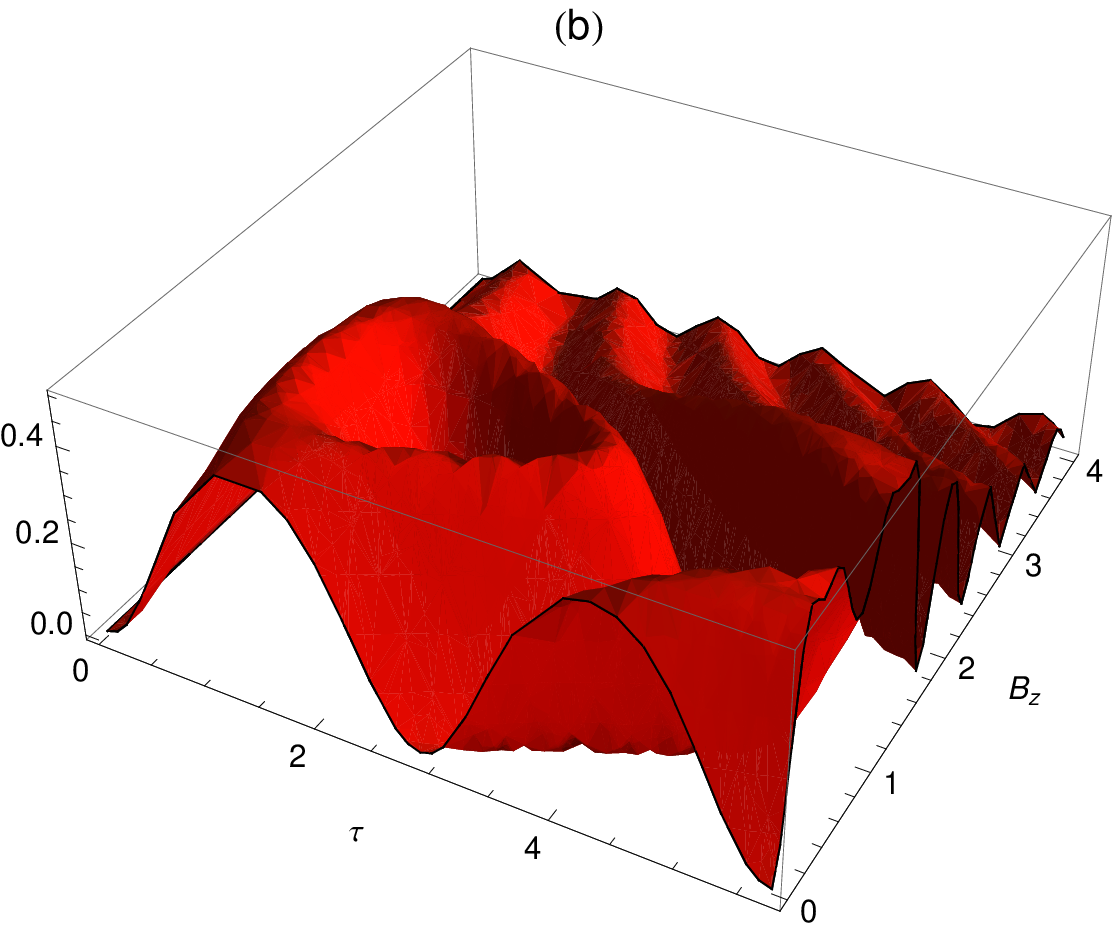}}
\hspace{0.3cm}
\subfigure{\label{psiaf1c}\includegraphics[width=6.5cm]{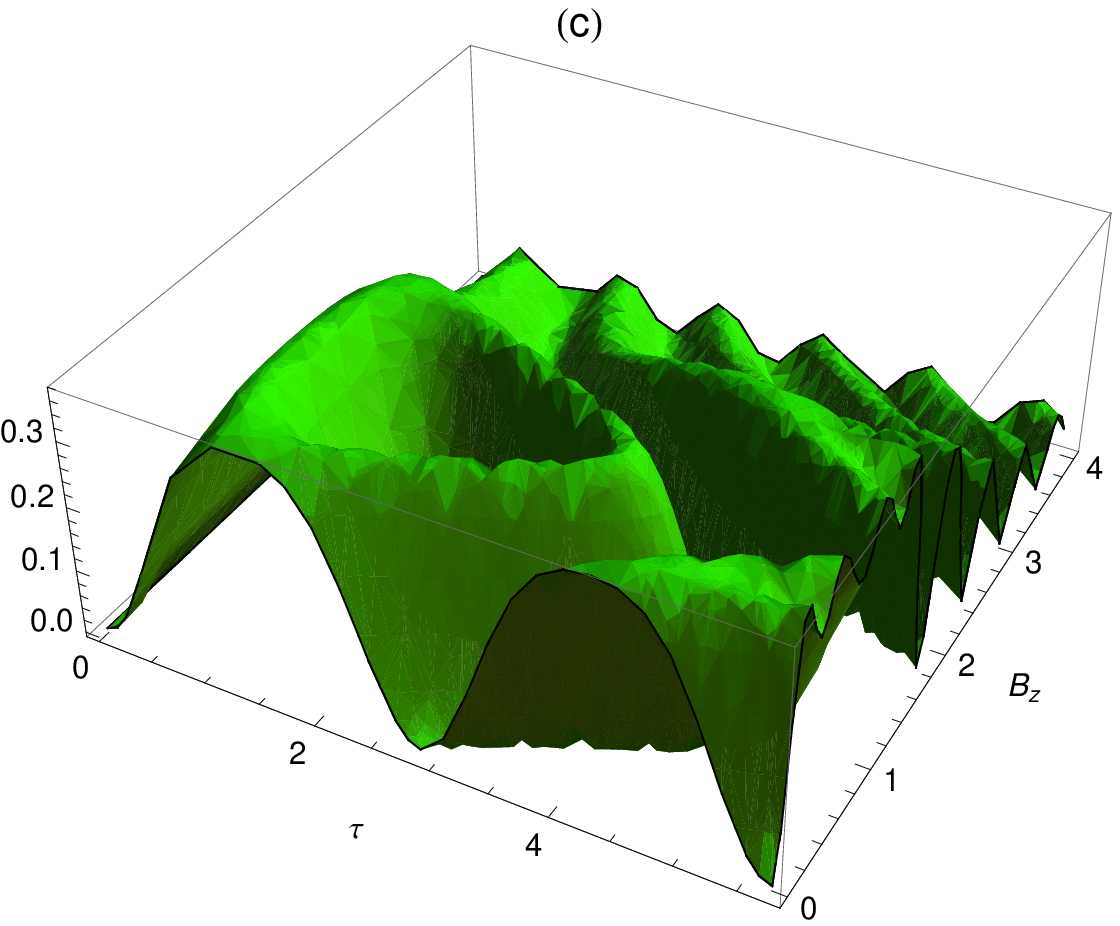}}
\hspace{0.3cm}
\subfigure{\label{psiaf1d}\includegraphics[width=6.5cm]{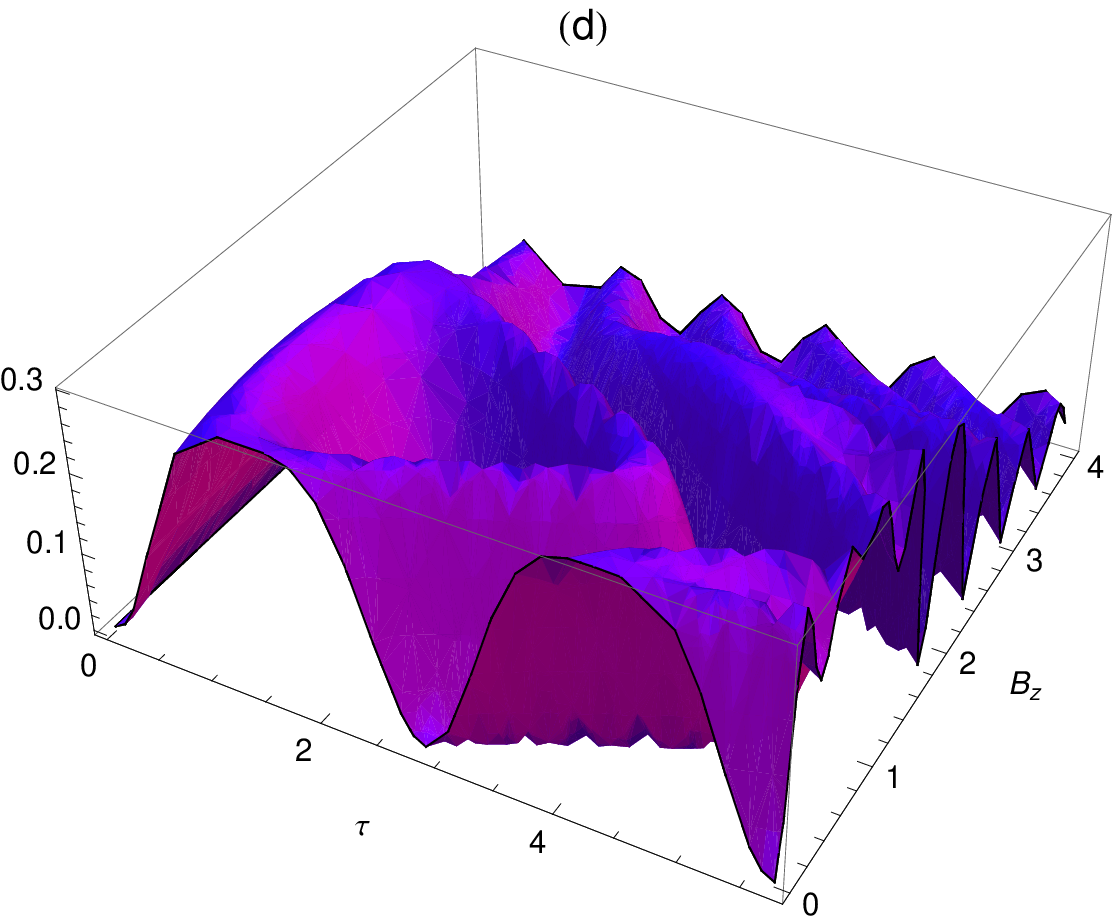}}
\hspace{0.3cm}
\subfigure{\label{psiaf1e}\includegraphics[width=6.5cm]{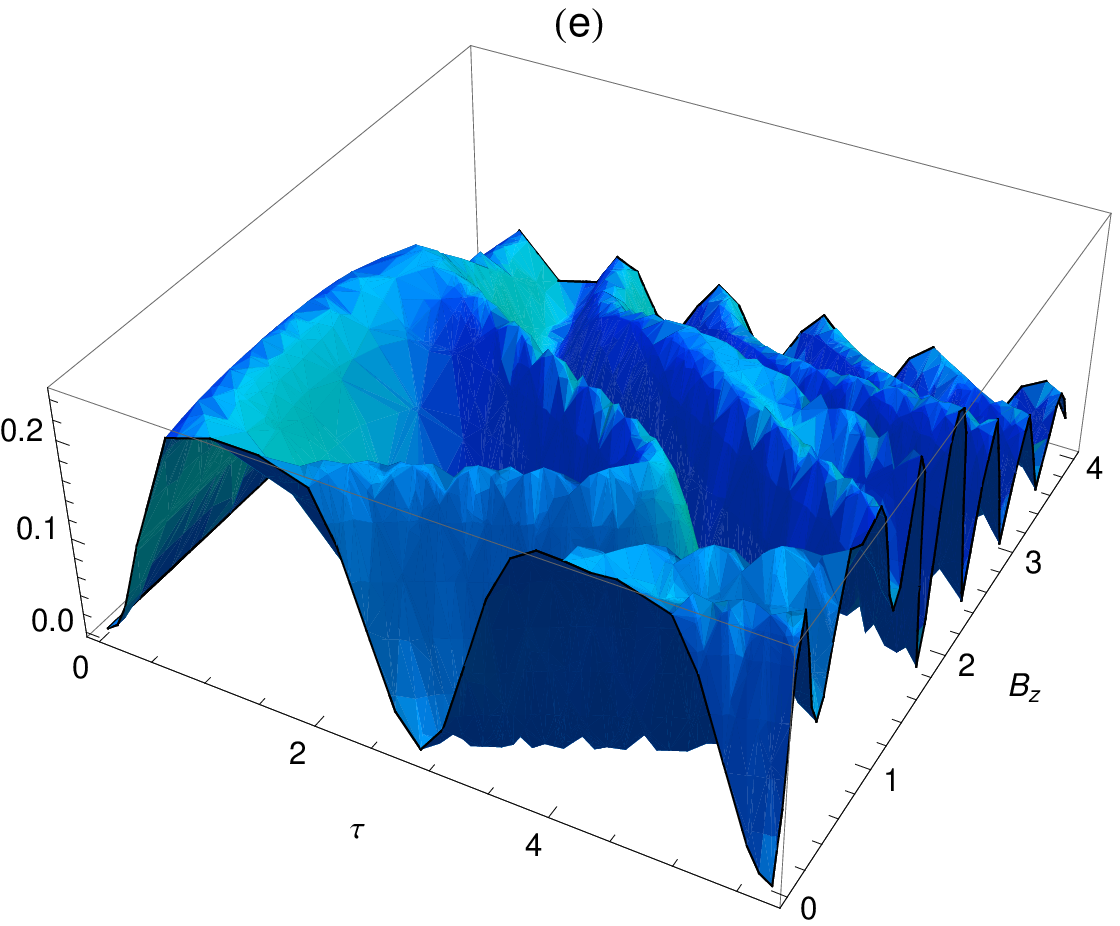}}
\hspace{0.3cm}
\subfigure{\label{psiaf1f}\includegraphics[width=6.5cm]{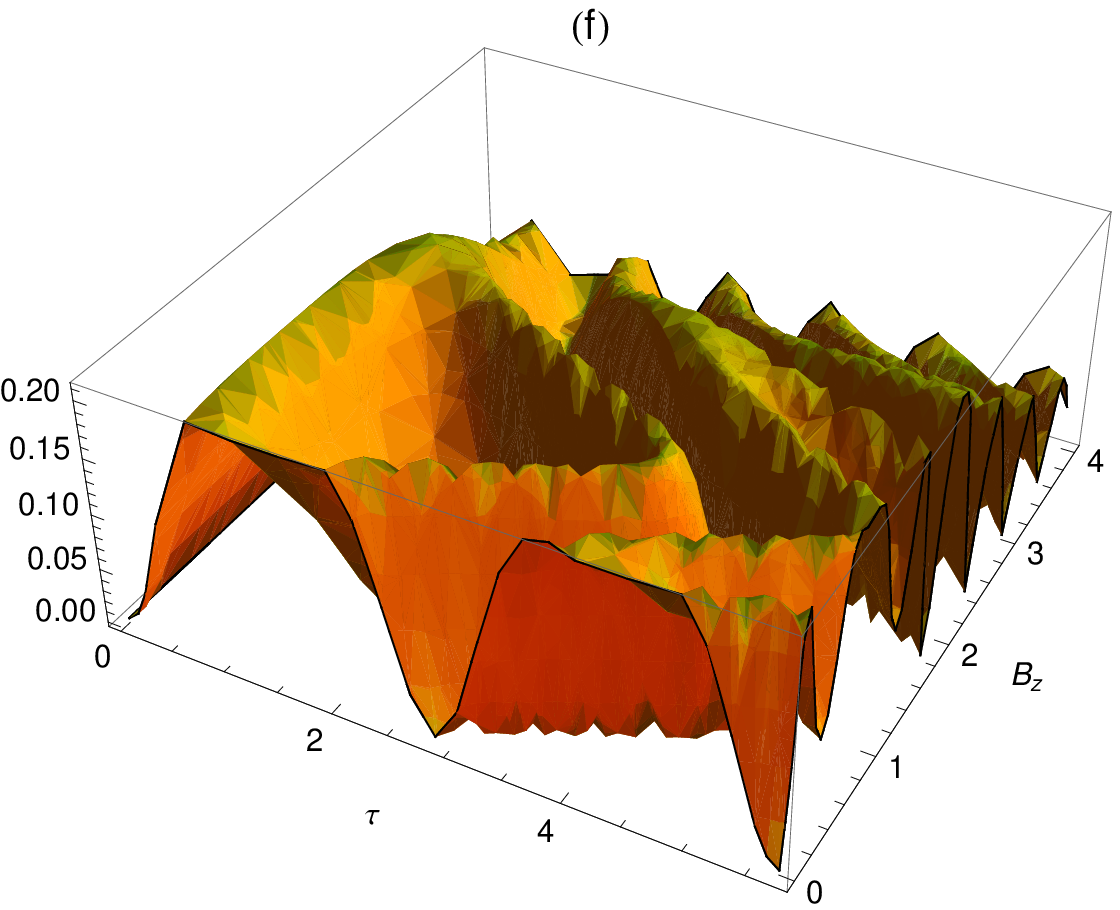}}
\end{center}
\caption{The concurrence $C_{1,n}$ in finite field for initial state $|\psi_0^A\rangle$. All quantities are expressed in natural units, with $\lambda=1$. (a) $n=2$; (b) $n=3$; (c) $n=4$; (d) $n=5$; (e) $n=6$; (f) $n=7$. }
\label{psiaf1}
\end{figure}

\begin{figure}[H]
\renewcommand{\captionfont}{\footnotesize}
\renewcommand{\captionlabelfont}{}
\begin{center}
\subfigure{\label{psiaf2a}\includegraphics[width=6.5cm]{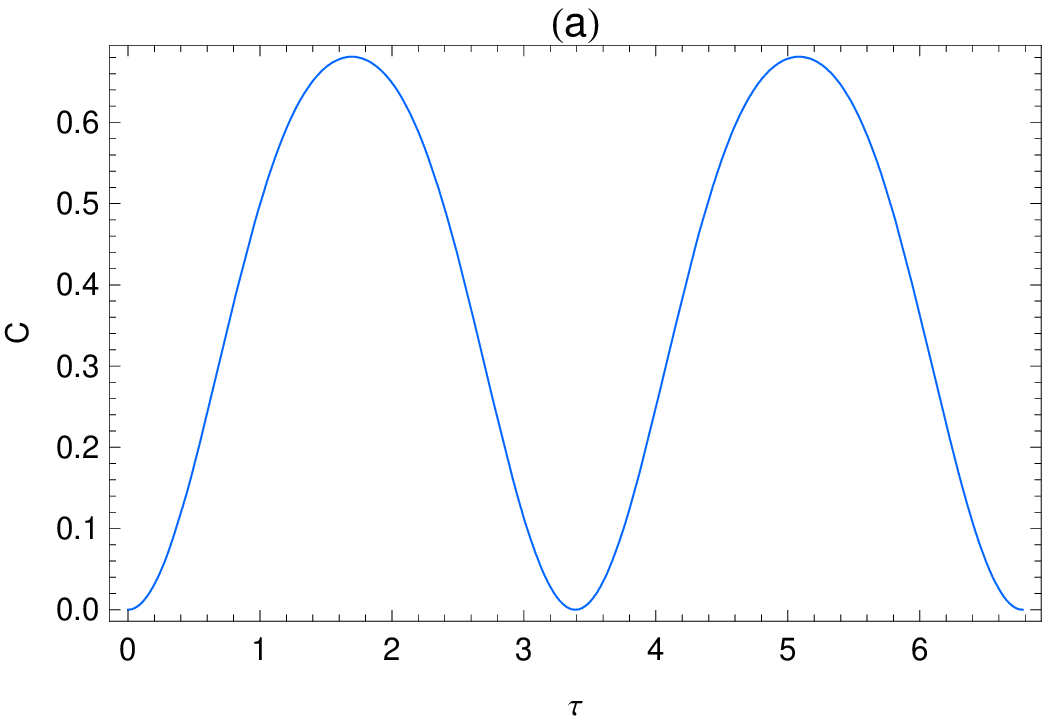}}
\hspace{0.3cm}
\subfigure{\label{psiaf2b}\includegraphics[width=6.5cm]{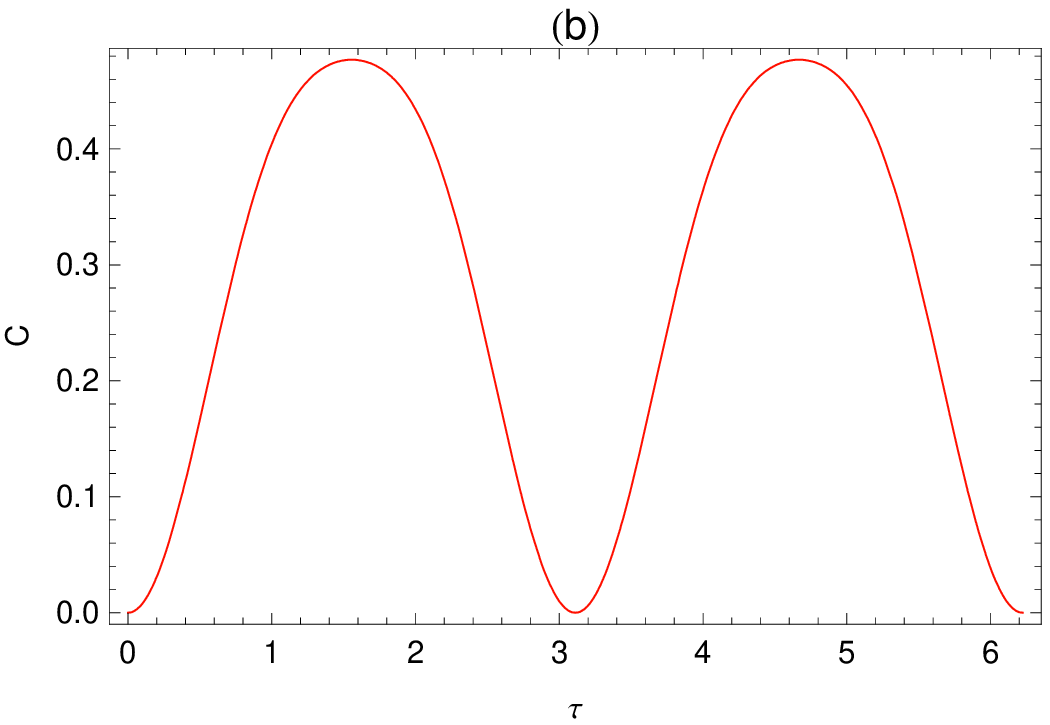}}
\hspace{0.3cm}
\subfigure{\label{psiaf2c}\includegraphics[width=6.5cm]{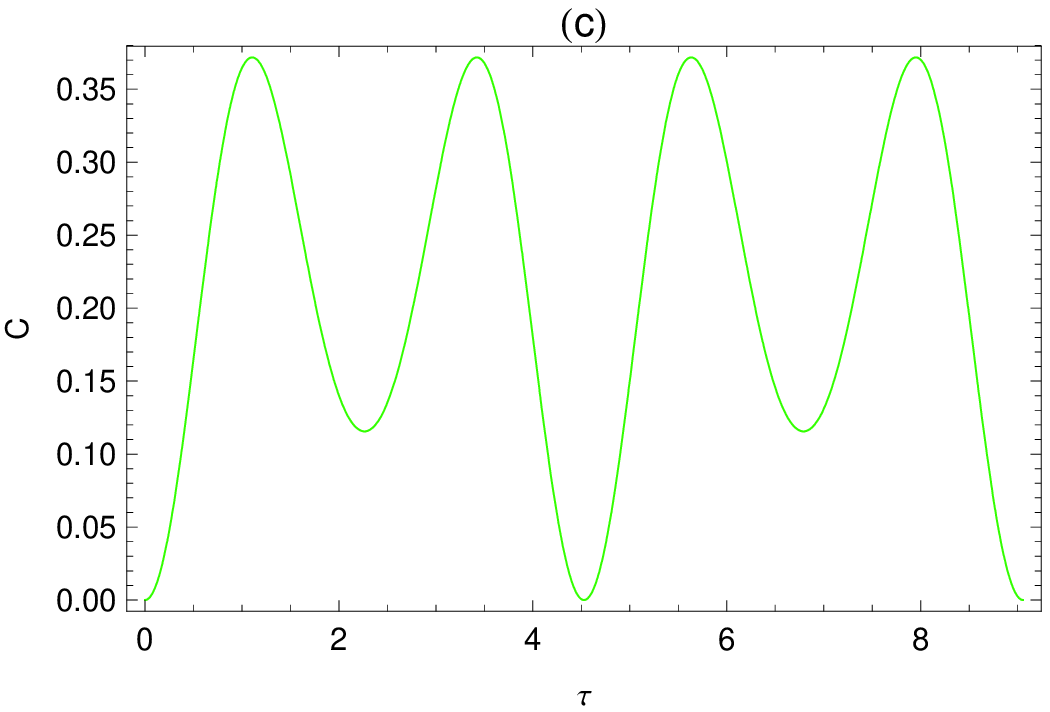}}
\hspace{0.3cm}
\subfigure{\label{psiaf2d}\includegraphics[width=6.5cm]{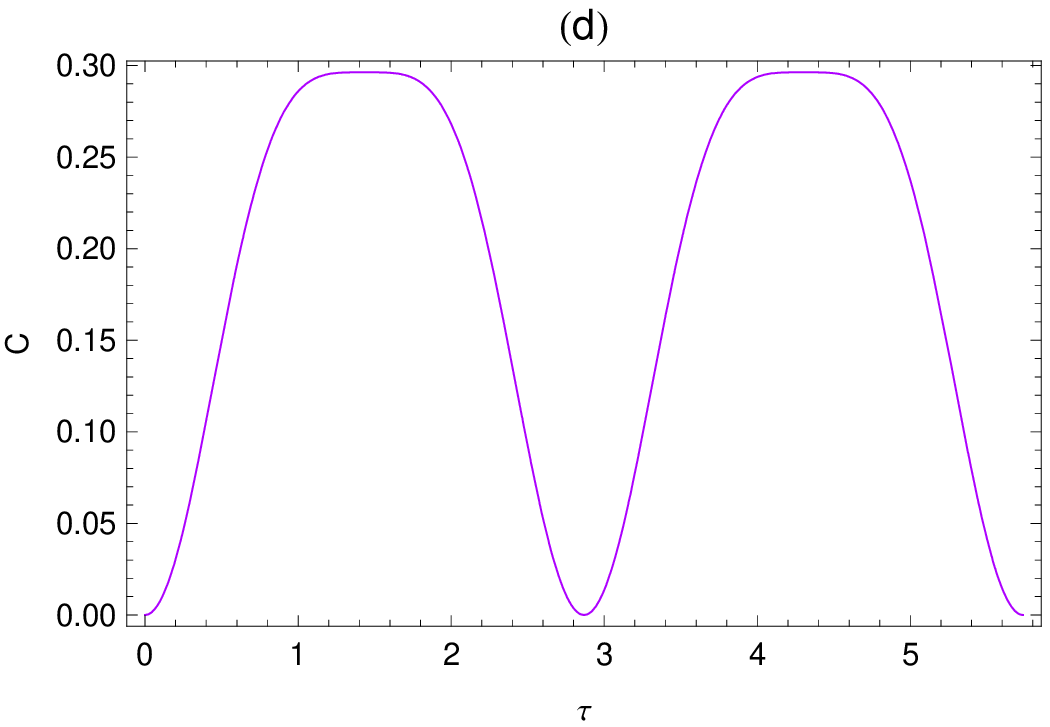}}
\hspace{0.3cm}
\subfigure{\label{psiaf2e}\includegraphics[width=6.5cm]{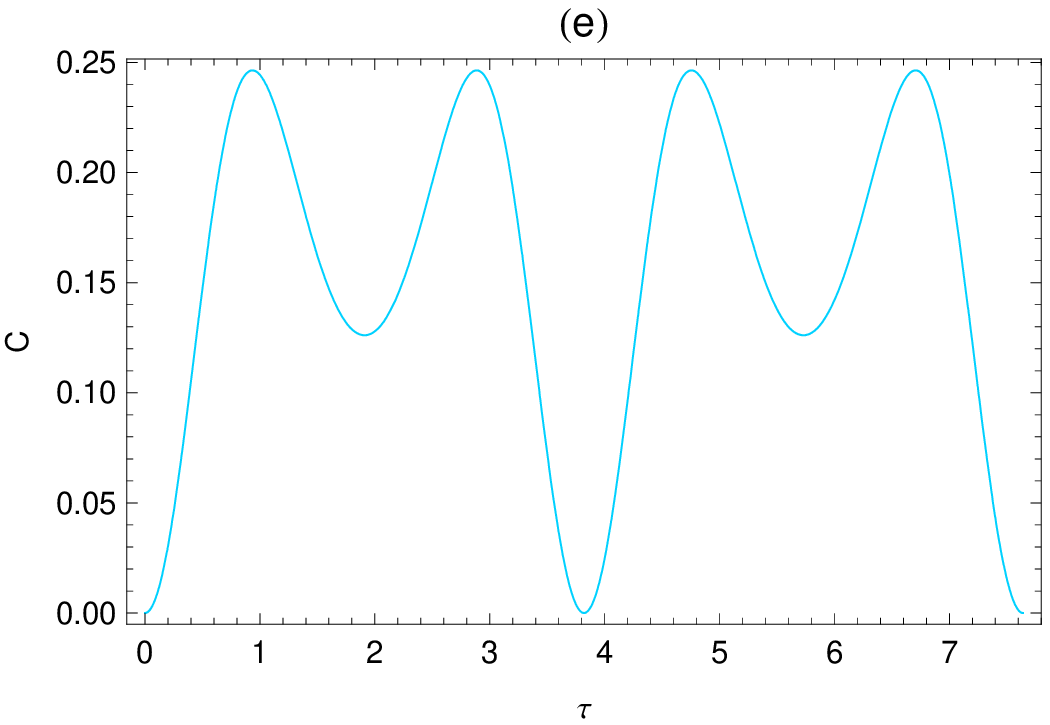}}
\hspace{0.3cm}
\subfigure{\label{psiaf2f}\includegraphics[width=6.5cm]{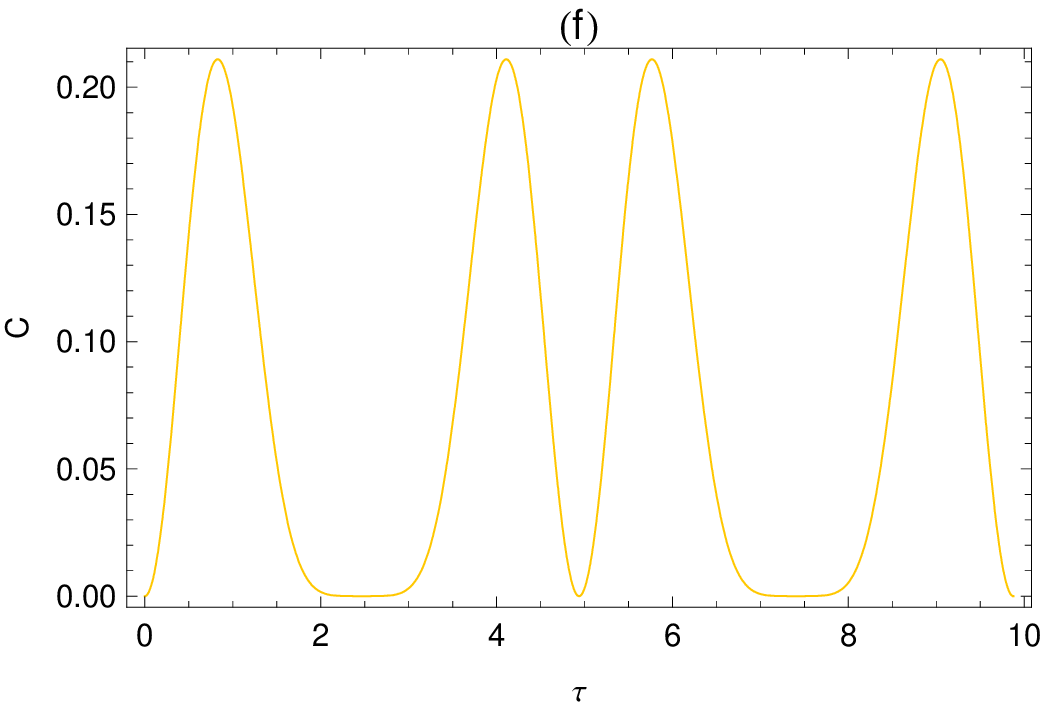}}
\end{center}
\caption{A slice through $C_{1,n}$ at randomly chosen values of $B_z$ (different for each plot) for initial state $|\psi_0^A\rangle$. All quantities are expressed in natural units, with $\lambda=1$. (a) $n=2$; (b) $n=3$; (c) $n=4$; (d) $n=5$; (e) $n=6$; (f) $n=7$. }
\label{psiaf2}
\end{figure}

In finite field also, the evolution of the concurrence of any neutron pair can be traced back to $C_{1,n}$. Let us then consider the behaviour of $C_{1,n}$ as a function of the applied field and the neutron interaction time, as shown in figure \ref{psiaf1}. A slice through these curves at any point along the $B_z$ axis illustrates the concurrence remains a periodic function of $\tau$ irrespective of the field (figure \ref{psiaf2}). $B_z$ does, however, determine the timescale of the oscillation, which is given as always by equation \eqref{T ex}. Comparing the shape of the curves for different values of $n$, one notes two main features. First, the peak concurrence falls with $n$. Second, the essential attributes of the curves do not change a great deal as $n$ is increased. In all cases, the concurrence at short times is characterized by a concave, approximately conical dip, which expands into a `valley' as $n$ becomes large. By inspection, maximum values of $C_{1,n}$ lie along the `mouth' of this valley, which is mirror-symmetric about $B_z=B_z^*$. We know this field maximizes the concurrence of the first two neutrons; therefore, from the form of $C_{1,n}$ and the trends shown in figure \ref{psiaf1}, we conclude that $B_z^*$ represents the optimal field value for all $m$ and $n$. Then, with the help of equation \eqref{c mn} and setting $B_z=B_z^*$, $C_{m,n}$ takes on the particularly simple form
\begin{equation}\label{c opt mn}
C_{m,n}=2\sin^2{\left(\frac{2\lambda\tau}{\sqrt{N}}\right)}\left|\cos^{m+n-2}{\left(\frac{2\lambda\tau}{\sqrt{N}}\right)}\right|,
\end{equation}
which peaks at time
\begin{equation}\label{tau star mn}
\tau^*=\frac{\sqrt{N}}{4\lambda}\cos^{-1}\left(\frac{m+n-4}{m+n}\right).
\end{equation}
As expected, both of these expressions reduce to \eqref{cp opt field} and \eqref{tau star} when $m=1$ and $n=2$. Finally, from \eqref{c opt mn} and \eqref{tau star mn}, the maximum concurrence of any neutron pair is given by
\begin{equation}\label{cp mn}
\mathcal{C}^p=4\sqrt{\frac{\left(\zeta-2\right)^{\zeta-2}}{\zeta^{\zeta}}},
\end{equation}
which, for ease of comparison, I have written in terms of $\zeta=m+n$. Indeed, a quick glance to section \ref{0f psia mn} shows equations \eqref{cp mn} and \eqref{cp0fp} are identical. This highlights an unexpected property of the system: applying a magnetic field can only improve the maximum concurrence of a given neutron pair with respect to its zero field value if the number of scattered neutrons is, roughly speaking, smaller than the number of spins in the sample. Above this, we run up against a fundamental limit, whereby the maximum concurrence is independent of the external parameters. Therefore, $\mathcal{C}^p$ represents an upper bound to the amount of entanglement any two neutrons can acquire. It is important to note that equations \eqref{tau star mn} and \eqref{cp mn} \emph{do not} in general describe the performance of the protocol in finite field; rather, they define how the concurrence of a given neutron pair \emph{would} scale if neutron interaction times were individually tuneable.

Let us then consider the concurrence of the pair $\lbrace m,n\rbrace$ at an arbitrary value of $B_z$. This has the form
\begin{equation}
C_{m,n}(N,\lambda,B_z,\tau)=\frac{32N\lambda^2\varphi^2\sin^2{\phi\tau}}{\left(4\lambda^2+N\varphi^2\right)^{\zeta}}
\sqrt{\left(16\lambda^4+N^2\varphi^4+8N\lambda^2\varphi^2\cos{2\phi\tau}\right)^{\zeta-2}},
\end{equation}
where $\phi$ and $\varphi$ are given by equations \eqref{phi} and \eqref{varphi}, respectively. This is the generalization of equation \eqref{conc psiA ex} for all $m$ and $n$. We know from section \ref{p0a zero mtm finite field} that within the interval $B_z\in\:[B_{z-},B_{z+}]$ the concurrence $C_{1,2}$ develops a double-peaked structure; if $m,n\neq\left[ 1,2\right]$, the limits of this interval are given by
\begin{align}
B_{z-}&=\lambda\left(1-\frac{1}{N}-\sqrt{\frac{2\left(\zeta-2\right)}{N}}\:\right)\label{bz-mn},\\
B_{z+}&=\lambda\left(1-\frac{1}{N}+\sqrt{\frac{2\left(\zeta-2\right)}{N}}\:\right)\label{bz+mn}.
\end{align}
Within this interval, $C_{m,n}$ peaks at time
\begin{equation}
\tau_B^{m,n}=\frac{1}{2\phi}\cos^{-1}\left[-\frac{16\lambda^4+N^2\varphi^4-4N\lambda^2\varphi^2\left(\zeta-2\right)}{4N\lambda^2\varphi^2\zeta}\right].
\end{equation}
Otherwise, maxima occur at $T_{\phi}/2$. Therefore, one finds
\begin{align}
C_o&=\frac{8N\lambda^2}{\left(\varphi-\phi\right)^2}\left[\frac{4\lambda^2-N\varphi^2}{4\lambda^2+N\varphi^2}\right]^{\zeta},\\
C_i&=4\sqrt{\frac{\left(\zeta-2\right)^{\zeta-2}}{\zeta^{\zeta}}}.
\end{align}
For a chosen neutron pair, these relations define the maximum achievable concurrence for any value of $N$ and $B_z$. As in the previous section, these limits can be phrased in terms of the number of scattered neutrons, rather than the applied field. Defining
\begin{equation}\label{zetab}
\zeta_B=1+\frac{1}{2N}+\frac{1}{2\lambda}\left[2B_z+\frac{N\left(B_z-\lambda\right)^2}{\lambda}\right],
\end{equation}
one has
\begin{equation}
\mathcal{C}^p_{m,n}=C_o
\end{equation}
if $\zeta<\zeta_B$, and
\begin{equation}
\mathcal{C}^p_{m,n}=C_i
\end{equation}
otherwise.

\begin{figure}[H]
\renewcommand{\captionfont}{\footnotesize}
\renewcommand{\captionlabelfont}{}
\begin{center}
\subfigure{\label{psiaf3a}\includegraphics[width=6.5cm]{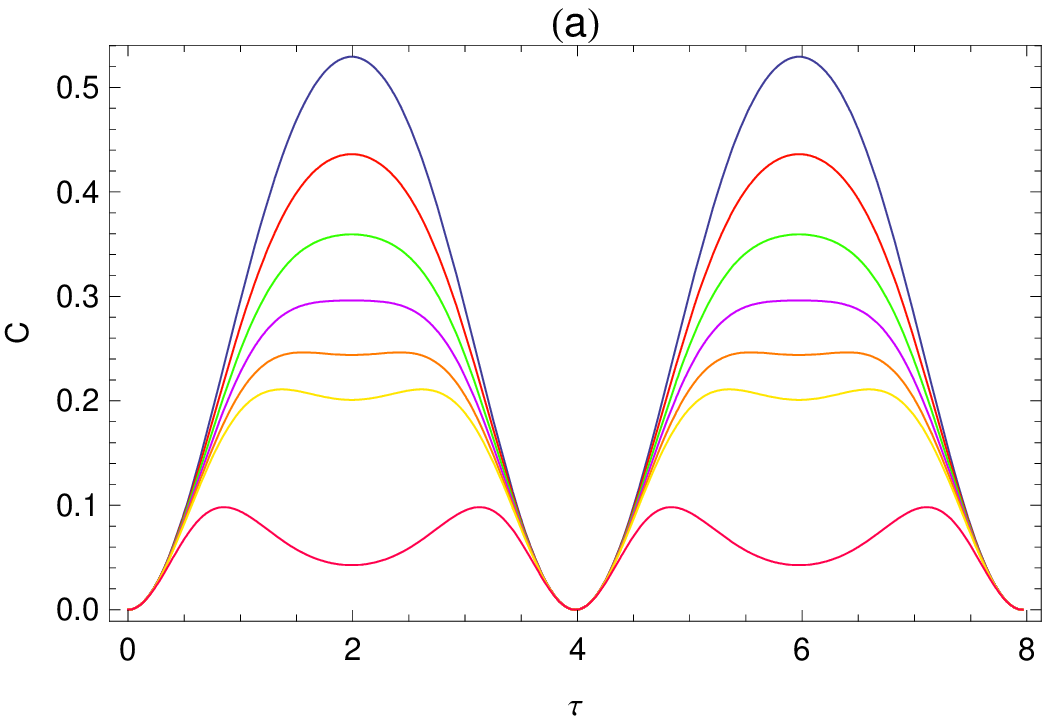}}
\hspace{0.3cm}
\subfigure{\label{psiaf3b}\includegraphics[width=6.5cm]{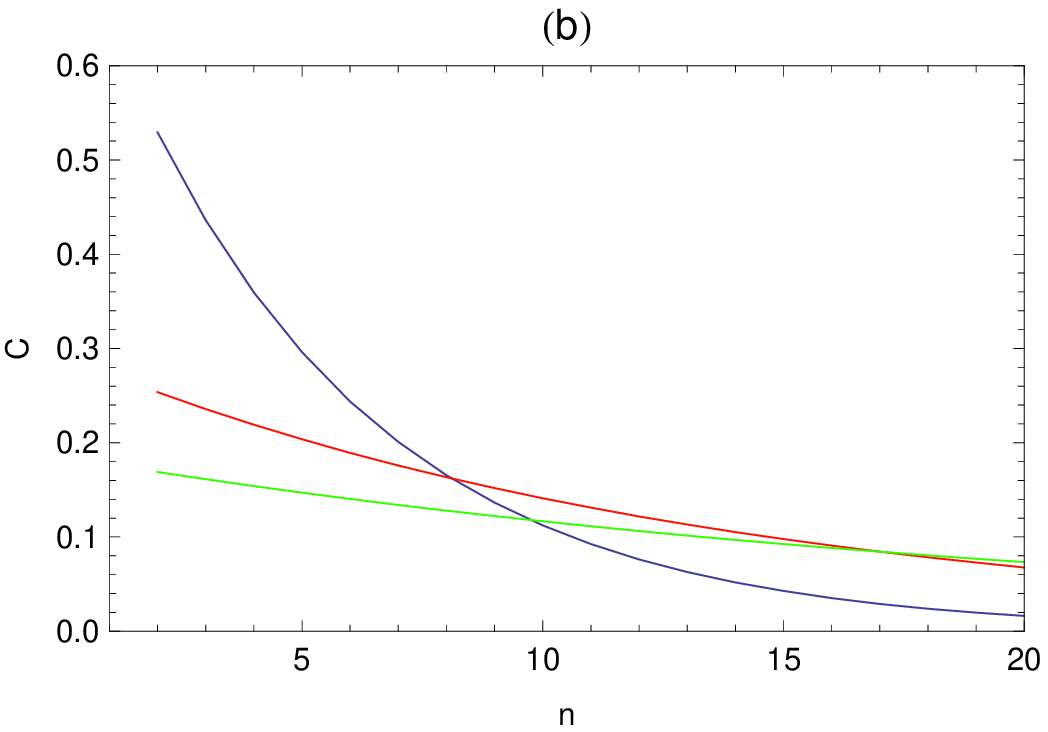}}
\hspace{0.3cm}
\subfigure{\label{psiaf3c}\includegraphics[width=6.5cm]{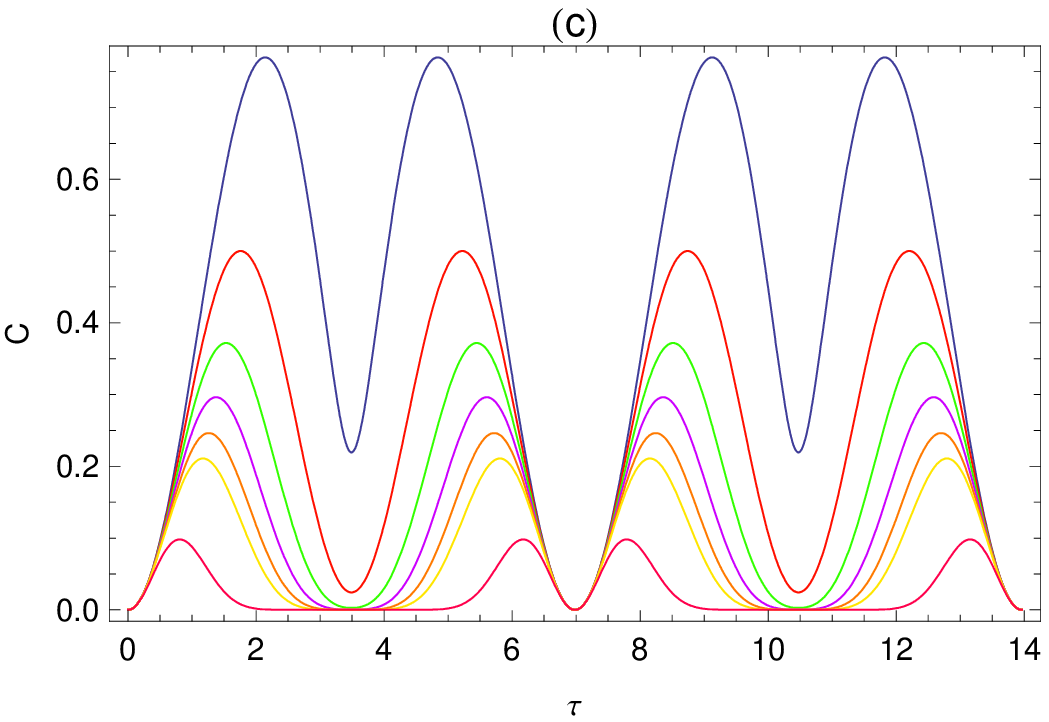}}
\hspace{0.3cm}
\subfigure{\label{psiaf3d}\includegraphics[width=6.5cm]{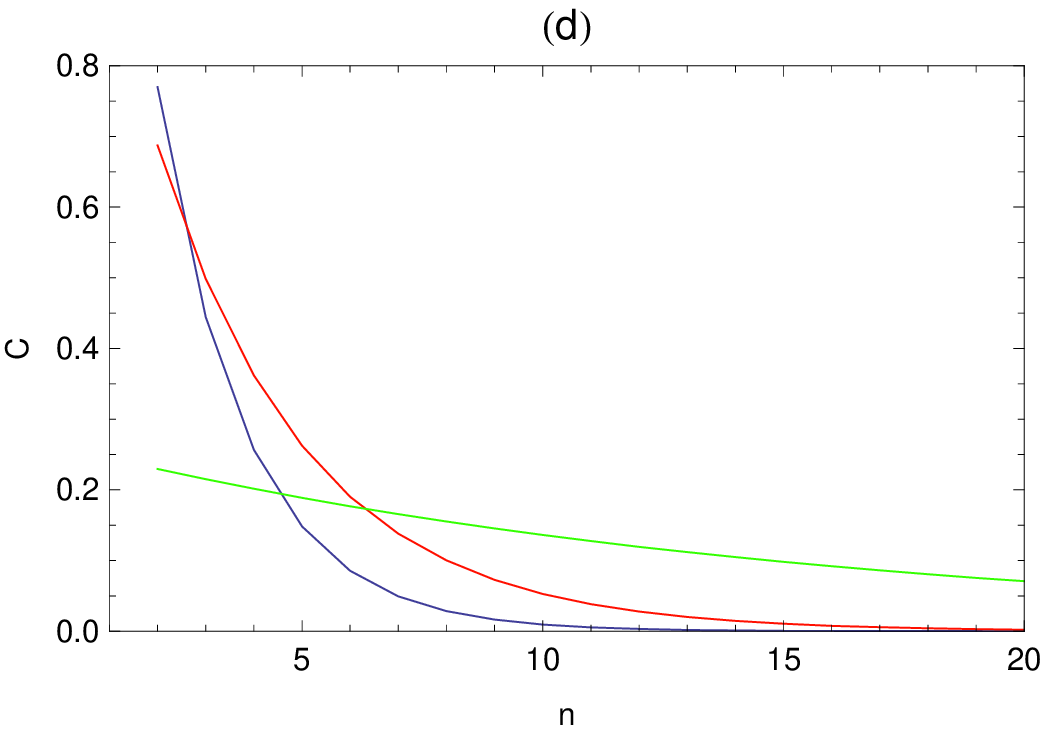}}
\hspace{0.3cm}
\subfigure{\label{psiaf3e}\includegraphics[width=6.5cm]{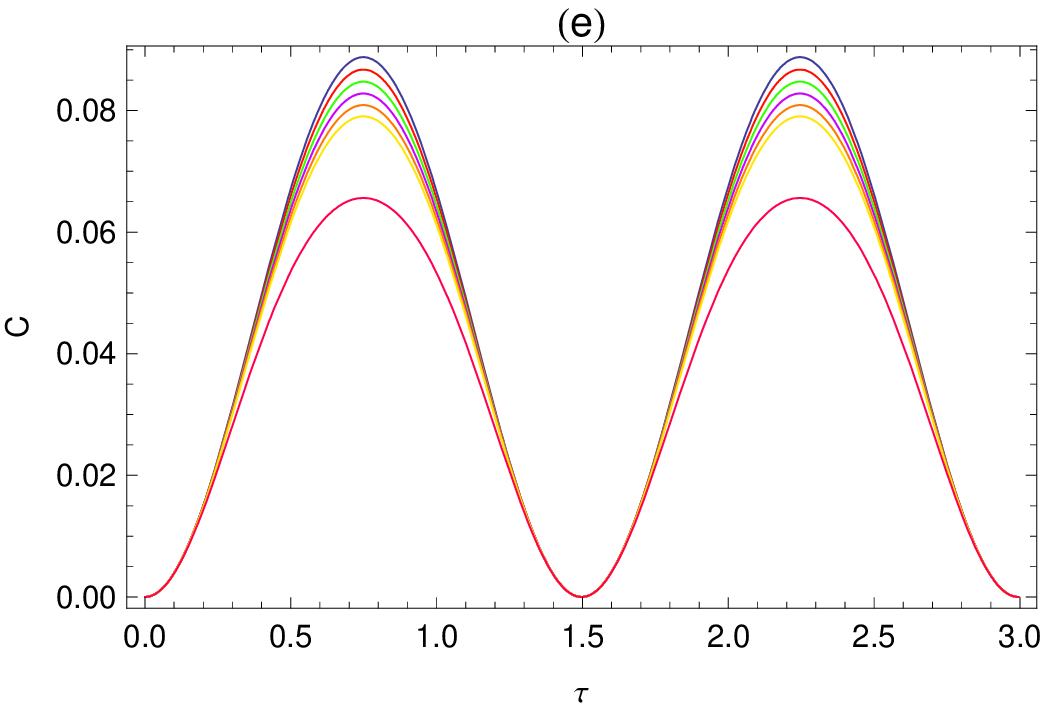}}
\hspace{0.3cm}
\subfigure{\label{psiaf3f}\includegraphics[width=6.5cm]{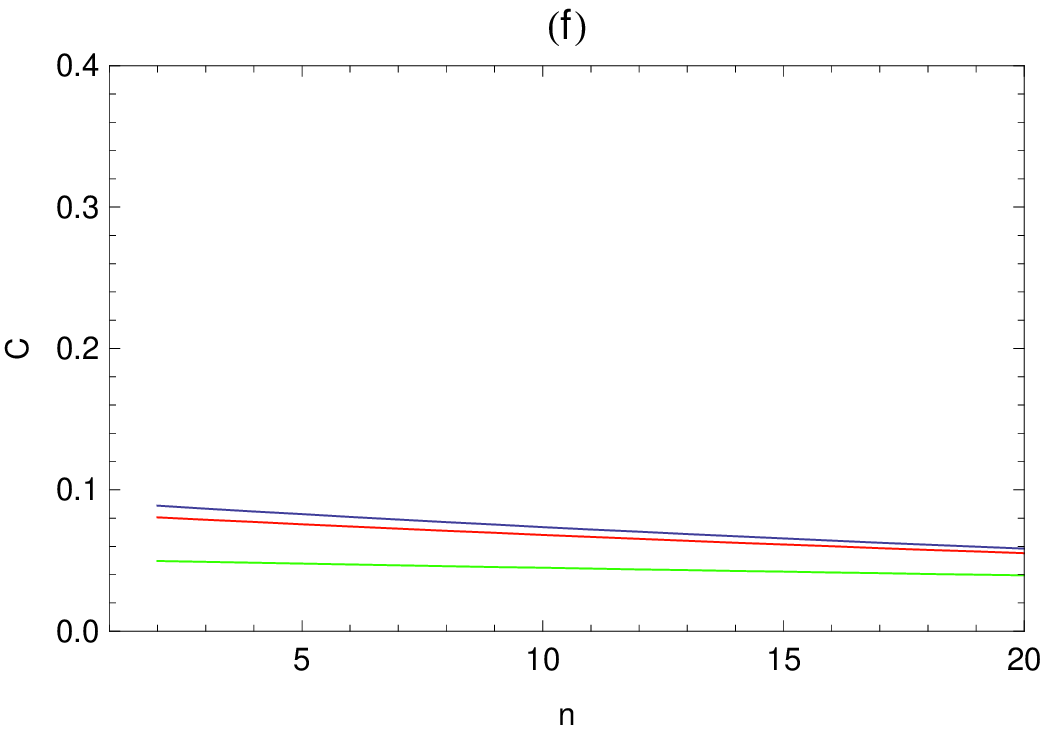}}
\end{center}
\caption{The concurrence $C_{1,n}$ in different field regimes for $N$=20 and $n=\lbrace1,2,3,4,5,6,7,15\rbrace$. Increasing values of $n$ are indicated by the blue, red, green, purple, orange, yellow and magenta curves, respectively. All quantities are expressed in natural units, with $\lambda=1$. (a) $C_{1,n}$ at $B_z=0.3$. (b) $C_{1,n}$ at $B_z=0.3$ and $\tau=T_{\phi}/2$ (blue curve), $\tau=0.9$ (red curve), and $\tau=0.7$ (green curve). (c) $C_{1,n}$ at $B_z=0.9$. (d) $C_{1,n}$ at $B_z=0.9$ and $\tau=\tau_B^{1,2}$ (blue curve), $\tau=1.7$ (red curve), and $\tau=0.8$ (green curve). (e) $C_{1,n}$ at $B_z=3$. (f) $C_{1,n}$ at $B_z=3$ and $\tau=T_{\phi}/2$ (blue curve), $\tau=0.6$ (red curve), and $\tau=0.4$ (green curve). }
\label{psiaf3}
\end{figure}

Finally, to assess the performance of the protocol in a finite field, let us consider how the concurrence scales with $m$ and $n$. If the field is weak, $\zeta_B$ is proportional to $N$. Therefore, the optimal interaction time at fixed $N$ will depend on the number of scattered neutrons relative to the number of spins in the sample. If this ratio is small, the optimal time is the same (i.e. $T_{\phi}/2$) for all $m$ and $n$. Otherwise, as discussed in section \ref{0f psia mn}, one must choose a time which yields appreciable values of $C_{m,n}$ for the broadest possible interval $|m-n|$. As the field is raised above $B_{z-}$, this optimization strategy becomes compulsory to avoid the concurrence plummeting after the first few scattering events. When $B_z$ exceeds $B_{z+}$, $\zeta_B$ becomes large and optimal measurement times come once more to coincide. However, the concurrence of any neutron pair is extremely low no matter how values of $\tau$ are chosen. These results are shown in figure \ref{psiaf3}.

In conclusion, if the sample is prepared in a single-magnon state the performance of the protocol can be optimized by applying a magnetic field within the range $[B_{z-},B_{z+}]$. In general the concurrence of an arbitrary neutron pair decreases with the total number of scattering events. However, by choosing the neutron velocity with care, one can ensure this decline is moderately slow. Therefore, the likelihood of detecting an entangled neutron pair is reasonably strong, though it remains desirable to detect neutrons scattered in quick succession. Peak values of the concurrence do not typically exceed 0.5 unless one of the neutrons detected is the first. Consequently, the yield of highly entangled pairs could be improved by periodically re-setting the sample, i.e re-initializing the system to the original input state.

\section{Scattering from a fully Polarized Sample}\label{mn psi0b}
Scattering from a fully polarized sample, with the system initialized to state $|\psi_0^B\rangle$ [see equation \eqref{psi0b}], cannot be described analytically, as the system wavefunction becomes increasingly convoluted with each step of the protocol. This occurs because the system can now accommodate as many spin flips as scattered neutrons, therefore the size of the working Hilbert space grows linearly in the total number of incoming neutrons. Consequently, the concurrence must be characterized using numerical methods. Owing to the increased computational demands as the number of scattered neutrons becomes large, I will limit my study to short interaction times bounded by $\tau=20$ (expressed in natural units, and for $\lambda=1$). As this interval suffices to observe the salient features of the concurrence of the first two neutrons, and the interaction of each neutron with the sample is the same, it is not unreasonable to assume that any interesting features of the function $C_{m,n}$ might emerge within this time frame.

At fixed $N$, we expect the concurrence of a given neutron pair to depend on three parameters
\begin{enumerate}
\item The interaction time;
\item The applied field;
\item The neutron polarization.
\end{enumerate}
Let us then observe how changing one or more of these parameters can affect the performance of the protocol.

\subsection{Zero-Field Evolution}
It is already known from section \ref{p0b zero mtm zero field} that when the applied field is zero the concurrence of the first two scattered neutrons is a decaying function of $N$. To study the concurrence of an arbitrary neutron pair, it is therefore convenient to work at fixed $N$, chosen such that the peak value of $C_{1,2}$ is non-zero within the allowed range of neutron polarizations $\left[\alpha_-,\alpha_+\right]$ (see figure \ref{aldep1}), and $N$ is large enough to accommodate several spin flips. I work with $N=10$, for which the peak concurrence of the first neutron pair falls just below 0.1 (see figure \ref{cexpsiB}).

\begin{figure}[H]
\renewcommand{\captionfont}{\footnotesize}
\renewcommand{\captionlabelfont}{}
\begin{center}
\subfigure{\label{mn0field0a}\includegraphics[width=6.5cm]{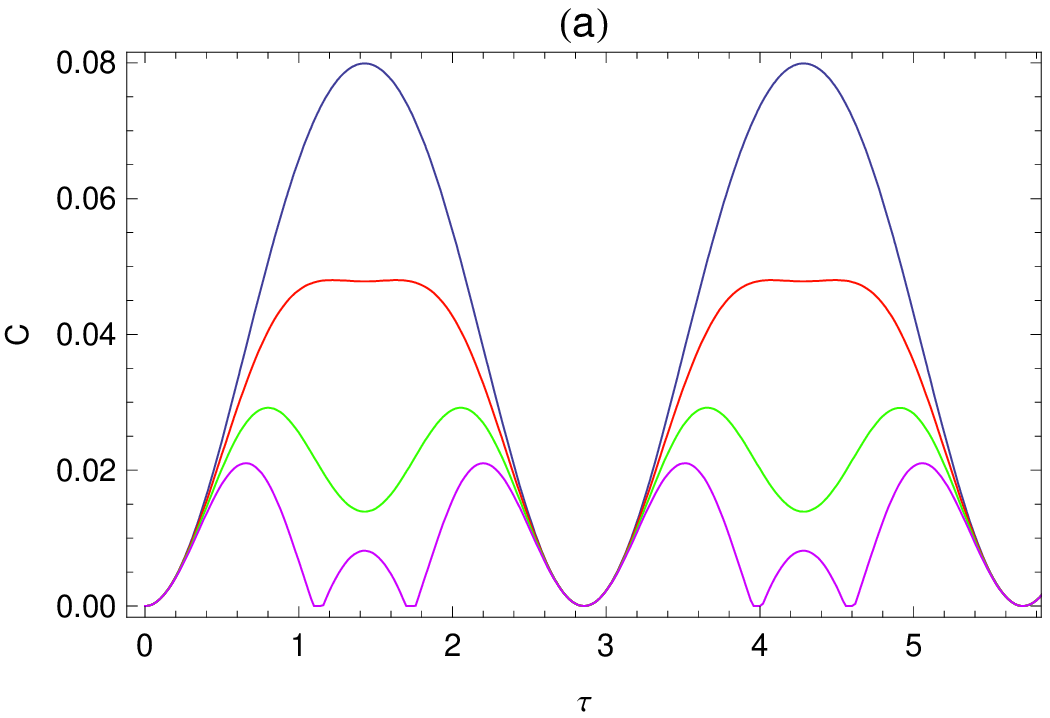}}
\hspace{0.3cm}
\subfigure{\label{mn0field0b}\includegraphics[width=6.5cm]{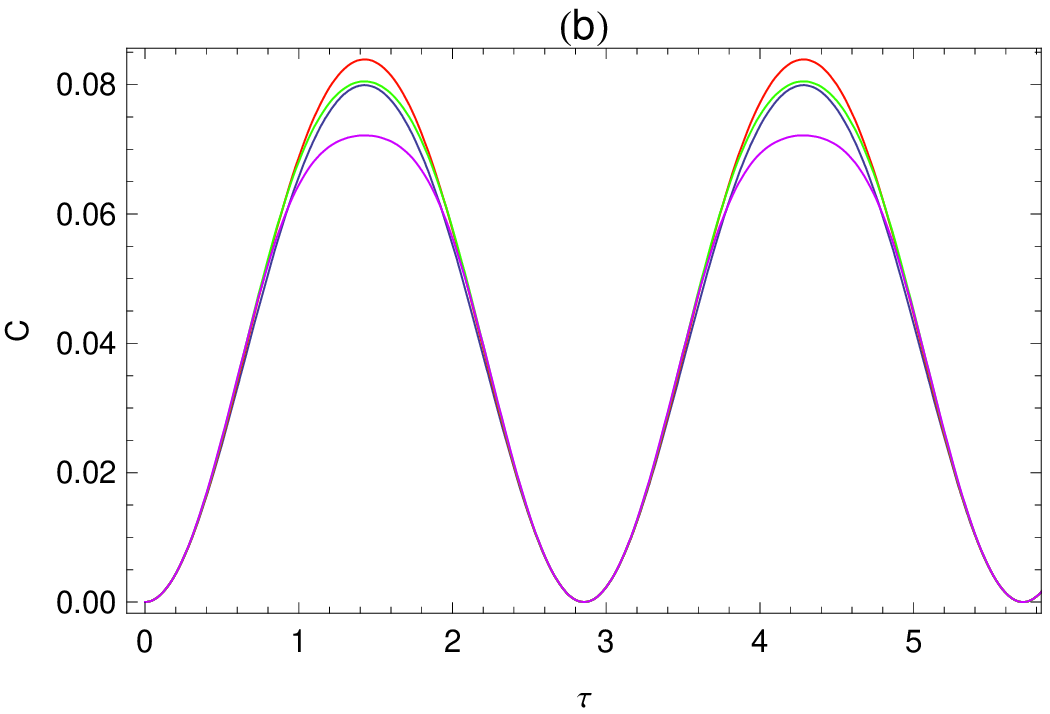}}
\hspace{0.3cm}
\subfigure{\label{mn0field0c}\includegraphics[width=6.5cm]{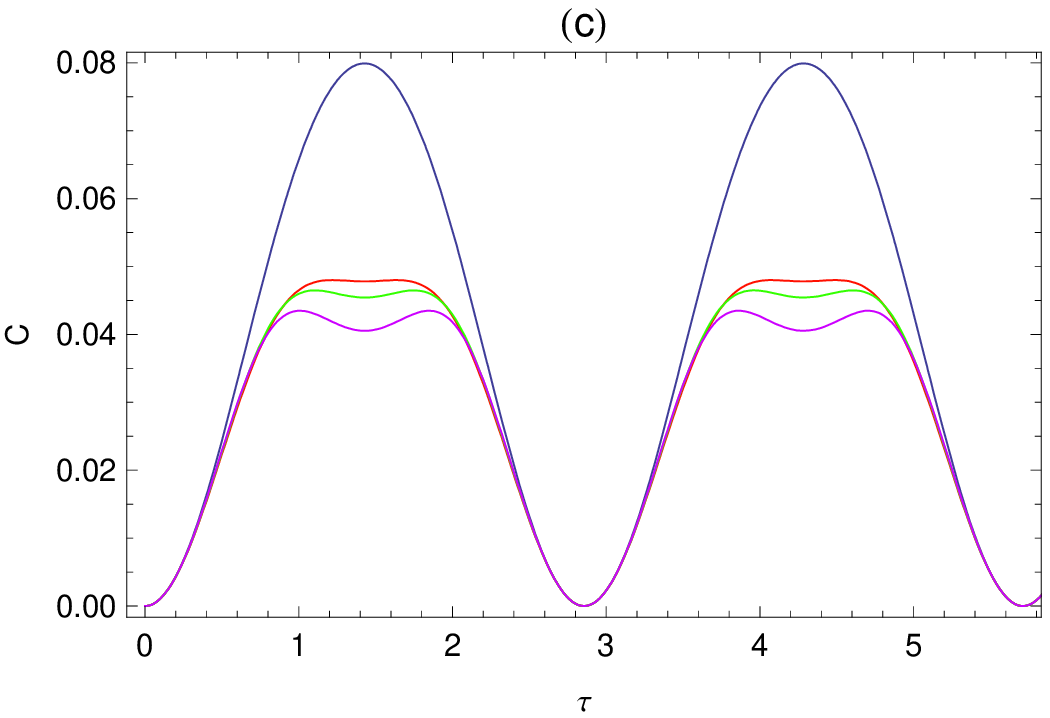}}
\hspace{0.3cm}
\subfigure{\label{mn0field0d}\includegraphics[width=6.5cm]{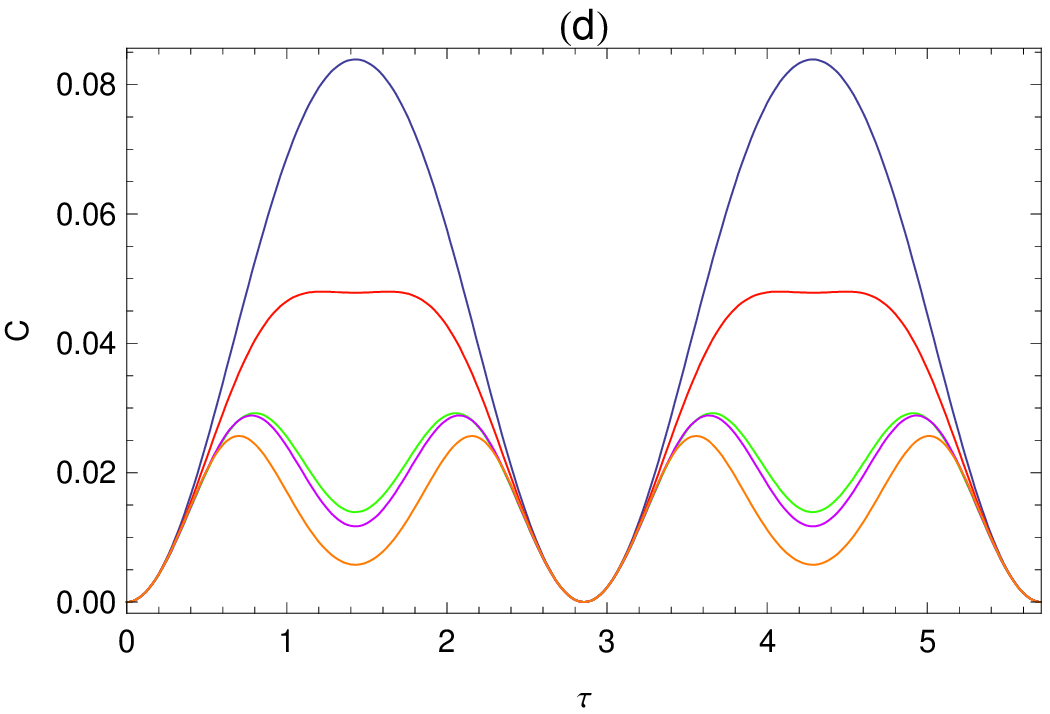}}
\hspace{0.3cm}
\subfigure{\label{mn0field0e}\includegraphics[width=6.5cm]{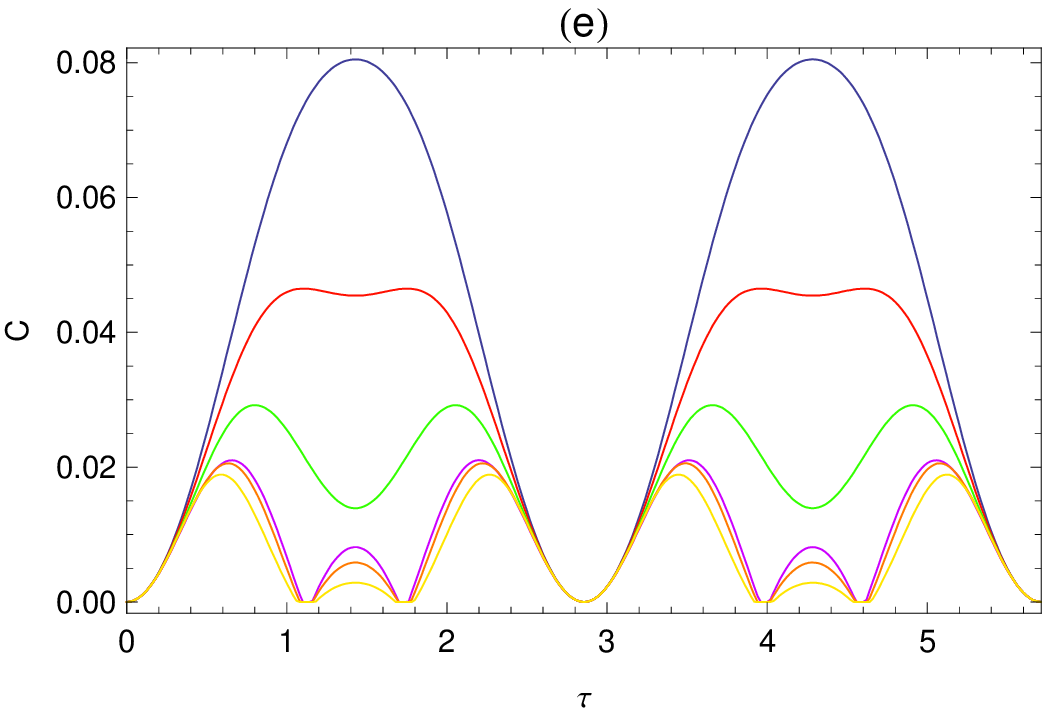}}
\end{center}
\caption{The concurrence of different neutron pairs for incoming neuron polarization $\alpha=\frac{\sqrt{3}}{2}$, $\beta=\frac{1}{2}$ and $B_z=0$. Increasing values of $n$ are indicated by the blue, red, green, purple, orange and yellow curves, respectively. (a) The concurrence of neutrons $m$ and $m+1$ for $m\leq4$. (b) The concurrence $C_{1n}$ for $1<n\leq5$. (c) The concurrence $C_{2n}$ for $n=\lbrace1,3,4,5\rbrace$. (d) The concurrence $C_{3n}$ for $n=\lbrace1,2,4,5,8\rbrace$. (e) The concurrence $C_{4n}$ for $n=\lbrace1,2,3,5,7,10\rbrace$. All quantities are expressed in natural units, with $\lambda=1$.}
\label{mn0field0}
\end{figure}

\begin{figure}[H]
\renewcommand{\captionfont}{\footnotesize}
\renewcommand{\captionlabelfont}{}
\begin{center}
\subfigure{\label{mn0field02a}\includegraphics[width=6.5cm]{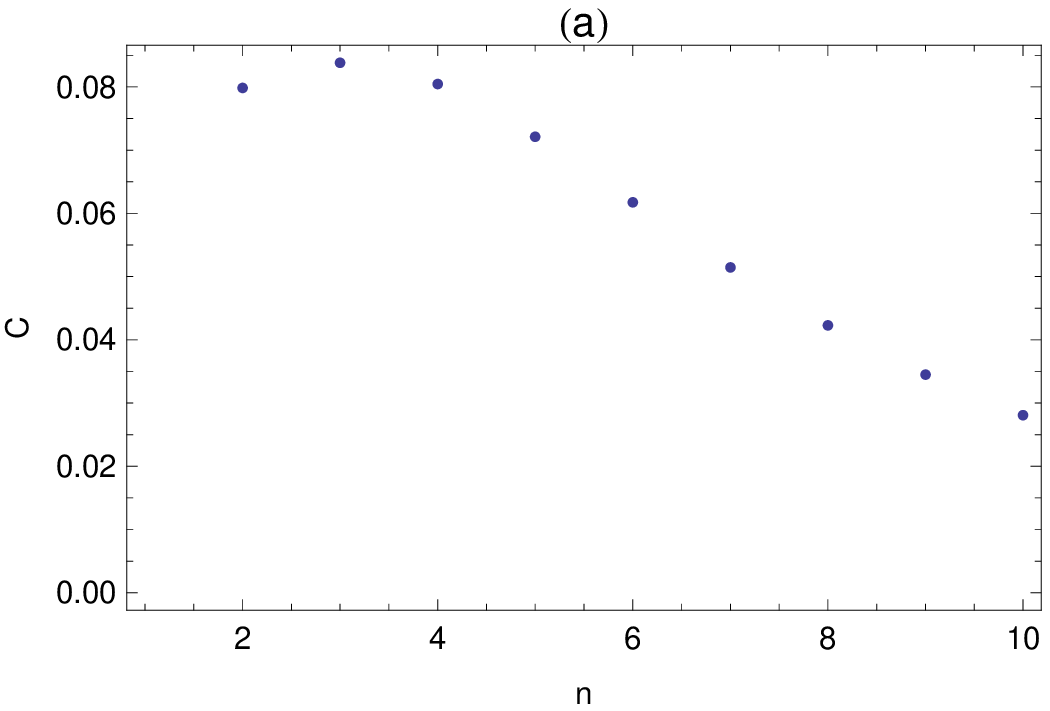}}
\hspace{0.3cm}
\subfigure{\label{mn0field02b}\includegraphics[width=6.5cm]{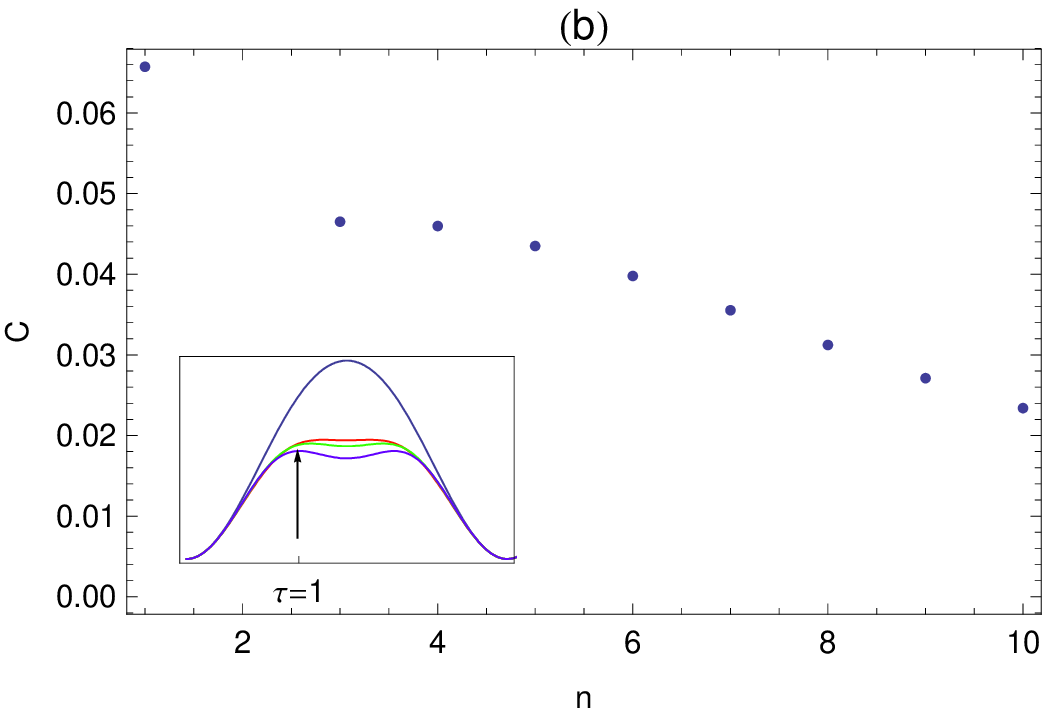}}
\hspace{0.3cm}
\subfigure{\label{mn0field02c}\includegraphics[width=6.5cm]{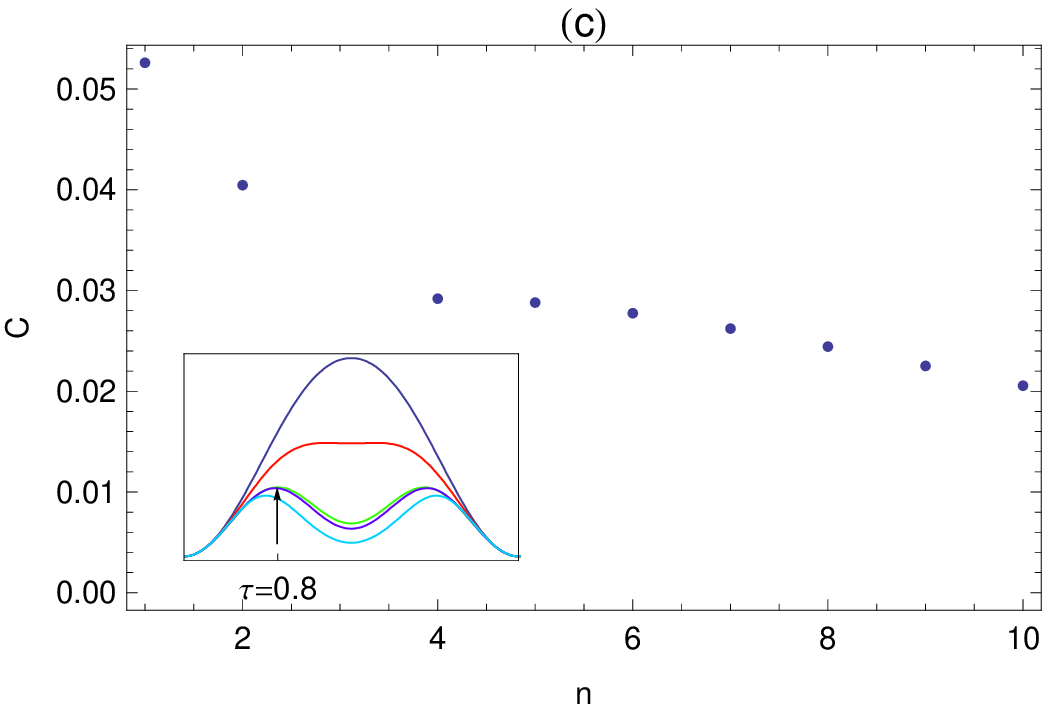}}
\hspace{0.3cm}
\subfigure{\label{mn0field02d}\includegraphics[width=6.5cm]{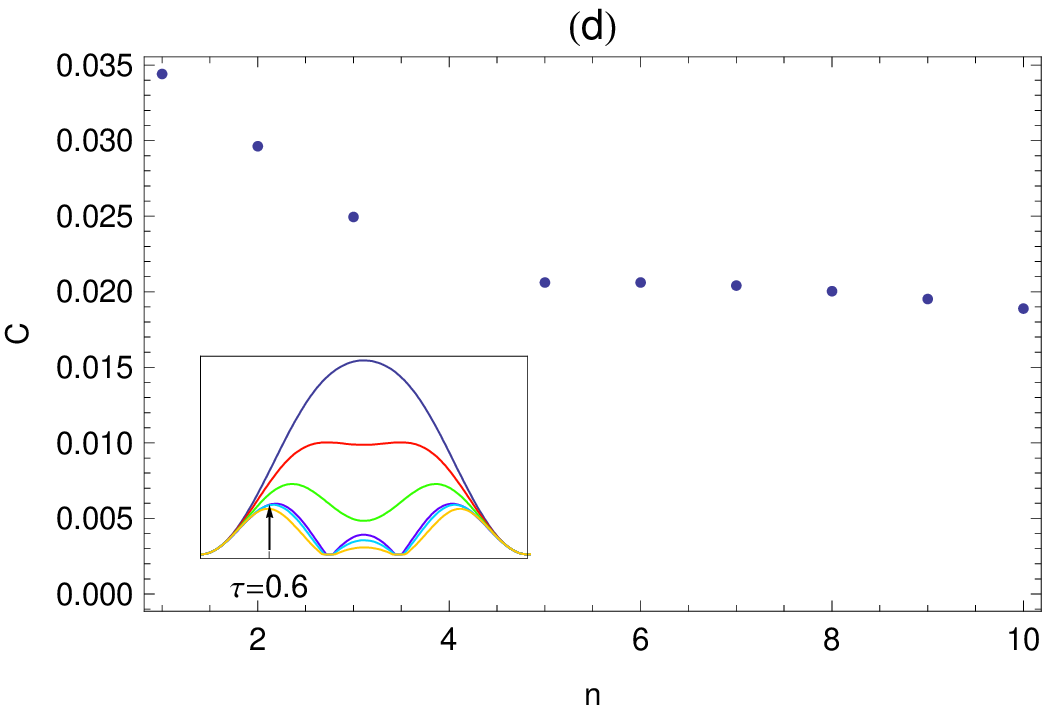}}
\end{center}
\caption{The concurrence $C_{mn}$ at fixed $\tau$ and $B_z=0$, for $m\in\left[1,4\right]$ and increasing $n$. Values of $\tau$ are inset, with the exception of figure (a), where $\tau=T_{\phi}/2$. All quantities are expressed in natural units, with $\lambda=1$. Results are for incoming neuron polarization $\alpha=\frac{\sqrt{3}}{2}$, $\beta=\frac{1}{2}$. (a) $m=1$. (b) $m=2$. (c) $m=3$. (d) $m=4$.}
\label{mn0field02}
\end{figure}

\begin{figure}[H]
\renewcommand{\captionfont}{\footnotesize}
\renewcommand{\captionlabelfont}{}
\begin{center}
\subfigure{\label{mn0field05a}\includegraphics[width=6.5cm]{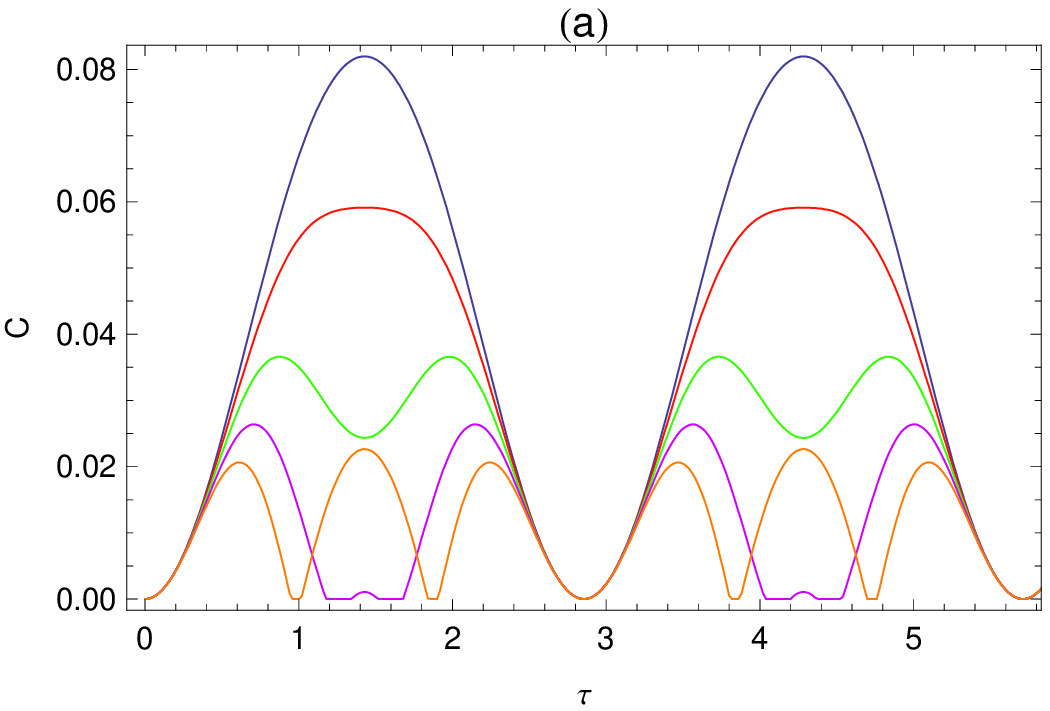}}
\hspace{0.3cm}
\subfigure{\label{mn0field05b}\includegraphics[width=6.5cm]{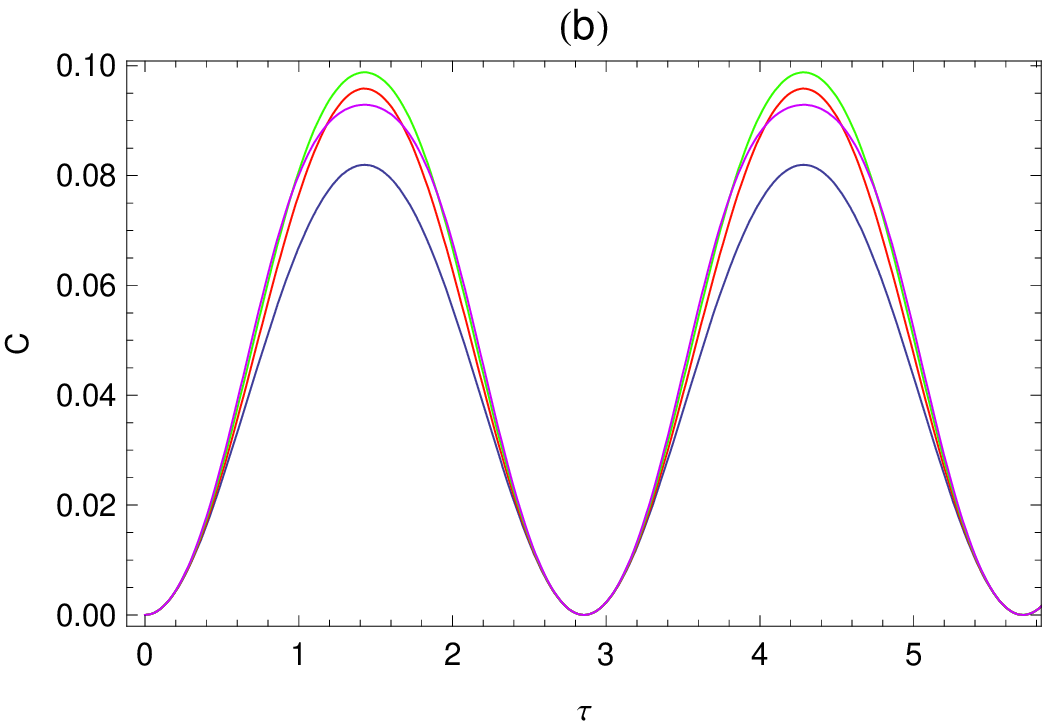}}
\hspace{0.3cm}
\subfigure{\label{mn0field05c}\includegraphics[width=6.5cm]{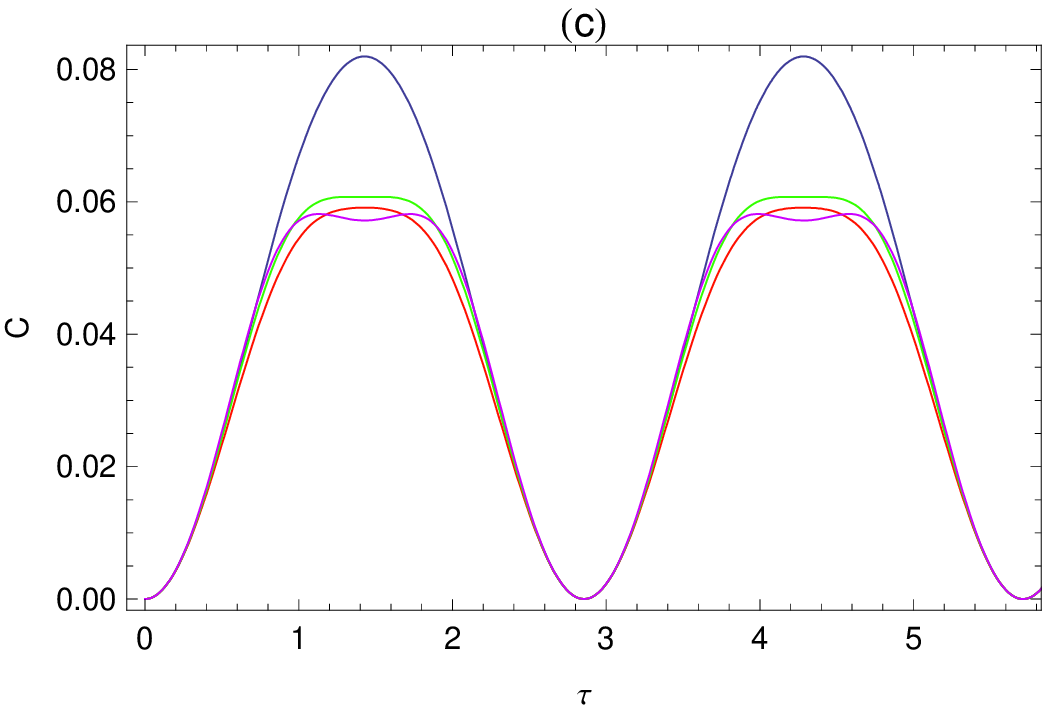}}
\hspace{0.3cm}
\subfigure{\label{mn0field05d}\includegraphics[width=6.5cm]{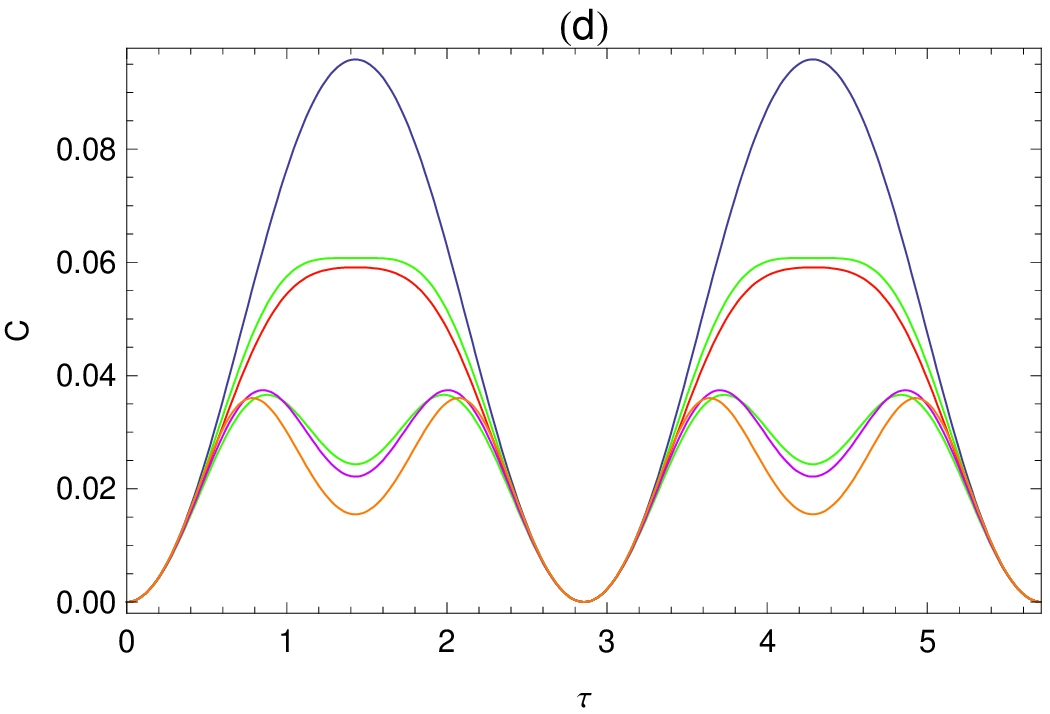}}
\hspace{0.3cm}
\subfigure{\label{mn0field05e}\includegraphics[width=6.5cm]{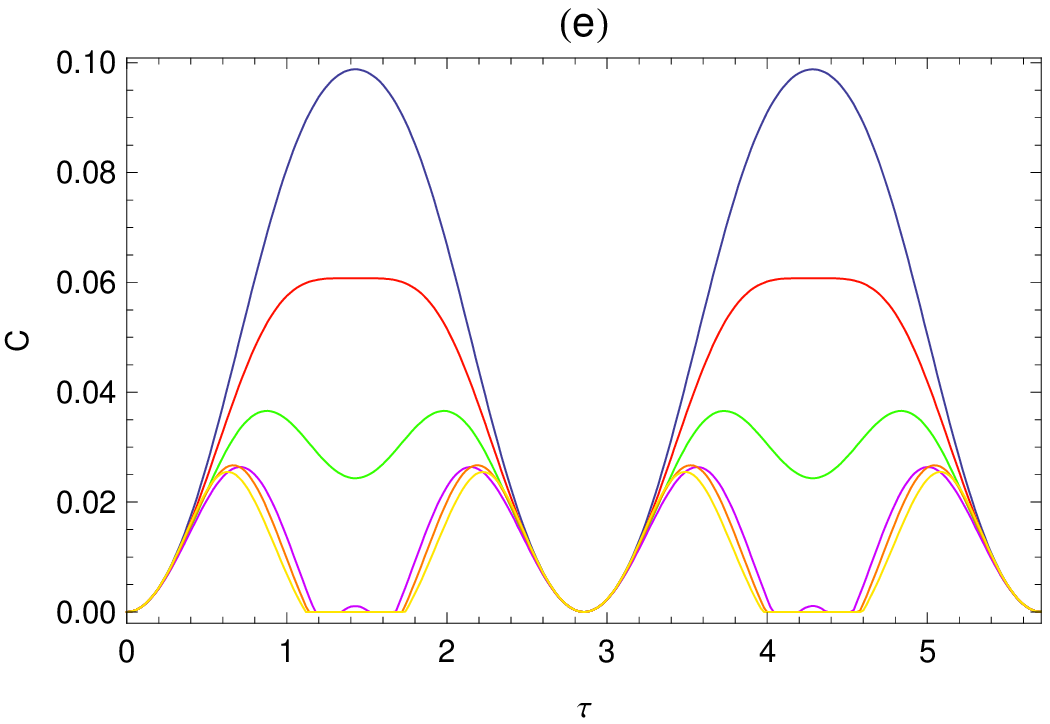}}
\end{center}
\caption{The concurrence of different neutron pairs for incoming neuron polarization $\alpha=\frac{\sqrt{2}}{\sqrt{3}}$, $\beta=\frac{1}{\sqrt{3}}$ and $B_z=0$. Increasing values of $n$ are indicated by the blue, red, green, purple, orange and yellow curves, respectively. (a) The concurrence of neutrons $m$ and $m+1$ for $m\leq5$. (b) $C_{1n}$ for $1<n\leq5$. (c) $C_{2n}$ for $n=\lbrace1,3,4,5\rbrace$. (d) $C_{3n}$ for $n=\lbrace1,2,4,5,7\rbrace$. (e) $C_{4n}$ for $n=\lbrace1,2,3,5,8,10\rbrace$. All quantities are expressed in natural units, with $\lambda=1$.}
\label{mn0field05}
\end{figure}

\begin{figure}[H]
\renewcommand{\captionfont}{\footnotesize}
\renewcommand{\captionlabelfont}{}
\begin{center}
\subfigure{\label{mn0field06a}\includegraphics[width=6.5cm]{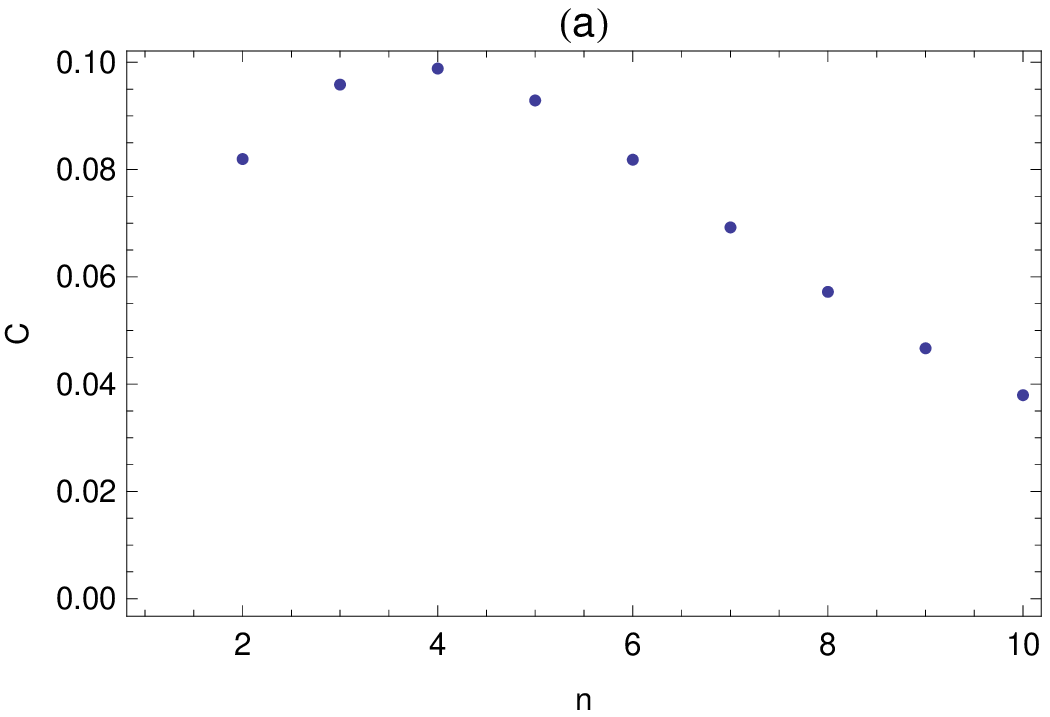}}
\hspace{0.3cm}
\subfigure{\label{mn0field06b}\includegraphics[width=6.5cm]{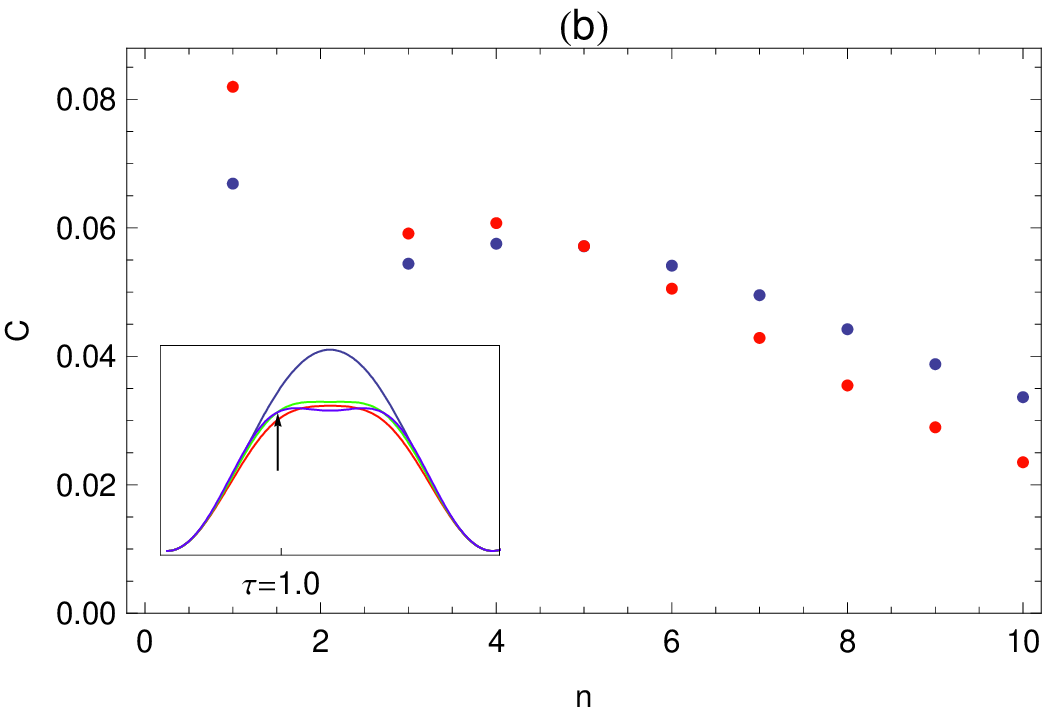}}
\hspace{0.3cm}
\subfigure{\label{mn0field06c}\includegraphics[width=6.5cm]{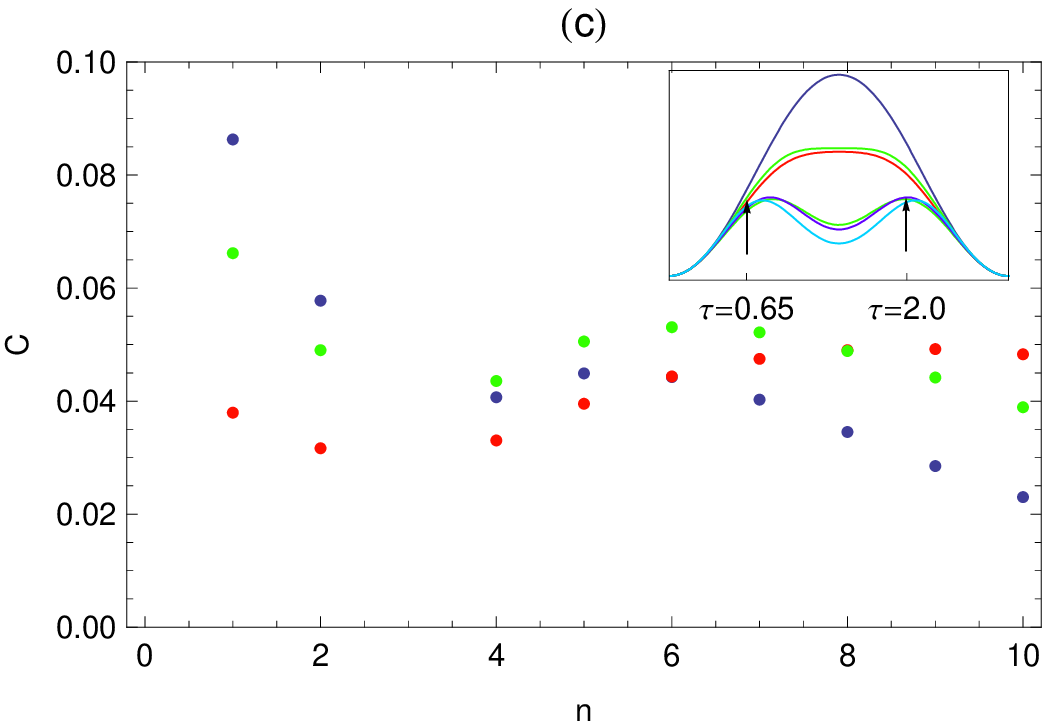}}
\hspace{0.3cm}
\subfigure{\label{mn0field06d}\includegraphics[width=6.5cm]{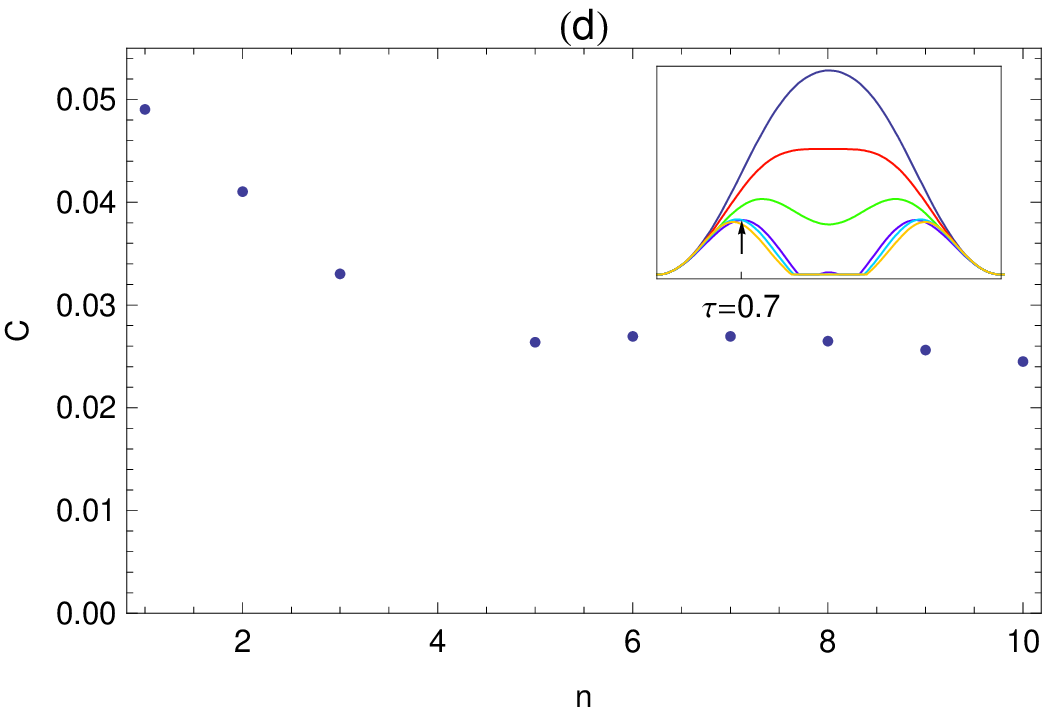}}
\end{center}
\caption{The concurrence $C_{mn}$ at fixed $\tau$ and $B_z=0$, for $m\in\left[1,4\right]$ and increasing $n$. Results are for incoming neuron polarization $\alpha=\frac{\sqrt{2}}{\sqrt{3}}$, $\beta=\frac{1}{\sqrt{3}}$. (a) $m=1$ at $\tau=T_{\phi}/2$. (b) $m=2$ at $\tau=1$ and $\tau=T_{\phi}/2$ (blue and red dots, respectively). (c) $m=3$ at $\tau=0.65$, $\tau=T_{\phi}/2$, and $\tau=2$ (blue, red and green dots, respectively). (d) $m=4$ at $\tau=0.7$. All quantities are expressed in natural units, with $\lambda=1$.}
\label{mn0field06}
\end{figure}

\begin{figure}[H]
\renewcommand{\captionfont}{\footnotesize}
\renewcommand{\captionlabelfont}{}
\begin{center}
\subfigure{\label{mn0field03a}\includegraphics[width=6.5cm]{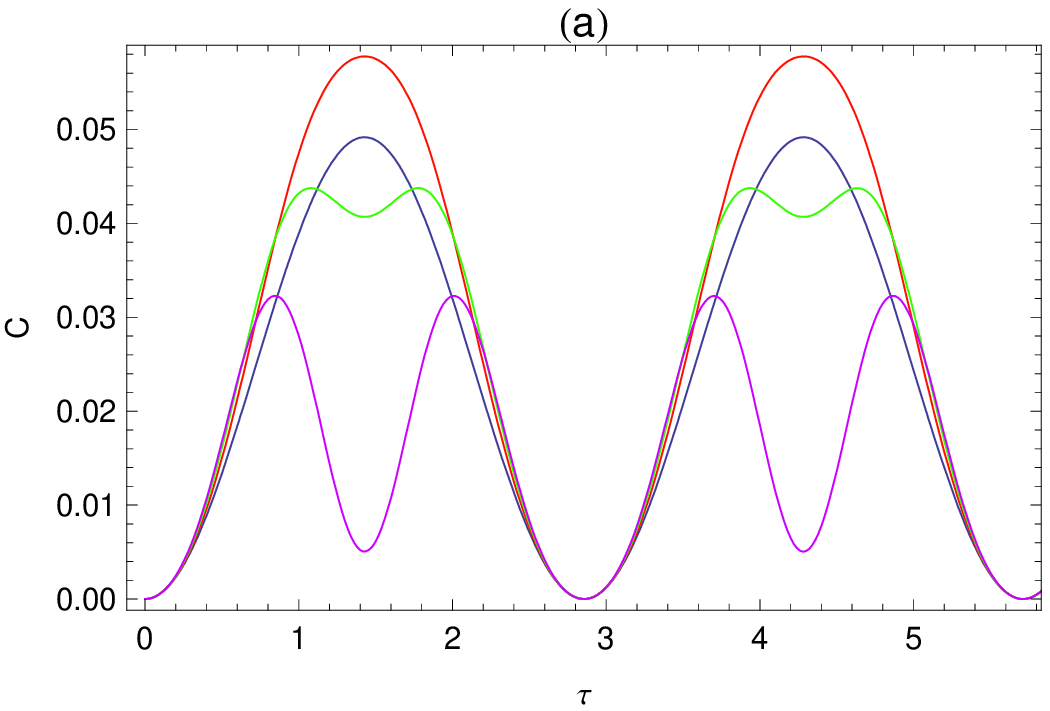}}
\hspace{0.3cm}
\subfigure{\label{mn0field03b}\includegraphics[width=6.5cm]{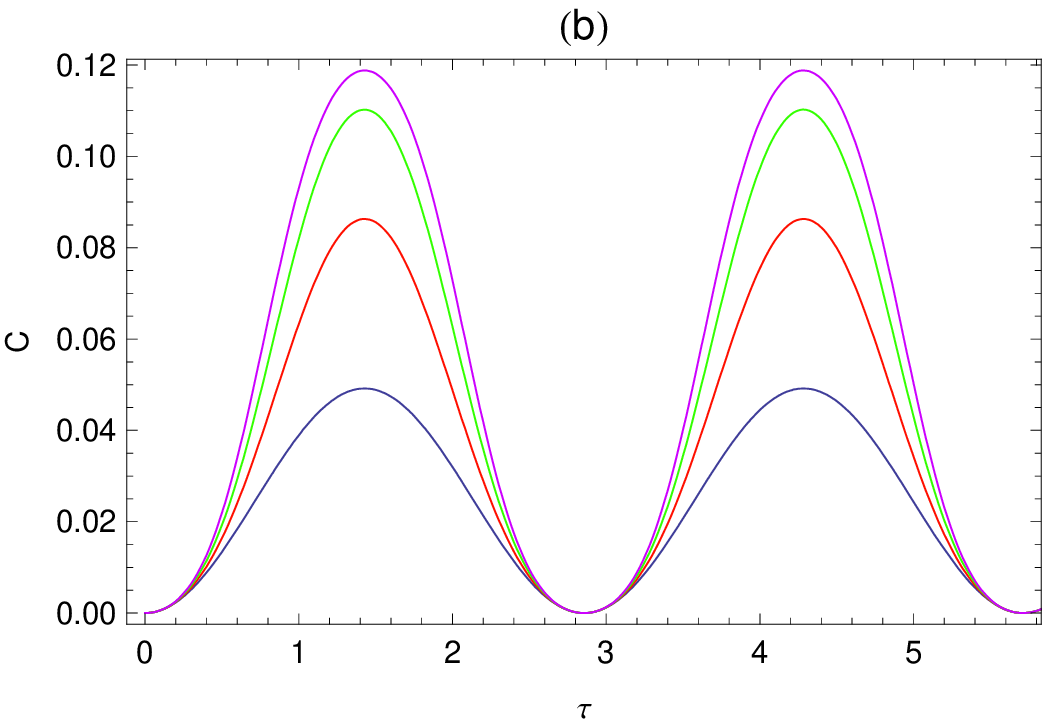}}
\hspace{0.3cm}
\subfigure{\label{mn0field03c}\includegraphics[width=6.5cm]{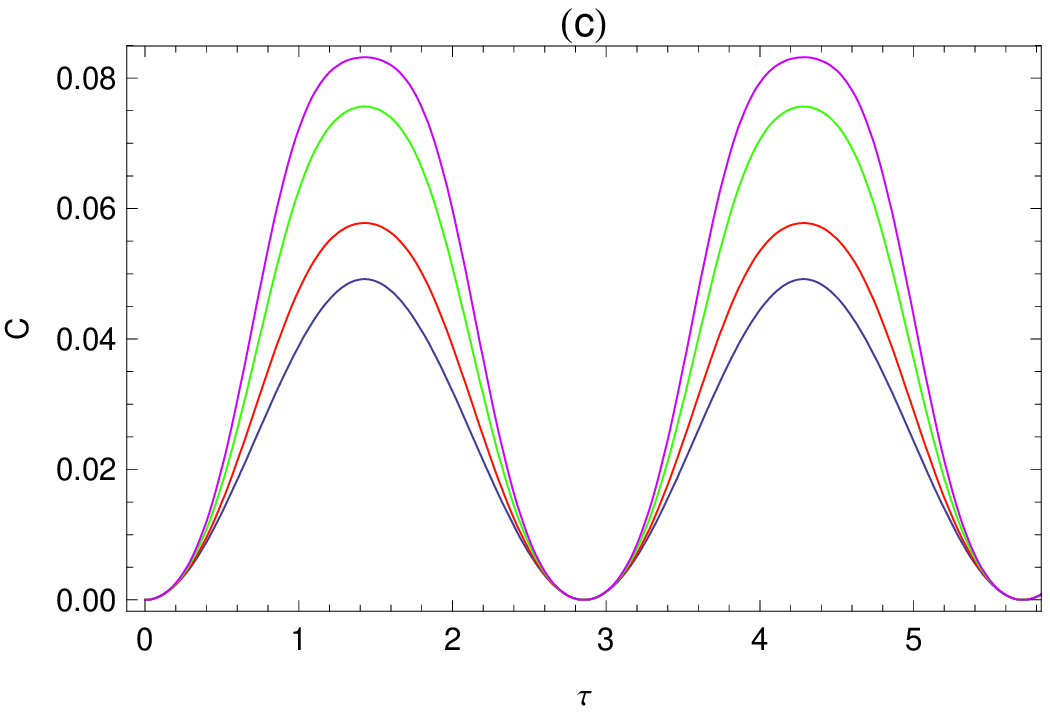}}
\hspace{0.3cm}
\subfigure{\label{mn0field03d}\includegraphics[width=6.5cm]{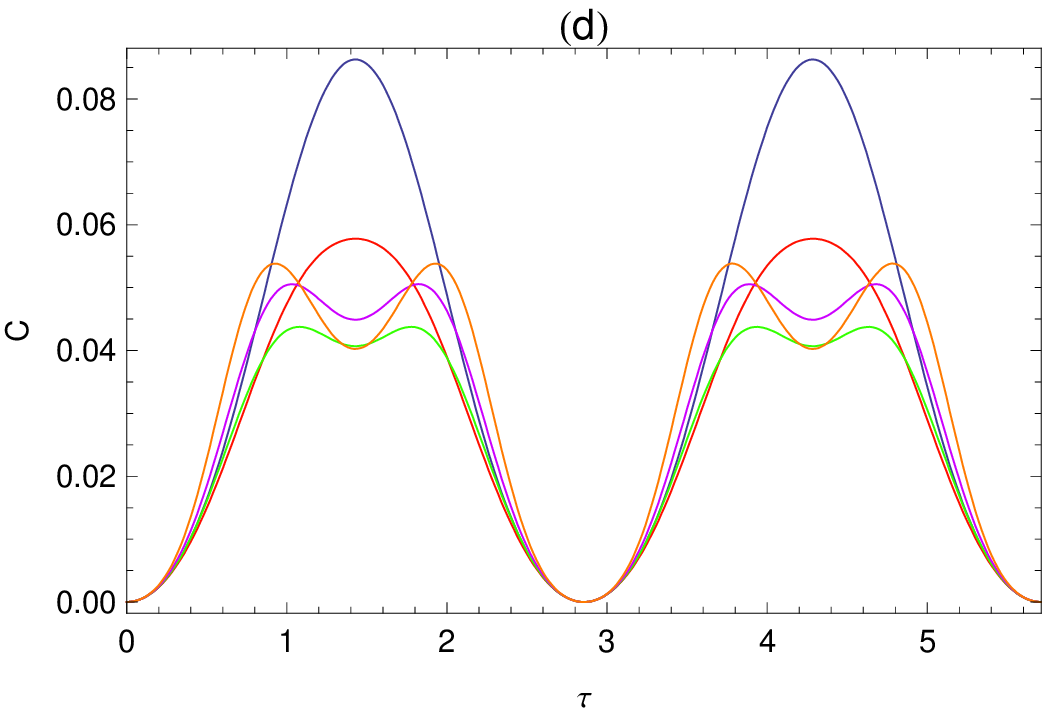}}
\hspace{0.3cm}
\subfigure{\label{mn0field03e}\includegraphics[width=6.5cm]{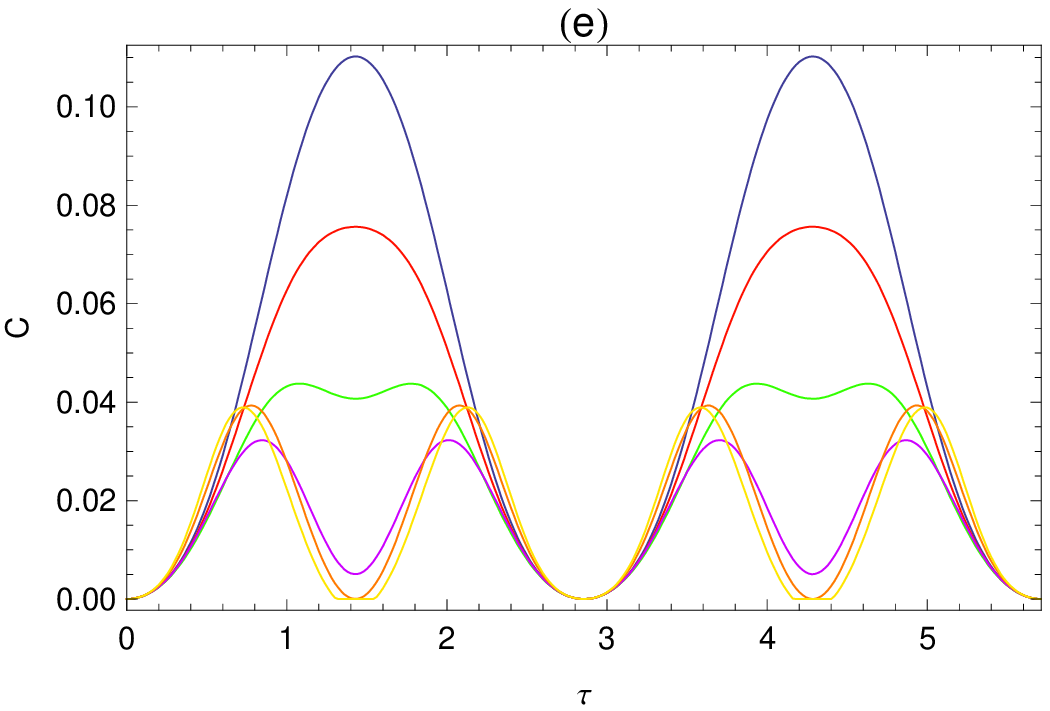}}
\end{center}
\caption{The concurrence of different neutron pairs for incoming neuron polarization $\alpha=\beta=\frac{1}{\sqrt{2}}$ and $B_z=0$. Increasing values of $n$ are indicated by the blue, red, green, purple, orange and yellow curves, respectively. (a) The concurrence of neutrons $m$ and $m+1$ for $m\leq4$. (b) $C_{1n}$ for $1<n\leq5$. (c) $C_{2n}$ for $n=\lbrace1,3,4,5\rbrace$. (d) $C_{3n}$ for $n=\lbrace1,2,4,5,7\rbrace$. (e) $C_{4n}$ for $n=\lbrace1,2,3,5,8,10\rbrace$. All quantities are expressed in natural units, with $\lambda=1$.}
\label{mn0field03}
\end{figure}

\begin{figure}[H]
\renewcommand{\captionfont}{\footnotesize}
\renewcommand{\captionlabelfont}{}
\begin{center}
\subfigure{\label{mn0field04a}\includegraphics[width=6.5cm]{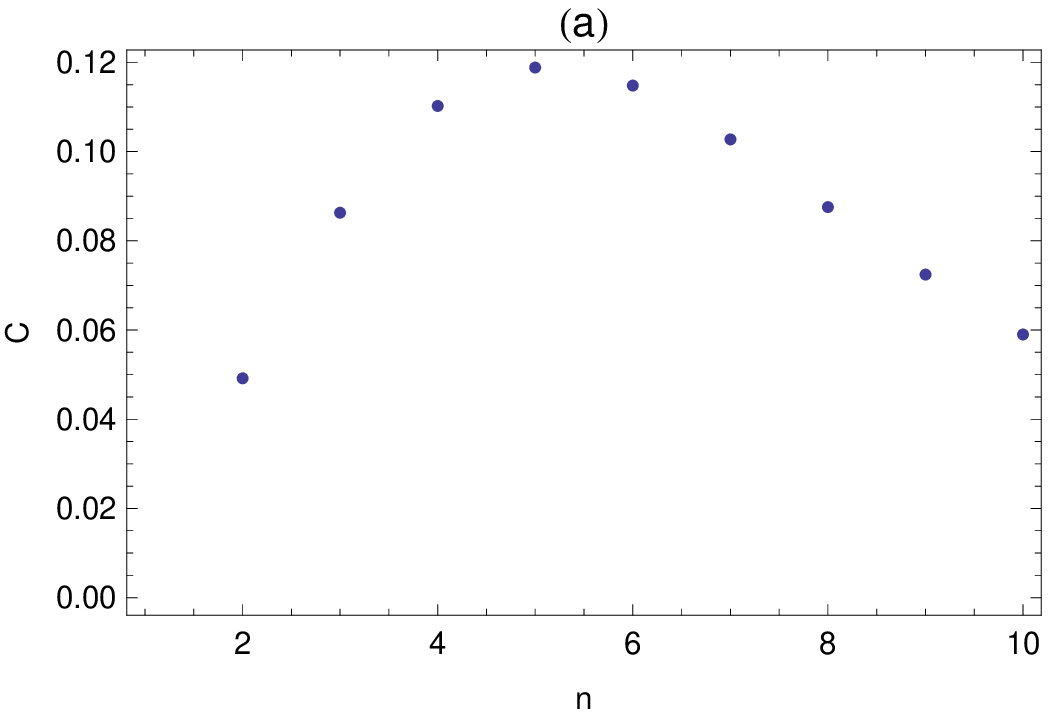}}
\hspace{0.3cm}
\subfigure{\label{mn0field04b}\includegraphics[width=6.5cm]{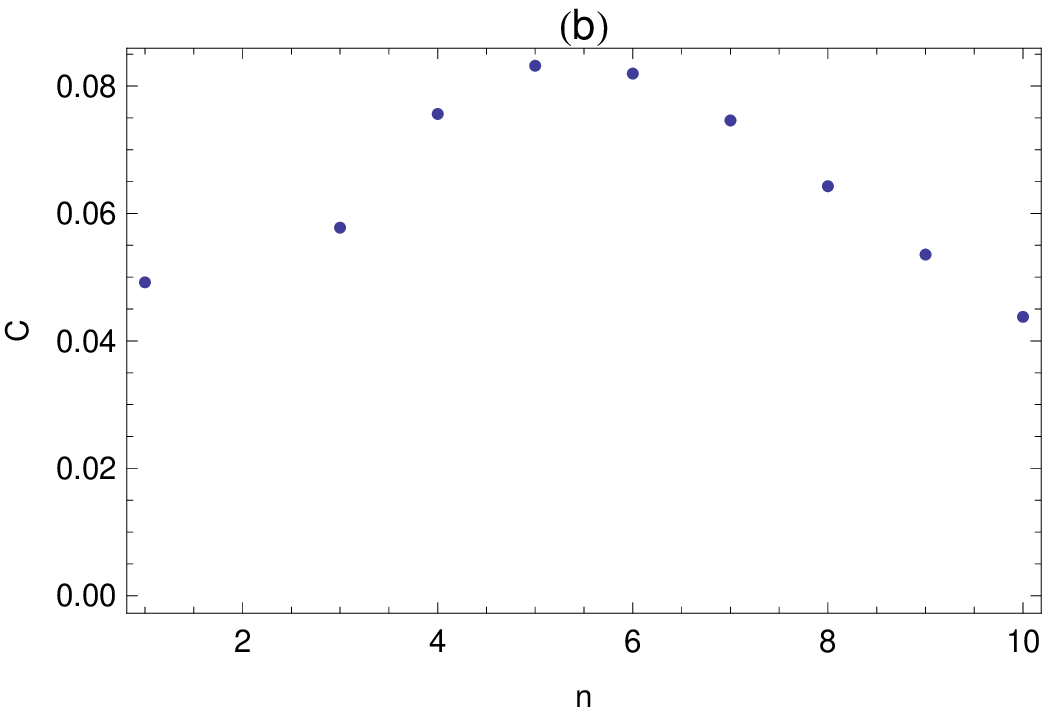}}
\hspace{0.3cm}
\subfigure{\label{mn0field04c}\includegraphics[width=6.5cm]{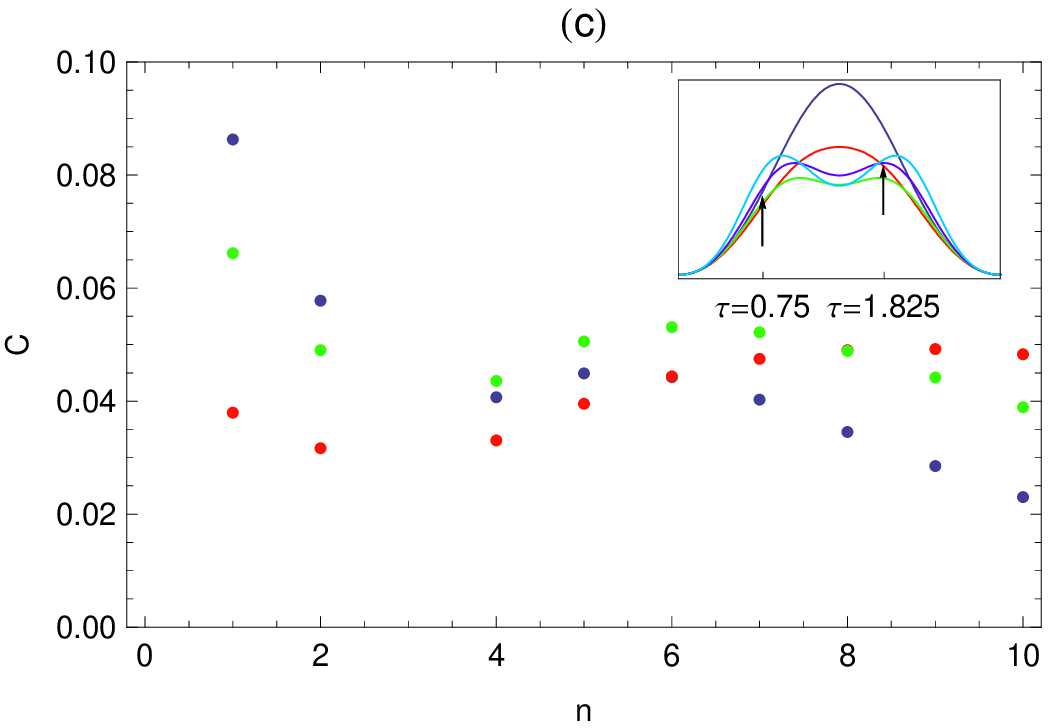}}
\hspace{0.3cm}
\subfigure{\label{mn0field04d}\includegraphics[width=6.5cm]{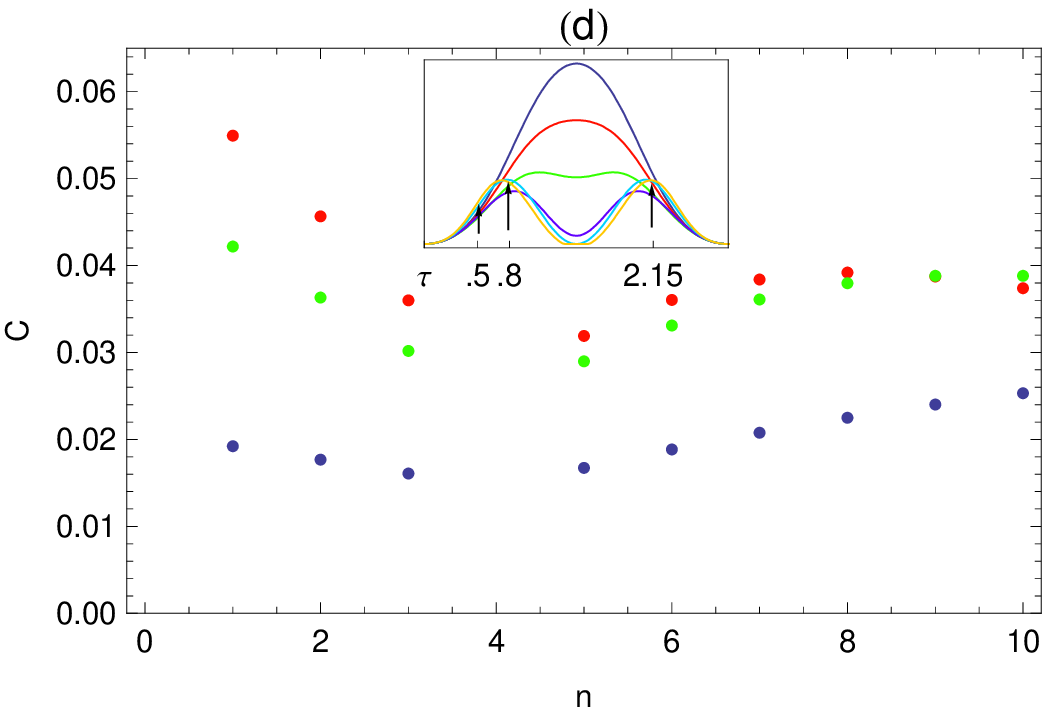}}
\end{center}
\caption{The concurrence $C_{mn}$ at fixed $\tau$ and $B_z=0$, for $m\in\left[1,4\right]$ and increasing $n$. Results are for incoming neuron polarization $\alpha=\beta=\frac{1}{\sqrt{2}}$. (a) $m=1$ at $\tau=T_{\phi}/2$. (b) $m=2$ at $\tau=T_{\phi}/2$. (c) $m=3$ at $\tau=T_{\phi}/2$, $\tau=0.75$ and $\tau=1.825$ (blue, red and green dots, respectively). (d) $m=4$ at $\tau=0.5$, $\tau=0.8$ and $\tau=2.15$ (blue, red and green dots, respectively). All quantities are expressed in natural units, with $\lambda=1$.}
\label{mn0field04}
\end{figure}

It is useful to begin with a brief recap of the properties of $C_{1,2}$. These are summarized in two points: (i) oscillatory behaviour as a function of interaction time; (ii) maximum amplitude for $\alpha=\frac{\sqrt{3}}{2}$. Let us then observe the concurrence of different neutron pairs at $\alpha=\frac{\sqrt{3}}{2}$; this is shown in figure \ref{mn0field0}. It is immediately evident that the concurrence maintains its characteristic periodicity irrespective of the indices $m$ and $n$, however these indices do affect the mode of oscillation. At fixed $n$, the concurrence falls with $m$ and, with the exception of $m=1$, the peak value of $C_{mn}$ falls with increasing $|m-n|$. More interesting features develop as $\alpha\rightarrow\frac{1}{\sqrt{2}}$. It now emerges that the concurrence of distant ($|m-n|>2$) neutrons can exceed that of neutrons scattered in succession (figures \ref{mn0field02}-\ref{mn0field04}). The most prominent example occurs at $m=1$ and $\alpha=\frac{1}{\sqrt{2}}$, for which the concurrence of the first and second neutrons represents a minimum value. This may seem promising, however the neutron concurrence remains low no matter which neutron pair we detect. Furthermore, even for an optimal choice of input parameters, the resulting state of the scattered neutrons is not compatible with the witness decomposition of equation \eqref{wit}.

\subsection{Finite-Field Evolution}
When a finite field is applied, the behaviour of the concurrence is highly unpredictable. Comparing the concurrence of an arbitrary pair $\lbrace m,n\rbrace$ with $C_{1,2}$ at $B_z<B_{z+}$, it is no longer a given that the peak value of $C_{m,n}$ will increase if $B_z$ is raised [see section \ref{p0b zero mtm finite field} and figure \ref{mn0field0b}]. As expected, the details of the time evolution of $C_{m,n}$ are determined by the strength of the field and the neutron polarization; however, exactly how these parameters affect the concurrence is \textit{a priori} unclear. To make our task more manageable, let us use the strength of the external field to define three regimes:
\begin{enumerate}
\item Weak field: $B_z\leq\frac{\lambda}{2}$;
\item Strong field: $B_z\geq3\lambda$;
\item Optimal field: $B_z=B_z^*$,
\end{enumerate}
and study the features of the concurrence in each case for different neutron polarizations.

In the weak field regime, the concurrence of the first and second neutrons shows a well-defined periodicity as a function of the interaction time (cfr. figure \ref{Bdep6}). As $m$ and $n$ are increased, this is no longer true unless $\alpha=0$. In general, the concurrence falls with increasing $|m-n|$, though when $m>2$ instances of $C_{m,n}>C_{m,m\pm1}$ become more common.

\begin{figure}[H]
\renewcommand{\captionfont}{\footnotesize}
\renewcommand{\captionlabelfont}{}
\begin{center}
\subfigure{\label{Bmin1a}\includegraphics[width=6.5cm]{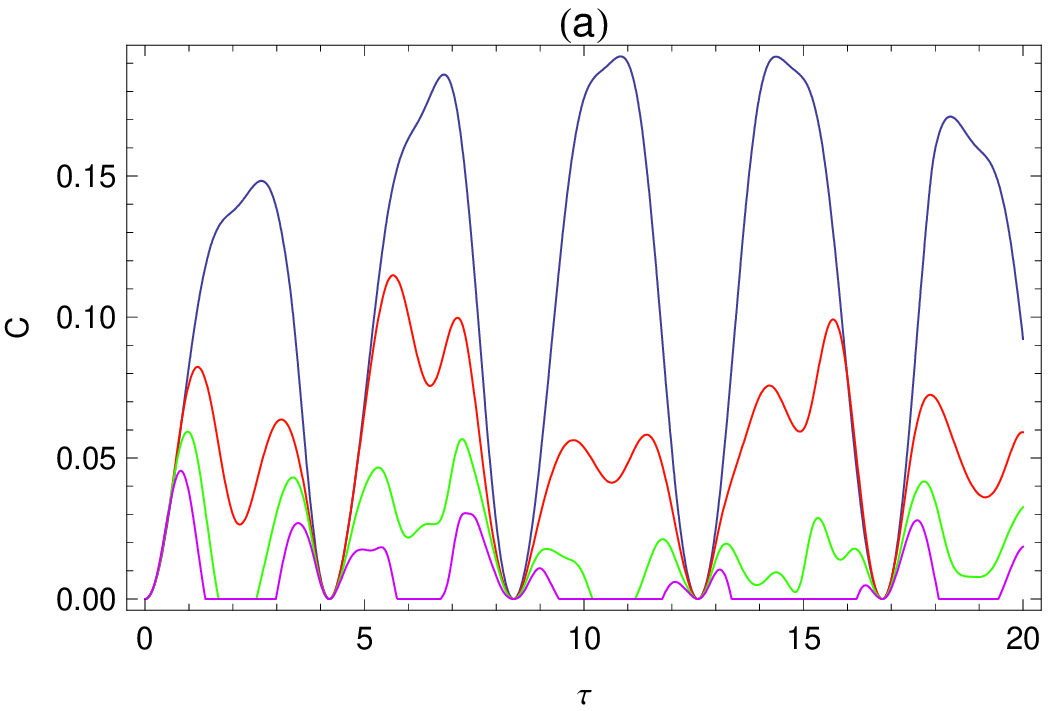}}
\hspace{0.3cm}
\subfigure{\label{Bmin1b}\includegraphics[width=6.5cm]{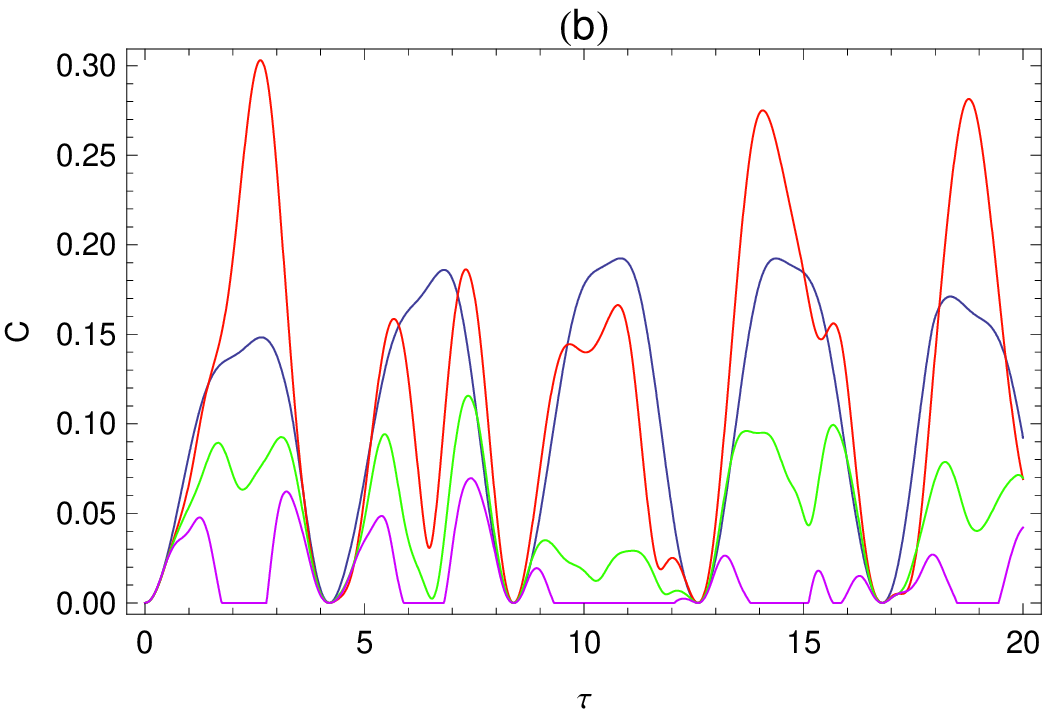}}
\end{center}
\caption{The concurrence $C_{mn}$ for $B_z=\lambda/2$ and $\alpha=\frac{\sqrt{3}}{2}$ at (a) $m=1$ and (b) $m=2$. Increasing values of $n$ are indicated by the blue, red, green and purple curves, respectively. All quantities are expressed in natural units, with $\lambda=1$.}
\label{Bmin1}
\end{figure}

\begin{figure}[H]
\renewcommand{\captionfont}{\footnotesize}
\renewcommand{\captionlabelfont}{}
\begin{center}
\subfigure{\label{Bmin2a}\includegraphics[width=6.5cm]{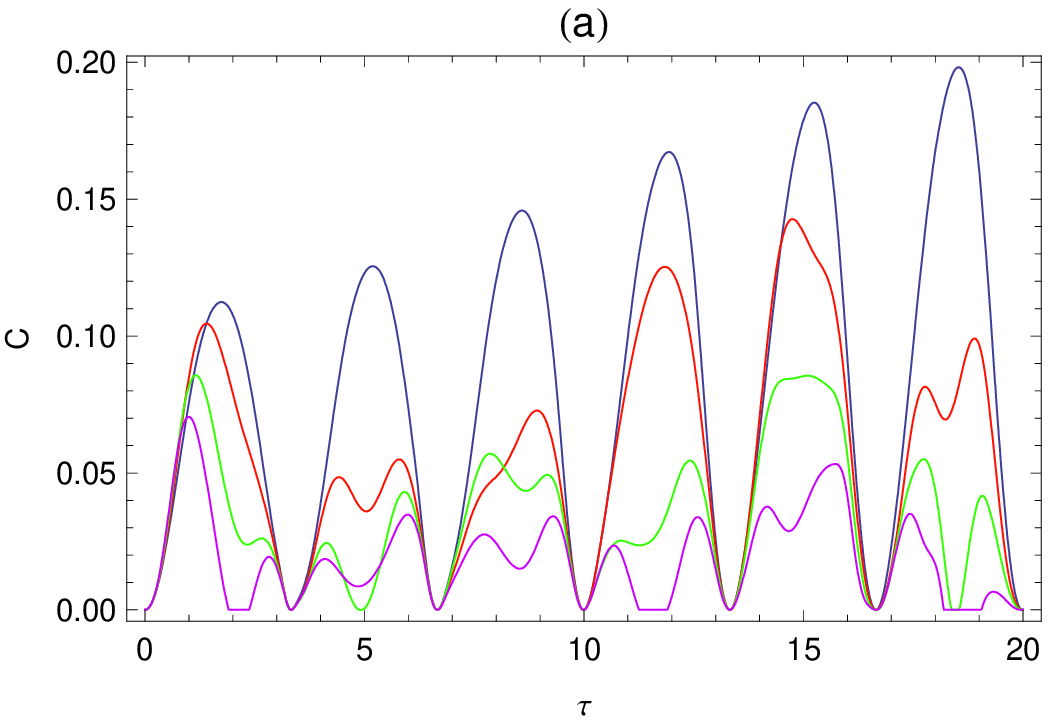}}
\hspace{0.3cm}
\subfigure{\label{Bmin2b}\includegraphics[width=6.5cm]{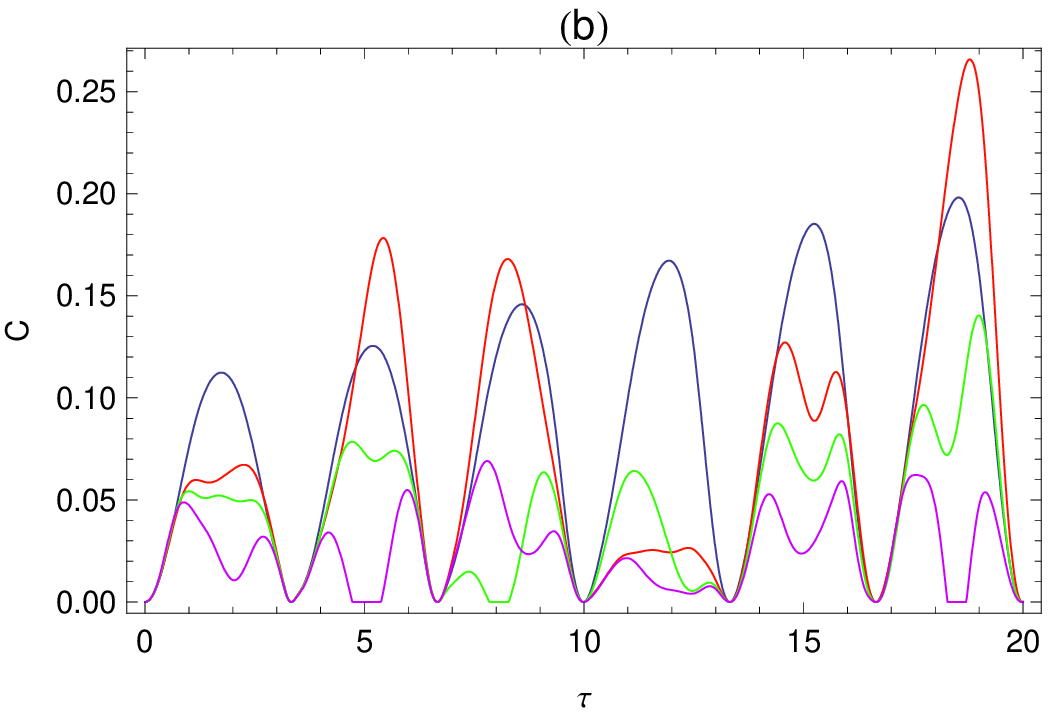}}
\hspace{0.3cm}
\subfigure{\label{Bmin2c}\includegraphics[width=6.5cm]{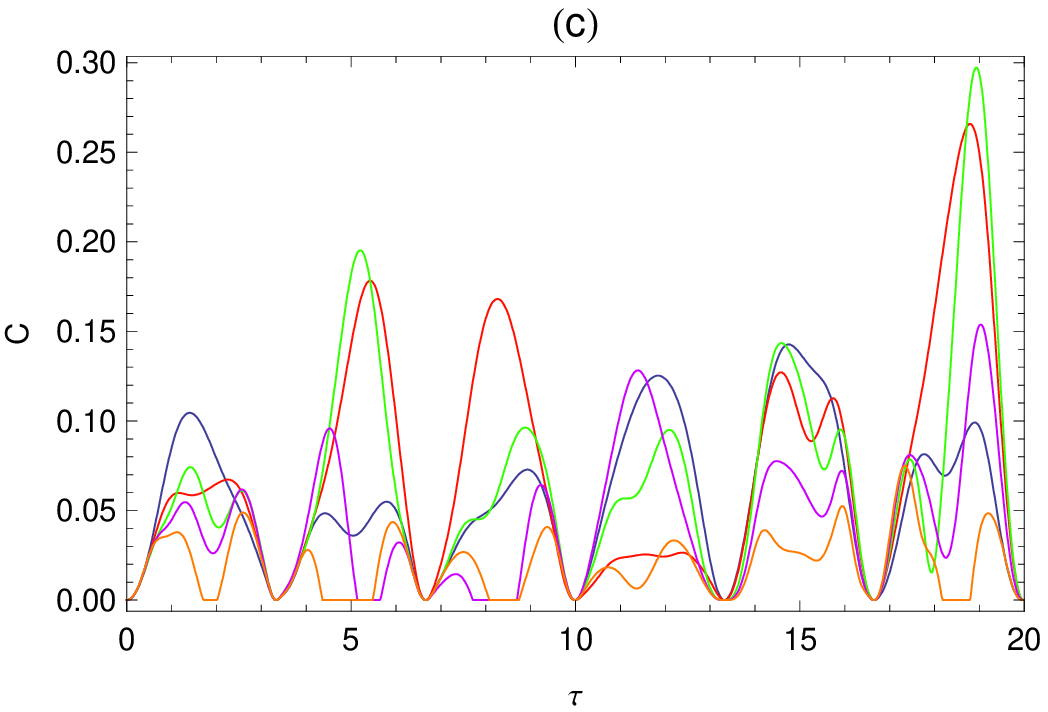}}
\hspace{0.3cm}
\subfigure{\label{Bmin2d}\includegraphics[width=6.5cm]{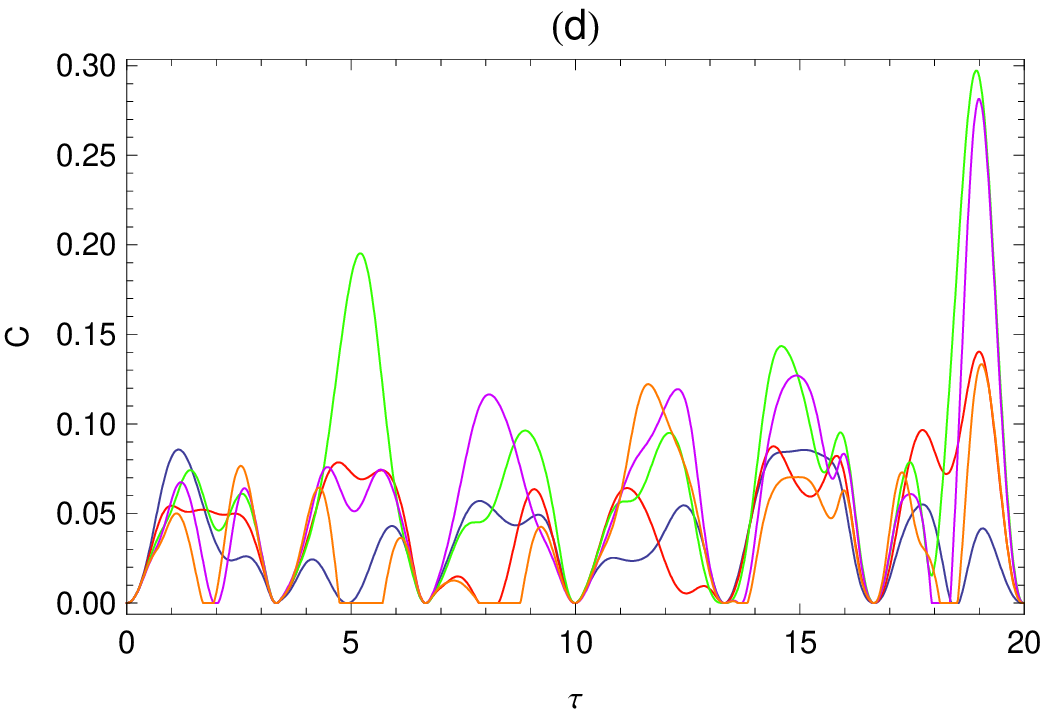}}
\end{center}
\caption{The concurrence $C_{mn}$ for $B_z=\lambda/5$ and $\alpha=\frac{\sqrt{2}}{\sqrt{3}}$ at (a) $m=1$, (b) $m=2$, (c) $m=3$ and (d) $m=4$. Increasing values of $n$ are indicated by the blue, red, green, purple and orange curves, respectively. All quantities are expressed in natural units, with $\lambda=1$.}
\label{Bmin2}
\end{figure}

\begin{figure}[H]
\renewcommand{\captionfont}{\footnotesize}
\renewcommand{\captionlabelfont}{}
\begin{center}
\subfigure{\label{Bmin3a}\includegraphics[width=6.5cm]{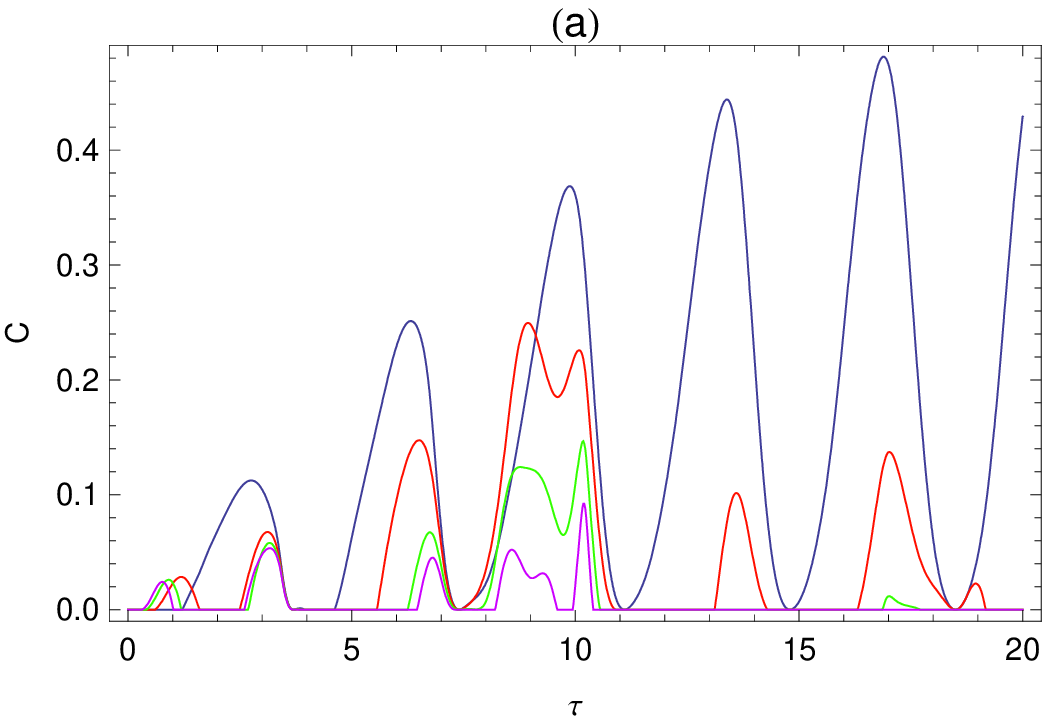}}
\hspace{0.3cm}
\subfigure{\label{Bmin3b}\includegraphics[width=6.5cm]{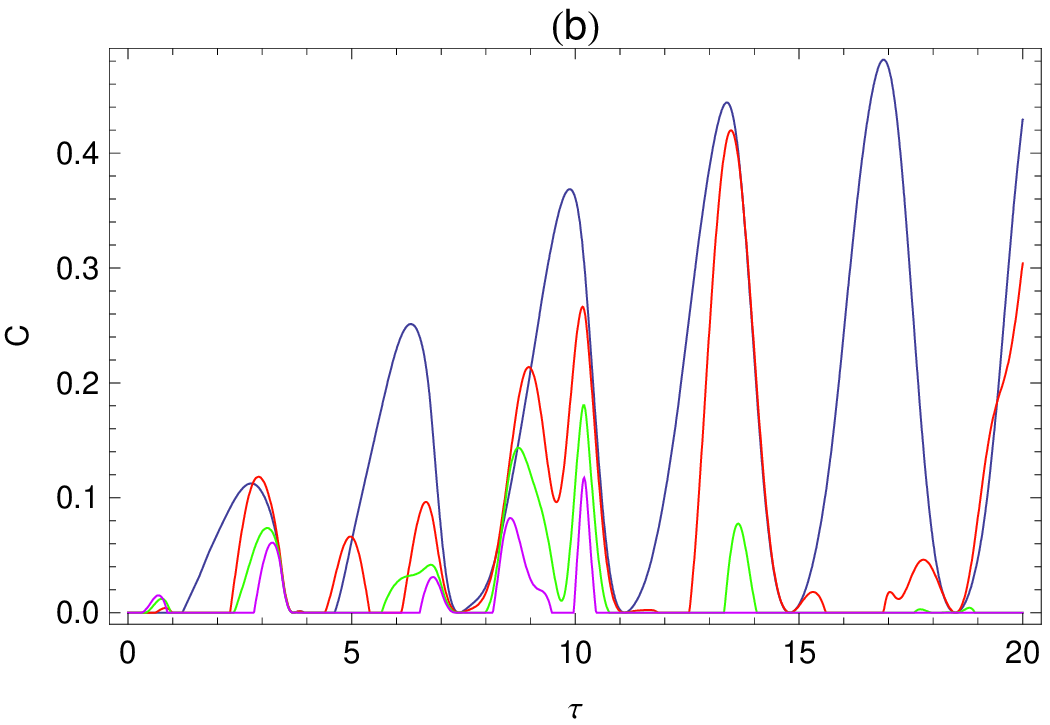}}
\hspace{0.3cm}
\subfigure{\label{Bmin3c}\includegraphics[width=6.5cm]{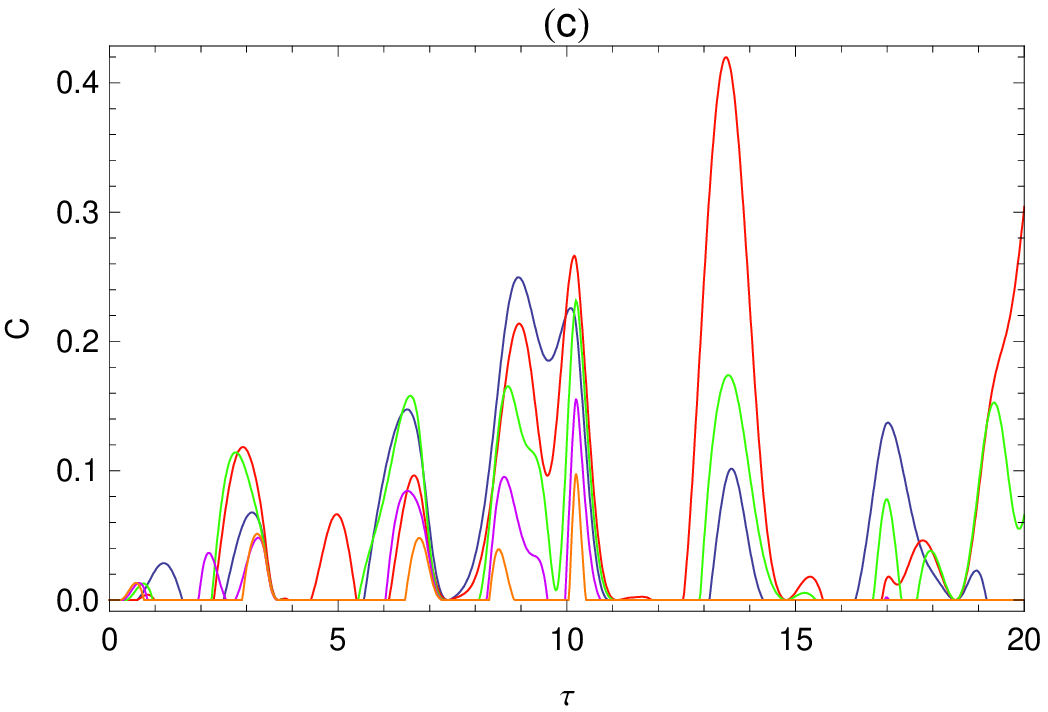}}
\end{center}
\caption{The concurrence $C_{mn}$ for $B_z=\lambda/3$ and $\alpha=\frac{1}{\sqrt{3}}$ at (a) $m=1$, (b) $m=2$ and (c) $m=3$. Increasing values of $n$ are indicated by the blue, red, green, purple and orange curves, respectively. All quantities are expressed in natural units, with $\lambda=1$.}
\label{Bmin3}
\end{figure}

\begin{figure}[H]
\renewcommand{\captionfont}{\footnotesize}
\renewcommand{\captionlabelfont}{}
\begin{center}
\subfigure{\label{Bmin4a}\includegraphics[width=6.5cm]{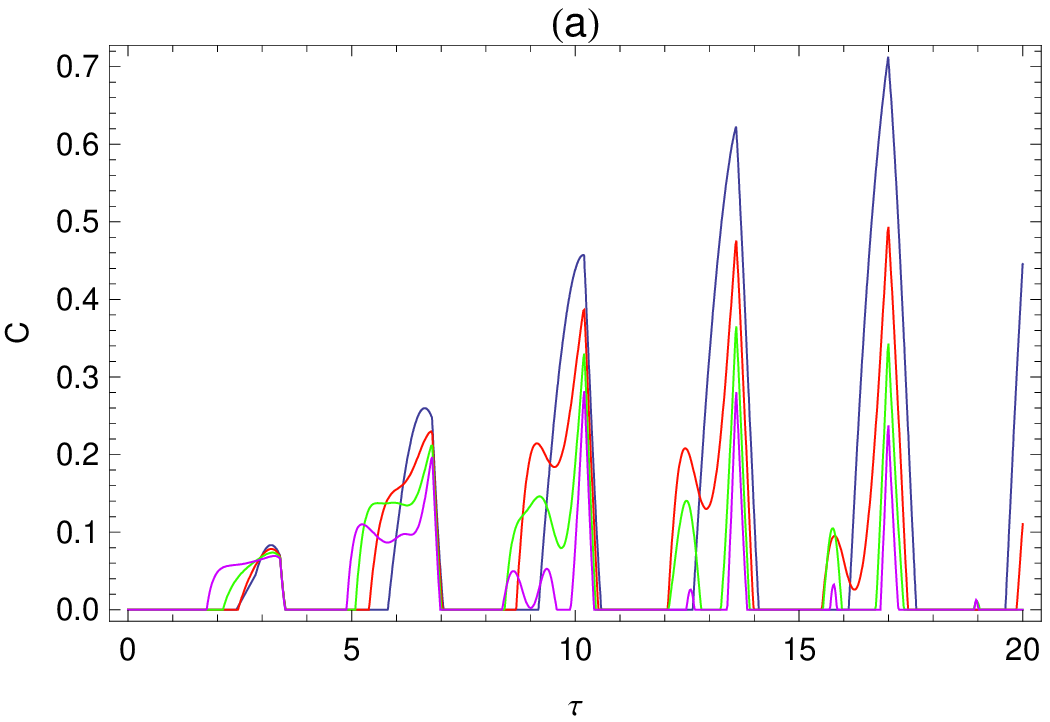}}
\hspace{0.3cm}
\subfigure{\label{Bmin4b}\includegraphics[width=6.5cm]{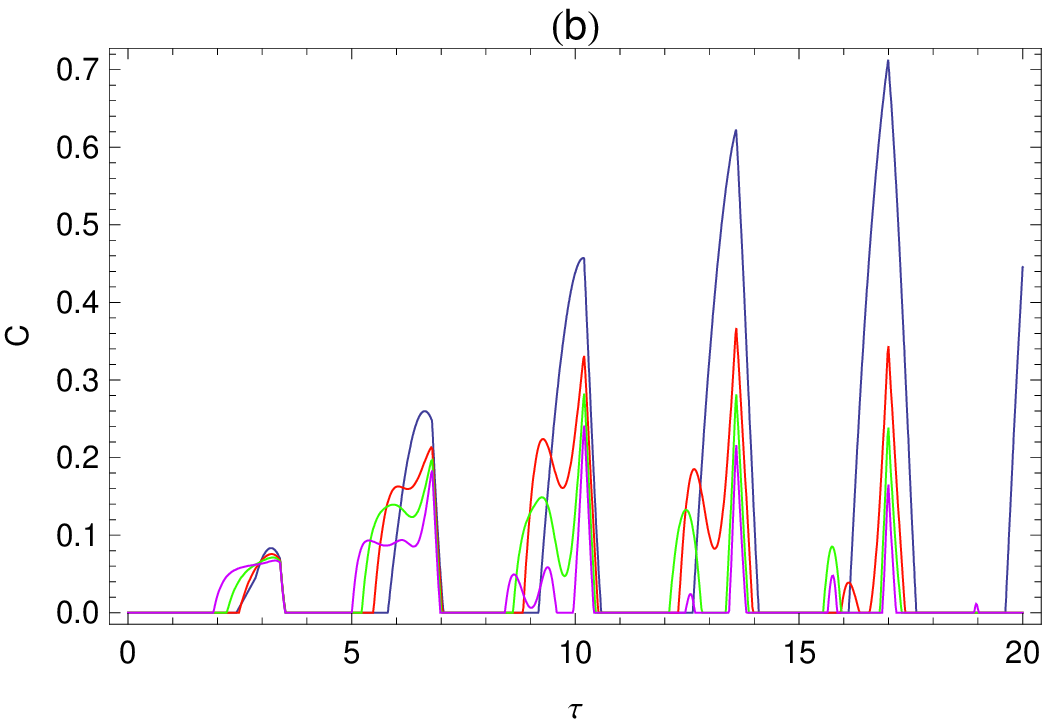}}
\hspace{0.3cm}
\subfigure{\label{Bmin4c}\includegraphics[width=6.5cm]{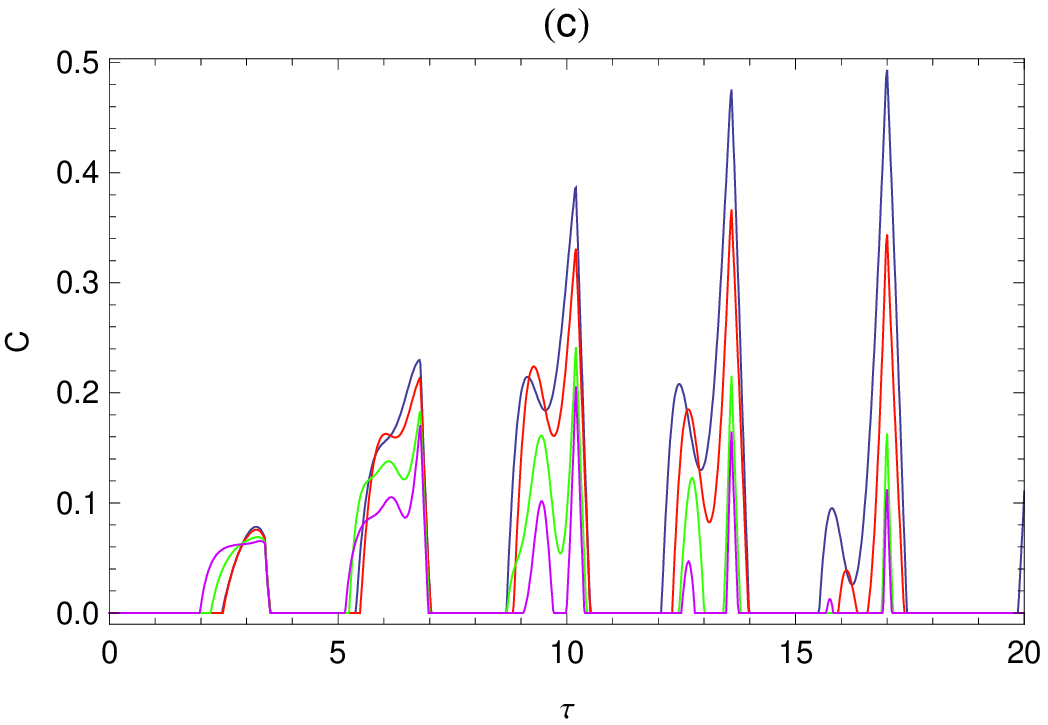}}
\hspace{0.3cm}
\subfigure{\label{Bmin4d}\includegraphics[width=6.5cm]{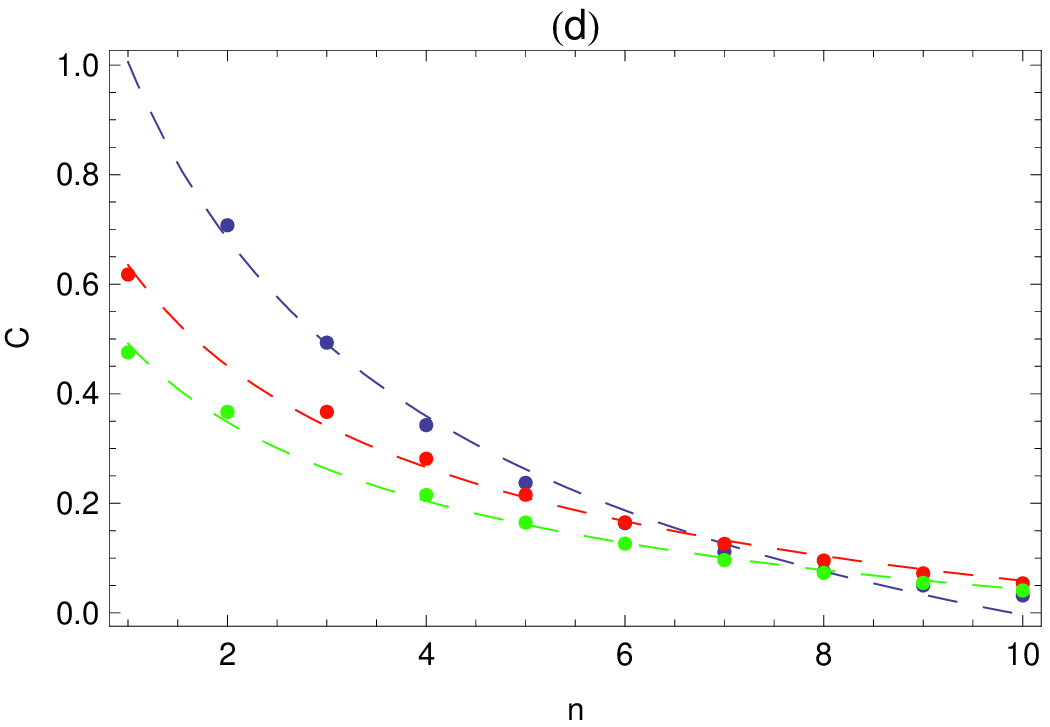}}
\end{center}
\caption{The concurrence for $B_z=\lambda/3$ and $\alpha=0$. Increasing values of $n$ are indicated by the blue, red, green and purple curves, respectively. All quantities are expressed in natural units, with $\lambda=1$. (a) $C_{1,n}$. (b) $C_{2,n}$. (c) $C_{3,n}$. (d) The peak concurrence for $m=\lbrace 1,2,3\rbrace$ (blue, red and green dots, respectively) at $\tau=17$ (blue dots) and $\tau=13.6$ (red and green dots). Dashed lines indicate the function $F(n)=\frac{a}{\sqrt{n+1}}+b$, evaluated at $a=2.49$ and $b=-0.75$ (blue curve), $a=1.42$ and $b=-0.37$ (red curve) and $a=1.11$ and $b=-0.29$ (green curve).}
\label{Bmin4}
\end{figure}

A strong dependence on the neutron polarization is also observed and, in analogy with the results of section \ref{p0b zero mtm finite field}, the performance of the system markedly improves as $\alpha\rightarrow0$ (figs. \ref{Bmin1}-\ref{Bmin4}). Indeed, the $\alpha=0$ case merits a separate discussion, as several interesting effects are observed. First, as noted above, the concurrence of any neutron pair shows traces of the periodic behaviour observed for $C_{1,2}$. Second, for any $m$ and $n$ the concurrence evolves through a series of peaks. At fixed $m$, peaks in $C_{m,n}$ for different $n$ are aligned [see figure \ref{Bmin4a}], though as $m$ increases the position of the strongest peak can change. In correspondence of this peak, the concurrence at fixed $m$ falls as $\left(n+1\right)^{-\frac{1}{2}}$. The scattered state of the neutrons at this time is always compatible with the witness decomposition of equation \eqref{wit}, however the concurrence of any non-successive neutron pair never exceeds 0.6. This is shown in figure \ref{Bmin4d}.

\begin{figure}[H]
\renewcommand{\captionfont}{\footnotesize}
\renewcommand{\captionlabelfont}{}
\begin{center}
\subfigure{\label{Bmax1a}\includegraphics[width=6.5cm]{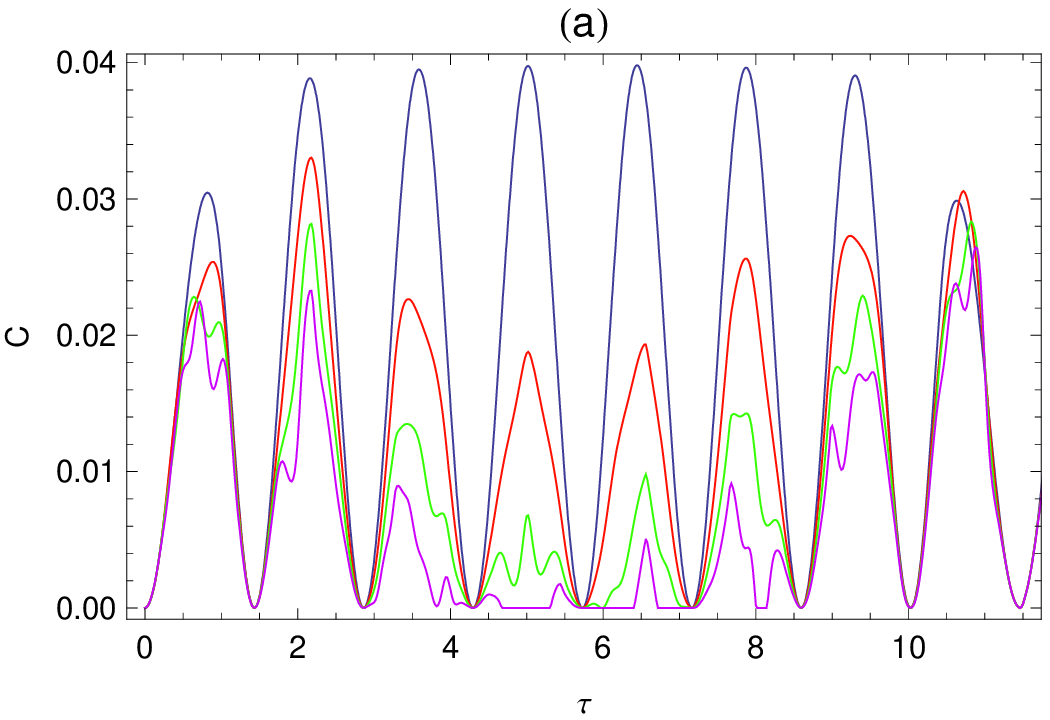}}
\hspace{0.3cm}
\subfigure{\label{Bmax1b}\includegraphics[width=6.5cm]{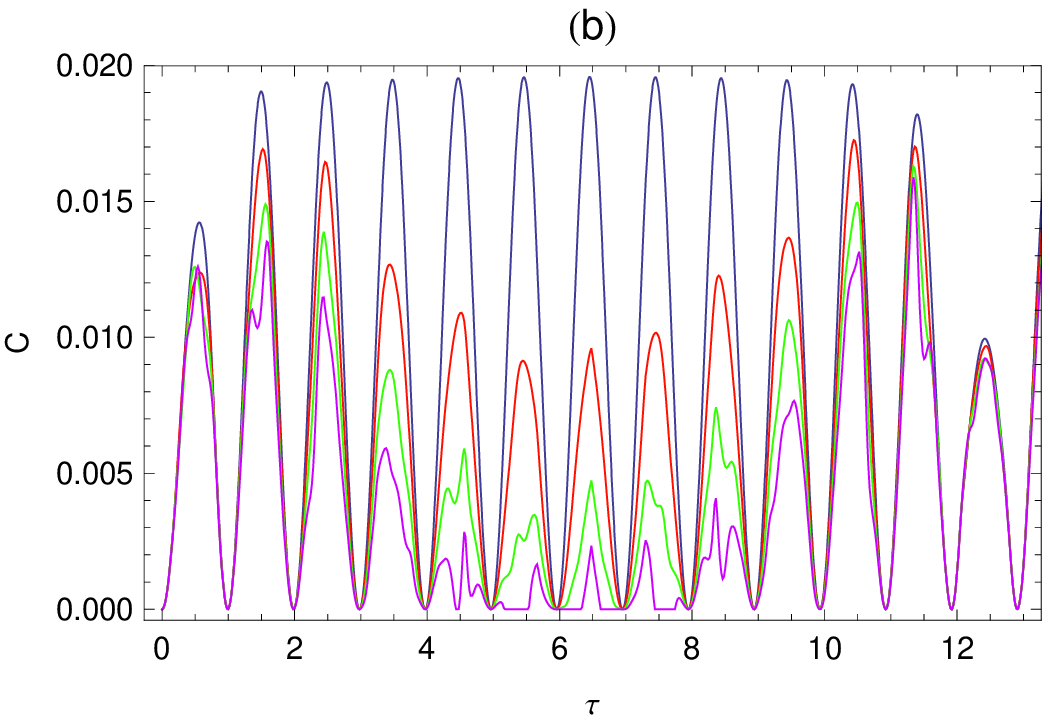}}
\hspace{0.3cm}
\subfigure{\label{Bmax1c}\includegraphics[width=6.5cm]{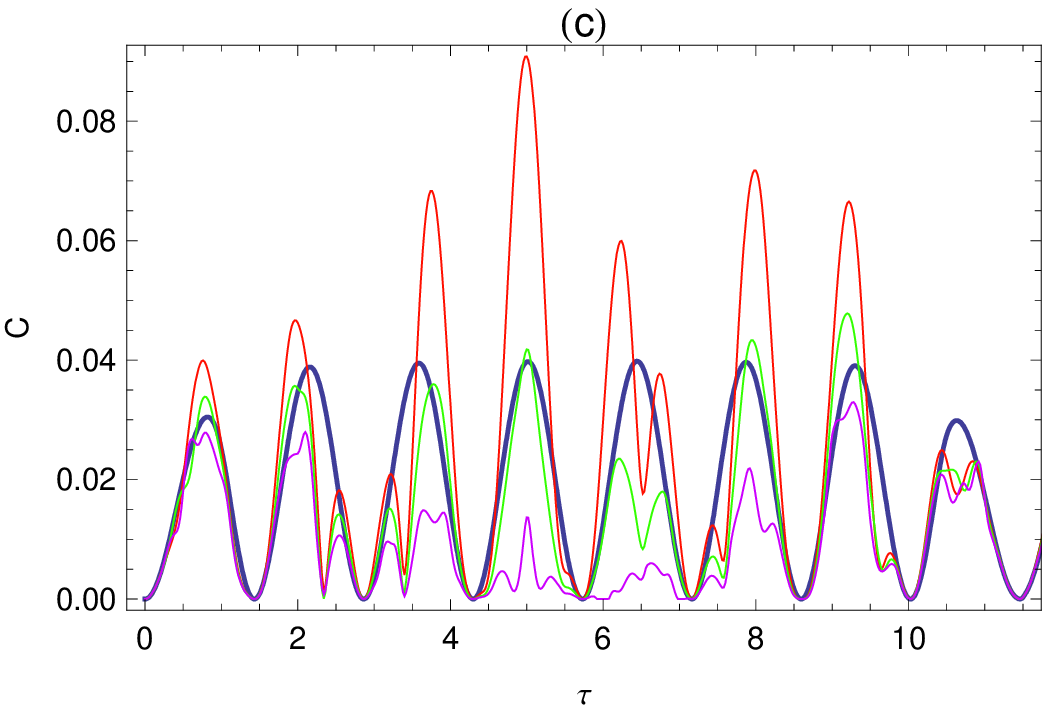}}
\hspace{0.3cm}
\subfigure{\label{Bmax1d}\includegraphics[width=6.5cm]{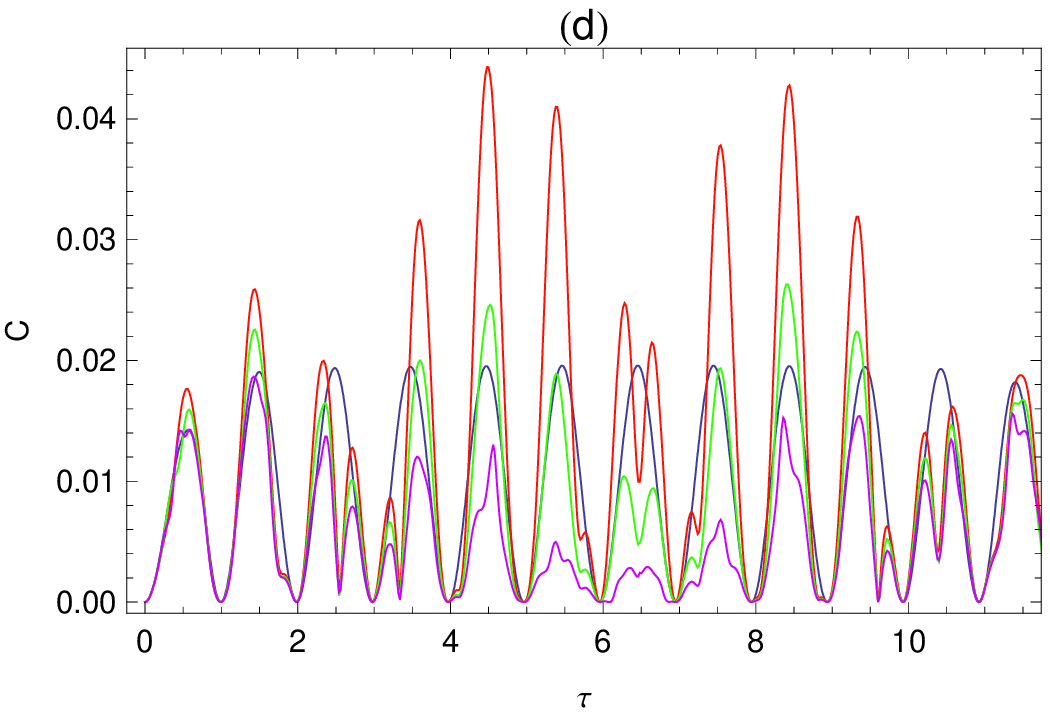}}
\hspace{0.3cm}
\subfigure{\label{Bmax1e}\includegraphics[width=6.5cm]{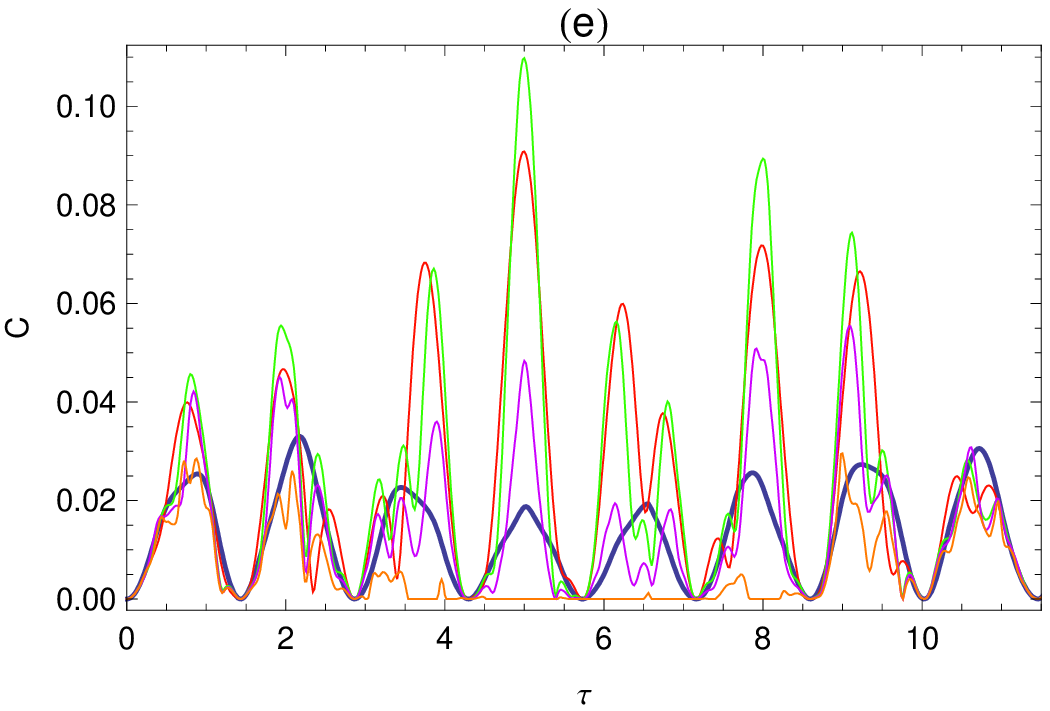}}
\hspace{0.3cm}
\subfigure{\label{Bmax1f}\includegraphics[width=6.5cm]{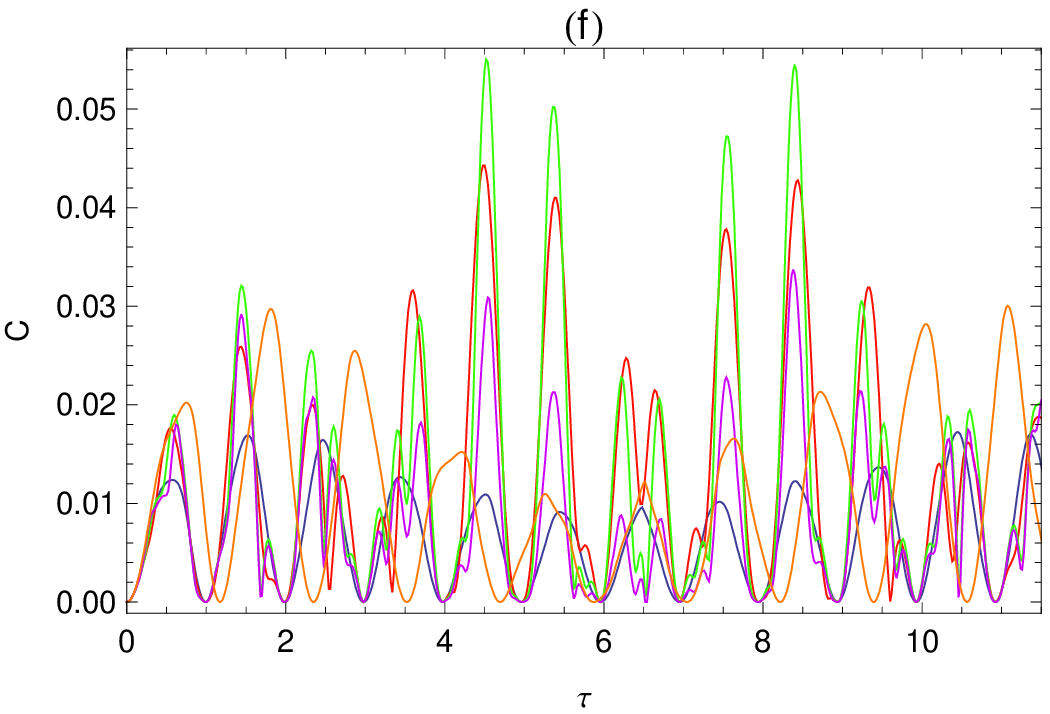}}
\end{center}
\caption{The concurrence of different neutron pairs for incoming neuron polarization $\alpha=\frac{\sqrt{3}}{2}$, $\beta=\frac{1}{2}$. Increasing values of $n$ are indicated by the blue, red, green, purple and orange curves, respectively. (a) The concurrence of neutrons $m$ and $m+1$ for $m\leq5$ and $B_z=3\lambda$. (b) As (a) for $B_z=4\lambda$. (c) $C_{2n}$ for $n=\lbrace1,3,4,5\rbrace$ and $B_z=3\lambda$. (d) As (c) for $B_z=4\lambda$. (e) $C_{3n}$ for $n=\lbrace1,2,4,5,7\rbrace$ and $B_z=3\lambda$. (f) As (e) for $B_z=4\lambda$. All quantities are expressed in natural units, with $\lambda=1$.}
\label{Bmax1}
\end{figure}

\begin{figure}[H]
\renewcommand{\captionfont}{\footnotesize}
\renewcommand{\captionlabelfont}{}
\begin{center}
\subfigure{\label{Bmax2a}\includegraphics[width=6.5cm]{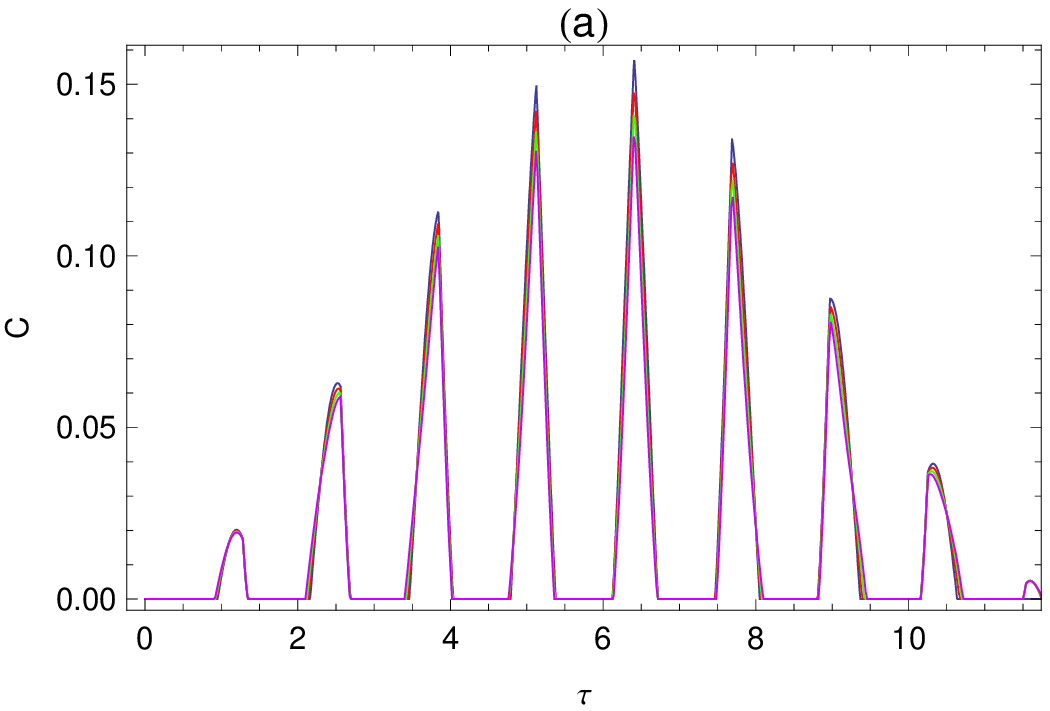}}
\hspace{0.3cm}
\subfigure{\label{Bmax2b}\includegraphics[width=6.5cm]{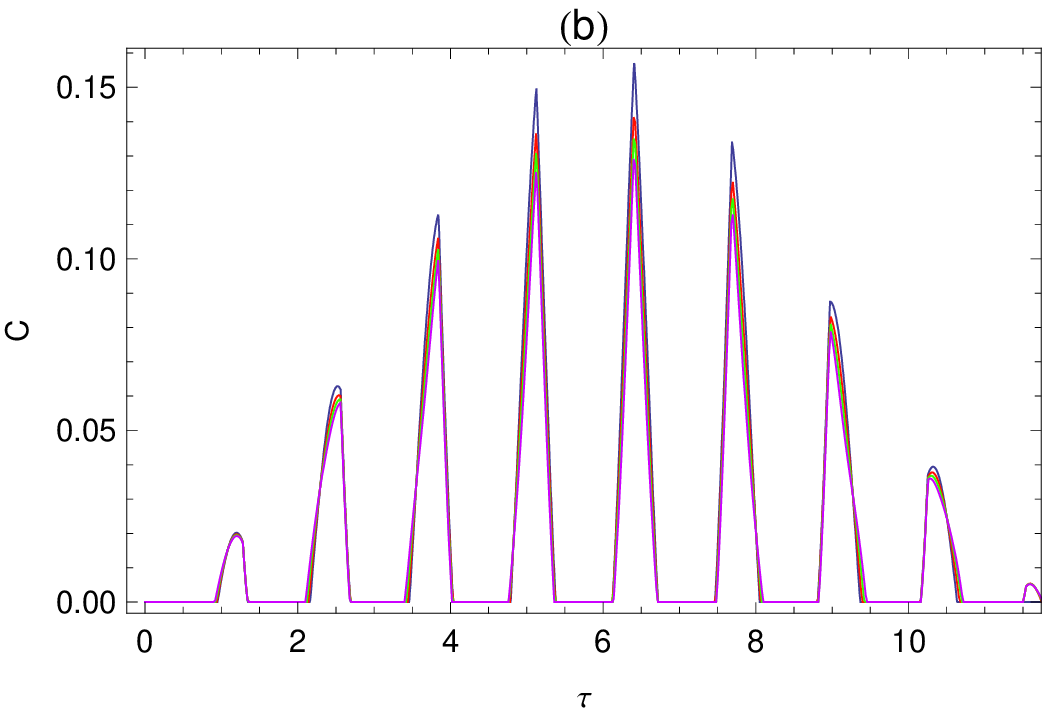}}
\hspace{0.3cm}
\subfigure{\label{Bmax2c}\includegraphics[width=6.5cm]{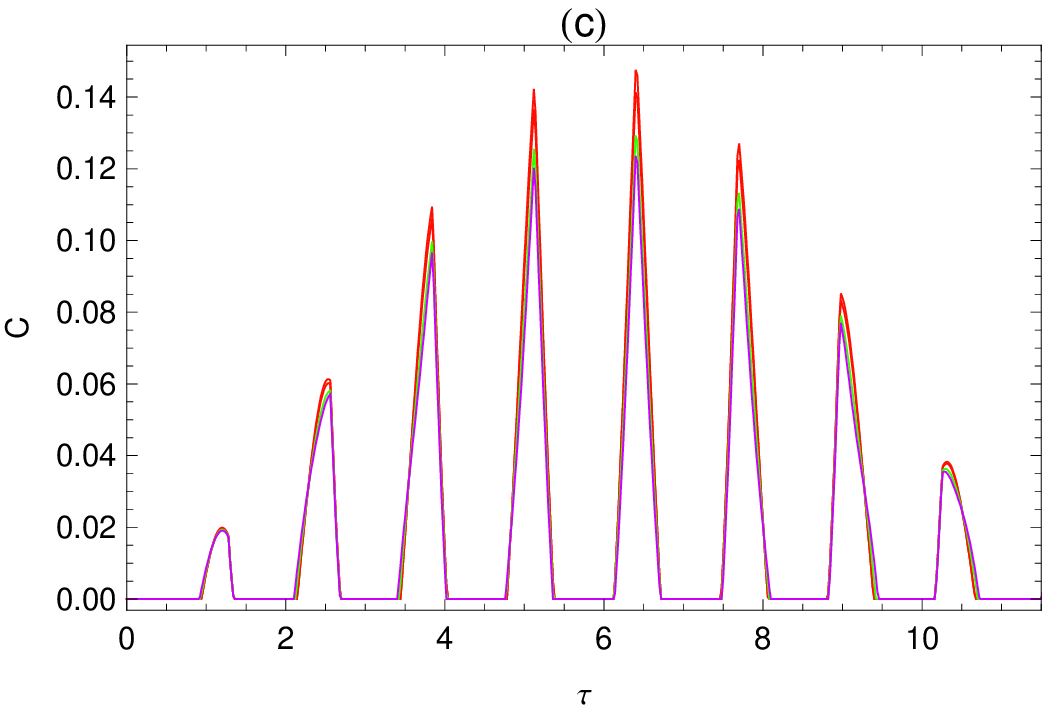}}
\hspace{0.3cm}
\subfigure{\label{Bmax2d}\includegraphics[width=6.5cm]{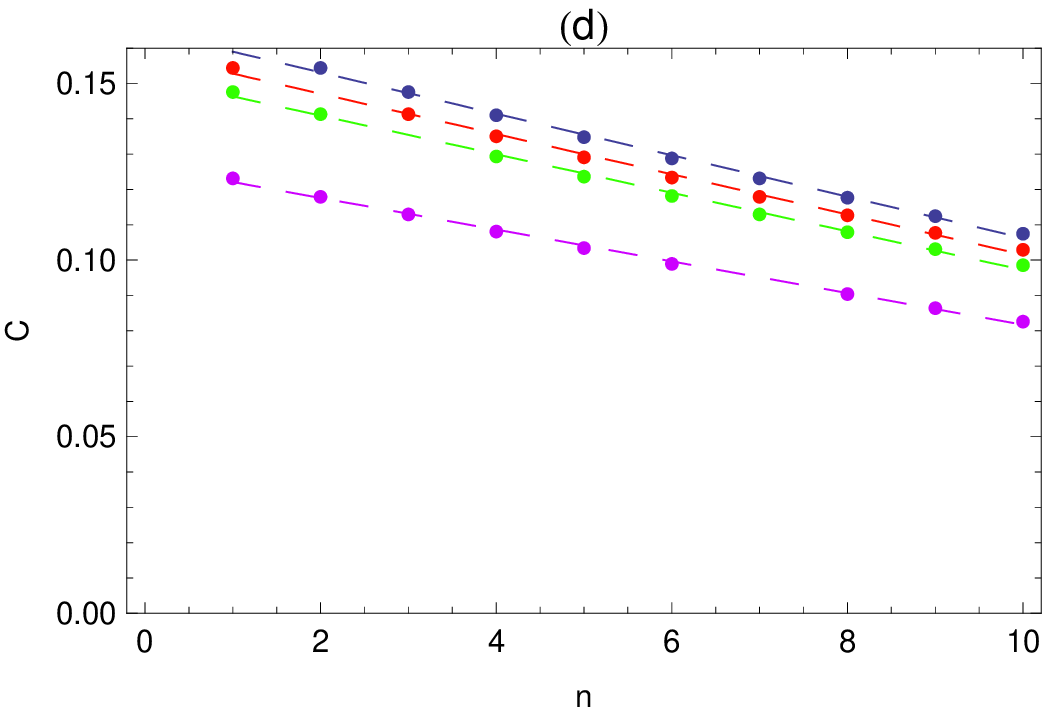}}
\end{center}
\caption{The concurrence for $B_z=3\lambda$ and $\alpha=0$. Increasing values of $n$ are indicated by the blue, red, green and purple curves, respectively. All quantities are expressed in natural units, with $\lambda=1$. (a) $C_{1,n}$. (b) $C_{2,n}$. (c) $C_{3,n}$. (d) The peak concurrence for $m=\lbrace 1,2,3,7\rbrace$ (blue, red, green and purple dots, respectively) at $\tau=6.4$. Dashed lines indicate the function $F(n)=an+b$, evaluated at $a=-0.006$ and $b=0.164$ (blue curve), $a=-0.005$ and $b=0.157$ (red curve), $a=-0.005$ and $b=0.152$ (green curve), and $a=-0.004$ and $b=0.127$ (purple curve).}
\label{Bmax2}
\end{figure}

If the field is strong, the concurrence is generally low - maximum values do not exceed 0.2. Excluding for the moment the case $\alpha=0$, for an arbitrary neutron polarization $C_{m,n}$ retains some traces of periodicity provided $m\leq2$. Otherwise, the concurrence is a disordered function of the interaction time. As in weak field, peak values of $C_{m,n}$ increase as $\alpha\rightarrow0$, but fall with $|m-n|$. However, is now extremely difficult to find instances of $\tau$ and $n$ for which the concurrence at fixed $m$ increases with respect to $C_{m,m\pm1}$. As opposed to $C_{1,2}$, if $m>1$ the concurrence of a given neutron pair can improve with respect to its zero-field value, even if $B_z$ exceeds the threshold $B_{z+}$. However, the concept of a limiting field still applies, as it is not possible to increase the peak concurrence indefinitely by making the field stronger (see figure \ref{Bmax1}).
As always, unusual features are observed at $\alpha=0$. In this case, $C_{m,n}$ retains the same periodicity as $C_{1,2}$ for all $m$ and $n$ (see figure \ref{Bmax2}). In analogy to its weak-field behaviour, the concurrence evolves through a series of peaks which, at fixed $m$, are aligned for different $n$. If $\tau$ is chosen to coincide with a peak, the concurrence at fixed $m$ falls with a very gentle gradient as $n^{-1}$. This gradient does not change significantly as $m$ is increased, therefore the concurrence of neutron $m$ with any other is generally a slowly-varying function of $n$. In other words, provided $|m-n|$ remains sufficiently small, the concurrence of any neutron pair is roughly constant (see figure \ref{Bmax2d}). Consequently, for a group of neutrons scattered in quick succession, the entanglement with respect to any bipartition is also approximately constant, and can be detected using the witness of equation \eqref{wit}.

\begin{figure}[H]
\renewcommand{\captionfont}{\footnotesize}
\renewcommand{\captionlabelfont}{}
\begin{center}
\subfigure{\label{Bopt1a}\includegraphics[width=6.5cm]{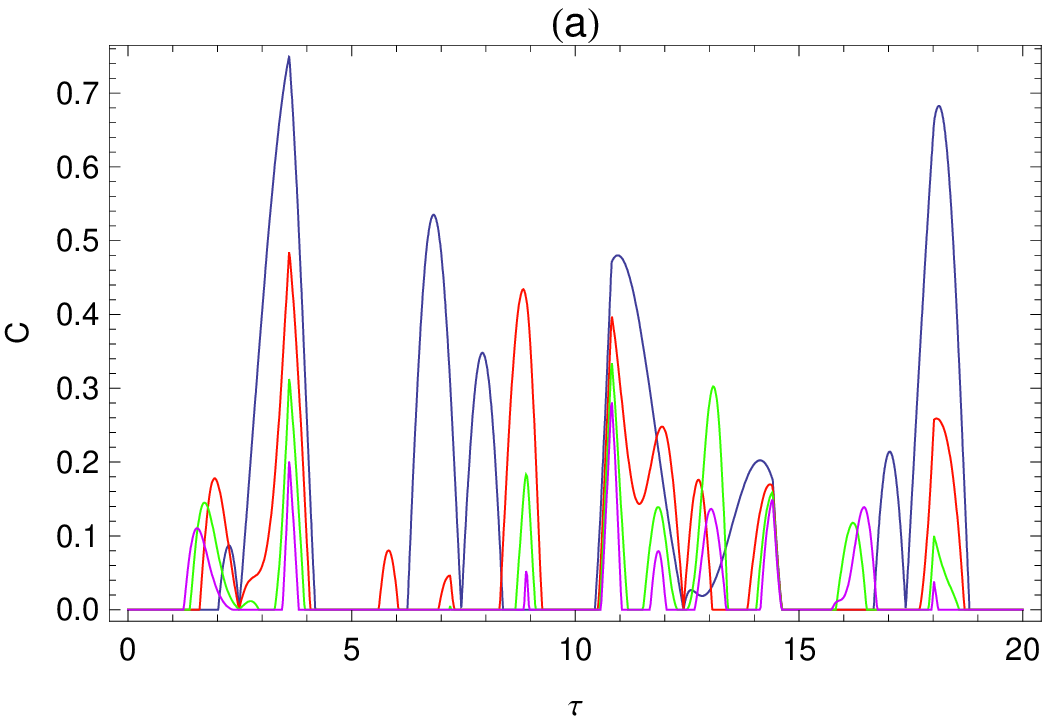}}
\hspace{0.3cm}
\subfigure{\label{Bopt1b}\includegraphics[width=6.5cm]{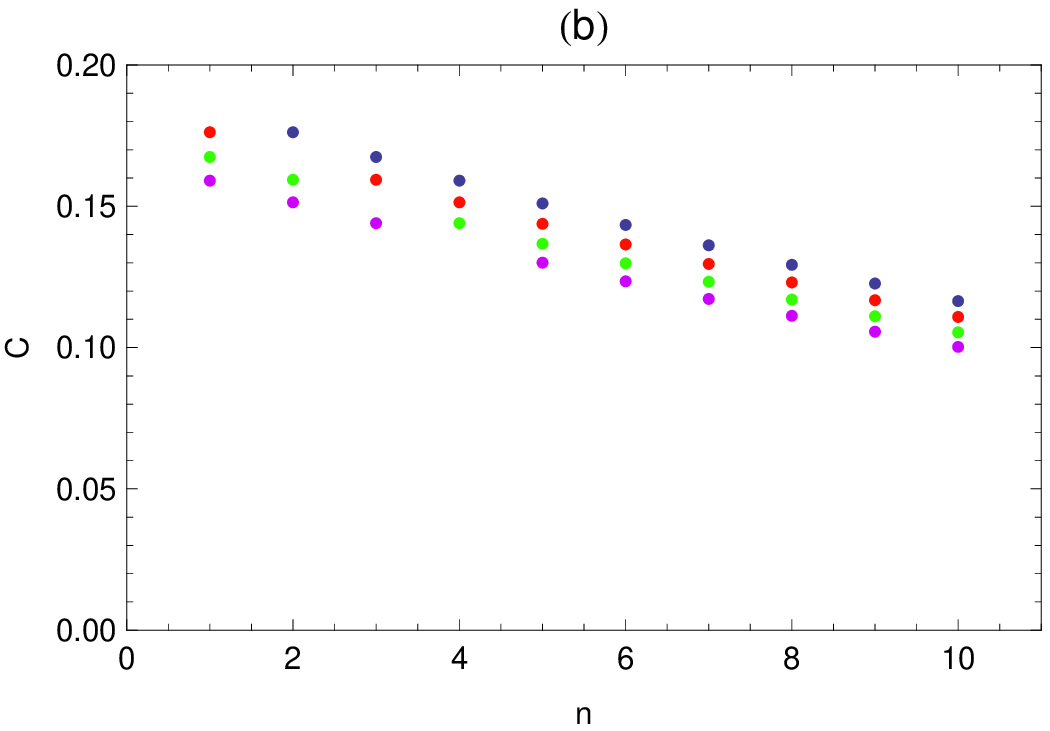}}
\hspace{0.3cm}
\subfigure{\label{Bopt1c}\includegraphics[width=6.5cm]{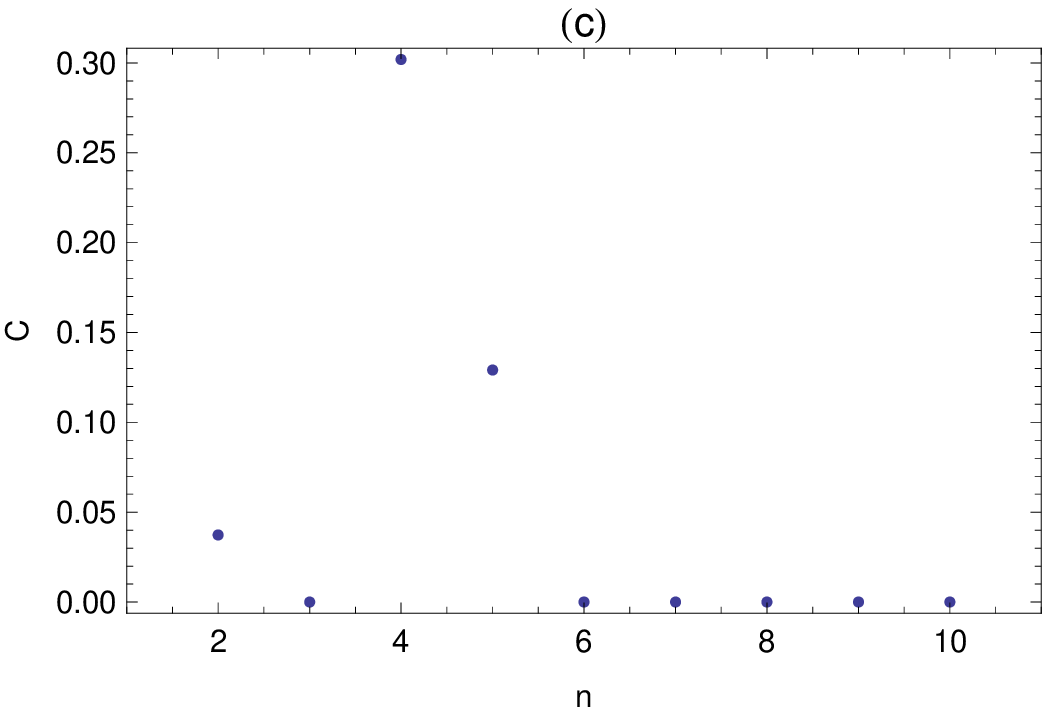}}
\end{center}
\caption{(a) The concurrence $C_{1,n}$ for $B_z=0.9\lambda$ and $\alpha=0$. Increasing values of $n$ are indicated by the blue, red, green and purple curves, respectively. All quantities are expressed in natural units, with $\lambda=1$. (b) The concurrence $C_{m,n}$ at $\tau=14.41$ for $m=1$ (blue dots), $m=2$ (red dots), $m=3$ (green dots), and $m=4$ (purple dots). Note the linear decrease with $n$ at fixed $m$. (c) $C_{1,n}$ at $\tau=13.1$. Note $C_{1,4}>C_{1,5}>C_{1,2}$.}
\label{B1opt}
\end{figure}

Finally, let us look to the properties of $C_{m,n}$ at $B_z=B_z^*\equiv\lambda\left(1-\frac{1}{N}\right)$. On the basis of our discussion so far, it is reasonable to assume the protocol will be most successful at $\alpha=0$, therefore I will focus on this limit. It was shown in section \ref{p0a zero mtm finite field} that the concurrence of the first two scattered neutrons is a maximum when the field is set to $B_z^*$. At general $m$ and $n$, this is no longer the case, because the concurrence of neutron $n$ with any neutron scattered before it will depend on the sample state it encounters. For the case of $C_{1,2}$, $B_z^*$ represents an optimum because it allows us to `deposit' entanglement in the system by creating a maximally entangled state of the first neutron and the sample. For all practical purposes, at this stage of the protocol the sample behaves like a two-level system; it is therefore conceivable that the second scattering event might swap - at least in part - the state of the second neutron and the state of the system. After all, the entanglement has few other places to go. As the sample acquires more spin-flips, several effects come into play. First, the entanglement we deposit acquires new channels over which to re-distribute itself. Second, the sample ceases to behave like a qubit, and becomes a \textit{bona fide} multilevel system. At that point, even if neutron $n$ were to encounter a fully entangled state of neutron $m$ and the sample, the protocol could only succeed if most of the sample wavefunction were concentrated in the single-excitation subspace. As the number of events grows, the chances of this happening become very slim. Therefore, in the context of this more general version of the protocol, $B_z^*$ simply defines an `intermediate' field regime. As one might expect, the concurrence in this regime shows a mixture of strong- and weak-field features. Among the former, we note the existence of interaction times for which the concurrence of any neutron pair is linearly dependent on $|m-n|$ [see figure \ref{Bopt1b}]. Among the latter, we find instances in which the concurrence of a non-consecutive pair is greater than that of two neutrons scattered in succession [see figure \ref{Bopt1c}].

In summary, if the sample is prepared in a fully polarized state, the concurrence of neutrons $m$ and $n$ falls with the number of scattering events between them. In some cases, it is possible to predict how $C_{m,n}$ scales with $|m-n|$, however the evolution of the concurrence is generally unpredictable, unless either the applied field is zero, or the neutrons are polarized in the negative $\hat{\bm{z}}$-direction. In the former case, the concurrence of any neutron pair is low; in the latter, the performance of the protocol depends on the strength of the applied field. Best results are obtained in the weak-field regime, where the concurrence at fixed $m$ scales roughly as $n^{-\frac{1}{2}}$. Maximum values of the concurrence of any non-consecutive pair do not exceed 0.5, hence in this respect a realization with input state $|\psi^B\rangle$ is equivalent to a realization with input state $|\psi^A\rangle$. What we lose is the ability to predict with certainty when the concurrence of a given pair will peak, although from the behaviour of $C_{m,n}$ with $\tau$ (cfr. figure \ref{Bmin4}) one could formulate a rough guess. Unfortunately, given the non-analytical form of $C_{m,n}$ in weak field, the margin of error is rather narrow.

\section{Average yield of Entangled Pairs}
I have shown so far that the entangling protocol of chapter \ref{neutron proposal} can be effective for a range of neutron indices $m$ and $n$, on the proviso that $|m-n|$ is not large compared to the number of spins in the sample, $N$. To assess the true performance of the protocol in a realistic situation, any reference to the indices of the scattered neutrons must now be removed. The question to ask becomes: will a pair of neutrons chosen at random from those which have scattered be entangled?

The results of sections \ref{mn psi0a} and \ref{mn psi0b} indicate that the entanglement of two neutrons $m$ and $n$ decreases with $|m-n|$. Let us therefore assume we apply a resetting pulse to the sample whenever the number of scattered neutrons approaches $N$, such that $|m-n|<N$. I introduce the quantities $t$ and $\bar{C}$ to represent the number of scattered neutrons on which measurements are performed, and the concurrence of the average state of any pair, respectively. For ease of comparison, I will compute the average concurrence relative to the situations presented in figures \ref{psia0f2a}, \ref{psia0f2b}, \ref{psiaf3b}, \ref{psiaf3d}, \ref{psiaf3e}, \ref{Bmin4}, \ref{Bmax2}, and \ref{B1opt}. These all correspond to optimal realizations of the protocol at fixed $N$ and $B_z$, either with input state $|\psi_0^A\rangle$ (first five cases) or with input state $|\psi_0^B\rangle$ (latter three cases). As expected, the average concurrence falls with the size of the sampling ensemble. However, some slightly surprising features are observed, according to the choice of input state and the strength of the applied field.

\begin{figure}[H]
\renewcommand{\captionfont}{\footnotesize}
\renewcommand{\captionlabelfont}{}
\begin{center}
\subfigure{\label{psia0av1}\includegraphics[width=6.5cm]{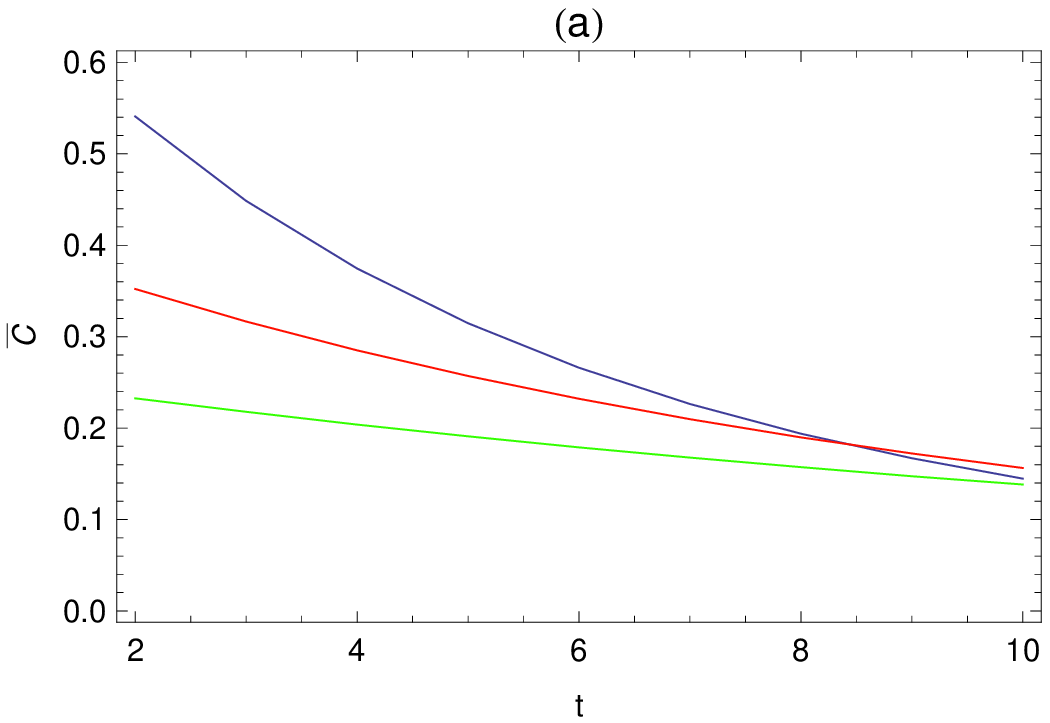}}
\hspace{0.3cm}
\subfigure{\label{psia0av2}\includegraphics[width=6.5cm]{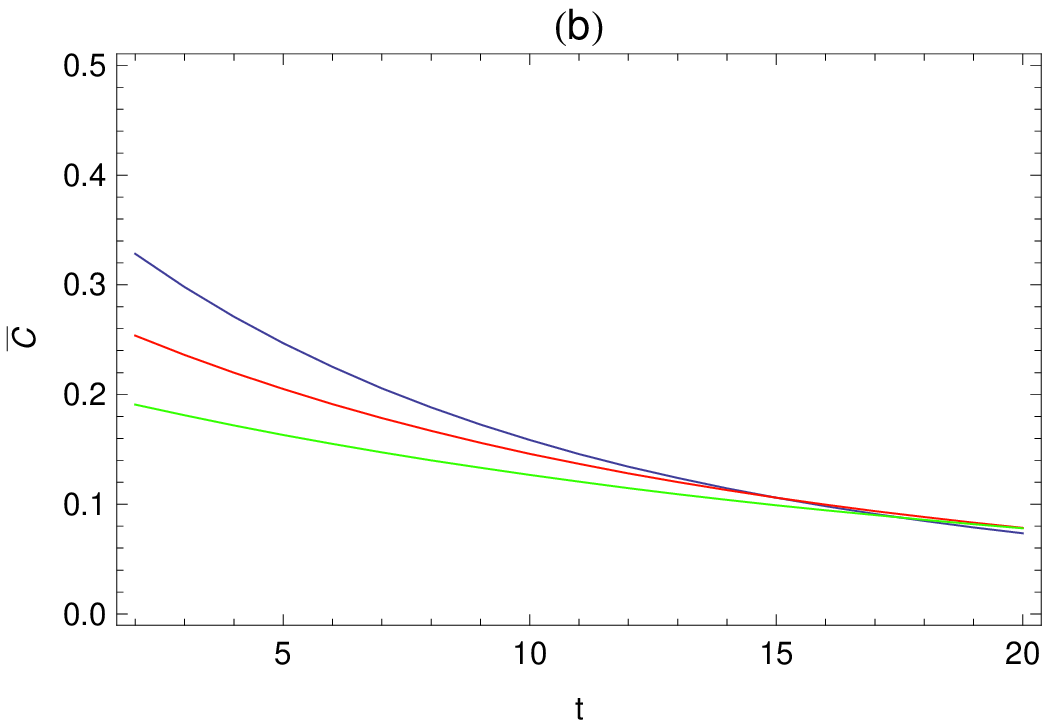}}
\end{center}
\caption{Input state $|\psi_0^A\rangle$. (a) The average concurrence of any neutron pair within the range $\lbrace m,n\rbrace\in\left[1,10\right]$ for $N=10$ and $B_z=0$, sampled at times $\tau=T_{\phi}/2$, $0.8$ and $0.6$ (blue, red and green curves, respectively). (b) The average concurrence of any neutron pair within the range $\lbrace m,n\rbrace\in\left[1,20\right]$ for $N=20$ and $B_z=0$, sampled at times $\tau=T_{\phi}/2$, $1$ and $0.8$. All quantities are expressed in natural units, with $\lambda=1$.}
\label{psia0av}
\end{figure}

\begin{figure}[H]
\renewcommand{\captionfont}{\footnotesize}
\renewcommand{\captionlabelfont}{}
\begin{center}
\subfigure{\label{psiafav3}\includegraphics[width=6.5cm]{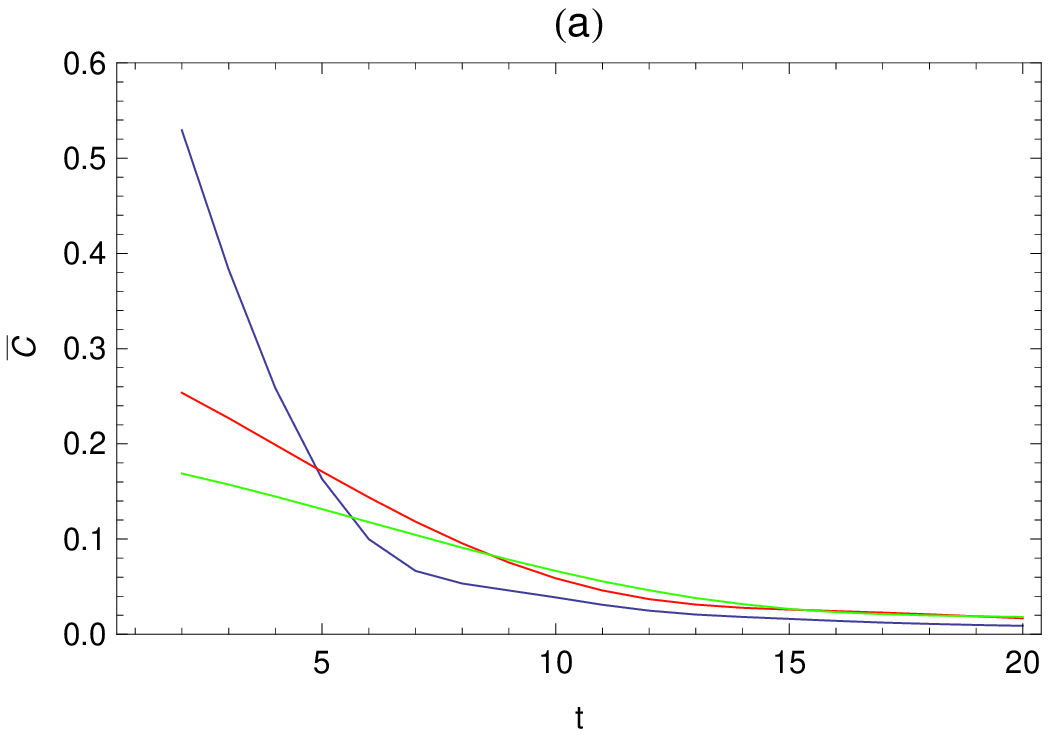}}
\hspace{0.3cm}
\subfigure{\label{psiafav4}\includegraphics[width=6.5cm]{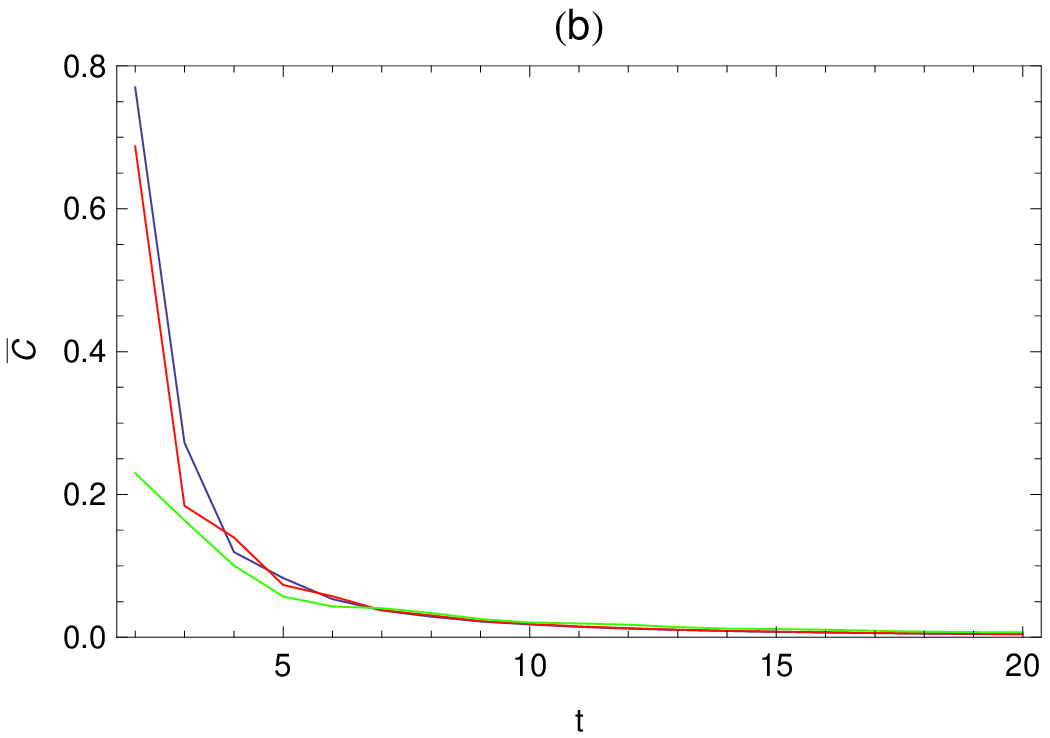}}
\hspace{0.3cm}
\subfigure{\label{psiafav5}\includegraphics[width=6.5cm]{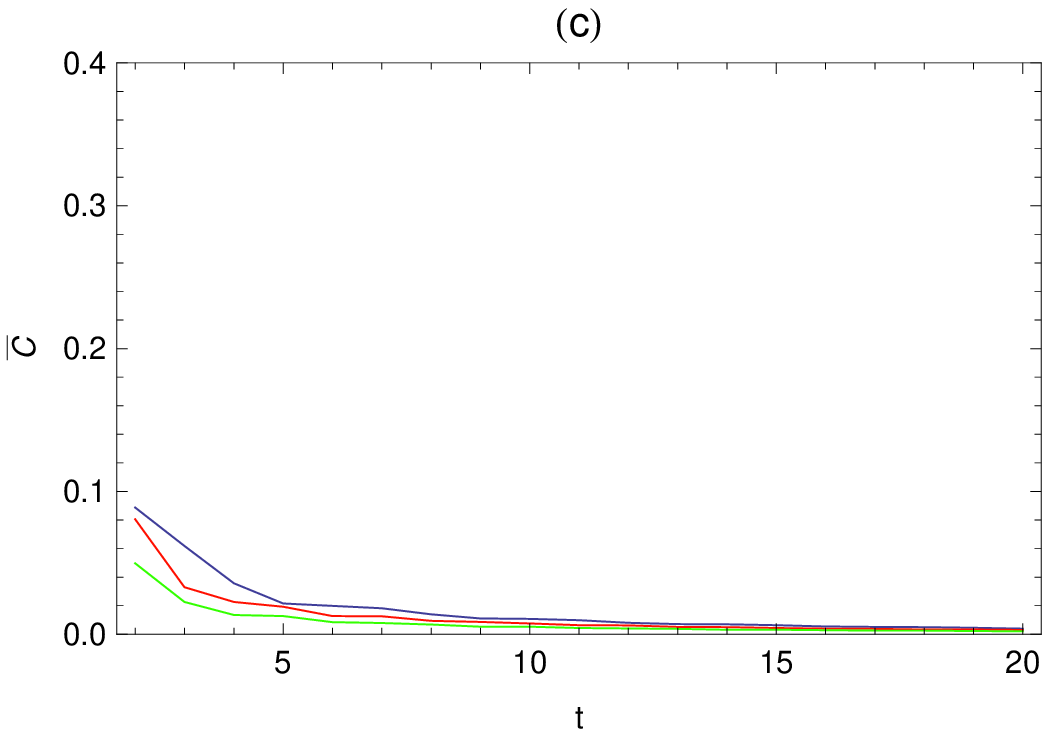}}
\end{center}
\caption{Input state $|\psi_0^A\rangle$. The average concurrence of any neutron pair within the range $\lbrace m,n\rbrace\in\left[1,20\right]$ for $N=20$. Values of $B_z$ and $\tau$ are as follows: (a) $B_z=0.3\lambda$, $\tau=T_{\phi}/2$, $0.9$ and $0.7$ (blue, red and green curves, respectively). (b) $B_z=0.9\lambda$, $\tau=\tau_B^{1,2}$, $1.7$ and $0.8$. (c)  $B_z=3\lambda$, $\tau=T_{\phi}/2$, $0.6$ and $0.4$. All quantities are expressed in natural units, with $\lambda=1$.}
\label{psiafav}
\end{figure}

If the sample is prepared in a single-magnon state (input state $|\psi_0^A\rangle$), the likelihood of detecting an entangled pair is strongly dependent on the value of $B_z$. If $B_z$ is zero, $\bar{C}$ is a slowly-varying function of $t$ - in fact, at certain values of $\tau$ the $\bar{C}$ vs. $t$ curve shows an almost linear trend (see figure \ref{psia0av}). Consequently, the average concurrence remains finite for an appreciable range of $t$, and shows signs of converging to a lower limit as $t\rightarrow N$. As the field is increased, the relationship between $\bar{C}$ and $t$ changes rather dramatically, approaching a near-exponential decay in the intermediate field regime [figure \ref{psiafav4}]. By the time $B_z$ reaches $B_{z+}$ the curve has somewhat flattened out, but values of $\bar{C}$ are extremely low, and fall to zero by $t=5$. Best results are therefore achieved at zero or small fields.

Conversely, if the sample is prepared in a fully polarized state, it makes little sense to consider an average concurrence \emph{unless} a field is applied. This is due to the fact that, in zero field, the concurrence is extremely low for all $m$ and $n$. The optimal operating mode for the system now seems to be the weak field regime, in which, for an optimal choice of $\tau$, the average concurrence falls as $t^{-\frac{1}{2}}$. As above, the situation deteriorates as the field is raised, and in the intermediate and strong field limits the behaviour of the system is similar to that observed for input state $|\psi_0^A\rangle$. We conclude that, for either input state, the probability of a measurement on two randomly chosen chosen neutrons yielding an entangled state is substantial, subject to a set of experimental conditions. First, the neutron interaction time must be chosen appropriately. Second, the field must remain below the threshold $B_{z-}$. Third, if input state $|\psi_0^B\rangle$ is chosen, the neutron spin must be antiparallel to the quantization axis.

\begin{figure}[H]
\renewcommand{\captionfont}{\footnotesize}
\renewcommand{\captionlabelfont}{}
\begin{center}
\subfigure{\label{avp1}\includegraphics[width=6.5cm]{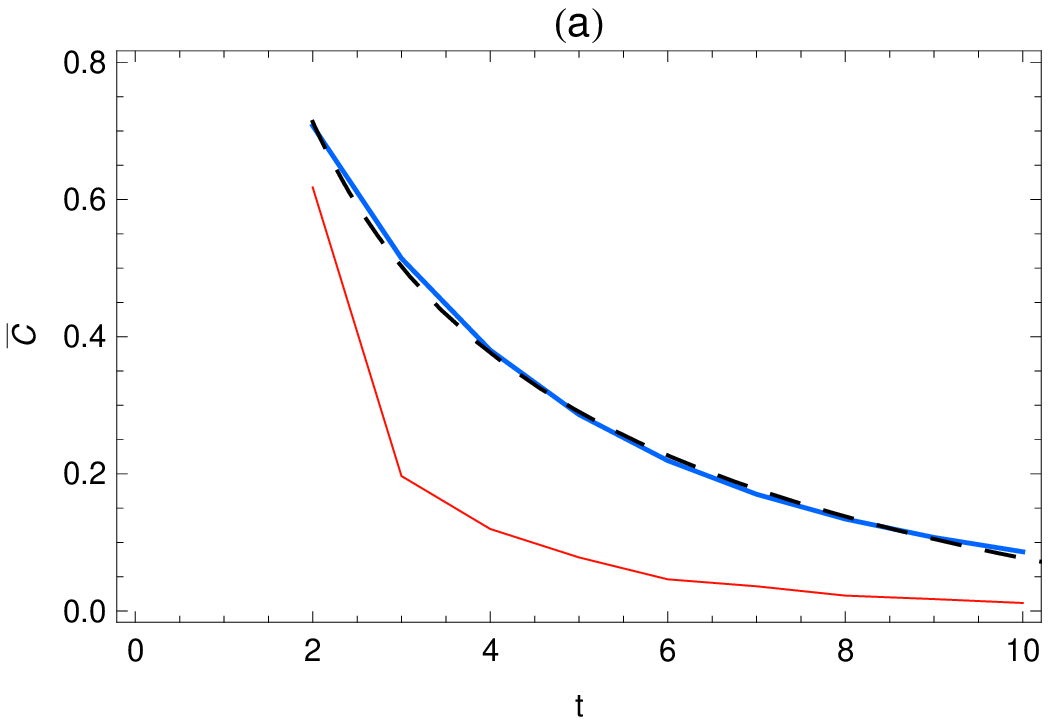}}
\hspace{0.3cm}
\subfigure{\label{avp2}\includegraphics[width=6.5cm]{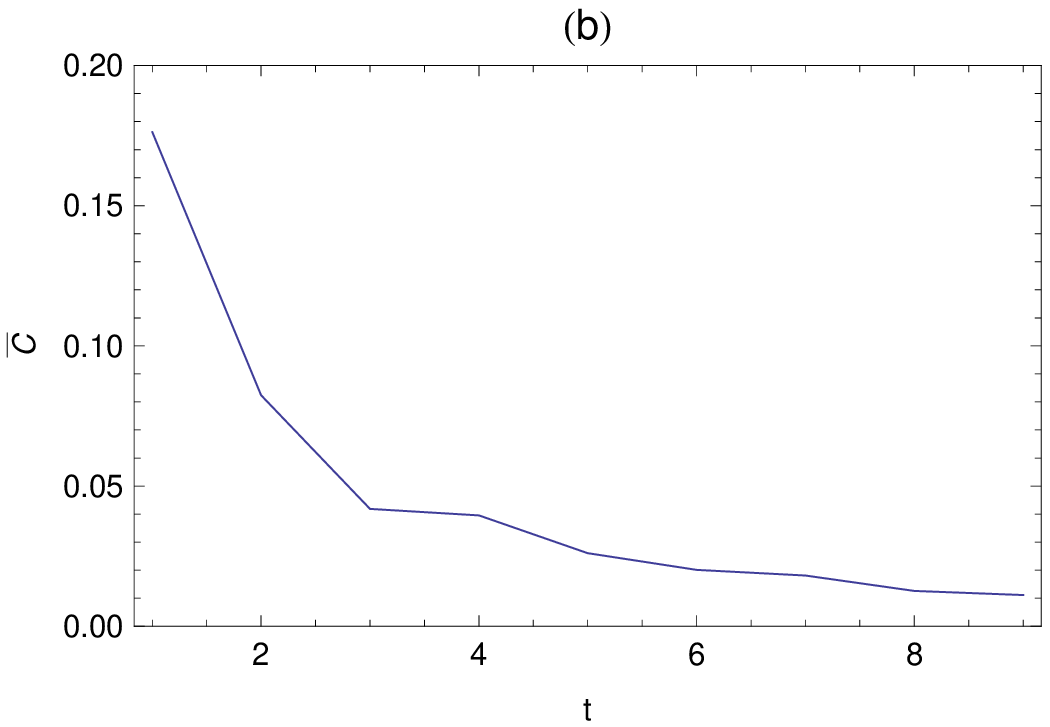}}
\hspace{0.3cm}
\subfigure{\label{avp3}\includegraphics[width=6.5cm]{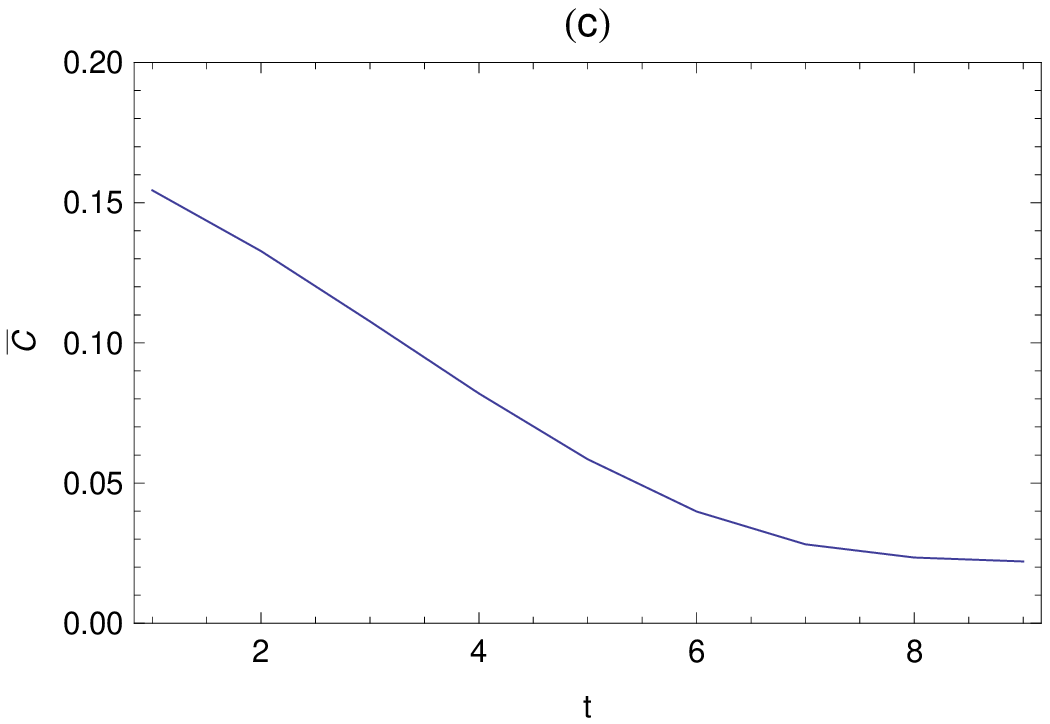}}
\end{center}
\caption{Input state $|\psi_0^B\rangle$. The average concurrence of any neutron pair within the range $\lbrace m,n\rbrace\in\left[1,10\right]$ for $N=10$. Values of $B_z$ and $\tau$ are as follows: (a) $B_z=\lambda/3$, $\tau=17$ and $13.6$ (blue and red curves, respectively). (b) $B_z=0.9\lambda$, $\tau=14.41$. (c) $B_z=3\lambda$, $\tau=6.4$. The black dashed line in figure (a) corresponds to the function $F(t)=at^{-\frac{1}{2}}+b$, with $a=1.63$ and $b=-0.44$. All quantities are expressed in natural units, with $\lambda=1$.}
\label{pav}
\end{figure}

\section{Conclusions}
It was shown in chapter \ref{two ns} of this thesis that two neutrons consecutively scattering from a sample prepared in a specific eigenstate of its internal Hamiltonian could come to share a substantial degree of entanglement, subject to an appropriate choice of the experimental parameters. Extending this protocol to include the ---more realistic--- possibility of multiple scattering events, it emerges that any two neutrons scattered within a certain time frame of each other can be measurably entangled in a similar fashion. The absolute value of this entanglement is generally low, and falls well short of the threshold $\mathcal{C}_{\mathcal{E}}=0.67$ defined in section \ref{exp feasibility}. However, the scheme remains of great interest from a fundamental viewpoint, providing as it does a simple, practical means to answer the as-yet unresolved question: can two distinct neutrons be entangled?

In general, the outcome of the protocol is determined by the number of scattered neutrons relative to the number of scatterers. Provided this ratio remains small, there is a strong possibility that measurements made on \emph{any} two scattered neutrons will yield an entangled state. Optimal performance can therefore be achieved by periodically resetting the sample to its initial state. In these circumstances, the average concurrence of any neutron pair is of order 0.2, corresponding to just under 0.1 ebits of entanglement. This can be boosted either by specifying which neutron pair to measure, or by increasing the frequency of the resetting pulses. Distillation may also be a possibility, but only if one could ensure detection of the same neutron pair subsequent to each re-initialization of the sample state.

A more detailed analysis of the distribution of entanglement within the scattered neutron bunch further shows the concurrence is finite with respect to several bi-partitions. Provided the number of scattered neutrons is small, one therefore creates genuine multi-partite entanglement. However, this particular aspect of the protocol has not been explored in great detail.

The experimental requirements of the scheme were assessed in section \ref{exp feasibility}, where it was concluded that measurement of the witness of equation \eqref{wit} may be extremely challenging, but not \textit{a priori} impossible. This is reinforced by the results of the present chapter. Therefore, the only outstanding problems remain the neutron velocity and the neutron coherence length. The former issue was discussed in detail in the previous chapter. As regards the neutron coherence volume, we recall this quantity depends on the degree to which a beam can be monochromated. A perfect crystal monochromator can achieve velocity resolutions of $\Delta v/v=10^{-5}$ \cite{mattoni04}, but for our purposes this is not enough. Higher resolutions may be achievable by selecting neutrons from the beam with the aid of an extremely high frequency chopper. This has not been reported to date, but should not be impossible, as there exists no fundamental lower limit to $\Delta v/v$. Hence, it is reasonable to conclude that meeting the final requirement of the protocol may simply be a matter of time.

%\begin{thebibliography}{9}
%\bibitem{mattoni04}
%C.E.H. Mattoni \textit{et al.}, Physica B \textbf{344}, 343 (2004).
%\end{thebibliography}
%\end{document} 

%% file: spin_chain_review.tex
%\documentclass[a4paper,12pt]{article}
%\usepackage{setspace}
%\usepackage{graphicx}
%\usepackage{subfigure}
%\usepackage{caption}
%\usepackage{color}
%\usepackage{amsmath}
%\usepackage{amssymb}
%\usepackage{caption}
%\usepackage{bm}
%\usepackage{float}
%\DeclareGraphicsExtensions{.epsi,.eps,.ps}
%
%\doublespacing
%\def\mathbi#1{\textbf{\em #1}}
%\def\hh{{\mathfrak h}}
%\begin{document}
%%\tableofcontents

\chapter{The Problem of Scalability and the Need for Communication}\label{spin chain review}
\textit{In this chapter, I introduce the topic of quantum state transfer with spin chains. The use of a spin chain as a communication channel was first suggested by Bose \cite{bose03}, in a bid to find an efficient means of transferring quantum states over short distances in solid-state devices. It was found that unmodulated spin chains are natural conduits of quantum information, but do not always allow high quality communication. Many improvements to the basic communication protocol have since been suggested, and these will be summarized. I will then explain the original communication scheme in detail, and review its performance. Possible experimental implementations will also be discussed, ending with a general overview of spin chain communication as a quantum computational technique.}

\section{Introduction}\label{spin chains part}
At the very beginning of this thesis, it was observed that communication represents the sharing of a correlation. In practice, this correlation arises when information flows from one party to another via some kind of mediating channel. There exist a wide variety of channels capable of transmitting classical information, but those same channels are not in general adequate for quantum state distribution. Indeed, it has been shown that transferring quantum information through a classical channel involves a loss of at least a third of the original information \cite{horodeki99}. This is unacceptable for protocols such as teleportation or quantum key distribution, which require faithful exchange of qubits. Currently, the most common means of transferring quantum information is the optical fibre, a choice born of necessity as most large-scale quantum computational tasks attempted to date have employed photons as qubits. This is down to the photon's low interaction cross-section, which enable it to remain in a coherent state over large distances both in air and in optical fibres. However, in recent years the problem of quantum state transfer in solid-state devices has attracted a great deal of attention; in fact, it has been suggested this may be the key to building quantum computers large enough to rival existing processors.

Just as the computing power of a classical machine is directly related to the number of transistors on its Central Processing Unit (CPU), the computing power of a quantum computer is directly proportional to the number of qubits it can access.
It has been calculated that a $10^{6}$-qubit quantum computer running with a clock speed of 100 MHz could, in just
one hour, solve a factorization problem that would keep the most
powerful existing supercomputer busy for a time comparable to the
age of the universe. Unfortunately, this is at least a million times larger than the machines
currently within our ability to build \cite{berggren04,preskill98}.

Scalability is therefore a key requirement for a feasible quantum computer architecture. Up-scaling current designs is not a straightforward task, for two main reasons. First, many-qubit systems acquire extra degrees of freedom, which are difficult to control and provide a path for spurious interactions with the environment. As a result, the system becomes more vulnerable to decoherence and, consequently, a less reliable quantum computer. A second problem relates to the implementation of two-qubit gates between distant qubits. To achieve this, the qubits must interact either directly or by a common bus mode, but neither approach is without its problems. If we aim for direct couplings, the qubits in question (or, alternatively, the information they encode) must be shuttled around the system until they are close enough to interact, and then returned to their original positions. This is clearly inefficient, and would slow the calculation down considerably. On the other hand, if one employs a common bus mode there is a physical limitation to the number of qubits that can be coupled. These issues become increasingly problematic as the size of the computer increases, thus rendering the construction of a many-qubit `processor unit' an extremely challenging task.

\section{A Review of Quantum Communication with Spin Chains}\label{review section}
How else could quantum processors be scaled up, if not by simply adding extra qubits? The past few years have witnessed the advent of a `divide and conquer' philosophy. This entails building a \emph{modular} processor composed of several interconnected few-qubit registers, which could either act as parallel processors, or perhaps have different functions (one register could be the CPU, another the memory etc.).

Such a `collaborative' approach clearly requires fast, efficient communication between the registers. The best known and most widely used carriers of quantum information are photons, because they travel naturally and are easily isolated from their environment. Photons can be successfully propagated over distances of hundreds of kilometers through optical fibres, free space - even through \emph{outer} space, as recently demonstrated by a team of researchers led by Zeilinger \cite{hiskett06,zeilinger07,zeilinger08}. One might then suggest constructing a simple quantum computer by connecting a series of solid-state processors by a mesh of optical fibres, but until very recently this approach seemed impractical, owing to the difficulty of interfacing photonic qubits with solid-state devices. It has now been shown that strong coupling can be achieved in a variety of systems \cite{reithmaier04,majer07}, which may open up the possibility of exploring this avenue. The option of a quantum computer with `all optical' components has been looked into, but such designs are complex and often involve non-linear processes, which are very challenging to implement
experimentally (see \cite{KLM01,obrien03,roadmap optical qc04} and references therein).

Recently, much effort has been dedicated to studying systems in
which quantum information is encoded in \emph{stationary} qubits and propagated from one part of the system to another by the natural
interaction between the system's components. One of the simplest
geometries in which this can be achieved is a one-dimensional chain
of interacting particles, where the qubit is encoded in the particles' spin degree of freedom. This is usually referred to as a \emph{spin chain}. The original proposal for quantum communication through spin chains was put forward by Bose \cite{bose03}, who studied a ferromagnetic, exchange-coupled chain of spins with a constant nearest-neighbour
interaction, in an external magnetic field. It was found that information
could be transferred with a fidelity exceeding the maximum classical
value in a time that grows polynomially with the length of the
chain. The importance of this result lies in the fact that the transfer protocol requires \emph{no external control} whatsoever. All the experimentalist must do is encode a quantum state into the first spin of the chain, wait for a well-defined amount of time and measure the last spin of the chain. Depending on the chain length, the fortunate experimentalist will have recovered a more or less perfect copy of the initial state without having to intervene in the transfer process in any way.

It was subsequently shown that slightly more complex systems allowed better performance. Osborne and Linden \cite{osborne04} demonstrated the viability of transmission fidelities close to unity if the qubit
is taken to be a carefully designed `wave packet', provided the
sending and receiving parties can access a sufficiently large
portion of the chain. In the absence of structural
imperfections, an XY Hamiltonian on a hyper-cubic lattice allows
perfect state transfer \cite{christandl04,christandl05,dechiara05},
as do parallel spin chains \cite{burgarth05}. Arbitrarily perfect state transfer can also be
achieved by applying a sequence of two-qubit gates at the receiving
end of a quantum chain \cite{burgarth07}, or by employing auxiliary
local memories \cite{he06}. Plenio \textit{et al.} \cite{plenio05}
have studied the situation for chains of harmonic oscillators (i.e.
where each particle on the lattice possesses a continuous, rather
than a discrete, degree of freedom), while Hartmann \textit{et al.}
\cite{hartmann05} have recently found that quantum information can
be made to propagate with arbitrarily high fidelity through both
oscillator and spin chains near a quantum phase transition. This transfer is exponentially slow; more rapid
transmission is possible at the quantum critical point, but at some
cost to the fidelity. High fidelities can also be attained by
engineering the strength and the nature of the interactions between
the spins \cite{yung05,yung06}, or by performing well-defined projective measurements on parts of the chain \cite{difranco08,monras08}. Perfect state transfer is even achievable in disordered or imperfect systems, and without state initialization, although extra resources or error correction may be required \cite{difranco08,allcock06,burrell08}.

The vast majority of these protocols assume only short-range couplings between the spins in the system. More recently, Kay
\cite{kay05} has made a detailed study of finite spin chains in which the total
Hamiltonian accounts for a long-range
magnetic dipole potential. It was found that state transfer can be optimized by fine-tuning the inter-particle spacings and applying local magnetic fields at each spin site.

Some effort has also been dedicated to examining the effect of thermal fluctuations on the transfer process \cite{bayat05}. For short chains, there exist analytical expressions which demonstrate that thermal fluctuations affect the fidelity of the transfer, but not its speed. For longer chains, the problem rapidly becomes intractable because the dimensions of the Hilbert space of the system blow up exponentially as the number of qubits increases. In most state transfer protocols suggested to date, this difficulty is avoided by choosing Hamiltonians with built-in symmetries, which allow one to work in a Hilbert space that grows linearly rather than exponentially in the number of spins (see section \ref{exch h}). These symmetries cannot be exploited if dissipative processes such as thermally induced excitations are present, therefore thermal effects are often neglected. This is not necessarily an unrealistic approximation, as it equates to considering thermal energy scales $kT$ which are negligible compared to the strength of, say, an external magnetic field or the inter-qubit couplings. One could certainly engineer such a situation in a laboratory environment.

\section{The Communication Protocol}\label{comm prot}
We now review the communication protocol originally suggested by Bose \cite{bose03}. The present discussion deals with a linear chain geometry, however the scheme is valid for an arbitrary graph of qubits.

A sending party, who in the name of tradition we call Alice, wishes to transmit an unknown quantum state to a receiving party, surprisingly named Bob, who is located at the opposite end of the chain. The transfer can be achieved with minimal effort following four very simple steps:
\begin{enumerate}
\item The chain is initialized to its ground state.
\item Alice prepares a qubit in the state she wishes to transmit and replaces the first spin in the chain with this qubit.
\item The chain is left to evolve for a finite amount of time under the effect of its internal Hamiltonian.
\item Bob measures the last spin in the chain and recovers a copy of the state originally prepared by Alice.
\end{enumerate}

The key stage of this protocol is point 3: the information encoded by Alice is transferred to Bob spontaneously by the inter-spin couplings.
As we will see in the following sections, information has flowed from Alice's end of the chain to Bob's. The chain therefore acts as a channel for  quantum information, albeit not a perfect one, as the state recovered by Bob is rarely an exact copy of that sent by Alice. This outcome is easily understood if one imagines Alice's quantum state as a wavepacket of finite width, initially centred on the first spin of the chain. The wavepacket spreads as it travels, hence the information it contains is no longer localized but \emph{dispersed} over several spins. To some extent, this effect is  related to the length of the chain: shorter chains often perform better as the initial wavepacket has less ``room'' to spread out. As a result of dispersion, when Bob measures the end spin he will only recover part of the original wavepacket, consequently receiving an incomplete sample of the information Alice sent.

To optimize the quality of the information he receives, Bob must measure his spin at the optimal moment. One might expect this to be the time at which the peak of the initial wavepacket reaches Bob's end of the chain; however, this is not always the case. Indeed, it is often advantageous to wait a little longer, thus allowing the wavepacket to propagate back and forth through the chain several times. As it is reflected from the ends, it interferes with itself and may experience a partial revival at the site of Bob's spin. If Bob is quick, he can measure at this time, but the occurrence of such revivals is unfortunately somewhat difficult to predict.

The realization of the protocol rests on three assumptions. First, it must be possible to initialize the chain to a known state, which in this case corresponds to the ground state. In a ferromagnetic chain, this is simply achieved by applying a strong magnetic field to polarize the spins, and is therefore not a great challenge. Second, Alice and Bob must be able to perform measurements on their spins in order to encode and extract the information. This is considerably trickier, owing to the difficulty of performing spin-selective measurements on the distance scales involved, though possible test systems have been suggested \cite{bose03}. Finally, these measurements must take place on a timescale shorter than the spin-lattice relaxation time $T_1$, as both the encoding and the decoding operations rely on the chain being in a specific state.

\subsection{The Hamiltonian}\label{exch h}
The great strength of the protocol we have just described lies in the fact that Alice and Bob do not need to intervene in the transfer process in any way: the state is transferred spontaneously by the interaction between the spins in the chain. In this respect, the transfer process can be thought of as the solid-state analogue of photon transmission through a vacuum. The protocol therefore exploits a natural property of the system, as opposed to the vast gamut of quantum computational tasks which require the experimentalist to keep a tight rein on any degrees of freedom their system may exhibit. One feels almost as if the much-fabled - and so far elusive - free lunch were within close reach.

The characteristics of the transfer process are governed by the Hamiltonian of the spin chain. Bose's original proposal sees the spins coupled by a ferromagnetic nearest-neighbour exchange interaction and immersed in an external magnetic field, such that the overall Hamiltonian of the system reads
\begin{equation}\label{h0}
H_0=-\sum_{\langle ij \rangle}J_{ij}\:\bm{\sigma}^i \cdot \bm{\sigma}^j-\sum_{j=1}^NB_j\mathbf{\sigma}_z^j,
\end{equation}
where the first sum runs over all nearest-neighbouring pairs $\langle ij \rangle$ and the second over all \textit{N} spins in the chain. Both $J_{ij}$ and $B_j$ are chosen to be positive numbers, hence the ground state of this Hamiltonian is fully polarized in the positive $\mathbf{\hat{z}}$ direction. Excitations from this ground state are represented by de-localized spin-flips, often described in terms of spin waves \cite{vanvleck58}.

The choice of Hamiltonian is determined by two main factors. First, it is physically realistic, as it accurately describes the energy level structure of many ferromagnets \cite{ashcroft}. Second, it conserves the number of excitations in the system; indeed, if the total magnetization is proportional to the number of spin flips, one finds
\begin{equation}\label{commutator1}
\left[H_0,\mathbi{S}_z\right]=0,
\end{equation}
where $\mathbi{S}_z$ is the sum of the individual spin components in the $\hat{\mathbi{z}}$-direction
\begin{equation}
\mathbi{S}_z=\sum_{j=1}^N \bm{\sigma}_z^j.
\end{equation}
As a result of (\ref{commutator1}), the dimensions of the Hilbert space needed to to describe the system is drastically reduced, as $\mathrm{dim}\mathcal{H}=\binom{N}{m}$, where \textit{m} is the number of excitations. In the simplest case of $m=1$, the dimensions of $\mathcal{H}$ grow linearly with the length of the chain. It is therefore possible to solve even very large systems analytically, an impossible task if one were obliged to work with the full Hilbert space.

\subsection{Notation}\label{notation}
Before proceeding with a more quantitative description of the communication protocol, a comment on notation is required. In keeping with the Bloch sphere representation of eigenkets of the Pauli spin operators, it has so far been assumed that states $|0\rangle$ and $|1\rangle$ represent the spin up and down eigenstates of $\sigma_z$, respectively. The convention adopted in \cite{bose03} reverses this labeling. Therefore, to avoid possible confusion, I will henceforth label the eigenstates of $\sigma_z$ explicitly as $|\uparrow\rangle$ and $|\downarrow\rangle$. The ground state of the chain can then be represented as $|\downarrow_1\downarrow_2\downarrow_3...\downarrow_N\rangle$, where the subscripts refer to the positions of the spins; I abbreviate this state as $|\Downarrow\rangle$. The ket $|\mathbi{j}\rangle$ will indicate a state with a spin flip on site $j$, the ket $|\mathbi{jk}\rangle$ a state with spin flips at sites $j$ and $k$, and so on. It will be assumed Alice encodes her information in the spin at position $s$ (standing for sender), and Bob chooses to measure the spin at position $r$ (standing for receiver). In all but very few cases it will be assumed that $s=1$ and $r=N$.

\subsection{The Fidelity as a Measure of Success}
With these tools in place, the steps of the protocol can be reiterated as follows:
\begin{enumerate}
\item The chain is initialized by applying a magnetic field and waiting until the ground state $|\Downarrow\rangle$ is attained.
\item Alice removes the first spin of the chain and replaces it with her encoded qubit $|a\rangle=\alpha|\downarrow\rangle+\beta|\uparrow\rangle$ to create the state $|\psi_i\rangle$
\begin{center}$|\psi_i\rangle=|a\rangle_1|\downarrow_2...\downarrow_N\rangle=\alpha|\Downarrow\rangle+\beta|\mathbf{1}\rangle$.\end{center}
\item The chain is left to evolve for time \textit{t} to output state $|\psi_f\rangle$ under the effect of (\ref{h0}):
\begin{center}$|\psi_f\rangle=\alpha|\Downarrow\rangle+\beta e^{-iH_0t}|\mathbf{1}\rangle=\alpha|\Downarrow\rangle+\beta \sum_{j=1}^N\langle\mathbi{j}|e^{-iH_0t}|\mathbf{1}\rangle|\mathbi{j}\rangle$.\end{center}
\item Bob measures the spin at his end of the chain and obtains the state $|b(t)\rangle$:
\begin{center}$|b(t)\rangle=\frac{1}{\sqrt{P(t)}}\left[\alpha |\downarrow\rangle + \beta f(t)|\uparrow\rangle\right]$,\end{center}
with \begin{center}$f(t)=\langle\mathbi{N}|e^{-iH_0t}|\mathbf{1}\rangle$\end{center}
\begin{center}$P(t)=|\alpha|^2+|\beta|^2|f(t)|^2.$\end{center}
\end{enumerate}
The function $f(t)$ is a propagator, and represents the transition amplitude from state $|\mathbf{1}\rangle$ to state $|\mathbf{N}\rangle$.
If the protocol is successful, at some time $t_0$ Bob will find $f(t_0)=1$, hence $|b(t_0)\rangle=|a\rangle$. The value of the propagator is therefore related to the quality of the transfer, and can be employed to measure of the success of the protocol via a quantity known as the  \textit{fidelity}
\begin{equation}\label{fidelity}
F(\rho,\rho^\prime)\equiv\left(Tr\sqrt{\rho^{\frac{1}{2}}\rho^\prime\rho^{\frac{1}{2}}}\right)^2,
\end{equation}
where $\rho$ and $\rho^\prime$ are the initial and final density operators. The fidelity evaluates how faithfully a quantum channel maps an input state $\rho$ to an output state $\rho^\prime$. It is a real-valued function spanning the interval $[0,1]$, which is non-zero provided there is some overlap between the input and output states, and unity if $\rho=\rho^\prime$. The definition (\ref{fidelity}) applies to both pure and mixed states, but when dealing with pure states it is more convenient to introduce a second quantity, known as the averaged fidelity
\begin{equation}\label{av fid}
\bar{F}=\frac{|f(t)|}{3}+\frac{|f(t)|^2}{6}+\frac{1}{2},
\end{equation}
where the average is calculated over all pure input states of the Bloch sphere. There are several other measures of the performance of a quantum channel, but these are more complex and will not be described here. As the present discussion is limited to pure states, the term `fidelity' should hereafter be understood as referring to (\ref{av fid}); definition (\ref{fidelity}) is not used anywhere in this work.

If nearest neighbour couplings are constant and the applied field is uniform, it is possible to derive simple expressions both for the spectrum of (\ref{h0}) and for the fidelity. Setting $J_{ij}\equiv \left(\frac{J}{2}\right)_{\delta_{i+1,j}}$ and $B_j \equiv B$, we quote the result obtained by Bose in \cite{bose03}
\begin{eqnarray}
|\mathbi{m}\rangle&=&a_m\sum_{j=1}^N\cos\left[\frac{\pi}{2N}(m-1)(2j-1)\right]|\mathbi{j}\rangle\label{m states}\\
E_m&=&2B+2J\bigg\lbrace1-\cos\left[\frac{\pi}{N}(m-1)\right]\bigg\rbrace\label{em s},
\end{eqnarray}
where $m=1...N$, $a_1=\frac{1}{\sqrt{N}}$, $a_{m\neq1}=\sqrt{\frac{2}{N}}$, and the ground state energy $E_0=-\frac{J(N-1)\:+\:2BN}{2}$ is set to zero. The propagator between sites $1$ and $N$ of the chain can then be expressed as
\begin{equation}\label{propagator}
f(t)=\sum_{m=1}^N\langle\mathbi{N}|\mathbi{m}\rangle \langle\mathbi{m}|\mathbf{1}\rangle e^{-iE_mt}.
\end{equation}
To calculate the fidelity, this is numerically evaluated and substituted into equation (\ref{av fid}).

\subsection{The Performance of the Protocol}
\begin{figure}[H]
\centering
\resizebox{9cm}{!}{\includegraphics{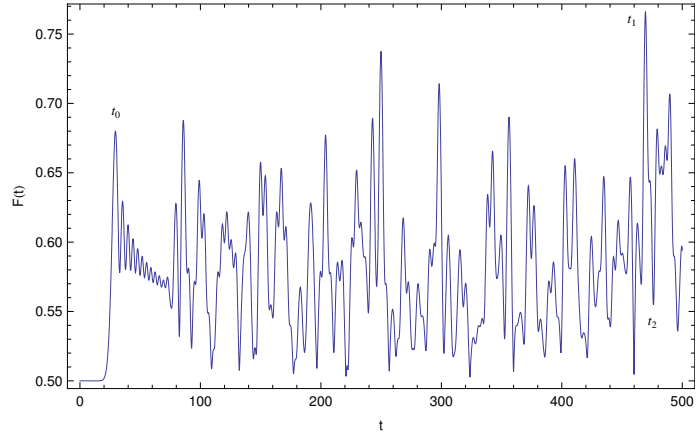}}
\renewcommand{\captionfont}{\footnotesize}
\renewcommand{\captionlabelfont}{\sffamily}
\caption{\label{fid_exch}The fidelity of state transfer $F(t)$ for a chain of length 27. In this example we have set $J=\frac{1}{2}$ and $B=0$. }
\end{figure}

For arbitrary chain length (with the exception of $N=2$), the evolution of the fidelity as a function of time is disordered and largely unpredictable (fig. \ref{fid_exch}). For $N<30$, with few exceptions, time-optimized fidelities exceed 0.8, but the channel becomes progressively worse as the length of the chain increases. Nevertheless, the optimal fidelity at $N=80$ is still greater than that achievable with a classical channel \cite{bose03,horodeki99}. The presence of several high `spikes' in the fidelity is encouraging, as it suggests there are several times at which Bob could measure his spin and obtain a reasonably good copy of Alice's input state. However, it also implies Bob has a very short window of opportunity in which to make his measurement. For the example shown in figure \ref{fid_exch}, for instance, if Bob measures his spin at time $t_1$ he will find state $|a\rangle$ with over $75\%$ fidelity; if, on the other hand, he measures at the not too distant time $t_2$ he will obtain a much poorer result.

One might expect the optimal measurement time to coincide with the time $t_0$ taken by the wavepacket to reach the end of the chain. Roughly speaking, this can be approximated as the timescale of a single nearest-neighbour coupling multiplied by the length of the chain. From the uncertainty principle
\begin{equation}\label{t0}
t_0\approx\frac{N-1}{2J},
\end{equation}
where $1/2J$ is the timescale of a single coupling, $N-1$ is the length of a chain of $N$ spins with unit spacing, and $h$ is set to 1.

The same result is obtained from equations \eqref{m states} and \eqref{em s}, recalling that a localized spin-flip can be written as a superposition of spin waves with different momenta
\begin{equation}\label{spin wave}
|\bm{j}\rangle=\sum_{m=1}^Na_m\cos {k_m}|\bm{k}_m\rangle,
\end{equation}
where I have replaced the argument of the cosine with $k_m$ to make explicit the momentum dependence of the $j$-states. Given the dispersion relation of equation \eqref{em s}, the group velocity of the individual spin waves
\begin{equation}\label{d omega}
v^g=\frac{\mathrm{d}\omega}{\mathrm{d}k_m}=2J\sin k_m,
\end{equation}
is itself a function of momentum. It then becomes clear that wavepacket of equation \eqref{spin wave} spreads because each component of the superposition propagates at a different velocity. As observed in \cite{osborne04}, spreading the spin flip over several sites counters the effect by reducing the number of momenta involved in the summation. The fastest-propagating component of \eqref{spin wave} will reach the end of a chain of length $L$ in time
\begin{equation}\label{t0 spin wave}
t_0=\frac{L}{2J}\equiv\frac{N-1}{2J},
\end{equation}
which is the result of equation \eqref{t0}.

The value of $t_0$ for a chain of $27$ spins with $J=\frac{1}{2}$ is shown in figure \ref{fid_exch}. We see that, although a local maximum is attained at this time, there are several instances with $t>t_0$ at which the fidelity of state transfer is greater than at $t_0$.

There have been many interesting studies regarding the relationship between the optimal fidelity and the length of the chain. It was already noted in \cite{bose03} that the fidelity seemed lower for chains whose length was divisible by 3. Subsequently, it was shown that chains of length $N>4$ cannot achieve unit fidelity \cite{christandl05}, and also that if $N$ is a prime number the fidelity approaches unity in exponential time \cite{burgarth thesis}. This may sound enticing, but ultimately one cannot afford to wait so long to achieve perfect state transfer. We will return to this point in section \ref{conclusions1}.

\section{Decoherence}
The discussion so far has been centred on the performance of an \emph{isolated} spin chain. However, to obtain a realistic description of the performance of the protocol in a realistic physical setting, interactions of the spin chain with the environment should be considered. With few exceptions \cite{giovannetti05,burgarth06,kay07}, the vast majority of existing spin chain protocols neglect the problem of \textit{decoherence}, and the work presented in this thesis is no exception. However, recent studies regarding the influence of dissipation and decoherence on the transfer process \cite{cai06,zhou06} allow us to estimate how several of the protocols described above may perform in less idealized settings.

The decoherence mechanisms most relevant to quantum information processing are dephasing and damping \cite{nielsen and chuang}. Dephasing takes place when interaction with the environment destroys the phase relationships between the components of a quantum state. It is a conservative interaction, because no energy is exchanged between the system and the environment, and the timescale on which it occurs is usually referred to as $T_2$. Damping, on the other hand, affects the populations of the eigenstates of the local Hamiltonian. Typical examples might be the relaxation to the ground state of a sample prepared in some high-energy configuration, or the appearance of thermal spin excitations in a perfectly ordered ferromagnetic state when the temperature is raised above zero. Damping is therefore a dissipative process, because energy is exchanged with the environment. The timescale associated with damping is usually referred to as $T_1$.

A detailed study of the effect of $T_1$ and $T_2$ processes on spin chain communication has recently been carried out in \cite{cai06}. Here, the authors assume two possible modes of interaction of the spin chain with its environment. The `global' mode involves each spin in the chain coupling to a single, `block' environment, whereas in the `local' mode each spin interacts with a local independent environment. In both cases, effects on the fidelity of state transfer are strongly dependent on the internal Hamiltonian of the spin chain. For a Heisenberg Hamiltonian, the fidelity in the presence of global noise is determined by the transfer time and the so called-environment parameter $\nu$, which quantifies the relative strength of spin-spin and spin-environment coupling. For a chain of given length, the quality of the transfer becomes steadily worse as $\nu$ increases. For a given $\nu$, the quality of the transfer is inversely proportional to the length of the chain, because the longer the chain, the longer it takes for the state to be transferred, hence the longer the coupling time to the environment. Similar results are found for the local environment model.

Interestingly, the situation changes if one considers the mirror-periodic Hamiltonian of \cite{christandl04}. In this model, in the absence of decoherence, the time at which perfect state transfer is achieved is constant. For a global dephasing or local damping environment, this yields an average state transfer fidelity which is independent of chain length, but decays with increasing environment parameter. On the other hand, in a local dephasing environment, the transfer fidelity falls both with increasing $\nu$ and with increasing chain length.

The effect of a pure local damping environment is analogous to that of a spin bath at zero temperature \cite{nielsen and chuang,carvalho04}. Recent work has been dedicated to studying the effects of thermal fluctuations in more detail. For short spin chains, the problem has been solved analytically \cite{bayat05}. It emerges that finite temperature effects can influence the fidelity of state transfer, but not the timescale. Recently, more detailed results have been obtained for longer chains globally coupled to a heat bath via dissipative exchange interactions \cite{zhou06}. The authors consider state transfer of a spinless fermion in a tightly bound lattice, which is analogous to an $XY$ spin model. At zero temperature, the fidelity of state transfer is found to decay exponentially with time. At fixed temperature, the same exponential decrease is observed; at fixed time, however, the performance of the system shows an interesting dependence on the coefficients of the state one sends. Specifically, for a state $\alpha|0\rangle+\beta|1\rangle$, the transfer fidelity decreases with temperature if $|\alpha|>|\beta|$, but increases with temperature if $|\alpha|<|\beta|$. However, this trend is not particularly strong, and becomes negligible at large times.

The above discussion allows us to conclude that, in all realistic systems, decoherence can seriously impair the efficiency of state transfer. The effect is especially pronounced when the transfer time is long, which is increasingly the case as the length of the chain grows. Strategies to combat decoherence might then include more robust state-encoding mechanisms or faster transfer times. The latter objective could perhaps be achieved by engineering nearest-neighbour couplings. In some systems, this might only be possible on a global scale; in others, however, locally tunable interactions may be a realistic possibility (see Section \ref{experiments1}).

\section{Experimental Quantum Communication with Spin Chains}\label{experiments1}
In the years immediately following Bose's proposal, research in quantum communication with spin chains was largely biased towards the theoretical. Practical achievements are now emerging, as progress in instrumentation begins to offer increased freedom in the systems we use as qubits. The majority of proposals for experimental implementation of existing communication schemes are centred on superconducting qubits. It has been shown that both charge and flux qubits allow high-fidelity quantum state transfer, even in the presence of manufacturing defects and decoherence \cite{romito05,lyakhov05}, though no experimental evidence has yet been presented. Quantum state mirror-inversion in a chain of three spins has recently been demonstrated in the field of NMR, which is well-suited to proof-of-principle trials, but fails for large systems, owing to the limitations of the state-preparation procedure \cite{fitzsimons06}. Promising candidates are also `carbon peapods', single-walled carbon nanotubes containing spin qubits trapped in fullerene cages \cite{luzzi98}. The properties of the peapod are strongly dependent on the nature of the trapped spin. Sc@C$_{60}$ peapods, for example, can be accurately modeled as Heisenberg antiferromagnetic spin chains with tunable nearest-neighbour couplings, as the spin qubit resides mainly on the surface of the fullerene cage \cite{ge07}. Conversely, in systems such as N@C$_60$, the spin is well-screened. Therefore, exchange couplings are more limited, but other mechanisms, such as dipolar couplings can come into play. In any case, the spin qubit is largely unaffected by the presence of the nanotube, which only serves as structural support, and a buffer against environment-induced decoherence. Atoms in coupled optical cavities have also been considered; recent results show these can simulate anisotropic $XXZ$ or isotropic $XXX$ chains of spin qubits with $s\geq \frac{1}{2}$ \cite{angelakis08}. Given an optimal cavity, all necessary qubit control is provided by laser fields; however, obstacles to practical implementation include decoherence effects and the requirement of strong atom-cavity coupling. Heisenberg and Ising Hamiltonians, as well as more exotic couplings, can also be realized in optical lattices \cite{duan03}. The idea here is that atoms can be made to interact controllably by adjusting the trapping potential. Several possible implementations have been suggested, however these are also not without their difficulties \cite{garcia04}. Finally, several proposals centred on quantum dot architectures have been advanced. These hold much promise, as it is now possible to grow regular two- and three- dimensional quantum dot arrays of controllable size and density, potentially allowing single-qubit addressability \cite{kiravittaya05}. Interactions between dots are provided by the F\"{o}rster effect, through which an excited donor dot transfers an exciton to an un-excited acceptor via an electric dipole coupling \cite{nazir05}. Previous work identified the main difficulties with the quantum dot approach in short decoherence times and manufacturing defects \cite{damico05}; however, recent findings suggest these difficulties can be largely overcome \cite{spiller07}, and with some luck, practical tests will soon bear out this result.

\section{Conclusions}\label{conclusions1}
Quantum communication with spin chains is currently one of the most active areas of research in quantum information processing. It has been widely shown that an isolated spin chain can afford versatile, high quality distribution of arbitrary quantum states with minimal control requirements, though optimal performance sometimes requires less flexibility over the operating conditions. A general survey of the communication protocols suggested to date indicates that perfect state transfer is achievable in a variety of different ways, provided one can engineer the geometry and the energetics of the system, and has access to ancillary registers. It is generally observed that spin chains allow more robust state transfer than sequences of $SWAP$ operations, because the use of a permanently-coupled system minimizes the possibility of accumulating errors. The issue of state-transfer time has also been raised, because an optimal protocol must allow for both quality and speed. Transfer times do not always bear an obvious relationship to the size of the system, though in several cases a rising trend is observed. In absolute terms, the transfer time is determined by the inter-spin couplings, which in a typical ferromagnet may be of order $10$ meV. We will see in the following chapter that for the original communication protocol of \cite{bose03} carried out with a chain of 20 spins, this might yield an optimal measurement time of order $10^{-15}$ s [this is shown in the following chapter, in figure \ref{Fidtimech}(a)]. Compared to the timescale of a single $SWAP$ gate, which hovers about the pico-second mark in a double quantum dot structure \cite{spiller07,petta05}, this figure is clearly advantageous; however, other protocols may not perform as efficiently. Practical demonstrations of spin chain communication are few, but recent advancements in experimental methods and instrumentation suggest the situation is poised to change.

%% file: dipolar_spin_chains.tex
\chapter{Quantum State Transfer with Long-Range Interactions}\label{dipolar chains}
\textit{This chapter examines the quality and efficiency of quantum
state transfer through closed- and open-ended chains of
spin-$\frac{1}{2}$ fermions coupled by long-range magnetic dipole interactions. The two geometries are treated separately because they are not physically equivalent; indeed, an open chain has end points, but a closed chain does not. The structural difference is reflected in the dynamics, which show interesting and perhaps slightly counter-intuitive features. We will see that the fidelity of state transfer in a chain has a clear periodicity and a predictable evolution in time, but the same cannot be said for a ring. This might seem unusual, given that a ring is a periodic structure and a chain is not; however, analysis of the respective eigenspectra reveals the origin of this behaviour. After considering a basic communication scheme analogous to that described in \cite{bose03}, I will explore possible optimization strategies, concluding with some remarks on the robustness of the protocol in the face of experimental uncertainties, fabrication defects and possible decoherence channels.}

\section{Introduction}\label{intro}
The majority of spin-chain communication protocols are designed for systems exhibiting only nearest-neighbour interactions. However, in recent years new effort has been dedicated to studying systems with anisotropic, long-range couplings. The possibility of implementing quantum-computational tasks in one- and two-dimensional spin lattices with tunable dipolar interactions has been known for some years \cite{porras04}; however, until the work of Kay \cite{kay05}, the performance of a spin chain with long-range couplings as a communication channel had never been investigated. The study detailed in this chapter represents the first rigorous attempt to analyze the dynamics of quantum state transfer in a chain of spins coupled by the magnetic dipole interaction. Thanks to recent advances in ion- and electron- trapping technology, several possible implementations of the present scheme have come to light. These, together with subsequent newly published extensions to the scheme, will be discussed in section \ref{experiments2}.

The protocol described here is, in essence, a mapping of Bose's original communication scheme to a system in which all the spins are coupled to each other via a long-range magnetic interaction that scales as $r^{-3}$. I examine a simplified system in which any external magnetic field is
constant and parallel to the axis joining the dipoles. This axis is
chosen to coincide with the $\mathbi{z}\:$-direction, so that the $z$-component of the total magnetization \textbf{M} is conserved (cfr. equation \ref{commutator1}). It is therefore possible to work in the single excitation sub-space, which grows linearly in the size of the system (see Section \ref{exch h}). Within this subspace, the effect of the magnetic field is to add a
constant to the energies. This constant will hereafter be omitted.
For simplicity, it will also be assumed that the system is at zero temperature, hence contains
no thermally excited spin-flips. Unless otherwise stated natural units will be used, such that $\mu_0 = \mu_B
= \hbar = m_e = 1$.

\section{The General Dipolar Hamiltonian}\label{dipole h}
The magnetic dipole coupling between two particles separated by a distance $r_{ij}\hat{\bm{z}}$ can be expressed as follows
\begin{equation}\label{h dip gen}
H_{ij} = \frac{C}{|\mathbi{r}_{ij}|^3}\:\lbrack\mathbi{S}_i \cdot \mathbi{S}_j - 3
\mathbi{S}_i^z \mathbi{S}_j^z\rbrack,
\end{equation}
where $\mathbi{S}_i$ and $\mathbi{S}_j$
are the total spin operators at sites $\mathbi{r}_i$ and $\mathbi{r}_j$, and
$\mathbi{S}_i^z$ and $\mathbi{S}_j^z$ are their respective \textit{z}-components. The value of \textit{C} is determined by the type of
particle one considers; for a system of spin-$\frac{1}{2}$ fermions such as electrons, \textit{C} is given by
\begin{equation}
C = \frac{\mu_0(\mu_B g_e)^2}{4\pi \hbar^2} ,
\end{equation}
where $g_e$ is the electron \textit{g}-factor. In natural units, the value of $C$ is then $C\equiv\frac{g_e^2}{4\pi}$. As the majority of the following discussion will be concerned with uniform systems, it is useful to define the interaction strength between anti-aligned nearest neighbours
\begin{equation}\label{eqnnint}
\epsilon=|\langle \mathbi{j}|H|\mathbi{j} \pm 1 \rangle| = \frac{C}{2a^3},
\end{equation}
where $a\equiv|\mathbi{r}_{j,\:j \pm 1}|$ is the nearest-neighbour spacing. Length, energy and time units are hereafter defined by
setting $a$ and $\epsilon$ to unity. This does not affect the qualitative behaviour of the system, as equation (\ref{h dip gen}) has an overall scaling factor of $a^3$ ; hence,
a uniform compression or expansion of the system simply re-scales all energies by
a constant. Consequently, for a fixed number of component spins, the energy and performance of a
system of any size can be extrapolated by adjusting the value
of \textit{a} as necessary.

The above considerations are entirely general. The specific form of the energy spectrum and Hamiltonian of the system are determined by the geometry. I will consider two cases: the closed and the open chain. These two systems differ in that the latter has end points, but the former does not. In other words, the closed chain is a linear system obeying periodic boundary conditions, whereas the open chain is a linear system \emph{not} obeying periodic boundary conditions. For brevity, I will refer to the closed chain as a ring, however it must be noted this does not in any way assume a circular geometry.

\section{Dipolar Rings: the Energy Spectrum and the Hamiltonian}\label{ring h}
Let us begin by calculating the Hamiltonian of a dipolar ring. From equation (\ref{h dip gen}), it is easily found that the off-diagonal matrix elements are
\begin{equation}\label{off diag}
\langle \mathbi{j}^{\:\prime}|H|\mathbi{j}\:\rangle=\frac{\epsilon}{|j-j^{\:\prime}|^3},
\end{equation}
where $\epsilon$ represents unit energy. For each pair $\lbrace j,\:j^{\:\prime}\rbrace$, equation (\ref{off diag}) must be evaluated in accordance with the minimum image convention, whereby if
\begin{equation}\label{min im}
|j-j^{\:\prime}|>
\begin{cases}
N/2 & \text{for even N}  \\
(N-1)/2 & \text{for odd N},
\end{cases}
\end{equation}
one sets
\begin{equation}\label{min image1}
\langle \mathbi{j}^{\:\prime}\:|H|\mathbi{j}\:\rangle=\langle |\mathbi{N}-\mathbi{j}^{\:\prime}|\:|H|\mathbi{j}\:\rangle.
\end{equation}
The diagonal elements correspond to the average energy of a single-spin-flip state. Given the ground-state energies for even and odd $N$
\begin{eqnarray}
E_0^e&=&-N\sum_{j=1}^{\frac{N}{2}-1}\frac{1}{j^{\:3}}-\left(\frac{2}{N}\right)^2\label{gs even}\\
E_0^o&=&-N\sum_{j=1}^{\frac{N-1}{2}}\frac{1}{j^{\:3}}\label{gs odd},
\end{eqnarray}
it is possible to show that the diagonal matrix elements of the corresponding Hamiltonians are related to the Euler Gamma function $\Gamma(z)$ and the Reimann Zeta function $\zeta(s)$, as
\begin{eqnarray}
\langle\mathbi{j}\:|H|\mathbi{j}\:\rangle^e&=&-\left(\frac{N-4}{2}\right)\left[\frac{8}{N^3}+\left(\frac{d^2}{dz^2}\right)\ln\Gamma\left(z\right)+2\zeta(3)\right]\label{dme}\\
\langle\mathbi{j}\:|H|\mathbi{j}\:\rangle^o&=&-\left(\frac{N-4}{2}\right)\left[\left(\frac{d^2}{dz^2}\right)\ln\Gamma\left(z\right)+2\zeta(3)\right]\label{dmo},
\end{eqnarray}
with $z=\frac{N}{2}$ and $z=\frac{N+1}{2}$, for even and odd $N$, respectively. Equations (\ref{dme}) and (\ref{dmo}) depend only on $N$; therefore, as required by symmetry, the energy needed to flip a spin with respect to the ground state is independent of position along the ring.

To calculate the eigenspectrum, we recall that the single excitation eigenvectors of a periodic system are single-magnon Bloch states. Labeling these states with $|\mathbi{m}\rangle$, one has
\begin{equation}\label{estates}
|\mathbi{m}\rangle=\frac{1}{\sqrt{N}}\sum_{j=1}^Ne^{i\mathbi{k}_m\cdot\mathbi{r}_j}|\mathbi{j}\:\rangle,
\end{equation}
where $\mathbi{k}_m$ is the magnitude of the Bloch wavevector, $\mathbi{r}_j$ is the position of the spin at site \textit{j}, and $|\mathbi{j}\:\rangle$ is the single spin-flip state.

If the nearest neighbour separation is set to unity, one has $|\mathbi{r}_j|=j-1$, as the origin $\mathbi{r}_j=0$ coincides with the first site of the ring. Since the excitations travel along the dipole axis, it follows that $\mathbi{k}_m\cdot\mathbi{r}_j=k_m(j-1)$. The eigenenergies of the ring can therefore be expressed as follows
\begin{equation}\label{em}
E_m=\langle \mathbi{m}|H|\mathbi{m}\rangle=\frac{1}{N}\sum_{j,\:j^{\:\prime}}e^{ik_m(j-j^{\:\prime} )}\langle \mathbi{j}^{\:\prime}|H|\mathbi{j}\:\rangle,
\end{equation}
where the sum runs over all pairs $j\neq\:j^{\:\prime}$.

According to condition (\ref{min im}), for a given $j$ there exist $\frac{N}{2}$ (even $N$) or $\frac{N-1}{2}$ (odd $N$) independent values of $|j-j^{\:\prime}|$. The sum over $j^{\:\prime}$ contains twice this number of terms, because it must account for positive and negative values of $\left(j-j^{\:\prime}\right)$. It follows that, for a given $j$ and even $N$\footnote{The term in $j=j^{\:\prime}$ is neglected.}
\begin{equation}\label{min image2}
\sum_{j,\:j^{\:\prime}}e^{ik_m(j-j^{\:\prime} )}\langle \mathbi{j}^{\:\prime}|H|\mathbi{j}\:\rangle=\sum_{\left(j-\:j^{\:\prime}\right)=-\frac{N}{2}}^{\frac{N}{2}}e^{ik_m(j-j^{\:\prime})} \frac{\epsilon}{|j-j^{\:\prime}|^3},
\end{equation}
where we have used the fact that interaction energies of distant neighbours $\lbrace j,\:j^{\:\prime}\rbrace$ are obtained by scaling the nearest-neighbour coupling by a factor $|j-j^{\:\prime}|^3$. Finally, considering all $N$ possible values of $j$
\begin{eqnarray}
E_m&=&2\epsilon\sum_{\left(j-\:j^{\:\prime}\right)=1}^{\frac{N}{2}}\frac{\cos{k_m(j-j^{\:\prime})}}{|j-j^{\:\prime}|^3}\nonumber\\
&=&2\sum_{j=1}^{\frac{N}{2}}\frac{\cos{k_mj}}{j^3},\label{em1}
\end{eqnarray}
as $\epsilon$ is set to unity. The corresponding expression for odd $N$ is obtained by simply changing the upper limit of the summation to $j=(N-1)/2$.

Equations (\ref{av fid}), (\ref{propagator}), (\ref{estates}) and (\ref{em1}) are now sufficient to provide a full characterization of the time-evolution of the fidelity of state transfer in a dipolar ring.

\section{The Performance of a Ring}\label{perf ring}
Thanks to the symmetries of the ring geometry, the fidelity of state transfer can be expressed in analytical form. To this end, it is convenient to write the propagator of equation (\ref{propagator}) in more general terms as
\begin{equation}\label{propgen}
f(t)_{r,\:s}=\sum_{m=1}^N\langle\mathbi{r}|\mathbi{m}\rangle \langle\mathbi{m}|\mathbi{s}\rangle e^{-iE_mt},
\end{equation}
where $|\mathbi s\rangle$ and $|\mathbi r\rangle$ are the sender and receiver sites, respectively. Equation (\ref{estates}) can be substituted into (\ref{propgen}), to obtain
\begin{equation}\label{ring prop1}
f(t)_{r,\:s}=\frac{1}{N}\sum_{m=1}^N\sum_{j,\:j^{\:\prime}=1}^N\langle\mathbi{r}|\mathbi{j}\rangle \langle\mathbi{j}^{\:\prime}|\mathbf{s}\rangle e^{ik_m(j-j^{\:\prime})}e^{-iE_mt}.
\end{equation}
For a ring of circumference $L=Na$, Bloch's theorem requires the phase factor $e^{ik_mNa}$ to be unity. This yields the allowed values of $k_m$:
\begin{equation}
k_m=n\:\frac{2 \pi}{Na}\equiv \frac{2\pi n}{N},
\end{equation}
with $a=1$. As the $|\mathbi{j}\:\rangle$ states form an orthonormal set, the propagator between sites $r$ and $s$ is found to be
\begin{equation}\label{ring prop2}
f(t)_{r,\:s}=\frac{1}{N}\sum_{n=0}^{N-1} e^{i\frac{2\pi n(r-s)}{N}}e^{-iE_nt},
\end{equation}
with
\begin{equation}\label{en}
E_n=2\:\sum_{j=1}^{\frac{N}{2}}\frac{\cos{\frac{2\pi n\: j}{N}}}{j^3}.
\end{equation}
To obtain the fidelity of state transfer, equations (\ref{ring prop2}) and (\ref{en}) are then substituted into (\ref{av fid}), taking $(r-s)=\frac{N}{2}$ or $\frac{N-1}{2}$, for even and odd $N$ respectively.

\begin{figure}[H]
\renewcommand{\captionfont}{\footnotesize}
\renewcommand{\captionlabelfont}{}
\begin{center}
\subfigure{\label{Fidsring1}\includegraphics[width=6.5cm]{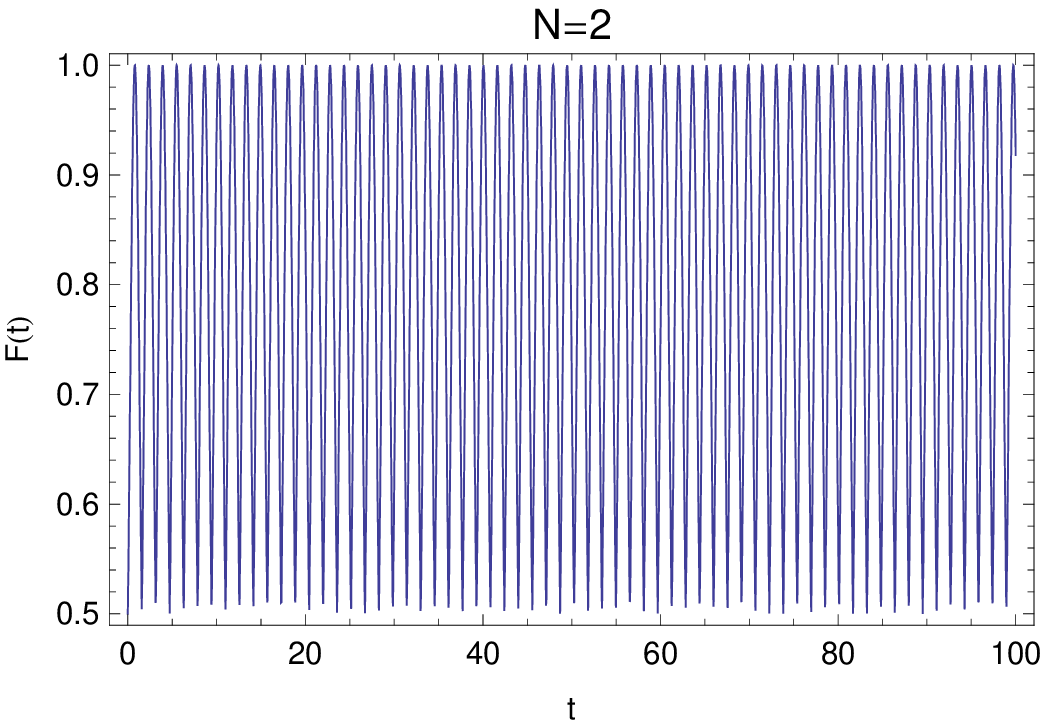}}
\hspace{0.3cm}
\subfigure{\label{Fidsring2}\includegraphics[width=6.5cm]{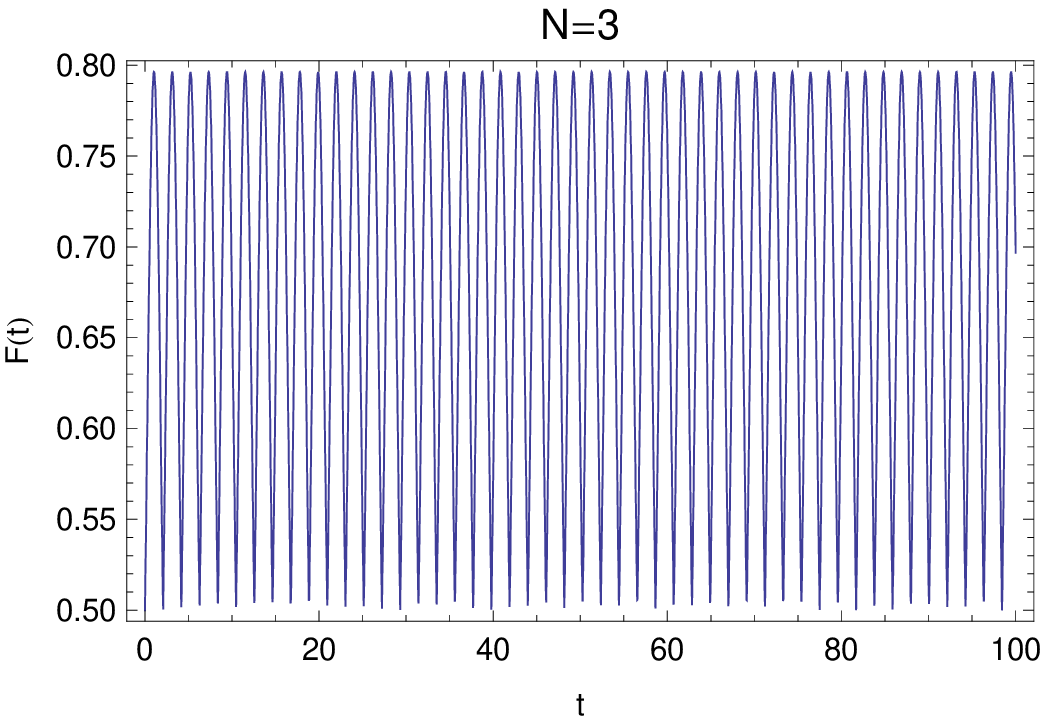}}
\hspace{0.3cm}
\subfigure{\label{Fidsring3}\includegraphics[width=6.5cm]{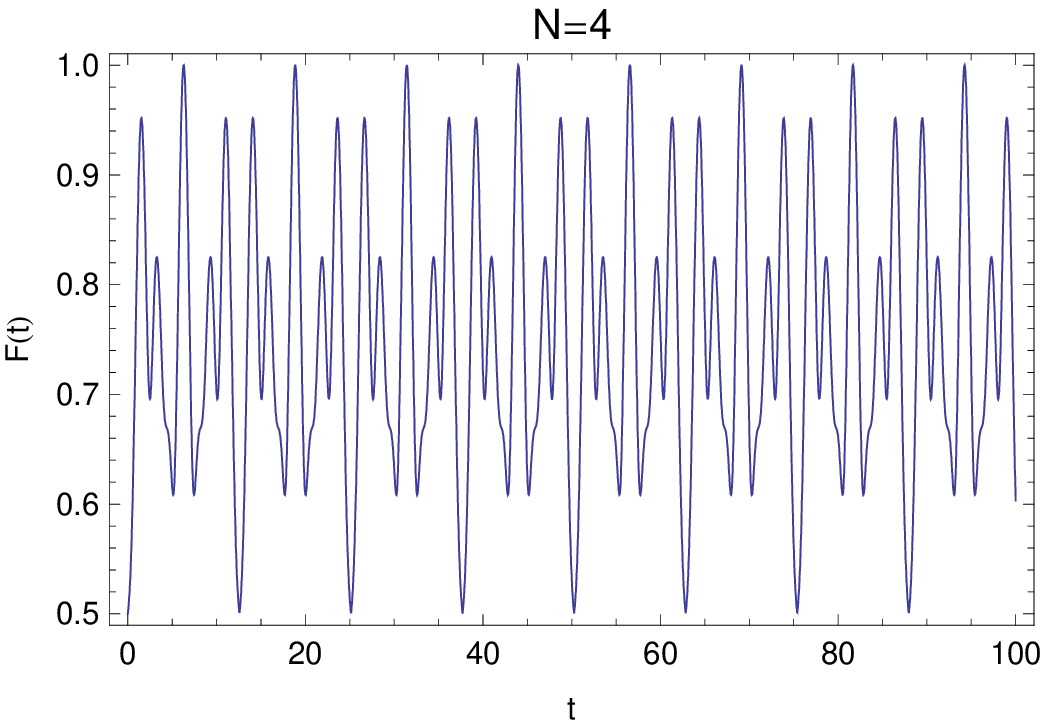}}
\hspace{0.3cm}
\subfigure{\label{Fidsring4}\includegraphics[width=6.5cm]{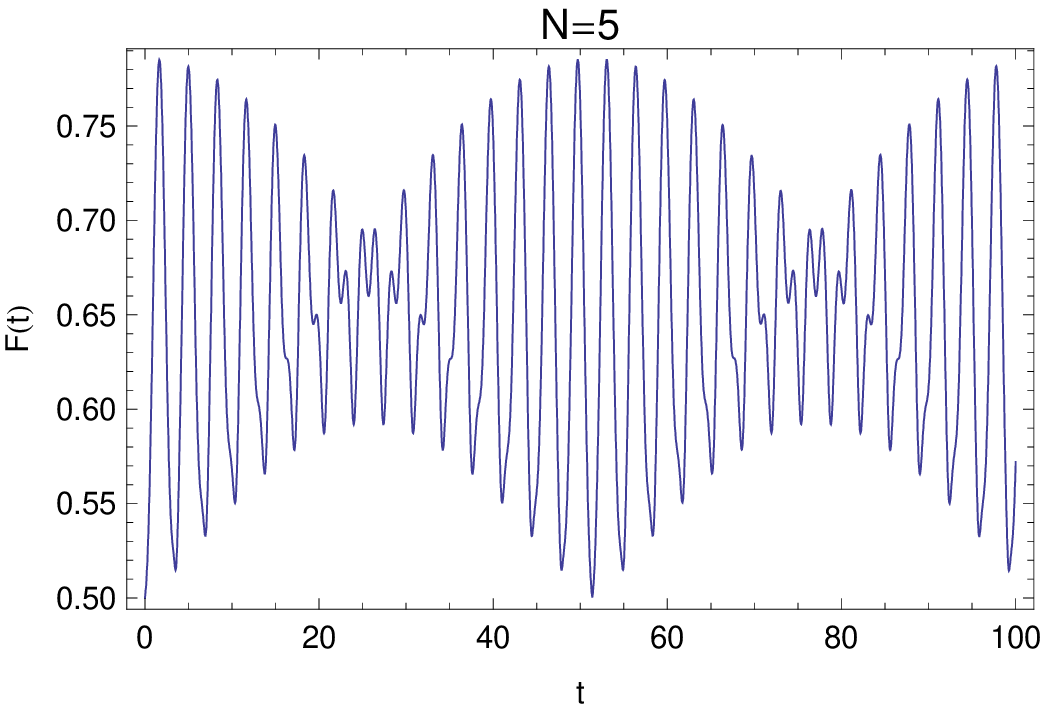}}
\hspace{0.3cm}
\subfigure{\label{Fidsring5}\includegraphics[width=6.5cm]{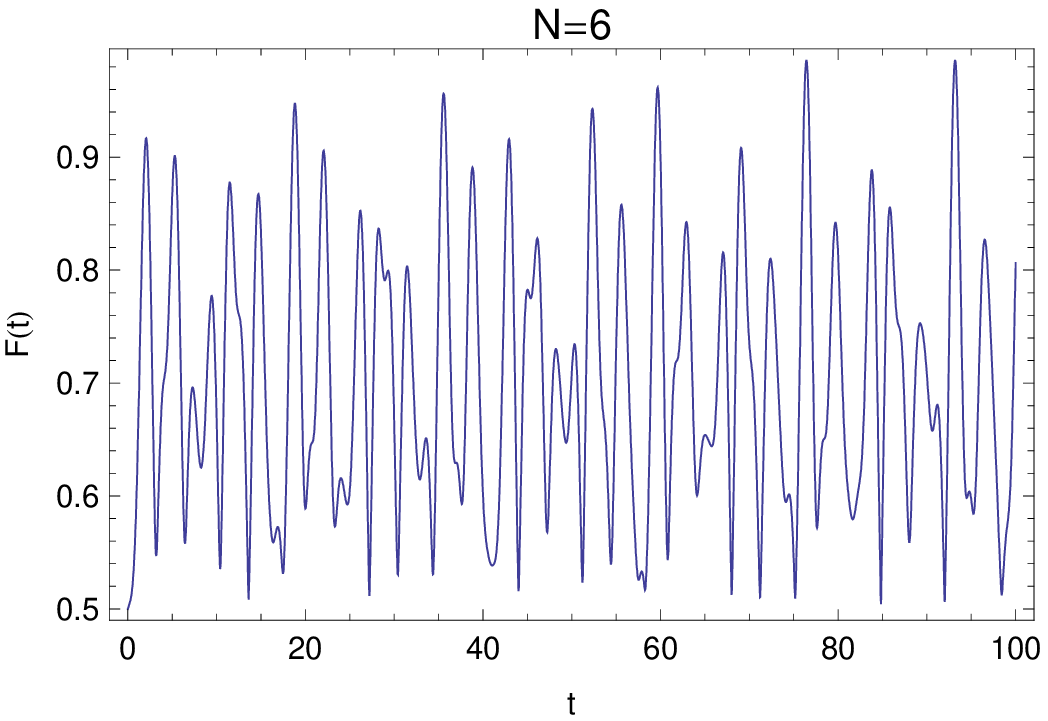}}
\hspace{0.3cm}
\subfigure{\label{Fidsring6}\includegraphics[width=6.5cm]{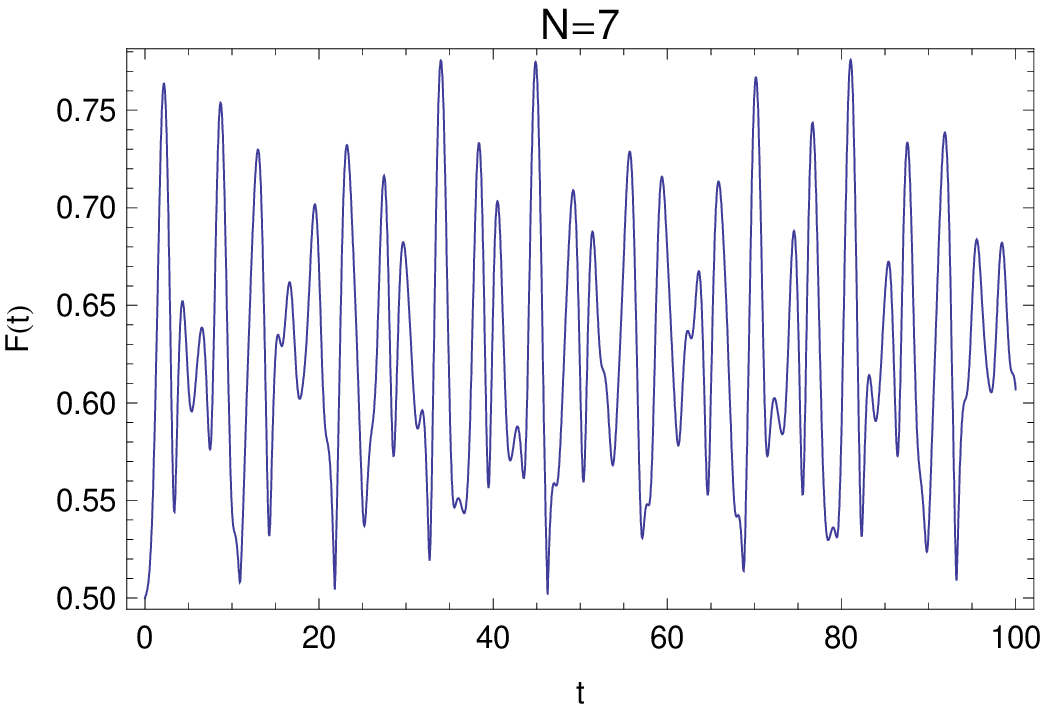}}
\end{center}
\caption{\label{fid1ring}The evolution of the fidelity at short times $t\leq\frac{100}{\epsilon}$ for $N=2-7$. }
\end{figure}

\begin{figure}[H]
\centering
\resizebox{11cm}{!}{\includegraphics{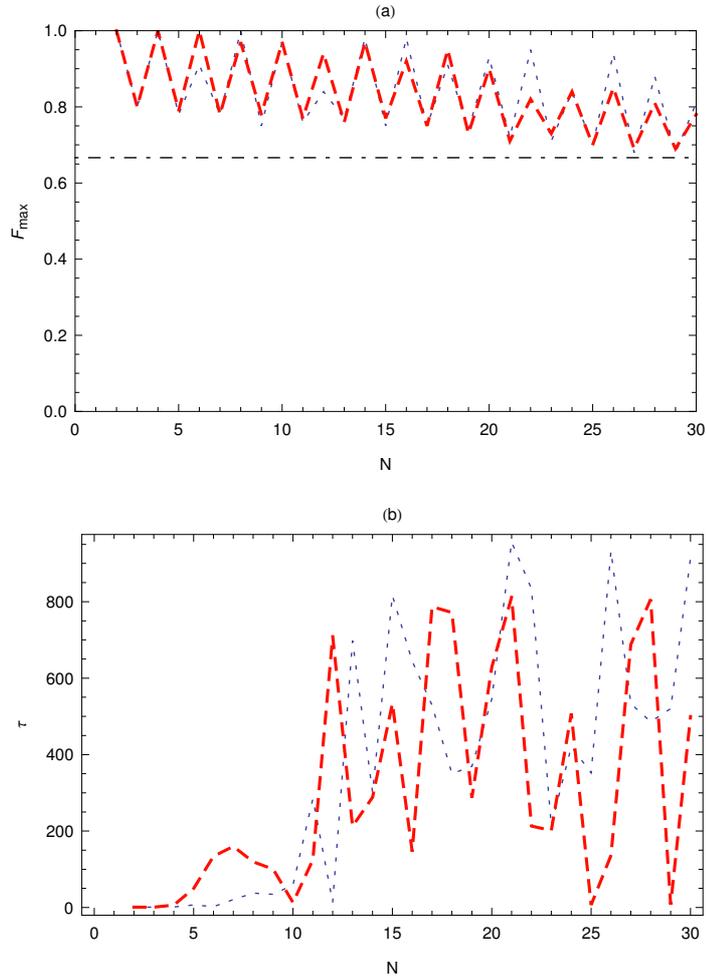}}
\renewcommand{\captionfont}{\footnotesize}
\renewcommand{\captionlabelfont}{\sffamily}
\caption{The maximal fidelity (a) and the timescale (b) of state transfer between diametrically opposite sites for a dipole-coupled (thick red dashed curve) and nearest-neighbour coupled (blue dotted curve) ring. The fidelity is optimized over a time period $T=\frac{1000}{\epsilon}$. In SI units, the value of $T$ depends on the inter-spin distance one chooses. For a chain of particles spaced at $3$ \AA, say, the dipolar interaction between nearest neighbours is approximately $4 \mu$eV, roughly corresponding to a timescale of $10^{-10}$ s. The chosen cutoff time is then of order $100$ ns, which is well within the decoherence time of, for example, N@C$_{60}$ \cite{tyrishkin03,spaeth96}. The black dot-dashed line indicates the maximum fidelity for classical transmission of a quantum state.\label{Fidtime}}
\end{figure}

The evolution of the fidelity in time for $N=2-7$ is illustrated in figure \ref{fid1ring}. Figure \ref{Fidtime} (a) represents instead the maximum fidelity $F_{max}$ that can be achieved within an arbitrary time period $T=\frac{1000}{\epsilon}$ for both a dipolar and a Heisenberg ring\footnote{For a Heisenberg ring, the fidelity is easily obtained from equations (\ref{av fid}) and (\ref{ring prop2}) by setting the eigenenergies $E_n^H=-2J\cos{\frac{2\pi n}{N}}$ (see Appendix \ref{app_c}).}. The definition of a cutoff time stems from the practical consideration that one cannot afford to wait indefinitely for the state to be transferred. To give a more complete description of the efficiency of the protocol, the time $\tau$ required to achieve fidelity $F_{max}$ is also shown, in Figure \ref{Fidtime} (b).

For $N>5$, the fidelity is an unpredictable function of time. The maximal fidelity falls rapidly with increasing $N$, though for even $N$ the fidelity is systematically higher. A comparison with a nearest-neighbour-coupled ring shows their average behaviour is very evenly matched in both quality and speed, although at large (odd) $N$ a Heisenberg ring has higher fidelity. The results illustrated in figure \ref{Fidtime} are complementary to those published in figure 1 of \cite{us06}, though not completely identical. In the majority of cases, this is owed to the fact that in \cite{us06} optimization of the fidelity was carried out over the much shorter timescale of $\frac{500}{\epsilon}$. However, there exist additional numerical inconsistencies between the results presented in figure \ref{Fidtime}(b) and those of figure 1(b) of \cite{us06}, which occur at $N=6$, $8$, $9$ and $16$ for the dipolar ring, and $N=8$, $12$, and $24$ for the Heisenberg ring. At $N=8$, $9$, and $12$ these arise from the fact that, upon repeating the analysis, it was found that the value of the fidelity at the (longer) time of figure 1(b) of \cite{us06} exceeds that at the (shorter) time of figure \ref{Fidtime}(b) by $\mathcal{O}\:(10^{-3})$. Given such a small advantage, it is more efficient overall to measure the output state at this shorter time, hence the result presented here. The discrepancies at $N=6$, $16$, and $24$ are unfortunately owed to a recently discovered indexing oversight, but this does not yield qualitative disagreement with previous findings except in the case $N=16$. Fortunately, this is not sufficient to affect the conclusions drawn so far, both in \cite{us06} and in the present thesis.

\section{State Transfer in a Dipolar Chain}
The following section is dedicated to the analysis of quantum state transfer in a dipolar chain. The lack of periodicity resulting from the finite structure yields important qualitative differences between the chain and ring geometries, which from many viewpoints render the chain the more efficient of the two quantum channels. As we will see, the main drawback of the chain configuration is the length of time required for a quantum state to be optimally transferred, which remains essentially impervious to most optimization techniques suggested so far. However, drawing also on more recent work published earlier this year \cite{gualdi08}, the performance of the dipolar chain will be critically evaluated, to establish whether such a system will ever be useful for anything other than proof-of-principle applications.

\subsection{The Hamiltonian}
\begin{figure}[H]
\centering
\resizebox{9cm}{!}{\includegraphics{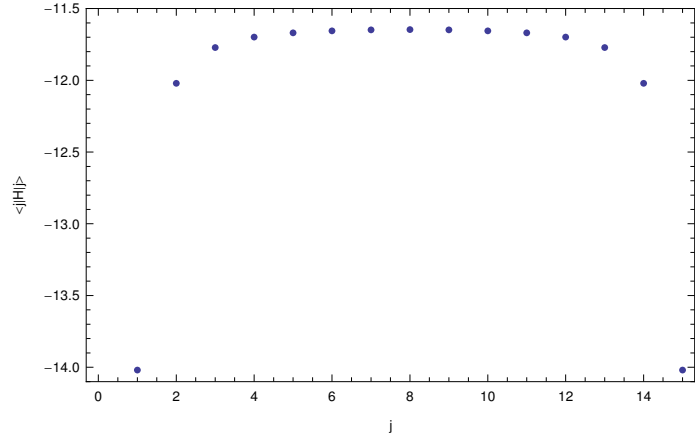}}
\renewcommand{\captionfont}{\footnotesize}
\renewcommand{\captionlabelfont}{\sffamily}
\caption{\label{diagelch}The on-site energy $\langle \mathbi{j}\:|H|\mathbi{j}\: \rangle$ as a function of \textit{j} for a dipolar chain of 15 spins. }
\end{figure}

It is instructive to begin our analysis by calculating the Hamiltonian of the uniform dipolar chain. It is easily shown that the off-diagonal matrix elements are given by equation (\ref{off diag}), but this time without recourse to condition (\ref{min im}). The quantity $|j-j^{\:\prime}|$ is therefore the linear distance between sites $j$ and $j^{\:\prime}$, irrespective of the values of these indices. The diagonal elements are calculated by evaluating the energy gap between the ground state and the spin-flipped state $|\mathbi{j}\:\rangle$. Given the ground-state energy
\begin{equation}
\langle\Downarrow|H|\Downarrow \rangle =
-\sum_{j=1}^{N-1}\frac{N-j}{j^3},
\end{equation}
it follows that
\begin{equation}\label{eqdiag}
\langle \mathbi{j}|H|\mathbi{j}\: \rangle = \langle\Downarrow|H|\Downarrow \rangle + 2\sum_{j^{\:\prime} \not=
j}\frac{1}{|j^{\:\prime} - j|^3}.
\end{equation}
Equation (\ref{eqdiag}) can be compared to (\ref{dme}) and (\ref{dmo}) to highlight an extremely important characteristic of the chain geometry: the energy required to flip a spin from the ground-state configuration is dependent on the site of the spin flip, and reaches a minimum at sites 1 and $N$ (Fig. \ref{diagelch}). The first and final sites of the chain are therefore the most favourable environments for a spin flip to exist, which bodes extremely well for the prospect of transferring a spin flip between these sites. The full implications of this result will be illustrated in greater detail in section \ref{performance chain}.

\subsection{The Performance of a Chain}\label{performance chain}
Equations (\ref{av fid}), (\ref{propagator}), (\ref{off diag}) and (\ref{eqdiag}) provide all the information required to calculate the fidelity of state transfer in a dipolar chain. The time evolution of the fidelity of state transfer between end sites is shown in figure \ref{fid1ch} for $N=2-7$.

It is immediately evident that, provided a large enough sampling time is chosen, the behaviour of a dipolar chain is substantially different from that of both a ring and a Heisenberg chain. The first thing we observe is the presence of a regular - though slightly noisy - oscillation, whose amplitude is close to unity. This shows there exist predictable times at which the transfer process is close to being perfect. Second, the period of the oscillation is long and increases with $N$, which suggests transfer time grows with the length of the chain. In some respects, this is a disadvantage; in others, one could argue that a slow variation of the fidelity in time provides the receiving party with a reasonably large window in which to measure their spin. Indeed, the fidelity remains at least above the classical threshold for roughly a quarter of a period either side of the peak.

\begin{figure}[H]
\centering
\resizebox{11cm}{!}{\includegraphics{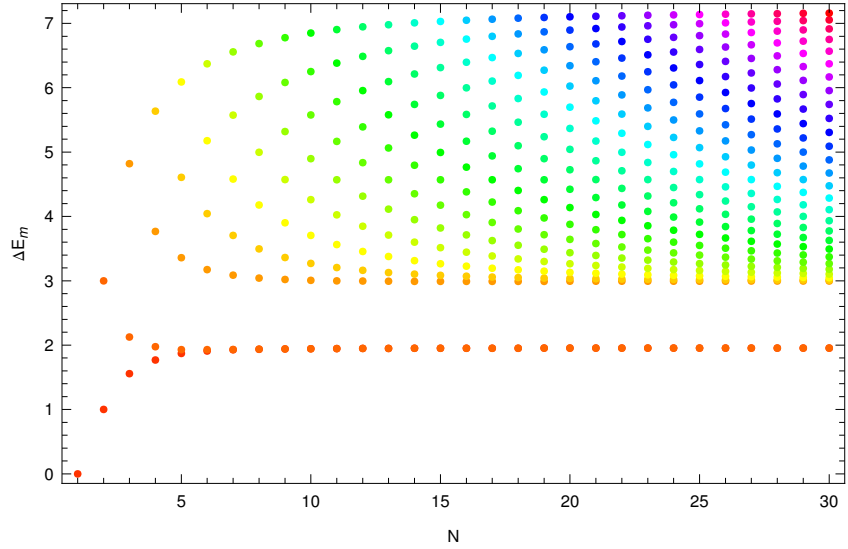}}
\renewcommand{\captionfont}{\footnotesize}
\renewcommand{\captionlabelfont}{\sffamily}
\caption{\label{spectrum}The energy splitting $\Delta E_m$ between
the ground state $|\Downarrow\:\rangle$ and states with a single flipped
spin, which shows the evolution
of the two bound states. Values of $\Delta E_m$ of the same index
$m$, counting from the bottom of the spectrum, are shown in
the same colour.}
\end{figure}

\begin{figure}[H]
\renewcommand{\captionfont}{\footnotesize}
\renewcommand{\captionlabelfont}{}
\begin{center}
\subfigure{\label{Fidsch1}\includegraphics[width=6.5cm]{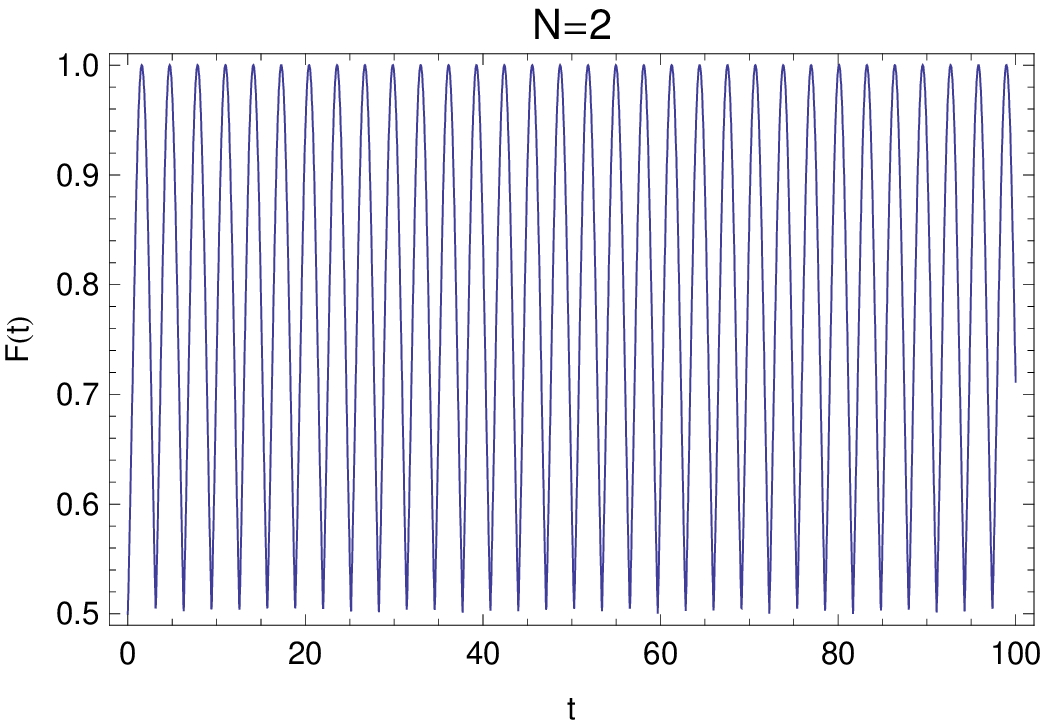}}
\hspace{0.3cm}
\subfigure{\label{Fidsch2}\includegraphics[width=6.5cm]{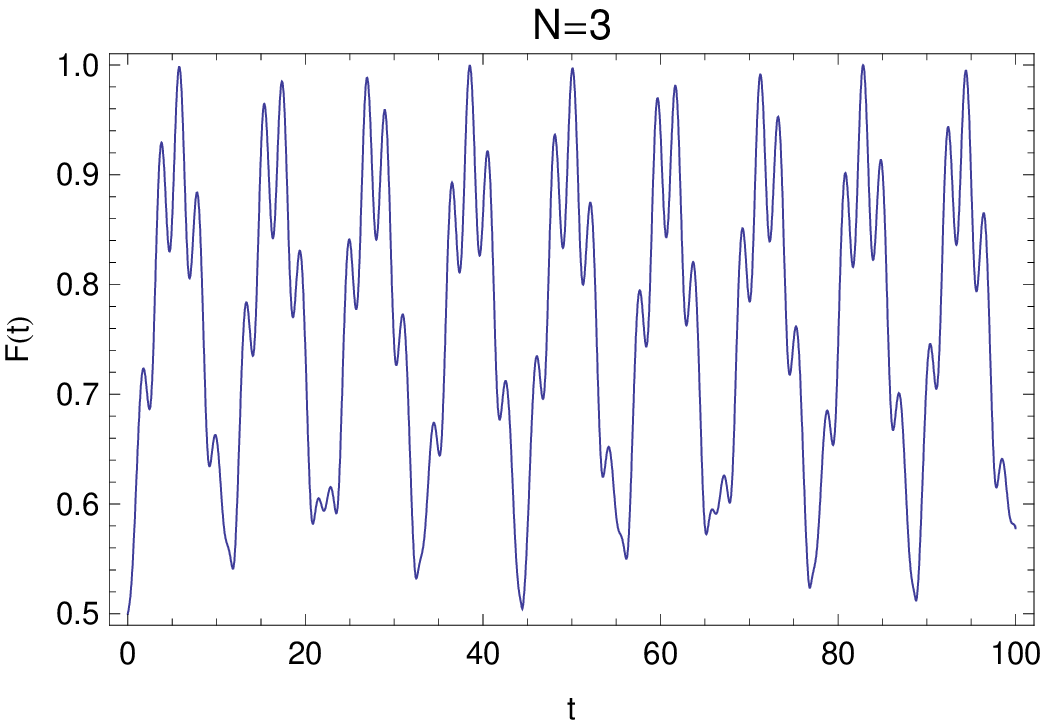}}
\hspace{0.3cm}
\subfigure{\label{Fidsch3}\includegraphics[width=6.5cm]{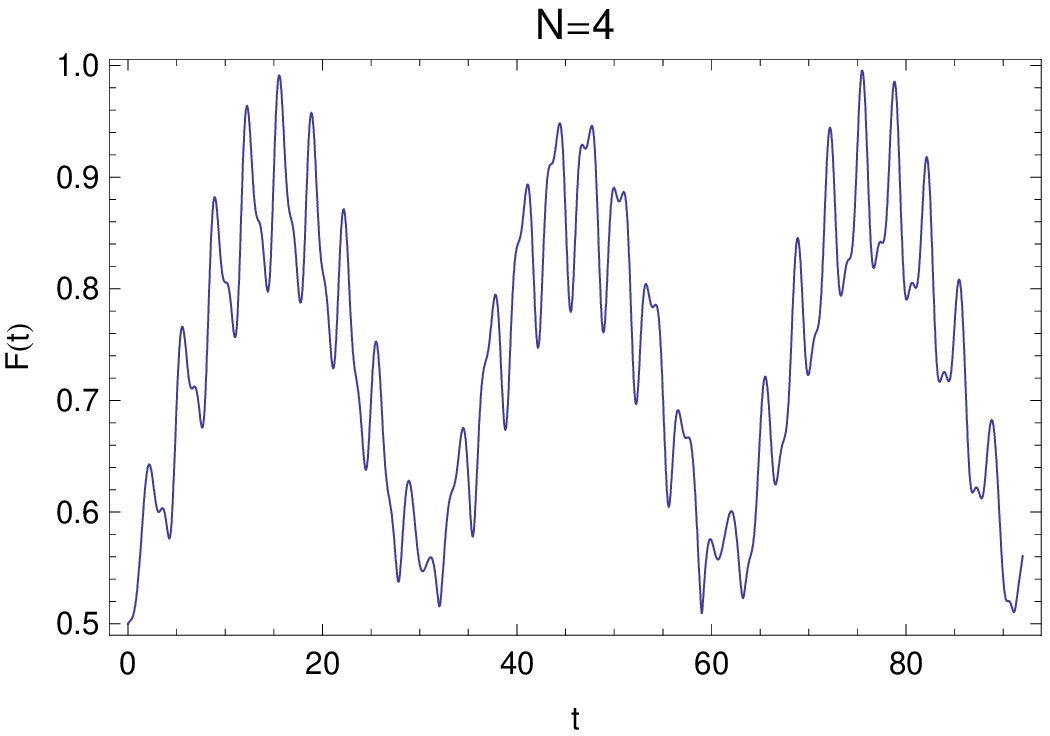}}
\hspace{0.3cm}
\subfigure{\label{Fidsch4}\includegraphics[width=6.5cm]{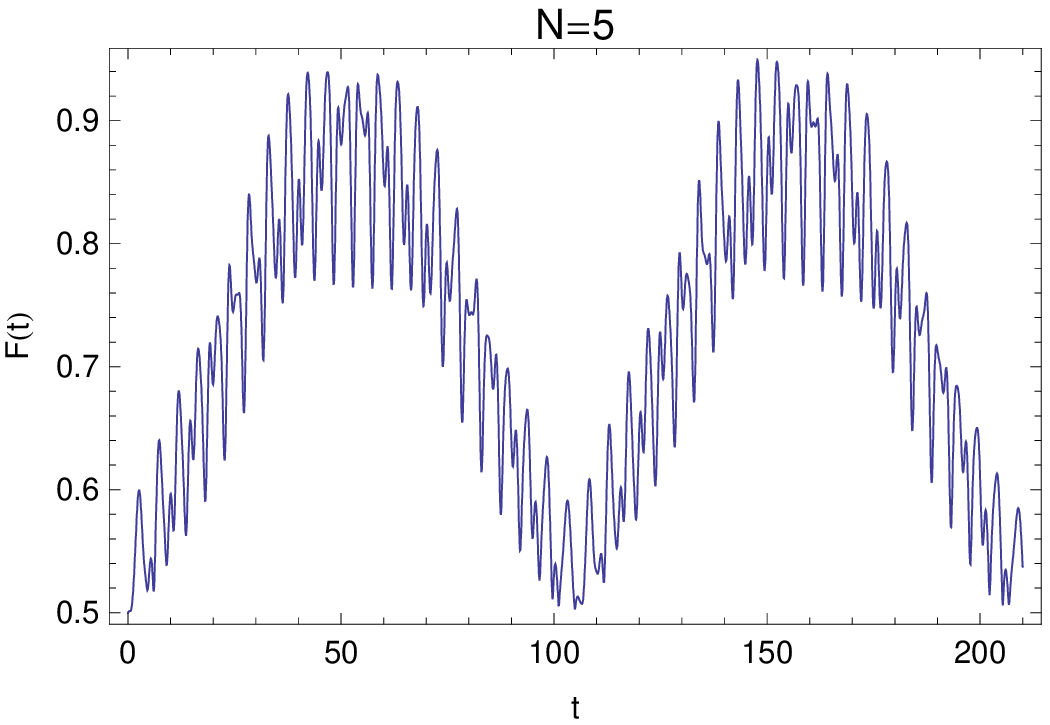}}
\hspace{0.3cm}
\subfigure{\label{Fidsch5}\includegraphics[width=6.5cm]{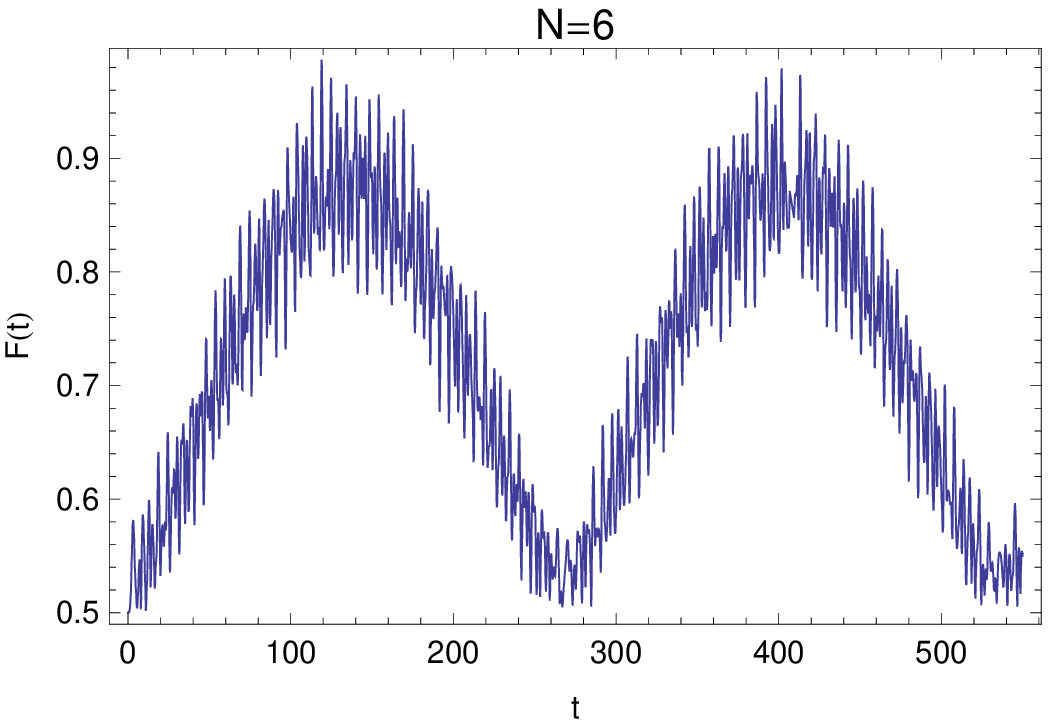}}
\hspace{0.3cm}
\subfigure{\label{Fidsch6}\includegraphics[width=6.5cm]{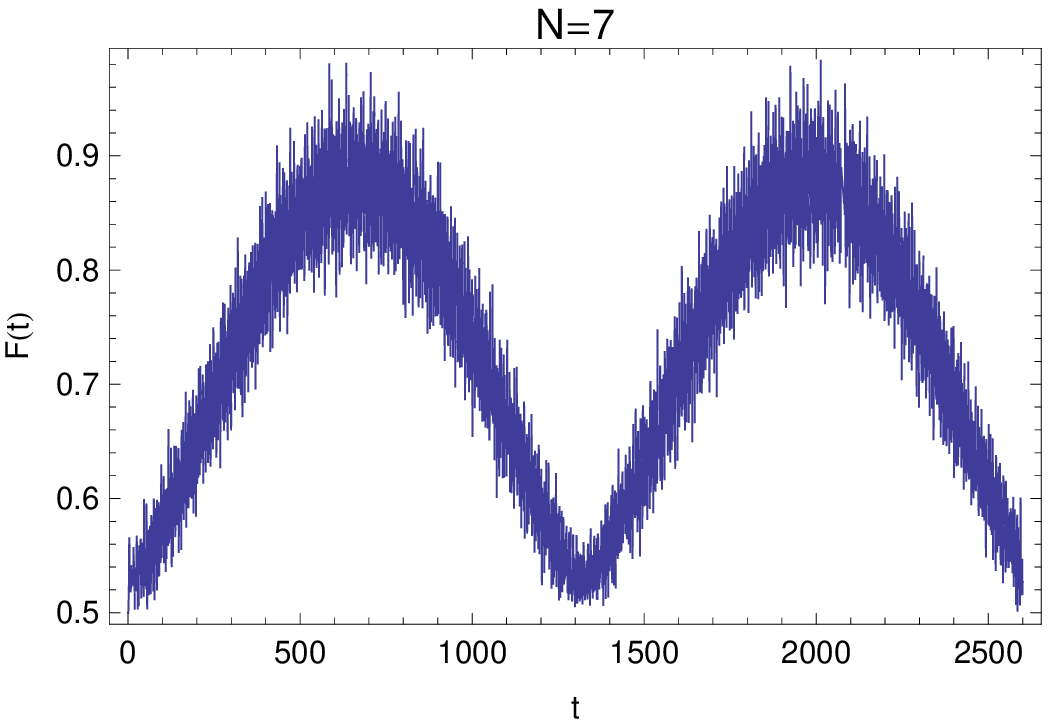}}
\end{center}
\caption{\label{fid1ch}The evolution of the fidelity for $N=2-7$. The range of abscissa values varies with $N$ in order to highlight the characteristic oscillation of the fidelity with time.}
\end{figure}

The period of oscillation of $F(t)$ is
uniquely defined by the energy splitting $\Delta \lambda$ between
the two lowest eigenvalues of the Hamiltonian. The transfer process is
therefore dominated by the beating of two nearly degenerate states
localized near the ends of the chain. As noted in the previous section, this behaviour is explained by
the variation of the on-site energies of the spins as a function of
$j$; it is clear from fig. \ref{diagelch} that the most favourable positions for a spin to flip are
sites 1 and \textit{N}. Consequently, states $|1 \rangle$ and $|N
\rangle$ are the most strongly coupled to the system's (two) bound
states, which are shown in fig.~\ref{spectrum}. In a dipolar chain, this
phenomenon is a natural consequence of the geometry, but systems in
which the spin flip energy is specifically chosen site by site have
also been studied \cite{santos05}.

The period of $F(t)$ is related to $\Delta \lambda$ by
\begin{equation}
\tau = \frac{2 \pi}{\Delta \lambda}
\end{equation}
Consequently, for each $N$ the time at which $F(t)$ first peaks is
\begin{equation}
t_0(N) = \frac{\tau}{2} = \frac{\pi}{\Delta \lambda}.
\end{equation}
This time grows with chain length, as the splitting $\Delta \lambda$
decreases with increasing \textit{N}.

\begin{figure}[H]
\centering
\resizebox{11cm}{!}{\includegraphics{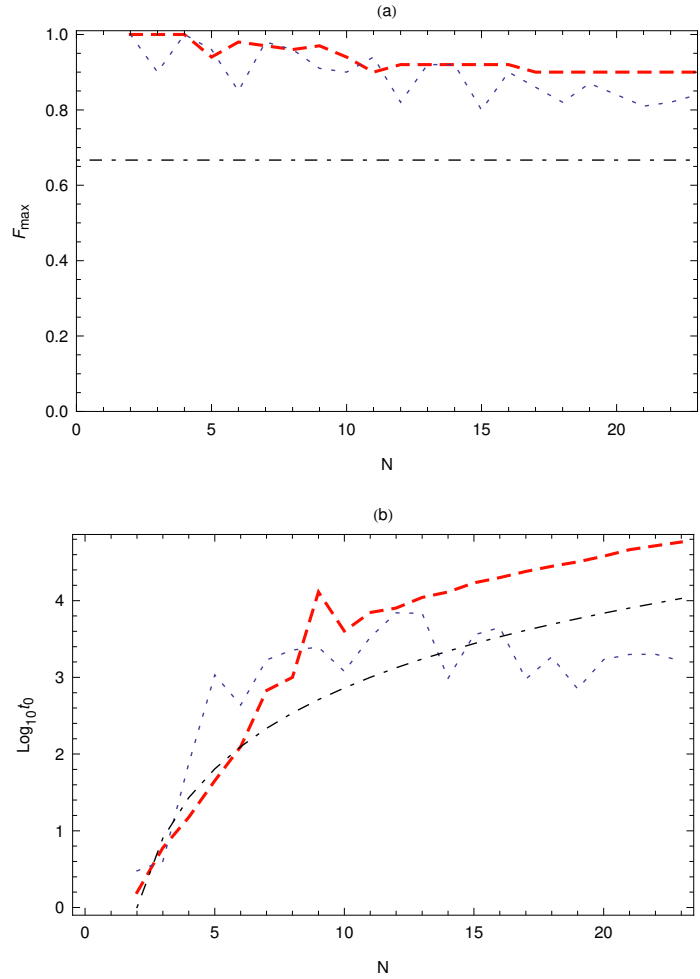}}
\renewcommand{\captionfont}{\footnotesize}
\renewcommand{\captionlabelfont}{\sffamily}
\caption{The maximal fidelity (a) and the timescale (b) of state transfer between sites 1 and $N$ of a dipole-coupled (thick red dashed curve) and nearest-neighbour coupled (blue dotted curve) chain. The black dot-dashed line in (a) indicates the maximum fidelity for classical transmission of a quantum state. Fig. (b) shows the times of optimal fidelity on a $\log_{10}$ scale for both types of chain. Here, the black dot-dashed line represents the function $y = (N-1)^3$. At large $N$, this is parallel to the red dotted curve, indicating that the transfer time in a dipolar chain scales as the cube of the chain
length. In absolute terms, for a chain of particles spaced by $3$ \AA, say, unit time corresponds to $\mathcal{O}(10^{-10})$ s. Transfer times for chains longer than 20 spins therefore exceed the micro-second mark. \label{Fidtimech}}
\end{figure}

Fig.~\ref{Fidtimech} compares the performance of dipolar and nearest-neighbour-coupled chains. We note that, in addition to $N = 2$, $N = 3$ and
$N = 4$ also give perfect transfer, and in general $F_{max} \geq
0.9$. This is a marked improvement on a
Heisenberg chain; specifically, it
is no longer the case that the fidelity is poor when $N$ is a multiple of 3
\cite{bose03}. These high fidelities are achieved at the cost of long transfer times; indeed, at large $N$, $t_0(N)$ is proportional to  the cube of the chain length (fig.~\ref{Fidtimech}). One is forced to conclude that, unless optimization strategies can be found, the time required to complete the protocol will become impractical as the chain length increases.
Following the work of Kay \cite{kay05} and Osborne and Linden \cite{osborne04}, two main optimization methods will be investigated. The first consists of changing the encoding and measurement convention so that the sending and receiving parties have access to more than one spin at the ends of the chain. I refer to this as input and measurement optimization. The second involves changing the geometric properties of the chain, such as the number of spins or their placement. I refer to this as structural optimization.

\subsection{Input and Measurement Optimization}
Let us begin by asking whether the performance of the protocol can be boosted by allowing Alice and Bob manipulate the first and last \emph{two} spins of the chain. For example, suppose Alice wishes to transmit the state $|1\rangle$. Rather than encoding this state in the first spin only, Alice encodes it in the first two spins, generating the initial state
\begin{equation}
|\mathbi{s}\rangle=c_1|\mathbi{1}\rangle+c_2|\mathbi{2}\rangle.
\end{equation}
Following the discussion of the previous section, it is advantageous to choose the coefficients $c_1$ and $c_2$ so as to maximize the overlap of $|\mathbi{s}\rangle$ with one of the system's bound states, as these dominate the transfer. The eigenvectors of the system are symmetrized and antisymmetrized combinations of $|\mathbi{j}\:\rangle$-states, having the form $|\mathbi{m}\:\rangle=\sum_{j=1}^Na_j^m|\mathbi{j}\:\rangle$, with $|a_i^m|=|a_{N-1+i}^m|$, and $a_i^m \in \Re$. For either of the bound states ($m=1$ and $m=2$), the amplitudes $\{a_3^m \cdots a_{N-2}^m\}$ decrease steadily with $N$; hence, most of the amplitude is concentrated on the first and last two sites. Suppose Alice takes the first of these bound states as her `reference' state; she should then choose the coefficients $c_1$ and $c_2$ to be as close as possible to $a_1^1$ and $a_2^1$, while still obeying the normalization condition $c_1^2+c_2^2=1$.

\begin{figure}[H]
\renewcommand{\captionfont}{\footnotesize}
\renewcommand{\captionlabelfont}{}
\begin{center}
\subfigure{\label{Fidsch1a}\includegraphics[width=6.5cm]{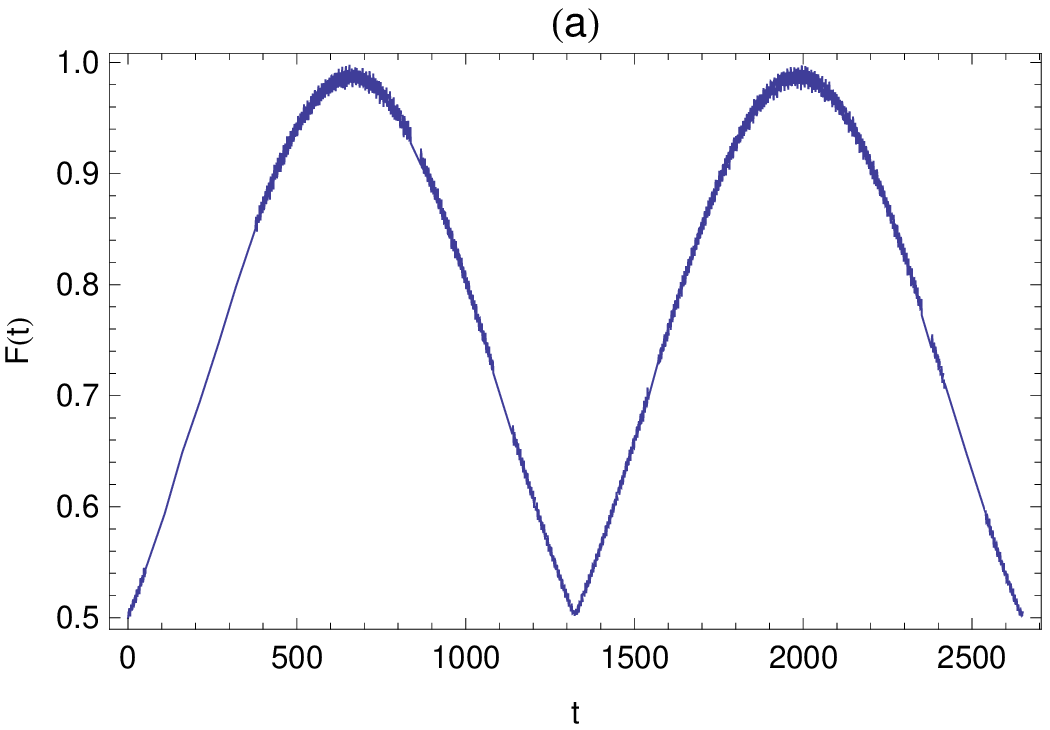}}
\hspace{0.3cm}
\subfigure{\label{Fidsch1b}\includegraphics[width=6.5cm]{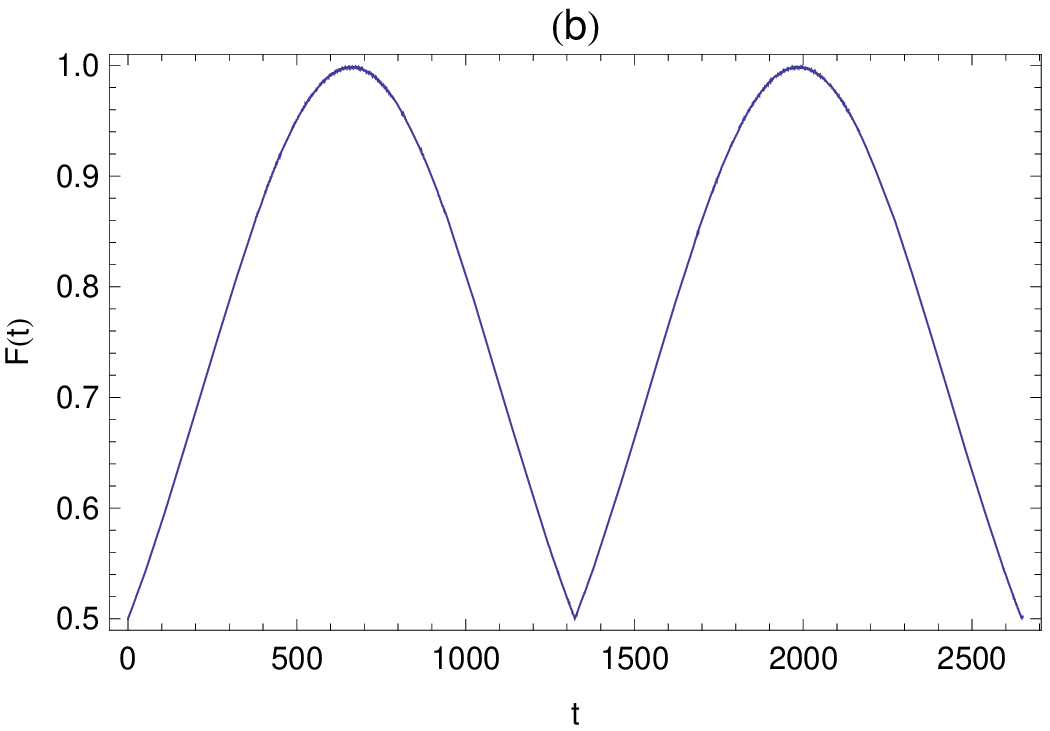}}
\end{center}
\caption{\label{superpositions}The evolution of the fidelity for $N=7$, when the initial and final states are distributed over (a) the first and last two, or (b) the first and last three spins of the chain.}
\end{figure}

Bob must now establish which output state $|\mathbi{r}\rangle$ to aim for. Knowing as he does the criterion Alice has adopted to fix her $c_1$ and $c_2$, and also that $|a_i^m|=|a_{N-1+i}^m|$, he chooses $|\mathbi{r}\rangle=c_2|\mathbi{N-1}\rangle+c_1|\mathbi{N}\rangle$. The propagator of equation (\ref{propagator}) can then be expressed as
\begin{equation}
f(t)=\sum_{m=1}^N\left(c_1a_1^m+c_2a_2^m\right)\left(c_2a_{N-1}^m+c_1a_N^m\right)e^{iE_mt}.
\end{equation}
The outcome of this optimization procedure for a chain of 7 spins is shown in figure \ref{superpositions}. Comparing with figure \ref{fid1ch}, it is evident that the transfer time has remained unaffected. However, we have been successful in smoothing out the oscillation and boosting its amplitude even closer to unity. This agrees completely with the result of \cite{osborne04}; indeed, the quality of the transfer can be further improved if Alice and Bob are have access to an even larger portion of the chain, as shown in figure \ref{superpositions}(b).

\begin{figure}[H]
\renewcommand{\captionfont}{\footnotesize}
\renewcommand{\captionlabelfont}{}
\begin{center}
\subfigure{\label{Fidsch2a}\includegraphics[width=6.5cm]{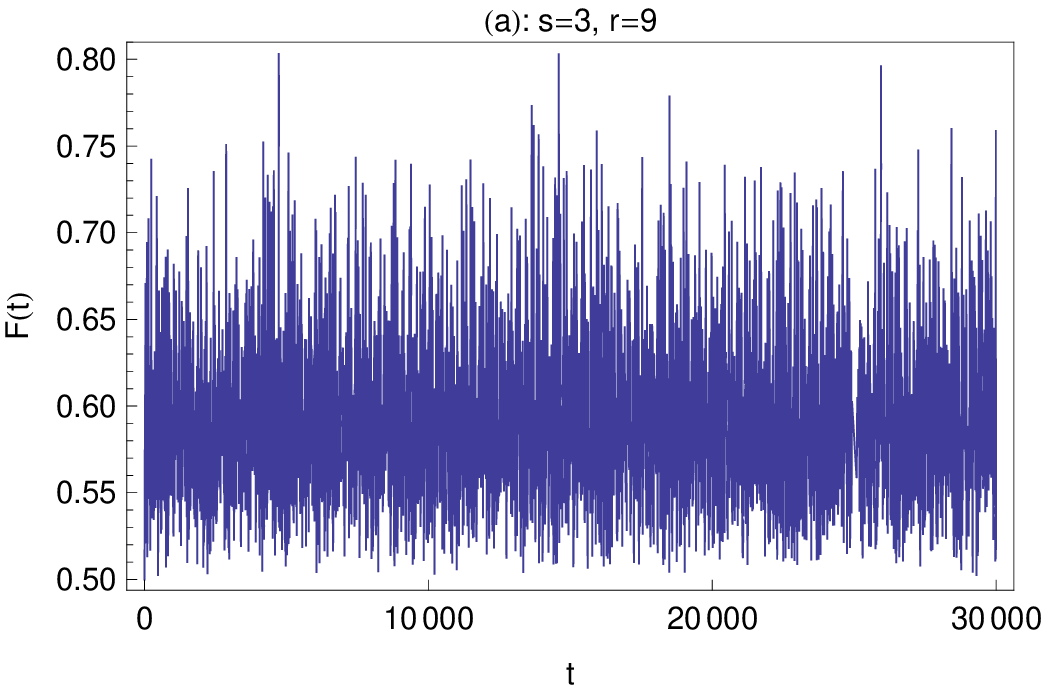}}
\hspace{0.3cm}
\subfigure{\label{Fidsch2b}\includegraphics[width=6.5cm]{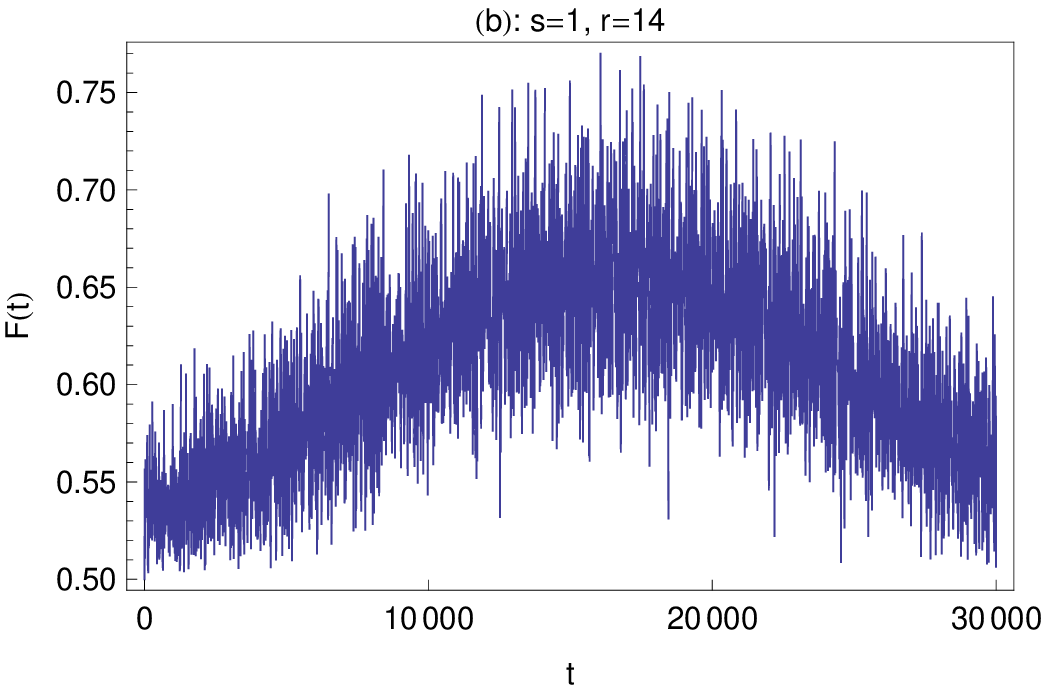}}
\hspace{0.3cm}
\subfigure{\label{Fidsch2c}\includegraphics[width=6.5cm]{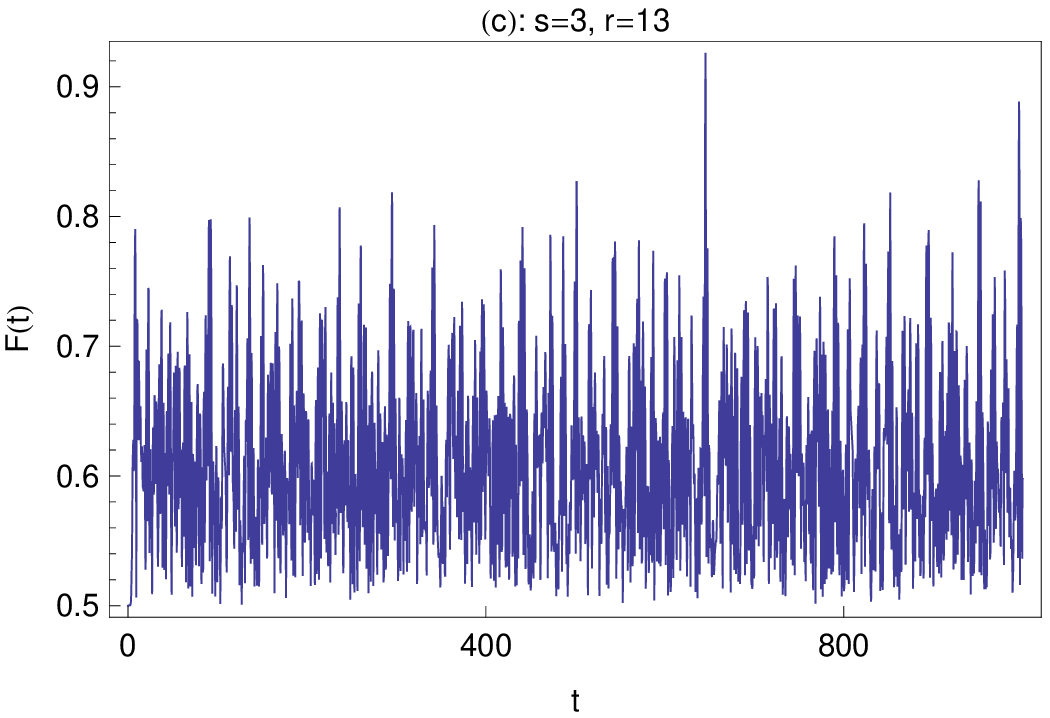}}
\end{center}
\caption{\label{inoutopt}The evolution of the fidelity for $N=15$, when the initial and final states are (a) chosen at random, (b) taken as $|\mathbi{s} \rangle = |1 \rangle$ and $|\mathbi{r} \rangle = |\mathbf{14}
\rangle$, and (c) symmetric ($|\mathbi{s} \rangle = |3 \rangle$ and $|\mathbi{r} \rangle = |\mathbi{13}
\rangle$). We note that in (c) the fidelity peaks within the time limit $T=\frac{1000}{\epsilon}$ imposed for the ring geometry.}
\end{figure}

It is also interesting to ask how well the system performs if the initial and final states are chosen at random; this is illustrated in figure \ref{inoutopt}. We find that the characteristic oscillation of $F(t)$ is lost,
unless either $|\mathbi{s} \rangle = |\mathbi{2} \rangle$ and $|\mathbi{r} \rangle = |\mathbi{N}
\rangle$, or $|\mathbi{s} \rangle = |\mathbi{1} \rangle$ and $|\mathbi{r} \rangle = |\mathbi{N - 1}
\rangle$  [figure \ref{inoutopt}(b)]. However, in both cases the signal is considerably noisier,
and the maximum fidelity is greatly reduced. This is a result of the
lesser efficiency of coupling to the bound states as one moves away
from the ends of the chain (cfr. fig.~\ref{spectrum}).

Conversely, fidelities in excess of 0.8 can be attained if sites $s$ and $r$ are symmetric\footnote{A mathematical justification of this result was recently published in \cite{gualdi08}.} (Figure \ref{inoutopt}). In this case also the oscillatory pattern is lost, as the eigenstates that dominate the transfer are well-separated in energy. In return, however, for $N<23$ the fidelity can peak at times shorter than $\tau/2$. The system is therefore more efficient, though less predictable. Furthermore, the choice of symmetric sites implies Alice and Bob must be capable of site-specific encoding and measurement at arbitrary positions in the chain; in other words, they must be able to act upon a more or less extended portion of the chain containing sites $s$ (Alice) or $r$ (Bob). From a practical viewpoint, this is much more demanding than end control only.

We conclude that by optimizing the encoding and decoding operations whilst remaining in control of the end spins only, Alice and Bob can ensure perfect transmission of quantum information. However, the problem of excessively long transfer times still remains. This is not surprising, as the transfer time is determined by the Hamiltonian rather than the details of the encoding. It is therefore clear that to act on the transfer time one must change the Hamiltonian, and as the Hamiltonian is proportional to the spacing of the spins, it can only be modified by changing the structural properties of the chain.

\subsection{Structural Optimization: Non-Uniform Chains}
It has already been suggested in \cite{kay05} that non-uniform chains may offer some advantages over uniform systems. To continue this line of investigation, we study a chain in which the placement of the spins is uneven, though not random. There are several ways to obtain such a structure; here, we imagine compressing or expanding some part of a uniform chain by a fixed amount. The system can then be described in terms of four parameters: (a) the number of spins in the chain, $N$; (b) the number of evenly spaced spins before the compressed (expanded) region, $i$; (c) the number of evenly spaced spins after the compressed (expanded) region, $f$; (d) the spacing of the spins in the compressed (expanded) region, $\delta$. The Hamiltonian remains that of equation (\ref{h dip gen}), except the overall scaling factor of $1/a^3$ is now lost.

\begin{figure}[H]
\centering
\resizebox{15cm}{!}{\includegraphics{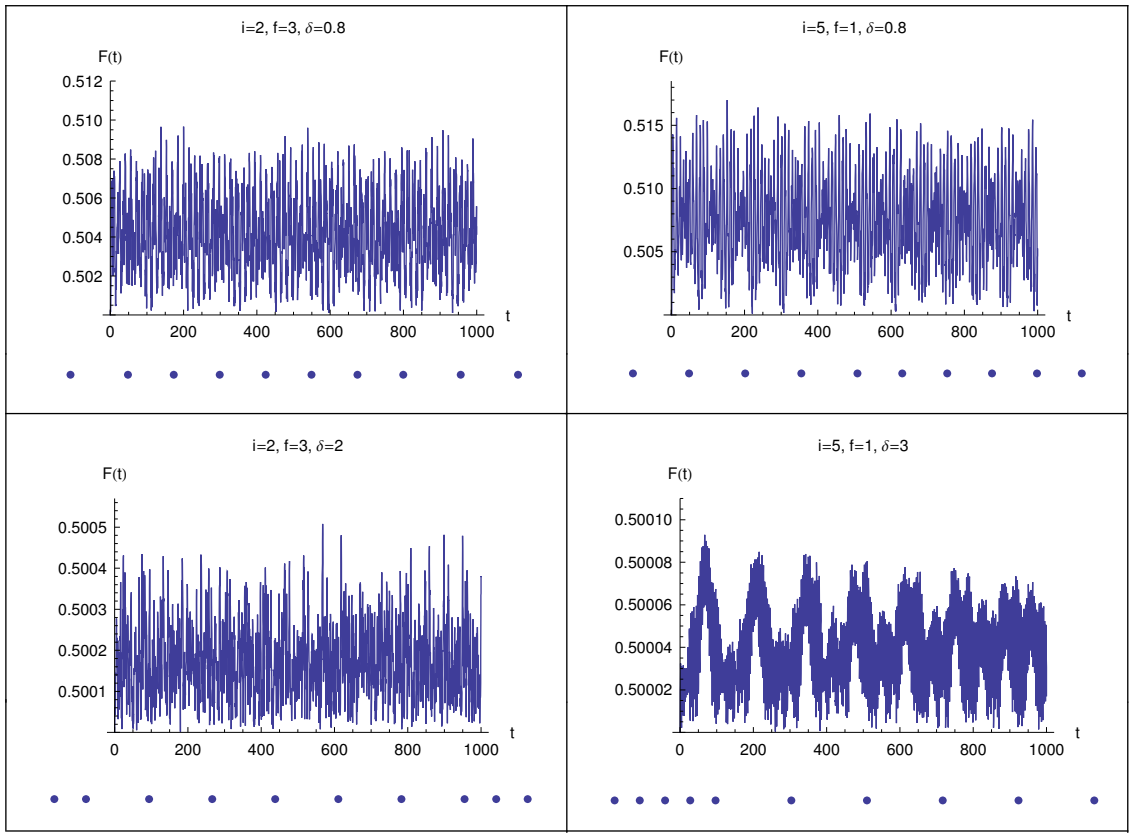}}
\renewcommand{\captionfont}{\footnotesize}
\renewcommand{\captionlabelfont}{\sffamily}
\caption{\label{uneven}The fidelity of state transfer between sites $1$ and $10$ of a ten-spin chain. Values of $\{N,i,f,\delta\}$ are indicated on each figure, with the corresponding chain structure inset.}
\end{figure}

\begin{figure}[H]
\centering
\resizebox{15cm}{!}{\includegraphics{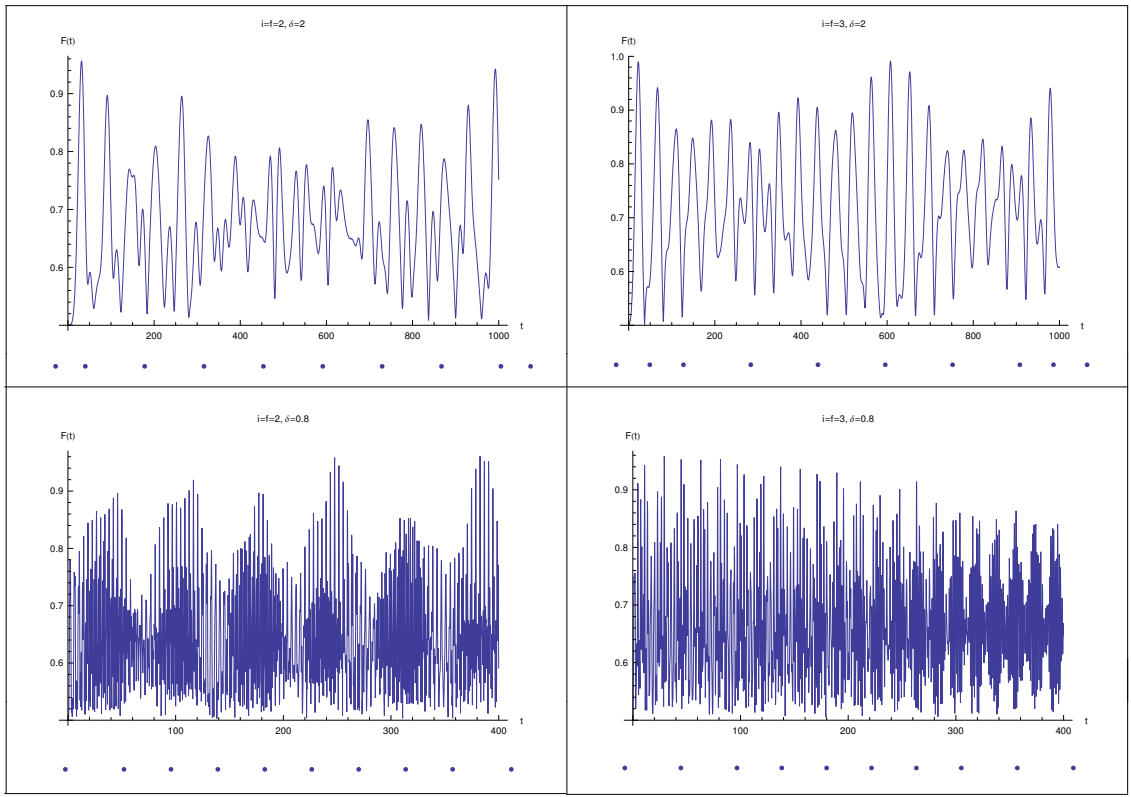}}
\renewcommand{\captionfont}{\footnotesize}
\renewcommand{\captionlabelfont}{\sffamily}
\caption{\label{msym}The fidelity of state transfer between symmetric sites of a non-uniform ten-spin chain. The left-hand figures take $s=3$ and $r=8$, the right-hand figures take $s=4$ and $r=7$. Values of $\{N,i,f,\delta\}$ are indicated on each figure, with the corresponding chain structure inset.}
\end{figure}

Figure \ref{uneven} shows the performance of a chain of 10 spins for different $\{N,i,f,\delta\}$ combinations with $i\neq f$. It has been assumed that the `default' spin spacing is $a$, hence $\delta$ is expressed in units of $a$. We observe that when the chain is not mirror-symmetric the fidelity is poor irrespective of the choice of parameters. Conversely, if the chain has mirror symmetry ($i=f$) it is again possible to achieve high-quality state transfer between all symmetric sites, as illustrated in figure \ref{msym}. The times of peak fidelity vary considerably; however, for a given $s$ and $r$, these times are shorter if the central section of the chain has been compressed (i.e. $\delta<1$). This is not surprising, as compressing the central section shortens the chain.

\begin{figure}[H]
\renewcommand{\captionfont}{\footnotesize}
\renewcommand{\captionlabelfont}{}
\begin{center}
\subfigure{\label{Fidsch2a}\includegraphics[width=6.5cm]{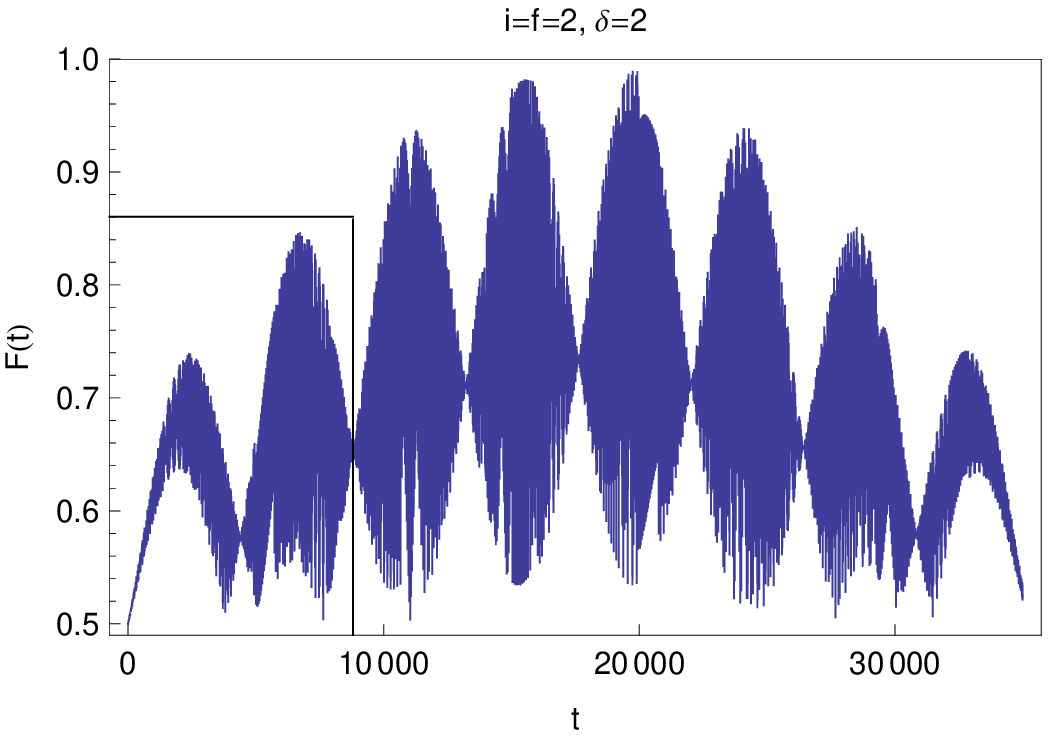}}
\hspace{0.3cm}
\subfigure{\label{Fidsch2b}\includegraphics[width=6.5cm]{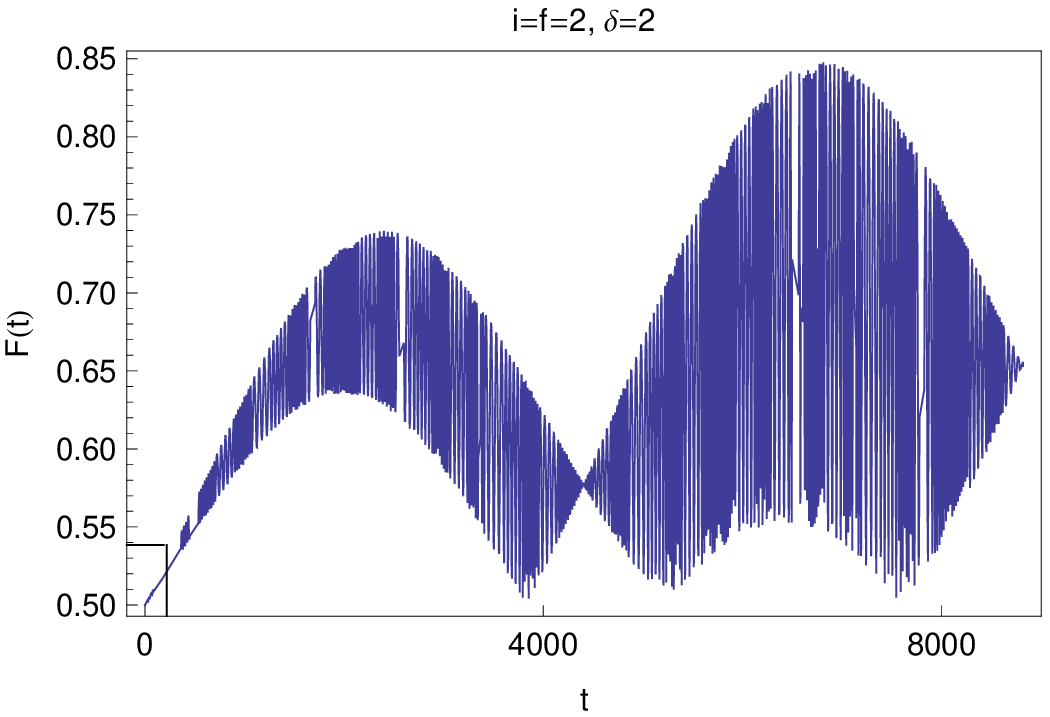}}
\hspace{0.3cm}
\subfigure{\label{Fidsch2c}\includegraphics[width=6.5cm]{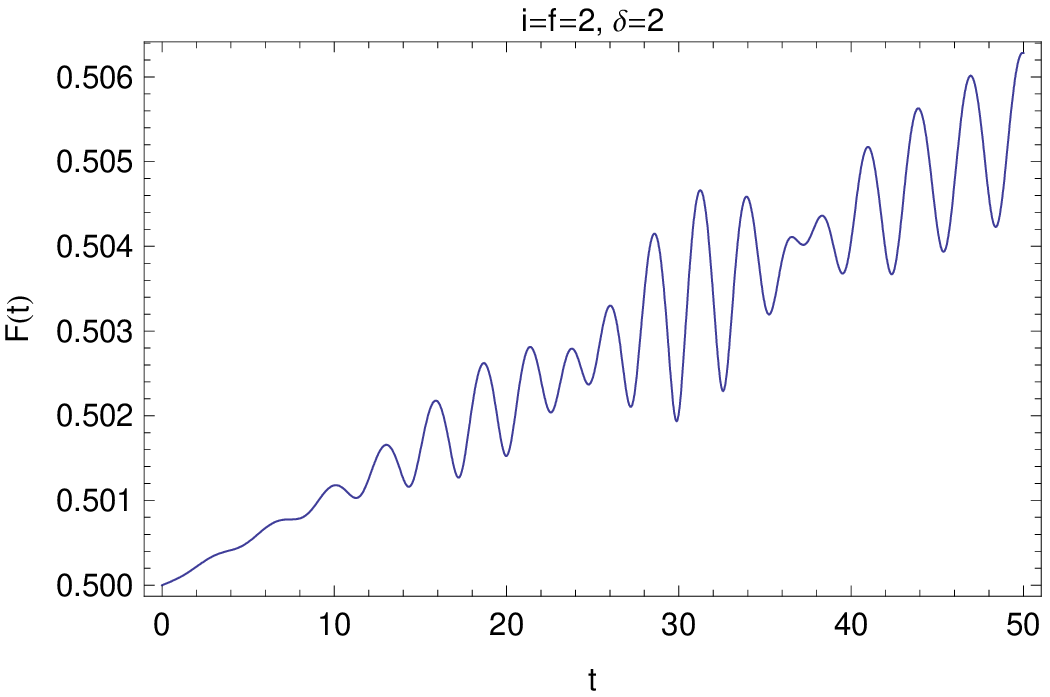}}
\end{center}
\caption{\label{msym1}The fidelity of state transfer between sites $1$ and $10$ of an expanded ten-spin chain, illustrating the different frequencies involved in the transfer. The plot to the right of each figure is a `zoom-in' of the boxed region.}
\end{figure}

If the chain is non uniform but mirror-symmetric, the time evolution of the fidelity of state transfer between the end sites exhibits a regular oscillatory behaviour, which is different from that of a uniform chain. The details of the oscillation depend on whether the chain has been expanded or compressed in the central region. If the chain has been expanded, the transfer is no longer dominated by the bound states but by the four upper-lying eigenstates of the system. This is a consequence of the geometry, as the end spins are now closer together (hence more strongly coupled) relative to the spins in the mid-section of the chain. As a result, there are now at least three clearly discernable frequencies beating against each other. The first is the frequency of the slowly varying envelope function, which is determined by the splitting of the third and fourth highest eigenenergies of the system. The second characterizes the rapid oscillation within the envelope, and is determined by the energy splitting of the highest and second-highest eigenstates. In turn, this serves as an envelope for an even faster oscillation, whose period is defined by the energy gap of the highest and third-highest eigenstates. These details are shown in figure \ref{msym1}. The peak fidelity is again close to unity, but the transfer now takes even longer than in a uniform chain, because expanding the central section of the chain increases its length.

Similar results are found for compressed chains with the exception that, as in uniform chains, we are concerned with the bottom of the energy spectrum rather than the top. The frequency of the outer envelope is therefore determined by the energy splitting of the two lowest-lying eigenstates, the frequency within this envelope is determined by the energy splitting of the third and fourth lowest-lying eigenstates, and so on. The transfer is quicker than in a uniform chain, as compressing the centre of the chain makes it shorter.

\begin{figure}[H]
\renewcommand{\captionfont}{\footnotesize}
\renewcommand{\captionlabelfont}{}
\begin{center}
\subfigure{\label{Fidsch2d}\includegraphics[width=6.5cm]{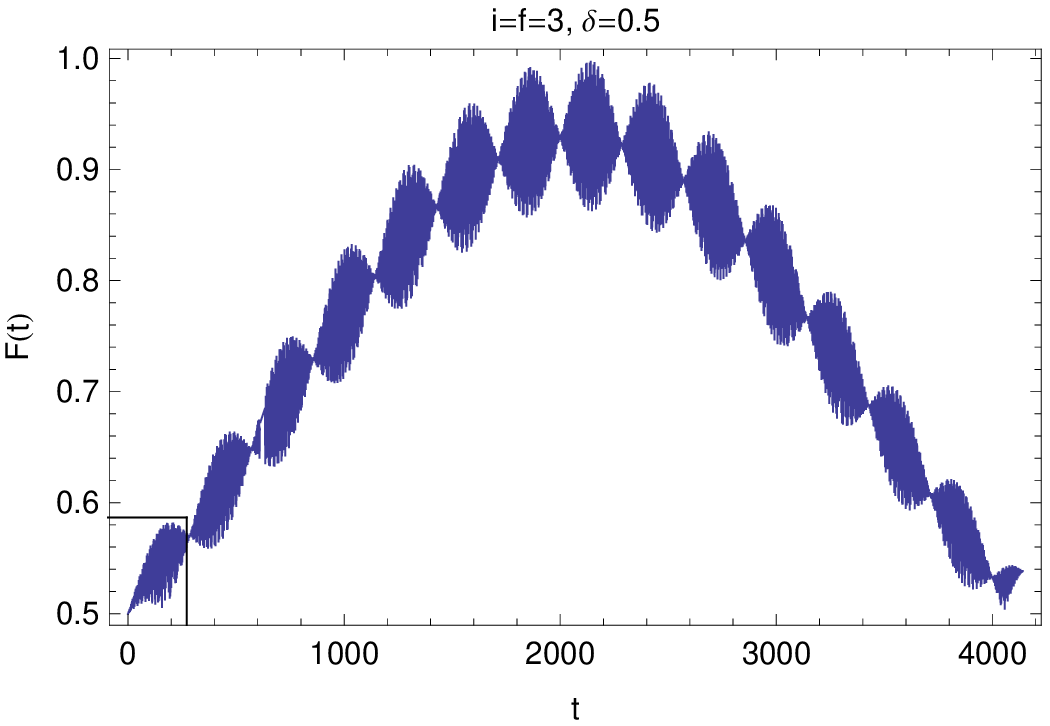}}
\hspace{0.3cm}
\subfigure{\label{Fidsch2e}\includegraphics[width=6.5cm]{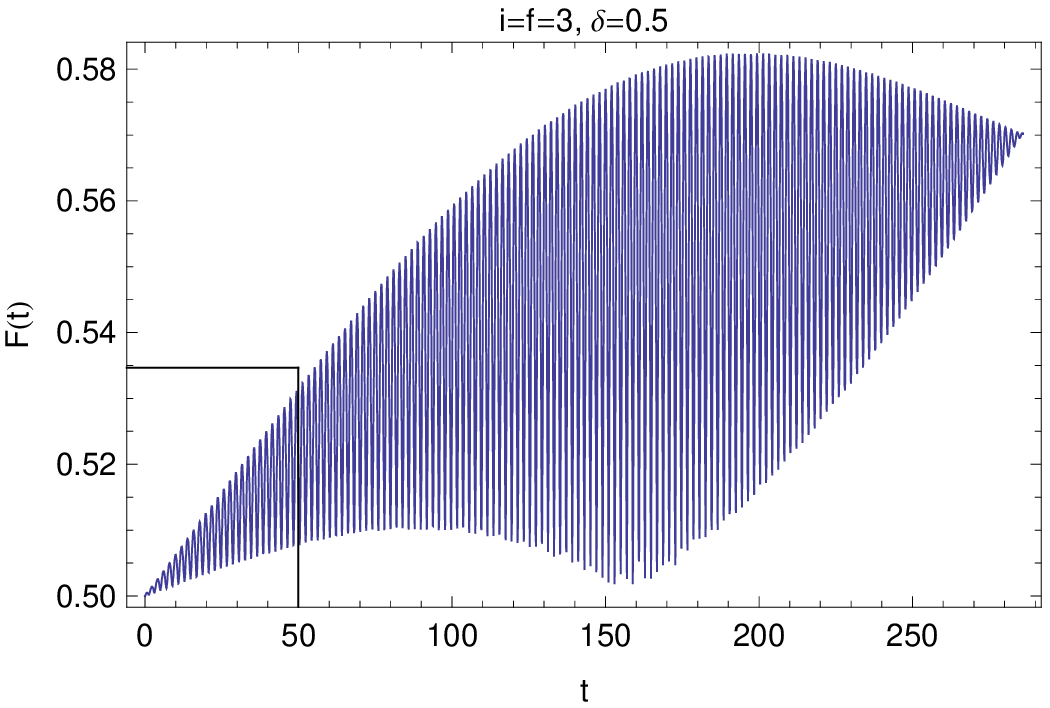}}
\hspace{0.3cm}
\subfigure{\label{Fidsch2f}\includegraphics[width=6.5cm]{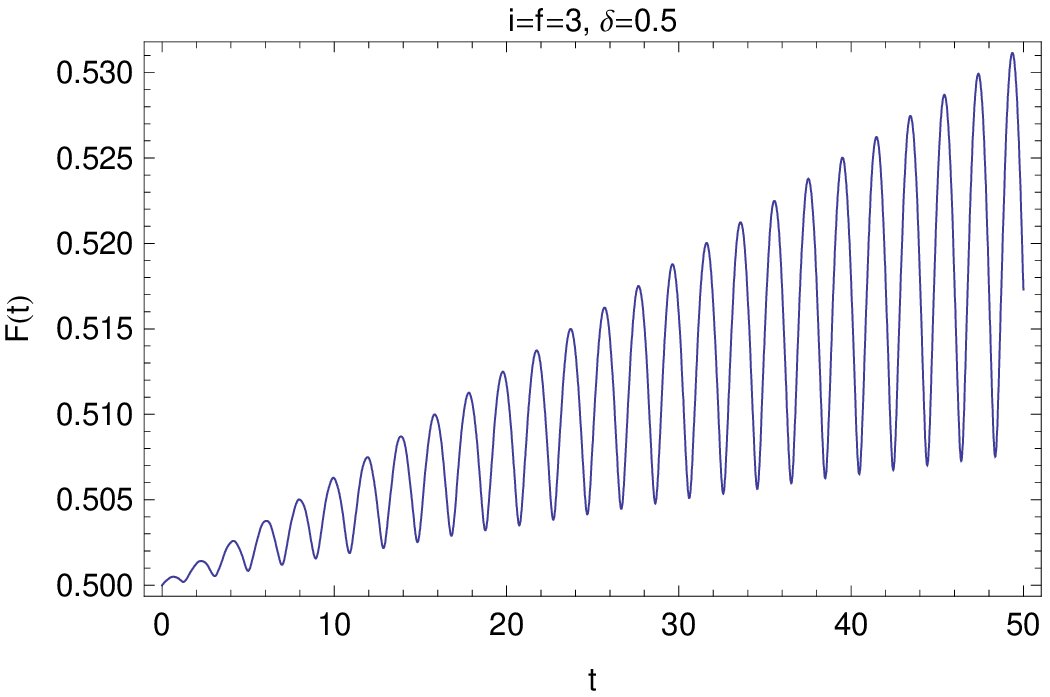}}
\end{center}
\caption{\label{msym2}The fidelity of state transfer between sites $1$ and $10$ of a compressed ten-spin chain, illustrating the different frequencies involved in the transfer. The plot to the right of each figure is a `zoom-in' of the boxed region.}
\end{figure}

In conclusion, we find that expanding or compressing the chain while keeping the number of spins fixed can offer some degree of speed-up over a uniform geometry provided the transfer takes place between symmetric spins \emph{not} located at the ends. However, this comes at the price of losing the regular oscillation which allows us to predict when the fidelity will peak. It would therefore be more desirable to find a way to minimize the transfer time while preserving the fidelity's periodic behaviour. This brings us to our last attempt at optimization, which will investigate whether the performance of a chain of a given length depends on the number of component spins.

\subsection{Structural Optimization: Fixed length Chains}
\begin{figure}[H]
\renewcommand{\captionfont}{\footnotesize}
\renewcommand{\captionlabelfont}{}
\begin{center}
 \includegraphics[width=7cm]{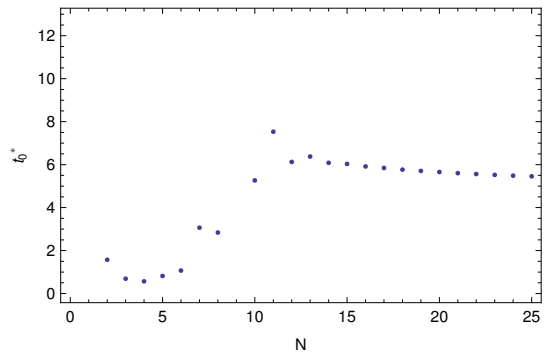}
 \end{center}
 \caption{The behaviour of $t_0^*$ as a function of the number
 of spins in the chain. Note the minimum at $N = 4$ and the flatness of
 the curve for $N > 15$. Units are as specified in Section
\ref{intro}.} \label{normtimech}
\end{figure}

In order to analyze the relationship between the
efficiency of state transfer and the number of spins for a fixed chain length $L$, we define $t_0^*$ as
the transmission time giving maximum fidelity at unit chain length. Hence
\begin{equation}
t_0^* = \frac{t_0(N)}{L^3},
\end{equation}
with $L=a(N-1)$. Fig. \ref{normtimech} shows the development of $t_0^*$ with increasing
$N$, and reveals two interesting features. The first is the
asymptotic behaviour of $t_0^*$ at large $N$; this suggests there exists a threshold length above which the central part of the chain ceases to play a role in the transfer. In other words, the evolution of the system is determined
almost exclusively by the coupling between the first and last $i$ spins, irrespective of the
number of spins that separate them.
To explore this hypothesis, and determine the behaviour of $t_0^*$
for large $N$, we work with states $|B \rangle$ and $|E
\rangle$ localized on the first and last $i$ spins of the chain, respectively. These are
the bound-state eigenfunctions of a semi-infinite chain extending to
the right and the left, and can be expressed as follows
\begin{eqnarray}\label{eqb}
|B \rangle&=&\sum_{n = 1}^ia_n |\mathbi{n} \rangle,\\
|E \rangle&=&\sum_{n = 1}^ia_n |\mathbi{N + 1 - n} \rangle.
\end{eqnarray}
Let us now assume the coefficients $a_n$ are non-zero only if $(j,j^{\:\prime})\leq i$ or $(j,j^{\:\prime})\geq N + 1 - i$. The states $|B \rangle$ and $|E\rangle$ can then be used as an alternative basis to represent the Hamiltonian as a $2\times2$ matrix, which takes the form
\[ \mathcal{H} = \left( \begin{array}{cc}
\langle B|H|B\rangle & \langle B|H|E\rangle \\
\langle B|H|E\rangle & \langle B|H|B\rangle \end{array} \right),\]
since by definition $\langle B|H|B\rangle=\langle E|H|E\rangle$ and $\langle B|H|E\rangle=\langle E|H|B\rangle$. The difference between the two bound-state energies of (\ref{h dip gen}) therefore corresponds to the difference between the eigenvalues of $\mathcal{H}$. This is given by
\begin{equation}
\Delta \lambda = 2 \langle B|\mathcal{H}|E \rangle.
\end{equation}
From (\ref{off diag}), we have
\begin{equation}
\langle \mathbi{j}^{\:\prime}|H|\mathbi{j}\:\rangle = H(|j - j^{\:\prime}|).
\end{equation}
Hence
\begin{equation}\label{eqtaylor}
\langle B|\mathcal{H}|E \rangle = \sum_{n,m = 1}^ia_n^*a_m \langle \mathbi{n}|H |\mathbi{N
+ 1 - m} \rangle = \sum_{n,m = 1}^ia_n^*a_m H(|N + 1 - m - n|)
\end{equation}
We adopt a `dummy' variable $X = |j - j^{\:\prime}|$, so that
\begin{equation}
H(X) = \frac{C}{2a^3X^3}.
\end{equation}
It follows that
\begin{equation}
\frac{\partial H(X)}{\partial X} = -\frac{3C}{2a^3X^4}.
\end{equation}
Using (\ref{eqtaylor}) and the fact that $L = a(N - 1)$, we can
expand $H(|N + 1 - m - n|)$ as a Taylor series to first order in
$\chi = m + n - 2$. Then
\begin{equation}
H(|L - \chi|) = H(L) - \chi \left.\frac{\partial
H}{\partial X} \right|_{X = L} = \frac{C}{2L^3} + \frac{3Ca (m + n
- 2)}{2L^4}.
\end{equation}
Therefore
\begin{eqnarray}
\langle B|\mathcal{H}|E \rangle &=& \frac{C}{2}\left[ \frac{1}{L^3}\sum_{n,m
= 1}^i a_n^*a_m + \frac{a}{L^4}\sum_{n,m = 1}^i 3a_n^*a_m(m + n -
2)\right]\\ &=& \frac{C}{2} \left[\frac{Q}{L^3} +
\frac{aR}{L^4}\right],
\end{eqnarray}
with
\begin{eqnarray}
Q&=&\sum_{n,m = 1}^i a_n^*a_m\label{eqnq},\\
R&=&\sum_{n,m = 1}^i 3a_n^*a_m (m + n - 2).
\end{eqnarray}
\begin{figure}[H]
\renewcommand{\captionfont}{\footnotesize}
\renewcommand{\captionlabelfont}{}
\begin{center}
 \includegraphics[width=7cm]{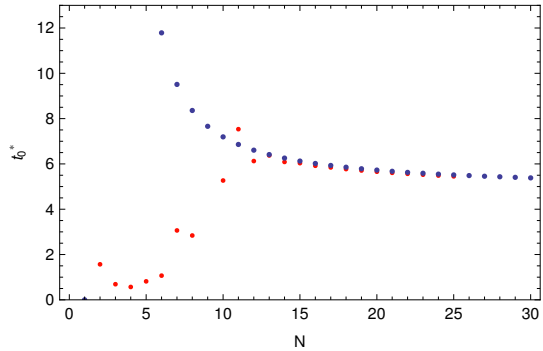}
 \end{center}
 \caption{Comparison between the predictions of a 14-spin model (blue points) and the
 data calculated from the treatment of the system in its entirety (red points).
 Note the model is only accurate at $N>13$.
 } \label{totfit}
\end{figure}

It emerges that the asymptotic behaviour of the
system can be modeled very accurately if the coefficients $a_n$ are taken to be the
amplitudes of the ground state eigenvector of the $i \times i$
sub-matrix of the original dipolar Hamiltonian. The values of
$a_n$ used in the fit of figure \ref{totfit} are thus obtained from the dipolar Hamiltonian of a 14-spin chain, with $i=4$. The corresponding values of $Q$ and $R$ are $Q \approx 0.325$ and $R \approx -0.957$. These
quantities show only a very weak dependence on $N$, so they
have been treated as constants. The fact that $Q < 1$ indicates that, at large $N$,
the transfer rate is always less
than that attained between two completely isolated spins; equation
(\ref{eqnq}) indicates this is a result of interference between
positive and negative components in the localized states $|B
\rangle$ and $|E \rangle$.

At small $N$, however, this is clearly not the case, as shown by the minimum in fig.~\ref{normtimech} at $N = 4$. The concavity of the curve between $N=2$ and $N=7$ indicates that only chains with less than 7 spins can improve on the
performance of a simple dipole pair; specifically, in a \emph{uniform}
chain of a given length, the best compromise between the quality and the speed of the
communication is obtained with 4 spins. This occurs because for small $N$ the bound states at the ends of the chain have a large overlap,
i.e. there exist terms in eqn. (\ref{eqtaylor}) which simultaneously
have significant positive values of $a_n^*a_m$ and small values of
$|N + 1 - m - n|$.

The uniform 4-spin chain can be optimized still further by symmetrically moving the inner spins slightly closer to the ends, so that $r_{1,2} = r_{3,4} \approx 0.312L$ and $r_{2,3}
\approx 0.375L$. The resulting transfer rate is then $t_0^{*\prime} \approx 0.512$, which is a 10\% improvement on the value one finds for a uniform system, $t_0^* \approx 0.568$. Considering the extra difficulty involved in such high precision placement of the spins, it is probably not useful to adopt this optimization technique as a rule. However, viewed from another angle, our result also indicates that if a symmetric error were made in the placement of the central spins, the protocol would not necessarily be compromised.

To explore in more detail the robustness of the protocol against structural imperfections in the chain, we can study the effect of introducing a random error on the placement of each spin. To evaluate whether there is a significant drop in performance, I have counted over 1000 iterations the number of times the fidelity of a 4-spin chain of arbitrary length, measured at the optimal time $t_0=Lt_0^*$, falls below $2/3$. It has been assumed the magnitude of the error lies in the interval $\lbrack-\frac{pL}{3}\leq \Delta \leq \frac{pL}{3}\rbrack $, with $p \in \lbrack0,1\rbrack$ so that the maximum allowed deviation of a spin is the distance to the next (or previous) site. As one might expect, the performance of the system is found to depend on the magnitude of $\Delta$. If $\Delta \leq 0.01L$, the fidelity at time $t_0$ rarely drops below the classical threshold. However, as $\Delta$ increases the number of instances of poor performance grow, and by the time $\Delta \approx 0.03L$ the system is failing more than 50\% of the time. Let us now be more demanding and raise the lower limit of the fidelity to 0.8; in this case, an error of $\Delta \leq 0.01L$ yields a 25\% failure rate, and at $\Delta \approx 0.07L$ the system can essentially be written off. These statistics may seem discouraging, but we must remember that as $L$ becomes large the absolute value of $\Delta$ also increases. Hence, in terms of precision requirements, long chains are more versatile than short ones because they allow more leeway for experimental error in the placement of the spins.

\section{Extensions and Experimental Implementations}\label{experiments2}
In section \ref{experiments1}, several methods for implementing quantum state transfer with short range interactions were reviewed. Among them, endohedral fullerenes are the leading candidates for the protocol discussed here, though natural magnetic dipolar systems, such as $\mathrm{LiHoF_4}$, should also be taken into account
\cite{bitko96}. Trapped ions are currently emerging as ideal subjects for one- and two-dimensional quantum manipulations. Indeed, as widely demonstrated by a variety of impressive experiments \cite{haffner05,barrett04,schaetz04,brickman05,chiaverini05}, ions can be efficiently cooled, initialized, entangled, read out and, most importantly in the present context, stored at fixed positions. Ion microtraps allow the highest degree of control over the qubits' placement, as each ion is individually confined. Linear traps can be used to create linear arrays, whereas confinement in two dimensions is achieved with Penning traps. In these systems, however, the single ionic positions cannot be fixed to the same degree of precision. In all three cases, the trapping potential is provided by a series of orthogonal electric and magnetic fields, whereas the inter-qubit interactions are mediated by common vibrational modes, usually tuned with directed laser pulses. It has recently been shown that by adjusting the trapping potential and the laser frequencies, it is possible to simulate a dipolar Hamiltonian similar to equation (\ref{h dip gen}). This involves accessing the so-called `stiff' vibrational modes, in which the Coulomb repulsion of the trapped qubits can be considered a perturbation on the trapping potential.

\begin{figure}[H]
\renewcommand{\captionfont}{\footnotesize}
\renewcommand{\captionlabelfont}{}
\begin{center}
\subfigure{\label{Fidsch4a}\includegraphics[width=6.5cm]{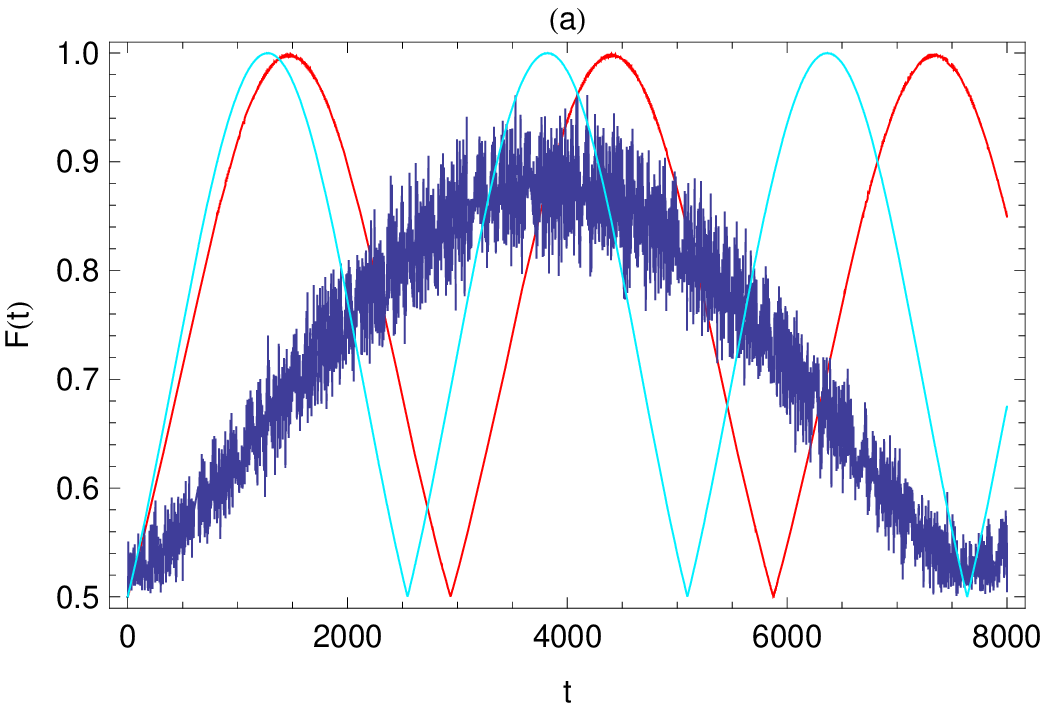}}
\hspace{0.3cm}
\subfigure{\label{Fidsch4b}\includegraphics[width=6.5cm]{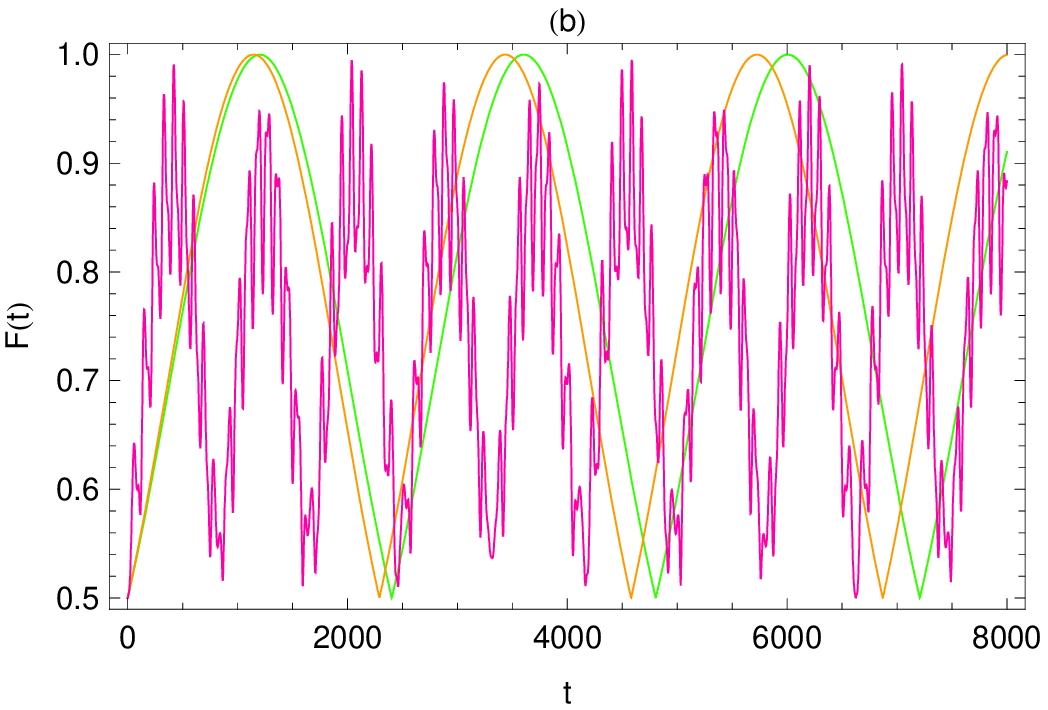}}

\end{center}
\caption{\label{th chain} The evolution of the fidelity of state transfer over a distance $L=9$ in (a) a complete, double-hole and four-hole chain (blue, red and cyan curves respectively), and (b) a six-hole chain, a uniform four-spin system, and two single spins (green, pink and orange curves, respectively). We note that a uniform chain of length $L=9$ is composed of ten spins, therefore a six-hole chain corresponds to a four-spin configuration, in which the two central spins have unit spacing and are separated from the end spins by a distance $L=4$. It is evident from figure (b) that the minimum transfer time is achieved in a uniform four-spin chain.}
\end{figure}

Motivated by these findings, new effort is being dedicated to optimizing quantum state transmission through dipolar chains \cite{gualdi08}. An interesting proposal, advanced earlier this year, involves a system known as the `double hole (DH) chain'. This is a linear chain of uniformly spaced, magnetic-dipole coupled spins, from which the particles at sites 2 and $N-1$ have been removed, isolating the end spins from the middle of the chain. As a result, the transfer process is faster than in a `complete' chain, and unit fidelities are achieved even in chains of over 80 spins. One can show that, by continuing to symmetrically remove spins, the performance of the DH chain can be further improved; however, no matter how many times the process is repeated, for a given chain length the lower bound for the transfer time is still that provided by a uniform four-spin system (figure \ref{th chain}). Nevertheless, a `many hole' chain may have other advantages; indeed, it is argued in \cite{gualdi08} that by splitting off the sending and receiving spins from the mid-section of the chain, perturbations in the channel are less likely to impair the protocol, making the DH chain more robust than its complete counterpart against possible decoherence.

A dipolar Hamiltonian can also be simulated in a solid-state spin system, using nuclear magnetic resonance \cite{cappellaro07,furman08}. Most recent work in this area relates to the appearance of multiple correlations between different parts of the spin chain, which can generate entanglement between distant spins. Interestingly, it is observed that long-distance entanglement is more likely to persist if the spins are coupled by nearest-neighbour interactions only.

\section{Conclusions}
Unmodulated spin chains with long-range dipolar interactions allow high quality transfer of quantum information. Periodic systems do not perform as well as finite chains, particularly if the component number of spins is even. Nevertheless, for systems of up to 30 spins, a dipolar ring often outdoes its nearest-neighbour coupled counterpart. In general, the fidelity of the channel is a decreasing function of chain length. If the chain is periodic, the fidelity falls quite rapidly with $N$; however, in a finite chain the optimal fidelity remains well above the classical threshold for chains of up to 100 spins \cite{gualdi08}. Furthermore, in a linear system the fidelity peaks at predictable times, and its dynamics are such to ensure the receiver has a reasonably ample window of opportunity in which to retrieve their spin. Both these factors are owed to the fact that the transfer process is dominated by the two end spins. The presence of the mid-section of the chain registers merely as noise; indeed, the more one separates the end spins from the mid-section, the better the fidelity. For a uniform chain of given length, the number of component spins is relevant to the overall performance of the system, which peaks when $N=4$.

The main drawback of the linear chain geometry is the time required for the state to be transferred, which grows as the cube of the chain length. Recent work has managed to reduce this figure by a factor of three, however in absolute terms this is not a sizeable advantage, because the transfer time in chains longer than $L=20$ still remains of micro-second order. Therefore, if all factors are taken into account, a dipolar chain is a \emph{less} efficient communication channel than a nearest-neighbour coupled chain. However, evaluating the effect of dipolar couplings on quantum state transfer is a valuable exercise, because any practical implementation of spin-chain communication would most likely be carried out in systems exhibiting both types of couplings, though perhaps not in equal measure. One could therefore imagine engineering the relative strengths of the nearest-neighbour and dipolar interactions so as to generate constructive interference, thus yielding a fidelity higher than that achievable by either effect taken separately.

%% file: conclusions.tex
\chapter{Summary and Future Work}\label{conclusions}
The past 50 years have witnessed the evolution of Quantum Information Technology from a theoretical construct to a tangible reality. Most recently, the field has been developing at a truly impressive rate, and seems on the verge of a leap of quality despite several outstanding issues, mainly related to preserving quantum properties for long enough to perform useful tasks. This thesis has explored means to achieve the extremely desirable goals of quantum communication over short distances in solid-state devices, and experimental entanglement generation. Novel techniques to attain these goals have been suggested, and related practical difficulties have been discussed.

\textbf{Chapter \ref{ent review}} was dedicated to an overview of quantum entanglement and its basic properties. Methods to quantify and measure entanglement were also analyzed.

\textbf{Chapter \ref{neutron proposal}} put forward a proposal for a scattering experiment designed to create entanglement between distinct neutrons. The protocol was illustrated in detail, together with the approximations underlying the mathematical model of the experiment. Given the possibility of describing the system in terms of pure states, it was decided that the scattering event should be treated using a time-dependent quantum mechanical analysis. In \textbf{Chapter \ref{s matrix}} it was verified that if no momentum is transferred between the incident neutron and the sample, the neutron interaction time can be identified with its free-flight time through the sample.

In \textbf{Chapter \ref{two ns}}, the performance of the scattering protocol at zero momentum transfer was assessed. It was found that the first pair of neutrons to scatter from a sample having the properties of a ferromagnetic insulator could come to share a substantial degree of \emph{measurable} entanglement, subject to the possibility of applying a static external magnetic field of tuneable strength. The experimental feasibility of the protocol was discussed, and it was concluded that the only true obstacle to a practical realization stemmed from the coherence properties of the neutron beam. These properties depend only on beam preparation apparatus and technique. Therefore, the protocol may indeed become feasible as neutron facilities become better equipped and more refined.

\textbf{Chapter \ref{many ns}} extended the scope of the entangling proposal to account for the possibility of many neutrons scattering from the sample. The aim of the chapter was to establish whether a measurement on any pair of neutrons detected within a certain time of each other would yield an entangled state. It was shown that the average entanglement of any pair is at best of order 0.1 ebits. This falls with the size of the sampling ensemble of outgoing neutrons. Consequently, to produce highly entangled states it is necessary to detect one of the very first scattered pairs. The emerging trends are interesting from a more fundamental viewpoint also, as they provide a practical means to establish whether or not two distinct neutrons can in fact be entangled.

In \textbf{Chapter \ref{spin chain review}}, the focus shifted from entanglement generation to quantum information transfer in the solid state. The chapter was dedicated to a review of quantum communication through spin chains, and a detailed analysis of the pioneering proposal of Bose \cite{bose03}. Recent experimental implementations of this proposal were also discussed.

Finally, \textbf{Chapter \ref{dipolar chains}} analyzed the performance of dipolar chains as conduits of quantum information. It was found that near-perfect transfer can be achieved without any special engineering of the system, but at the cost of long transfer times. Possible optimization methods were explored, and it was concluded that for a given distance the optimal balance between quality and speed was obtained with a uniform chain of four spins. Recent theoretical and experimental work on dipolar systems was also reported.

Much of this work has been concerned with simulating possible experimental conditions. As with many theoretical models, this leaves a great deal of scope for refining underlying approximations to attain a more faithful description of reality. Consequently, there are several directions in which future work might proceed.

In the context of neutron entanglement, a first question to ask might be how the protocol is affected when the temperature is raised above zero. Simple preliminary calculations suggest that if scattering takes place from a thermal state of the sample the outcome of the protocol is determined by the relative magnitude of the thermal energy to all other energy scales in the problem. Due to heavy computational demands, only small systems have been considered so far. It would be interesting to look at larger systems, and in greater detail. One could also examine the effect of changing the properties of the sample, and in this respect the possibilities are clearly boundless. Samples with different electrical properties (metals, semi-conductors, superconductors...) or more realistic anisotropies could be considered.

As regards quantum state transfer through dipolar spin chains, it may be interesting to re-asses the performance of the protocol in the presence of mixed short- and long-range interactions. A slightly more ambitious goal might be to scale up from a single spin chain, and study the flow of information around some form of `dipolar quantum network'.

The problem of time in scattering theory also offers huge scope for further investigation. From a very brief survey of the field, it would seem the matter has been `on the table' for the past fifty years at least, and still no unique resolution exists. As commented in chapter \ref{s matrix}, this may be due to a dependence on the context. Indeed, much of the literature tends to explore very specific scenarios. One could therefore begin with an extensive survey of work done to date, to identify results of general validity and gain insight on the most efficient way to tackle the problem. Some research into the dynamical theory of neutron scattering may also prove enlightening.

\textbf{As a final word}, I would like to say that the past three years have been an incredible experience, which has taught me a great many things about myself in both a professional and a personal capacity. I have often felt the only thing that could have prepared me to do a PhD would have been... to do a PhD! Paraphrasing slightly the words of Cervantes' legendary \textit{Don Quixote}: I have struggled, I have made mistakes, but I have done this work as best I can according to the world as I see it.

One of the most precious things I have learnt is what it takes for me to truly understand a concept and make it mine. I discovered this rather late in my PhD - perhaps I was not in a position to see it sooner, but that is another story. What matters is that I have rediscovered the joy of questioning, and the freedom to question what I wish. I hope to carry that feeling of wonder and adventure into any work I may do in the future, conveying it to those around me always. Thank you for listening.

%\end{document} 